\documentclass{article}%
\usepackage{amsmath}
\usepackage{amsfonts}
\usepackage{amssymb}
\usepackage{graphicx}%
\providecommand{\U}[1]{\protect\rule{.1in}{.1in}}
\newtheorem{theorem}{Theorem}
\newtheorem{acknowledgement}[theorem]{Acknowledgement}

\newenvironment{proof}[1][Proof]{\noindent\textbf{#1.} }{\ \rule{0.5em}{0.5em}}
\newcommand{\bpartial}{\mathop{\partial\kern -4pt\raisebox{.8pt}{$|$}}}
\newcommand{\bra}{\mathopen{[\kern-1.6pt[}}
\newcommand{\ket}{\mathclose{]\kern-1.5pt]}}
\newcommand{\bbra}{\mathopen{[\kern-2.2pt[\kern-2.3pt[}}
\newcommand{\bket}{\mathclose{]\kern-2.1pt]\kern-2.3pt]}}
\newcommand{\sle}{\mbox{\bfseries\slshape e}}
\newcommand{\slg}{\mbox{\bfseries\slshape g}}
\newcommand{\sslg}{\mbox{\tiny \bfseries\slshape g}}
\newcommand{\slh}{\mbox{\bfseries\slshape h}}
\newcommand{\sslh}{\mbox{\tiny \bfseries\slshape h}}
\newcommand{\sslm}{\mbox{\tiny \bfseries\slshape m}}
\newcommand{\itg}{\mbox{\bfseries\itshape g}}
\newcommand{\sitg}{\mbox{\tiny\bfseries\itshape g}}
\newcommand{\sll}{\mbox{\bfseries\slshape l}}
\newcommand{\slx}{\mbox{\bfseries\slshape x}}

\newcommand{\slD}{\mbox{\bfseries\slshape D}}
\newcommand{\slG}{\mbox{\bfseries\slshape G}}

\newcommand{\slR}{\mbox{\bfseries\slshape R}}

\makeindex
\begin{document}

\title{Gravitation as a Plastic Distortion of the Lorentz Vacuum\thanks{Some (odd)
misprints and typos have been corrected, some sentences have been changed for
better intelligibility and Appendix F has new important remarks which result
from discussions that W.A .R. had with A. Lasenby at ICCA10 (Tartu) in August
2014.}}
\author{Virginia V. Fern\'{a}ndez$^{(1)}$ and Waldyr A. Rodrigues Jr.$^{(2)}$
\and $^{(1)}${\footnotesize Instituto de Ingenier\'{\i}a Biom\'{e}dica, Facultad de
Ingenier\'{\i}a}
\and {\footnotesize Universidad de Buenos Aires, Av. Paseo Col\'{o}n 850,}
\and {\footnotesize C1063ACV, Buenos Aires -- Argentina}
\and $^{(2)}${\footnotesize Institute of Mathematics, Statistics and Scientific
Computation}
\and {\footnotesize IMECC-UNICAMP CP 6065 \ \ \ \ 13083-859 Campinas, SP, Brazil }\\{\footnotesize walrod@ime.unicamp.br \ \ \ virginvel@gmail.com}}
\maketitle

\begin{abstract}
In this paper we present a theory of the gravitational field where this field,
represented by a $(1,1)$-extensor field $%
%TCIMACRO{\TeXButton{h}{\slh}}%
%BeginExpansion
\slh
%EndExpansion
$ describing a plastic distortion of the Lorentz vacuum (a real substance that
lives in a Minkowski spacetime) due to the presence of matter. The field $%
%TCIMACRO{\TeXButton{h}{\slh}}%
%BeginExpansion
\slh
%EndExpansion
$ distorts the Minkowski metric extensor $\eta$ generating what may be
interpreted as an effective Lorentzian metric\ extensor $%
%TCIMACRO{\TeXButton{itg}{\itg}}%
%BeginExpansion
\itg
%EndExpansion
=%
%TCIMACRO{\TeXButton{h}{\slh}}%
%BeginExpansion
\slh
%EndExpansion
^{\dagger}\eta%
%TCIMACRO{\TeXButton{h}{\slh}}%
%BeginExpansion
\slh
%EndExpansion
$ and also it permits the introduction of different kinds of parallelism rules
on the world manifold, which may be interpreted as distortions of the
parallelism structure of Minkowski spacetime and which may have non null
curvature and/or torsion and/or non metricity tensors. We thus have different
possible effective geometries\ which may be associated to the gravitational
field and thus its description by a Lorentzian geometry is only a possibility,
not an imposition from Nature. Moreover, we developed with enough details the
theory of multiform functions and multiform functionals that permitted us to
successfully write a Lagrangian for $%
%TCIMACRO{\TeXButton{h}{\slh}}%
%BeginExpansion
\slh
%EndExpansion
$ and to obtain its equations of motion, that results equivalent to Einstein
field equations of General Relativity (for all those solutions where the
manifold $M$ is diffeomorphic to $\mathbb{R}^{4}$). \ However, in our theory,
differently from \ the case of General Relativity, trustful energy-momentum
and angular momentum conservation laws exist. We express also the results of
our theory in terms of the gravitational potentials $\mathfrak{g}^{\mu}=%
%TCIMACRO{\TeXButton{h}{\slh}}%
%BeginExpansion
\slh
%EndExpansion
^{\dagger}(\vartheta^{\mu})$ where $\{\vartheta^{\mu}\}$ is an orthonormal
basis of Minkowski spacetime in order to have results which may be easily
expressed with the theory of differential forms. The Hamiltonian formalism for
our theory (formulated in terms of the potentials $\mathfrak{g}^{\mu}$) is
also discussed. The paper contains also several important Appendices that
complete the material in the main text.

\end{abstract}
\tableofcontents

\section{Introduction}

In this article we present a theory of the gravitational field where this
field is described by a distortion\ field $%
%TCIMACRO{\TeXButton{h}{\slh}}%
%BeginExpansion
\slh
%EndExpansion
$ which lives and interacts with the matter fields in Minkowski spacetime. It
describe a plastic deformation of the Lorentz vacuum understood as a real
physical medium\footnote{We recall that deformations of a medium may be of the
elastic or plastic type \cite{Zorawski}. Later we explain the difference
between those two types of deformation.}. The distortion field is represented
in the mathematical formalism of this paper---called the \emph{multiform and
extensor calculus on manifolds }(\textit{MECM})--- by a $(1,1)$-extensor field
and as we are going to see $%
%TCIMACRO{\TeXButton{h}{\slh}}%
%BeginExpansion
\slh
%EndExpansion
$ it is a kind of square root of a $(1,1)$-extensor field $%
%TCIMACRO{\TeXButton{itg}{\itg}}%
%BeginExpansion
\itg
%EndExpansion
$ which is directly associated to the metric tensor $%
%TCIMACRO{\TeXButton{g}{\slg}}%
%BeginExpansion
\slg
%EndExpansion
$ which as well known represents important aspects of the gravitational field
in Einstein's General Relativity Theory (\textit{GRT}). Before going into the
details we make a digression in order to present the main motivations for our enterprise.

We start by recalling that in \textit{GRT} the gravitational field has a
status which is completely different from the other physical fields. Indeed in
\textit{GRT} the gravitational field is interpreted as aspects of a
\textit{geometrical} structure of the spacetime manifold. More precisely we
have that each gravitational field generated by a given matter distribution
(represented by a given energy-momentum tensor) is represented by a pentuple
$(M,%
%TCIMACRO{\TeXButton{g}{\slg}}%
%BeginExpansion
\slg
%EndExpansion
,D,\tau_{%
%TCIMACRO{\TeXButton{g}{\sslg}}%
%BeginExpansion
\sslg
%EndExpansion
},\uparrow)$, where\footnote{In fact a gravitational field is defined by an
equivalence class of pentuples, where $(M,%
%TCIMACRO{\TeXButton{g}{\slg}}%
%BeginExpansion
\slg
%EndExpansion
,D,\tau_{%
%TCIMACRO{\TeXButton{g}{\sslg}}%
%BeginExpansion
\sslg
%EndExpansion
},\uparrow)$ \ and $(M^{\prime},%
%TCIMACRO{\TeXButton{g}{\slg}}%
%BeginExpansion
\slg
%EndExpansion
^{\prime},D^{\prime},\tau_{%
%TCIMACRO{\TeXButton{g}{\sslg}}%
%BeginExpansion
\sslg
%EndExpansion
}^{\prime},\uparrow^{\prime})$ are said equivalent if there is a
diffeomorphism $\mathtt{h}:M\rightarrow M^{\prime}$, such that $%
%TCIMACRO{\TeXButton{g}{\slg}}%
%BeginExpansion
\slg
%EndExpansion
^{\prime}=\mathtt{h}^{\ast}%
%TCIMACRO{\TeXButton{g}{\slg}}%
%BeginExpansion
\slg
%EndExpansion
$, $D^{\prime}=\mathtt{h}^{\ast}D$, $\tau_{%
%TCIMACRO{\TeXButton{g}{\sslg}}%
%BeginExpansion
\sslg
%EndExpansion
}^{\prime}=\mathtt{h}^{\ast}\tau_{%
%TCIMACRO{\TeXButton{g}{\sslg}}%
%BeginExpansion
\sslg
%EndExpansion
},\uparrow^{\prime}=\mathtt{h}^{\ast}\uparrow,$ (where $\mathtt{h}^{\ast}$
here denotes the pullback mapping). For more details, see, e.g., \cite{sawu,
rodrosa,rodcap2007}.}:

\textbf{(mi)} $M$ is a $4$-dimensional Hausdorff manifold which paracompact,
connected and noncompact,

\textbf{(mii)} $%
%TCIMACRO{\TeXButton{g}{\slg}}%
%BeginExpansion
\slg
%EndExpansion
\in\sec T_{2}^{0}M$ \ is a tensor of signature $-2$, called a
\textit{Lorentzian metric}\footnote{The field $%
%TCIMACRO{\TeXButton{g}{\slg}}%
%BeginExpansion
\slg
%EndExpansion
$ \ in \textit{GRT} obeys Einstein's field equation, to be recalled below.} on
$M$,

\textbf{(miii)} $D$ is the Levi-Civita connection of $%
%TCIMACRO{\TeXButton{g}{\slg}}%
%BeginExpansion
\slg
%EndExpansion
,$

\textbf{(miv)} $M$ is an oriented by the volume element $\tau_{%
%TCIMACRO{\TeXButton{g}{\sslg}}%
%BeginExpansion
\sslg
%EndExpansion
}\in\sec%
%TCIMACRO{\dbigwedge \nolimits^{4}}%
%BeginExpansion
{\displaystyle\bigwedge\nolimits^{4}}
%EndExpansion
T^{\ast}M$ and the symbol $\uparrow$ means that $M$ is also time oriented.

\textit{GRT} supposes that particles and fields are some special
configurations on the manifold $M$. Particles are described by \ triples
$\langle(m,q),S,\sigma\rangle$ where $m$ ($0\leq m<\infty$) is the particle
mass, $q$ ($-\infty<q<\infty$) is a parameter called the electric charge, $S$
is the particle's spin\footnote{For details of how to characterize the
particle's spin in the Clifford bundle formalism see e.g., \cite{rvqp2003}.},
and $\sigma$ is a regular curve called the world line of the particle.

If $m>0$ \ the particle is called a \textit{bradyon} and in this case $\sigma$
is a \textit{timelike} curve pointing into the future.

If $m=0$ the particle is said to be a luxon and in this case $\sigma$ is a
\textit{lightlike} curve.

The different physical fields are modelled by special sections of the tensor
and spinor bundles over the basic manifold $M$.

The world lines that represent the history of particles as well as the tensor
and spinor fields which describe the physical fields are mathematically
described by some system of differential equations (in general called the
\textit{equations of motion}) in which the metric tensor $%
%TCIMACRO{\TeXButton{g}{\slg}}%
%BeginExpansion
\slg
%EndExpansion
$ appears.

The energy-momentum content of a system of particles and fields is described
by their respective \emph{energy-momentum} $\mathbf{T}_{p}\in\sec T_{0}^{2}M$
and $\mathbf{T}_{f}\in\sec T_{0}^{2}M$. As is well known \cite{mtw,sawu} those
objects appear in the second member of Einstein's field equation, where

as in the first member of that equation appears a tensor called the
\textit{Einstein tensor}.

Einstein's equation may be written (after a convenient choice of a local chart
on $U\subset M$) as a system of ten non linear partial differential equations
which contains a certain combination of the first and second order partial
derivatives of the components $g_{\mu\nu}$ of the metric tensor.

Einstein's equation is sometimes described in a pictorially way \cite{mtw} by
saying that \textquotedblleft the geometry says how the matter must move and
by its turn matter says how the geometry must curve\textquotedblright. This
pictorial description induces pedestrians on the subject to imagine that the
\textit{manifold} $M$ which models spacetime is some kind of \textit{curved
}hypersurface (a $4$-brane living in a $(4+p)$-dimensional pseudo-euclidean
space). It is then necessary in order to appreciate the developments of this
paper to deconstruct such an idea by explaining in a rigorous way first what
the concept of curvature used in \textit{GRT} really means and second to
realize that the use of this concept in certain gravitational theories is no
more than a coincidence.

We recall that from a mathematical point of view a general differential
manifold $M$ by itself possess only a topological structure and a differential
structure of the charts of its maximal atlas. It must be clear to start that
the topological structure of $M$ is not fixed in \textit{GRT} and indeed the
topology of $M$ is introduced\ `by hand' in specific problems and situations
\cite{eddington,sawu}. On the other hand any given $M$ may support in general
many distinct \textit{geometrical structures}.

In what follows we say that a\emph{ pair }$(M,\nabla)$ where $M$ is a general
differential manifold and $\mathbf{\nabla}$ is an arbitrary connection on $M$
\emph{is a space endowed with a parallelism rule}. We say that the
\emph{triple }$(M,%
%TCIMACRO{\TeXButton{g}{\slg}}%
%BeginExpansion
\slg
%EndExpansion
,\nabla)$\emph{ is a geometric space structure (GSS)} and that $%
%TCIMACRO{\TeXButton{g}{\slg}}%
%BeginExpansion
\slg
%EndExpansion
$ is a general metric tensor on $M$. The signature of a $%
%TCIMACRO{\TeXButton{g}{\slg}}%
%BeginExpansion
\slg
%EndExpansion
$ is arbitrary. So, even in the case\ where $(M,%
%TCIMACRO{\TeXButton{g}{\slg}}%
%BeginExpansion
\slg
%EndExpansion
,\mathbf{\nabla})$ is part of a Lorentzian spacetime structure and $%
%TCIMACRO{\TeXButton{g}{\slg}}%
%BeginExpansion
\slg
%EndExpansion
$ has signature $-2$ we can have another \textit{GSS }$(M,%
%TCIMACRO{\TeXButton{g}{\slg}}%
%BeginExpansion
\slg
%EndExpansion
^{\prime},\mathbf{\nabla})$ where $%
%TCIMACRO{\TeXButton{g}{\slg}}%
%BeginExpansion
\slg
%EndExpansion
^{\prime}$ has an another signature\footnote{In particular in our theory we
will see that we shall need to introduce a $%
%TCIMACRO{\TeXButton{g}{\slg}}%
%BeginExpansion
\slg
%EndExpansion
^{\prime}$ with signature $+4$.}.

Now the objects that characterize a \textit{GSS} are the following tensor
fields:\footnote{The definitions of those objects will be recalled in Chapter
3.}:

\textbf{(egi)} \emph{nonmetricity of} $\mathbf{\nabla},$\emph{\ }%
$\mathfrak{A}\in T_{3}^{0}M,$
\begin{equation}
\mathfrak{A}=\mathbf{\nabla}%
%TCIMACRO{\TeXButton{g}{\slg}}%
%BeginExpansion
\slg
%EndExpansion
. \label{0.1}%
\end{equation}

\textbf{(egii)} \emph{torsion of }$\mathbf{\nabla}$, represented by a tensor
field $\Theta\in T_{2}^{1}M$,

\textbf{(egiii)} the \textit{R}\emph{iemann curvature of} $\mathbf{\nabla}$,
represented by a tensor field $\mathfrak{R}\in T_{3}^{1}M\mathfrak{.}$

A geometric space structures is said to be a:

\textbf{(a)} Riemann-Cartan- Weyl \textit{GSS} iff $\mathfrak{A}\neq
0,\Theta\neq0,\mathfrak{R}\neq0$,

\textbf{(b)} Riemann-Cartan \textit{GSS} iff $\mathfrak{A}=0,\Theta
\neq0,\mathfrak{R}\neq0$,

\textbf{(c)} Riemann \textit{GSS} iff $\mathfrak{A}=\Theta=0,\mathfrak{R}%
\neq0$.\medskip

A general \textit{GSS }$(M,%
%TCIMACRO{\TeXButton{g}{\slg}}%
%BeginExpansion
\slg
%EndExpansion
,\mathbf{\nabla})$ such that $\mathbf{\nabla}%
%TCIMACRO{\TeXButton{g}{\slg}}%
%BeginExpansion
\slg
%EndExpansion
=0$ will be called a\emph{ metric compatible geometric space structure
(MCGSS)}

\subsection{Flat Spaces and Affine Spaces}

A \textit{MCGSS} $(M,%
%TCIMACRO{\TeXButton{g}{\slg}}%
%BeginExpansion
\slg
%EndExpansion
,\mathbf{\nabla})$ such that $\mathfrak{A}=\Theta=\mathfrak{R}=0$ is said to
be a \textit{globally flat} Riemann \textit{MCGSS.}

A \textit{MCGSS} where the parallelism rule is such that\ $\Theta
=\mathfrak{R}=0$ and where $M\simeq\mathbb{R}^{n}$ is isomorphic to a (real)
affine space of the same dimension, denoted by $\mathcal{A}^{n}$.

As it is well known \cite{trautman1} a (real) affine space has as a
fundamental property an operation called the difference between two given
points, such that if $p,$ $q\in\mathcal{A}^{n}$ (or $p,q\in\mathbb{E}^{n}$)
then $(p-q)\in\mathbf{V}$, where $\mathbf{V}$ is a real $n$-dimensional vector
space. So, we may say in a pictorially way that an affine space is a vector
space from where the origin has been stolen. Given an arbitrary point
$o\in\mathcal{A}^{n}$ (said to be the \textit{origin}) the object
$(p-o)\in\mathbf{V}$ is appropriately called the position vector of $p$
relative to $o$. This operation permit us to define an \textit{absolute
parallelism} rule (denoted $\upuparrows)$ in an affine space. Given two
arbitrary points $o$,$o^{\prime}\in\mathcal{A}^{n}$ consider the set of all
tangent vectors at those two points, i.e., $\{(p-o),$ $p\in\mathcal{A}^{n}\}$
and $\{(p^{\prime}-o^{\prime})$, $p^{\prime}\in\mathcal{A}^{n}\}$. Then we say
that a tangent vector at $o$, say $(p-o)$ is parallel to a tangent vector at
$o^{\prime}$, say $(p^{\prime}-o^{\prime})$ if being $(p-o)$ $=\mathbf{v\in
V}$ and $(p^{\prime}-o^{\prime})=\mathbf{v}^{\prime}\mathbf{\in V}$ we have
$\mathbf{v=v}^{\prime}$. This parallelism rule permit us to identify the pair
$(\mathcal{A}^{n},\upuparrows)$ with a parallelism structure ($M\simeq
\mathbb{R}^{n},\nabla^{\upuparrows}$) where $\nabla^{\upuparrows}$ is a
connection, called the \emph{Euclidean connection} and defined as follows. Let
$\{\mathtt{x}^{i}\}$ $i=1,2,...,n$ \ be standard Cartesian coordinates for
$M\simeq\mathbb{R}^{n}$. Consider the global coordinate basis $\{e_{i}\}$ for
$TM$ where $e_{i}=\partial/\partial\mathtt{x}^{i}$. Then, put
\begin{equation}
\nabla_{e_{j}}^{\upuparrows}e_{i}=0.
\end{equation}
We define an \emph{Euclidean MCGSS} as a triple $\mathbb{E}^{n}=(\mathbb{R}%
^{n},%
%TCIMACRO{\TeXButton{g}{\slg}}%
%BeginExpansion
\slg
%EndExpansion
,\nabla^{\upuparrows})$ where $%
%TCIMACRO{\TeXButton{g}{\slg}}%
%BeginExpansion
\slg
%EndExpansion
$ defines the standard Euclidean scalar product in each tangent space $T_{p}M$
by $\left.
%TCIMACRO{\TeXButton{g}{\slg}}%
%BeginExpansion
\slg
%EndExpansion
\right\vert _{p}(\left.  e_{i}\right\vert _{p},\left.  e_{j}\right\vert
_{p})=\delta_{ij}$. For future reference we say that if the
signature\footnote{A metric with signature $(n-1,1)$ is also called
pseudo-euclidean. Note also that mathematical textbooks define the signature
of a metric $(p,q)$ as the number $s=p-q$.} of $%
%TCIMACRO{\TeXButton{g}{\slg}}%
%BeginExpansion
\slg
%EndExpansion
$ is $(1,n-1)$, i.e. $\left.
%TCIMACRO{\TeXButton{g}{\slg}}%
%BeginExpansion
\slg
%EndExpansion
\right\vert _{p}(\left.  e_{i}\right\vert _{p},\left.  e_{j}\right\vert
_{p})=\eta_{ij}$ where the matrix $(\eta_{ij})$ with entries $\eta_{ij}$ is
the diagonal matrix $\mathrm{diag}(1,-1,...,-1)$.

Now, from the classification of the \textit{GSS} structures given above it is
clear that the Riemannian curvature does not refer to any \textit{intrinsic}
property of $M$, but it refers to a \textit{particular} parallelism structure
(i.e., a property of a particular connection $\mathbf{\nabla}$) that has been
defined on $M$. To keep this in mind is crucial, because very often we observe
notable confusions between the concept of \emph{Riemannian curvature} and the
concept of \emph{bending}\footnote{We recall that\textit{ }the bending of a
$r$-dimensional hypersurface $S$ considered as a subset of points of an
euclidean (or pseudo-eucldean) \textit{GSS} of appropriated dimension
\cite{clarke} $n$ is characterized by the so called \textit{shape tensor.}
\cite{hesob84,oneill}} of hypersurfaces embedded in a higher dimensional
euclidean \textit{MCGSS}, something that may lead to equivocated and even
psychedelic interpretations concerning the physical and mathematical contents
of \textit{GRT. }

So, before proceeding we recall some examples which we hope will clarify the
difference between \textit{bending} and \textit{curvature}.

Consider a $2$-dimensional cylindrical surface $\mathcal{C}$ embedded in
$\mathbb{E}^{3}$. Using\ as metric on $\mathcal{C}$ the pullback of the
Euclidean metric of $\mathbb{E}^{3}$ and as its connection the pullback of the
connection $\nabla^{\upuparrows}$ of $\mathbb{E}^{3}$ (which happens to be the
Levi-Civita connection of the induced metric) it is possible to verify without
difficulties that the Riemannian curvature of $\mathcal{C}$ is \textit{null},
i.e., the cylinder $\mathcal{C}$ is \textit{flat} according to its Riemannian
curvature. However a cylinder $S^{1}\times\mathbb{R}$ living in $\mathbb{E}%
^{3}$ is very different topologically from any plane $\mathbb{R}^{2}$ which
lives in $\mathbb{E}^{3}$ and which \textit{may} have zero curvature tensor
\textit{if} it is part of a \textit{MCGSS \ }(say $^{p}\mathbb{E}^{2}%
$)\textit{ }where the metric on it\ is the pullback of the Euclidean metric of
$\mathbb{E}^{3}$ and its connection is the pullback of the connection
$\nabla^{\upuparrows}$ of $\mathbb{E}^{3}$ (which happens to be the
Levi-Civita connection of the induced metric). As we already said (Footnote
8)\textbf{ }it is the \textit{shape tensor }the mathematical object that helps
to characterize distinct $r$-dimensional differential manifolds when they are
viewed as hypersurfaces embedded in an Euclidean (or pseudo-euclidean) GSS of
appropriate dimension\textit{ }$n>r$, and of course, the shape tensors of
$\mathcal{C}$ and $\mathbb{R}^{2}$ (as part of the \textit{MCGSS}
$^{p}\mathbb{E}^{2}$) are very different \cite{oneill}.

Another very instructive example showing the crucial distinction between
curvature and bending is the following $\ $\cite{dodson}:

\textbf{(A)} Consider the torus $\mathcal{T}_{1}$ as a surface embedded in
$\mathbb{E}^{3}$ and represented, e.g., in the canonical coordinates
of\ $\mathbb{E}^{3}$ by the equation
\begin{equation}
(x^{2}+y^{2}+z^{2}+3)^{2}=16(x^{2}+y^{2}). \label{0.2}%
\end{equation}

Of course, $\mathcal{T}_{1}$ is a $2$-dimensional manifold. If we consider on
$\mathcal{T}_{1}$ a \textit{MCGSS} structure where the metric on
$\mathcal{T}_{1}$ is the pullback of the Euclidean metric of $\mathbb{E}^{3}$
and its connection is the pullback of the connection $\nabla^{\upuparrows}$ of
$\mathbb{E}^{3}$ (which results to be the Levi-Civita connection of the
induced metric) then the Riemann curvature tensor of that pullback connection
on $\mathcal{T}_{1}$ is \textit{non null}, as it is easy to verify.

\textbf{(B)} Now, the subset $\mathcal{T}_{2}$ de $\mathbb{E}^{4}$ given by ,
\begin{equation}
\mathcal{T}_{2}=\{x\in\mathbf{E}^{4}\text{ }/\text{ }(x^{1})^{2}+(x^{2}%
)^{2}=(x^{3})^{2}+(x^{4})^{2}=1\}, \label{0.3}%
\end{equation}
is diffeomorphic to the torus $\mathcal{T}_{1}$ defined in \textbf{(A)}.
Consider the \textit{MCGSS structure defined on\ }$\mathcal{T}_{2}$ where the
metric\ is the pullback of the Euclidean metric of $\mathbb{E}^{4}$ and its
connection is the pullback of the connection $\nabla^{\upuparrows}$ of
$\mathbb{E}^{4}$. It can be easily verified that such \textit{MCGSS} is
globally flat, i.e., $\Theta=$ $\mathfrak{R}=0$.

Those two examples leave clear that we cannot confound curvature with bending.
Indeed, those examples show that a given manifold when interpreted as a
$r$-dimensional hypersurface embedded in an euclidean \textit{MCGSS }of
appropriate dimension $n>r$ may be or may be not curved (even if that
\textit{MCGSS} heritages as metric tensor and connection the pullback of the
metric tensor and the connection of $\mathbb{E}^{n}$). In particular keep in
mind that a $2$-dimensional torus in order to be part of a flat \textit{MCGSS}
(with a Levi-Civita connection of the pullback metric) must be embedded in
a\ euclidean \textit{GSS }with more dimensions than in the case it is part of
a non flat \textit{MCGSS }(with a Levi-Civita connection of the pullback
metric).\footnote{We can also put on a torus living in $\mathbb{E}^{3}$ a non
metric compatible connection with null torsion and curvature tensors or yet a
metric compatible connection with non null torsion tensor and null curvature
tensor. See Appendix A.}

We give yet another example (which Cartan showed to Einstein in 1922 when he
visited Paris \cite{debewer,goenner}) to clarify (we hope) the question of the
curvature of a given \textit{MCGSS}. Consider the punctured sphere
$\mathring{S}^{2}=\{S^{2}-$ north and south poles$\}\subset\mathbb{E}^{3}$.
$\mathring{S}^{2}$ is clearly a bent surface embedded in $\mathbb{E}^{3}$. As
well known it is a surface with non zero Riemann curvature tensor when
interpreted as part of a \textit{MCGSS }$\{\mathring{S}^{2},%
%TCIMACRO{\TeXButton{g}{\slg}}%
%BeginExpansion
\slg
%EndExpansion
,D\}$, where the metric $%
%TCIMACRO{\TeXButton{g}{\slg}}%
%BeginExpansion
\slg
%EndExpansion
$ and the connection $D$ are respectively the pullback of the metric and the
connection of the \textit{MCGSS} $\mathbb{E}^{3}$.

Besides the \textit{MCGSS }$\{\mathring{S}^{2},%
%TCIMACRO{\TeXButton{g}{\slg}}%
%BeginExpansion
\slg
%EndExpansion
,D\}$ it is possible to define on $\mathring{S}^{2}$ a Riemann-Cartan
\textit{MCGSS }$\{\mathring{S}^{2},%
%TCIMACRO{\TeXButton{g}{\slg}}%
%BeginExpansion
\slg
%EndExpansion
,\nabla\}$ where $%
%TCIMACRO{\TeXButton{g}{\slg}}%
%BeginExpansion
\slg
%EndExpansion
$ is as before and $\nabla$ is a metric compatible Riemann-Cartan connection
such that its Riemann curvature is null, but its torsion tensor is\textit{ non
null}. $\nabla$ is called in \cite{nakahara} the navigator connection and the
denomination Columbus connection is also sometimes used. In \cite{rodcap2007}
$\nabla$ has been\ appropriately called the Nunes connection\footnote{Pedro
Salacience Nunes (1502--1578) was one of the leading mathematicians and
cosmographers of Portugal during the Age of Discoveries. He is well known for
his studies in Cosmography, Spherical Geometry, Astronomic Navigation, and
Algebra, and particularly known for his discovery of loxodromic curves and the
nonius. Loxodromic curves, also called rhumb lines, are spirals that converge
to the poles. They are lines that maintain a fixed angle with the meridians.
In other words, loxodromic curves directly related to the construction of the
Nunes connection. A ship following a fixed compass direction travels along a
loxodromic, this being the reason why Nunes connection is also known as
navigator connection. Nunes discovered the loxodromic lines and advocated the
drawing of maps in which loxodromic spirals would appear as straight lines.
This led to the celebrated Mercator projection, constructed along these
recommendations. Nunes invented also the Nonius scales which allow a more
precise reading of the height of stars on a quadrant. The device was used and
perfected at the time by several people, including Tycho Brahe, Jacob Kurtz,
Christopher Clavius and further by Pierre Vernier who in 1630 constructed a
practical device for navigation. For some centuries, this device was called
nonius. During the 19th century, many countries, most notably France, started
to call it vernier. More details in
http://www.mlahanas.de/Stamps/Data/Mathematician/N.htm.}. The parallelism rule
defining $\nabla$ is very simple. Given a vector $\mathbf{v}$ at $p$, let
$\alpha$ be the angle it makes with the tangent vector to the latitude lines
that pass through $p$. Then $\mathbf{v}$ is said to be parallel transported
according to the Nunes connection from $p$ to $q$ along any curve containing
those points if at point in the curve the \ transported vector makes the same
angle $\alpha$ with \ the tangent vector to the latitude line at that point.
It is then possible to verify that indeed the Riemann curvature of is null and
its torsion is non null. Details are given in Appendix B where a comparison of
the transport rules on $\mathring{S}^{2}$. according to the Levi-Civita and
the Nunes connection is also given. From Appendix B it is clear that the
orthonormal basis $\{\mathbf{e}_{\mathbf{1}}=\frac{1}{r}\partial
/\partial\theta,\mathbf{e}_{\mathbf{2}}=\frac{1}{r\sin\theta}\partial
/\partial\phi\}$ (where $r$, $\theta$,$\phi$ are the usual spherical
coordinates) used in the calculations are not defined on the poles. This is
due to the well known fact that $S^{2}$ does not admit two linear independent
tangent vector fields in all its point. Any given vector field on $S^{2}$
necessarily is zero at some point of $S^{2\text{ }}$(see, e.g., \cite{eise}).

A $n$-dimensional manifold $M$ which admits $n$ linearly independent \ vector
fields $\{e_{i}\in\sec TM$, $i=1,2,...,n\}$ at all its points is said to be
\textit{parallelizable}.

A Riemann-Cartan \textit{MCGSS }$\{M,%
%TCIMACRO{\TeXButton{g}{\slg}}%
%BeginExpansion
\slg
%EndExpansion
,\mathbf{\nabla}\}$ which is also parallelizable and such that\ there exists a
set of $n$ linearly independent \ vector fields $\{e_{i}\in\sec TM$,
$i=1,2,...,n\}$ on $M$ such that%

\begin{equation}
\mathbf{\nabla}_{e_{j}}e_{i}=0,\text{ }\forall i,j=1,2,...,n \label{0.6}%
\end{equation}
is said to be a \textit{teleparallel} \textit{MCGSS}. As it may be verified
without difficulty any teleparallel \textit{MCGSS} has null Riemannian
curvature tensor but non null torsion tensor.

It is also an easy task to invent examples where a manifold $M$ diffeomorphic
to $\mathbb{R}^{n}$ is part of a \textit{MCGSS'} such that its connection has
non zero curvature and/or torsion. A simple example is the following. Take an
arbitrary global $%
%TCIMACRO{\TeXButton{g}{\slg}}%
%BeginExpansion
\slg
%EndExpansion
$-orthonormal non coordinate basis for $TM$ ($M\simeq\mathbb{R}^{n}$) given by
$\{\mathbf{f}_{i}\}$, with $\mathbf{f}_{i}=f_{i}^{j}\partial/\partial
\mathtt{x}^{j}$, and introduce on $M\simeq\mathbb{R}^{n}$ a connection
$\mathbf{\nabla}$ such that $\mathbf{\nabla}_{\mathbf{f}_{j}}\mathbf{f}%
_{i}=0.$ It is then easy to verify that the \emph{nonmetricity} of that
connection, i.e., $\mathbf{\nabla}%
%TCIMACRO{\TeXButton{g}{\slg}}%
%BeginExpansion
\slg
%EndExpansion
=0$, the Riemann curvature tensor of $\mathbf{\nabla}$ is null but the torsion
tensor of $\mathbf{\nabla}$ is non null. This last example is important
regarding the \textit{interpretation} of the gravitational field in theories
that can be shown to mathematically equivalent to \textit{GRT} (in a precise
sense to be disclosed in due course), \textit{not }as an element describing a
particular geometrical property of a given \textit{MCGSS}, but as a
\textit{physical field} in the sense of Faraday (i.e., a field with ontology
similar to the electromagnetic field) living on Minkowski spacetime. But a
reader may ask: is there any \textit{serious} reason for trying such an
interpretation for the gravitational field? The answer is yes and will be
briefly discussed now.

\subsection{Killing Vector Fields, Symmetries and Conservation Laws}

We start this section by given some specialized names for structures that may
represent a spacetime in \textit{GRT} and some of its more known
generalizations. Let then $M$ be a $4$-dimensional differentiable manifold
satisfying the properties required in the definition given above for
a\ (Lorentzian) spacetime of \textit{GRT} Consider the structure given by the
pentuple $\{M,%
%TCIMACRO{\TeXButton{g}{\slg}}%
%BeginExpansion
\slg
%EndExpansion
,\mathbf{\nabla,}\tau_{%
%TCIMACRO{\TeXButton{g}{\sslg}}%
%BeginExpansion
\sslg
%EndExpansion
},\uparrow\}$, where $%
%TCIMACRO{\TeXButton{g}{\slg}}%
%BeginExpansion
\slg
%EndExpansion
$ is a Lorentzian metric (with signature $-2$) on $M$ and $\mathbf{\nabla}$ is
an arbitrary connection on $M$.

We say that the pentuple $(M,%
%TCIMACRO{\TeXButton{g}{\slg}}%
%BeginExpansion
\slg
%EndExpansion
,\mathbf{\nabla,}\tau_{%
%TCIMACRO{\TeXButton{g}{\sslg}}%
%BeginExpansion
\sslg
%EndExpansion
},\uparrow)$ is:

(a Lorentz-Cartan-Weyl spacetime \emph{iff} $\mathfrak{A}\neq0,\Theta\neq0,$
$\mathfrak{R}\neq0$,

(b) Lorentz-Cartan spacetime \emph{iff} $\mathfrak{A}=0,\Theta\neq0,$
$\mathfrak{R}\neq0$,

(c) Lorentzian spacetime \emph{iff} $\mathfrak{A}=\Theta=0,$ $\mathfrak{R}%
\neq0$,

(d) Minkowski spacetime \textit{iff }$M\simeq\mathbb{R}^{4},$\textit{\ }%
$\mathfrak{A}=\Theta=$ $\mathfrak{R}=0.$

In a Minkowski spacetime we will denote the metric tensor by $%
%TCIMACRO{\TeXButton{eta}{\mbox{\boldmath{$\eta$}}}}%
%BeginExpansion
\mbox{\boldmath{$\eta$}}%
%EndExpansion
$. Moreover, we recall that \ in \ such spacetime there exists an equivalence
class of global coordinate systems for $M\simeq\mathbb{R}^{4}$ such that if
$\{\mathtt{x}^{\alpha}\}$ is one of these systems then%

\begin{equation}%
%TCIMACRO{\TeXButton{eta}{\mbox{\boldmath{$\eta$}}}}%
%BeginExpansion
\mbox{\boldmath{$\eta$}}%
%EndExpansion
=\eta_{\alpha\beta}d\mathtt{x}^{\alpha}\otimes\mathtt{x}^{\beta},
\label{0.66a}%
\end{equation}
with $\eta_{\alpha\beta}=%
%TCIMACRO{\TeXButton{eta}{\mbox{\boldmath{$\eta$}}}}%
%BeginExpansion
\mbox{\boldmath{$\eta$}}%
%EndExpansion
(\frac{\partial}{\partial\mathtt{x}^{\alpha}},\frac{\partial}{\partial
\mathtt{x}^{\beta}})$, where $\alpha,\beta=0,1,2,3$ and the matrix
$(\eta_{\alpha\beta})$ is:
\begin{equation}
(\eta_{\alpha\beta})=\mathrm{diag}(1,-1,-1,-1). \label{0.66}%
\end{equation}
\medskip

Now, consider a Lorentzian spacetime. Let $\mathbf{T\in}\sec T_{s}^{r}M$ be an
arbitrary differentiable tensor field. We say that a diffeomorphism
$l:U\rightarrow$ $U$ ($U\subset M$) generated by a one-parameter group of
diffeomorphisms characterized by the vector field $\xi$ is a \textit{symmetry}
of $\mathbf{T}$ iff%

\begin{equation}
\pounds _{\xi}\mathbf{T}=0, \label{0.7'}%
\end{equation}
where $\pounds _{\xi}$ is the Lie derivative \footnote{For details, see, e.g.,
\cite{choquet,rodcap2007}.} in the direction of $\xi$.

This means that the pullback field $l^{\ast}\mathbf{T}$\ satisfies
\begin{equation}
l^{\ast}\mathbf{T}=\mathbf{T.} \label{0.8}%
\end{equation}

For Lorentzian spacetimes (where $\dim M=4$) the symmetries of the metric
tensor\emph{ }$%
%TCIMACRO{\TeXButton{g}{\slg}}%
%BeginExpansion
\slg
%EndExpansion
$ play a very important role. Indeed, it can be shown that the equation. (see,
e.g., \cite{logunov1})
\begin{equation}
\pounds _{\xi}%
%TCIMACRO{\TeXButton{g}{\slg}}%
%BeginExpansion
\slg
%EndExpansion
=0, \label{0.9}%
\end{equation}
called \emph{Killing equation} can have the maximum number of ten Killing
vector fields, and that maximum number ten only occurs for Lorentzian
spacetimes with have constant \textit{scalar curvature. }There are only three
distinct Lorentzian spacetimes with the maximum number of Killing vector
fields, which are Minkowski spacetime, de Sitter spacetime and anti de Sitter spacetime.

As well known Minkowski spacetime is the mathematical structure used as
the\ `arena' for the classical theories of fields and particles and also for
the so called relativistic quantum field theories \cite{bolog}.

In the classical relativistic theories of fields and particles it can be shown
that there exists trustful conservation laws for the energy-momentum, angular
momentum and conservation of the center of mass for any system of particles
and fields\footnote{In the relativisitc quantum filed theories there are
trustful conservation laws for any system of interacting fields.}. The proof
\footnote{See, e.g., \cite{logunov1,rodcap2007}.} of that statement depends
crucially on the existence of the ten Killing vector fields of the Minkowski
metric tensor $%
%TCIMACRO{\TeXButton{eta}{\mbox{\boldmath{$\eta$}}}}%
%BeginExpansion
\mbox{\boldmath{$\eta$}}%
%EndExpansion
$ .

This fact is very important because given a general Lorentzian spacetime
modelling a given gravitational field according to \textit{GRT} and which has
a non constant scalar curvature in general there are not a sufficient number
of Killing vector fields for formulating trustful conservation laws for the
energy-momentum and angular momentum of a given systems of particles and
fields. Many tentatives have been done in order to establish energy-momentum
and angular momentum conservation laws in \textit{GRT}. All tentatives are
based on the fact that although the field $%
%TCIMACRO{\TeXButton{g}{\slg}}%
%BeginExpansion
\slg
%EndExpansion
$ is part of the spacetime structure the `geometry' must also have an object
that makes the role of its \ `energy-momentum tensor' in order to warrant the
conservation of the energy-momentum of the gravitational and matter fields
However all tentatives resulted (until now) deceptive (unless some additional
hypothesis are postulated for the Lorentzian spacetime structure, as e.g.,
that such structure is parallelizable) because all that have been attained was
the association of a series of \textit{pseudo-energy momentum tensors} to the
gravitational field \footnote{The first one to introduce an energy-momentum
pseudo-tensor for the gravitational field was \cite{einstein0}. However, it is
possible to introduce an infinity of distinct energy-momentum pseudo-tensors
for the gravitational field in \textit{GRT}. The mathematical reasons for this
manifold possibility may be found, e.g., in
\cite{tw78,rodcap2007,notterod2009}.} and without additional commentaries from
our part we quote here what the authors of \cite{sawu} have to say on this
issue:\medskip

\textquotedblleft\emph{It is a shame to loose the special relativistic total
energy conservation in General Relativity. Many of the tentatives to resurrect
it are quite interesting, many are simply garbage}\textquotedblright.\medskip

It can be shown that the non existence of trustful energy-momentum and angular
momentum for the system of the gravitational field and the matter fields lead
to serious inconsistencies. This has already been noted long ago by several
scientists, in particular by Levi-Civita (already in 1919!) \cite{levi1} and
discussed by Logunov and collaborators \cite{logunov1} in the eights of the
last century\footnote{A simple way to understand the origin of the issue has
already been clearly stated by Schr\"{o}ndinger, who in \cite{schrodinger}
observed that in a general Lorentzian spacetime you can not even define the
momentum of a pair of particles when they are at events say, $\mathfrak{e}%
_{1}$ and $\mathfrak{e}_{2}$ because vectors at the tangent spaces
$T_{\mathfrak{e}_{1}}M$ an $TM_{\mathfrak{e}_{2}}$ cannot be summed.}. A
modern discussion may be found in \cite{benn, rodcap2007,notterod2009}\ and
the continuous tentatives for finding a solution for the problem within
\textit{GRT} (including a today's popular approach called quasi-local energy)
may be found in \cite{szabados}. We will briefly discuss such an approach in
Section 7 where we study the Hamiltonian formulation of \textit{our} theory.

After more than 85 years of deceptive tentatives many physicists,
\footnote{Besides Logunov and collaborators we quote here also Feynmann
\cite{feynman}, Schwinger \cite{schw}, Weinberg \cite{weinberg}, Rosen
\cite{rosen} and \cite{gupta,thirring,grish1,grish2}.} think that \textit{GRT}
needs a revision where the main emphasis must be given to the fact that the
gravitational field is a physical field in the \textit{sense} of Faraday
living and interacting with the other physical fields in Minkowski spacetime.
However, the formulation of such a theory using as unique ingredients the
geometric objects of Minkowski spacetime and the correct representative of the
gravitational field (the distortion field $%
%TCIMACRO{\TeXButton{h}{\slh}}%
%BeginExpansion
\slh
%EndExpansion
$\textbf{$)$} living in that spacetime and in such\ way that the resulting
theory becomes equivalent to \textit{GRT} in some well defined sense resulted
to be a nontrivial task.

In \cite{rodsouza} (using the Clifford bundle formalism \cite{rodcap2007})\ a
theory of that kind, where the gravitational field is described by a physical
field\ living in Minkowski spacetime and has its dynamics described by a well
defined Lagrangian density such that its equations of motion result equivalent
to Einstein's field equations (at least for models of \textit{GRT }where the
manifold $M$ can be taken as $\mathbb{R}^{4}$) has been proposed. In that
\ theory it is suggested that the gravitational field $%
%TCIMACRO{\TeXButton{g}{\slg}}%
%BeginExpansion
\slg
%EndExpansion
$ is to be interpreted as a deformation of the Minkowski metric $%
%TCIMACRO{\TeXButton{eta}{\mbox{\boldmath{$\eta$}}}}%
%BeginExpansion
\mbox{\boldmath{$\eta$}}%
%EndExpansion
$ product by a special $(1,1)$-extensor field $%
%TCIMACRO{\TeXButton{h}{\slh}}%
%BeginExpansion
\slh
%EndExpansion
:\sec%
%TCIMACRO{\dbigwedge \nolimits^{1}}%
%BeginExpansion
{\displaystyle\bigwedge\nolimits^{1}}
%EndExpansion
T^{\ast}M\rightarrow\sec%
%TCIMACRO{\dbigwedge \nolimits^{1}}%
%BeginExpansion
{\displaystyle\bigwedge\nolimits^{1}}
%EndExpansion
T^{\ast}M$ said to be the distortion (or deformation) \emph{tensor field.
}However, in that opportunity those authors were not able to produce a
mathematical formalism to work directly with $%
%TCIMACRO{\TeXButton{h}{\slh}}%
%BeginExpansion
\slh
%EndExpansion
$, i.e., writing a Lagrangian density for that field and deducing its
equations of motion.

The necessary mathematical formalism to do that job now exists, it is the
\textit{MECM }mentioned above\textit{. }The \textit{MECM} has its origin, by
the best of our knowledge in some mathematical ideas first proposed by
Hestenes and Sobczyc in their book \textquotedblleft\emph{Clifford Algebra to
Geometrical Calculus}\textquotedblright\ \cite{hesob84}. Those ideas have been
investigated in \cite{vfernandez, moya,quin} and attained maturity in
\cite{fmr011,fmr012,fmr013,fmr014,fmr015,fmr016,fmr017,fmr071,fmr072,fmr073,fmr074}%
. In particular in \ \cite{fmr017} the theory of derivatives of functionals of
extensor fields necessary for the developments of the present paper has been
introduced in a rigorous way. That crucial concept will be recalled (with
several examples) in Section 3. Utilizing \textit{MECM,} presented with enough
details in Section 4 we will show that it is indeed possible to present a
theory of the gravitational field as the theory of a field $%
%TCIMACRO{\TeXButton{h}{\slh}}%
%BeginExpansion
\slh
%EndExpansion
$\textbf{ }living on Minkowski spacetime and describing a \textit{plastic}
\textit{deformation }of the Lorentz vacuum. The Lorentz vacuum is in our view
a real and very complex physical substance\footnote{For some interesting ideas
on the nature of such a substance see
\cite{laughlin,unzicker0,unzicker1,volovik}.} which lives in an arena
mathematically described\ by Minkowski spacetime\footnote{Eventually future
experimental developments will show that this hypothesis must be modified,
e.g., the substance describing the vacuum may live in a spacetime with more
dimensions that Minkowski spacetime.}. When the Lorentz vacuum is in its
ground state it is described by a trivial distortion field $%
%TCIMACRO{\TeXButton{h}{\slh}}%
%BeginExpansion
\slh
%EndExpansion
=\mathrm{id}$, but when it is disturbed due to the presence of what we call
matter fields it is described by a nontrivial distortion field $%
%TCIMACRO{\TeXButton{h}{\slh}}%
%BeginExpansion
\slh
%EndExpansion
$\textbf{ }of the \textit{plastic type} which as we shall see (Section 5)
satisfies a well defined field equation, which are (Section 6) equivalent to
Einstein field equation for $%
%TCIMACRO{\TeXButton{g}{\slg}}%
%BeginExpansion
\slg
%EndExpansion
$ (at least for models of \textit{GRT }where the manifold $M$ can be taken as
$\mathbb{R}^{4}$). In Section 6 we introduce also genuine energy-momentum and
angular momentum conservation laws for the system composed of the matter and
gravitational fields.

One of the features of the \textit{general }mathematical formalism used in the
present article is a formulation of a theory of covariant derivatives on
manifolds, where some novel concepts appear, like for example the rotation
extensor field $\Omega$ defined in a $4$-dimensional vector manifold
$U_{o}\subset U$, where $U_{o}$ represents the points of a given
$\mathcal{U\subset}M$ and $U$ is called the canonical space.

Once $U$ is constructed we introduce an Euclidean Clifford algebra denoted
$\mathcal{C\ell(}U,\cdot\mathcal{)}$ on $U_{o}\subset U$ whose \textit{main
purpose} is to define an algorithm to perform very sophisticated and indeed
nontrivial calculations. Next we introduce on $U_{0}$ a constant Minkowski
metric extensor field $\eta$ and the distortion field $%
%TCIMACRO{\TeXButton{h}{\slh}}%
%BeginExpansion
\slh
%EndExpansion
$ such that the extensor field $%
%TCIMACRO{\TeXButton{itg}{\itg}}%
%BeginExpansion
\itg
%EndExpansion
=%
%TCIMACRO{\TeXButton{h}{\slh}}%
%BeginExpansion
\slh
%EndExpansion
^{\dagger}\eta%
%TCIMACRO{\TeXButton{h}{\slh}}%
%BeginExpansion
\slh
%EndExpansion
$ represents in $U_{o}$ the metric tensor $%
%TCIMACRO{\TeXButton{g}{\slg}}%
%BeginExpansion
\slg
%EndExpansion
$ in $\mathcal{U\subset}M$. Several different\textit{ MCGSS }are introduced
having $U_{o}$ as the first member of the triple and it is shown that some of
those structures may be clearly interpreted as a distortion of another one
through the action of the distortion extensor $%
%TCIMACRO{\TeXButton{h}{\slh}}%
%BeginExpansion
\slh
%EndExpansion
$. In that way it will be shown that when $M\simeq\mathbb{R}^{4}$ in which
case we may identify $\mathbb{R}^{4}\simeq U_{0}\simeq U$, the rotation
extensor field that appears in the representative in $U$ of the Levi-Civita
connection of $%
%TCIMACRO{\TeXButton{g}{\slg}}%
%BeginExpansion
\slg
%EndExpansion
$ can be written as a functional of $%
%TCIMACRO{\TeXButton{h}{\slh}}%
%BeginExpansion
\slh
%EndExpansion
^{\clubsuit}$ $=%
%TCIMACRO{\TeXButton{h}{\slh}}%
%BeginExpansion
\slh
%EndExpansion
^{-1\dagger}$ and its \textit{vector derivatives} $\cdot\partial%
%TCIMACRO{\TeXButton{h}{\slh}}%
%BeginExpansion
\slh
%EndExpansion
^{\clubsuit}$ and $\cdot\partial\cdot\partial%
%TCIMACRO{\TeXButton{h}{\slh}}%
%BeginExpansion
\slh
%EndExpansion
^{\clubsuit}$. This permits to write the usual Einstein-Hilbert Lagrangian as
a functional of $(%
%TCIMACRO{\TeXButton{h}{\slh}}%
%BeginExpansion
\slh
%EndExpansion
^{\clubsuit},\partial%
%TCIMACRO{\TeXButton{h}{\slh}}%
%BeginExpansion
\slh
%EndExpansion
^{\clubsuit},\cdot\partial\cdot\partial%
%TCIMACRO{\TeXButton{h}{\slh}}%
%BeginExpansion
\slh
%EndExpansion
^{\clubsuit})$ and to find directly the equation of motion for $%
%TCIMACRO{\TeXButton{h}{\slh}}%
%BeginExpansion
\slh
%EndExpansion
$, a result that for the best of our knowledge has not been obtained before in
a consistent way. Indeed, it must be said that tentatives of using the
rudiments of the multiform and extensor calculus to formulated a theory of the
gravitational field as a gauge theory (but where the world gauge does not have
the meaning it has in those field theories which use gauge fields defined as
sections of some principal bundles) has been developed\ in \cite{ladogu,dola}.

The gravitational theory in \cite{ladogu,dola} has been recently reviewed by
\cite{hestenes2005}.However, despite some very good ideas presented there,
those works which do not specify clearly (in our opinion) the relationship
between their gauge covariant derivative $\mathcal{D}_{\mu}$ and the standard
Levi-Civita \ covariant derivative $D_{e_{\mu}}$ of $%
%TCIMACRO{\TeXButton{g}{\slg}}%
%BeginExpansion
\slg
%EndExpansion
$ implies in confusion, for the formalism as presented in
\cite{lagodu,dola,hestenes2005} does not permit to deduce any relation between
$\mathcal{D}_{\mu}\ $and $D_{e_{\mu}}$. The confusion, now clarified is
described in Appendix F which has additional remarks (concerning the
originally written in 2010) and resulted from a discussion of W.A.R. with A.
Lasenby at ICCA10\footnote{See http://icca10.ut.ee/.} in August 2014.

For completeness we also present in Appendix C the derivation of the equations
of motion for a theory where the gravitational field is described by two
independent fields, a distortion field $%
%TCIMACRO{\TeXButton{h}{\slh}}%
%BeginExpansion
\slh
%EndExpansion
$ and a rotation field $\Omega$ \ This is in line of ideas (first present in
\cite{hehl} where it is claimed that $\Omega$ is related to the source of
spin.\footnote{However, it must be said here that as it is clear form the the
application of the Lagrangian formalism for mltiform fields (including spinor
fields) using the \ multiform and extensor calculus, as developed e.g., in
\cite{rodcap2007} the source of spins is to be found in the antisymmetric part
of the canonical energy-momentum tensor.}. Appendix D contains a proof of the
equivalence of two apparently very different expressions for the
Einstein-Hilbert Lagrangian density and Appendix E recall the derivation of
the field equations for the potential fields $\mathfrak{g}^{\mathbf{\alpha}}=%
%TCIMACRO{\TeXButton{h}{\slh}}%
%BeginExpansion
\slh
%EndExpansion
^{\dagger}(%
%TCIMACRO{\TeXButton{vt}{\mbox{\boldmath{$\vartheta$}}}}%
%BeginExpansion
\mbox{\boldmath{$\vartheta$}}%
%EndExpansion
^{\mathbf{\alpha}})$.

Finally in Section 8 we present our conclusions.

\section{Multiforms and Extensors}

Let $\mathbf{V}$ be a vector space ($\dim\mathbf{V}=n$) over the real field
$\mathbb{R}$. The dual space of $\mathbf{V}$ is usually denoted by
$\mathbf{V}^{\ast}$, but in what follows it will be denoted by $V$. The space
of $k$-forms ($0\leq k\leq n$) will be denoted by $%
%TCIMACRO{\dbigwedge \nolimits^{k}}%
%BeginExpansion
{\displaystyle\bigwedge\nolimits^{k}}
%EndExpansion
V$ and we have the usual identifications: $%
%TCIMACRO{\dbigwedge \nolimits^{0}}%
%BeginExpansion
{\displaystyle\bigwedge\nolimits^{0}}
%EndExpansion
V=\mathbb{R}$ and $%
%TCIMACRO{\dbigwedge \nolimits^{1}}%
%BeginExpansion
{\displaystyle\bigwedge\nolimits^{1}}
%EndExpansion
V=V$.

\subsection{Multiforms}

A \textit{formal} sum $X=X_{0}+X_{1}+\cdots+X_{k}+\cdots+X_{n}$, where
$X_{k}\in%
%TCIMACRO{\dbigwedge \nolimits^{k}}%
%BeginExpansion
{\displaystyle\bigwedge\nolimits^{k}}
%EndExpansion
V$ will be called a \textit{nonhomogeneous multiform}.\emph{ }The set of
all\emph{ }nonhomogeneous multiforms\emph{ }has a natural vector space
structure\footnote{The sum of multiforms and the multiplicatin of multiforms
by scalars are defined by: \textbf{(i)} if $X=X_{0}+X_{1}+\cdots+X_{n}$ e
$Y=Y_{0}+Y_{1}+\cdots+Y_{n}$ then $X+Y=(X_{0}+Y_{0})+(X_{1}+Y_{1}%
)+\cdots+(X_{n}+Y_{n})$, \textbf{(ii)} if $\alpha\in R$ and $X=X_{0}%
+X_{1}+\cdots+X_{n}$ then $\alpha X=\alpha X_{0}+\alpha X_{1}+\cdots+\alpha
X_{n}$.}\ over $\mathbb{R}$ and will be denoted by $%
%TCIMACRO{\dbigwedge }%
%BeginExpansion
{\displaystyle\bigwedge}
%EndExpansion
V$.

Let $\left\{  \mathbf{e}_{k}\right\}  $ and $\left\{  \varepsilon^{k}\right\}
$ be respectively basis of $\mathbf{V}$ and $V$ where the second is the
\textit{dual} of the first, i.e., $\varepsilon^{k}(\mathbf{e}_{j})=\delta
_{j}^{k}$. Then $\{1,\varepsilon^{j_{1}},\ldots,\varepsilon^{j_{1}}%
\wedge\ldots\wedge\varepsilon^{j_{k}},\ldots,\varepsilon^{j_{1}}\wedge
\ldots\wedge\varepsilon^{j_{n}}\}$ is a basis for $%
%TCIMACRO{\dbigwedge }%
%BeginExpansion
{\displaystyle\bigwedge}
%EndExpansion
V$, and since $\dim\mathbf{V=}n$ we have that $\dim%
%TCIMACRO{\dbigwedge }%
%BeginExpansion
{\displaystyle\bigwedge}
%EndExpansion
V=\binom{n}{0}+\binom{n}{1}+\cdots+\binom{n}{k}+\cdots+\binom{n}{n}=2^{n}$.

\subsection{The $k$-Part Operator and Involutions}

We introduce on $%
%TCIMACRO{\dbigwedge }%
%BeginExpansion
{\displaystyle\bigwedge}
%EndExpansion
V$ a fundamental operator. Let $0\leq k\leq n$, the linear operator
$\left\langle {}\right\rangle _{k}$ defined by $X\mapsto\left\langle
X\right\rangle _{k}:=X_{k}$ ($X_{k}\in%
%TCIMACRO{\dbigwedge \nolimits^{k}}%
%BeginExpansion
{\displaystyle\bigwedge\nolimits^{k}}
%EndExpansion
V$) will be called the $k$-part operator and $\left\langle X\right\rangle
_{k}$ is read as the $k$\emph{-part} of $X$.

The $k$-forms are called \textit{homogeneous} \emph{multiforms of grade} $k$
(or $k$\emph{-homogeneous multiforms}). It is obvious that for any
$k$-homogeneous multiform $X$ we have: $\left\langle X\right\rangle _{l}=X$ if
$l=k,$ or $\left\langle X\right\rangle _{l}=0$ if $l\neq k$. Also, any
multiform may be expressed as a sum of its $k$-parts, i.e.,
$X=\overset{n}{\underset{k=0}{\sum}}\left\langle X\right\rangle _{k}.$

The $0$-forms (i.e.,the real numbers), the $1$-forms, the $2$-forms$\ldots$
are called scalars, forms, biforms, etc., and the $n$-forms and the
($n-1)$-forms are sometimes called pseudo-scalars and pseudo-forms
\footnote{Such a nomenclature must be used with care, since\ it may lead to
serious confusions if mislead with the concept of de Rham's pair and impair
forms, those latter objects also called by some authors pseudo-forms or
twisted forms. See \cite{droro2009} for a discussion of the issue.}\emph{.}

We now introduce two fundamental involutions on $%
%TCIMACRO{\dbigwedge }%
%BeginExpansion
{\displaystyle\bigwedge}
%EndExpansion
V$.

The linear operator$\ \symbol{94},$ defined by $X\mapsto\hat{X}$ such that
$\left\langle \hat{X}\right\rangle _{k}=(-1)^{k}\left\langle X\right\rangle
_{k},$ is called the \textit{conjugation operator} and $\hat{X}$ \emph{is read
as the conjugate of} $X$.

The linear operator $\underset{}{\sim},$ defined by $X\mapsto\widetilde{X}$
such that $\left\langle \widetilde{X}\right\rangle _{k}=(-1)^{\frac{1}%
{2}k(k-1)}\left\langle X\right\rangle _{k},$ is called the reversion operator
and $\widetilde{X}$ \emph{is read as the reverse of }$X.$

Both \ operators are involutions, i.e., $\overset{\wedge}{\hat{X}}=X$ and
$\widetilde{\widetilde{X}}=X.$

\subsection{Exterior Product}

The exterior product (or Grassmann product) $\wedge:%
%TCIMACRO{\dbigwedge }%
%BeginExpansion
{\displaystyle\bigwedge}
%EndExpansion
V\times%
%TCIMACRO{\dbigwedge }%
%BeginExpansion
{\displaystyle\bigwedge}
%EndExpansion
V\rightarrow%
%TCIMACRO{\dbigwedge }%
%BeginExpansion
{\displaystyle\bigwedge}
%EndExpansion
V$ is defined for arbitrary $X,Y\in%
%TCIMACRO{\dbigwedge }%
%BeginExpansion
{\displaystyle\bigwedge}
%EndExpansion
V$ as the element $X\wedge Y\in%
%TCIMACRO{\dbigwedge }%
%BeginExpansion
{\displaystyle\bigwedge}
%EndExpansion
V$ such that
\begin{equation}
\left\langle X\wedge Y\right\rangle _{k}=\overset{k}{\underset{j=1}{\sum}%
}\left\langle X\right\rangle _{j}\wedge\left\langle Y\right\rangle _{k-j}.
\label{1.1}%
\end{equation}

On the right side of Eq.(\ref{1.1}) appears a sum of the exterior product of
\footnote{The exterior product of $k$-formas is defined as follows:
\textbf{(i)\ } $\alpha,\beta\in\mathbb{R}:\alpha\wedge\beta=\alpha\beta$;
\textbf{(ii)\ }$\alpha\in\mathbb{R}$ and $x\in%
%TCIMACRO{\dbigwedge \nolimits^{k}}%
%BeginExpansion
{\displaystyle\bigwedge\nolimits^{k}}
%EndExpansion
V$ with $k\geq1:\alpha\wedge x=x\wedge\alpha=\alpha x,$ \textbf{(iii)\ } if
$X_{j}\in%
%TCIMACRO{\dbigwedge \nolimits^{j}}%
%BeginExpansion
{\displaystyle\bigwedge\nolimits^{j}}
%EndExpansion
V$ and $X_{j}\in%
%TCIMACRO{\dbigwedge \nolimits^{k}}%
%BeginExpansion
{\displaystyle\bigwedge\nolimits^{k}}
%EndExpansion
V$ with $j,k\geq1$ then $X_{j}\wedge Y_{k}=\frac{(j+k)!}{j!k!}\mathcal{A}%
(x\otimes y)$ where $\mathcal{A}$ is the well known antisymmetrization
operator of ordinary linear algebra. Keep in mind when reading texts on the
subject that eventually another definition of the exterior product (not
equivalent to the one used here) may be being used, as e.g., in
\cite{rodcap2007}.} $j$-form by a $(k-j)$-form. The exterior product (like the
sum) is a an internal law in the space $%
%TCIMACRO{\dbigwedge }%
%BeginExpansion
{\displaystyle\bigwedge}
%EndExpansion
V$. It satisfy distributive laws on the left and on the right\ and an
associative laws. The first ones are consequence of the distributive laws of
the the exterior product of $k$-forms and the second ones may be shown to be
true without difficulty). The linear space $%
%TCIMACRO{\dbigwedge }%
%BeginExpansion
{\displaystyle\bigwedge}
%EndExpansion
V$ equipped with the exterior product is an associative algebra called the
\textit{exterior algebra of multiforms}.

We present now the main properties of the exterior product of multiforms.

\textbf{(i)\ }For all $\alpha,\beta\in\mathbb{R}$ and $X\in%
%TCIMACRO{\dbigwedge }%
%BeginExpansion
{\displaystyle\bigwedge}
%EndExpansion
V$
\begin{subequations}
\label{1.2}%
\begin{align}
\alpha\wedge\beta &  =\alpha\beta,\label{1.2a}\\
\alpha\wedge X  &  =X\wedge\alpha=\alpha X, \label{1.2b}%
\end{align}

\textbf{(ii)\ }For all $X_{j}\in%
%TCIMACRO{\dbigwedge \nolimits^{j}}%
%BeginExpansion
{\displaystyle\bigwedge\nolimits^{j}}
%EndExpansion
V$ and $Y_{k}\in%
%TCIMACRO{\dbigwedge \nolimits^{k}}%
%BeginExpansion
{\displaystyle\bigwedge\nolimits^{k}}
%EndExpansion
V$%
\end{subequations}
\begin{equation}
X_{j}\wedge Y_{k}=(-1)^{jk}Y_{k}\wedge X_{j}. \label{1.3}%
\end{equation}

\textbf{(iii)\ }For all $a\in%
%TCIMACRO{\dbigwedge \nolimits^{1}}%
%BeginExpansion
{\displaystyle\bigwedge\nolimits^{1}}
%EndExpansion
V$ and $X\in%
%TCIMACRO{\dbigwedge }%
%BeginExpansion
{\displaystyle\bigwedge}
%EndExpansion
V$%
\begin{equation}
a\wedge X=\hat{X}\wedge a. \label{1.4}%
\end{equation}

\textbf{(iv)\ }For all $X,Y\in%
%TCIMACRO{\dbigwedge }%
%BeginExpansion
{\displaystyle\bigwedge}
%EndExpansion
V$
\begin{subequations}
\label{1.5}%
\begin{align}
(X\wedge Y)\symbol{94}  &  =\hat{X}\wedge\hat{Y},\label{1.5a}\\
\widetilde{X\wedge Y}  &  =\widetilde{Y}\wedge\widetilde{X}. \label{1.5b}%
\end{align}
\vspace*{0.3in}

\subsection{The Scalar Canonical Product}

Let us fix on $\mathbf{V}$ an arbitrary basis $\{\mathbf{b}_{j}\}$. A scalar
product of multiforms $X,Y\in%
%TCIMACRO{\dbigwedge }%
%BeginExpansion
{\displaystyle\bigwedge}
%EndExpansion
V$ may be defined by
\end{subequations}
\begin{equation}
X\cdot Y=\left\langle X\right\rangle _{0}\left\langle Y\right\rangle
_{0}+\frac{1}{k!}\left\langle X\right\rangle _{k}(\mathbf{b}^{j_{1}}%
,\ldots,\mathbf{b}^{j_{k}})\left\langle Y\right\rangle _{k}(b_{j_{1}}%
,\ldots,b_{j_{k}}), \label{1.6}%
\end{equation}
where, in order to utilize Einstein's sum convention rule, we wrote $b_{j_{i}%
}:=\mathbf{b}^{j_{i}}$, $i=1,2,...,k$.

The product defined by Eq.(\ref{1.6}) (obviously associated to the basis
$\{\mathbf{b}_{j}\}$) is a well defined scalar product on the vector space$%
%TCIMACRO{\dbigwedge }%
%BeginExpansion
{\displaystyle\bigwedge}
%EndExpansion
V$.\textbf{ }It is symmetric ($X\cdot Y=Y\cdot X$), is linear, i.e.,
($X\cdot(Y+Z)=X\cdot Y+X\cdot Z$), possess mist associativity ($(\alpha
X)\cdot Y=X\cdot(\alpha Y)=\alpha(X\cdot Y)$) and is non degenerated (i.e.,
$X\cdot Y=0$ for all $X$, implies $Y=0$). Even more that product is positive
definite, i.e., $X\cdot X\geq0$ for all $X$, and $X\cdot X=0$ iff $X=0$.

The scalar product of multiforms which we just introduced will be called the
\textit{canonical scalar product}, despite the fact that for its definition it
is necessary to choice an arbitrary basis $\{\mathbf{b}_{j}\}$ of $\mathbf{V}$.

Some important properties of the canonical scalar product are:

\textbf{(i)\ }For all scalars $\alpha,\beta\in\mathbb{R}$%
\begin{equation}
\alpha\cdot\beta=\alpha\beta, \label{1.7}%
\end{equation}
\ \ 

\textbf{(ii)\ }For all $X_{j}\in%
%TCIMACRO{\dbigwedge \nolimits^{j}}%
%BeginExpansion
{\displaystyle\bigwedge\nolimits^{j}}
%EndExpansion
V$ and $Y_{k}\in%
%TCIMACRO{\dbigwedge \nolimits^{k}}%
%BeginExpansion
{\displaystyle\bigwedge\nolimits^{k}}
%EndExpansion
V$%
\begin{equation}
X_{j}\cdot Y_{k}=0,\text{ if }j\neq k. \label{1.8}%
\end{equation}

\textbf{(iii)\ }For any two simple $k$-forms, say $v_{1}\wedge\ldots\wedge
v_{k}$ and $w_{1}\wedge\ldots\wedge w_{k},$ we have:
\begin{equation}
(v_{1}\wedge\ldots\wedge v_{k})\cdot(w_{1}\wedge\ldots\wedge w_{k}%
)=\left\vert
\begin{array}
[c]{ccc}%
v_{1}\cdot w_{1} & \cdots & v_{1}\cdot w_{k}\\
\vdots &  & \vdots\\
v_{k}\cdot w_{1} & \cdots & v_{k}\cdot w_{k}%
\end{array}
\right\vert , \label{1.9}%
\end{equation}
where $\left\vert {}\right\vert $ denote, here the classical determinant.

\textbf{(iv)}\emph{\ }For all $X,Y\in%
%TCIMACRO{\dbigwedge }%
%BeginExpansion
{\displaystyle\bigwedge}
%EndExpansion
V$
\begin{subequations}
\label{1.10}%
\begin{align}
\hat{X}\cdot Y  &  =X\cdot\hat{Y},\label{1.10a}\\
\widetilde{X}\cdot Y  &  =X\cdot\widetilde{Y}. \label{1.10b}%
\end{align}

\textbf{(v) }Let $(\{\varepsilon^{j}\}$, $\{\varepsilon_{j}\})$ be a pair of
reciprocal bases for $V$ (i.e., $\varepsilon^{j}\cdot\varepsilon_{k}%
=\delta_{k}^{j}$). \ We can then construct exactly two natural basis for the
space $%
%TCIMACRO{\dbigwedge }%
%BeginExpansion
{\displaystyle\bigwedge}
%EndExpansion
V$, respectively
\end{subequations}
\begin{subequations}
\label{1.11}%
\begin{align}
&  \{1,\varepsilon^{j_{1}},\ldots,\varepsilon^{j_{1}}\wedge\ldots
\wedge\varepsilon^{j_{k}},\ldots,\varepsilon^{j_{1}}\wedge\ldots
\wedge\varepsilon^{j_{n}}\},\label{1.11a}\\
&  \text{and}\nonumber\\
&  \{1,\varepsilon_{j_{1}},\ldots,\varepsilon_{j_{1}}\wedge\ldots
\wedge\varepsilon_{j_{k}},\ldots,\varepsilon_{j_{1}}\wedge\ldots
\wedge\varepsilon_{j_{n}}\}. \label{1.11b}%
\end{align}

Then, any $X\in%
%TCIMACRO{\dbigwedge }%
%BeginExpansion
{\displaystyle\bigwedge}
%EndExpansion
V$ \ may be expressed using these basis as :
\end{subequations}
\begin{subequations}
\label{1.12}%
\begin{align}
X  &  =X\cdot1+\overset{n}{\underset{k=1}{\sum}}\frac{1}{k!}X\cdot
(\varepsilon^{j_{1}}\wedge\ldots\wedge\varepsilon^{j_{k}})(\varepsilon_{j_{1}%
}\wedge\ldots\wedge\varepsilon_{j_{k}}),\label{1.12a}\\
&  =X\cdot1+\overset{n}{\underset{k=1}{\sum}}\frac{1}{k!}X\cdot(\varepsilon
_{j_{1}}\wedge\ldots\wedge\varepsilon_{j_{k}})(\varepsilon^{j_{1}}\wedge
\ldots\wedge\varepsilon^{j_{k}}), \label{1.12b}%
\end{align}
or in a more compact notation using the collective index\footnote{The
collective index $J$ take values over the follwing set of indices: $\emptyset$
(the null set), $j_{1},\ldots,j_{1}\ldots j_{k},\ldots,j_{1}\ldots j_{n}$. The
symbol $\varepsilon^{J}$ denots then a scalar basis, $\varepsilon^{\emptyset
}=1$, \ a basis of $1$-forms $\varepsilon^{j_{1}},\ldots$, and $\varepsilon
^{j_{1}\ldots j_{k}}=\varepsilon^{j_{1}}\wedge\ldots\wedge\varepsilon^{j_{k}}$
abasis for $%
%TCIMACRO{\dbigwedge \nolimits^{k}}%
%BeginExpansion
{\displaystyle\bigwedge\nolimits^{k}}
%EndExpansion
V.$ We will use also when convenient the notation $\varepsilon^{j_{1}\ldots
j_{n}}=\varepsilon^{j_{1}}\wedge\ldots\wedge\varepsilon^{j_{n}}$. Finally we
register that $\nu(J)$ is the number of indices, defined by $\nu
(J)=0,1,\ldots,k,\ldots,n$ with $J$ \ taking values over the set just
introduced.} $J$,
\end{subequations}
\begin{equation}
X=\underset{J}{\sum}\frac{1}{\nu(J)!}(X\cdot\varepsilon^{J})\varepsilon
_{J}=\underset{J}{\sum}\frac{1}{\nu(J)!}(X\cdot\varepsilon_{J})\varepsilon
^{J}. \label{1.13}%
\end{equation}

\subsection{Canonical Contract Products}

We now introduce two other important products on $X,Y\in%
%TCIMACRO{\dbigwedge }%
%BeginExpansion
{\displaystyle\bigwedge}
%EndExpansion
V$ \ by utilizing only the properties of the canonical scalar product.

Given $X,Y\in%
%TCIMACRO{\dbigwedge }%
%BeginExpansion
{\displaystyle\bigwedge}
%EndExpansion
V$ the \ \textit{canonical left contraction} $(X,Y)\mapsto X\lrcorner Y$ is
defined by
\begin{equation}
(X\lrcorner Y)\cdot Z:=Y\cdot(\widetilde{X}\wedge Z), \label{1.14}%
\end{equation}
for arbitrary $Z\in%
%TCIMACRO{\dbigwedge }%
%BeginExpansion
{\displaystyle\bigwedge}
%EndExpansion
V$

Given $X,Y\in%
%TCIMACRO{\dbigwedge }%
%BeginExpansion
{\displaystyle\bigwedge}
%EndExpansion
V$ the \ canonical right contraction $(X,Y)\mapsto X\lrcorner Y$ is defined
by
\begin{equation}
(X\llcorner Y)\cdot Z:=X\cdot(Z\wedge\widetilde{Y}), \label{1.15}%
\end{equation}
or arbitrary $Z\in%
%TCIMACRO{\dbigwedge }%
%BeginExpansion
{\displaystyle\bigwedge}
%EndExpansion
V$

The space $%
%TCIMACRO{\dbigwedge }%
%BeginExpansion
{\displaystyle\bigwedge}
%EndExpansion
V$ equipped with anyone of those\ canonical contractions $\lrcorner$ or
$\llcorner$ \ possess a natural structure of non associative algebras. The
triple $(%
%TCIMACRO{\dbigwedge }%
%BeginExpansion
{\displaystyle\bigwedge}
%EndExpansion
V,\lrcorner,\llcorner)$ will be called the canonical interior algebra of multiforms

.Some important properties of \ $\lrcorner$ and $\llcorner$ are:

\textbf{(i)} For all $\alpha,\beta\in\mathbb{R}$ and $X\in%
%TCIMACRO{\dbigwedge }%
%BeginExpansion
{\displaystyle\bigwedge}
%EndExpansion
V$
\begin{subequations}
\label{1.16}%
\begin{align}
\alpha\lrcorner\beta &  =\alpha\llcorner\beta=\alpha\beta,\label{1.16a}\\
\alpha\lrcorner X  &  =X\llcorner\alpha=\alpha X, \label{1.16b}%
\end{align}

\textbf{(ii)}\emph{\ }For all $X_{j}\in%
%TCIMACRO{\dbigwedge \nolimits^{j}}%
%BeginExpansion
{\displaystyle\bigwedge\nolimits^{j}}
%EndExpansion
V$ and $Y_{k}\in%
%TCIMACRO{\dbigwedge \nolimits^{k}}%
%BeginExpansion
{\displaystyle\bigwedge\nolimits^{k}}
%EndExpansion
V$%
\end{subequations}
\begin{align}
X_{j}\lrcorner Y_{k}  &  =0,\text{ if }j>k,\label{1.17a}\\
X_{j}\llcorner Y_{k}  &  =0,\text{ if }j<k. \label{1.17b}%
\end{align}

\textbf{(iii)} For all $X_{j}\in%
%TCIMACRO{\dbigwedge \nolimits^{j}}%
%BeginExpansion
{\displaystyle\bigwedge\nolimits^{j}}
%EndExpansion
V$ and $Y_{k}\in%
%TCIMACRO{\dbigwedge \nolimits^{k}}%
%BeginExpansion
{\displaystyle\bigwedge\nolimits^{k}}
%EndExpansion
V,$ with $j\leq k,$ we have that $X_{j}\lrcorner Y_{k}\in%
%TCIMACRO{\dbigwedge \nolimits^{k-j}}%
%BeginExpansion
{\displaystyle\bigwedge\nolimits^{k-j}}
%EndExpansion
V$, $Y_{k}\llcorner X_{j}\in%
%TCIMACRO{\dbigwedge \nolimits^{k-j}}%
%BeginExpansion
{\displaystyle\bigwedge\nolimits^{k-j}}
%EndExpansion
V$ and
\begin{equation}
X_{j}\lrcorner Y_{k}=(-1)^{j(k-j)}Y_{k}\llcorner X_{j}. \label{1.18}%
\end{equation}

\textbf{(iv)}\emph{\ }For all $X_{k},Y_{k}\in%
%TCIMACRO{\dbigwedge \nolimits^{k}}%
%BeginExpansion
{\displaystyle\bigwedge\nolimits^{k}}
%EndExpansion
V$%
\begin{equation}
X_{k}\lrcorner Y_{k}=X_{k}\llcorner Y_{k}=\widetilde{X_{k}}\cdot Y_{k}%
=X_{k}\cdot\widetilde{Y_{k}}. \label{1.19}%
\end{equation}

\textbf{(v)}\emph{\ }For all $a\in%
%TCIMACRO{\dbigwedge \nolimits^{1}}%
%BeginExpansion
{\displaystyle\bigwedge\nolimits^{1}}
%EndExpansion
V$ and $X\in%
%TCIMACRO{\dbigwedge }%
%BeginExpansion
{\displaystyle\bigwedge}
%EndExpansion
V$%
\begin{equation}
a\lrcorner X=-\hat{X}\llcorner a. \label{1.20}%
\end{equation}

\textbf{(vi)}\emph{\ }For all $a\in%
%TCIMACRO{\dbigwedge \nolimits^{1}}%
%BeginExpansion
{\displaystyle\bigwedge\nolimits^{1}}
%EndExpansion
V$ and $X,Y\in%
%TCIMACRO{\dbigwedge }%
%BeginExpansion
{\displaystyle\bigwedge}
%EndExpansion
V$%
\begin{equation}
a\lrcorner(X\wedge Y)=(a\lrcorner X)\wedge Y+\hat{X}\wedge(a\lrcorner Y).
\label{1.21}%
\end{equation}

\textbf{(vii)}\emph{\ }For all $X,Y,Z\in%
%TCIMACRO{\dbigwedge }%
%BeginExpansion
{\displaystyle\bigwedge}
%EndExpansion
V$%
\begin{align}
X\lrcorner(Y\lrcorner Z)  &  =(X\wedge Y)\lrcorner Z,\tag{1.22a}\\
(X\llcorner Y)\llcorner Z  &  =X\llcorner(Y\wedge Z). \tag{1.22b}%
\end{align}
\vspace*{0.3in}

\subsection{The Canonical Clifford Product}

We define now on $%
%TCIMACRO{\dbigwedge }%
%BeginExpansion
{\displaystyle\bigwedge}
%EndExpansion
V$ the \textit{canonical Clifford product} (or \textit{geometrical product})
\ utilizing the canonical contractions and the exterior product. For any
$X,Z\in%
%TCIMACRO{\dbigwedge }%
%BeginExpansion
{\displaystyle\bigwedge}
%EndExpansion
V$ the Clifford product of $X$ by $Y$ will be denoted by juxtaposition of
symbols and obeys the following axiomatic:

\textbf{(i)} For all $\alpha\in\mathbb{R}$ and $X\in%
%TCIMACRO{\dbigwedge }%
%BeginExpansion
{\displaystyle\bigwedge}
%EndExpansion
V$%
\begin{equation}
\alpha X=X\alpha\label{1.22aa}%
\end{equation}

\textbf{(ii)} For all $a\in%
%TCIMACRO{\dbigwedge \nolimits^{1}}%
%BeginExpansion
{\displaystyle\bigwedge\nolimits^{1}}
%EndExpansion
V$ an any $X\in%
%TCIMACRO{\dbigwedge }%
%BeginExpansion
{\displaystyle\bigwedge}
%EndExpansion
V$
\begin{subequations}
\label{1.22a}%
\begin{align}
aX  &  =a\lrcorner X+a\wedge X,\label{1.22aaa}\\
Xa  &  =X\llcorner a+X\wedge a. \label{1.22aab}%
\end{align}

\textbf{(iii)} For all $X,Y,Z\in%
%TCIMACRO{\dbigwedge }%
%BeginExpansion
{\displaystyle\bigwedge}
%EndExpansion
V$%
\end{subequations}
\begin{equation}
X(YZ)=(XY)Z. \label{1.22aac}%
\end{equation}

The canonical Clifford product is distributive and associative (precisely
axiom \textbf{(iii)}). Its distributive law follows directly from the
corresponding distributive laws of the canonical contractions \ and of the
exterior product.

The linear space $%
%TCIMACRO{\dbigwedge }%
%BeginExpansion
{\displaystyle\bigwedge}
%EndExpansion
V$ equipped with the canonical Clifford product is a (real) associative
algebra. It will be called here the canonical Clifford algebra of multiforms
\emph{and denoted by} $\mathcal{C}\ell(V,\cdot).$

We list now the main properties of the canonical Clifford product.

\textbf{(i) }For all $X,Y\in%
%TCIMACRO{\dbigwedge }%
%BeginExpansion
{\displaystyle\bigwedge}
%EndExpansion
V$
\begin{subequations}
\label{1.23}%
\begin{align}
(XY)\symbol{94}  &  =\hat{X}\text{ }\hat{Y},\label{1.23a}\\
\widetilde{XY}  &  =\widetilde{Y}\widetilde{X}. \label{1.23b}%
\end{align}

\textbf{(ii)} For all $a\in%
%TCIMACRO{\dbigwedge \nolimits^{1}}%
%BeginExpansion
{\displaystyle\bigwedge\nolimits^{1}}
%EndExpansion
V$ and $X\in%
%TCIMACRO{\dbigwedge }%
%BeginExpansion
{\displaystyle\bigwedge}
%EndExpansion
V$
\end{subequations}
\begin{subequations}
\label{1.24}%
\begin{align}
a\lrcorner X  &  =\frac{1}{2}(aX-\hat{X}\text{ }a).\label{1.24a}\\
a\wedge X  &  =\frac{1}{2}(aX+\hat{X}a). \label{1.24b}%
\end{align}

\textbf{(iii)} For all $X,Y\in%
%TCIMACRO{\dbigwedge }%
%BeginExpansion
{\displaystyle\bigwedge}
%EndExpansion
V$%
\end{subequations}
\begin{equation}
X\cdot Y=\left\langle \widetilde{X}Y\right\rangle _{0}=\left\langle
X\widetilde{Y}\right\rangle _{0}. \label{1.25}%
\end{equation}

\textbf{(iv)} Let $\tau\in%
%TCIMACRO{\dbigwedge \nolimits^{n}}%
%BeginExpansion
{\displaystyle\bigwedge\nolimits^{n}}
%EndExpansion
V,$ $a\in%
%TCIMACRO{\dbigwedge \nolimits^{1}}%
%BeginExpansion
{\displaystyle\bigwedge\nolimits^{1}}
%EndExpansion
V$ and $X\in%
%TCIMACRO{\dbigwedge }%
%BeginExpansion
{\displaystyle\bigwedge}
%EndExpansion
V$, then
\begin{equation}
\tau(a\wedge X)=(-1)^{n-1}a\lrcorner(\tau X). \label{1.26}%
\end{equation}
This notable formula is sometimes referred in the mathematical literature as
the \textit{duality identity.}

\subsection{Extensors}

\subsubsection{The Space $extV$}

A linear mapping between two arbitrary parts\footnote{A direct sum of any
vector spaces $%
%TCIMACRO{\dbigwedge \nolimits^{k}}%
%BeginExpansion
{\displaystyle\bigwedge\nolimits^{k}}
%EndExpansion
V$\textbf{\ } is clearly a subspace of $%
%TCIMACRO{\dbigwedge }%
%BeginExpansion
{\displaystyle\bigwedge}
%EndExpansion
V$ and is said to be a part of\emph{ } $%
%TCIMACRO{\dbigwedge }%
%BeginExpansion
{\displaystyle\bigwedge}
%EndExpansion
V.$ A convenenient notation for a part of $%
%TCIMACRO{\dbigwedge }%
%BeginExpansion
{\displaystyle\bigwedge}
%EndExpansion
V$ is $%
%TCIMACRO{\dbigwedge \nolimits^{\diamond}}%
%BeginExpansion
{\displaystyle\bigwedge\nolimits^{\diamond}}
%EndExpansion
V$.} $%
%TCIMACRO{\dbigwedge \nolimits_{1}^{\diamond}}%
%BeginExpansion
{\displaystyle\bigwedge\nolimits_{1}^{\diamond}}
%EndExpansion
V,%
%TCIMACRO{\dbigwedge \nolimits_{2}^{\diamond}}%
%BeginExpansion
{\displaystyle\bigwedge\nolimits_{2}^{\diamond}}
%EndExpansion
V$ of $%
%TCIMACRO{\dbigwedge }%
%BeginExpansion
{\displaystyle\bigwedge}
%EndExpansion
V$, i.e., $t:%
%TCIMACRO{\dbigwedge \nolimits_{1}^{\diamond}}%
%BeginExpansion
{\displaystyle\bigwedge\nolimits_{1}^{\diamond}}
%EndExpansion
V\rightarrow%
%TCIMACRO{\dbigwedge \nolimits_{2}^{\diamond}}%
%BeginExpansion
{\displaystyle\bigwedge\nolimits_{2}^{\diamond}}
%EndExpansion
V$ such that for all $\alpha,\alpha^{\prime}\in\mathbb{R}$ and $X,X^{\prime
}\in%
%TCIMACRO{\dbigwedge \nolimits_{1}^{\diamond}}%
%BeginExpansion
{\displaystyle\bigwedge\nolimits_{1}^{\diamond}}
%EndExpansion
V$ we have%
\begin{equation}
t(\alpha X+\alpha^{\prime}X^{\prime})=\alpha t(X)+\alpha^{\prime}t(X^{\prime
}),
\end{equation}
will be called an \textit{extensor over} $V$. The set of all extensor over $V$
whose domain and codomain are $%
%TCIMACRO{\dbigwedge \nolimits_{1}^{\diamond}}%
%BeginExpansion
{\displaystyle\bigwedge\nolimits_{1}^{\diamond}}
%EndExpansion
V$ and $%
%TCIMACRO{\dbigwedge \nolimits_{2}^{\diamond}}%
%BeginExpansion
{\displaystyle\bigwedge\nolimits_{2}^{\diamond}}
%EndExpansion
V$ possess a natural structure of vector space over\emph{ }$\mathbb{R}%
$\emph{.}

When $%
%TCIMACRO{\dbigwedge \nolimits_{1}^{\diamond}}%
%BeginExpansion
{\displaystyle\bigwedge\nolimits_{1}^{\diamond}}
%EndExpansion
V$ $=%
%TCIMACRO{\dbigwedge \nolimits_{2}^{\diamond}}%
%BeginExpansion
{\displaystyle\bigwedge\nolimits_{2}^{\diamond}}
%EndExpansion
V=%
%TCIMACRO{\dbigwedge }%
%BeginExpansion
{\displaystyle\bigwedge}
%EndExpansion
V$ the space of extensors $t:%
%TCIMACRO{\dbigwedge }%
%BeginExpansion
{\displaystyle\bigwedge}
%EndExpansion
V\rightarrow%
%TCIMACRO{\dbigwedge }%
%BeginExpansion
{\displaystyle\bigwedge}
%EndExpansion
V$ will be denoted by $extV$.

\subsubsection{\noindent{\protect\Large \ }The Space $(p,q)$-$extV$ of the
$(p,q)$-Extensors}

Let $p,q$ two integers with $0\leq p,q\leq n.$ An extensor with domain $%
%TCIMACRO{\dbigwedge \nolimits^{p}}%
%BeginExpansion
{\displaystyle\bigwedge\nolimits^{p}}
%EndExpansion
V$ and codomain $%
%TCIMACRO{\dbigwedge \nolimits^{q}}%
%BeginExpansion
{\displaystyle\bigwedge\nolimits^{q}}
%EndExpansion
V$ is said to be a $(p,q)$\emph{-extensor over }$V$\emph{. }The real vector
space of the $(p,q)$-extensors will be denoted $(p,q)$-$extV$. If
$\dim(\mathbf{V})=n$, then $\dim((p,q)$-$extV)=\binom{n}{p}\binom{n}{q}$.

\subsubsection{The Adjoint Operator}

Let $(\{\varepsilon^{j}\},\{\varepsilon_{j}\})$ be a pair of arbitrary
reciprocal basis for $V$ (i.e., $\varepsilon^{j}\cdot\varepsilon_{i}%
=\delta_{i}^{j}$). The linear operator acting on the space $(p,q)$-$extV\ni
t\mapsto t^{\dagger}\in(q,p)$-$extV$ such that
\begin{subequations}
\label{1.27}%
\begin{align}
t^{\dagger}(X)  &  =\frac{1}{p!}(t(\varepsilon^{j_{1}}\wedge\ldots
\wedge\varepsilon^{j_{p}})\cdot X)\varepsilon_{j_{1}}\wedge\ldots
\wedge\varepsilon_{j_{p}}\label{1.27a}\\
&  =\frac{1}{p!}(t(\varepsilon_{j_{1}}\wedge\ldots\wedge\varepsilon_{j_{p}%
})\cdot X)\varepsilon^{j_{1}}\wedge\ldots\wedge\varepsilon^{j_{p}},
\label{1.27b}%
\end{align}
is called the \textit{adjoint operator} and $t^{\dagger}$ \textit{is read as
the adjoint of}\emph{ }$t.$

The adjoint operator is well defined since the sums appearing in
Eqs.(\ref{1.27a}),(\ref{1.27b}) do not depend on the choice of the pair of
reciprocal basis $(\{\varepsilon^{j}\},\{\varepsilon_{j}\}).$

We give now some important properties of the adjoint operator.

\textbf{(i)} The adjoint operator is involutive, i.e.,
\end{subequations}
\begin{equation}
(t^{\dagger})^{\dagger}=t. \label{1.28}%
\end{equation}

\textbf{(ii)}\emph{\ }Let $t$ $\in$ $(p,q)$-$extV$\textbf{. } Then for all
$X\in%
%TCIMACRO{\dbigwedge \nolimits^{p}}%
%BeginExpansion
{\displaystyle\bigwedge\nolimits^{p}}
%EndExpansion
V$, $Y\in%
%TCIMACRO{\dbigwedge \nolimits^{q}}%
%BeginExpansion
{\displaystyle\bigwedge\nolimits^{q}}
%EndExpansion
V$ we have that
\begin{equation}
t(X)\cdot Y=X\cdot t^{\dagger}(Y). \label{1.29}%
\end{equation}

\textbf{(iii)}\emph{\ }Let $t\in(q,r)$-$extV$ and $u\in(p,q)$-$extV$. The
extensor $t\circ u$ (composition of $u$ with $t$), which clearly belongs to
$(p,r)$-$extV,$ satisfies
\begin{equation}
(t\circ u)^{\dagger}=u^{\dagger}\circ t^{\dagger}. \label{1.30}%
\end{equation}

\subsubsection{$(1,1)$-Extensors}

\paragraph{Symmetric and Antisymmetric parts of $(1,1)$-Extensors}

An extensor $t\in(1,1)$-$extV$ is said to be \textit{adjoint symmetrical}
\ (respectively \textit{adjoint antisymmetric}) iff $t=t^{\dagger}$
(respectively $t=-t^{\dagger}$).

The following result is important. For any $t\in(1,1)$-$extV$, there exist
exactly two $(1,1)$-extensors over $V$\textbf{,} say $t_{+}$ and $t_{-}$, such
that $t_{+}$ is adjoint symmetric (i.e., $t_{+}=t_{+}^{\dagger}$) and $t_{-}$
is adjoint antisymmetric (i.e., $t_{-}=-t_{-}^{\dagger}$) and the following
decomposition is valid:
\begin{equation}
t(a)=t_{+}(a)+t_{-}(a). \label{1.31}%
\end{equation}
Those $(1,1)$-extensors are given by
\begin{equation}
t_{\pm}(a)=\frac{1}{2}(t(a)\pm t^{\dagger}(a)). \label{1.32}%
\end{equation}

The extensors$t_{+}$ and $t_{-}$ are called the adjoint symmetric and the
adjoint antisymmetric parts \emph{of }$t$.

\paragraph{The extension of $(1,1)$-Extensors}

Let $(\{\varepsilon^{j}\},\{\varepsilon_{j}\})$ be an arbitrary pair of
reciprocal basis for$V$ (i.e., $\varepsilon^{j}\cdot\varepsilon_{i}=\delta
_{i}^{j}$) and let $t\in(1,1)$-$extV$. The \textit{extension operator} is the
linear mapping%
\begin{align}
\underset{-}{}  &  :(1,1)\text{-}extV\rightarrow extV,\nonumber\\
t  &  \mapsto\underline{t}\text{,} \label{extensao}%
\end{align}
such that
\begin{align}
\underline{t}(X)  &  =1\cdot X+\overset{n}{\underset{k=1}{\sum}}\frac{1}%
{k!}((\varepsilon^{j_{1}}\wedge\ldots\wedge\varepsilon^{j_{k}})\cdot
X)t(\varepsilon_{j_{1}})\wedge\ldots\wedge t(\varepsilon_{j_{k}}%
)\label{1.33a}\\
&  =1\cdot X+\overset{n}{\underset{k=1}{\sum}}\frac{1}{k!}((\varepsilon
_{j_{1}}\wedge\ldots\wedge\varepsilon_{j_{k}})\cdot X)t(\varepsilon^{j_{1}%
})\wedge\ldots\wedge t(\varepsilon^{j_{k}}), \label{1.33b}%
\end{align}
In what follows we say that $\underline{t}$ is the \emph{extension of }$t$
\footnote{Some authors call $\underline{t}$ the exterior algebra extension of
$t$.}.

The extension operator is well defined since the sums appearing in
Eqs.(\ref{1.33a}) and \ref{1.33b} do not depend on the pair $(\{\varepsilon
^{j}\},\{\varepsilon_{j}\})$. Moreover the extension operator preserves the
graduation, i.e., if $X\in%
%TCIMACRO{\dbigwedge }%
%BeginExpansion
{\displaystyle\bigwedge}
%EndExpansion
V$, then $\underline{t}(X)\in%
%TCIMACRO{\dbigwedge \nolimits^{k}}%
%BeginExpansion
{\displaystyle\bigwedge\nolimits^{k}}
%EndExpansion
V$.

We\ present now some important properties of the $(1,1)$-extensors

\textbf{(i) }Let $t\in(1,1)$-$extV$, for all $\alpha\in\mathbb{R},$ $v\in%
%TCIMACRO{\dbigwedge }%
%BeginExpansion
{\displaystyle\bigwedge}
%EndExpansion
V$ and $v_{1},\ldots,v_{k}\in%
%TCIMACRO{\dbigwedge }%
%BeginExpansion
{\displaystyle\bigwedge}
%EndExpansion
V$ we have that
\begin{subequations}
\label{1.43}%
\begin{align}
\underline{t}(\alpha)  &  =\alpha,\label{1.43a}\\
\underline{t}(v)  &  =t(v),\label{1.43b}\\
\underline{t}(v_{1}\wedge\ldots\wedge v_{k})  &  =t(v_{1})\wedge\ldots\wedge
t(v_{k}). \label{1.43c}%
\end{align}
The last property possess an immediate corollary,
\end{subequations}
\begin{equation}
\underline{t}(X\wedge Y)=\underline{t}(X)\wedge\underline{t}(Y), \label{1.35}%
\end{equation}
for all $X,Y\in%
%TCIMACRO{\dbigwedge }%
%BeginExpansion
{\displaystyle\bigwedge}
%EndExpansion
V.$

\textbf{(ii)} For all $t,u\in(1,1)$-$extV$ the following identity holds;
\begin{equation}
\underline{t\circ u}=\underline{t}\circ\underline{u}. \label{1.36}%
\end{equation}

\textbf{(iii)} Let $t\in(1,1)$-$extV$ with inverse $t^{-1}\in(1,1)$-$extV$
(i.e., $t\circ t^{-1}=t^{-1}\circ t=i_{d},$ where $i_{d}\in(1,1)$-$extV$ is
the identity extensor, then%
\begin{equation}
(\underline{t})^{-1}=\underline{(t^{-1})}, \label{1.48}%
\end{equation}
and we shall use the symbol $\underline{t}^{-1}$ to denote both extensors
$(\underline{t})^{-1}$ and $\underline{(t^{-1})}.$

\textbf{(iv)} For any $(1,1)$-extensor over $V\mathbf{,}$ the extension
operator commutes with the adjoint operator, i.e.,
\begin{equation}
\underline{(t^{\dagger})}=(\underline{t})^{\dagger}, \label{1.39}%
\end{equation}

and we shall use the symbol $\underline{t}^{\dagger}$ to denote both extensors
$\underline{(t^{\dagger})}$ and $(\underline{t})^{\dagger}$.

\textbf{(v)} Ley $t\in(1,1)$-$extV$, for all $X,Y\in%
%TCIMACRO{\dbigwedge }%
%BeginExpansion
{\displaystyle\bigwedge}
%EndExpansion
V$ we have
\begin{equation}
X\lrcorner\underline{t}(Y)=\underline{t}(\underline{t}^{\dagger}(X)\lrcorner
Y). \label{1.40}%
\end{equation}

\paragraph{The Characteristic Scalars $tr[t]$ and $\det[t]$}

Let $(\{\varepsilon^{j}\},\{\varepsilon_{j}\})$ \ be an arbitrary pair of
reciprocal basis for $V$ (i.e., $\varepsilon^{j}\cdot\varepsilon_{i}%
=\delta_{i}^{j}$).The \textit{trace mapping}
\begin{align}
\mathrm{tr}  &  :(1,1)\text{-}extV\rightarrow\mathbb{R},\nonumber\\
t  &  \mapsto tr[t]
\end{align}
is such that
\begin{equation}
\mathrm{tr}[t]=t(\varepsilon^{j})\cdot\varepsilon_{j}=t(\varepsilon_{j}%
)\cdot\varepsilon^{j}. \label{1.41}%
\end{equation}

Observe that $\mathrm{tr}$ does not depend on the pair $(\{\varepsilon
^{j}\},\{\varepsilon_{j}\})$ used in Eq.(\ref{1.41}). We have immediately that
\ for any $t\in(1,1)$-$extV$,%
\begin{equation}
\mathrm{tr}[t^{\dagger}]=\mathrm{tr}[t]. \tag{1.42}%
\end{equation}

The \textit{determinant mapping}
\begin{align*}
\det[t]  &  :(1,1)\text{-}extV\rightarrow\mathbb{R},\\
t  &  \mapsto\det[t],
\end{align*}
is such that
\begin{subequations}
\begin{align}
\det[t]  &  :=\frac{1}{n!}\underline{t}(\varepsilon^{j_{1}}\wedge\ldots
\wedge\varepsilon^{j_{n}})\cdot(\varepsilon_{j_{1}}\wedge\ldots\wedge
\varepsilon_{j_{n}})\label{1.43aa}\\
&  =\frac{1}{n!}\underline{t}(\varepsilon_{j_{1}}\wedge\ldots\wedge
\varepsilon_{j_{n}})\cdot(\varepsilon^{j_{1}}\wedge\ldots\wedge\varepsilon
^{j_{n}}). \label{1.43ab}%
\end{align}

Like the trace the determinant also does not depend on the pair of arbitrary
reciprocal basis $(\{\varepsilon^{j}\},\{\varepsilon_{j}\})$.

By using the combinatorial formulas $v^{j_{1}}\wedge\ldots\wedge v^{j_{n}%
}=\epsilon^{j_{1}\ldots j_{n}}v^{1}\wedge\ldots\wedge v^{n}$ and $v_{j_{1}%
}\wedge\ldots\wedge v_{j_{n}}=\epsilon_{j_{1}\ldots j_{n}}v_{1}\wedge
\ldots\wedge v_{n},$ where $\epsilon^{j_{1}\ldots j_{n}}$ and $\epsilon
_{j_{1}\ldots j_{n}}$ are symbols of permutation\footnote{The permutation
symbols of order $n$ are defined by
\par
$\epsilon^{j_{1}\ldots j_{n}}=\epsilon_{j_{1}\ldots j_{n}}=\left\{
\begin{array}
[c]{cc}%
1, & \text{if }j_{1}\ldots j_{n}\text{ is an even permutation of }1,\ldots,n\\
-1, & \text{if }j_{1}\ldots j_{n}\text{ is an odd permutation of }1,\ldots,n\\
0, & \text{in all other cases}%
\end{array}
\right.  .$
\par
Moreover, recall that $\epsilon^{j_{1}\ldots j_{n}}\epsilon_{j_{1}\ldots
j_{n}}=n!$} of order $n$ and $v^{1},\ldots,v^{n}$ and $v_{1},\ldots,v_{n}$ are
$1$-forms, \ we can write formulas more convenient for $\det t$. Indeed,we
have:
\end{subequations}
\begin{subequations}
\label{1.44}%
\begin{align}
\det[t]  &  =\underline{t}(\varepsilon^{1}\wedge\ldots\wedge\varepsilon
^{n})\cdot(\varepsilon_{1}\wedge\ldots\wedge\varepsilon_{n})=\underline{t}%
(\overline{\varepsilon})\cdot\underline{\varepsilon}\label{1.44a}\\
&  =\underline{t}(\varepsilon_{1}\wedge\ldots\wedge\varepsilon_{n}%
)\cdot(\varepsilon^{1}\wedge\ldots\wedge\varepsilon^{n})=\underline{t}%
(\underline{\varepsilon})\cdot\overline{\varepsilon}, \label{1.44b}%
\end{align}
where we used the short notations: $\overline{\varepsilon}=\varepsilon
^{1}\wedge\ldots\wedge\varepsilon^{n}$ and $\underline{\varepsilon
}=\varepsilon_{1}\wedge\ldots\wedge\varepsilon_{n}$.

The mapping $\det$\ possess the following important properties.

\textbf{(i)} For all $t\in(1,1)$-$extV,$%
\end{subequations}
\begin{equation}
\det[t^{\dagger}]=\det[t]. \label{1.45}%
\end{equation}

\textbf{(ii)} Let $t\in(1,1)$-$extV$, for any $\tau\in%
%TCIMACRO{\dbigwedge }%
%BeginExpansion
{\displaystyle\bigwedge}
%EndExpansion
V$ \ we have
\begin{equation}
\underline{t}(\tau)=\det[t]\tau. \label{1.46}%
\end{equation}

\textbf{(iii)} For all $t,u\in(1,1)$-$extV,$
\begin{equation}
\det[t\circ u]=\det[t]\circ\det[u]. \label{1.57}%
\end{equation}

\textbf{(iv)} Let $t\in(1,1)$-$extV$ with inverse $t^{-1}\in(1,1)$-$extV$
(i.e., $t\circ t^{-1}=t^{-1}\circ t=i_{d},$ where $i_{d}\in(1,1)$-$extV$ is
the identity extensor ), then:
\begin{equation}
\det[t^{-1}]=(\det[t])^{-1}. \label{1.48a}%
\end{equation}

We shall use the notation$\left.  \det\right.  ^{-1}[t]$ to denote both
$\det[t^{-1}]$ and $(\det[t])^{-1}$.

\textbf{(v)} If $t\in(1,1)$-$extV$ is non degenerated (i.e., $\det[t]\neq0$),
then its inverse $t^{-1}\in(1,1)$-$extV$exists and is given by
\begin{equation}
t^{-1}(a)=\left.  \det\right.  ^{-1}[t]\underline{t}^{\dagger}(a\tau)\tau
^{-1}, \label{1.49}%
\end{equation}
where $\tau\in%
%TCIMACRO{\dbigwedge ^{n}}%
%BeginExpansion
{\displaystyle\bigwedge^{n}}
%EndExpansion
V$, $\tau\neq0$.

\paragraph{The Characteristic Biform\ Mapping $\mathrm{bif}$}

Let $(\{\varepsilon^{j}\},\{\varepsilon_{j}\})$ be an arbitrary pair of
reciprocal basis for $V$ (i.e., $\varepsilon^{j}\cdot\varepsilon_{i}%
=\delta_{i}^{j}$). The \textit{biform mapping}%
\[
\mathrm{bif}:(1,1)\text{-}extV\rightarrow%
%TCIMACRO{\dbigwedge \nolimits^{2}}%
%BeginExpansion
{\displaystyle\bigwedge\nolimits^{2}}
%EndExpansion
V,
\]
$\ $ is such that
\begin{equation}
\mathrm{bif}[t]=t(\varepsilon^{j})\wedge\varepsilon_{j}=t(\varepsilon
_{j})\wedge\varepsilon^{j}. \label{1.50}%
\end{equation}

Note that $\mathrm{bif}[t]$ is a $2$-form characteristic of the extensor $t$
(since it does not depend on the pair of reciprocal basis) $(\{\varepsilon
^{j}\},\{\varepsilon_{j}\})$, and $\mathrm{bif}[t]$ is read as the\textit{
biform} of $t$.

We give now some important properties of \textrm{bif}..

\textbf{(i)} Let $t\in(1,1)$-$extV$, then
\begin{equation}
\mathrm{bif}[t^{\dagger}]=-\mathrm{bif}[t]. \label{1.51}%
\end{equation}

\textbf{(ii)} The antisymmetric adjoint of any $t\in(1,1)$-$extV$ may be
factored by the following formula
\begin{equation}
t_{-}(a)=\frac{1}{2}\mathrm{bif}[t]\times a, \label{1.52}%
\end{equation}
where $\times$ means here the canonical commutator defined for any $A,B\in%
%TCIMACRO{\dbigwedge }%
%BeginExpansion
{\displaystyle\bigwedge}
%EndExpansion
V$ by%
\begin{equation}
A\times B=\frac{1}{2}(AB-BA). \label{comut}%
\end{equation}

\subsubsection{The Generalization Operator of $(1,1)$-Extensors}

Let $(\{\varepsilon^{j}\},\{\varepsilon_{j}\})$ be an arbitrary pair of
reciprocal basis for $V$ (i.e., $\varepsilon^{j}\cdot\varepsilon_{i}%
=\delta_{i}^{j}$).The linear operator whose domain is the space $(1,1)$-$extV$
and whose codomain is the space $extV$, given by%
\begin{align}
(1,1)\text{-}extV  &  \ni t\mapsto T\in extV,\label{1.53}\\
T(X)  &  =t(\varepsilon^{j})\wedge(\varepsilon_{j}\lrcorner X)=t(\varepsilon
_{j})\wedge(\varepsilon^{j}\lrcorner X),\nonumber
\end{align}
is called the \textit{generalization operator}, and $T$ is read as the
\textit{generalized of}\emph{ }$t.$

The generalization operator is well defined since the sums in Eq.(\ref{1.53})
do not depend on $(\{\varepsilon^{j}\},\{\varepsilon_{j}\})$. The
generalization operators preserves the grade, i.e., if $X\in%
%TCIMACRO{\dbigwedge \nolimits^{k}}%
%BeginExpansion
{\displaystyle\bigwedge\nolimits^{k}}
%EndExpansion
V$, then $T(X)\in%
%TCIMACRO{\dbigwedge \nolimits^{k}}%
%BeginExpansion
{\displaystyle\bigwedge\nolimits^{k}}
%EndExpansion
V$.

We list some important properties of the generalization of a $(1,1)$-extensor
$t$:

\textbf{(i)} For all $\alpha\in\mathbb{R}$ and $v\in%
%TCIMACRO{\dbigwedge \nolimits^{1}}%
%BeginExpansion
{\displaystyle\bigwedge\nolimits^{1}}
%EndExpansion
V$,
\begin{subequations}
\label{1.54}%
\begin{align}
T(\alpha)  &  =0,\label{1.54a}\\
T(v)  &  =t(v). \label{1.54b}%
\end{align}

\textbf{(ii)} For all $X,Y\in%
%TCIMACRO{\dbigwedge }%
%BeginExpansion
{\displaystyle\bigwedge}
%EndExpansion
V$ it is:
\end{subequations}
\begin{equation}
T(X\wedge Y)=T(X)\wedge Y+X\wedge T(Y) \label{1.55}%
\end{equation}

\textbf{(iii)} The generalization operator commutes with\ the adjoint
operator. Then $T^{\dagger}$ denotes both the adjoint of the generalized and
the generalized of the\emph{ adjoint.}

\textbf{(iv)} The antisymmetric part of the adjoint of the generalized
coincides with the generalized of the adjoint antisymmetric part of $t$ and
can then be factored as
\begin{equation}
T_{-}(X)=\frac{1}{2}\mathrm{bif}[t]\times X, \label{1.56}%
\end{equation}
for all $X\in%
%TCIMACRO{\dbigwedge }%
%BeginExpansion
{\displaystyle\bigwedge}
%EndExpansion
V$.

\paragraph{Normal $(1,1)$-Extensors}

A extensor$\ t\in(1,1)$-$extV$ is said to be \textit{normal} if its adjoint
symmetric and antisymmetric parts commutes, i.e., for any $a\in V$,%

\begin{equation}
t_{+}(t_{-}(a))=t_{-}(t_{+}(a)) \label{1.n1}%
\end{equation}

We can show without difficulties \cite{hesob84} that Eq.(\ref{1.n1}) is
equivalent to any one of the following two equations
\begin{align}
\underline{t}(\underline{t}^{\dagger}(A))  &  =\underline{t}^{\dagger
}(\underline{t}(A)),\nonumber\\
\langle\underline{t}(A)\underline{t}(B)\rangle_{0}  &  =\langle\underline{t}%
^{\dagger}(A)\underline{t}^{\dagger}(B)\rangle_{0}, \label{1.n2}%
\end{align}
for any $A,B\in%
%TCIMACRO{\dbigwedge }%
%BeginExpansion
{\displaystyle\bigwedge}
%EndExpansion
V$

\subsection{The Metric Clifford Algebra $C\ell(V,%
%TCIMACRO{\TeXButton{itg}{\itg}}%
%BeginExpansion
\itg
%EndExpansion
)$}

A $(1,1)$-extensor over $V$, say $%
%TCIMACRO{\TeXButton{itg}{\itg}}%
%BeginExpansion
\itg
%EndExpansion
$ which is symmetric adjoint (i.e., $%
%TCIMACRO{\TeXButton{itg}{\itg}}%
%BeginExpansion
\itg
%EndExpansion
=%
%TCIMACRO{\TeXButton{itg}{\itg}}%
%BeginExpansion
\itg
%EndExpansion
^{\dagger}$) and non degenerated (i.e., $\det%
%TCIMACRO{\TeXButton{itg}{\itg}}%
%BeginExpansion
\itg
%EndExpansion
\neq0$) is said to be a metric extensor on $V$. Under these conditions $%
%TCIMACRO{\TeXButton{itg}{\itg}}%
%BeginExpansion
\itg
%EndExpansion
^{-1}$ exists and is also a metric extensor on $V$, and as we are going to see
both $%
%TCIMACRO{\TeXButton{itg}{\itg}}%
%BeginExpansion
\itg
%EndExpansion
$ and $%
%TCIMACRO{\TeXButton{itg}{\itg}}%
%BeginExpansion
\itg
%EndExpansion
^{-1}$ have equally important roles in our theory. The reason is that we are
going to use $%
%TCIMACRO{\TeXButton{itg}{\itg}}%
%BeginExpansion
\itg
%EndExpansion
$ to represent the metric tensor $%
%TCIMACRO{\TeXButton{slg}{\slg}}%
%BeginExpansion
\slg
%EndExpansion
\in T_{2}^{0}\mathbf{V}$\textbf{\ }whereas $%
%TCIMACRO{\TeXButton{itg}{\itg}}%
%BeginExpansion
\itg
%EndExpansion
^{-1}$ will represent \texttt{g}$\in T_{0}^{2}\mathbf{V}$. This is done as
follows. Given an arbitrary basis $\{\mathbf{e}_{i}\}$ of $\mathbf{V}$ and the
corresponding dual basis $\{\varepsilon^{i}\}$ de $V$ and the reciprocal basis
of $\{\varepsilon^{i}\}$, i.e., $\varepsilon_{i}\cdot\varepsilon^{j}%
=\delta_{i}^{j}$, if $%
%TCIMACRO{\TeXButton{slg}{\slg}}%
%BeginExpansion
\slg
%EndExpansion
(\mathbf{e}_{i},\mathbf{e}_{j})=g_{ij}$ and \texttt{g}$(\varepsilon
^{i},\varepsilon^{j})=g^{ij}$ with $g_{ij}g^{jk}=\delta_{i}^{k}$ then:%

\begin{align}%
%TCIMACRO{\TeXButton{itg}{\itg}}%
%BeginExpansion
\itg
%EndExpansion
^{-1}(\varepsilon^{i})\cdot\varepsilon^{j}  &  =g^{ij},\nonumber\\%
%TCIMACRO{\TeXButton{itg}{\itg}}%
%BeginExpansion
\itg
%EndExpansion
(\varepsilon_{i})\cdot\varepsilon_{j}  &  =g_{ij}. \label{g and g}%
\end{align}
\ 

\subsubsection{The Metric Scalar Products}

Given a metric extensor $%
%TCIMACRO{\TeXButton{itg}{\itg}}%
%BeginExpansion
\itg
%EndExpansion
\in(1,1)$-$extV$ we may define two new scalar products on $%
%TCIMACRO{\dbigwedge }%
%BeginExpansion
{\displaystyle\bigwedge}
%EndExpansion
V$, respectively $\underset{%
%TCIMACRO{\TeXButton{sig}{\sitg}}%
%BeginExpansion
\sitg
%EndExpansion
}{\cdot}$ and $\underset{%
%TCIMACRO{\TeXButton{sig}{\sitg}}%
%BeginExpansion
\sitg
%EndExpansion
^{-1}}{\cdot}$ such that:%
\begin{align}
X\underset{%
%TCIMACRO{\TeXButton{sig}{\sitg}}%
%BeginExpansion
\sitg
%EndExpansion
}{\cdot}Y  &  =\underline{%
%TCIMACRO{\TeXButton{itg}{\itg}}%
%BeginExpansion
\itg
%EndExpansion
}(X)\cdot Y,\label{1.57a}\\
X\underset{%
%TCIMACRO{\TeXButton{sig}{\sitg}}%
%BeginExpansion
\sitg
%EndExpansion
^{-1}}{\cdot}Y  &  =\underline{%
%TCIMACRO{\TeXButton{itg}{\itg}}%
%BeginExpansion
\itg
%EndExpansion
}^{-1}(X)\cdot Y \label{1.57b}%
\end{align}
Both $\underset{%
%TCIMACRO{\TeXButton{sig}{\sitg}}%
%BeginExpansion
\sitg
%EndExpansion
}{\cdot}$ and $\underset{%
%TCIMACRO{\TeXButton{sig}{\sitg}}%
%BeginExpansion
\sitg
%EndExpansion
^{-1}}{\cdot}$ will be called a \textit{metric scalar product.}

The mappings $\underset{%
%TCIMACRO{\TeXButton{sig}{\sitg}}%
%BeginExpansion
\sitg
%EndExpansion
}{\cdot}$ and $\underset{%
%TCIMACRO{\TeXButton{sig}{\sitg}}%
%BeginExpansion
\sitg
%EndExpansion
^{-1}}{\cdot}$ are indeed well defined scalar products on $%
%TCIMACRO{\dbigwedge }%
%BeginExpansion
{\displaystyle\bigwedge}
%EndExpansion
V$. The scalar product $\underset{%
%TCIMACRO{\TeXButton{sig}{\sitg}}%
%BeginExpansion
\sitg
%EndExpansion
}{\cdot}$ (and obviously also $\underset{%
%TCIMACRO{\TeXButton{sig}{\sitg}}%
%BeginExpansion
\sitg
%EndExpansion
^{-1}}{\cdot}$ ) is symmetric (i.e., $X\underset{%
%TCIMACRO{\TeXButton{sig}{\sitg}}%
%BeginExpansion
\sitg
%EndExpansion
}{\cdot}Y=Y\underset{%
%TCIMACRO{\TeXButton{sig}{\sitg}}%
%BeginExpansion
\sitg
%EndExpansion
}{\cdot}X$), is linear (i.e., $X\underset{%
%TCIMACRO{\TeXButton{sig}{\sitg}}%
%BeginExpansion
\sitg
%EndExpansion
}{\cdot}(Y+Z)=X\underset{%
%TCIMACRO{\TeXButton{sig}{\sitg}}%
%BeginExpansion
\sitg
%EndExpansion
}{\cdot}Y+X\underset{%
%TCIMACRO{\TeXButton{sig}{\sitg}}%
%BeginExpansion
\sitg
%EndExpansion
}{\cdot}Z$), possess mist associativity (i.e., $(\alpha X)\underset{%
%TCIMACRO{\TeXButton{sig}{\sitg}}%
%BeginExpansion
\sitg
%EndExpansion
}{\cdot}Y=X\underset{%
%TCIMACRO{\TeXButton{sig}{\sitg}}%
%BeginExpansion
\sitg
%EndExpansion
}{\cdot}(\alpha Y)=\alpha(X\underset{%
%TCIMACRO{\TeXButton{sig}{\sitg}}%
%BeginExpansion
\sitg
%EndExpansion
}{\cdot}Y$ $\forall a\in\mathbb{R})$) and is non degenerated (i.e.,
$X\underset{%
%TCIMACRO{\TeXButton{sig}{\sitg}}%
%BeginExpansion
\sitg
%EndExpansion
}{\cdot}Y=0$ for all $X,$ implies $Y=0$). In what follows we list the main
properties of $\underset{%
%TCIMACRO{\TeXButton{sig}{\sitg}}%
%BeginExpansion
\sitg
%EndExpansion
}{\cdot}$, since $\underset{%
%TCIMACRO{\TeXButton{sig}{\sitg}}%
%BeginExpansion
\sitg
%EndExpansion
^{-1}}{\cdot}$ obey similar properties.

\textbf{(i)} For $\alpha,\beta\in\mathbb{R}$
\begin{equation}
\alpha\underset{%
%TCIMACRO{\TeXButton{sig}{\sitg}}%
%BeginExpansion
\sitg
%EndExpansion
}{\cdot}\beta=\alpha\beta, \label{1.58}%
\end{equation}

\textbf{(ii)} For all $X_{j}\in%
%TCIMACRO{\dbigwedge \nolimits^{j}}%
%BeginExpansion
{\displaystyle\bigwedge\nolimits^{j}}
%EndExpansion
V$ and $Y_{k}\in%
%TCIMACRO{\dbigwedge \nolimits^{k}}%
%BeginExpansion
{\displaystyle\bigwedge\nolimits^{k}}
%EndExpansion
V$
\begin{equation}
X_{j}\underset{%
%TCIMACRO{\TeXButton{sig}{\sitg}}%
%BeginExpansion
\sitg
%EndExpansion
}{\cdot}Y_{k}=0,\text{ if }j\neq k. \label{1.59}%
\end{equation}

\textbf{(iii)} For all simple $k$-forms $v_{1}\wedge\ldots\wedge v_{k}\in%
%TCIMACRO{\dbigwedge \nolimits^{k}}%
%BeginExpansion
{\displaystyle\bigwedge\nolimits^{k}}
%EndExpansion
V$ and $w_{1}\wedge\ldots\wedge w_{k}\in%
%TCIMACRO{\dbigwedge \nolimits^{k}}%
%BeginExpansion
{\displaystyle\bigwedge\nolimits^{k}}
%EndExpansion
V$, we have
\begin{equation}
(v_{1}\wedge\ldots\wedge v_{k})\underset{%
%TCIMACRO{\TeXButton{sig}{\sitg}}%
%BeginExpansion
\sitg
%EndExpansion
}{\cdot}(w_{1}\wedge\ldots\wedge w_{k})=\left\vert
\begin{array}
[c]{ccc}%
v_{1}\underset{%
%TCIMACRO{\TeXButton{sig}{\sitg}}%
%BeginExpansion
\sitg
%EndExpansion
}{\cdot}w_{1} & \ldots & v_{1}\underset{%
%TCIMACRO{\TeXButton{sig}{\sitg}}%
%BeginExpansion
\sitg
%EndExpansion
}{\cdot}w_{k}\\
\vdots &  & \vdots\\
v_{k}\underset{%
%TCIMACRO{\TeXButton{sig}{\sitg}}%
%BeginExpansion
\sitg
%EndExpansion
}{\cdot}w_{1} & \ldots & v_{k}\underset{%
%TCIMACRO{\TeXButton{sig}{\sitg}}%
%BeginExpansion
\sitg
%EndExpansion
}{\cdot}w_{k}%
\end{array}
\right\vert , \label{1.60}%
\end{equation}
where as before $\left\vert {}\right\vert $ denotes the classical determinant
of the matrix with the entries $v_{i}\underset{%
%TCIMACRO{\TeXButton{sig}{\sitg}}%
%BeginExpansion
\sitg
%EndExpansion
}{\cdot}w_{j}$ .

\textbf{(iv)} For all $X,Y\in%
%TCIMACRO{\dbigwedge }%
%BeginExpansion
{\displaystyle\bigwedge}
%EndExpansion
V$
\begin{subequations}
\label{1.61}%
\begin{align}
\hat{X}\underset{%
%TCIMACRO{\TeXButton{sig}{\sitg}}%
%BeginExpansion
\sitg
%EndExpansion
}{\cdot}Y  &  =X\underset{%
%TCIMACRO{\TeXButton{sig}{\sitg}}%
%BeginExpansion
\sitg
%EndExpansion
}{\cdot}\hat{Y}.\label{1.61a}\\
\widetilde{X}\underset{%
%TCIMACRO{\TeXButton{sig}{\sitg}}%
%BeginExpansion
\sitg
%EndExpansion
}{\cdot}Y  &  =X\underset{%
%TCIMACRO{\TeXButton{sig}{\sitg}}%
%BeginExpansion
\sitg
%EndExpansion
}{\cdot}\widetilde{Y}. \label{1.61b}%
\end{align}
\vspace*{0.3in}

\subsubsection{The Metric Left and Right Contraction}

The metric left and right contractions of $X,Y\in%
%TCIMACRO{\dbigwedge }%
%BeginExpansion
{\displaystyle\bigwedge}
%EndExpansion
V$ are given by
\end{subequations}
\begin{align}
X\underset{%
%TCIMACRO{\TeXButton{sig}{\sitg}}%
%BeginExpansion
\sitg
%EndExpansion
}{\lrcorner}Y  &  :=\underline{%
%TCIMACRO{\TeXButton{itg}{\itg}}%
%BeginExpansion
\itg
%EndExpansion
}(X)\lrcorner Y,\label{1.62}\\
X\underset{%
%TCIMACRO{\TeXButton{sig}{\sitg}}%
%BeginExpansion
\sitg
%EndExpansion
}{\llcorner}Y  &  :=X\llcorner\underline{%
%TCIMACRO{\TeXButton{itg}{\itg}}%
%BeginExpansion
\itg
%EndExpansion
}(Y). \label{1.63}%
\end{align}

The linear space $%
%TCIMACRO{\dbigwedge }%
%BeginExpansion
{\displaystyle\bigwedge}
%EndExpansion
V$ equipped with any one of those contractions has a structure of non
associative algebra, and ($%
%TCIMACRO{\dbigwedge }%
%BeginExpansion
{\displaystyle\bigwedge}
%EndExpansion
V,\underset{%
%TCIMACRO{\TeXButton{sig}{\sitg}}%
%BeginExpansion
\sitg
%EndExpansion
}{\lrcorner}$,$\underset{%
%TCIMACRO{\TeXButton{sig}{\sitg}}%
%BeginExpansion
\sitg
%EndExpansion
}{\llcorner})$ will be called metric interior algebra of multiforms

The most important properties of the metric left and right contractions are:

\textbf{(i)} For all $X,Y$, $Z\in%
%TCIMACRO{\dbigwedge }%
%BeginExpansion
{\displaystyle\bigwedge}
%EndExpansion
V$,
\begin{subequations}
\label{1.64}%
\begin{align}
(X\underset{%
%TCIMACRO{\TeXButton{sig}{\sitg}}%
%BeginExpansion
\sitg
%EndExpansion
}{\lrcorner}Y)\underset{%
%TCIMACRO{\TeXButton{sig}{\sitg}}%
%BeginExpansion
\sitg
%EndExpansion
}{\cdot}Z  &  =Y\underset{%
%TCIMACRO{\TeXButton{sig}{\sitg}}%
%BeginExpansion
\sitg
%EndExpansion
}{\cdot}(\widetilde{X}\wedge Z),\label{1.64a}\\
(X\underset{%
%TCIMACRO{\TeXButton{sig}{\sitg}}%
%BeginExpansion
\sitg
%EndExpansion
}{\llcorner}Y)\underset{%
%TCIMACRO{\TeXButton{sig}{\sitg}}%
%BeginExpansion
\sitg
%EndExpansion
}{\cdot}Z  &  =X\underset{%
%TCIMACRO{\TeXButton{sig}{\sitg}}%
%BeginExpansion
\sitg
%EndExpansion
}{\cdot}(Z\wedge\widetilde{Y}). \label{1.64b}%
\end{align}

\textbf{(ii)} For all $\alpha,\beta\in\mathbb{R}$ and $X\in%
%TCIMACRO{\dbigwedge }%
%BeginExpansion
{\displaystyle\bigwedge}
%EndExpansion
V$
\end{subequations}
\begin{subequations}
\label{1.65}%
\begin{align}
\alpha\underset{%
%TCIMACRO{\TeXButton{sig}{\sitg}}%
%BeginExpansion
\sitg
%EndExpansion
}{\lrcorner}\beta &  =\alpha\underset{%
%TCIMACRO{\TeXButton{sig}{\sitg}}%
%BeginExpansion
\sitg
%EndExpansion
}{\llcorner}\beta=\alpha\beta,\label{1.65a}\\
\alpha\underset{%
%TCIMACRO{\TeXButton{sig}{\sitg}}%
%BeginExpansion
\sitg
%EndExpansion
}{\lrcorner}X  &  =X\underset{%
%TCIMACRO{\TeXButton{sig}{\sitg}}%
%BeginExpansion
\sitg
%EndExpansion
}{\llcorner}\alpha=\alpha X, \label{1.65b}%
\end{align}

\textbf{(iii)} For all $X_{j}\in%
%TCIMACRO{\dbigwedge \nolimits^{j}}%
%BeginExpansion
{\displaystyle\bigwedge\nolimits^{j}}
%EndExpansion
V$ and $Y_{k}\in%
%TCIMACRO{\dbigwedge \nolimits^{k}}%
%BeginExpansion
{\displaystyle\bigwedge\nolimits^{k}}
%EndExpansion
V$
\end{subequations}
\begin{subequations}
\label{1.66}%
\begin{align}
X_{j}\underset{v}{\lrcorner}Y_{k}  &  =0,\text{ if }j>k.\label{1.66a}\\
X_{j}\underset{%
%TCIMACRO{\TeXButton{sig}{\sitg}}%
%BeginExpansion
\sitg
%EndExpansion
}{\llcorner}Y_{k}  &  =0,\text{ if }j<k. \label{1.66b}%
\end{align}

\textbf{(iv)} For all $X_{j}\in%
%TCIMACRO{\dbigwedge \nolimits^{j}}%
%BeginExpansion
{\displaystyle\bigwedge\nolimits^{j}}
%EndExpansion
V$ and $Y_{k}\in%
%TCIMACRO{\dbigwedge \nolimits^{k}}%
%BeginExpansion
{\displaystyle\bigwedge\nolimits^{k}}
%EndExpansion
V$, with $j\leq k,$ we have: $X_{j}\underset{%
%TCIMACRO{\TeXButton{sig}{\sitg}}%
%BeginExpansion
\sitg
%EndExpansion
}{\lrcorner}Y_{k}\in%
%TCIMACRO{\dbigwedge \nolimits^{k-j}}%
%BeginExpansion
{\displaystyle\bigwedge\nolimits^{k-j}}
%EndExpansion
V$ and $Y_{k}\underset{%
%TCIMACRO{\TeXButton{sig}{\sitg}}%
%BeginExpansion
\sitg
%EndExpansion
}{\llcorner}X_{j}\in%
%TCIMACRO{\dbigwedge \nolimits^{k-j}}%
%BeginExpansion
{\displaystyle\bigwedge\nolimits^{k-j}}
%EndExpansion
V$ \ and
\end{subequations}
\begin{equation}
X_{j}\underset{%
%TCIMACRO{\TeXButton{sig}{\sitg}}%
%BeginExpansion
\sitg
%EndExpansion
}{\lrcorner}Y_{k}=(-1)^{j(k-j)}Y_{k}\underset{%
%TCIMACRO{\TeXButton{sig}{\sitg}}%
%BeginExpansion
\sitg
%EndExpansion
}{\llcorner}X_{j}. \label{1.67}%
\end{equation}

\textbf{(v)} For all $X_{k},Y_{k}\in%
%TCIMACRO{\dbigwedge \nolimits^{k}}%
%BeginExpansion
{\displaystyle\bigwedge\nolimits^{k}}
%EndExpansion
V$%
\begin{equation}
X_{k}\underset{%
%TCIMACRO{\TeXButton{sig}{\sitg}}%
%BeginExpansion
\sitg
%EndExpansion
}{\lrcorner}Y_{k}=X_{k}\underset{%
%TCIMACRO{\TeXButton{sig}{\sitg}}%
%BeginExpansion
\sitg
%EndExpansion
}{\llcorner}Y_{k}=\widetilde{X}_{k}\underset{%
%TCIMACRO{\TeXButton{sig}{\sitg}}%
%BeginExpansion
\sitg
%EndExpansion
}{\cdot}Y_{k}=X_{k}\underset{%
%TCIMACRO{\TeXButton{sig}{\sitg}}%
%BeginExpansion
\sitg
%EndExpansion
}{\cdot}\widetilde{Y}_{k}. \label{1.68}%
\end{equation}

\textbf{(vi)} For all $a\in%
%TCIMACRO{\dbigwedge \nolimits^{1}}%
%BeginExpansion
{\displaystyle\bigwedge\nolimits^{1}}
%EndExpansion
V$ and $X\in%
%TCIMACRO{\dbigwedge }%
%BeginExpansion
{\displaystyle\bigwedge}
%EndExpansion
V$%
\begin{equation}
a\underset{%
%TCIMACRO{\TeXButton{sig}{\sitg}}%
%BeginExpansion
\sitg
%EndExpansion
}{\lrcorner}X=-\hat{X}\underset{%
%TCIMACRO{\TeXButton{sig}{\sitg}}%
%BeginExpansion
\sitg
%EndExpansion
}{\llcorner}a. \label{1.69}%
\end{equation}

\textbf{(vii)}$.$For all $a\in%
%TCIMACRO{\dbigwedge \nolimits^{1}}%
%BeginExpansion
{\displaystyle\bigwedge\nolimits^{1}}
%EndExpansion
V$ and $X,Y\in%
%TCIMACRO{\dbigwedge }%
%BeginExpansion
{\displaystyle\bigwedge}
%EndExpansion
V$%
\begin{equation}
a\underset{%
%TCIMACRO{\TeXButton{sig}{\sitg}}%
%BeginExpansion
\sitg
%EndExpansion
}{\lrcorner}(X\wedge Y)=(a\underset{%
%TCIMACRO{\TeXButton{sig}{\sitg}}%
%BeginExpansion
\sitg
%EndExpansion
}{\lrcorner}X)\wedge Y+\hat{X}\wedge(a\underset{%
%TCIMACRO{\TeXButton{sig}{\sitg}}%
%BeginExpansion
\sitg
%EndExpansion
}{\lrcorner}Y). \label{1.70}%
\end{equation}

\subsubsection{The Metric Clifford Product}

The metric Clifford product $\underset{%
%TCIMACRO{\TeXButton{sig}{\sitg}}%
%BeginExpansion
\sitg
%EndExpansion
}{}$ of two elements of $%
%TCIMACRO{\dbigwedge }%
%BeginExpansion
{\displaystyle\bigwedge}
%EndExpansion
V$ is given by the following axiomatic:

\textbf{(i)} For all $\alpha\in\mathbb{R}$ and $X\in%
%TCIMACRO{\dbigwedge }%
%BeginExpansion
{\displaystyle\bigwedge}
%EndExpansion
V$%
\[
\alpha\underset{%
%TCIMACRO{\TeXButton{sig}{\sitg}}%
%BeginExpansion
\sitg
%EndExpansion
}{}X=X\underset{%
%TCIMACRO{\TeXButton{sig}{\sitg}}%
%BeginExpansion
\sitg
%EndExpansion
}{}\alpha=\alpha X,
\]

\textbf{(ii)} For all $a\in%
%TCIMACRO{\dbigwedge \nolimits^{1}}%
%BeginExpansion
{\displaystyle\bigwedge\nolimits^{1}}
%EndExpansion
V$ and $X\in%
%TCIMACRO{\dbigwedge }%
%BeginExpansion
{\displaystyle\bigwedge}
%EndExpansion
V$%
\begin{align*}
a\underset{%
%TCIMACRO{\TeXButton{sig}{\sitg}}%
%BeginExpansion
\sitg
%EndExpansion
}{}X  &  =a\underset{%
%TCIMACRO{\TeXButton{sig}{\sitg}}%
%BeginExpansion
\sitg
%EndExpansion
}{\lrcorner}X+a\wedge X,\\
X\underset{%
%TCIMACRO{\TeXButton{sig}{\sitg}}%
%BeginExpansion
\sitg
%EndExpansion
}{}a  &  =X\underset{%
%TCIMACRO{\TeXButton{sig}{\sitg}}%
%BeginExpansion
\sitg
%EndExpansion
}{\llcorner}a+X\wedge a.
\end{align*}

\textbf{(iii)} For all $X,Y,Z\in%
%TCIMACRO{\dbigwedge }%
%BeginExpansion
{\displaystyle\bigwedge}
%EndExpansion
V$%
\[
X\underset{%
%TCIMACRO{\TeXButton{sig}{\sitg}}%
%BeginExpansion
\sitg
%EndExpansion
}{}(Y\underset{%
%TCIMACRO{\TeXButton{sig}{\sitg}}%
%BeginExpansion
\sitg
%EndExpansion
}{}Z)=(X\underset{%
%TCIMACRO{\TeXButton{sig}{\sitg}}%
%BeginExpansion
\sitg
%EndExpansion
}{}Y)\underset{%
%TCIMACRO{\TeXButton{sig}{\sitg}}%
%BeginExpansion
\sitg
%EndExpansion
}{}Z.
\]

The metric Clifford product is distributive and associative Its distributive
law follows from the correspondent distributive laws satisfied by the left and
right contractions and the exterior product. Its associative law is just the
axiom \textbf{(iii)}.

$%
%TCIMACRO{\dbigwedge }%
%BeginExpansion
{\displaystyle\bigwedge}
%EndExpansion
V$ equipped with the metric Clifford product is an associative algebra, called
a \textit{metric Clifford algebra over} $V$. It will be denoted by
$\mathcal{C\ell}(V,%
%TCIMACRO{\TeXButton{itg}{\itg}}%
%BeginExpansion
\itg
%EndExpansion
)$. In a totally similar way we may define the algebra $\mathcal{C\ell}(V,%
%TCIMACRO{\TeXButton{itg}{\itg}}%
%BeginExpansion
\itg
%EndExpansion
^{-1})$.

We conclude this section by listing some useful properties of the metric
Clifford product which will be used in the calculations of the following sections.

\textbf{(i)} For all $a\in%
%TCIMACRO{\dbigwedge \nolimits^{1}}%
%BeginExpansion
{\displaystyle\bigwedge\nolimits^{1}}
%EndExpansion
V$ and $X,Y,Z\in%
%TCIMACRO{\dbigwedge }%
%BeginExpansion
{\displaystyle\bigwedge}
%EndExpansion
V:$%
\begin{align}
\widehat{X\underset{%
%TCIMACRO{\TeXButton{sig}{\sitg}}%
%BeginExpansion
\sitg
%EndExpansion
}{}Y}  &  =\hat{X}\underset{%
%TCIMACRO{\TeXButton{sig}{\sitg}}%
%BeginExpansion
\sitg
%EndExpansion
}{}\hat{Y},\nonumber\\
\widetilde{X\underset{%
%TCIMACRO{\TeXButton{sig}{\sitg}}%
%BeginExpansion
\sitg
%EndExpansion
}{}Y}  &  =\widetilde{X}\underset{%
%TCIMACRO{\TeXButton{sig}{\sitg}}%
%BeginExpansion
\sitg
%EndExpansion
}{}\widetilde{Y}.\nonumber\\
a\underset{%
%TCIMACRO{\TeXButton{sig}{\sitg}}%
%BeginExpansion
\sitg
%EndExpansion
}{\lrcorner}X  &  =\frac{1}{2}(a\underset{%
%TCIMACRO{\TeXButton{sig}{\sitg}}%
%BeginExpansion
\sitg
%EndExpansion
}{}X-\hat{X}\underset{%
%TCIMACRO{\TeXButton{sig}{\sitg}}%
%BeginExpansion
\sitg
%EndExpansion
}{}a),\nonumber\\
a\wedge X  &  =\frac{1}{2}(a\underset{%
%TCIMACRO{\TeXButton{sig}{\sitg}}%
%BeginExpansion
\sitg
%EndExpansion
}{}X+\hat{X}\underset{%
%TCIMACRO{\TeXButton{itg}{\itg}}%
%BeginExpansion
\itg
%EndExpansion
}{}a).\label{1.71}\\
X\underset{%
%TCIMACRO{\TeXButton{sig}{\sitg}}%
%BeginExpansion
\sitg
%EndExpansion
}{\cdot}Y  &  =\left\langle \widetilde{X}\underset{%
%TCIMACRO{\TeXButton{sig}{\sitg}}%
%BeginExpansion
\sitg
%EndExpansion
}{}Y\right\rangle _{0}=\left\langle X\underset{%
%TCIMACRO{\TeXButton{sig}{\sitg}}%
%BeginExpansion
\sitg
%EndExpansion
}{}\widetilde{Y}\right\rangle _{0}.\nonumber\\
X\underset{%
%TCIMACRO{\TeXButton{sig}{\sitg}}%
%BeginExpansion
\sitg
%EndExpansion
}{\lrcorner}(Y\underset{%
%TCIMACRO{\TeXButton{sig}{\sitg}}%
%BeginExpansion
\sitg
%EndExpansion
}{\lrcorner}Z)  &  =(X\wedge Y)\underset{%
%TCIMACRO{\TeXButton{sig}{\sitg}}%
%BeginExpansion
\sitg
%EndExpansion
}{\lrcorner}Z,\nonumber\\
(X\underset{%
%TCIMACRO{\TeXButton{sig}{\sitg}}%
%BeginExpansion
\sitg
%EndExpansion
}{\llcorner}Y)\underset{%
%TCIMACRO{\TeXButton{sig}{\sitg}}%
%BeginExpansion
\sitg
%EndExpansion
}{\llcorner}Z  &  =X\underset{%
%TCIMACRO{\TeXButton{sig}{\sitg}}%
%BeginExpansion
\sitg
%EndExpansion
}{\llcorner}(Y\wedge Z).\nonumber
\end{align}

\textbf{(ii)} Let $\tau\in%
%TCIMACRO{\dbigwedge \nolimits^{n}}%
%BeginExpansion
{\displaystyle\bigwedge\nolimits^{n}}
%EndExpansion
V$ then for all $a\in%
%TCIMACRO{\dbigwedge \nolimits^{1}}%
%BeginExpansion
{\displaystyle\bigwedge\nolimits^{1}}
%EndExpansion
V$ and $X\in%
%TCIMACRO{\dbigwedge }%
%BeginExpansion
{\displaystyle\bigwedge}
%EndExpansion
V:$%
\begin{equation}
\tau\underset{%
%TCIMACRO{\TeXButton{sig}{\sitg}}%
%BeginExpansion
\sitg
%EndExpansion
}{}(a\wedge X)=(-1)^{n-1}a\underset{%
%TCIMACRO{\TeXButton{sig}{\sitg}}%
%BeginExpansion
\sitg
%EndExpansion
}{\lrcorner}(\tau\underset{%
%TCIMACRO{\TeXButton{sig}{\sitg}}%
%BeginExpansion
\sitg
%EndExpansion
}{}X). \label{1.72}%
\end{equation}
Eq.(\ref{1.72}) will be called \textit{metric dual identity}.

\subsubsection{Pseudo-Euclidean Metric Extensors on $V$}

\paragraph{The metric extensor $\eta$}

Consider the vector spaces $\mathbf{V}$ and $V$ \ ($\dim\mathbf{V}=\dim V=n)$
and the dual bases $\{\mathbf{b}_{\mu}\}$ and $\{%
%TCIMACRO{\TeXButton{beta}{\mbox{\boldmath{$\beta$}}}}%
%BeginExpansion
\mbox{\boldmath{$\beta$}}%
%EndExpansion
^{\mu}\}$ for $\mathbf{V}$ and $V$, $%
%TCIMACRO{\TeXButton{beta}{\mbox{\boldmath{$\beta$}}}}%
%BeginExpansion
\mbox{\boldmath{$\beta$}}%
%EndExpansion
^{\mu}(\mathbf{b}_{\nu})=\delta_{\nu}^{\mu}$. Moreover, denote the reciprocal
basis of the basis $\{%
%TCIMACRO{\TeXButton{beta}{\mbox{\boldmath{$\beta$}}}}%
%BeginExpansion
\mbox{\boldmath{$\beta$}}%
%EndExpansion
^{\mu}\}$ by $\{\beta_{\mu}\}$, with $%
%TCIMACRO{\TeXButton{beta}{\mbox{\boldmath{$\beta$}}}}%
%BeginExpansion
\mbox{\boldmath{$\beta$}}%
%EndExpansion
^{\mu}\cdot\beta_{\nu}=\delta_{\nu}^{\mu}$, and $\mu,\nu=0,1,2...,n$.

The extensor on $V$, say $\eta\in(1,1)$-$extV$ defined for any $a\in%
%TCIMACRO{\dbigwedge ^{1}}%
%BeginExpansion
{\displaystyle\bigwedge^{1}}
%EndExpansion
V$ by $a\mapsto\eta(a)\in%
%TCIMACRO{\dbigwedge ^{1}}%
%BeginExpansion
{\displaystyle\bigwedge^{1}}
%EndExpansion
V$, such that%
\begin{equation}
\eta(a)=%
%TCIMACRO{\TeXButton{beta}{\mbox{\boldmath{$\beta$}}}}%
%BeginExpansion
\mbox{\boldmath{$\beta$}}%
%EndExpansion
^{0}a%
%TCIMACRO{\TeXButton{beta}{\mbox{\boldmath{$\beta$}}}}%
%BeginExpansion
\mbox{\boldmath{$\beta$}}%
%EndExpansion
^{0}, \label{defeta}%
\end{equation}
is a well defined metric extensor on $V$ (i.e., $\eta$ is symmetric and non
degenerated) and is called a \textit{pseudo-Euclidean metric extensor for
}$\mathbf{V}$

The main properties of $\eta$ are:

\textbf{(i)} $%
%TCIMACRO{\TeXButton{beta}{\mbox{\boldmath{$\beta$}}}}%
%BeginExpansion
\mbox{\boldmath{$\beta$}}%
%EndExpansion
^{0}$ is an eigenvector with eigenvalue $1,$ i.e., $\eta(%
%TCIMACRO{\TeXButton{beta}{\mbox{\boldmath{$\beta$}}}}%
%BeginExpansion
\mbox{\boldmath{$\beta$}}%
%EndExpansion
^{0})=%
%TCIMACRO{\TeXButton{beta}{\mbox{\boldmath{$\beta$}}}}%
%BeginExpansion
\mbox{\boldmath{$\beta$}}%
%EndExpansion
^{0},$ and $%
%TCIMACRO{\TeXButton{beta}{\mbox{\boldmath{$\beta$}}}}%
%BeginExpansion
\mbox{\boldmath{$\beta$}}%
%EndExpansion
^{k}$ $(k=1,2,...,n-1)$ are eigenvectors with eigenvalues $-1,$ i.e., $\eta(%
%TCIMACRO{\TeXButton{beta}{\mbox{\boldmath{$\beta$}}}}%
%BeginExpansion
\mbox{\boldmath{$\beta$}}%
%EndExpansion
^{k})=-%
%TCIMACRO{\TeXButton{beta}{\mbox{\boldmath{$\beta$}}}}%
%BeginExpansion
\mbox{\boldmath{$\beta$}}%
%EndExpansion
^{k}.$

\textbf{(ii)} The 1-forms of the canonical basis $\{%
%TCIMACRO{\TeXButton{beta}{\mbox{\boldmath{$\beta$}}}}%
%BeginExpansion
\mbox{\boldmath{$\beta$}}%
%EndExpansion
^{\mu}\}$ are $\eta$ orthonormal i.e.,%
\begin{equation}%
%TCIMACRO{\TeXButton{beta}{\mbox{\boldmath{$\beta$}}}}%
%BeginExpansion
\mbox{\boldmath{$\beta$}}%
%EndExpansion
^{\mu}\underset{\eta}{\cdot}%
%TCIMACRO{\TeXButton{beta}{\mbox{\boldmath{$\beta$}}}}%
%BeginExpansion
\mbox{\boldmath{$\beta$}}%
%EndExpansion
^{\nu}:=\eta(%
%TCIMACRO{\TeXButton{beta}{\mbox{\boldmath{$\beta$}}}}%
%BeginExpansion
\mbox{\boldmath{$\beta$}}%
%EndExpansion
^{\mu})\cdot%
%TCIMACRO{\TeXButton{beta}{\mbox{\boldmath{$\beta$}}}}%
%BeginExpansion
\mbox{\boldmath{$\beta$}}%
%EndExpansion
^{\nu}=\eta^{\mu\nu},
\end{equation}
where the matrix with entries $(\eta^{\mu\nu})$ is the diagonal matrix
$(1,-1,...,-1)$.

Then, $\eta$ has signature\footnote{Or as some mathematicians say, signature
$(1-n)$.}\emph{ }$(1,n).$

\textbf{(iii) }$\mathrm{tr}[\eta]=2-n$ and $\det[\eta]=(-1)^{n-1}.$

\textbf{(iv)} The extended of $\eta$ has the same generator of $\eta,$ i.e.,
for any $X\in%
%TCIMACRO{\dbigwedge }%
%BeginExpansion
{\displaystyle\bigwedge}
%EndExpansion
U$, $\underline{\eta}(X)=%
%TCIMACRO{\TeXButton{beta}{\mbox{\boldmath{$\beta$}}}}%
%BeginExpansion
\mbox{\boldmath{$\beta$}}%
%EndExpansion
^{0}X%
%TCIMACRO{\TeXButton{beta}{\mbox{\boldmath{$\beta$}}}}%
%BeginExpansion
\mbox{\boldmath{$\beta$}}%
%EndExpansion
^{0}$.

\textbf{(v)} $\eta$ is orthogonal canonical, i.e., since%
\[
\eta(%
%TCIMACRO{\TeXButton{beta}{\mbox{\boldmath{$\beta$}}}}%
%BeginExpansion
\mbox{\boldmath{$\beta$}}%
%EndExpansion
^{\mu})\cdot%
%TCIMACRO{\TeXButton{beta}{\mbox{\boldmath{$\beta$}}}}%
%BeginExpansion
\mbox{\boldmath{$\beta$}}%
%EndExpansion
^{\nu}=%
%TCIMACRO{\TeXButton{beta}{\mbox{\boldmath{$\beta$}}}}%
%BeginExpansion
\mbox{\boldmath{$\beta$}}%
%EndExpansion
^{\mu}\cdot\eta(%
%TCIMACRO{\TeXButton{beta}{\mbox{\boldmath{$\beta$}}}}%
%BeginExpansion
\mbox{\boldmath{$\beta$}}%
%EndExpansion
^{\nu}%
\]
it is $\eta=\eta^{\diamond}$ (recall that $\eta^{\diamond}=(\eta^{\dagger
})^{-1}=(\eta^{-1})^{\dagger}$).Then, $\eta^{2}=\mathrm{id}$.\medskip

\textbf{Remark 2.1 }When $\dim\mathbf{V}=4$ and the signature of $\eta$ is
$(1,3)$ the pair $(\mathbf{V,}\eta)$ is called a \textit{Minkowski vector
space} and $\eta$ is called a Minkowski metric.\medskip

\paragraph{The Metric Extensor $%
%TCIMACRO{\TeXButton{itg}{\itg}}%
%BeginExpansion
\itg
%EndExpansion
$ with the Same Signature of $\eta$}

Let $%
%TCIMACRO{\TeXButton{itg}{\itg}}%
%BeginExpansion
\itg
%EndExpansion
$) be a \ general metric extensor on $V$ with the same signature of the $\eta$
extensor defined by Eq.(\ref{defeta}). Then we have the following fundamental
theorem\footnote{Originally proved in \cite{fmr013}.}, which will play a
crucial role in the considerations that follows.\medskip

\textbf{Theorem 2.1}\emph{.}For any pseudo-euclidean metric extensor $%
%TCIMACRO{\TeXButton{itg}{\itg}}%
%BeginExpansion
\itg
%EndExpansion
$ for $V$ there exists a invertible $(1,1)$-extensor $%
%TCIMACRO{\TeXButton{h}{\slh}}%
%BeginExpansion
\slh
%EndExpansion
$ \emph{(non unique!)}, such that
\begin{equation}%
%TCIMACRO{\TeXButton{itg}{\itg}}%
%BeginExpansion
\itg
%EndExpansion
=%
%TCIMACRO{\TeXButton{h}{\slh}}%
%BeginExpansion
\slh
%EndExpansion
^{\dagger}\eta%
%TCIMACRO{\TeXButton{h}{\slh}}%
%BeginExpansion
\slh
%EndExpansion
, \label{4.1}%
\end{equation}
where $\eta$ is defined by Eq.(\ref{defeta}). The $(1,1)$-extensor field $%
%TCIMACRO{\TeXButton{h}{\slh}}%
%BeginExpansion
\slh
%EndExpansion
$ is given by:
\begin{equation}%
%TCIMACRO{\TeXButton{h}{\slh}}%
%BeginExpansion
\slh
%EndExpansion
(a)=\overset{3}{\underset{\mu=0}{\sum}}\sqrt{\left\vert \lambda^{\mu
}\right\vert }(a\cdot v^{\mu})%
%TCIMACRO{\TeXButton{beta}{\mbox{\boldmath{$\beta$}}}}%
%BeginExpansion
\mbox{\boldmath{$\beta$}}%
%EndExpansion
^{\mu}, \label{4.2}%
\end{equation}
where the $\lambda^{\mu}$are the eigenvalue fields of $%
%TCIMACRO{\TeXButton{itg}{\itg}}%
%BeginExpansion
\itg
%EndExpansion
$) and $v^{\mu}\in%
%TCIMACRO{\dbigwedge ^{1}}%
%BeginExpansion
{\displaystyle\bigwedge^{1}}
%EndExpansion
U$ are the associated eigenvector of $%
%TCIMACRO{\TeXButton{itg}{\itg}}%
%BeginExpansion
\itg
%EndExpansion
$ The 1-form $\{v^{\mu}\}$ defines an\emph{ }\textit{euclidean orthonormal
basis }for $V$, i.e., $v^{\mu}\cdot v^{\nu}=\delta^{\mu\nu}$.\medskip

\textbf{Proof \ }We must prove that $%
%TCIMACRO{\TeXButton{h}{\slh}}%
%BeginExpansion
\slh
%EndExpansion
$ in Eq.(\ref{4.2}) satisfy the equation for composition of extensors $%
%TCIMACRO{\TeXButton{h}{\slh}}%
%BeginExpansion
\slh
%EndExpansion
^{\dagger}\eta%
%TCIMACRO{\TeXButton{h}{\slh}}%
%BeginExpansion
\slh
%EndExpansion
=%
%TCIMACRO{\TeXButton{itg}{\itg}}%
%BeginExpansion
\itg
%EndExpansion
$.

First we need to calculate the adjoint of $%
%TCIMACRO{\TeXButton{h}{\slh}}%
%BeginExpansion
\slh
%EndExpansion
$. Take two arbitrary $a,$ $b\in%
%TCIMACRO{\dbigwedge \nolimits^{1}}%
%BeginExpansion
{\displaystyle\bigwedge\nolimits^{1}}
%EndExpansion
U$. Then using Eq.(\ref{1.29}) and Eq.(\ref{4.2}) we can write
\begin{align}%
%TCIMACRO{\TeXButton{h}{\slh}}%
%BeginExpansion
\slh
%EndExpansion
^{\dagger}(a)\cdot b  &  =a\cdot%
%TCIMACRO{\TeXButton{h}{\slh}}%
%BeginExpansion
\slh
%EndExpansion
(b)\nonumber\\
&  =a\cdot(\overset{3}{\underset{\mu=0}{\sum}}\sqrt{\left\vert \lambda^{\mu
}\right\vert }(b\cdot v^{\mu})%
%TCIMACRO{\TeXButton{beta}{\mbox{\boldmath{$\beta$}}}}%
%BeginExpansion
\mbox{\boldmath{$\beta$}}%
%EndExpansion
^{\mu})\nonumber\\
&  =(\overset{3}{\underset{\mu=0}{\sum}}\sqrt{\left\vert \lambda^{\mu
}\right\vert }(a\cdot%
%TCIMACRO{\TeXButton{beta}{\mbox{\boldmath{$\beta$}}}}%
%BeginExpansion
\mbox{\boldmath{$\beta$}}%
%EndExpansion
^{\mu})v^{\mu})\cdot b, \label{4.2iii}%
\end{align}
and from the non degeneracy of the scalar product we have
\begin{equation}%
%TCIMACRO{\TeXButton{h}{\slh}}%
%BeginExpansion
\slh
%EndExpansion
^{\dagger}(a)=\overset{3}{\underset{\mu=0}{\sum}}\sqrt{\left\vert \lambda
^{\mu}\right\vert }(a\cdot%
%TCIMACRO{\TeXButton{beta}{\mbox{\boldmath{$\beta$}}}}%
%BeginExpansion
\mbox{\boldmath{$\beta$}}%
%EndExpansion
^{\mu})v^{\mu}. \label{4.2iv}%
\end{equation}

Now, using Eqs.(\ref{4.2}) and (\ref{4.2iv}) we get
\begin{align*}%
%TCIMACRO{\TeXButton{h}{\slh}}%
%BeginExpansion
\slh
%EndExpansion
^{\dagger}\eta%
%TCIMACRO{\TeXButton{h}{\slh}}%
%BeginExpansion
\slh
%EndExpansion
(a)  &  =\overset{3}{\underset{\mu=0}{\sum}}\overset{3}{\underset{\nu=0}{\sum
}}\sqrt{\left\vert \lambda^{\mu}\lambda^{\nu}\right\vert }\eta(%
%TCIMACRO{\TeXButton{beta}{\mbox{\boldmath{$\beta$}}}}%
%BeginExpansion
\mbox{\boldmath{$\beta$}}%
%EndExpansion
^{\mu})\cdot%
%TCIMACRO{\TeXButton{beta}{\mbox{\boldmath{$\beta$}}}}%
%BeginExpansion
\mbox{\boldmath{$\beta$}}%
%EndExpansion
^{\nu}(a\cdot v^{\mu})v^{\nu}\\
&  =\sqrt{\left\vert \lambda^{0}\right\vert ^{2}}\eta(%
%TCIMACRO{\TeXButton{beta}{\mbox{\boldmath{$\beta$}}}}%
%BeginExpansion
\mbox{\boldmath{$\beta$}}%
%EndExpansion
^{0})\cdot%
%TCIMACRO{\TeXButton{beta}{\mbox{\boldmath{$\beta$}}}}%
%BeginExpansion
\mbox{\boldmath{$\beta$}}%
%EndExpansion
^{0}(a\cdot v^{0})v^{0}+\overset{3}{\underset{j=1}{\sum}}\sqrt{\left\vert
\lambda^{j}\lambda^{0}\right\vert }\eta(%
%TCIMACRO{\TeXButton{beta}{\mbox{\boldmath{$\beta$}}}}%
%BeginExpansion
\mbox{\boldmath{$\beta$}}%
%EndExpansion
^{j})\cdot%
%TCIMACRO{\TeXButton{beta}{\mbox{\boldmath{$\beta$}}}}%
%BeginExpansion
\mbox{\boldmath{$\beta$}}%
%EndExpansion
^{0}(a\cdot v^{j})v^{0}\\
&  +\overset{3}{\underset{k=1}{\sum}}\sqrt{\left\vert \lambda^{0}\lambda
^{k}\right\vert }\eta(%
%TCIMACRO{\TeXButton{beta}{\mbox{\boldmath{$\beta$}}}}%
%BeginExpansion
\mbox{\boldmath{$\beta$}}%
%EndExpansion
^{0})\cdot%
%TCIMACRO{\TeXButton{beta}{\mbox{\boldmath{$\beta$}}}}%
%BeginExpansion
\mbox{\boldmath{$\beta$}}%
%EndExpansion
^{k}(a\cdot v^{0})v^{k}+\overset{3}{\underset{j=1}{\sum}}%
\overset{3}{\underset{k=1}{\sum}}\sqrt{\left\vert \lambda^{j}\lambda
^{k}\right\vert }\eta(%
%TCIMACRO{\TeXButton{beta}{\mbox{\boldmath{$\beta$}}}}%
%BeginExpansion
\mbox{\boldmath{$\beta$}}%
%EndExpansion
^{j})\cdot%
%TCIMACRO{\TeXButton{beta}{\mbox{\boldmath{$\beta$}}}}%
%BeginExpansion
\mbox{\boldmath{$\beta$}}%
%EndExpansion
^{k}(a\cdot v^{j})v^{k},
\end{align*}
and since $\eta(%
%TCIMACRO{\TeXButton{beta}{\mbox{\boldmath{$\beta$}}}}%
%BeginExpansion
\mbox{\boldmath{$\beta$}}%
%EndExpansion
^{0})=%
%TCIMACRO{\TeXButton{beta}{\mbox{\boldmath{$\beta$}}}}%
%BeginExpansion
\mbox{\boldmath{$\beta$}}%
%EndExpansion
^{0}$ and $\eta(%
%TCIMACRO{\TeXButton{beta}{\mbox{\boldmath{$\beta$}}}}%
%BeginExpansion
\mbox{\boldmath{$\beta$}}%
%EndExpansion
^{k})=-%
%TCIMACRO{\TeXButton{beta}{\mbox{\boldmath{$\beta$}}}}%
%BeginExpansion
\mbox{\boldmath{$\beta$}}%
%EndExpansion
^{k}$ ($k=1,2,...,n-1$), and $%
%TCIMACRO{\TeXButton{beta}{\mbox{\boldmath{$\beta$}}}}%
%BeginExpansion
\mbox{\boldmath{$\beta$}}%
%EndExpansion
^{\mu}\cdot%
%TCIMACRO{\TeXButton{beta}{\mbox{\boldmath{$\beta$}}}}%
%BeginExpansion
\mbox{\boldmath{$\beta$}}%
%EndExpansion
^{\nu}=\delta^{\mu\nu}$ we get
\begin{align}%
%TCIMACRO{\TeXButton{h}{\slh}}%
%BeginExpansion
\slh
%EndExpansion
^{\dagger}\eta%
%TCIMACRO{\TeXButton{h}{\slh}}%
%BeginExpansion
\slh
%EndExpansion
(a)  &  =\left\vert \lambda^{0}\right\vert (a\cdot v^{0})v^{0}%
-\overset{3}{\underset{j=1}{\sum}}\overset{3}{\underset{k=1}{\sum}}%
\sqrt{\left\vert \lambda^{j}\lambda^{k}\right\vert }\delta^{jk}(a\cdot
v^{j})v^{k}\nonumber\\%
%TCIMACRO{\TeXButton{h}{\slh}}%
%BeginExpansion
\slh
%EndExpansion
^{\dagger}\eta%
%TCIMACRO{\TeXButton{h}{\slh}}%
%BeginExpansion
\slh
%EndExpansion
(a)  &  =\left\vert \lambda^{0}\right\vert (a\cdot v^{0})v^{0}%
-\overset{3}{\underset{j=1}{\sum}}\left\vert \lambda^{j}\right\vert (a\cdot
v^{j})v^{j}. \label{4.2v}%
\end{align}

Finally using Eqs.(\ref{4.2iv}) and (\ref{4.2iii}) in Eq.(\ref{4.2v}), we
have
\begin{align*}%
%TCIMACRO{\TeXButton{h}{\slh}}%
%BeginExpansion
\slh
%EndExpansion
^{\dagger}\eta%
%TCIMACRO{\TeXButton{h}{\slh}}%
%BeginExpansion
\slh
%EndExpansion
(a)  &  =(a\cdot v^{0})%
%TCIMACRO{\TeXButton{itg}{\itg}}%
%BeginExpansion
\itg
%EndExpansion
(v^{0})+\overset{3}{\underset{j=1}{\sum}}(a\cdot v^{j})%
%TCIMACRO{\TeXButton{itg}{\itg}}%
%BeginExpansion
\itg
%EndExpansion
(v^{j})\\
&  =\overset{3}{\underset{\mu=0}{\sum}}(a\cdot v^{\mu})%
%TCIMACRO{\TeXButton{itg}{\itg}}%
%BeginExpansion
\itg
%EndExpansion
(v^{\mu})=%
%TCIMACRO{\TeXButton{itg}{\itg}}%
%BeginExpansion
\itg
%EndExpansion
\overset{3}{(\underset{\mu=0}{\sum}}(a\cdot v^{\mu})v^{\mu})\\%
%TCIMACRO{\TeXButton{h}{\slh}}%
%BeginExpansion
\slh
%EndExpansion
^{\dagger}\eta%
%TCIMACRO{\TeXButton{h}{\slh}}%
%BeginExpansion
\slh
%EndExpansion
(a)  &  =%
%TCIMACRO{\TeXButton{itg}{\itg}}%
%BeginExpansion
\itg
%EndExpansion
(a).%
%TCIMACRO{\TeXButton{End Proof}{\endproof}}%
%BeginExpansion
\endproof
%EndExpansion
\end{align*}

The $(1,1)$-extensor $%
%TCIMACRO{\TeXButton{h}{\slh}}%
%BeginExpansion
\slh
%EndExpansion
$ given by Eq.(\ref{4.2} )is indeed invertible
\begin{align}%
%TCIMACRO{\TeXButton{h}{\slh}}%
%BeginExpansion
\slh
%EndExpansion
^{\dagger}\eta%
%TCIMACRO{\TeXButton{h}{\slh}}%
%BeginExpansion
\slh
%EndExpansion
=%
%TCIMACRO{\TeXButton{itg}{\itg}}%
%BeginExpansion
\itg
%EndExpansion
&  \Rightarrow\det[%
%TCIMACRO{\TeXButton{h}{\slh}}%
%BeginExpansion
\slh
%EndExpansion
^{\dagger}]\det[\eta]\det[%
%TCIMACRO{\TeXButton{h}{\slh}}%
%BeginExpansion
\slh
%EndExpansion
]=\det[%
%TCIMACRO{\TeXButton{itg}{\itg}}%
%BeginExpansion
\itg
%EndExpansion
]\nonumber\\
&  \Rightarrow(\det[%
%TCIMACRO{\TeXButton{h}{\slh}}%
%BeginExpansion
\slh
%EndExpansion
])^{2}(-1)=\det[%
%TCIMACRO{\TeXButton{itg}{\itg}}%
%BeginExpansion
\itg
%EndExpansion
],
\end{align}
and this implies that $\det[%
%TCIMACRO{\TeXButton{h}{\slh}}%
%BeginExpansion
\slh
%EndExpansion
]\neq0$ i.e., $%
%TCIMACRO{\TeXButton{h}{\slh}}%
%BeginExpansion
\slh
%EndExpansion
$ is non degenerated and thus is indeed invertible.

On the other hand the $(1,1)$-extensor $%
%TCIMACRO{\TeXButton{h}{\slh}}%
%BeginExpansion
\slh
%EndExpansion
$ which satisfies Eq.(\ref{4.1}) is certainly not unique. There is a
\textit{gauge freedom} which is represented by a local pseudo orthogonal
transformation. Indeed, consider the $(1,1)$-extensor $%
%TCIMACRO{\TeXButton{h}{\slh}}%
%BeginExpansion
\slh
%EndExpansion
^{\prime}\equiv%
%TCIMACRO{\TeXButton{l}{\sll}}%
%BeginExpansion
\sll
%EndExpansion%
%TCIMACRO{\TeXButton{h}{\slh}}%
%BeginExpansion
\slh
%EndExpansion
,$ where $%
%TCIMACRO{\TeXButton{l}{\sll}}%
%BeginExpansion
\sll
%EndExpansion
$ is an $\eta$-orthogonal $(1,1)$-extensor $%
%TCIMACRO{\TeXButton{l}{\sll}}%
%BeginExpansion
\sll
%EndExpansion
=%
%TCIMACRO{\TeXButton{l}{\sll}}%
%BeginExpansion
\sll
%EndExpansion
^{(\eta)}$,(or equivalently, $%
%TCIMACRO{\TeXButton{l}{\sll}}%
%BeginExpansion
\sll
%EndExpansion
^{\dagger}\eta%
%TCIMACRO{\TeXButton{l}{\sll}}%
%BeginExpansion
\sll
%EndExpansion
=\eta$) \footnote{I.e., a \emph{transformation } $l(a)\underset{\eta}{\cdot
}l(b)=a\underset{\eta}{\cdot}b$.} and $%
%TCIMACRO{\TeXButton{h}{\slh}}%
%BeginExpansion
\slh
%EndExpansion
$ is any $(1,1)$-extensor that satisfies Eq.(\ref{4.1}) (e.g., the$%
%TCIMACRO{\TeXButton{h}{\slh}}%
%BeginExpansion
\slh
%EndExpansion
$ given by Eq.(\ref{4.2})), then
\[
(%
%TCIMACRO{\TeXButton{h}{\slh}}%
%BeginExpansion
\slh
%EndExpansion
^{\prime})^{\dagger}\eta%
%TCIMACRO{\TeXButton{h}{\slh}}%
%BeginExpansion
\slh
%EndExpansion
^{\prime}=(%
%TCIMACRO{\TeXButton{l}{\sll}}%
%BeginExpansion
\sll
%EndExpansion%
%TCIMACRO{\TeXButton{h}{\slh}}%
%BeginExpansion
\slh
%EndExpansion
)^{\dagger}\eta%
%TCIMACRO{\TeXButton{l}{\sll}}%
%BeginExpansion
\sll
%EndExpansion%
%TCIMACRO{\TeXButton{h}{\slh}}%
%BeginExpansion
\slh
%EndExpansion
=%
%TCIMACRO{\TeXButton{h}{\slh}}%
%BeginExpansion
\slh
%EndExpansion
^{\dagger}l^{\dagger}\eta%
%TCIMACRO{\TeXButton{l}{\sll}}%
%BeginExpansion
\sll
%EndExpansion%
%TCIMACRO{\TeXButton{h}{\slh}}%
%BeginExpansion
\slh
%EndExpansion
=%
%TCIMACRO{\TeXButton{h}{\slh}}%
%BeginExpansion
\slh
%EndExpansion
^{\dagger}\eta%
%TCIMACRO{\TeXButton{h}{\slh}}%
%BeginExpansion
\slh
%EndExpansion
=%
%TCIMACRO{\TeXButton{itg}{\itg}}%
%BeginExpansion
\itg
%EndExpansion
.
\]
i.e., $%
%TCIMACRO{\TeXButton{h}{\slh}}%
%BeginExpansion
\slh
%EndExpansion
^{\prime}$ also satisfies Eq.(\ref{4.1}).

\textbf{Remark 2.2 }When $\dim V=4$ and the signature of $%
%TCIMACRO{\TeXButton{itg}{\itg}}%
%BeginExpansion
\itg
%EndExpansion
$ (and, of course, of $\eta$) is ($1,3$) $%
%TCIMACRO{\TeXButton{itg}{\itg}}%
%BeginExpansion
\itg
%EndExpansion
$ is said a \textit{Lorentz metric} \textit{extensor.}

\subsubsection{Some Remarkable Results}

We recall now some remarkable results that will be need in Section 6.

\paragraph{Golden Rule}

Let$%
%TCIMACRO{\TeXButton{itg}{\itg}}%
%BeginExpansion
\itg
%EndExpansion
=%
%TCIMACRO{\TeXButton{h}{\slh}}%
%BeginExpansion
\slh
%EndExpansion
^{\dagger}\eta%
%TCIMACRO{\TeXButton{h}{\slh}}%
%BeginExpansion
\slh
%EndExpansion
$ be defined as in Eq.(\ref{4.1}). Then for any $X,Y\in%
%TCIMACRO{\dbigwedge }%
%BeginExpansion
{\displaystyle\bigwedge}
%EndExpansion
U$ it holds the following remarkable identity known as the \textit{golden
rule} \cite{fmr014}:%

\begin{equation}
\underline{%
%TCIMACRO{\TeXButton{h}{\slh}}%
%BeginExpansion
\slh
%EndExpansion
}(X\underset{%
%TCIMACRO{\TeXButton{sig}{\sitg}}%
%BeginExpansion
\sitg
%EndExpansion
}{\ast}Y)=\underline{%
%TCIMACRO{\TeXButton{h}{\slh}}%
%BeginExpansion
\slh
%EndExpansion
}(X)\underset{\eta}{\ast}\underline{%
%TCIMACRO{\TeXButton{h}{\slh}}%
%BeginExpansion
\slh
%EndExpansion
}(Y), \label{4.4a}%
\end{equation}
where $\ast$ denotes here as previously agreed any one of the multiform products.

\paragraph{Hodge Star Operators}

The canonical Hodge star operator is defined by%
\begin{align}
\star &  :%
%TCIMACRO{\dbigwedge \nolimits^{p}}%
%BeginExpansion
{\displaystyle\bigwedge\nolimits^{p}}
%EndExpansion
U\rightarrow%
%TCIMACRO{\dbigwedge \nolimits^{4-p}}%
%BeginExpansion
{\displaystyle\bigwedge\nolimits^{4-p}}
%EndExpansion
U,\nonumber\\
X  &  \mapsto\star X:=\tilde{X}\lrcorner\tau=\tilde{X}\tau, \label{4.4b}%
\end{align}
where $\tau$ is the canonical volume element\ $\tau=%
%TCIMACRO{\TeXButton{beta}{\mbox{\boldmath{$\beta$}}}}%
%BeginExpansion
\mbox{\boldmath{$\beta$}}%
%EndExpansion
^{0}\wedge%
%TCIMACRO{\TeXButton{beta}{\mbox{\boldmath{$\beta$}}}}%
%BeginExpansion
\mbox{\boldmath{$\beta$}}%
%EndExpansion
^{1}\wedge%
%TCIMACRO{\TeXButton{beta}{\mbox{\boldmath{$\beta$}}}}%
%BeginExpansion
\mbox{\boldmath{$\beta$}}%
%EndExpansion
^{2}\wedge%
%TCIMACRO{\TeXButton{beta}{\mbox{\boldmath{$\beta$}}}}%
%BeginExpansion
\mbox{\boldmath{$\beta$}}%
%EndExpansion
^{3}$. Then the Hodge star operators associated to the metric extensors $\eta$
and $%
%TCIMACRO{\TeXButton{itg}{\itg}}%
%BeginExpansion
\itg
%EndExpansion
$ are defined by:
\begin{align}
\underset{\eta}{\star}  &  :%
%TCIMACRO{\dbigwedge \nolimits^{p}}%
%BeginExpansion
{\displaystyle\bigwedge\nolimits^{p}}
%EndExpansion
U\rightarrow%
%TCIMACRO{\dbigwedge \nolimits^{4-p}}%
%BeginExpansion
{\displaystyle\bigwedge\nolimits^{4-p}}
%EndExpansion
U\text{, \ \ }\underset{%
%TCIMACRO{\TeXButton{sig}{\sitg}}%
%BeginExpansion
\sitg
%EndExpansion
}{\star}:%
%TCIMACRO{\dbigwedge \nolimits^{p}}%
%BeginExpansion
{\displaystyle\bigwedge\nolimits^{p}}
%EndExpansion
U\rightarrow%
%TCIMACRO{\dbigwedge \nolimits^{4-p}}%
%BeginExpansion
{\displaystyle\bigwedge\nolimits^{4-p}}
%EndExpansion
U,\label{4.4c}\\
\underset{\eta}{\star}X  &  :=\tilde{X}\lrcorner\tau_{\eta}=\tilde{X}%
\tau_{\eta}\text{, \ \ \ \ \ \ \ \ \ \ \ \ \ \ }\underset{%
%TCIMACRO{\TeXButton{sig}{\sitg}}%
%BeginExpansion
\sitg
%EndExpansion
}{\star}X:=\tilde{X}\lrcorner\tau_{%
%TCIMACRO{\TeXButton{itg}{\sitg}}%
%BeginExpansion
\sitg
%EndExpansion
}=\tilde{X}\tau_{%
%TCIMACRO{\TeXButton{itg}{\sitg}}%
%BeginExpansion
\sitg
%EndExpansion
},\nonumber
\end{align}
where
\begin{equation}
\tau_{\eta}=\sqrt{\left\vert \det\eta\right\vert }\tau\text{,\ }\tau_{%
%TCIMACRO{\TeXButton{itg}{\sitg}}%
%BeginExpansion
\sitg
%EndExpansion
}=\sqrt{\left\vert \det%
%TCIMACRO{\TeXButton{itg}{\itg}}%
%BeginExpansion
\itg
%EndExpansion
\right\vert }\tau. \label{4.4d}%
\end{equation}
\textbf{ }

\paragraph{Relation Between the Hodge Star Operators of $%
%TCIMACRO{\TeXButton{itg}{\itg}}%
%BeginExpansion
\itg
%EndExpansion
$ and the Canonical Hodge Star Operator}

We then have the following nontrivial results relating $\underset{%
%TCIMACRO{\TeXButton{sig}{\sitg}}%
%BeginExpansion
\sitg
%EndExpansion
}{\star}$ and $\star$. For any $X\in%
%TCIMACRO{\dbigwedge }%
%BeginExpansion
{\displaystyle\bigwedge}
%EndExpansion
V,$%
\begin{equation}
\underset{%
%TCIMACRO{\TeXButton{sig}{\sitg}}%
%BeginExpansion
\sitg
%EndExpansion
}{\star}X=\frac{1}{\sqrt{\left\vert \det%
%TCIMACRO{\TeXButton{itg}{\itg}}%
%BeginExpansion
\itg
%EndExpansion
\right\vert }}\underline{%
%TCIMACRO{\TeXButton{itg}{\itg}}%
%BeginExpansion
\itg
%EndExpansion
}(\tilde{X}\lrcorner\tau)=\frac{1}{\sqrt{\left\vert \det%
%TCIMACRO{\TeXButton{itg}{\itg}}%
%BeginExpansion
\itg
%EndExpansion
\right\vert }}\underline{%
%TCIMACRO{\TeXButton{itg}{\itg}}%
%BeginExpansion
\itg
%EndExpansion
}\circ\star X, \label{4.4.ee}%
\end{equation}

\paragraph{Relation Between the Hodge Star Operators of $%
%TCIMACRO{\TeXButton{itg}{\itg}}%
%BeginExpansion
\itg
%EndExpansion
$ and $\eta$}

Moreover, putting%
\begin{equation}%
%TCIMACRO{\TeXButton{h}{\slh}}%
%BeginExpansion
\slh
%EndExpansion
^{\clubsuit}:=%
%TCIMACRO{\TeXButton{h}{\slh}}%
%BeginExpansion
\slh
%EndExpansion
^{\dagger-1}, \label{4.4hh}%
\end{equation}
the following result is valid when $%
%TCIMACRO{\TeXButton{itg}{\itg}}%
%BeginExpansion
\itg
%EndExpansion
$ and $\eta$ has the same signature
\begin{equation}
\underset{%
%TCIMACRO{\TeXButton{sig}{\sitg}}%
%BeginExpansion
\sitg
%EndExpansion
}{\star}=\mathrm{sgn}(\det%
%TCIMACRO{\TeXButton{h}{\slh}}%
%BeginExpansion
\slh
%EndExpansion
)\underline{%
%TCIMACRO{\TeXButton{h}{\slh}}%
%BeginExpansion
\slh
%EndExpansion
}^{\dagger}\circ\underset{\eta}{\star}\circ\underline{%
%TCIMACRO{\TeXButton{h}{\slh}}%
%BeginExpansion
\slh
%EndExpansion
}^{\clubsuit}, \label{4.4e}%
\end{equation}
which we denoted for simplicity by $\underset{%
%TCIMACRO{\TeXButton{sig}{\sitg}}%
%BeginExpansion
\sitg
%EndExpansion
}{\star}=%
%TCIMACRO{\TeXButton{h}{\slh}}%
%BeginExpansion
\slh
%EndExpansion
^{\dagger}\underset{\eta}{\star}%
%TCIMACRO{\TeXButton{h}{\slh}}%
%BeginExpansion
\slh
%EndExpansion
^{\clubsuit}$. Moreover if we suppose that $%
%TCIMACRO{\TeXButton{h}{\slh}}%
%BeginExpansion
\slh
%EndExpansion
$ is continuously connected to the identity extensor which has determinant
equal to $1$ we can write
\begin{equation}
\underset{%
%TCIMACRO{\TeXButton{sig}{\sitg}}%
%BeginExpansion
\sitg
%EndExpansion
}{\star}=\underline{%
%TCIMACRO{\TeXButton{h}{\slh}}%
%BeginExpansion
\slh
%EndExpansion
}^{\dagger}\underset{\eta}{\star}\underline{%
%TCIMACRO{\TeXButton{h}{\slh}}%
%BeginExpansion
\slh
%EndExpansion
}^{\clubsuit} \label{4.4ee}%
\end{equation}

\paragraph{Useful Identities}

We shall list some identities involving contractions, exterior product and the
Hodge star operator that will be need for some derivations in Section 6 and
Appendix D and E. These are

For any metric $%
%TCIMACRO{\TeXButton{sig}{\sitg}}%
%BeginExpansion
\sitg
%EndExpansion
$ and for any $A_{r}\in\bigwedge^{r}U$ and $B_{s}\in\bigwedge^{s}U$,
$r,s\geq0$:
\begin{equation}%
\begin{array}
[c]{l}%
A_{r}\wedge\text{\ }\underset{%
%TCIMACRO{\TeXButton{sig}{\sitg}}%
%BeginExpansion
\sitg
%EndExpansion
}{\star}B_{s}=B_{s}\wedge\text{\ }\underset{%
%TCIMACRO{\TeXButton{sig}{\sitg}}%
%BeginExpansion
\sitg
%EndExpansion
}{\star}A_{r};\quad r=s\\
A_{r}\underset{%
%TCIMACRO{\TeXButton{sig}{\sitg}}%
%BeginExpansion
\sitg
%EndExpansion
^{-1}}{\cdot}\text{\ }\underset{%
%TCIMACRO{\TeXButton{sig}{\sitg}}%
%BeginExpansion
\sitg
%EndExpansion
}{\star}B_{s}=B_{s}\cdot\text{\ }\underset{%
%TCIMACRO{\TeXButton{sig}{\sitg}}%
%BeginExpansion
\sitg
%EndExpansion
}{\star}A_{r};\quad r+s=n\\
A_{r}\wedge\text{\ }\underset{%
%TCIMACRO{\TeXButton{sig}{\sitg}}%
%BeginExpansion
\sitg
%EndExpansion
}{\star}B_{s}=(-1)^{r(s-1)}\text{\ }\underset{%
%TCIMACRO{\TeXButton{sig}{\sitg}}%
%BeginExpansion
\sitg
%EndExpansion
}{\star}(\tilde{A}_{r}\underset{%
%TCIMACRO{\TeXButton{sig}{\sitg}}%
%BeginExpansion
\sitg
%EndExpansion
^{-1}}{\lrcorner}B_{s});\quad r\leq s\\
A_{r}\lrcorner\text{\ }\underset{%
%TCIMACRO{\TeXButton{sig}{\sitg}}%
%BeginExpansion
\sitg
%EndExpansion
}{\star}B_{s}=(-1)^{rs}\text{\ }\underset{%
%TCIMACRO{\TeXButton{sig}{\sitg}}%
%BeginExpansion
\sitg
%EndExpansion
}{\star}(\tilde{A}_{r}\wedge B_{s});\quad r+s\leq n\\
\underset{%
%TCIMACRO{\TeXButton{sig}{\sitg}}%
%BeginExpansion
\sitg
%EndExpansion
}{\star}A_{r}=\tilde{A}_{r}\underset{%
%TCIMACRO{\TeXButton{sig}{\sitg}}%
%BeginExpansion
\sitg
%EndExpansion
^{-1}}{\lrcorner}\tau_{\overset{\circ}{%
%TCIMACRO{\TeXButton{sig}{\sitg}}%
%BeginExpansion
\sitg
%EndExpansion
}}=\tilde{A}_{r}\;\tau_{%
%TCIMACRO{\TeXButton{sig}{\sitg}}%
%BeginExpansion
\sitg
%EndExpansion
}\\
\underset{%
%TCIMACRO{\TeXButton{sig}{\sitg}}%
%BeginExpansion
\sitg
%EndExpansion
}{\star}\tau_{%
%TCIMACRO{\TeXButton{sig}{\sitg}}%
%BeginExpansion
\sitg
%EndExpansion
}=\mathrm{sgn}%
%TCIMACRO{\TeXButton{itg}{\itg}}%
%BeginExpansion
\itg
%EndExpansion
;\quad\underset{%
%TCIMACRO{\TeXButton{sig}{\sitg}}%
%BeginExpansion
\sitg
%EndExpansion
}{\star}1=\tau_{%
%TCIMACRO{\TeXButton{sig}{\sitg}}%
%BeginExpansion
\sitg
%EndExpansion
}.
\end{array}
\label{440new}%
\end{equation}

\bigskip

Finally, we shall need also the following result. Let $\{\varepsilon^{\mu}\}$
be an arbitrary basis of $U$ such that $\varepsilon^{\mu}\underset{%
%TCIMACRO{\TeXButton{sig}{\sitg}}%
%BeginExpansion
\sitg
%EndExpansion
^{-1}}{\cdot}$ $\varepsilon^{\mu}=g^{\mu\nu}$. Then, if
\[
\varepsilon^{\mu_{1}...\mu_{p}}=\varepsilon^{\mu_{1}}\wedge...\wedge
\varepsilon^{\mu_{p}},\varepsilon^{\nu_{p+1}...\nu_{n}}=\varepsilon^{\nu
_{p+1}}\wedge...\wedge\varepsilon^{\nu_{n}}%
\]
we have, with $g^{\mu\nu}g_{\mu\alpha}=\delta_{\alpha}^{\nu},$
\begin{equation}
\underset{%
%TCIMACRO{\TeXButton{sig}{\sitg}}%
%BeginExpansion
\sitg
%EndExpansion
}{\star}\varepsilon^{\mu_{1}...\mu_{p}}=\frac{1}{(n-p)!}\sqrt{\left\vert
\det(g_{\mu\nu})\right\vert }g^{\mu_{1}\nu_{1}}...g^{\mu_{p}\nu_{p}%
}\varepsilon_{\nu_{1}...\nu_{n}}\varepsilon^{\nu_{p+1}...\nu_{n}}, \label{fhd}%
\end{equation}
where $\det(g_{\mu\nu})$ is the determinant of the matrix with entries
$g_{\mu\nu}$.

\section{Multiform Functions and Multiform Functionals}

\subsection{Multiform Functions of Real Variable}

A mapping%
\begin{equation}
X:S\rightarrow%
%TCIMACRO{\dbigwedge }%
%BeginExpansion
{\displaystyle\bigwedge}
%EndExpansion
V\text{, }(S\subseteq\mathbb{R}),
\end{equation}
is called a \textit{multiform function of real variable}.

For simplicity reasons\ when the image of $X$ is a scalar, a $1$-form, a
biform, etc., $X$ is said a scalar function, $1$-form function, biform
function, etc.

\subsubsection{Limit and Continuity}

The concepts of limit and continuity may be easily introduced following a path
analog to the one used in the theory of ordinary functions

Then, a multiform $L\in%
%TCIMACRO{\dbigwedge }%
%BeginExpansion
{\displaystyle\bigwedge}
%EndExpansion
V$ is said to be the limit of\emph{ }$X(\lambda)$\emph{\ for }$\lambda\in
S$\emph{\ }approaching\emph{ }$\lambda_{0}\in S$ iff for any real number
$\varepsilon>0$ there exists as real number $\delta>0$ such that\footnote{For
any $A\in%
%TCIMACRO{\dbigwedge }%
%BeginExpansion
{\displaystyle\bigwedge}
%EndExpansion
V$, $\left\Vert X\right\Vert :=\sqrt{\langle X\cdot X\rangle_{0}}$}
$\left\Vert X(\lambda)-L\right\Vert <\varepsilon,$ if $0<\left\vert
\lambda-\lambda_{0}\right\vert <\delta.$ As usual, such a concept will be
denoted by $\underset{\lambda\rightarrow\lambda_{0}}{\lim}X(\lambda)=L$.

The theorems for the limits of multiform functions are completely analogous to
the ones of the theory of ordinary functions, i.e.,
\begin{align}
\underset{\lambda\rightarrow\lambda_{0}}{\lim}X(\lambda)+Y(\lambda)  &
=\underset{\lambda\rightarrow\lambda_{0}}{\lim}X(\lambda)+\underset{\lambda
\rightarrow\lambda_{0}}{\lim}Y(\lambda).\label{2.1}\\
\underset{\lambda\rightarrow\lambda_{0}}{\lim}X(\lambda)\ast Y(\lambda)  &
=\underset{\lambda\rightarrow\lambda_{0}}{\lim}X(\lambda)\ast\underset{\lambda
\rightarrow\lambda_{0}}{\lim}Y(\lambda), \label{2.2}%
\end{align}
where $\ast$ denotes any one of the multiform products, i.e., $(\wedge),$
$(\cdot),$ $(\lrcorner,\llcorner)$ or $($\emph{Clifford product}$).$

$X(\lambda)$ is said to be \textit{continuous} at \emph{ }$\lambda_{0}\in S$
iff $\underset{\lambda\rightarrow\lambda_{0}}{\lim}X(\lambda)=X(\lambda_{0})$
and sum and product of continuous multiform functions are continuous.

\subsubsection{Derivative}

The notion of derivative of a multiform function of a real variable is also
formulated in completely analogy to the one used in the theory of ordinary
functions. Then the derivative of $X(\lambda)$\emph{\ at }$\lambda_{0}%
$\emph{\ }is defined by
\begin{equation}
X^{\prime}(\lambda_{0}):=\underset{\lambda\rightarrow\lambda_{0}}{\lim}%
\dfrac{X(\lambda)-X(\lambda_{0})}{\lambda-\lambda_{0}}. \label{2.3}%
\end{equation}

The rules for derivations as expected are completely analogous to the ones
valid for ordinary functions.
\begin{align}
(X+Y)^{\prime}  &  =X^{\prime}+Y^{\prime},\label{2.4}\\
(X\ast Y)^{\prime}  &  =X^{\prime}\ast Y+X\ast Y^{\prime}\text{ (Leibniz's
rule),}\label{2.5}\\
(X\circ\phi)^{\prime}  &  =(X^{\prime}\circ\phi)\phi^{\prime}\text{ (chain
rule),} \label{2.6}%
\end{align}
where $\ast$, as above means $(\wedge),$ $(\cdot),$ $(\lrcorner,\llcorner)$ or
(\emph{Clifford product}) and $\phi$ is a real ordinary function
($\phi^{\prime}$ is the derivative of $\phi$).

In the space of derivable multiform functions of real variable, we introduce
the symbol $\dfrac{d}{d\lambda}$ (derivative operator), by $\dfrac{d}%
{d\lambda}X(\lambda)=X^{\prime}(\lambda).$

We can easily generalize all the above rules for multiform functions of
several real variables, so that needs no additional comments. Now we will
introduce the real important objects need in this paper.

\subsection{Multiform Functions of Multiform Variables}

A mapping $F:%
%TCIMACRO{\dbigwedge \nolimits^{\diamond}}%
%BeginExpansion
{\displaystyle\bigwedge\nolimits^{\diamond}}
%EndExpansion
V\rightarrow%
%TCIMACRO{\dbigwedge }%
%BeginExpansion
{\displaystyle\bigwedge}
%EndExpansion
V$ is called a \textit{multiform function of a multiform variable} and when
$F(X)$ is a scalar, a $1$-form, a biform, etc., then $F$ is said to be \emph{a
scalar function, a }$1$\emph{-form function, a biform function, etc.}

\subsubsection{Limit and Continuity}

Let $F:%
%TCIMACRO{\dbigwedge \nolimits^{\diamond}}%
%BeginExpansion
{\displaystyle\bigwedge\nolimits^{\diamond}}
%EndExpansion
V\rightarrow%
%TCIMACRO{\dbigwedge }%
%BeginExpansion
{\displaystyle\bigwedge}
%EndExpansion
V$. A \emph{multiform }$M\in$\emph{\ }$%
%TCIMACRO{\dbigwedge }%
%BeginExpansion
{\displaystyle\bigwedge}
%EndExpansion
V$ is said to be the limit of\emph{ }$F(X)$\emph{\ for }$X\in%
%TCIMACRO{\dbigwedge }%
%BeginExpansion
{\displaystyle\bigwedge}
%EndExpansion
V$ \emph{approaching }$X_{0}\in%
%TCIMACRO{\dbigwedge }%
%BeginExpansion
{\displaystyle\bigwedge}
%EndExpansion
V$ iff for any real $\varepsilon>0$ there exists a real $\delta>0$ such that
$\left\Vert F(X)-M\right\Vert <\varepsilon,$ if $0<\left\Vert X-X_{0}%
\right\Vert <\delta$. This will be denoted by $\underset{X\rightarrow
X_{0}}{\lim}F(X)=M$.

When we are considering scalar functions of multiform variables $%
%TCIMACRO{\dbigwedge \nolimits^{\diamond}}%
%BeginExpansion
{\displaystyle\bigwedge\nolimits^{\diamond}}
%EndExpansion
V\ni X\mapsto\Phi(X)\in\mathbb{R},$ the definition $\underset{X\rightarrow
X_{0}}{\lim}\Phi(X)=\alpha$ is simply the statement: for any real
$\varepsilon>0$ there exists a real $\delta>0$ such that $\left\vert
\Phi(X)-\alpha\right\vert <\varepsilon,$ if $0<\left\Vert X-X_{0}\right\Vert
<\delta$.

The limit theorems are analogous to the ones of the theory of ordinary
functions, i.e.,
\begin{align}
\underset{X\rightarrow X_{0}}{\lim}F(X)+G(X)  &  =\underset{X\rightarrow
X_{0}}{\lim}F(X)+\underset{X\rightarrow X_{0}}{\lim}G(X).\label{2.7}\\
\underset{X\rightarrow X_{0}}{\lim}F(X)\ast G(X)  &  =\underset{X\rightarrow
X_{0}}{\lim}F(X)\ast\underset{X\rightarrow X_{0}}{\lim}G(X), \label{2.8}%
\end{align}
where $\ast$ means as now usual $(\wedge),$ $(\cdot),$ $(\lrcorner,\llcorner)$
or the $($\emph{Clifford product}$)$.

Let $F:%
%TCIMACRO{\dbigwedge \nolimits^{\diamond}}%
%BeginExpansion
{\displaystyle\bigwedge\nolimits^{\diamond}}
%EndExpansion
V\rightarrow%
%TCIMACRO{\dbigwedge }%
%BeginExpansion
{\displaystyle\bigwedge}
%EndExpansion
V$ \ be a multiform function of multiform variable. We say that $F$ is
continuous at $X_{0}\in%
%TCIMACRO{\dbigwedge }%
%BeginExpansion
{\displaystyle\bigwedge}
%EndExpansion
V$ iff $\underset{X\rightarrow X_{0}}{\lim}F(X)=F(X_{0})$.

The sum $X\mapsto(F+G)(X)=F(X)+G(X)$ and any product of \ continuous multiform
functions of multiform variable $X\mapsto(F\ast G)(X)=F(X)\ast G(X)$ are also
continuous multiform functions of multiform variable.

\subsubsection{Differentiability}

A multiform function of multiform variable $F:%
%TCIMACRO{\dbigwedge \nolimits^{\diamond}}%
%BeginExpansion
{\displaystyle\bigwedge\nolimits^{\diamond}}
%EndExpansion
V\rightarrow%
%TCIMACRO{\dbigwedge }%
%BeginExpansion
{\displaystyle\bigwedge}
%EndExpansion
V$ \ is said differentiable at $X_{0}\in%
%TCIMACRO{\dbigwedge }%
%BeginExpansion
{\displaystyle\bigwedge}
%EndExpansion
V$ iff it exists $f_{X_{0}}\in extV$, such that
\begin{equation}
\underset{X\rightarrow X_{0}}{\lim}\dfrac{F(X)-F(X_{0})-f_{X_{0}}(X-X_{0}%
)}{\left\Vert X-X_{0}\right\Vert }=0.
\end{equation}

If such $f_{X_{0}}$ \emph{exists}, \ it must be unique and will be called the
\textit{differential }of \emph{ }$F$\emph{\ at }$X_{0}.$

\subsubsection{The Directional Derivative $A\cdot\partial_{X}$}

Let $F:%
%TCIMACRO{\dbigwedge \nolimits^{\diamond}}%
%BeginExpansion
{\displaystyle\bigwedge\nolimits^{\diamond}}
%EndExpansion
V\rightarrow%
%TCIMACRO{\dbigwedge }%
%BeginExpansion
{\displaystyle\bigwedge}
%EndExpansion
V$ \ be differentiable at $X$. Take a real number $\lambda\neq0$ and an
arbitrary multiform $A$. Then the limit%
\begin{equation}
\underset{\lambda\rightarrow0}{\lim}\dfrac{F(X+\lambda\left\langle
A\right\rangle _{X})-F(X)}{\lambda}%
\end{equation}
exists. It will be denoted by $F_{A}^{\prime}(X)$ (or, $A\cdot\partial
_{X}F(X)$)\footnote{For some few special cases we use yet some special
symbols.} and called the directional derivative of\emph{ }$F$\emph{\ at }%
$X$\emph{\ in the direction of the multiform} $A$. The operator $A\cdot
\partial_{X}$ is said to be the \emph{directional derivative operator in the
direction of} $A$. We have:
\begin{equation}
F_{A}^{\prime}(X)=A\cdot\partial_{X}F(X)=\underset{\lambda\rightarrow0}{\lim
}\dfrac{F(X+\lambda\left\langle A\right\rangle _{X})-F(X)}{\lambda},
\label{2.9}%
\end{equation}
or in an equivalent form
\begin{equation}
F_{A}^{\prime}(X)=A\cdot\partial_{X}F(X)=\left.  \frac{d}{d\lambda}%
F(X+\lambda\left\langle A\right\rangle _{X})\right\vert _{\lambda=0}.
\tag{2.10}%
\end{equation}

The directional derivative of a differentiable multiform function $F:%
%TCIMACRO{\dbigwedge \nolimits^{\diamond}}%
%BeginExpansion
{\displaystyle\bigwedge\nolimits^{\diamond}}
%EndExpansion
V\rightarrow%
%TCIMACRO{\dbigwedge }%
%BeginExpansion
{\displaystyle\bigwedge}
%EndExpansion
V$ is linear with respect to the direction multiform, i.e., for all
$\alpha,\beta\in\mathbb{R}$ and $A,B\in%
%TCIMACRO{\dbigwedge }%
%BeginExpansion
{\displaystyle\bigwedge}
%EndExpansion
V$
\begin{equation}
F_{\alpha A+\beta B}^{\prime}(X)=\alpha F_{A}^{\prime}(X)+\beta F_{B}^{\prime
}(X), \label{2.11}%
\end{equation}
or yet
\begin{equation}
(\alpha A+\beta B)\cdot\partial_{X}F(X)=\alpha A\cdot\partial_{X}F(X)+\beta
A\cdot\partial_{X}F(X). \label{2.12}%
\end{equation}
\medskip

We give now the main properties of the directional derivative.

Let $F,G:%
%TCIMACRO{\dbigwedge \nolimits^{\diamond}}%
%BeginExpansion
{\displaystyle\bigwedge\nolimits^{\diamond}}
%EndExpansion
V\rightarrow%
%TCIMACRO{\dbigwedge }%
%BeginExpansion
{\displaystyle\bigwedge}
%EndExpansion
V$ be differentiable, the sum $X\mapsto(F+G)(X)=F(X)+G(X)$ and the products
$X\mapsto(F\ast G)(X)=F(X)\ast G(X)$, where $\ast$ means $(\wedge),$
$(\cdot),(\lrcorner,\llcorner)$ or (\emph{Clifford product}), are also
differentiable and we have
\begin{align}
(F+G)_{A}^{\prime}(X)  &  =F_{A}^{\prime}(X)+G_{A}^{\prime}(X),\label{2.13}\\
(F\ast G)_{A}^{\prime}(X)  &  =F_{A}^{\prime}(X)\ast G(X)+F(X)\ast
G_{A}^{\prime}(X)\text{ (Leibniz's rule).} \label{2.14}%
\end{align}
\medskip

\subsubsection{Chain Rules}

Let $F,G:%
%TCIMACRO{\dbigwedge \nolimits^{\diamond}}%
%BeginExpansion
{\displaystyle\bigwedge\nolimits^{\diamond}}
%EndExpansion
V\rightarrow%
%TCIMACRO{\dbigwedge }%
%BeginExpansion
{\displaystyle\bigwedge}
%EndExpansion
V$ be differentiable. The composition $X\mapsto(F\circ G)(X)=F(\left\langle
G(X)\right\rangle _{Y})$, with $Y\in\mathrm{dom}F$ is differentiable and
\begin{equation}
(F\circ G)_{A}^{\prime}(X)=F_{G_{A}^{\prime}(X)}^{\prime}(\left\langle
G(X)\right\rangle _{Y}). \label{2.15}%
\end{equation}

Let $X\mapsto F(X)$ and $\lambda\mapsto X(\lambda)$ be differentiable
multiform functions, the first one of multiform variable and the second of
real variable. The composition $\lambda\mapsto(F\circ X)(\lambda
)=F(\left\langle X(\lambda)\right\rangle _{Y})$, with $Y\in\mathrm{dom}F$ is a
differentiable multiform function of real variable and
\begin{equation}
(F\circ X)^{\prime}(\lambda)=F_{X^{\prime}(\lambda)}^{\prime}(\left\langle
X(\lambda)\right\rangle _{Y}). \label{2.16}%
\end{equation}

Let $\phi:\mathbb{R}\rightarrow\mathbb{R}$ and $\Psi:%
%TCIMACRO{\dbigwedge \nolimits^{\diamond}}%
%BeginExpansion
{\displaystyle\bigwedge\nolimits^{\diamond}}
%EndExpansion
V\rightarrow\mathbb{R}$ \ respectively an ordinary differentiable function and
a scalar differentiable multiform function of multiform variable. The
composition $\phi\circ\Psi:%
%TCIMACRO{\dbigwedge \nolimits^{\diamond}}%
%BeginExpansion
{\displaystyle\bigwedge\nolimits^{\diamond}}
%EndExpansion
V\rightarrow R$ such that $(\phi\circ\Psi)(X)=\phi(\Psi(X))$ is differentiable
multiform function of multiform variable and
\begin{equation}
(\phi\circ\Psi)_{A}^{\prime}(X)=\phi^{\prime}(\Psi(X))\Psi_{A}^{\prime}(X).
\label{2.17}%
\end{equation}

In resume we have:
\begin{align}
A\cdot\partial_{X}(F+G)(X)  &  =A\cdot\partial_{X}F(X)+A\cdot\partial
_{X}G(X).\label{2.18}\\
A\cdot\partial_{X}(F\ast G)(X)  &  =A\cdot\partial_{X}F(X)\ast G(X)+F(X)\ast
A\cdot\partial_{X}G(X).\label{2.19}\\
A\cdot\partial_{X}(F\circ G)(X)  &  =A\cdot\partial_{X}G(X)\cdot\partial
_{Y}F(\left\langle G(X)\right\rangle _{Y}),\text{ with }Y\in domF.
\label{2.20}\\
\frac{d}{d\lambda}(F\circ X)(\lambda)  &  =\frac{d}{d\lambda}X(\lambda
)\cdot\partial_{Y}F(\left\langle X(\lambda)\right\rangle _{Y}),\text{ with
}Y\in domF.\label{2.21}\\
A\cdot\partial_{X}(\phi\circ\Psi)(X)  &  =\frac{d}{d\mu}\phi(\Psi
(X))A\cdot\partial_{X}\Psi(X). \label{2.22}%
\end{align}
\vspace*{0.3in}

\subsection{The Derivative Mapping $\partial_{X}$}

Let $F:%
%TCIMACRO{\dbigwedge \nolimits^{\diamond}}%
%BeginExpansion
{\displaystyle\bigwedge\nolimits^{\diamond}}
%EndExpansion
V\rightarrow%
%TCIMACRO{\dbigwedge }%
%BeginExpansion
{\displaystyle\bigwedge}
%EndExpansion
V$ \ be differentiable at $X,$ Take any pair of arbitrary reciprocal basis for
$V$, say $(\{\varepsilon^{j}\},\{\varepsilon_{j}\})$, ( $\varepsilon^{j}%
\cdot\varepsilon_{i}=\delta_{i}^{j}$). The it exists a well defined multiform
function (i.e., it is independent of the pair $(\{\varepsilon^{j}%
\},\{\varepsilon_{j}\})$ depending only on $F$) called the derivative of
$F\emph{\ at}$ $X_{0}$ and given by%

\begin{equation}
F^{\prime}(X)=\partial_{X}F(X):=\underset{J}{%
%TCIMACRO{\dsum }%
%BeginExpansion
{\displaystyle\sum}
%EndExpansion
}\dfrac{1}{\nu(J)!}\varepsilon^{J}F_{\varepsilon_{J}}^{\prime}%
(X)=\underset{J}{%
%TCIMACRO{\dsum }%
%BeginExpansion
{\displaystyle\sum}
%EndExpansion
}\dfrac{1}{\nu(J)!}\varepsilon_{J}F_{\varepsilon^{J}}^{\prime}(X).
\label{2.23}%
\end{equation}

The main properties of $\partial_{X}F$ are.

\textbf{(i)} Let $X\mapsto\Phi(X)$ be a scalar multiform function, then
\begin{equation}
A\cdot\partial_{X}\Phi(X)=A\cdot(\partial_{X}\Phi(X)). \label{2.24}%
\end{equation}

\textbf{(ii)} Let $X\mapsto F(X),$ $X\mapsto G(X)$ and $X\mapsto\Phi(X)$ be
differentiable multiform functions. Then
\begin{align}
\partial_{X}(F+G)(X)  &  =\partial_{X}F(X)+\partial_{X}G(X).\label{2.25}\\
\partial_{X}(\Phi G)(X)  &  =\partial_{X}\Phi(X)G(X)+\Phi(X)\partial
_{X}G(X)\text{ (Leibniz's rule).} \label{2.26}%
\end{align}

\subsection{Examples}

We present now some example which illustrate some of the most important
formulas of the calculus of multiform functions of multiform variables and
which will be used in the following sections.

\textbf{Example 3.1}\textit{\ }Let $%
%TCIMACRO{\dbigwedge \nolimits^{\diamond}}%
%BeginExpansion
{\displaystyle\bigwedge\nolimits^{\diamond}}
%EndExpansion
V\ni X\mapsto X\cdot X\in\mathbb{R}$. $%
%TCIMACRO{\dbigwedge }%
%BeginExpansion
{\displaystyle\bigwedge}
%EndExpansion
V.$Let us calculate $A\cdot\partial_{X}(X\cdot X)$ and $\partial_{X}(X\cdot
X).$
\begin{subequations}
\label{2.27a}%
\begin{align}
A\cdot\partial_{X}(X\cdot X)  &  =\left.  \frac{d}{d\lambda}(X+\lambda
\left\langle A\right\rangle _{X})\cdot(X+\lambda\left\langle A\right\rangle
_{X})\right\vert _{\lambda=0}\nonumber\\
&  =\left.  \frac{d}{d\lambda}(X\cdot X+2\lambda\left\langle A\right\rangle
_{X}\cdot X+\lambda^{2}\left\langle A\right\rangle _{X}\cdot\left\langle
A\right\rangle _{X})\right\vert _{\lambda=0},\nonumber\\
A\cdot\partial_{X}(X\cdot X)  &  =2\left\langle A\right\rangle _{X}\cdot
X=2A\cdot X.\label{2.27aa}\\
\partial_{X}(X\cdot X)  &  =\underset{J}{%
%TCIMACRO{\tsum }%
%BeginExpansion
{\textstyle\sum}
%EndExpansion
}\frac{1}{\nu(J)!}\varepsilon^{J}\varepsilon_{J}\cdot\partial_{X}(X\cdot
X)=\underset{J}{%
%TCIMACRO{\tsum }%
%BeginExpansion
{\textstyle\sum}
%EndExpansion
}\frac{1}{\nu(J)!}\varepsilon^{J}2(\varepsilon_{J}\cdot X)=2X. \label{2.27ab}%
\end{align}

\textbf{Example 3.2} Let $%
%TCIMACRO{\dbigwedge \nolimits^{\diamond}}%
%BeginExpansion
{\displaystyle\bigwedge\nolimits^{\diamond}}
%EndExpansion
V\ni X\mapsto B\cdot X\in\mathbb{R}$, with $B\in%
%TCIMACRO{\dbigwedge }%
%BeginExpansion
{\displaystyle\bigwedge}
%EndExpansion
V$. Let us calculate $A\cdot\partial_{X}(B\cdot X)$ and $\partial_{X}(B\cdot
X)$.
\end{subequations}
\begin{subequations}
\label{2.28a}%
\begin{align}
A\cdot\partial_{X}(B\cdot X)  &  =\left.  \frac{d}{d\lambda}B\cdot
(X+\lambda\left\langle A\right\rangle _{X})\right\vert _{\lambda=0}%
=B\cdot\left\langle A\right\rangle _{X}=A\cdot\left\langle B\right\rangle
_{X},\label{2.28aa}\\
\partial_{X}(B\cdot X)  &  =\underset{J}{%
%TCIMACRO{\tsum }%
%BeginExpansion
{\textstyle\sum}
%EndExpansion
}\frac{1}{\nu(J)!}\varepsilon^{J}\varepsilon_{J}\cdot\partial_{X}(B\cdot
X)=\underset{J}{%
%TCIMACRO{\tsum }%
%BeginExpansion
{\textstyle\sum}
%EndExpansion
}\frac{1}{\nu(J)!}\varepsilon^{J}(\varepsilon_{J}\cdot\left\langle
B\right\rangle _{X})=\left\langle B\right\rangle _{X}. \label{2.28ab}%
\end{align}

\textbf{Example 3.3}\textit{\ }Let $%
%TCIMACRO{\dbigwedge \nolimits^{\diamond}}%
%BeginExpansion
{\displaystyle\bigwedge\nolimits^{\diamond}}
%EndExpansion
V\ni X\mapsto(BXC)\cdot X\in\mathbb{R}$, with $B,C\in%
%TCIMACRO{\dbigwedge }%
%BeginExpansion
{\displaystyle\bigwedge}
%EndExpansion
V$. Let us calculate: \newline$A\cdot\partial_{X}((BXC)\cdot X)$ and
$\partial_{X}((BXC)\cdot X)$.
\end{subequations}
\begin{subequations}
\label{2.29a}%
\begin{align}
A\cdot\partial_{X}((BXC)\cdot X)  &  =\left.  \frac{d}{d\lambda}%
(B(X+\lambda\left\langle A\right\rangle _{X})C)\cdot(X+\lambda\left\langle
A\right\rangle _{X})\right\vert _{\lambda=0}\nonumber\\
&  =%
\begin{array}
[c]{c}%
\left.  \dfrac{d}{d\lambda}((BXC)\cdot X+\lambda(B\left\langle A\right\rangle
_{X}C)\cdot X+\lambda(BXC)\cdot\left\langle A\right\rangle _{X})\right. \\
+\left.  \lambda^{2}(B\left\langle A\right\rangle _{X}C)\cdot\left\langle
A\right\rangle _{X})\right\vert _{\lambda=0}%
\end{array}
\nonumber\\
&  =(B\left\langle A\right\rangle _{X}C)\cdot X+(BXC)\cdot\left\langle
A\right\rangle _{X}=\left\langle A\right\rangle _{X}\cdot(BXC+\widetilde{B}%
X\widetilde{C}),\label{2.29aa}\\
A\cdot\partial_{X}((BXC)\cdot X)  &  =A\cdot\left\langle BXC+\widetilde{B}%
X\widetilde{C}\right\rangle _{X}.\nonumber\\
\partial_{X}((BXC)\cdot X)  &  =\underset{J}{%
%TCIMACRO{\tsum }%
%BeginExpansion
{\textstyle\sum}
%EndExpansion
}\frac{1}{\nu(J)!}\varepsilon^{J}\varepsilon_{J}\cdot\partial_{X}(BXC\cdot
X)\nonumber\\
&  =\underset{J}{%
%TCIMACRO{\tsum }%
%BeginExpansion
{\textstyle\sum}
%EndExpansion
}\frac{1}{\nu(J)!}\varepsilon^{J}(\varepsilon_{J}\cdot\left\langle
BXC+\widetilde{B}X\widetilde{C}\right\rangle _{X}),\nonumber\\
\partial_{X}((BXC)\cdot X)  &  =\left\langle BXC+\widetilde{B}X\widetilde{C}%
\right\rangle _{X}. \label{2.29ab}%
\end{align}

In this example we essentially utilized the nontrivial multiform identity
$(AXB)\cdot Y=X\cdot(\widetilde{A}Y\widetilde{B})$.

\textbf{Example 3.4}\textit{\ }Consider $x\wedge B$, with $x\in$ $%
%TCIMACRO{\dbigwedge }%
%BeginExpansion
{\displaystyle\bigwedge}
%EndExpansion
V$ and $B\in$ $%
%TCIMACRO{\dbigwedge \nolimits^{2}}%
%BeginExpansion
{\displaystyle\bigwedge\nolimits^{2}}
%EndExpansion
V$. Then $x\wedge B\in%
%TCIMACRO{\dbigwedge \nolimits^{3}}%
%BeginExpansion
{\displaystyle\bigwedge\nolimits^{3}}
%EndExpansion
V$. Let us calculate the directional derivative $a\cdot\partial_{x}(x\wedge
B)$, the \textit{divergent} $\partial_{x}\lrcorner(x\wedge B)$, the
\textit{rotational} $\partial_{x}\wedge(x\wedge B)$ and the gradient
$\partial_{x}(x\wedge B)$.
\end{subequations}
\begin{subequations}
\label{2.30a}%
\begin{align}
a\cdot\partial_{x}(x\wedge B)  &  =\left.  \frac{d}{d\lambda}(x+\lambda
a)\wedge B\right\vert _{\lambda=0}=a\wedge B,\label{2.30aa}\\
\partial_{x}\lrcorner(x\wedge B)  &  =\overset{n}{\underset{j=1}{%
%TCIMACRO{\tsum }%
%BeginExpansion
{\textstyle\sum}
%EndExpansion
}}\varepsilon^{j}\lrcorner\varepsilon_{j}\cdot\partial_{x}(x\wedge
B)=\overset{n}{\underset{j=1}{%
%TCIMACRO{\tsum }%
%BeginExpansion
{\textstyle\sum}
%EndExpansion
}}\varepsilon^{j}\lrcorner(\varepsilon_{j}\wedge B)=(n-2)B.\label{2.30ab}\\
\partial_{x}\wedge(x\wedge B)  &  =\overset{n}{\underset{j=1}{%
%TCIMACRO{\tsum }%
%BeginExpansion
{\textstyle\sum}
%EndExpansion
}}\varepsilon^{j}\wedge\varepsilon_{j}\cdot\partial_{x}(x\wedge
B)=\overset{n}{\underset{j=1}{%
%TCIMACRO{\tsum }%
%BeginExpansion
{\textstyle\sum}
%EndExpansion
}}\varepsilon^{j}\wedge(\varepsilon_{j}\wedge B_{k})=0.\label{2.30ac}\\
\partial_{x}(x\wedge B)  &  =\overset{n}{\underset{j=1}{%
%TCIMACRO{\tsum }%
%BeginExpansion
{\textstyle\sum}
%EndExpansion
}}\varepsilon^{j}\varepsilon_{j}\cdot\partial_{x}(x\wedge
B)=\overset{n}{\underset{j=1}{%
%TCIMACRO{\tsum }%
%BeginExpansion
{\textstyle\sum}
%EndExpansion
}}\varepsilon^{j}(\varepsilon_{j}\wedge B)\nonumber\\
&  =\overset{n}{\underset{j=1}{%
%TCIMACRO{\tsum }%
%BeginExpansion
{\textstyle\sum}
%EndExpansion
}}(\varepsilon^{j}\lrcorner(\varepsilon_{j}\wedge B)+\varepsilon^{j}%
\wedge(\varepsilon_{j}\wedge B)),\label{2.30ad}\\
\partial_{x}(x\wedge B)  &  =(n-2)B,\hspace{0.5in}(n=\dim V). \label{2.30af}%
\end{align}

\textbf{Example 3.5}\textit{\ }Consider $x\lrcorner$ $B\in%
%TCIMACRO{\dbigwedge \nolimits^{1}}%
%BeginExpansion
{\displaystyle\bigwedge\nolimits^{1}}
%EndExpansion
V$. Let us calculate $a\cdot\partial_{x}(x\lrcorner B)$, $\partial
_{x}\lrcorner(x\lrcorner B)$, $\partial_{x}\wedge(x\lrcorner B)$ and
$\partial_{x}(x\lrcorner B)$
\end{subequations}
\begin{subequations}
\label{2.31a}%
\begin{align}
a\cdot\partial_{x}(x\lrcorner B)  &  =\left.  \frac{d}{d\lambda}(x+\lambda
a)\lrcorner B\right\vert _{\lambda=0}=a\lrcorner B.\label{2.31aa}\\
\partial_{x}\lrcorner(x\lrcorner B)  &  =\overset{n}{\underset{j=1}{%
%TCIMACRO{\tsum }%
%BeginExpansion
{\textstyle\sum}
%EndExpansion
}}\varepsilon^{j}\lrcorner\varepsilon_{j}\cdot\partial_{x}(x\lrcorner
B)=\overset{n}{\underset{j=1}{%
%TCIMACRO{\tsum }%
%BeginExpansion
{\textstyle\sum}
%EndExpansion
}}\varepsilon^{j}\lrcorner(\varepsilon_{j}\lrcorner
B)=\overset{n}{\underset{j=1}{%
%TCIMACRO{\tsum }%
%BeginExpansion
{\textstyle\sum}
%EndExpansion
}(}\varepsilon^{j}\wedge\varepsilon_{j})\lrcorner\text{ }B_{k}%
=0,\label{2.31ab}\\
\partial_{x}\wedge(x\lrcorner B)  &  =\overset{n}{\underset{j=1}{%
%TCIMACRO{\tsum }%
%BeginExpansion
{\textstyle\sum}
%EndExpansion
}}\varepsilon^{j}\wedge\varepsilon_{j}\cdot\partial_{x}(x\lrcorner
B)=\overset{n}{\underset{j=1}{%
%TCIMACRO{\tsum }%
%BeginExpansion
{\textstyle\sum}
%EndExpansion
}}\varepsilon^{j}\wedge(\varepsilon_{j}\lrcorner B)=2B.\label{2.31ac}\\
\partial_{x}(x\lrcorner B)  &  =\overset{n}{\underset{j=1}{%
%TCIMACRO{\tsum }%
%BeginExpansion
{\textstyle\sum}
%EndExpansion
}}\varepsilon^{j}\varepsilon_{j}\cdot\partial_{x}(x\lrcorner
B)=\overset{n}{\underset{j=1}{%
%TCIMACRO{\tsum }%
%BeginExpansion
{\textstyle\sum}
%EndExpansion
}}\varepsilon^{j}(\varepsilon_{j}\lrcorner B)\label{2.31ad}\\
\partial_{x}(x\lrcorner B)  &  =\overset{n}{\underset{j=1}{%
%TCIMACRO{\tsum }%
%BeginExpansion
{\textstyle\sum}
%EndExpansion
}}(\varepsilon^{j}\lrcorner(\varepsilon_{j}\lrcorner B)+\varepsilon^{j}%
\wedge(\varepsilon_{j}\lrcorner B))=2B.\hspace{0.5in}\text{(}n=\dim V\text{)}
\label{2.31ae}%
\end{align}

\textbf{Example 3.6 Let }$X\in%
%TCIMACRO{\dbigwedge \nolimits^{r}}%
%BeginExpansion
{\displaystyle\bigwedge\nolimits^{r}}
%EndExpansion
V$ and consider the multiform function $F:%
%TCIMACRO{\dbigwedge \nolimits^{r}}%
%BeginExpansion
{\displaystyle\bigwedge\nolimits^{r}}
%EndExpansion
\ni X\mapsto X\wedge\star X\in%
%TCIMACRO{\dbigwedge \nolimits^{n}}%
%BeginExpansion
{\displaystyle\bigwedge\nolimits^{n}}
%EndExpansion
V$. Let us calculate the directional derivative $W\cdot\partial_{X}F$, with
$W\in%
%TCIMACRO{\dbigwedge \nolimits^{r}}%
%BeginExpansion
{\displaystyle\bigwedge\nolimits^{r}}
%EndExpansion
V$ and $\partial_{X}F$. We have
\end{subequations}
\begin{align}
W\cdot\partial_{X}F(X)  &  =\lim_{\lambda\rightarrow0}\frac{(X+\lambda
W)\wedge\star(X+\lambda W)-X\wedge\star X}{\lambda}\nonumber\\
&  =2W\wedge\star X. \label{2.31n}%
\end{align}
On the other hand using Eq.(\ref{2.23}), we have:
\begin{align*}
\partial_{X}F(X)  &  =%
%TCIMACRO{\dsum }%
%BeginExpansion
{\displaystyle\sum}
%EndExpansion
(\frac{1}{r!}\varepsilon^{j_{1}}\wedge...\wedge\varepsilon^{j_{r}%
})[(\varepsilon_{j_{1}}\wedge...\wedge\varepsilon_{j_{r}})\cdot\partial
_{X}F(X)]\\
&  =2%
%TCIMACRO{\dsum }%
%BeginExpansion
{\displaystyle\sum}
%EndExpansion
\frac{1}{r!}\varepsilon^{j_{1}}\wedge...\wedge\varepsilon^{j_{r}}%
[(\varepsilon_{j_{1}}\wedge...\wedge\varepsilon_{j_{r}})\wedge\star X]\\
&  =2%
%TCIMACRO{\dsum }%
%BeginExpansion
{\displaystyle\sum}
%EndExpansion
\frac{1}{r!}\varepsilon^{j_{1}}\wedge...\wedge\varepsilon^{j_{r}}%
(-1)^{r(r-1)}[\star(\widetilde{(\varepsilon_{j_{1}}\wedge...\wedge
\varepsilon_{j_{r}})}\lrcorner X))]\\
&  =2%
%TCIMACRO{\dsum }%
%BeginExpansion
{\displaystyle\sum}
%EndExpansion
\frac{1}{r!}\varepsilon^{j_{1}}\wedge...\wedge\varepsilon^{j_{r}}%
[(\varepsilon_{j_{1}}\wedge...\wedge\varepsilon_{j_{r}})\cdot X)\tau]\\
&  =2X\tau=2(-1)^{\frac{r}{2}(r-1)}\tilde{X}\lrcorner\tau=2(-1)^{\frac{r}%
{2}(r-1)}\star X.
\end{align*}
\medskip

\textbf{Remark 3.1 }Sometimes the directional derivative of a multiform
function $F:%
%TCIMACRO{\dbigwedge \nolimits^{r}}%
%BeginExpansion
{\displaystyle\bigwedge\nolimits^{r}}
%EndExpansion
\ni X\mapsto X\wedge\star X\in%
%TCIMACRO{\dbigwedge \nolimits^{n}}%
%BeginExpansion
{\displaystyle\bigwedge\nolimits^{n}}
%EndExpansion
$ in the direction of $W=%
%TCIMACRO{\TeXButton{bdelta}{\mbox{\boldmath{$\delta$}}}}%
%BeginExpansion
\mbox{\boldmath{$\delta$}}%
%EndExpansion
X\in%
%TCIMACRO{\dbigwedge \nolimits^{r}}%
%BeginExpansion
{\displaystyle\bigwedge\nolimits^{r}}
%EndExpansion
V$ is written as%
\begin{equation}%
%TCIMACRO{\TeXButton{bdelta}{\mbox{\boldmath{$\delta$}}}}%
%BeginExpansion
\mbox{\boldmath{$\delta$}}%
%EndExpansion
F:=%
%TCIMACRO{\TeXButton{bdelta}{\mbox{\boldmath{$\delta$}}}}%
%BeginExpansion
\mbox{\boldmath{$\delta$}}%
%EndExpansion
X\wedge\frac{\partial F}{\partial X} \label{2.31nn}%
\end{equation}
and called the \textit{variational derivative} of $F$ and $\frac{\partial
F}{\partial X}$ is called the algebraic derivative of $F$. Now, since $%
%TCIMACRO{\TeXButton{bdelta}{\mbox{\boldmath{$\delta$}}}}%
%BeginExpansion
\mbox{\boldmath{$\delta$}}%
%EndExpansion
F=W\cdot\partial_{X}$ we see that we have the identification for our multiform
function%
\begin{equation}%
%TCIMACRO{\TeXButton{bdelta}{\mbox{\boldmath{$\delta$}}}}%
%BeginExpansion
\mbox{\boldmath{$\delta$}}%
%EndExpansion
X\cdot\partial_{X}F=(-1)^{\frac{r}{2}(r-1)}%
%TCIMACRO{\TeXButton{bdelta}{\mbox{\boldmath{$\delta$}}}}%
%BeginExpansion
\mbox{\boldmath{$\delta$}}%
%EndExpansion
X\wedge\frac{\partial F}{\partial X}. \label{2.31nnn}%
\end{equation}

\subsection{The{\protect\Large \ }Operators $\partial_{X}\ast$\ and
$\protect\underline{t}(\partial_{X})\ast$}

We introduce now the linear differential operators $\partial_{X}\ast$ defined
by
\begin{equation}
\partial_{X}\ast F(X):=\underset{J}{%
%TCIMACRO{\dsum }%
%BeginExpansion
{\displaystyle\sum}
%EndExpansion
}\dfrac{1}{\nu(J)!}\varepsilon^{J}\ast\varepsilon_{J}\cdot\partial_{X}F(X),
\label{2.27}%
\end{equation}
where $\ast$ denotes any one of the multiform products $(\wedge),$ $(\cdot),$
$(\lrcorner,\llcorner)$ or $($\emph{Clifford product}$).$

Those linear differentiable operators are well defined since the multiforms in
the second member of Eq.(\ref{2.27}) depends only on the multiform function
$F$ sand do not depend on the pair of arbitrary reciprocal basis used.

Note that if $\ast$ refers to the Clifford product then the operator
$\partial_{X}\ast$ is precisely the operator $\partial_{X}$ (i.e., the
derivative operator). Sometimes, $\partial_{X}$ is called the \textit{gradient
operator }and $\partial_{X}F$ is said to be the \emph{\ gradient of }$F.$

Also, the operator $\partial_{X}\wedge$ is called the\emph{ rotational}
\textit{operator} and $\partial_{X}\wedge F$ is said to be the
\emph{rotational of }$F.$

The operators $\partial_{X}\cdot$ and $\partial_{X}\lrcorner$ are called
\emph{divergent} operators, $\partial_{X}\cdot F$ is said to be the
\emph{scalar divergent }and\emph{ } $\partial_{X}\lrcorner F$ is said to be
the \emph{contracted divergent.}

Let $t\in(1,1)$-$extV$\textbf{.} We define the linear differential operators
$\underline{t}(\partial_{X})\ast$ by
\begin{equation}
\underline{t}(\partial_{X})\ast F(X):=\underset{J}{\sum}\frac{1}{\nu
(J)!}\underline{t}(\varepsilon^{J})\ast\varepsilon_{J}\cdot\partial_{X}F(X),
\label{2.28bis}%
\end{equation}
where $\ast$ is any one of the multiform products $(\wedge),$ $(\cdot),$
$(\lrcorner,\llcorner)$ or $($\emph{Clifford product}$)$.

Those linear differential operators are well defined since they depends only
on $F$ and do not depend on the pair of reciprocal basis $(\{\varepsilon
^{j}\},\{\varepsilon_{j}\})$.

We call $\underline{t}(\partial_{X})\wedge$ the $t$-rotational\emph{ operator,
}$\underline{t}(\partial_{X})\wedge F$ is read as the $t$-\emph{rotational of
}$F.$

Also, $\underline{t}(\partial_{X})\cdot$ and $\underline{t}(\partial
_{X})\lrcorner$ \ are the $t$\emph{-divergent operators,} $\underline{t}%
(\partial_{X})\cdot F$ is the $t$\emph{-scalar divergent of }$F$\emph{\ }and
$\underline{t}(\partial_{X})\lrcorner F$ is $t$\emph{-contracted divergent of
}$F$. Finally, $\underline{t}(\partial_{X})$ is the $t$\emph{- gradient
operator and} $\underline{t}(\partial_{X})F$ is read as the $t$\emph{-gradient
of }$F$.

\subsection{Multiform Functionals $\mathcal{F}_{(X^{1},\ldots,X^{k})}[t]$}

Let%

\begin{align}
F  &  :\underset{k\text{ copies}}{\underbrace{%
%TCIMACRO{\dbigwedge \nolimits^{q}}%
%BeginExpansion
{\displaystyle\bigwedge\nolimits^{q}}
%EndExpansion
V\times\cdots\times%
%TCIMACRO{\dbigwedge \nolimits^{q}}%
%BeginExpansion
{\displaystyle\bigwedge\nolimits^{q}}
%EndExpansion
V}}\rightarrow%
%TCIMACRO{\dbigwedge }%
%BeginExpansion
{\displaystyle\bigwedge}
%EndExpansion
V,\nonumber\\
(X^{1},\ldots,X^{k})  &  \mapsto F(X^{1},\ldots,X^{k})=F_{(X^{1},\ldots
,X^{k})}.
\end{align}

A mapping%
\begin{align}
\mathcal{F}_{(X^{1},\ldots,X^{k})}  &  :(p,q)\text{-}extV\rightarrow%
%TCIMACRO{\dbigwedge }%
%BeginExpansion
{\displaystyle\bigwedge}
%EndExpansion
V,\nonumber\\
t  &  \mapsto\mathcal{F}_{(X^{1},\ldots,X^{k})}[t]:=F[t(X^{1}),\ldots
,t(X^{k})]. \label{fmult}%
\end{align}
will be called a \textit{multiform functional} $(p,q)$\emph{-extensor variable
induced by the function }$F$. We say also that $F$ is the \textit{generator
of} $\mathcal{F}_{(X^{1},\ldots,X^{k})}$.

When $\mathcal{F}[t]$ is scalar valued, 1-form valued, biform valued,
etc...,$\mathcal{F}$ it is said \ to be a \textit{scalar} \emph{functional ,
}1-\textit{form functional, biform functional, etc...,}\emph{\ of a }%
$(p,q)$\emph{-extensor variable.}

\subsection{Derivatives of Induced Multiform Functionals}

Let $(X^{1},\ldots,X^{k})\mapsto F(X^{1},\ldots,X^{k})$ \ be a differentiable
multiform function, i.e., differentiable with respect to each one of its
$q$-multiform variables $(X^{1},\ldots,X^{k}).$ Then the partial derivatives
of $F,$ $\partial_{X^{1}}F,\ldots,\partial_{X^{k}}F$ exist. We can then
construct exactly $k$ multiform functionals of a $(p,q)$-extensor variable
$t\in(p,q)$-$extV$ .which are induced by them, namely%
\begin{align}
t  &  \mapsto(\partial_{X^{1}}F)_{(X^{1},\ldots,X^{k})}[t]:=\partial
_{t(X^{1})}F[t(X^{1}),\ldots,t(X^{k})],\ldots,\nonumber\\
t  &  \mapsto(\partial_{X^{k}}F)_{(X^{1},\ldots,X^{k})}[t]:=\partial
_{t(X^{k})}F[t(X^{1}),\ldots,t(X^{k})]. \label{fm1}%
\end{align}

\subsubsection{The $A$-Directional $A\cdot\partial_{t}$\ Derivative of a
Multiform Functional}

Let $A\in%
%TCIMACRO{\dbigwedge }%
%BeginExpansion
{\displaystyle\bigwedge}
%EndExpansion
V$ \ be an arbitrary multiform. The multiform functional of $(p,q)$-extensor
variable $t$,%
\begin{equation}
(p,q)\text{-}extV\ni t\mapsto A\cdot\partial_{t}\mathcal{F}_{(X^{1}%
,\ldots,X^{k})}[t]\in%
%TCIMACRO{\dbigwedge }%
%BeginExpansion
{\displaystyle\bigwedge}
%EndExpansion
V \label{fm3}%
\end{equation}
such
\begin{align}
A\cdot\partial_{t}\mathcal{F}_{(X^{1},\ldots,X^{k})}[t]  &  :=A\cdot
X^{1}(\partial_{X^{1}}F)_{(X^{1},\ldots,X^{k})}[t]+\cdots+A\cdot
X^{k}(\partial_{X^{k}})_{(X^{1},\ldots,X^{k})}[t]\nonumber\\
&  =%
%TCIMACRO{\dsum \limits_{i=1}^{k}}%
%BeginExpansion
{\displaystyle\sum\limits_{i=1}^{k}}
%EndExpansion
A\cdot X^{i}(\partial_{X^{i}}F)_{(X^{1},\ldots,X^{k})}[t]\nonumber\\
&  =%
%TCIMACRO{\dsum \limits_{i=1}^{k}}%
%BeginExpansion
{\displaystyle\sum\limits_{i=1}^{k}}
%EndExpansion
A\cdot X^{i}\partial_{t(X^{i})}F[t(X^{1}),\ldots,t(X^{k})] \label{2.29}%
\end{align}
is said to be \footnote{Note that in \cite{fmr017} we used the notation\emph{
}$\partial_{t(A)}$ in the place of $A\cdot\partial_{t}$ \ and called that
operator the directional derivative of the functional $\mathcal{F}%
_{(X^{1},\ldots,X^{k})}[t]$ in the direction of $A$. We think now that the new
denomination is more appropriated, something that will become clear when we
calculate some examples.} the\emph{ }$A$-\emph{directional derivative of the
functional }$\mathcal{F}_{(X^{1},\ldots,X^{k})}[t]$. A convenient notation in
order to write short formulas is $\mathcal{F}_{(X^{1},\ldots,X^{k})A}^{\prime
}[t]$, i.e., we have%
\begin{equation}
\mathcal{F}_{(X^{1},\ldots,X^{k})A}^{\prime}[t]\equiv A\cdot\partial
_{t}\mathcal{F}_{(X^{1},\ldots,X^{k})}[t]
\end{equation}

Observe that $A\cdot\partial_{t}$\ is linear with respect to the direction
multiform $A$, i.e., for all $\alpha,\beta\in\mathbb{R}$ and $A,B\in%
%TCIMACRO{\dbigwedge }%
%BeginExpansion
{\displaystyle\bigwedge}
%EndExpansion
V$,
\begin{equation}
(\alpha A+\beta B)\cdot\partial_{t}=\alpha A\cdot\partial_{t}+\beta
B\cdot\partial_{t}. \label{2.30}%
\end{equation}

The operator $A\cdot\partial_{t}$ possess three important properties:
\begin{align}
A\cdot\partial_{t}\ (\mathcal{F}_{(X^{1},\ldots,X^{k})}+\mathcal{G}%
_{(X^{1},\ldots,X^{k})})[t]  &  =A\cdot\partial_{t}\ \mathcal{F}%
_{(X^{1},\ldots,X^{k})}[t]+A\cdot\partial_{t}\ \mathcal{G}_{(X^{1}%
,\ldots,X^{k})}[t].\label{2.31}\\
A\cdot\partial_{t}\ (\alpha\mathcal{F}_{(X^{1},\ldots,X^{k})})[t]  &
=\alpha(A\cdot\partial_{t}\ \mathcal{F}_{(X^{1},\ldots,X^{k})}[t])\label{2.32}%
\\
A\cdot\partial_{t}(\mathcal{F}_{(X^{1},\ldots,X^{k})}B)[t]  &  =(A\cdot
\partial_{t}\ \mathcal{F}_{(X^{1},\ldots,X^{k})}[t])B, \label{2.33}%
\end{align}
where $\mathcal{F}_{(X^{1},\ldots,X^{k})}$ and $\mathcal{G}_{(X^{1}%
,\ldots,X^{k})}$ are the multiform functionals induced by $F$ and $G$, and
$\alpha\in\mathbb{R}$ and $B\in%
%TCIMACRO{\dbigwedge }%
%BeginExpansion
{\displaystyle\bigwedge}
%EndExpansion
V$.

\subsubsection{The Operators $\partial_{t}\ast$}

Let $(\{\varepsilon^{j}\},\{\varepsilon_{j}\})$ be a pair of arbitrary
reciprocal basis for $V$ and let $\ast$ \ be any one of the multiform products
as before. We define%
\begin{equation}
\partial_{t}\ast\mathcal{F}_{(X^{1},\ldots,X^{k})}[t]:=\frac{1}{p!}%
\varepsilon^{j_{1}}\wedge...\wedge\varepsilon^{j_{p}}\ast\mathcal{F}%
_{(X^{1},\ldots,X^{k})\varepsilon_{j_{1}}\wedge...\wedge\varepsilon_{j_{p}}%
}^{\prime}[t] \label{2.33a}%
\end{equation}
Observe first that $\partial_{t}\ast\mathcal{F}_{(A^{1},\ldots,A^{k})}[t]$
does not depend on the pair of reciprocal basis and that taking into account
Eq.(\ref{2.29}) and Eq.(\ref{fm1})we have%
\begin{align}
\partial_{t}\ast\mathcal{F}_{(X^{1},\ldots,X^{k})}[t]  &  =\frac{1}%
{p!}\varepsilon^{j_{1}}\wedge...\wedge\varepsilon^{j_{p}}\ast%
%TCIMACRO{\dsum \limits_{i=1}^{k}}%
%BeginExpansion
{\displaystyle\sum\limits_{i=1}^{k}}
%EndExpansion
(\varepsilon_{j_{1}}\wedge...\wedge\varepsilon_{j_{p}})\cdot X^{i}%
(\partial_{X^{i}}F)_{(X^{1},\ldots,X^{k})}[t]\nonumber\\
&  =%
%TCIMACRO{\dsum \limits_{i=1}^{k}}%
%BeginExpansion
{\displaystyle\sum\limits_{i=1}^{k}}
%EndExpansion
\frac{1}{p!}(\varepsilon_{j_{1}}\wedge...\wedge\varepsilon_{j_{p}})\cdot
X^{i}(\varepsilon^{j_{1}}\wedge...\wedge\varepsilon^{j_{p}})\ast
(\partial_{X^{i}}F)_{(A^{1},\ldots,A^{k})}[t]\nonumber\\
&  =%
%TCIMACRO{\dsum \limits_{i=1}^{k}}%
%BeginExpansion
{\displaystyle\sum\limits_{i=1}^{k}}
%EndExpansion
X^{i}\ast(\partial_{X^{i}}F)_{(X^{1},\ldots,X^{k})}[t]\nonumber\\
&  =%
%TCIMACRO{\dsum \limits_{i=1}^{k}}%
%BeginExpansion
{\displaystyle\sum\limits_{i=1}^{k}}
%EndExpansion
X^{i}\ast\partial_{t(X^{k})}F[t(X^{1}),\ldots,t(X^{k})]. \label{2.33b}%
\end{align}

We call the objects $\partial_{t}\wedge\mathcal{F}_{(A^{1},\ldots,A^{k})}[t]$,
$\partial_{t}\cdot\mathcal{F}_{(A^{1},\ldots,A^{k})}[t]$, $\partial
_{t}\lrcorner\mathcal{F}_{(A^{1},\ldots,A^{k})}[t]$ and $\partial
_{t}\mathcal{F}_{(A^{1},\ldots,A^{k})}[t]$ respectively the
\textit{rotational}, the divergence, the left contracted divergence
\textit{and the gradient} (or simply the derivative) of $\mathcal{F}%
_{(X^{1},\ldots,X^{k})}$ with respect to $t$. \cite{fmr017}.

\subsubsection{Examples}

We present now some illustrative examples of derivatives of multiform
functionals that will be need in Section 6 where we shall derive the equations
of motion of the gravitational field from the variational principle.\medskip

\textbf{Example 3.7 }\emph{\ }Let $%
%TCIMACRO{\TeXButton{h}{\slh}}%
%BeginExpansion
\slh
%EndExpansion
\in(1,1)$-$extV$ \ and consider the scalar functional $%
%TCIMACRO{\TeXButton{h}{\slh}}%
%BeginExpansion
\slh
%EndExpansion
\mapsto%
%TCIMACRO{\TeXButton{h}{\slh}}%
%BeginExpansion
\slh
%EndExpansion
(b)\cdot%
%TCIMACRO{\TeXButton{h}{\slh}}%
%BeginExpansion
\slh
%EndExpansion
(c)$ and the biform functional $%
%TCIMACRO{\TeXButton{l}{\sll}}%
%BeginExpansion
\sll
%EndExpansion
\mapsto%
%TCIMACRO{\TeXButton{l}{\sll}}%
%BeginExpansion
\sll
%EndExpansion
(b)\wedge%
%TCIMACRO{\TeXButton{l}{\sll}}%
%BeginExpansion
\sll
%EndExpansion
(c)$, with $b,c\in%
%TCIMACRO{\dbigwedge \nolimits^{1}}%
%BeginExpansion
{\displaystyle\bigwedge\nolimits^{1}}
%EndExpansion
V$. The first has as generator function the scalar function of two 1-form
variables $(x,y)\mapsto x\cdot y$ and the generator of the second functional
is the biform function of two 1-form variables $(x,y)\mapsto x\wedge y$. Let
us calculate the derivative of those functionals,with respect to $%
%TCIMACRO{\TeXButton{h}{\slh}}%
%BeginExpansion
\slh
%EndExpansion
$ in the direction of $a\in%
%TCIMACRO{\dbigwedge \nolimits^{1}}%
%BeginExpansion
{\displaystyle\bigwedge\nolimits^{1}}
%EndExpansion
V$.%
\begin{align}
a\cdot\partial_{%
%TCIMACRO{\TeXButton{sh}{\sslh}}%
%BeginExpansion
\sslh
%EndExpansion
}(%
%TCIMACRO{\TeXButton{h}{\slh}}%
%BeginExpansion
\slh
%EndExpansion
(b)\cdot%
%TCIMACRO{\TeXButton{h}{\slh}}%
%BeginExpansion
\slh
%EndExpansion
(c))  &  =a\cdot b\partial_{%
%TCIMACRO{\TeXButton{sh}{\sslh}}%
%BeginExpansion
\sslh
%EndExpansion
(b)}(%
%TCIMACRO{\TeXButton{h}{\slh}}%
%BeginExpansion
\slh
%EndExpansion
(b)\cdot%
%TCIMACRO{\TeXButton{h}{\slh}}%
%BeginExpansion
\slh
%EndExpansion
(c))+a\cdot c\partial_{%
%TCIMACRO{\TeXButton{sh}{\sslh}}%
%BeginExpansion
\sslh
%EndExpansion
(c)}(%
%TCIMACRO{\TeXButton{h}{\slh}}%
%BeginExpansion
\slh
%EndExpansion
(b)\cdot%
%TCIMACRO{\TeXButton{h}{\slh}}%
%BeginExpansion
\slh
%EndExpansion
(c))\nonumber\\
&  =a\cdot b%
%TCIMACRO{\TeXButton{h}{\slh}}%
%BeginExpansion
\slh
%EndExpansion
(c)+a\cdot c%
%TCIMACRO{\TeXButton{h}{\slh}}%
%BeginExpansion
\slh
%EndExpansion
(b), \label{ef1}%
\end{align}
were we utilized $\partial_{x}(x\cdot y)=y.$

Also
\begin{align}
a\cdot\partial_{%
%TCIMACRO{\TeXButton{sh}{\sslh}}%
%BeginExpansion
\sslh
%EndExpansion
}(%
%TCIMACRO{\TeXButton{h}{\slh}}%
%BeginExpansion
\slh
%EndExpansion
(b)\wedge%
%TCIMACRO{\TeXButton{h}{\slh}}%
%BeginExpansion
\slh
%EndExpansion
(c))  &  =a\cdot b\partial_{%
%TCIMACRO{\TeXButton{sh}{\sslh}}%
%BeginExpansion
\sslh
%EndExpansion
(b)}(%
%TCIMACRO{\TeXButton{h}{\slh}}%
%BeginExpansion
\slh
%EndExpansion
(b)\wedge%
%TCIMACRO{\TeXButton{h}{\slh}}%
%BeginExpansion
\slh
%EndExpansion
(c))+a\cdot c\partial_{%
%TCIMACRO{\TeXButton{sh}{\sslh}}%
%BeginExpansion
\sslh
%EndExpansion
(c)}(%
%TCIMACRO{\TeXButton{h}{\slh}}%
%BeginExpansion
\slh
%EndExpansion
(b)\wedge%
%TCIMACRO{\TeXButton{h}{\slh}}%
%BeginExpansion
\slh
%EndExpansion
(c))\nonumber\\
&  =a\cdot b(n-1)%
%TCIMACRO{\TeXButton{h}{\slh}}%
%BeginExpansion
\slh
%EndExpansion
(c)-a\cdot c(n-1)%
%TCIMACRO{\TeXButton{h}{\slh}}%
%BeginExpansion
\slh
%EndExpansion
(b)\nonumber\\
&  =(n-1)(a\cdot b%
%TCIMACRO{\TeXButton{h}{\slh}}%
%BeginExpansion
\slh
%EndExpansion
(c)-a\cdot c%
%TCIMACRO{\TeXButton{h}{\slh}}%
%BeginExpansion
\slh
%EndExpansion
(b))\nonumber\\
&  =(n-1)%
%TCIMACRO{\TeXButton{h}{\slh}}%
%BeginExpansion
\slh
%EndExpansion
(a\cdot bc-a\cdot cb)\nonumber\\
&  =(n-1)%
%TCIMACRO{\TeXButton{h}{\slh}}%
%BeginExpansion
\slh
%EndExpansion
(a\lrcorner(b\wedge c)), \label{ef2}%
\end{align}
where we used $\partial_{x}(x\wedge y)=(n-1)y.$

Eq.(\ref{ef2}) has a very useful generalization. The derivative of the
$k$-form functional $%
%TCIMACRO{\TeXButton{h}{\slh}}%
%BeginExpansion
\slh
%EndExpansion
\mapsto\underline{%
%TCIMACRO{\TeXButton{h}{\slh}}%
%BeginExpansion
\slh
%EndExpansion
}(a^{1}\wedge\ldots\wedge a^{k})=%
%TCIMACRO{\TeXButton{h}{\slh}}%
%BeginExpansion
\slh
%EndExpansion
(a^{1})\wedge\ldots\wedge%
%TCIMACRO{\TeXButton{h}{\slh}}%
%BeginExpansion
\slh
%EndExpansion
(a^{k})$, with $a^{1},\ldots,a^{k}\in%
%TCIMACRO{\dbigwedge \nolimits^{1}}%
%BeginExpansion
{\displaystyle\bigwedge\nolimits^{1}}
%EndExpansion
V$ is
\begin{equation}
a\cdot\partial_{%
%TCIMACRO{\TeXButton{sh}{\sslh}}%
%BeginExpansion
\sslh
%EndExpansion
}\underline{%
%TCIMACRO{\TeXButton{h}{\slh}}%
%BeginExpansion
\slh
%EndExpansion
}(a^{1}\wedge\ldots\wedge a^{k})=(n-k+1)\underline{%
%TCIMACRO{\TeXButton{h}{\slh}}%
%BeginExpansion
\slh
%EndExpansion
}(a\lrcorner(a^{1}\wedge\ldots\wedge a^{k})). \label{ef3}%
\end{equation}
\medskip

\textbf{Example 3.8 } Let $t\in(1,1)$-$extV$. The trace of $t$ (i.e.,
$t\mapsto\mathrm{tr}[t]=t(\varepsilon^{j})\cdot\varepsilon_{j}$) is a scalar
functional of $t$ whose generator is the scalar function of $n$ 1-form
variables, $(x^{1},\ldots,x^{n})\mapsto x^{1}\cdot\varepsilon_{1}+\cdots
+x^{n}\cdot\varepsilon_{n},$ and \textrm{bif}$[t]$ $=t(\varepsilon^{j}%
)\wedge\varepsilon_{j}$) is the biform functional of $t$ whose generator is
the biform function of $n$ 1-form variables, $(x^{1},\ldots,x^{n})\mapsto
x^{1}\wedge\varepsilon_{1}+\cdots+x^{n}\wedge\varepsilon_{n}$. Let us
calculate $a\cdot\partial_{t}\mathrm{tr}[t]$ and $\partial_{%
%TCIMACRO{\TeXButton{h}{\slh}}%
%BeginExpansion
\slh
%EndExpansion
(a)}\mathrm{bif}[t]$. We have%
\begin{align}
a\cdot\partial_{%
%TCIMACRO{\TeXButton{sh}{\sslh}}%
%BeginExpansion
\sslh
%EndExpansion
}\mathrm{tr}[t]  &  =a\cdot\varepsilon^{1}\partial_{t(\varepsilon^{1}%
)}(t(\varepsilon^{1})\cdot\varepsilon_{1}+\cdots+t(\varepsilon^{n}%
)\cdot\varepsilon_{n})+\cdots\nonumber\\
&  +a\cdot\varepsilon^{n}\partial_{t(\varepsilon^{n})}(t(\varepsilon^{1}%
)\cdot\varepsilon_{1}+\cdots+t(\varepsilon^{n})\cdot\varepsilon_{n}%
)\nonumber\\
&  =a\cdot\varepsilon^{1}\varepsilon_{1}+\cdots+a\cdot\varepsilon
^{n}\varepsilon_{n}=a, \label{ef4}%
\end{align}

and
\begin{align}
a\cdot\partial_{%
%TCIMACRO{\TeXButton{sh}{\sslh}}%
%BeginExpansion
\sslh
%EndExpansion
}\mathrm{bif}[t]  &  =a\cdot\varepsilon^{1}\partial_{t(\varepsilon^{1}%
)}(t(\varepsilon^{1})\wedge\varepsilon_{1}+\cdots+t(\varepsilon^{n}%
)\wedge\varepsilon_{n})+\cdots\nonumber\\
&  +a\cdot\varepsilon^{n}\partial_{t(\varepsilon^{n})}(t(\varepsilon
^{1})\wedge\varepsilon_{1}+\cdots+t(\varepsilon^{n})\wedge\varepsilon
_{n})\nonumber\\
&  =a\cdot\varepsilon^{1}(n-1)\varepsilon_{1}+\cdots+a\cdot\varepsilon
^{n}(n-1)\varepsilon_{n}=(n-1)a. \label{ef5}%
\end{align}
\medskip

\textbf{Example 3.9}\emph{\ } Let $%
%TCIMACRO{\TeXButton{h}{\slh}}%
%BeginExpansion
\slh
%EndExpansion
\in(1,1)$-$extV$, and consider the 1-form functional $%
%TCIMACRO{\TeXButton{h}{\slh}}%
%BeginExpansion
\slh
%EndExpansion
\mapsto%
%TCIMACRO{\TeXButton{h}{\slh}}%
%BeginExpansion
\slh
%EndExpansion
^{\dagger}(b)$, with $b\in%
%TCIMACRO{\dbigwedge \nolimits^{1}}%
%BeginExpansion
{\displaystyle\bigwedge\nolimits^{1}}
%EndExpansion
V$, and the pseudo-scalar functional $%
%TCIMACRO{\TeXButton{h}{\slh}}%
%BeginExpansion
\slh
%EndExpansion
\mapsto\underline{%
%TCIMACRO{\TeXButton{h}{\slh}}%
%BeginExpansion
\slh
%EndExpansion
}(\tau)$, with $\tau\in%
%TCIMACRO{\dbigwedge \nolimits^{n}}%
%BeginExpansion
{\displaystyle\bigwedge\nolimits^{n}}
%EndExpansion
V$. Let us calculate the derivatives of those functionals.
\begin{align}
a\cdot\partial_{%
%TCIMACRO{\TeXButton{sh}{\sslh}}%
%BeginExpansion
\sslh
%EndExpansion
}%
%TCIMACRO{\TeXButton{h}{\slh}}%
%BeginExpansion
\slh
%EndExpansion
^{\dagger}(b)  &  =a\cdot\partial_{%
%TCIMACRO{\TeXButton{sh}{\sslh}}%
%BeginExpansion
\sslh
%EndExpansion
}(%
%TCIMACRO{\TeXButton{h}{\slh}}%
%BeginExpansion
\slh
%EndExpansion
^{\dagger}(b)\cdot\varepsilon^{k}\varepsilon_{k})=a\cdot\partial_{%
%TCIMACRO{\TeXButton{sh}{\sslh}}%
%BeginExpansion
\sslh
%EndExpansion
}(b\cdot%
%TCIMACRO{\TeXButton{h}{\slh}}%
%BeginExpansion
\slh
%EndExpansion
(\varepsilon^{k})\varepsilon_{k})\nonumber\\
&  =a\cdot\varepsilon^{1}\partial_{%
%TCIMACRO{\TeXButton{sh}{\sslh}}%
%BeginExpansion
\sslh
%EndExpansion
(\varepsilon^{1})}(b\cdot%
%TCIMACRO{\TeXButton{h}{\slh}}%
%BeginExpansion
\slh
%EndExpansion
(\varepsilon^{k})\varepsilon_{k})+\cdots+a\cdot\varepsilon^{n}\partial_{%
%TCIMACRO{\TeXButton{sh}{\sslh}}%
%BeginExpansion
\sslh
%EndExpansion
(\varepsilon^{n})}(b\cdot%
%TCIMACRO{\TeXButton{h}{\slh}}%
%BeginExpansion
\slh
%EndExpansion
(\varepsilon^{k})\varepsilon_{k})\nonumber\\
&  =a\cdot\varepsilon^{1}(b\varepsilon_{1})+\cdots+a\cdot\varepsilon
^{n}(b\varepsilon_{n})\nonumber\\
&  =b(a\cdot\varepsilon^{1}\varepsilon_{1}+\cdots+a\cdot\varepsilon
^{n}\varepsilon_{n})=ba, \label{ef6}%
\end{align}
where we utilized a formula for the expansion of 1-forms in order to get $%
%TCIMACRO{\TeXButton{h}{\slh}}%
%BeginExpansion
\slh
%EndExpansion
^{\dagger}(b)$ as a 1-form functional of $%
%TCIMACRO{\TeXButton{h}{\slh}}%
%BeginExpansion
\slh
%EndExpansion
$, the scalar product condition involving $%
%TCIMACRO{\TeXButton{h}{\slh}}%
%BeginExpansion
\slh
%EndExpansion
$ and $%
%TCIMACRO{\TeXButton{h}{\slh}}%
%BeginExpansion
\slh
%EndExpansion
^{\dagger}$ (i.e., $%
%TCIMACRO{\TeXButton{h}{\slh}}%
%BeginExpansion
\slh
%EndExpansion
^{\dagger}(x)\cdot y=x\cdot%
%TCIMACRO{\TeXButton{h}{\slh}}%
%BeginExpansion
\slh
%EndExpansion
(y)$) and the formula $\partial_{x}(b\cdot x)c=bc.$

Also,%
\begin{align}
a\cdot\partial_{%
%TCIMACRO{\TeXButton{sh}{\sslh}}%
%BeginExpansion
\sslh
%EndExpansion
}\underline{%
%TCIMACRO{\TeXButton{h}{\slh}}%
%BeginExpansion
\slh
%EndExpansion
}(\tau)  &  =a\cdot\partial_{%
%TCIMACRO{\TeXButton{sh}{\sslh}}%
%BeginExpansion
\sslh
%EndExpansion
}\underline{%
%TCIMACRO{\TeXButton{h}{\slh}}%
%BeginExpansion
\slh
%EndExpansion
}((\tau\cdot\varepsilon_{1}\wedge\ldots\wedge\varepsilon_{n})\varepsilon
^{1}\wedge\ldots\wedge\varepsilon^{n})\nonumber\\
&  =(\tau\cdot\varepsilon_{1}\wedge\ldots\wedge\varepsilon_{n})a\cdot
\partial_{%
%TCIMACRO{\TeXButton{sh}{\sslh}}%
%BeginExpansion
\sslh
%EndExpansion
}\underline{%
%TCIMACRO{\TeXButton{h}{\slh}}%
%BeginExpansion
\slh
%EndExpansion
}(\varepsilon^{1}\wedge\ldots\wedge\varepsilon^{n})\nonumber\\
&  =(\tau\cdot\varepsilon_{1}\wedge\ldots\wedge\varepsilon_{n})\underline{%
%TCIMACRO{\TeXButton{h}{\slh}}%
%BeginExpansion
\slh
%EndExpansion
}(a\lrcorner(\varepsilon^{1}\wedge\ldots\wedge\varepsilon^{n}))\nonumber\\
&  =\underline{%
%TCIMACRO{\TeXButton{h}{\slh}}%
%BeginExpansion
\slh
%EndExpansion
}(a\lrcorner(\tau\cdot\varepsilon_{1}\wedge\ldots\wedge\varepsilon
_{n})\varepsilon^{1}\wedge\ldots\wedge\varepsilon^{n})\nonumber\\
&  =\underline{%
%TCIMACRO{\TeXButton{h}{\slh}}%
%BeginExpansion
\slh
%EndExpansion
}(a\lrcorner\tau)=\underline{%
%TCIMACRO{\TeXButton{h}{\slh}}%
%BeginExpansion
\slh
%EndExpansion
}(a\tau), \label{ef7}%
\end{align}
where we utilized the formula for expansion of pseudo-scalars, Eq.(\ref{2.32})
and Eq.(\ref{ef3}).\medskip

\textbf{Example 3.10} Let $%
%TCIMACRO{\TeXButton{h}{\slh}}%
%BeginExpansion
\slh
%EndExpansion
\in(1,1)$-$extV$, the determinant of $%
%TCIMACRO{\TeXButton{h}{\slh}}%
%BeginExpansion
\slh
%EndExpansion
$ is a well defined scalar functional, $%
%TCIMACRO{\TeXButton{h}{\slh}}%
%BeginExpansion
\slh
%EndExpansion
\mapsto\det[%
%TCIMACRO{\TeXButton{h}{\slh}}%
%BeginExpansion
\slh
%EndExpansion
]=\underline{h}(\varepsilon^{1}\wedge\ldots\wedge\varepsilon^{n}%
)\cdot(\varepsilon_{1}\wedge\ldots\wedge\varepsilon_{n})$. Let us calculate
$\partial_{%
%TCIMACRO{\TeXButton{sh}{\sslh}}%
%BeginExpansion
\sslh
%EndExpansion
(a)}\det[%
%TCIMACRO{\TeXButton{h}{\slh}}%
%BeginExpansion
\slh
%EndExpansion
]$.%
\begin{align}
a\cdot\partial_{%
%TCIMACRO{\TeXButton{sh}{\sslh}}%
%BeginExpansion
\sslh
%EndExpansion
}\det[%
%TCIMACRO{\TeXButton{h}{\slh}}%
%BeginExpansion
\slh
%EndExpansion
]  &  =a\cdot\partial_{%
%TCIMACRO{\TeXButton{sh}{\sslh}}%
%BeginExpansion
\sslh
%EndExpansion
}(\underline{%
%TCIMACRO{\TeXButton{h}{\slh}}%
%BeginExpansion
\slh
%EndExpansion
}(\tau)\tau)=(a\cdot\partial_{%
%TCIMACRO{\TeXButton{h}{\slh}}%
%BeginExpansion
\slh
%EndExpansion
}\underline{%
%TCIMACRO{\TeXButton{h}{\slh}}%
%BeginExpansion
\slh
%EndExpansion
}(\tau))\tau^{-1}\nonumber\\
&  =\underline{%
%TCIMACRO{\TeXButton{h}{\slh}}%
%BeginExpansion
\slh
%EndExpansion
}(a\tau)\tau^{-1}=\det[%
%TCIMACRO{\TeXButton{h}{\slh}}%
%BeginExpansion
\slh
%EndExpansion
]%
%TCIMACRO{\TeXButton{h}{\slh}}%
%BeginExpansion
\slh
%EndExpansion
^{\clubsuit}(a), \label{ef8}%
\end{align}
where we utilized Eq.(\ref{2.33}), Eq.(\ref{ef7}) and the following identities
involving $(1,1)$-extensors: $\det[t]\tau=\underline{t}(\tau)$ and
$t^{-1}(a)=\left.  \det\right.  ^{-1}[t]\underline{t}^{\dagger}(a\tau
)\tau^{-1},$ (recall that $%
%TCIMACRO{\TeXButton{h}{\slh}}%
%BeginExpansion
\slh
%EndExpansion
^{\clubsuit}=(%
%TCIMACRO{\TeXButton{h}{\slh}}%
%BeginExpansion
\slh
%EndExpansion
^{\dagger})^{-1}=(%
%TCIMACRO{\TeXButton{h}{\slh}}%
%BeginExpansion
\slh
%EndExpansion
^{-1})^{\dagger}$).

\subsection{The Variational Operator $%
%TCIMACRO{\TeXButton{delta}{\mbox{\boldmath{$\delta$}}}}%
%BeginExpansion
\mbox{\boldmath{$\delta$}}%
%EndExpansion
_{t}^{w}$}

Let $\mathcal{F}_{(X^{1},\ldots,X^{k})}[t]$ be a scalar functional of the
extensor variable $t:%
%TCIMACRO{\dbigwedge \nolimits^{1}}%
%BeginExpansion
{\displaystyle\bigwedge\nolimits^{1}}
%EndExpansion
V\rightarrow%
%TCIMACRO{\dbigwedge \nolimits^{1}}%
%BeginExpansion
{\displaystyle\bigwedge\nolimits^{1}}
%EndExpansion
V$. Let also $w:$ $%
%TCIMACRO{\dbigwedge \nolimits^{1}}%
%BeginExpansion
{\displaystyle\bigwedge\nolimits^{1}}
%EndExpansion
V\rightarrow%
%TCIMACRO{\dbigwedge \nolimits^{1}}%
%BeginExpansion
{\displaystyle\bigwedge\nolimits^{1}}
%EndExpansion
V$. Construct next the real variable function $f_{w}\mathcal{(\lambda)=}$
$\mathcal{F}_{(X^{1},\ldots,X^{k})}[t+\lambda w]$. Then according to the mean
value theorem we have%
\begin{equation}
f_{w}\mathcal{(\lambda)=}f_{w}\mathcal{(}0\mathcal{)+}\frac{d}{d\lambda
}\left.  f_{w}\mathcal{(}\lambda\mathcal{)}\right\vert _{\lambda=0}%
\lambda+\frac{1}{2!}\frac{d^{2}}{d\lambda^{2}}\left.  f_{w}\mathcal{(}%
\lambda\mathcal{)}\right\vert _{\lambda=\lambda_{1}}\lambda^{2}\text{,
\ \ \ \ \ }0<\left\vert \lambda_{1}\right\vert <\left\vert \lambda\right\vert
.
\end{equation}
The functional variation of $\mathcal{F}_{(X^{1},\ldots,X^{k})}[t]$ is by
definition:
\begin{equation}%
%TCIMACRO{\TeXButton{delta}{\mbox{\boldmath{$\delta$}}}}%
%BeginExpansion
\mbox{\boldmath{$\delta$}}%
%EndExpansion
_{t}^{w}\mathcal{F}_{(X^{1},\ldots,X^{k})}[t]:=\frac{d}{d\lambda}\left.
\mathcal{F}_{w}\mathcal{(}\lambda\mathcal{)}\right\vert _{\lambda=0}
\label{vf}%
\end{equation}

Since the Lagrangian for the field theory\ of gravitation (to be developed in
Section 5) is a scalar functional of a $(1,1)$-extensor field $%
%TCIMACRO{\TeXButton{h}{\slh}}%
%BeginExpansion
\slh
%EndExpansion
$ , its functional variation will play an important role in this paper. In
particular we shall need the following result.

\textbf{Example 3.11} Calculate $%
%TCIMACRO{\TeXButton{delta}{\mbox{\boldmath{$\delta$}}}}%
%BeginExpansion
\mbox{\boldmath{$\delta$}}%
%EndExpansion
_{t}^{w}\det[t]$. We first construct the real variable function%
\begin{equation}
f_{w}\mathcal{(\lambda)=}\det[t+\lambda w]
\end{equation}
Next we define $g=tt^{+}$ and take a pair of reciprocal basis $(\{\varepsilon
^{i}\},\{\varepsilon_{i}\})$ for $V$ satisfying the condition
\begin{equation}
g(\varepsilon^{i})\cdot\varepsilon^{j}=\delta^{ij},
\end{equation}
which implies that
\begin{equation}
t(\varepsilon^{i})\cdot t(\varepsilon^{j})=\delta^{ij}.
\end{equation}
Recalling Eq.(\ref{1.43aa}) we can write
\begin{equation}
f_{w}\mathcal{(\lambda)=}\frac{1}{n!}\mathcal{\{(}t(\varepsilon^{1})+\lambda
w(\varepsilon^{1})]\wedge...\wedge\lbrack t(\varepsilon^{1})+\lambda
w(\varepsilon^{1})]\}\lrcorner(\varepsilon_{1}\wedge...\wedge\varepsilon_{n}).
\end{equation}
Then we have%
\begin{align}
\frac{d}{d\lambda}\left.  f_{w}\mathcal{(}\lambda\mathcal{)}\right\vert
_{\lambda=0}  &  =\frac{1}{n!}\{[w(\varepsilon^{1})\wedge t(\varepsilon
^{2})\wedge...t(\varepsilon^{n})]\lrcorner(\varepsilon_{1}\wedge
...\wedge\varepsilon_{n})\nonumber\\
&  +.....+[t(\varepsilon^{1})\wedge t(\varepsilon^{2})\wedge...\wedge
w(\varepsilon^{n})]\lrcorner(\varepsilon_{1}\wedge...\wedge\varepsilon_{n})\}
\label{int}%
\end{align}
and recalling the identity giving by Eq.(\ref{1.22a}) which says that for nay
$A,B,C\in%
%TCIMACRO{\dbigwedge }%
%BeginExpansion
{\displaystyle\bigwedge}
%EndExpansion
V$. it is $(A\wedge B)\lrcorner C$ $=A\lrcorner(B\lrcorner C)$, and
Eq.(\ref{1.49}) ($t^{-1}(a)=\left.  \det\right.  ^{-1}[t]\underline{t}%
^{\dagger}(a\tau)\tau^{-1}$), we can write the second member of Eq.(\ref{int})
as%
\begin{align}
&  \frac{1}{n!}\{w(\varepsilon^{1})\lrcorner(t(\varepsilon^{2})\wedge...\wedge
t(\varepsilon^{n}))\lrcorner(\varepsilon_{1}\wedge...\varepsilon
_{n})+...\nonumber\\
&  +(-1)^{n-1}w(\varepsilon^{n})\lrcorner(t(\varepsilon^{2})\wedge...\wedge
t(\varepsilon^{n-1}))\lrcorner(\varepsilon_{1}\wedge...\wedge\varepsilon
_{n})\}\nonumber\\
&  =\frac{1}{n!}\{w(\varepsilon^{1})\lrcorner\lbrack t(\varepsilon
^{1})\lrcorner(t(\varepsilon^{1})(t(\varepsilon^{2})\wedge...\wedge
t(\varepsilon^{n}))\lrcorner(\varepsilon_{1}\wedge...\wedge\varepsilon
_{n})...\nonumber\\
&  +w(\varepsilon^{n})\lrcorner\lbrack t(\varepsilon^{n})\lrcorner
(t(\varepsilon^{1})(t(\varepsilon^{2})\wedge...\wedge t(\varepsilon
^{n}))\lrcorner(\varepsilon_{1}\wedge...\wedge\varepsilon_{n})\}\nonumber\\
&  =%
%TCIMACRO{\dsum \limits_{j=1}^{n}}%
%BeginExpansion
{\displaystyle\sum\limits_{j=1}^{n}}
%EndExpansion
w(\varepsilon^{i})\lrcorner\lbrack t(\varepsilon^{i})\lrcorner(t(\varepsilon
^{1})(t(\varepsilon^{2})\wedge...\wedge t(\varepsilon^{n}))\lrcorner
(\varepsilon_{1}\wedge...\wedge\varepsilon_{n})\nonumber\\
&  =%
%TCIMACRO{\dsum \limits_{j=1}^{n}}%
%BeginExpansion
{\displaystyle\sum\limits_{j=1}^{n}}
%EndExpansion
w(\varepsilon^{i})\lrcorner t(\varepsilon^{i}\tau)\tau^{-1}\nonumber\\
&  =%
%TCIMACRO{\dsum \limits_{j=1}^{n}}%
%BeginExpansion
{\displaystyle\sum\limits_{j=1}^{n}}
%EndExpansion
w(\varepsilon^{i})\lrcorner t^{\clubsuit}(\varepsilon^{i})\det[t]\nonumber\\
&  =w(\partial_{a})\lrcorner t^{\clubsuit}(a)\det[t].
\end{align}
Taking moreover into account that for any $a\in%
%TCIMACRO{\dbigwedge \nolimits^{1}}%
%BeginExpansion
{\displaystyle\bigwedge\nolimits^{1}}
%EndExpansion
V$,%
\begin{equation}%
%TCIMACRO{\dsum \limits_{j=1}^{n}}%
%BeginExpansion
{\displaystyle\sum\limits_{j=1}^{n}}
%EndExpansion
w(\varepsilon^{i})\lrcorner t^{\clubsuit}(\varepsilon^{i})\det[t]=w(\partial
_{a})\lrcorner t^{\clubsuit}(a)\det[t],
\end{equation}
we finally get
\begin{equation}%
%TCIMACRO{\TeXButton{delta}{\mbox{\boldmath{$\delta$}}}}%
%BeginExpansion
\mbox{\boldmath{$\delta$}}%
%EndExpansion
_{t}^{w}\det[t]=w(\partial_{a})\lrcorner t^{\clubsuit}(a)\det[t]. \label{vdet}%
\end{equation}

\section{Multiform and Extensor Calculus on Manifolds}

\subsection{Canonical Space}

Let $M$ be a smooth (i.e., $C^{\infty})$ differential manifold, $\dim M=n$.
\ Take an arbitrary point $O\in M$ , a local chart $(\mathcal{U},\phi)_{o}$ of
the atlas of such that $O\in\mathcal{U}.$

Any point $p\in\mathcal{U}$ \ is then localized by a $n$-uple of real numbers
$\phi(p)\in\mathbb{R}^{n}$, say $\phi(p)=(\xi^{1}(p),\ldots,\xi^{n}(p))$. As
usual both $\mathcal{U}\ni p\mapsto\xi^{\mu}(p)\in\mathbb{R}$, both the $\mu
$-th coordinate function as well $\xi^{\mu}(p)$, the $\mu$-th coordinate, are
denoted by the same notation $\xi^{\mu}$, with $\mu=1,\ldots,n$.

At $p\in\mathcal{U},$ the set of coordinate tangent vectors $\left\{  \left.
\dfrac{\partial}{\partial\xi^{1}}\right\vert _{(p)},\ldots,\left.
\dfrac{\partial}{\partial\xi^{n}}\right\vert _{(p)}\right\}  $ is a natural
basis for the tangent space $T_{p}M$, and the set of tangent coordinate
$1$-forms $\left\{  \left.  d\xi^{1}\right\vert _{(p)},\ldots,\left.  d\xi
^{n}\right\vert _{(p)}\right\}  $ is a natural basis for the cotangent (or
dual) space $T_{p}^{\ast}M.$

We introduce an equivalence relation on the (sub)tangent bundle $T\mathcal{U=}%
$ $\underset{p\in\mathcal{U}}{\bigcup}T_{p}M$ as follows. Let $\mathbf{v}%
_{p}\in T_{p}M$ and $\mathbf{v}_{q}\in T_{p}M$. We say that $\ \mathbf{v}%
_{p}\sim\mathbf{v}_{q}$ iff $\mathbf{v}_{p}\xi^{\mu}=\mathbf{v}_{q}\xi^{\mu}$,
with $\mu=1,\ldots,n$.

This equivalence relation is well defined and, of course, it is not void,
since the coordinate tangent vectors at any two points, $p,q\in\mathcal{U}$
are equivalent. Indeed, we have%
\begin{equation}
\left.  \dfrac{\partial}{\partial\xi^{\alpha}}\right\vert _{(p)}\xi^{\mu
}=\left.  \dfrac{\partial\xi^{\mu}}{\partial\xi^{\alpha}}\right\vert
_{(p)}=\delta_{\alpha}^{\mu}=\left.  \dfrac{\partial\xi^{\mu}}{\partial
\xi^{\alpha}}\right\vert _{(q)}=\left.  \dfrac{\partial}{\partial\xi^{\alpha}%
}\right\vert _{(q)}\xi^{\mu},\text{ \ }\mu,\alpha=1,\ldots,n,
\end{equation}
and thus $\left.  \dfrac{\partial}{\partial\xi^{\alpha}}\right\vert _{(p)}%
\sim\left.  \dfrac{\partial}{\partial\xi^{\alpha}}\right\vert _{(q)}.$

For the point $O\in\mathcal{U}$, the equivalence class of a tangent vector
$\mathbf{v}_{o}\in T_{o}M$ is given by $[\mathbf{v}_{o}]=\{\mathbf{v}_{p}%
\sim\mathbf{v}_{o}$ $\mid$ $p\in\mathcal{U}\}.$

The set of the equivalence classes of each one of the tangent vectors
$\mathbf{v}_{o}\in T_{o}M$, i.e., $\mathbf{U=\{}[\mathbf{v}_{o}]$ $/$
$\mathbf{v}_{o}\in T_{o}M\mathcal{\}}$, equipped with the \textit{sum} of
equivalence classes and the \textit{multiplication} of equivalence classes by
scalars (real numbers), defined by
\begin{equation}
\lbrack\mathbf{v}_{o}]+[\mathbf{w}_{o}]=[\mathbf{v}_{o}+\mathbf{w}_{o}]\text{;
\ \ \ \ }\alpha\lbrack\mathbf{v}_{o}]=[\alpha\mathbf{v}_{o}],
\end{equation}
has a natural structure of vector space over $\mathbb{R}$.

The equivalence class of the null vector $\mathbf{0}_{o}\in\mathcal{T}_{o}M,$
i.e., $\mathbf{0\equiv}[\mathbf{0}_{o}]\in\mathbf{U}$, i.e., the set of all
null vectors in $T\mathcal{U}$ is the null vector of $\mathbf{U}.$

The set of the equivalence classes of the $n$-tangent coordinate vectors
$\left\{  \left.  \dfrac{\partial}{\partial\xi^{i}}\right\vert _{(o)}\right\}
$, with $\left.  \dfrac{\partial}{\partial\xi^{i}}\right\vert _{(o)}\in
T_{o}M$, say $\{\mathbf{b}_{1},\ldots,\mathbf{b}_{n}\}$ with
\begin{equation}
\mathbf{b}_{1}\equiv\lbrack\left.  \dfrac{\partial}{\partial\xi^{1}%
}\right\vert _{(o)}],\ldots,\mathbf{b}_{n}\equiv\lbrack\left.  \dfrac
{\partial}{\partial\xi^{n}}\right\vert _{(o)}]\in\mathbf{U,}%
\end{equation}
is a basis for $\mathbf{U}$ called here \emph{fiducial basis} and, of course,
$\dim\mathbf{U}=n.$

$\mathbf{U}$ is a vector space obviously associated to the chart
$(\mathcal{U},\phi)_{o}$, but nevertheless it will be said to be the canonical
space since it will play a fundamental role in our theory of multiform and
extensor fields on manifolds. To continue we recall that from the above
identifications we have immediately the identification of $%
%TCIMACRO{\dbigwedge }%
%BeginExpansion
{\displaystyle\bigwedge}
%EndExpansion
T_{p}^{\ast}M$ with $%
%TCIMACRO{\dbigwedge }%
%BeginExpansion
{\displaystyle\bigwedge}
%EndExpansion
T_{q}^{\ast}M$ for all $p.q\in\mathcal{U}$.

Taking into account the notations introduced in Section 2 the dual space of
$\mathbf{U}$ is denoted $U$, the space of $k$-forms is denoted by $%
%TCIMACRO{\dbigwedge \nolimits^{k}}%
%BeginExpansion
{\displaystyle\bigwedge\nolimits^{k}}
%EndExpansion
U$ and the space of multiforms is denoted by$%
%TCIMACRO{\dbigwedge }%
%BeginExpansion
{\displaystyle\bigwedge}
%EndExpansion
U$.

The dual basis of the fiducial basis \{$\mathbf{b}_{\mu}\}$ is said to be a
fiducial basis for $U$ and is denoted $\{%
%TCIMACRO{\TeXButton{beta}{\mbox{\boldmath{$\beta$}}}}%
%BeginExpansion
\mbox{\boldmath{$\beta$}}%
%EndExpansion
^{\mu}\}$, i.e., $%
%TCIMACRO{\TeXButton{beta}{\mbox{\boldmath{$\beta$}}}}%
%BeginExpansion
\mbox{\boldmath{$\beta$}}%
%EndExpansion
^{\mu}(\mathbf{b}_{\nu})=\delta_{\nu}^{\mu}$. To be able to use Einstein's sum
convention we introduce also the notation,%
\begin{equation}
b^{\mu}=\mathbf{b}_{\mu}\text{ }e\text{ }\beta_{\mu}=%
%TCIMACRO{\TeXButton{beta}{\mbox{\boldmath{$\beta$}}}}%
%BeginExpansion
\mbox{\boldmath{$\beta$}}%
%EndExpansion
^{\mu}.
\end{equation}

The canonical scalar product is constructed as in Section 2 using
$\{\mathbf{b}_{\mu}\}$ and in what follows we use for our calculations the
canonical Clifford $\mathcal{C\ell(}U,\cdot)$.

\subsubsection{The Position $1$-Form}

To any $p\in\mathcal{U}$ there correspond exactly $n$ real numbers $\xi
^{1},\ldots,\xi^{n}$, its position coordinates in the local chart
$(\mathcal{U},\phi)_{o}$. It is thus possible to define a $1$-form on $U$,
\begin{equation}
x=\xi^{\mu}\beta_{\mu}, \label{3.1}%
\end{equation}
associated to $p\in\mathcal{U}$. The $1$-form $x\in U$ \emph{localize }$p$ in
the vector space $U$ and we call $x$ the\emph{\ position 1-form }(\emph{or
position vector})\emph{ of }$p.$

Observe that given an arbitrary $1$-form on $U$\ it will not necessarily will
be the\ position vector of some point of the open set $\mathcal{U}$, since the
components of an arbitrary element of $U$\ does not need to be necessarily the
coordinates of some point of $\mathcal{U}$.

The set of all position vectors of the points of $\mathcal{U}$ is denoted
$U_{o}\subset U$. This subset of the vector space $U$ is not necessarily a
vector subspace of $U$.

Observe also that we have immediately for any $x,y\in U_{o}$ the
identification of the tangent spaces $%
%TCIMACRO{\dbigwedge }%
%BeginExpansion
{\displaystyle\bigwedge}
%EndExpansion
T_{x}^{\ast}U$ and $%
%TCIMACRO{\dbigwedge }%
%BeginExpansion
{\displaystyle\bigwedge}
%EndExpansion
T_{y}^{\ast}U$ with $%
%TCIMACRO{\dbigwedge }%
%BeginExpansion
{\displaystyle\bigwedge}
%EndExpansion
U$.

\subsection{Multiform Fields}

From what has been said it is obvious that any $\mathcal{X\in}\sec%
%TCIMACRO{\dbigwedge }%
%BeginExpansion
{\displaystyle\bigwedge}
%EndExpansion
t^{\clubsuit}M$ when restricted to $\mathcal{U}$ will be represented by a
multiform function $X$ of the position vector, which we will write as%
\begin{equation}
X\mathcal{\in}\sec%
%TCIMACRO{\dbigwedge }%
%BeginExpansion
{\displaystyle\bigwedge}
%EndExpansion
T^{\ast}U
\end{equation}
meaning that for each $x\in U_{o}$, $X(x)=X_{(x)}\in%
%TCIMACRO{\dbigwedge }%
%BeginExpansion
{\displaystyle\bigwedge}
%EndExpansion
T_{x}^{\ast}U\simeq%
%TCIMACRO{\dbigwedge }%
%BeginExpansion
{\displaystyle\bigwedge}
%EndExpansion
U$. Eventually we also use the notation $X:U_{0}\rightarrow%
%TCIMACRO{\dbigwedge }%
%BeginExpansion
{\displaystyle\bigwedge}
%EndExpansion
U.$

If $X\mathcal{\in}\sec%
%TCIMACRO{\dbigwedge }%
%BeginExpansion
{\displaystyle\bigwedge}
%EndExpansion
T^{\ast}U$ is $C^{\infty}$-differentiable on the set $U_{0},$ then the
multiform $\mathcal{X}\sec%
%TCIMACRO{\dbigwedge }%
%BeginExpansion
{\displaystyle\bigwedge}
%EndExpansion
t^{\clubsuit}M$ is differentiable on $\mathcal{U}$.

\subsubsection{Extensor Fields}

Any $(p,q)$-extensor field $\mathfrak{t}$ on $M$, when restricted to
$\mathcal{U}$ is represented by an extensor function $t$ of the position
vector. We write
\begin{equation}
t\in\sec(p,q)\text{-}extU,
\end{equation}
and eventually also write $t:U_{0}\rightarrow(p,q)$-$extU$.

If $t$ is $C^{\infty}$-differentiable\footnote{A $(p,q)$-extensorial function
$t$ is said to be $C^{\infty}$-differentiable on $U_{0}$ iff for an any
$C^{\infty}$-differentiable $p$-form function on $U_{0}$, say $x\mapsto X(x)$,
the $q$-form function $x\mapsto t_{(x)}(X(x))$ is $C^{\infty}$-differentiable
on $U_{0}$.} on $U_{o},$ then the $(p,q)$-extensor field $\mathfrak{t}$ is
smooth on $\mathcal{U}$. For a general extensor field, say $\Delta$ we use the
notation $\Delta\in\sec extU$.

\subsection{Parallelism Structure $(U_{0},\lambda)$ and Covariant Derivatives}

Given a smooth manifold $M$ ($\dim M=n)$ and an \textit{arbitrary} linear
connection $\mathbf{\nabla}$ on $M$, the pair $(M,\mathbf{\nabla)}$ is said to
be a \textit{parallelism structure}. We will now analyze some important
parallelism structures that will appear in the next sections.

\subsubsection{The Connection $2$\textit{-extensor field }$\gamma$ on $U_{o}$
and Associated Extensors Covariant Derivative of Multiform Fields Associated
to $(U_{0},\gamma)$}

The parallelism structure $(M,\mathbf{\nabla)}$ is represented on $U_{o}$ by a
pair $(U_{0},\gamma)$ (also called a parallelism structure) where $\gamma
\in\sec extU$ is called a smooth \textit{connection }$2$\textit{-extensor
field} on $U_{o}$, i.e., for each $x\in U_{0}$,
\begin{equation}
\gamma_{(x)}:%
%TCIMACRO{\dbigwedge \nolimits^{1}}%
%BeginExpansion
{\displaystyle\bigwedge\nolimits^{1}}
%EndExpansion
T_{x}U\times%
%TCIMACRO{\dbigwedge \nolimits^{1}}%
%BeginExpansion
{\displaystyle\bigwedge\nolimits^{1}}
%EndExpansion
T_{x}U\rightarrow%
%TCIMACRO{\dbigwedge \nolimits^{1}}%
%BeginExpansion
{\displaystyle\bigwedge\nolimits^{1}}
%EndExpansion
T_{x}U.
\end{equation}

We also define the following fields associated to $\gamma$. First, given
$a\in\sec%
%TCIMACRO{\dbigwedge \nolimits^{1}}%
%BeginExpansion
{\displaystyle\bigwedge\nolimits^{1}}
%EndExpansion
T^{\ast}U$, define the smooth $(1,1)$-extensor field on $U_{o}$, $\gamma
_{a}\in\sec(1,1)$-$extU$ by
\begin{equation}
\gamma_{a}(b)=\gamma(a,b)
\end{equation}
for all $b\in\sec%
%TCIMACRO{\dbigwedge \nolimits^{1}}%
%BeginExpansion
{\displaystyle\bigwedge\nolimits^{1}}
%EndExpansion
T^{\ast}U$. The field $\gamma_{a}$ is called a \textit{connection}\emph{
}$(1,1)$\emph{-extensor field on }$U_{o}$.

Next we introduce a smooth $(1,2)$-extensor field on $U_{0}$, say $\omega
\in\sec(1,2)$-$extU$ such that for $a\in\sec%
%TCIMACRO{\dbigwedge \nolimits^{1}}%
%BeginExpansion
{\displaystyle\bigwedge\nolimits^{1}}
%EndExpansion
T^{\ast}U$ $\omega(a)\in\sec%
%TCIMACRO{\dbigwedge \nolimits^{2}}%
%BeginExpansion
{\displaystyle\bigwedge\nolimits^{2}}
%EndExpansion
T^{\ast}U$ is given by
\begin{equation}
\omega(a)=\frac{1}{2}\mathrm{bif}[\gamma_{a}]
\end{equation}
The field $\omega$ will be called (for reasons that will become clear in
awhile) a\emph{ rotation gauge field. }

Finally we define the smooth extensor field on $U_{o}$, $\Gamma_{a}\in\sec
extU$ such that for each $x\in U_{o}$, $\Gamma_{(x)a}$ is the generalized of
$\gamma_{(x)a}$ (recall Eq.(\ref{1.53})).

\subsubsection{Covariant Derivative of Multiform Fields Associated to
$(U_{0},\gamma)$}

The connection $2$-extensor field $\gamma$ is used in the following way. For
any smooth $a\in\sec%
%TCIMACRO{\dbigwedge \nolimits^{1}}%
%BeginExpansion
{\displaystyle\bigwedge\nolimits^{1}}
%EndExpansion
T^{\ast}U$ which is a representative on $U$ of a vector field $\mathfrak{a}%
\in\sec T\mathcal{U}$ we introduce two covariant operators $\nabla_{a}$ and
$\nabla_{a}^{-},$ acting the module of smooth multiform fields on $U_{o}$.

First, if $b\in\sec%
%TCIMACRO{\dbigwedge \nolimits^{1}}%
%BeginExpansion
{\displaystyle\bigwedge\nolimits^{1}}
%EndExpansion
T^{\ast}U$
\begin{subequations}
\label{3.77}%
\begin{align}
\nabla_{a}b  &  =a\cdot\partial b+\gamma_{a}(b),\label{3.77a}\\
\nabla_{a}b  &  =a\cdot\partial b-\gamma_{a}(b) \label{3.77b}%
\end{align}
For any smooth $X\in\sec%
%TCIMACRO{\dbigwedge }%
%BeginExpansion
{\displaystyle\bigwedge}
%EndExpansion
T^{\ast}U$ we define
\end{subequations}
\begin{subequations}
\label{3.7}%
\begin{align}
\nabla_{a}X  &  =a\cdot\partial X+\Gamma_{a}(X),\label{3.7a}\\
\nabla_{a}^{-}X  &  =a\cdot\partial X-\Gamma_{a}^{\dagger}(X), \label{3.7b}%
\end{align}
where $\Gamma_{a}$ is the generalized of $\gamma_{a}$ and $\Gamma_{a}%
^{\dagger}$ is the adjoint of $\Gamma_{a}$.

The operators $\nabla_{a}$ and $\nabla_{a}^{-}$ \ are well defined covariant
derivatives on $U_{o}$ sand thus satisfy all properties of a covariant
derivative operator, i.e., they are linear with respect to the direction
1-form field, i.e., we have
\end{subequations}
\begin{equation}
\nabla_{\alpha a+\beta b}^{\pm}X=\alpha\nabla_{a}^{\pm}X+\beta\nabla_{b}^{\pm
}X, \label{3.8}%
\end{equation}
for $\alpha,\beta\in\mathbb{R}$ and $a,b:\in\sec%
%TCIMACRO{\dbigwedge ^{1}}%
%BeginExpansion
{\displaystyle\bigwedge^{1}}
%EndExpansion
T^{\ast}U$ and
\begin{equation}
\nabla_{a}f=a\cdot\partial f,\text{ \ }\nabla_{a}(X+Y)=\nabla_{a}X+\nabla
_{a}Y\text{, \ }\nabla_{a}(fX)=a\cdot\partial f+f\nabla_{a}X, \label{3.88}%
\end{equation}
with analog equations for $\nabla_{a}^{-}.$

To prove, e.g., that $\nabla_{a}f=a\cdot\partial f$ it is enough to recall a
property of the generalized $T$ of a given $(1,1)$-extensor $t$, explicitly
$T(\alpha)=0$, with $\alpha\in\mathbb{R}$. We then have using Eq.(\ref{3.7a})
that for any smooth scalar field $f$ indeed $\nabla_{a}f=a\cdot\partial
f+\Gamma_{a}(f)=a\cdot\partial f.$

The proofs of the remaining formulas in Eq.(\ref{3.88}) are equally elementary.

We can prove also that $\nabla_{a}$ and $\nabla_{a}^{-}$ satisfy
\emph{Leibniz's rule} for the exterior product of smooth multiform fields,
i.e.,
\begin{equation}
\nabla_{a}(X\wedge Y)=(\nabla_{a}X)\wedge Y+X\wedge(\nabla_{a}Y),
\label{3.888}%
\end{equation}
with an analogous formula for $\nabla_{a}^{-}$. Indeed, take two smooth
multiform fields $X$ and $Y$. Utilizing the Leibniz's rule satisfied by the
directional derivative operator $a\cdot\partial$ when applied to the exterior
product of smooth multiform fields and a property of the generalized of a
$(1,1)$-extensor $t$ (namely $T(A\wedge B)=T(A)\wedge B+A\wedge T(B)$, for any
$A,B\in\sec%
%TCIMACRO{\dbigwedge }%
%BeginExpansion
{\displaystyle\bigwedge}
%EndExpansion
T^{\ast}U$), we get
\begin{align}
\nabla_{a}(X\wedge Y)  &  =a\cdot\partial(X\wedge Y)+\Gamma_{a}(X\wedge
Y)\nonumber\\
&  =(a\cdot\partial X)\wedge Y+X\wedge(a\cdot\partial Y)+\Gamma_{a}(X)\wedge
Y+X\wedge\Gamma_{a}(Y)\nonumber\\
\nabla_{a}(X\wedge Y)  &  =(\nabla_{a}X)\wedge Y+X\wedge(\nabla_{a}Y),
\label{3.11}%
\end{align}
which proves Eq.(\ref{3.888})

On the other hand the operator of ordinary directional derivative
$a\cdot\partial$\ and the operators $\nabla_{a}$ and $\nabla_{a}^{-}$ are
related by a notable identity:%
\begin{equation}
a\cdot\partial(X\cdot Y)=(\nabla_{a}X)\cdot Y+X\cdot(\nabla_{a}^{-}%
Y)=(\nabla_{a}^{-}X)\cdot Y+X\cdot(\nabla_{a}Y). \label{inot}%
\end{equation}

Indeed, take two smooth multiform fields $X$ and $Y$. Utilizing the Leibniz's
rule for $a\cdot\partial$ when applied to the exterior product of smooth
multiform fields and that $T(A)\cdot B=A\cdot T^{\dagger}(B)$, for all
$A,B\in\sec%
%TCIMACRO{\dbigwedge }%
%BeginExpansion
{\displaystyle\bigwedge}
%EndExpansion
T^{\ast}U$we see immediately the validity of the formulas
\begin{align}
(\nabla_{a}X)\cdot Y+X\cdot(\nabla_{a}^{-}Y)  &  =(a\cdot\partial X)\cdot
Y+\Gamma_{a}(X)\cdot Y+X\cdot(a\cdot\partial Y)-X\cdot\Gamma_{a}^{\dagger
}(Y)\nonumber\\
&  =(a\cdot\partial X)\cdot Y+X\cdot(a\cdot\partial Y)\nonumber\\
(\nabla_{a}X)\cdot Y+X\cdot(\nabla_{a}^{-}Y)  &  =a\cdot\partial(X\cdot Y).
\label{3.12}%
\end{align}

\subsubsection{Covariant Derivative of Extensor Fields Associated to
$(U_{0},\gamma)$}

The covariant derivative operators $\nabla_{a}$ and $\nabla_{a}^{-}$ \ may be
extended to act on the \textit{module} of smooth $(p,q)$-extensor fields on
$U_{o}$ (which represent smooth $(p,q)$-extensor field on $\mathcal{U\subset
}M$). We define for $t\in\sec(p,q)extU$, $\nabla_{a}t$ and $\nabla_{a}^{-}t$
as the smooth $(p,q)$-extensor fields on $U_{o}$ such that for any smooth
$X\in\in\sec%
%TCIMACRO{\dbigwedge ^{p}}%
%BeginExpansion
{\displaystyle\bigwedge^{p}}
%EndExpansion
T^{\ast}U,$
\begin{subequations}
\label{3.13}%
\begin{align}
(\nabla_{a}t)(X)  &  =\nabla_{a}t(X)-t(\nabla_{a}^{-}X),\label{3.13a}\\
(\nabla_{a}^{-}t)(X)  &  =\nabla_{a}^{-}t(X)-t(\nabla_{a}X). \label{3.13b}%
\end{align}

Observe that in the above formulas $\nabla_{a}t(X)$ denotes the covariant
derivative $\nabla_{a}$ of the smooth multiform field $t(X)\ $and $\nabla
_{a}^{-}X$ is the covariant derivative $\nabla_{a}^{-}$ of the smooth
multiform field $X.$

Certainly those properties are consistent with the linearity property of the
smooth $(p,q)$-extensor fields. Indeed, take two smooth $p$-form fields $X$
and $Y,$ and a smooth scalar field $f$. We have immediately
\end{subequations}
\begin{align}
(\nabla_{a}t)(X+Y)  &  =\nabla_{a}t(X+Y)-t(\nabla_{a}^{-}(X+Y))\nonumber\\
&  =\nabla_{a}(t(X)+t(Y))-t(\nabla_{a}^{-}X+\nabla_{a}^{-}Y)\nonumber\\
&  =\nabla_{a}t(X)+\nabla_{a}t(Y)-t(\nabla_{a}^{-}X)-t(\nabla_{a}%
^{-}Y)=(\nabla_{a}t)(X)+(\nabla_{a}t)(Y),
\end{align}
and
\begin{align}
(\nabla_{a}t)(fX)  &  =\nabla_{a}t(fX)-t(\nabla_{a}^{-}(fX))=\nabla
_{a}(ft(X))-t((a\cdot\partial f)X+f\nabla_{a}^{-}X)\nonumber\\
&  =(a\cdot\partial f)t(X)+f\nabla_{a}t(X)-(a\cdot\partial f)t(X)-ft(\nabla
_{a}^{-}X)=f(\nabla_{a}t)(X).
\end{align}

Each one of the covariant derivatives $t\mapsto\nabla_{a}t$ and $t\mapsto
\nabla_{a}^{-}t$ possess the property of linearity with relation to the
direction 1-form, i.e., $\nabla_{\alpha a+\beta b}^{\pm}t=\alpha\nabla
_{a}^{\pm}t+\beta\nabla_{b}^{\pm}t$, where $\alpha,\beta\in\mathbb{R}$ and
$x\mapsto a,b\in\sec%
%TCIMACRO{\dbigwedge ^{1}}%
%BeginExpansion
{\displaystyle\bigwedge^{1}}
%EndExpansion
T^{\ast}U$

Moreover, $t\mapsto\nabla_{a}t,$and $t\mapsto\nabla_{a}^{-}t$ possess also the
following properties:
\begin{align}
\nabla_{a}^{\pm}(t+u)  &  =\nabla_{a}^{\pm}t+\nabla_{a}^{\pm}u,\label{3.14}\\
\nabla_{a}^{\pm}(ft)  &  =(a\cdot\partial f)t+f\nabla_{a}^{\pm}t, \label{3.15}%
\end{align}
where $t$ and $u$ are smooth $(p,q)$-extensor fields and $f$ is a scalar field.

Let us prove the identities for $\nabla_{a}.$Let $X$ be an arbitrary smooth
$p$-form field,then
\begin{align}
(\nabla_{a}(t+u))(X)  &  =\nabla_{a}(t+u)(X)-(t+u)(\nabla_{a}^{-}X)\nonumber\\
&  =\nabla_{a}(t(X)+u(X))-t(\nabla_{a}^{-}X)-u(\nabla_{a}^{-}X)\nonumber\\
&  =\nabla_{a}t(X)+\nabla_{a}u(X)-t(\nabla_{a}^{-}X)-u(\nabla_{a}%
^{-}X)\nonumber\\
&  =(\nabla_{a}t)(X)+(\nabla_{a}u)(X)=(\nabla_{a}t+\nabla_{a}u)(X),
\end{align}
i.e., $\nabla_{a}(t+u)=\nabla_{a}t+\nabla_{a}u$. Also,
\begin{align}
(\nabla_{a}(ft))(X)  &  =\nabla_{a}(ft)(X)-ft(\nabla_{a}X)=\nabla
_{a}ft(X)-ft(\nabla_{a}X)\nonumber\\
&  =(a\cdot\partial f)t(X)+f\nabla_{a}t(X)-ft(\nabla_{a}X)=(a\cdot\partial
f)t(X)+f(\nabla_{a}t)(X),
\end{align}
i.e.,, $\nabla_{a}(ft)=(a\cdot\partial f)t+f(\nabla_{a}t)$.

\subsubsection{Notable Identities}

We end this section presenting two notable identities, which are:

\textbf{(i)} Let $X$ be a smooth $p$-form field and $Y$ a smooth $q$-form
field, then
\begin{equation}
(\nabla_{a}t)(X)\cdot Y=a\cdot\partial(t(X)\cdot Y)-t(\nabla_{a}^{-}X)\cdot
Y-t(X)\cdot\nabla_{a}^{-}Y. \label{3.16}%
\end{equation}

Indeed, we have
\begin{align*}
(\nabla_{a}t)(X)\cdot Y  &  =\nabla_{a}t(X)\cdot Y-t(\nabla_{a}^{-}X)\cdot Y\\
&  =a\cdot\partial t(X)\cdot Y+\Gamma_{a}(t(X))\cdot Y-t(\nabla_{a}^{-}X)\cdot
Y\\
&  =a\cdot\partial t(X)\cdot Y+t(X)\cdot\Gamma_{a}^{\dagger}(Y)-t(\nabla
_{a}^{-}X)\cdot Y\\
&  =a\cdot\partial t(X)\cdot Y+t(X)\cdot a\cdot\partial Y-t(X)\cdot
a\cdot\partial Y\\
&  +t(X)\cdot\Gamma_{a}^{\dagger}(Y)-t(\nabla_{a}^{-}X)\cdot Y\\
&  =a\cdot\partial t(X)\cdot Y+t(X)\cdot a\cdot\partial Y-t(\nabla_{a}%
^{-}X)\cdot Y\\
&  -t(X)\cdot(a\cdot\partial Y-\Gamma_{a}^{\dagger}(Y))\\
&  =a\cdot\partial(t(X)\cdot Y)-t(\nabla_{a}^{-}X)\cdot Y-t(X)\cdot\nabla
_{a}^{-}Y.
\end{align*}

\textbf{(ii)} For any smooth $(p,q)$-extensor field $t,$ it is
\begin{equation}
\nabla_{a}t^{\dagger}=(\nabla_{a}t)^{\dagger}. \label{3.17}%
\end{equation}

Take the smooth fields $X\in\sec%
%TCIMACRO{\dbigwedge ^{p}}%
%BeginExpansion
{\displaystyle\bigwedge^{p}}
%EndExpansion
T^{\ast}U,$ and $Y\in\sec%
%TCIMACRO{\dbigwedge ^{q}}%
%BeginExpansion
{\displaystyle\bigwedge^{q}}
%EndExpansion
T^{\ast}U$, .\ Utilizing twice Eq.(\ref{3.16}) and the algebraic property
$t(X)\cdot Y=X\cdot t^{\dagger}(Y)$, we have
\begin{align}
(\nabla_{a}t^{\dagger})(Y)\cdot X  &  =a\cdot\partial(t^{\dagger}(Y)\cdot
X)-t^{\dagger}(\nabla_{a}^{-}Y)\cdot X-t^{\dagger}(Y)\cdot\nabla_{a}%
^{-}X\nonumber\\
&  =a\cdot\partial(t(X)\cdot Y)-t(X)\cdot\nabla_{a}^{-}Y-t(\nabla_{a}%
^{-}X)\cdot Y\nonumber\\
&  =(\nabla_{a}t)(X)\cdot Y=X\cdot(\nabla_{a}t)^{\dagger}(Y),
\end{align}
which implies $(\nabla_{a}t^{\dagger})(Y)=(\nabla_{a}t)^{\dagger}(Y)$, i.e.,
$\nabla_{a}t^{\dagger}=(\nabla_{a}t)^{\dagger}$.\medskip

For $\nabla_{a}^{-}$ completely analog properties hold, i.e., for any smooth
$p$-form field $X$ and all smooth $q$-form field $Y$ it is
\begin{equation}
(\nabla_{a}^{-}t)(X)\cdot Y=a\cdot\partial(t(X)\cdot Y)-t(\nabla_{a}X)\cdot
Y-t(X)\cdot\nabla_{a}Y. \label{3.18}%
\end{equation}
Also, for any smooth $(p,q)$-extensor field $t$,
\begin{equation}
\nabla_{a}^{-}t^{\dagger}=(\nabla_{a}^{-}t)^{\dagger}. \label{3.19}%
\end{equation}

\subsubsection{The $2$-Exform Torsion Field of the Structure $(U_{o},\gamma)$}

Given a parallelism structure $(M,\mathbf{\nabla})$ we know that the torsion
and curvature operators (used for the definition of the torsion and Riemann
curvature tensors) characterize $\mathbf{\nabla}$ completely. Let
$(U_{o},\gamma)$ be the representative of the structure $(M,\mathbf{\nabla})$
restricted to $\mathcal{U}$. We now introduce the \textit{representatives} of
the torsion and curvature operators of $(M,\mathbf{\nabla})$ on $U_{o}$.

First we introduce the smooth torsion\emph{ }$2$\emph{-exform}\footnote{A
$2$-exform on $U$ is an antisymmetric $2$-extensor on $U$ , i.e., a linear
mapping $\theta:%
%TCIMACRO{\dbigwedge \nolimits^{1}}%
%BeginExpansion
{\displaystyle\bigwedge\nolimits^{1}}
%EndExpansion
U\times%
%TCIMACRO{\dbigwedge \nolimits^{1}}%
%BeginExpansion
{\displaystyle\bigwedge\nolimits^{1}}
%EndExpansion
U\rightarrow%
%TCIMACRO{\dbigwedge }%
%BeginExpansion
{\displaystyle\bigwedge}
%EndExpansion
U$ \ such that for all $a,b\in%
%TCIMACRO{\dbigwedge \nolimits^{1}}%
%BeginExpansion
{\displaystyle\bigwedge\nolimits^{1}}
%EndExpansion
U$ it is $\theta(a,b)=-\theta(b,a)$.} field $\overset{\gamma}{\tau}\in\sec
extU$, such for smooth $a,b\in\sec%
%TCIMACRO{\dbigwedge \nolimits^{1}}%
%BeginExpansion
{\displaystyle\bigwedge\nolimits^{1}}
%EndExpansion
T^{\ast}U$ we have $(a,b)\mapsto\overset{\gamma}{\tau}(a,b)\in\sec%
%TCIMACRO{\dbigwedge \nolimits^{1}}%
%BeginExpansion
{\displaystyle\bigwedge\nolimits^{1}}
%EndExpansion
T^{\ast}U$, given by
\begin{equation}
\overset{\gamma}{\tau}(a,b):=\nabla_{a}b-\nabla_{b}a-[a,b]=\gamma
_{a}(b)-\gamma_{b}(a), \label{3.20}%
\end{equation}

Next we introduce the $(2,1)$-extensor field $\overset{\gamma}{T}\in\sec
(2,1)$-$extU$ ,such that for any smooth $B\in\sec%
%TCIMACRO{\dbigwedge ^{1}}%
%BeginExpansion
{\displaystyle\bigwedge^{1}}
%EndExpansion
T^{\ast}U$, $B\mapsto\overset{\gamma}{T}(B)\in\sec%
%TCIMACRO{\dbigwedge ^{1}}%
%BeginExpansion
{\displaystyle\bigwedge^{1}}
%EndExpansion
T^{\ast}U$ we have:
\begin{equation}
\overset{\gamma}{T}(B):=\frac{1}{2}B\cdot(\partial_{a}\wedge\partial
_{b})\overset{\gamma}{\tau}(a,b). \label{3.21}%
\end{equation}
$\overset{\gamma}{T}$ is called the \textit{torsion}\emph{ }$(2,1)$%
\emph{-extensor field}

Observe that if the covariant derivative $\nabla_{a}$ is \emph{symmetric}
(i.e., $\nabla_{a}b-\nabla_{b}a=[a,b],$ for all smooth 1-form fields $a$ and
$b$),then $\overset{\gamma}{\tau}(a,b)=0$ and $\overset{\gamma}{T}(B)=0$.

\subsection{Curvature Operator and Curvature Extensor Fields of the Structure
$(U_{o},\gamma)$}

First we introduce a linear operator that acts on the Lie algebra of smooth
$1$-form fields on $U_{o}$. Given $a,b,c\in\sec%
%TCIMACRO{\dbigwedge ^{1}}%
%BeginExpansion
{\displaystyle\bigwedge^{1}}
%EndExpansion
T^{\ast}U$ the curvature operator is the mapping%
\[%
\begin{array}
[c]{ccccc}%
\overset{\gamma}{\rho} & : & \sec(%
%TCIMACRO{\dbigwedge \nolimits^{1}}%
%BeginExpansion
{\displaystyle\bigwedge\nolimits^{1}}
%EndExpansion
T^{\ast}U\times%
%TCIMACRO{\dbigwedge \nolimits^{1}}%
%BeginExpansion
{\displaystyle\bigwedge\nolimits^{1}}
%EndExpansion
T^{\ast}U\times%
%TCIMACRO{\dbigwedge \nolimits^{1}}%
%BeginExpansion
{\displaystyle\bigwedge\nolimits^{1}}
%EndExpansion
T^{\ast}U) & \rightarrow & \sec%
%TCIMACRO{\dbigwedge \nolimits^{1}}%
%BeginExpansion
{\displaystyle\bigwedge\nolimits^{1}}
%EndExpansion
T^{\ast}U,\\
&  & (a,b,c) & \mapsto & \overset{\gamma}{\rho}(a,b,c),
\end{array}
\]
such that
\begin{equation}
\overset{\gamma}{\rho}(a,b,c):=[\nabla_{a},\nabla_{b}]c-\nabla_{\lbrack
a,b]}c. \label{3.22}%
\end{equation}
The operator $\overset{\gamma}{\rho}$ characterizes the curvature of the
parallelism structure\emph{ }$(M,\mathbf{\nabla})$ on $U_{0}$\emph{. }Its main
properties are:

\textbf{(i)} $\overset{\gamma}{\rho}$ \emph{is antisymmetric} with respect to
the first and second variables, i.e.,
\begin{equation}
\overset{\gamma}{\rho}(a,b,c)=-\overset{\gamma}{\rho}(b,a,c). \label{3.23a}%
\end{equation}

\textbf{(ii)} If the \emph{covariant derivative} $\nabla_{a}$\emph{ is
symmetric} (i.e., $\nabla_{a}b-\nabla_{b}a=[a,b]$ (or equivalently $\gamma
_{a}(b)=\gamma_{b}(a),$ for all smooth $1$-form fields $a$ and $b$) then
$\overset{\gamma}{\rho}$ has a cyclic property\emph{,} i.e.,
\begin{equation}
\overset{\gamma}{\rho}(a,b,c)+\overset{\gamma}{\rho}(b,c,a)+\overset{\gamma
}{\rho}(c,a,b)=0. \label{3.23b}%
\end{equation}

The smooth scalar $4$-extensor field\footnote{A linear mapping $t:\sec(%
%TCIMACRO{\dbigwedge \nolimits^{1}}%
%BeginExpansion
{\displaystyle\bigwedge\nolimits^{1}}
%EndExpansion
U\times%
%TCIMACRO{\dbigwedge \nolimits^{1}}%
%BeginExpansion
{\displaystyle\bigwedge\nolimits^{1}}
%EndExpansion
U\times%
%TCIMACRO{\dbigwedge \nolimits^{1}}%
%BeginExpansion
{\displaystyle\bigwedge\nolimits^{1}}
%EndExpansion
U\times%
%TCIMACRO{\dbigwedge \nolimits^{1}}%
%BeginExpansion
{\displaystyle\bigwedge\nolimits^{1}}
%EndExpansion
U)\rightarrow\mathbb{R}$ is called a scalar $4$-extensor.}, $(w,a,b,c)\mapsto
\overset{\gamma}{\mathbf{R}}_{1}(w,a,b,c)$ such that
\begin{equation}
\overset{\gamma}{\mathbf{R}}_{1}(w,a,b,c)=w\cdot\overset{\gamma}{\rho}(b,c,a),
\label{3.24}%
\end{equation}
for all smooth $1$-form fields $w,a,b$ and $c,$ is called the \emph{curvature
}$4$\emph{-extensor field}

The main properties of $\overset{\gamma}{\mathbf{R}}_{1}$ are:

\textbf{(i)} $\overset{\gamma}{\mathbf{R}}_{1}$ \emph{is antisymmetric} with
respect to the third and fourth variables, i.e.,
\begin{equation}
\overset{\gamma}{\mathbf{R}}_{1}(w,a,b,c)=-\overset{\gamma}{\mathbf{R}}%
_{1}(w,a,c,b). \label{3.25a}%
\end{equation}

\textbf{(iI)} \emph{If} the covariant derivative $\nabla_{a}$ is symmetric
then $\overset{\gamma}{\mathbf{R}}$ possess a \emph{cyclic property}, i.e.,
\begin{equation}
\overset{\gamma}{\mathbf{R}_{1}}(w,a,b,c)+\overset{\gamma}{\mathbf{R}_{1}%
}(w,b,c,a)+\overset{\gamma}{\mathbf{R}_{1}}(w,c,a,b)=0. \label{3.25b}%
\end{equation}

The smooth scalar $2$-extensor field $(a,b)\mapsto\overset{\gamma}{\mathbf{R}%
}_{2}(a,b)$ such that
\begin{equation}
\overset{\gamma}{\mathbf{R}}_{2}(a,b)=\overset{\gamma}{\mathbf{R}}%
_{2}(\partial_{w},a,w,b), \label{3.26}%
\end{equation}
for all smooth $1$-form fields $a$ and $b,$ is called \emph{Ricci }%
$2$\emph{-extensor field}. Note that $\overset{\gamma}{\mathbf{R}}_{2}$ is an
\textit{internal contraction}\emph{\ }of $\overset{\gamma}{\mathbf{R}}_{1}$
between the first and third variables and defined as follows. Let
$(\{\varepsilon^{\mu}\},\{\varepsilon_{\mu}\})$ be a pair of reciprocal basis
on $U_{0}$. We define an internal contraction by%
\[
\overset{\gamma}{\mathbf{R}}_{1}(\partial_{w},a,w,b):=\frac{\partial}{\partial
w_{\mu}}\overset{\gamma}{\mathbf{R}}_{1}(\varepsilon^{\mu}%
,a,w,b)=\overset{\gamma}{\mathbf{R}}_{1}(\varepsilon^{\mu},a,\varepsilon_{\mu
},b)=\overset{\gamma}{\mathbf{R}}_{1}(\varepsilon_{\mu},a,\varepsilon^{\mu
},b),
\]
with $\mu$ summed from $0$ to $3.$

The smooth $(1,1)$-extensor field $b\mapsto\overset{\gamma}{%
%TCIMACRO{\TeXButton{R}{\slR}}%
%BeginExpansion
\slR
%EndExpansion
}_{1}(b)$ such that
\begin{equation}
a\cdot\overset{\gamma}{%
%TCIMACRO{\TeXButton{R}{\slR}}%
%BeginExpansion
\slR
%EndExpansion
}_{1}(b)=\overset{\gamma}{\mathbf{R}}_{2}(a,b), \label{3.27}%
\end{equation}
for all smooth $1$-form fields $a$ and $b,$ is called the \emph{Ricci }%
$(1,1)$\emph{-extensor field.}

Note that we can write $\overset{\gamma}{%
%TCIMACRO{\TeXButton{R}{\slR}}%
%BeginExpansion
\slR
%EndExpansion
}_{1}(b)$ as\footnote{once we recall that $\partial_{a}\mathbf{R}%
_{2}(a,b)\equiv\varepsilon^{\mu}(\varepsilon_{\mu}\cdot\partial_{a}%
\mathbf{R}_{2}(a,b))$, i.e., $\partial_{a}\mathbf{R}_{2}(a,b)=\varepsilon
^{\mu}\mathbf{R}_{2}(\varepsilon_{\mu},b)$, oncew we utilize the linearity of
$\mathbf{R}_{2}$ with respect to the first variable.}
\begin{equation}
\overset{\gamma}{%
%TCIMACRO{\TeXButton{R}{\slR}}%
%BeginExpansion
\slR
%EndExpansion
}_{1}(b)=\partial_{a}\overset{\gamma}{\mathbf{R}}_{2}(a,b), \label{3.27a}%
\end{equation}
once we recall that $\partial_{a}\overset{\gamma}{\mathbf{R}}_{2}%
(a,b)\equiv\varepsilon^{\mu}(\varepsilon_{\mu}\cdot\partial_{a}\overset{\gamma
}{\mathbf{R}}_{2}(a,b))$, i.e., $\partial_{a}\overset{\gamma}{\mathbf{R}}%
_{2}(a,b)=\varepsilon^{\mu}\mathbf{R}_{2}(\varepsilon_{\mu},b)$ due to the
linearity of $\overset{\gamma}{\mathbf{R}}_{2}$ with respect to the first variable.

\subsection{Covariant Derivatives Associated to Metric Structures $(U_{o},%
%TCIMACRO{\TeXButton{itg}{\itg}}%
%BeginExpansion
\itg
%EndExpansion
)$}

\subsubsection{Metric Structures}

Let $\mathbf{g}$\textbf{ }be a smooth $(1,1)$-extensor field (associated to a
metric tensor $%
%TCIMACRO{\TeXButton{slg}{\slg}}%
%BeginExpansion
\slg
%EndExpansion
$ $\in\sec T_{2}^{0}M$ ). Its representative $%
%TCIMACRO{\TeXButton{itg}{\itg}}%
%BeginExpansion
\itg
%EndExpansion
$ on a given $U_{o}$ is symmetric (i.e., $%
%TCIMACRO{\TeXButton{itg}{\itg}}%
%BeginExpansion
\itg
%EndExpansion
_{(x)}=%
%TCIMACRO{\TeXButton{itg}{\itg}}%
%BeginExpansion
\itg
%EndExpansion
_{(x)}^{\dagger}$ for all $x\in U_{0}$), is non degenerated (i.e., $\det[%
%TCIMACRO{\TeXButton{itg}{\itg}}%
%BeginExpansion
\itg
%EndExpansion
_{(x)}]\neq0$ for $x\in U_{0}$). In what follows we say that $%
%TCIMACRO{\TeXButton{itg}{\itg}}%
%BeginExpansion
\itg
%EndExpansion
$ is a metric extensor field on $U_{0}$.

A pair $(M,%
%TCIMACRO{\TeXButton{slg}{\slg}}%
%BeginExpansion
\slg
%EndExpansion
)$ is said to be a \textit{metric structure} and its restriction
to$\emph{\ }\mathcal{U}\subset M$\emph{ }is represented by $(U_{o},%
%TCIMACRO{\TeXButton{itg}{\itg}}%
%BeginExpansion
\itg
%EndExpansion
)$, and is also called a \emph{metric structure.}

\subsubsection{Christofell Operators for the Metric Structure $(U_{o},%
%TCIMACRO{\TeXButton{itg}{\itg}}%
%BeginExpansion
\itg
%EndExpansion
)$}

Let $%
%TCIMACRO{\TeXButton{itg}{\itg}}%
%BeginExpansion
\itg
%EndExpansion
$ be the representative of the extensor field $\mathbf{g}$ (associated to $%
%TCIMACRO{\TeXButton{slg}{\slg}}%
%BeginExpansion
\slg
%EndExpansion
$) on $U_{o}$.

The two classical Christofell operators on $\mathcal{U\subset}M$ are
represented on the structure $(U_{o},%
%TCIMACRO{\TeXButton{itg}{\itg}}%
%BeginExpansion
\itg
%EndExpansion
)$\ by two operators that have the same name and that act on the Lie algebra
of smooth $1$-form fields on $U_{0}$. If $a,b,c\in\sec%
%TCIMACRO{\dbigwedge ^{1}}%
%BeginExpansion
{\displaystyle\bigwedge^{1}}
%EndExpansion
T^{\ast}U$ are such fields, we have

\textbf{(i)} \emph{The first Christoffel operator} is the smooth mapping
$(a,b,c)\mapsto\lbrack a,b,c]$ such that
\begin{align}
\lbrack a,b,c]  &  =\frac{1}{2}(a\cdot\partial(%
%TCIMACRO{\TeXButton{itg}{\itg}}%
%BeginExpansion
\itg
%EndExpansion
(b)\cdot c)+b\cdot\partial(%
%TCIMACRO{\TeXButton{itg}{\itg}}%
%BeginExpansion
\itg
%EndExpansion
(c)\cdot a)-c\cdot\partial(%
%TCIMACRO{\TeXButton{itg}{\itg}}%
%BeginExpansion
\itg
%EndExpansion
(a)\cdot b)\nonumber\\
&  +%
%TCIMACRO{\TeXButton{itg}{\itg}}%
%BeginExpansion
\itg
%EndExpansion
(c)\cdot\lbrack a,b]+%
%TCIMACRO{\TeXButton{itg}{\itg}}%
%BeginExpansion
\itg
%EndExpansion
(b)\cdot\lbrack c,a]-%
%TCIMACRO{\TeXButton{itg}{\itg}}%
%BeginExpansion
\itg
%EndExpansion
(a)\cdot\lbrack b,c]), \label{3.2}%
\end{align}
where $[a,b]$ is the Lie bracket of $a$ and $b$ defined by:%

\begin{equation}
\lbrack a,b]=a\cdot\partial b-b\cdot\partial a. \label{clie}%
\end{equation}

\textbf{(ii)} \emph{The second Christoffel operator}, $(a,b,c)\mapsto%
%TCIMACRO{\QDATOPD{\{}{\}}{c}{a,b}}%
%BeginExpansion
\genfrac{\{}{\}}{0pt}{0}{c}{a,b}%
%EndExpansion
$is defined by
\begin{equation}%
%TCIMACRO{\QDATOPD{\{}{\}}{c}{a,b}}%
%BeginExpansion
\genfrac{\{}{\}}{0pt}{0}{c}{a,b}%
%EndExpansion
=[a,b,%
%TCIMACRO{\TeXButton{itg}{\itg}}%
%BeginExpansion
\itg
%EndExpansion
^{-1}(c)]. \label{3.3}%
\end{equation}

Note that $%
%TCIMACRO{\QDATOPD{\{}{\}}{c}{a,b}}%
%BeginExpansion
\genfrac{\{}{\}}{0pt}{0}{c}{a,b}%
%EndExpansion
$is also a smooth $1$-form field since has been defined algebraically from
$[a,b,c]$.

For all smooth form fields $a,a^{\prime},b,b^{\prime},c,c^{\prime}$ and smooth
scalar field $f$, the first Christoffel operator satisfies the elementary
properties:
\begin{align}
\lbrack a+a^{\prime},b,c]  &  =[a,b,c]+[a^{\prime},b,c].\nonumber\\
\lbrack fa,b,c]  &  =f[a,b,c].\nonumber\\
\lbrack a,b+b^{\prime},c]  &  =[a,b,c]+[a,b^{\prime},c].\nonumber\\
\lbrack a,fb,c]  &  =f[a,b,c]+(a\cdot\partial f)%
%TCIMACRO{\TeXButton{itg}{\itg}}%
%BeginExpansion
\itg
%EndExpansion
(b)\cdot c.\nonumber\\
\lbrack a,b,c+c^{\prime}]  &  =[a,b,c]+[a,b,c^{\prime}].\nonumber\\
\lbrack a,b,fc]  &  =f[a,b,c]. \label{3.4}%
\end{align}

Eq.(\ref{3.4}) says that $[a,b,c]$ is linear with respect to the first and
third arguments, but it is \textit{not} linear with respect to the second argument.

Given the smooth 1-form fields $a,b$ and $c$, another relevant properties of
$[a,b,c]$ \ are:
\begin{subequations}
\label{3.5}%
\begin{align}
\lbrack a,b,c]+[b,a,c]  &  =a\cdot\partial(%
%TCIMACRO{\TeXButton{itg}{\itg}}%
%BeginExpansion
\itg
%EndExpansion
(b)\cdot c)+b\cdot\partial(%
%TCIMACRO{\TeXButton{itg}{\itg}}%
%BeginExpansion
\itg
%EndExpansion
(c)\cdot a)-c\cdot\partial(%
%TCIMACRO{\TeXButton{itg}{\itg}}%
%BeginExpansion
\itg
%EndExpansion
(a)\cdot b)\nonumber\\
&  +%
%TCIMACRO{\TeXButton{itg}{\itg}}%
%BeginExpansion
\itg
%EndExpansion
(b)\cdot\lbrack c,a]-%
%TCIMACRO{\TeXButton{itg}{\itg}}%
%BeginExpansion
\itg
%EndExpansion
(a)\cdot\lbrack b,c]\label{3.5a}\\
\lbrack a,b,c]-[b,a,c]  &  =%
%TCIMACRO{\TeXButton{itg}{\itg}}%
%BeginExpansion
\itg
%EndExpansion
(c)\cdot\lbrack a,b].\label{3.5b}\\
\lbrack a,b,c]+[a,c,b]  &  =a\cdot\partial(%
%TCIMACRO{\TeXButton{itg}{\itg}}%
%BeginExpansion
\itg
%EndExpansion
(b)\cdot c)\label{3.5c}\\
\lbrack a,b,c]-[a,c,b]  &  =b\cdot\partial(%
%TCIMACRO{\TeXButton{itg}{\itg}}%
%BeginExpansion
\itg
%EndExpansion
(c)\cdot a)-c\cdot\partial(%
%TCIMACRO{\TeXButton{itg}{\itg}}%
%BeginExpansion
\itg
%EndExpansion
(a)\cdot b)+%
%TCIMACRO{\TeXButton{itg}{\itg}}%
%BeginExpansion
\itg
%EndExpansion
(c)\cdot\lbrack a,b]\nonumber\\
&  +%
%TCIMACRO{\TeXButton{itg}{\itg}}%
%BeginExpansion
\itg
%EndExpansion
(b)\cdot\lbrack c,a]-%
%TCIMACRO{\TeXButton{itg}{\itg}}%
%BeginExpansion
\itg
%EndExpansion
(a)\cdot\lbrack b,c].\label{3.5d}\\
\lbrack a,b,c]+[c,b,a]  &  =b\cdot\partial(%
%TCIMACRO{\TeXButton{itg}{\itg}}%
%BeginExpansion
\itg
%EndExpansion
(c)\cdot a)+%
%TCIMACRO{\TeXButton{itg}{\itg}}%
%BeginExpansion
\itg
%EndExpansion
(c)\cdot\lbrack a,b]-%
%TCIMACRO{\TeXButton{itg}{\itg}}%
%BeginExpansion
\itg
%EndExpansion
(a)\cdot\lbrack b,c]\label{3.5e}\\
\lbrack a,b,c]-[c,b,a]  &  =a\cdot\partial(%
%TCIMACRO{\TeXButton{itg}{\itg}}%
%BeginExpansion
\itg
%EndExpansion
(b)\cdot c)-c\cdot\partial(%
%TCIMACRO{\TeXButton{itg}{\itg}}%
%BeginExpansion
\itg
%EndExpansion
(a)\cdot b)+%
%TCIMACRO{\TeXButton{itg}{\itg}}%
%BeginExpansion
\itg
%EndExpansion
(b)\cdot\lbrack c,a]. \label{3.5f}%
\end{align}

\subsubsection{The $2$-Extensor field $\lambda$}

Given the metric structure $(M,%
%TCIMACRO{\TeXButton{slg}{\slg}}%
%BeginExpansion
\slg
%EndExpansion
)$ and $(U_{o},%
%TCIMACRO{\TeXButton{itg}{\itg}}%
%BeginExpansion
\itg
%EndExpansion
)$, the representative on $U_{o}$ of its restriction to $\mathcal{U}$, we can
construct a connection 2-extensor field $\lambda$ on $U_{0}$, $\lambda\in\sec
extU$, such that $a,b\in\sec%
%TCIMACRO{\dbigwedge \nolimits^{1}}%
%BeginExpansion
{\displaystyle\bigwedge\nolimits^{1}}
%EndExpansion
T^{\ast}U,$
\end{subequations}
\begin{align}
\lambda &  :\sec(%
%TCIMACRO{\dbigwedge \nolimits^{1}}%
%BeginExpansion
{\displaystyle\bigwedge\nolimits^{1}}
%EndExpansion
T^{\ast}U\times%
%TCIMACRO{\dbigwedge \nolimits^{1}}%
%BeginExpansion
{\displaystyle\bigwedge\nolimits^{1}}
%EndExpansion
T^{\ast}U)\rightarrow\sec%
%TCIMACRO{\dbigwedge \nolimits^{1}}%
%BeginExpansion
{\displaystyle\bigwedge\nolimits^{1}}
%EndExpansion
T^{\ast}U,\nonumber\\
\lambda(a,b)  &  =\partial_{c}%
%TCIMACRO{\QATOPD{\{}{\}}{c}{a,b}}%
%BeginExpansion
\genfrac{\{}{\}}{0pt}{}{c}{a,b}%
%EndExpansion
-a\cdot\partial b. \label{3.6}%
\end{align}
\medskip\medskip\textbf{Remark 4.1 }In completely analogy to the case of
connection 2-extensor field $\gamma$, we also introduce here the extensor
fields $\lambda_{a}$ and $\Lambda_{a}$. Those object $\lambda$ and some others
constructed from it (see next subsection) enter in the calculations of
covariant derivatives of multiform fields and extensor fields defined on
$U_{o}$. The field $\lambda$ will be called a \textit{Levi-Civita connection}
on $U_{o}$.\medskip

\subsubsection{Riemann-Cartan and Lorentz-Cartan \textit{MGSS's }$(U_{o},%
%TCIMACRO{\TeXButton{itg}{\itg}}%
%BeginExpansion
\itg
%EndExpansion
,\gamma)$}

Consider a triple $(M,%
%TCIMACRO{\TeXButton{slg}{\slg}}%
%BeginExpansion
\slg
%EndExpansion
,\nabla)$ where $M$ is a smooth manifold ($\dim M=n)$, $%
%TCIMACRO{\TeXButton{slg}{\slg}}%
%BeginExpansion
\slg
%EndExpansion
\in\sec T_{2}^{0}M$ is a Riemannian or a Lorentzian metric tensor and
$\mathbf{\nabla}$ is an arbitrary \textit{metric compatible connection}. Let
$\mathbf{g}$ be the metric extensor corresponding to $%
%TCIMACRO{\TeXButton{slg}{\slg}}%
%BeginExpansion
\slg
%EndExpansion
$ and let $%
%TCIMACRO{\TeXButton{itg}{\itg}}%
%BeginExpansion
\itg
%EndExpansion
$ the representative of $\mathbf{g}$ on $U_{o}$. As we already know the
connection $\mathbf{\nabla}$ is characterized on $U_{o}$ by a smooth
$2$-extensor field $\gamma$ (and other two extensor fields defined from
$\gamma$). The metric compatibility means
\begin{equation}
\nabla%
%TCIMACRO{\TeXButton{slg}{\slg}}%
%BeginExpansion
\slg
%EndExpansion
=0, \label{mc1}%
\end{equation}
and such condition is obviously expressed on $U_{0}$ by%
\begin{equation}
\nabla_{a}^{-}%
%TCIMACRO{\TeXButton{itg}{\itg}}%
%BeginExpansion
\itg
%EndExpansion
=0. \label{mc2}%
\end{equation}
In what follows when $%
%TCIMACRO{\TeXButton{slg}{\slg}}%
%BeginExpansion
\slg
%EndExpansion
$ has Euclidean signature ($(n,0)$ or ($0,n$)) we call $(M,%
%TCIMACRO{\TeXButton{slg}{\slg}}%
%BeginExpansion
\slg
%EndExpansion
,\nabla)$ a \textit{Riemann-Cartan} \textit{MCGSS. When }$%
%TCIMACRO{\TeXButton{slg}{\slg}}%
%BeginExpansion
\slg
%EndExpansion
$ has pseudo-euclidean signature $(1,n-1)$\textit{ }we call $(M,%
%TCIMACRO{\TeXButton{slg}{\slg}}%
%BeginExpansion
\slg
%EndExpansion
,\nabla)$ a \textit{Lorentz-Cartan MCGSS\footnote{The \textit{GSS }where $%
%TCIMACRO{\TeXButton{itg}{\itg}}%
%BeginExpansion
\itg
%EndExpansion
$ has other possible signatures may be called semi-riemannian.}}. We use also
such denominations for the triple $(U_{o},%
%TCIMACRO{\TeXButton{itg}{\itg}}%
%BeginExpansion
\itg
%EndExpansion
,\gamma)$.

\subsubsection{Existence Theorem of the $\gamma%
%TCIMACRO{\TeXButton{itg}{\itg}}%
%BeginExpansion
\itg
%EndExpansion
$-gauge Rotation Extensor of the \textit{MCGSS} $(U_{o},%
%TCIMACRO{\TeXButton{itg}{\itg}}%
%BeginExpansion
\itg
%EndExpansion
,\gamma)$}

We present now a theorem whose proof may be found in \cite{fmr074} and which
pays\ a crucial role in our theory of the gravitational field in Section
6.\medskip

\textbf{Theorem 4.1 \cite{fmr073} }On the \textit{MCGSS} $(U_{o},%
%TCIMACRO{\TeXButton{itg}{\itg}}%
%BeginExpansion
\itg
%EndExpansion
,\gamma)$ there exists a $(1,2)$-extensor field $\overset{\gamma%
%TCIMACRO{\TeXButton{sig}{\sitg}}%
%BeginExpansion
\sitg
%EndExpansion
}{\omega}(a)=\frac{1}{2}\overset{%
%TCIMACRO{\TeXButton{sig}{\sitg}}%
%BeginExpansion
\sitg
%EndExpansion
}{\mathrm{bif}}[\gamma_{a}]$, such that for all smooth $1$-form fields $a$ and
$b$%
\begin{equation}
\gamma_{a}(b)=\frac{1}{2}%
%TCIMACRO{\TeXButton{itg}{\itg}}%
%BeginExpansion
\itg
%EndExpansion
^{-1}(a\cdot\partial%
%TCIMACRO{\TeXButton{itg}{\itg}}%
%BeginExpansion
\itg
%EndExpansion
)(b)+\overset{\gamma%
%TCIMACRO{\TeXButton{sig}{\sitg}}%
%BeginExpansion
\sitg
%EndExpansion
}{\omega}(a)\underset{%
%TCIMACRO{\TeXButton{sig}{\sitg}}%
%BeginExpansion
\sitg
%EndExpansion
}{\times}b, \label{3.28}%
\end{equation}
where
\begin{equation}
\overset{\gamma%
%TCIMACRO{\TeXButton{sig}{\sitg}}%
%BeginExpansion
\sitg
%EndExpansion
}{\omega}(a)\underset{%
%TCIMACRO{\TeXButton{sig}{\sitg}}%
%BeginExpansion
\sitg
%EndExpansion
}{\times}b:=\overset{\gamma%
%TCIMACRO{\TeXButton{sig}{\sitg}}%
%BeginExpansion
\sitg
%EndExpansion
}{\omega}(a)\underset{%
%TCIMACRO{\TeXButton{sig}{\sitg}}%
%BeginExpansion
\sitg
%EndExpansion
}{}b-b\underset{%
%TCIMACRO{\TeXButton{sig}{\sitg}}%
%BeginExpansion
\sitg
%EndExpansion
}{}\overset{\gamma%
%TCIMACRO{\TeXButton{sig}{\sitg}}%
%BeginExpansion
\sitg
%EndExpansion
}{\omega}(a).
\end{equation}

The field $\overset{\gamma%
%TCIMACRO{\TeXButton{sig}{\sitg}}%
%BeginExpansion
\sitg
%EndExpansion
}{\omega}(a)$ is called the\emph{ }$\gamma%
%TCIMACRO{\TeXButton{itg}{\itg}}%
%BeginExpansion
\itg
%EndExpansion
$-\emph{gauge rotation field}.\medskip

\textbf{Corollary 4.1 \cite{fmr073}} Let $a\mapsto%
%TCIMACRO{\TeXButton{itg}{\itg}}%
%BeginExpansion
\itg
%EndExpansion
(a)$ be a metric extensor field defined on $U_{o}$ and let $a\mapsto
\overset{\gamma%
%TCIMACRO{\TeXButton{sig}{\sitg}}%
%BeginExpansion
\sitg
%EndExpansion
}{\omega}(a)$ a smooth $(1,2)$-extensor field defined on $U_{o}$. Define a
smooth connection $2$-extensor field on $U_{o}$, say $(a,b)\mapsto\gamma
_{a}(b)$ by
\begin{equation}
\gamma_{a}(b)=\frac{1}{2}%
%TCIMACRO{\TeXButton{itg}{\itg}}%
%BeginExpansion
\itg
%EndExpansion
^{-1}(a\cdot\partial%
%TCIMACRO{\TeXButton{itg}{\itg}}%
%BeginExpansion
\itg
%EndExpansion
)(b)+\overset{\gamma%
%TCIMACRO{\TeXButton{sig}{\sitg}}%
%BeginExpansion
\sitg
%EndExpansion
}{\omega}(a)\underset{%
%TCIMACRO{\TeXButton{sig}{\sitg}}%
%BeginExpansion
\sitg
%EndExpansion
}{\times}b, \label{3.29}%
\end{equation}
for all smooth $1$-form fields $a$ and $b$. Then the triple $(U_{o},%
%TCIMACRO{\TeXButton{itg}{\itg}}%
%BeginExpansion
\itg
%EndExpansion
,\gamma)$ is a \emph{Riemann-Cartan or Lorentz-Cartan }\textit{MCGSS
}(depending on the signature of $%
%TCIMACRO{\TeXButton{itg}{\itg}}%
%BeginExpansion
\itg
%EndExpansion
$).\textit{ }

\subsubsection{Some Important Properties of a Metric Compatible Connection}

Given the \textit{MCGSS }$(U_{o},%
%TCIMACRO{\TeXButton{itg}{\itg}}%
%BeginExpansion
\itg
%EndExpansion
,\gamma)$ the operators $\nabla_{a}$ and $\nabla_{a}^{-}$ satisfy the
following properties \cite{fmr073}.

\textbf{(i)} \textit{The\emph{ Ricci Theorem} }\emph{ }For all smooth 1-form
fields $a,b$, $c$ \ we have
\begin{equation}
a\cdot\partial(%
%TCIMACRO{\TeXButton{itg}{\itg}}%
%BeginExpansion
\itg
%EndExpansion
(b)\cdot c)=%
%TCIMACRO{\TeXButton{itg}{\itg}}%
%BeginExpansion
\itg
%EndExpansion
(\nabla_{a}b)\cdot c+%
%TCIMACRO{\TeXButton{itg}{\itg}}%
%BeginExpansion
\itg
%EndExpansion
(b)\cdot\nabla_{a}c, \label{3.30}%
\end{equation}

Ricci theorem may be generalized for arbitrary smooth multiform fields $X$ and
$Y$, i.e.,
\begin{equation}
a\cdot\partial(\underline{%
%TCIMACRO{\TeXButton{itg}{\itg}}%
%BeginExpansion
\itg
%EndExpansion
}(X)\cdot Y)=\underline{%
%TCIMACRO{\TeXButton{itg}{\itg}}%
%BeginExpansion
\itg
%EndExpansion
}(\nabla_{a}X)\cdot Y+\underline{%
%TCIMACRO{\TeXButton{itg}{\itg}}%
%BeginExpansion
\itg
%EndExpansion
}(X)\cdot\nabla_{a}Y. \label{3.30aa}%
\end{equation}

\textbf{(ii)}\emph{ Leibniz's rule f}or the metric scalar product (i.e.,
$X\underset{%
%TCIMACRO{\TeXButton{sig}{\sitg}}%
%BeginExpansion
\sitg
%EndExpansion
}{\cdot}Y\equiv\underline{g}(X)\cdot Y$),
\begin{equation}
a\cdot\partial(X\underset{%
%TCIMACRO{\TeXButton{sig}{\sitg}}%
%BeginExpansion
\sitg
%EndExpansion
}{\cdot}Y)=(\nabla_{a}^{-}X)\underset{%
%TCIMACRO{\TeXButton{sig}{\sitg}}%
%BeginExpansion
\sitg
%EndExpansion
}{\cdot}Y+X\underset{%
%TCIMACRO{\TeXButton{sig}{\sitg}}%
%BeginExpansion
\sitg
%EndExpansion
}{\cdot}(\nabla_{a}^{-}Y), \label{3.30b}%
\end{equation}
for all smooth multiform fields $X$ and $Y.$

\textbf{(iii)} For any smooth multiform field $X,$
\begin{equation}
\nabla_{a}^{-}X=\underline{%
%TCIMACRO{\TeXButton{itg}{\itg}}%
%BeginExpansion
\itg
%EndExpansion
}(\nabla_{a}\underline{%
%TCIMACRO{\TeXButton{itg}{\itg}}%
%BeginExpansion
\itg
%EndExpansion
}^{-1}(X)). \label{3,.30c}%
\end{equation}

\textbf{(iv)} The action of the operators $\nabla_{a}$ and $\nabla_{a}^{-}$,
on smooth extensor fields\ permit us to write the compatibility between the
metric and the parallelism as
\begin{align}
\nabla_{a}^{-}%
%TCIMACRO{\TeXButton{itg}{\itg}}%
%BeginExpansion
\itg
%EndExpansion
&  =0,\label{3.30d}\\
\nabla_{a}%
%TCIMACRO{\TeXButton{itg}{\itg}}%
%BeginExpansion
\itg
%EndExpansion
^{-1}  &  =0. \label{3.30e}%
\end{align}

\subsubsection{The Riemann $4$-Extensor Field of a \textit{MCGSS} $(U_{o},%
%TCIMACRO{\TeXButton{itg}{\itg}}%
%BeginExpansion
\itg
%EndExpansion
,\gamma)$}

The smooth scalar $4$-extensor field on $U_{o}$, $(a,b,c,w)\mapsto
\overset{\gamma%
%TCIMACRO{\TeXButton{sig}{\sitg}}%
%BeginExpansion
\sitg
%EndExpansion
}{\mathbf{R}}_{3}(a,b,c,w)$ such that
\begin{equation}
\overset{\gamma%
%TCIMACRO{\TeXButton{sig}{\sitg}}%
%BeginExpansion
\sitg
%EndExpansion
}{\mathbf{R}}_{3}(a,b,c,w)=-\overset{\gamma%
%TCIMACRO{\TeXButton{sig}{\sitg}}%
%BeginExpansion
\sitg
%EndExpansion
}{\rho}(a,b,c)\cdot%
%TCIMACRO{\TeXButton{itg}{\itg}}%
%BeginExpansion
\itg
%EndExpansion
(w), \label{3.31}%
\end{equation}
for all smooth $1$-form fields $a,b,c$ and $w,$ is called the \emph{Riemann
(curvature) }$4$\emph{-extensor field of the structure }$(U_{o},%
%TCIMACRO{\TeXButton{itg}{\itg}}%
%BeginExpansion
\itg
%EndExpansion
,\gamma)$.

$\overset{\gamma%
%TCIMACRO{\TeXButton{sig}{\sitg}}%
%BeginExpansion
\sitg
%EndExpansion
}{\mathbf{R}}_{3}$ satisfy the following important properties: \cite{fmr074}:

\textbf{(i)} On the parallelism structure $(U_{o},\gamma)$, $\overset{\gamma%
%TCIMACRO{\TeXButton{sig}{\sitg}}%
%BeginExpansion
\sitg
%EndExpansion
}{\mathbf{R}}_{3}$ is antisymmetric with respect to the first and second
variables, i.e.,
\begin{equation}
\overset{\gamma%
%TCIMACRO{\TeXButton{sig}{\sitg}}%
%BeginExpansion
\sitg
%EndExpansion
}{\mathbf{R}}_{3}(a,b,c,w)=-\overset{\gamma%
%TCIMACRO{\TeXButton{sig}{\sitg}}%
%BeginExpansion
\sitg
%EndExpansion
}{\mathbf{R}}_{3}(b,a,c,w). \label{3.32a}%
\end{equation}

\textbf{(ii) }On the parallelism structure\ $(U_{o},\gamma)$ if the covariant
derivative $\nabla_{a}$ is symmetric (i.e., $\nabla_{a}b-\nabla_{b}a=[a,b]$),
then $\overset{\gamma%
%TCIMACRO{\TeXButton{sig}{\sitg}}%
%BeginExpansion
\sitg
%EndExpansion
}{\mathbf{R}}_{3}$ possess a cyclic property, i.e.,
\begin{equation}
\overset{\gamma%
%TCIMACRO{\TeXButton{sig}{\sitg}}%
%BeginExpansion
\sitg
%EndExpansion
}{\mathbf{R}}_{3}(a,b,c,w)+\overset{\gamma%
%TCIMACRO{\TeXButton{sig}{\sitg}}%
%BeginExpansion
\sitg
%EndExpansion
}{\mathbf{R}}_{3}(b,c,a,w)+\overset{\gamma%
%TCIMACRO{\TeXButton{sig}{\sitg}}%
%BeginExpansion
\sitg
%EndExpansion
}{\mathbf{R}}_{3}(c,a,b,w)=0. \label{3.32b}%
\end{equation}

\textbf{(iii)} On the \textit{MCGSS} $(U_{o},%
%TCIMACRO{\TeXButton{itg}{\itg}}%
%BeginExpansion
\itg
%EndExpansion
,\gamma)$, $\overset{\gamma%
%TCIMACRO{\TeXButton{sig}{\sitg}}%
%BeginExpansion
\sitg
%EndExpansion
}{\mathbf{R}}_{3}$ is antisymmetric with respect to the third and fourth
variables, i.e.,
\begin{equation}
\overset{\gamma%
%TCIMACRO{\TeXButton{sig}{\sitg}}%
%BeginExpansion
\sitg
%EndExpansion
}{\mathbf{R}}_{3}(a,b,c,w)=-\overset{\gamma%
%TCIMACRO{\TeXButton{sig}{\sitg}}%
%BeginExpansion
\sitg
%EndExpansion
}{\mathbf{R}}_{3}(a,b,w,c). \label{3.32c}%
\end{equation}

\textbf{(iv)}$.$On the \textit{MCGSS} $(U_{o},%
%TCIMACRO{\TeXButton{itg}{\itg}}%
%BeginExpansion
\itg
%EndExpansion
,\gamma)$, if the covariant derivative $\nabla_{a}$ is symmetric (i.e.,
$\nabla_{a}b-\nabla_{b}a=[a,b]$), then $\overset{\gamma%
%TCIMACRO{\TeXButton{sig}{\sitg}}%
%BeginExpansion
\sitg
%EndExpansion
}{\mathbf{R}}_{3}$
\begin{equation}
\overset{\gamma%
%TCIMACRO{\TeXButton{sig}{\sitg}}%
%BeginExpansion
\sitg
%EndExpansion
}{\mathbf{R}}_{3}(a,b,c,w)=\overset{\gamma%
%TCIMACRO{\TeXButton{sig}{\sitg}}%
%BeginExpansion
\sitg
%EndExpansion
}{\mathbf{R}}_{3}(c,w,a,b). \label{3.32d}%
\end{equation}

Note that from the Eqs.(\ref{3.32a}) and (\ref{3.32c}) we have
$\overset{\gamma%
%TCIMACRO{\TeXButton{sig}{\sitg}}%
%BeginExpansion
\sitg
%EndExpansion
}{\mathbf{R}}_{3}(a,b,c,w)=\overset{\gamma%
%TCIMACRO{\TeXButton{sig}{\sitg}}%
%BeginExpansion
\sitg
%EndExpansion
}{\mathbf{R}}_{3}(b,a,w,c)$ and under those conditions, given a pair of
reciprocal basis $(\{\varepsilon^{\mu}\},\{\varepsilon_{\mu}\})$ on $U_{o}$ we
agree in writing \cite{choquet,rodcap2007}%
\begin{equation}
\overset{\gamma%
%TCIMACRO{\TeXButton{sig}{\sitg}}%
%BeginExpansion
\sitg
%EndExpansion
}{\mathbf{R}}_{1}(\varepsilon_{\beta},\varepsilon^{\alpha},\varepsilon_{\rho
},\varepsilon_{\sigma})=R_{\beta\;\rho\sigma}^{\;\alpha}, \label{3.32e}%
\end{equation}
for the \emph{components of the Riemann tensor.} Also, under the same
conditions Eq.(\ref{3.26}) may be written as%
\begin{equation}
\overset{\gamma%
%TCIMACRO{\TeXButton{sig}{\sitg}}%
%BeginExpansion
\sitg
%EndExpansion
}{\mathbf{R}}_{2}(a,b)=\overset{\gamma%
%TCIMACRO{\TeXButton{sig}{\sitg}}%
%BeginExpansion
\sitg
%EndExpansion
}{\mathbf{R}}_{1}(\partial_{w},a,w,b)=\overset{\gamma%
%TCIMACRO{\TeXButton{sig}{\sitg}}%
%BeginExpansion
\sitg
%EndExpansion
}{\mathbf{R}}_{1}(a,\partial_{w},b,w)
\end{equation}
and we have%
\begin{equation}
R_{\beta\alpha}=\overset{\gamma%
%TCIMACRO{\TeXButton{sig}{\sitg}}%
%BeginExpansion
\sitg
%EndExpansion
}{\mathbf{R}}_{2}(\varepsilon_{\beta},\varepsilon_{\rho})=\overset{\gamma%
%TCIMACRO{\TeXButton{sig}{\sitg}}%
%BeginExpansion
\sitg
%EndExpansion
}{\mathbf{R}}_{1}(\varepsilon_{\beta},\varepsilon^{\alpha},\varepsilon_{\rho
},\varepsilon_{\alpha})=R_{\beta\;\rho\alpha}^{\;\alpha} \label{ricci}%
\end{equation}
for the \textit{components of the Ricci tensor}.

The scalar field%
\begin{equation}
\overset{\gamma%
%TCIMACRO{\TeXButton{sig}{\sitg}}%
%BeginExpansion
\sitg
%EndExpansion
}{R}=\overset{\gamma%
%TCIMACRO{\TeXButton{sig}{\sitg}}%
%BeginExpansion
\sitg
%EndExpansion
}{\mathbf{R}}_{2}(%
%TCIMACRO{\TeXButton{itg}{\itg}}%
%BeginExpansion
\itg
%EndExpansion
^{-1}(\partial_{a}),a), \label{escurv}%
\end{equation}
i.e., a $%
%TCIMACRO{\TeXButton{itg}{\itg}}%
%BeginExpansion
\itg
%EndExpansion
^{-1}$-contraction between the first and second variables of $\overset{\gamma%
%TCIMACRO{\TeXButton{sig}{\sitg}}%
%BeginExpansion
\sitg
%EndExpansion
}{\mathbf{R}}_{2}$ is called the\emph{ Ricci scalar field }(or scalar
curvature)\emph{.}

\subsubsection{Existence Theorem for the Riemann $(2,2)$-extensor Field on
$(U_{o},%
%TCIMACRO{\TeXButton{itg}{\itg}}%
%BeginExpansion
\itg
%EndExpansion
,\gamma)$}

\textbf{Proposition 4.1} \cite{fmr074} There exists an unique smooth
$(2,2)$-extensor field, say $B\mapsto\overset{\gamma%
%TCIMACRO{\TeXButton{sig}{\sitg}}%
%BeginExpansion
\sitg
%EndExpansion
}{%
%TCIMACRO{\TeXButton{R}{\slR}}%
%BeginExpansion
\slR
%EndExpansion
}_{2}(B)$, such that
\begin{equation}
\overset{\gamma%
%TCIMACRO{\TeXButton{sig}{\sitg}}%
%BeginExpansion
\sitg
%EndExpansion
}{\mathbf{R}}_{3}(a,b,c,w)=\overset{\gamma%
%TCIMACRO{\TeXButton{sig}{\sitg}}%
%BeginExpansion
\sitg
%EndExpansion
}{%
%TCIMACRO{\TeXButton{R}{\slR}}%
%BeginExpansion
\slR
%EndExpansion
}_{2}(a\wedge b)\cdot(c\wedge w). \label{3.33}%
\end{equation}

This field $B\mapsto\overset{\gamma%
%TCIMACRO{\TeXButton{sig}{\sitg}}%
%BeginExpansion
\sitg
%EndExpansion
}{%
%TCIMACRO{\TeXButton{R}{\slR}}%
%BeginExpansion
\slR
%EndExpansion
}_{2}(B)$ is called the Riemann\emph{ }$(2,2)$\emph{-extensor field }and we
see that\emph{ }Eq.(\ref{3.33}) may be though as a factorization of\emph{
}$\overset{\gamma%
%TCIMACRO{\TeXButton{sig}{\sitg}}%
%BeginExpansion
\sitg
%EndExpansion
}{\mathbf{R}}_{3}$\emph{.\medskip}

\textbf{Proposition 4.2} The Ricci $(1,1)$-extensor field $b\mapsto
\overset{\gamma%
%TCIMACRO{\TeXButton{sig}{\sitg}}%
%BeginExpansion
\sitg
%EndExpansion
}{%
%TCIMACRO{\TeXButton{R}{\slR}}%
%BeginExpansion
\slR
%EndExpansion
}_{1}(b)$ and the Ricci scalar field $R$ may be written as $%
%TCIMACRO{\TeXButton{itg}{\itg}}%
%BeginExpansion
\itg
%EndExpansion
^{-1}$-divergences of the Riemann $(2,2)$-extensor field $B\mapsto
\overset{\gamma%
%TCIMACRO{\TeXButton{sig}{\sitg}}%
%BeginExpansion
\sitg
%EndExpansion
}{%
%TCIMACRO{\TeXButton{R}{\slR}}%
%BeginExpansion
\slR
%EndExpansion
}_{2}(B)$, i.e.,
\begin{align}
\overset{\gamma%
%TCIMACRO{\TeXButton{sig}{\sitg}}%
%BeginExpansion
\sitg
%EndExpansion
}{%
%TCIMACRO{\TeXButton{R}{\slR}}%
%BeginExpansion
\slR
%EndExpansion
}_{1}(b)  &  =%
%TCIMACRO{\TeXButton{itg}{\itg}}%
%BeginExpansion
\itg
%EndExpansion
^{-1}(\partial_{a})\lrcorner\overset{\gamma%
%TCIMACRO{\TeXButton{sig}{\sitg}}%
%BeginExpansion
\sitg
%EndExpansion
}{%
%TCIMACRO{\TeXButton{R}{\slR}}%
%BeginExpansion
\slR
%EndExpansion
}_{2}(a\wedge b),\\
\overset{\gamma%
%TCIMACRO{\TeXButton{sig}{\sitg}}%
%BeginExpansion
\sitg
%EndExpansion
}{R}  &  =%
%TCIMACRO{\TeXButton{itg}{\itg}}%
%BeginExpansion
\itg
%EndExpansion
^{-1}(\partial_{b})\cdot\overset{\gamma%
%TCIMACRO{\TeXButton{sig}{\sitg}}%
%BeginExpansion
\sitg
%EndExpansion
}{%
%TCIMACRO{\TeXButton{R}{\slR}}%
%BeginExpansion
\slR
%EndExpansion
}(b)=\underline{%
%TCIMACRO{\TeXButton{itg}{\itg}}%
%BeginExpansion
\itg
%EndExpansion
}^{-1}(\partial_{a}\wedge\partial_{b})\cdot\overset{\gamma%
%TCIMACRO{\TeXButton{sig}{\sitg}}%
%BeginExpansion
\sitg
%EndExpansion
}{%
%TCIMACRO{\TeXButton{R}{\slR}}%
%BeginExpansion
\slR
%EndExpansion
}_{2}(a\wedge b). \label{3.35}%
\end{align}

\textbf{Proof }\cite{fmr074} A simple algebraic manipulation of Eq.(\ref{3.24}%
) and Eq. (\ref{3.31}) gives the extensor identity
\begin{equation}
\overset{\gamma%
%TCIMACRO{\TeXButton{sig}{\sitg}}%
%BeginExpansion
\sitg
%EndExpansion
}{\mathbf{R}}_{1}(w,a,b,c)=\overset{\gamma%
%TCIMACRO{\TeXButton{sig}{\sitg}}%
%BeginExpansion
\sitg
%EndExpansion
}{\mathbf{R}}_{3}(c,b,a,%
%TCIMACRO{\TeXButton{itg}{\itg}}%
%BeginExpansion
\itg
%EndExpansion
^{-1}(w)). \label{4i}%
\end{equation}

Now, according to Eq.(\ref{3.26}) and Eq. (\ref{3.27}), and taking also in
account Eq.(\ref{4i}), we have
\begin{align}
a\cdot\overset{%
%TCIMACRO{\TeXButton{sig}{\sitg}}%
%BeginExpansion
\sitg
%EndExpansion
}{%
%TCIMACRO{\TeXButton{R}{\slR}}%
%BeginExpansion
\slR
%EndExpansion
}_{1}(b)  &  =\overset{%
%TCIMACRO{\TeXButton{sig}{\sitg}}%
%BeginExpansion
\sitg
%EndExpansion
}{\mathbf{R}}_{2}(a,b)\nonumber\\
&  =\overset{\gamma%
%TCIMACRO{\TeXButton{sig}{\sitg}}%
%BeginExpansion
\sitg
%EndExpansion
}{\mathbf{R}}_{1}(\varepsilon^{\mu},a,\varepsilon_{\mu},b),\text{ (}\mu\text{
summed from }0\text{ to }3)\nonumber\\
&  =\overset{\gamma%
%TCIMACRO{\TeXButton{sig}{\sitg}}%
%BeginExpansion
\sitg
%EndExpansion
}{\mathbf{R}}_{3}(b,\varepsilon_{\mu},a,%
%TCIMACRO{\TeXButton{itg}{\itg}}%
%BeginExpansion
\itg
%EndExpansion
^{-1}(\varepsilon^{\mu})),
\end{align}
and utilizing essentially the factorization given by Eq.(\ref{3.33}) we have
\begin{align*}
a\cdot\overset{\gamma%
%TCIMACRO{\TeXButton{sig}{\sitg}}%
%BeginExpansion
\sitg
%EndExpansion
}{%
%TCIMACRO{\TeXButton{R}{\slR}}%
%BeginExpansion
\slR
%EndExpansion
}_{1}(b)  &  =\overset{\gamma%
%TCIMACRO{\TeXButton{sig}{\sitg}}%
%BeginExpansion
\sitg
%EndExpansion
}{%
%TCIMACRO{\TeXButton{R}{\slR}}%
%BeginExpansion
\slR
%EndExpansion
}_{2}(b\wedge\varepsilon_{\mu})\cdot(a\wedge%
%TCIMACRO{\TeXButton{itg}{\itg}}%
%BeginExpansion
\itg
%EndExpansion
^{-1}(\varepsilon^{\mu}))=(%
%TCIMACRO{\TeXButton{itg}{\itg}}%
%BeginExpansion
\itg
%EndExpansion
^{-1}(\varepsilon^{\mu})\wedge a)\cdot\overset{\gamma%
%TCIMACRO{\TeXButton{sig}{\sitg}}%
%BeginExpansion
\sitg
%EndExpansion
}{%
%TCIMACRO{\TeXButton{R}{\slR}}%
%BeginExpansion
\slR
%EndExpansion
}_{2}(\varepsilon_{\mu}\wedge b)\\
&  =a\cdot(%
%TCIMACRO{\TeXButton{itg}{\itg}}%
%BeginExpansion
\itg
%EndExpansion
^{-1}(\varepsilon^{\mu})\lrcorner\overset{%
%TCIMACRO{\TeXButton{sig}{\sitg}}%
%BeginExpansion
\sitg
%EndExpansion
}{%
%TCIMACRO{\TeXButton{R}{\slR}}%
%BeginExpansion
\slR
%EndExpansion
}_{2}(\varepsilon_{\mu}\wedge b)),
\end{align*}
i.e., $\overset{\gamma%
%TCIMACRO{\TeXButton{sig}{\sitg}}%
%BeginExpansion
\sitg
%EndExpansion
}{%
%TCIMACRO{\TeXButton{R}{\slR}}%
%BeginExpansion
\slR
%EndExpansion
}_{1}(b)=%
%TCIMACRO{\TeXButton{itg}{\itg}}%
%BeginExpansion
\itg
%EndExpansion
^{-1}(\varepsilon^{\mu})\lrcorner\overset{\gamma%
%TCIMACRO{\TeXButton{sig}{\sitg}}%
%BeginExpansion
\sitg
%EndExpansion
}{%
%TCIMACRO{\TeXButton{R}{\slR}}%
%BeginExpansion
\slR
%EndExpansion
}_{2}(\varepsilon_{\mu}\wedge b)=%
%TCIMACRO{\TeXButton{itg}{\itg}}%
%BeginExpansion
\itg
%EndExpansion
^{-1}(\partial_{a})\lrcorner\overset{\gamma%
%TCIMACRO{\TeXButton{sig}{\sitg}}%
%BeginExpansion
\sitg
%EndExpansion
}{%
%TCIMACRO{\TeXButton{R}{\slR}}%
%BeginExpansion
\slR
%EndExpansion
}_{2}(a\wedge b)$.%
%TCIMACRO{\TeXButton{End Proof}{\endproof}}%
%BeginExpansion
\endproof
%EndExpansion

\subsubsection{The Einstein $(1,1)$-Extensor Field}

The smooth $(1,1)$-extensor field, $a\mapsto\overset{\gamma%
%TCIMACRO{\TeXButton{sig}{\sitg}}%
%BeginExpansion
\sitg
%EndExpansion
}{%
%TCIMACRO{\TeXButton{G}{\slG}}%
%BeginExpansion
\slG
%EndExpansion
}(a)$, given by
\begin{equation}
\overset{\gamma%
%TCIMACRO{\TeXButton{sig}{\sitg}}%
%BeginExpansion
\sitg
%EndExpansion
}{%
%TCIMACRO{\TeXButton{G}{\slG}}%
%BeginExpansion
\slG
%EndExpansion
}(a)=\overset{\gamma%
%TCIMACRO{\TeXButton{sig}{\sitg}}%
%BeginExpansion
\sitg
%EndExpansion
}{%
%TCIMACRO{\TeXButton{R}{\slR}}%
%BeginExpansion
\slR
%EndExpansion
}_{1}(a)-\frac{1}{2}%
%TCIMACRO{\TeXButton{itg}{\itg}}%
%BeginExpansion
\itg
%EndExpansion
(a)\overset{\gamma%
%TCIMACRO{\TeXButton{sig}{\sitg}}%
%BeginExpansion
\sitg
%EndExpansion
}{R}, \label{3.36}%
\end{equation}
is called the \textit{Einstein}\emph{ }$(1,1)$\emph{-extensor field.}

\subsection{Riemann and Lorentz \textit{MCGSS's} $(U_{o},%
%TCIMACRO{\TeXButton{itg}{\itg}}%
%BeginExpansion
\itg
%EndExpansion
,\lambda)$}

\subsubsection{Levi-Civita Covariant Derivative}

It is important to have in mind that the unique \textit{MCGSS} $(U_{o},%
%TCIMACRO{\TeXButton{itg}{\itg}}%
%BeginExpansion
\itg
%EndExpansion
,\gamma)$ where the covariant derivative $\nabla_{a}$ is symmetric is the one
where the connection $2$-extensor field $\gamma$ is precisely the Levi-Civita
one,\ that we already denoted by $\lambda$ (recall Eq.(\ref{3.6})). In this
case the \textit{MCGSS} $(M,%
%TCIMACRO{\TeXButton{slg}{\slg}}%
%BeginExpansion
\slg
%EndExpansion
,\mathbf{\nabla}=D)$ is said a Riemann or \textit{Lorentz MCGSS }depending on
the signature of $%
%TCIMACRO{\TeXButton{slg}{\slg}}%
%BeginExpansion
\slg
%EndExpansion
$\textit{.}\emph{ }We also use this denomination for the structure\emph{\ }%
$(U_{o},%
%TCIMACRO{\TeXButton{itg}{\itg}}%
%BeginExpansion
\itg
%EndExpansion
,\lambda)$ The covariant derivatives defined by $\lambda$ are precisely the
Levi-Civita covariant derivatives $D_{a}$ and $D_{a}^{-}$. Some properties of
the curvature extensors of the \textit{MCGSS }$(U_{o},%
%TCIMACRO{\TeXButton{itg}{\itg}}%
%BeginExpansion
\itg
%EndExpansion
,\gamma)$\ which are important for this paper will be given below, and for
that important case \textit{simplified} notations will be used.

The covariant derivative operators associated to the connection $2$-extensor
field $\lambda$ will be denoted $D_{a}$ and $D_{a}^{-}$, and%

\begin{subequations}
\label{3.37}%
\begin{align}
D_{a}X  &  =a\cdot\partial X+\Lambda_{a}(X),\label{3.37a}\\
D_{a}^{-}X  &  =a\cdot\partial X-\Lambda_{a}^{\dagger}(X), \label{3.37b}%
\end{align}
for all smooth multiform field $X\in\sec%
%TCIMACRO{\dbigwedge }%
%BeginExpansion
{\displaystyle\bigwedge}
%EndExpansion
T^{\ast}U$ and are called \textit{Levi-Civita covariant derivative operators.
}Note that $\Lambda_{a}$ is the generalized of $\lambda_{a}$ and $\Lambda
_{a}^{\dagger}$ is the adjoint of $\Lambda_{a}.$

\subsubsection{Properties of $D_{a}$}

\textbf{(i) }For all smooth $1$-form fields $a,b$ and $c$ it is:
\end{subequations}
\begin{equation}
(D_{a}b)\cdot c=%
%TCIMACRO{\QATOPD{\{}{\}}{c}{a,b}}%
%BeginExpansion
\genfrac{\{}{\}}{0pt}{}{c}{a,b}%
%EndExpansion
. \label{3.38}%
\end{equation}

Indeed, using Eq.(\ref{3.37a}) and Eq.(\ref{3.6}) we have immediately
\[
D_{a}b=a\cdot\partial b+\lambda_{a}(b)=a\cdot\partial b+\partial_{c}%
%TCIMACRO{\QATOPD{\{}{\}}{c}{a,b}}%
%BeginExpansion
\genfrac{\{}{\}}{0pt}{}{c}{a,b}%
%EndExpansion
-a\cdot\partial b=\partial_{c}%
%TCIMACRO{\QATOPD{\{}{\}}{c}{a,b}}%
%BeginExpansion
\genfrac{\{}{\}}{0pt}{}{c}{a,b}%
%EndExpansion
.
\]

Now, utilizing the formula for multiform differentiation $c\cdot\partial
_{n}\phi(n)=c\cdot(\partial_{n}\phi(n))$, with $\phi$ a scalar function of
$1$-form variable, the formula $c\cdot\partial_{n}f(n)=f(c)$, where $f$ is a
multiform function of $1$-form variable and taking into account the linearity
of the Christoffel operator with respect to the superior argument, we have:
\[
(D_{a}b)\cdot c=c\cdot(\partial_{n}%
%TCIMACRO{\QATOPD{\{}{\}}{n}{a,b}}%
%BeginExpansion
\genfrac{\{}{\}}{0pt}{}{n}{a,b}%
%EndExpansion
)=c\cdot\partial_{n}%
%TCIMACRO{\QATOPD{\{}{\}}{n}{a,b}}%
%BeginExpansion
\genfrac{\{}{\}}{0pt}{}{n}{a,b}%
%EndExpansion
=%
%TCIMACRO{\QATOPD{\{}{\}}{c}{a,b}}%
%BeginExpansion
\genfrac{\{}{\}}{0pt}{}{c}{a,b}%
%EndExpansion
.%
%TCIMACRO{\TeXButton{End Proof}{\endproof}}%
%BeginExpansion
\endproof
%EndExpansion
\]

\textbf{(ii)} In order to get acquainted with the algebraic manipulations of
the multiform and extensor calculus we give the details in the derivation of
the \emph{Ricci} \emph{theorem for the} Levi-Civita covariant derivative
$D_{a}$, i.e.,
\begin{equation}
a\cdot\partial(%
%TCIMACRO{\TeXButton{itg}{\itg}}%
%BeginExpansion
\itg
%EndExpansion
(b)\cdot c)=%
%TCIMACRO{\TeXButton{itg}{\itg}}%
%BeginExpansion
\itg
%EndExpansion
(D_{a}b)\cdot c+%
%TCIMACRO{\TeXButton{itg}{\itg}}%
%BeginExpansion
\itg
%EndExpansion
(b)\cdot D_{a}c, \label{3.39}%
\end{equation}
where $a,b,c\in\sec%
%TCIMACRO{\dbigwedge \nolimits^{1}}%
%BeginExpansion
{\displaystyle\bigwedge\nolimits^{1}}
%EndExpansion
T^{\ast}U$ are smooth $1$-form fields.

Utilizing Eq.(\ref{3.38}), Eq.(\ref{3.2}) and Eq.(\ref{3.5c}), we have
\begin{align*}%
%TCIMACRO{\TeXButton{itg}{\itg}}%
%BeginExpansion
\itg
%EndExpansion
(D_{a}b)\cdot c+%
%TCIMACRO{\TeXButton{itg}{\itg}}%
%BeginExpansion
\itg
%EndExpansion
(b)\cdot D_{a}c  &  =(D_{a}b)\cdot%
%TCIMACRO{\TeXButton{itg}{\itg}}%
%BeginExpansion
\itg
%EndExpansion
(c)+D_{a}c\cdot%
%TCIMACRO{\TeXButton{itg}{\itg}}%
%BeginExpansion
\itg
%EndExpansion
(b)\\
&  =%
%TCIMACRO{\QATOPD{\{}{\}}{\TeXButton{itg}{\itg}(c)}{a,b}}%
%BeginExpansion
\genfrac{\{}{\}}{0pt}{}{\itg(c)}{a,b}%
%EndExpansion
+%
%TCIMACRO{\QATOPD{\{}{\}}{\TeXButton{itg}{\itg}(b)}{a,c}}%
%BeginExpansion
\genfrac{\{}{\}}{0pt}{}{\itg(b)}{a,c}%
%EndExpansion
\\
&  =[a,b,%
%TCIMACRO{\TeXButton{itg}{\itg}}%
%BeginExpansion
\itg
%EndExpansion
^{-1}(%
%TCIMACRO{\TeXButton{itg}{\itg}}%
%BeginExpansion
\itg
%EndExpansion
(c))]+[a,c,%
%TCIMACRO{\TeXButton{itg}{\itg}}%
%BeginExpansion
\itg
%EndExpansion
^{-1}(%
%TCIMACRO{\TeXButton{itg}{\itg}}%
%BeginExpansion
\itg
%EndExpansion
(b))]\\
&  =[a,b,c]+[a,c,b]=a\cdot\partial(%
%TCIMACRO{\TeXButton{itg}{\itg}}%
%BeginExpansion
\itg
%EndExpansion
(b)\cdot c).
\end{align*}
\medskip

\textbf{(iii) }This theorem is also valid for smooth multiform fields $X,Y$,
i.e.,
\begin{equation}
a\cdot\partial(X\cdot Y)=\underline{%
%TCIMACRO{\TeXButton{itg}{\itg}}%
%BeginExpansion
\itg
%EndExpansion
}(D_{a}X)\cdot Y+\underline{%
%TCIMACRO{\TeXButton{itg}{\itg}}%
%BeginExpansion
\itg
%EndExpansion
}(X)\cdot D_{a}Y, \label{3.40}%
\end{equation}

\textbf{(iv) }The Levi-Civita covariant derivative $D_{a}$ is symmetric,
i.e.,
\begin{equation}
D_{a}b-D_{b}a=[a,b], \label{3.41}%
\end{equation}
for all smooth $1$-form fields $a$ and $b.$

Take three smooth $1$-form fields $a,b$ and $c.$ Utilizing Eq.(\ref{3.38}),
Eq.(\ref{3.3}) and Eq.(\ref{3.5b}), we can write
\begin{align*}
(D_{a}b)\cdot c-(D_{b}a)\cdot c  &  =%
%TCIMACRO{\QATOPD{\{}{\}}{c}{a,b}}%
%BeginExpansion
\genfrac{\{}{\}}{0pt}{}{c}{a,b}%
%EndExpansion
-%
%TCIMACRO{\QATOPD{\{}{\}}{c}{b,a}}%
%BeginExpansion
\genfrac{\{}{\}}{0pt}{}{c}{b,a}%
%EndExpansion
=[a,b,%
%TCIMACRO{\TeXButton{itg}{\itg}}%
%BeginExpansion
\itg
%EndExpansion
^{-1}(c)]-[b,a,%
%TCIMACRO{\TeXButton{itg}{\itg}}%
%BeginExpansion
\itg
%EndExpansion
^{-1}(c)]\\
&  =%
%TCIMACRO{\TeXButton{itg}{\itg}}%
%BeginExpansion
\itg
%EndExpansion
(%
%TCIMACRO{\TeXButton{itg}{\itg}}%
%BeginExpansion
\itg
%EndExpansion
^{-1}(c))\cdot\lbrack a,b]=[a,b]\cdot c,
\end{align*}
which from the non degeneracy of the scalar product gives
\[
D_{a}b-D_{b}a=[a,b].%
%TCIMACRO{\TeXButton{End Proof}{\endproof}}%
%BeginExpansion
\endproof
%EndExpansion
\]

\textbf{(v) }The Levi-Civita $\ $covariant derivative $D_{a}$ is $%
%TCIMACRO{\TeXButton{itg}{\itg}}%
%BeginExpansion
\itg
%EndExpansion
$\emph{-compatible,} i.e.,
\begin{equation}
D_{a}^{-}%
%TCIMACRO{\TeXButton{itg}{\itg}}%
%BeginExpansion
\itg
%EndExpansion
=0. \label{3.42}%
\end{equation}

Take three smooth $1$-form fields $a,b$ and $c$. Utilizing Eq.(\ref{3.18}) and
the Ricci theorem\ (Eq.(\ref{3.39})), we have
\begin{align*}
(D_{a}^{-}%
%TCIMACRO{\TeXButton{itg}{\itg}}%
%BeginExpansion
\itg
%EndExpansion
)(b)\cdot c  &  =a\cdot\partial(%
%TCIMACRO{\TeXButton{itg}{\itg}}%
%BeginExpansion
\itg
%EndExpansion
(b)\cdot c)-%
%TCIMACRO{\TeXButton{itg}{\itg}}%
%BeginExpansion
\itg
%EndExpansion
(D_{a}b)\cdot c-%
%TCIMACRO{\TeXButton{itg}{\itg}}%
%BeginExpansion
\itg
%EndExpansion
(b)\cdot D_{a}c,\\
&  =0,
\end{align*}
which implies that $(D_{a}^{-}%
%TCIMACRO{\TeXButton{itg}{\itg}}%
%BeginExpansion
\itg
%EndExpansion
)(b)=0$, i.e., $D_{a}^{-}%
%TCIMACRO{\TeXButton{itg}{\itg}}%
%BeginExpansion
\itg
%EndExpansion
=0$.$%
%TCIMACRO{\TeXButton{End Proof}{\endproof}}%
%BeginExpansion
\endproof
%EndExpansion
$

We emphasize that \ the symmetry property and the $%
%TCIMACRO{\TeXButton{itg}{\itg}}%
%BeginExpansion
\itg
%EndExpansion
$-compatibility uniquely characterizes the Levi-Civita covariant derivative,
i.e., there exists an unique pair of covariant derivative operators
$\nabla_{a}$ and $\nabla_{a}^{-}$ such that $\nabla_{a}$ is symmetric (i.e.,
$\nabla_{a}b-\nabla_{b}a=[a,b]$) and $\nabla_{a}^{-}%
%TCIMACRO{\TeXButton{itg}{\itg}}%
%BeginExpansion
\itg
%EndExpansion
=0$ (i.e., $\nabla_{a}^{-}$ is $%
%TCIMACRO{\TeXButton{itg}{\itg}}%
%BeginExpansion
\itg
%EndExpansion
$-compatible). Those $\nabla_{a}$ and $\nabla_{a}^{-}$ are precisely $D_{a}$
and $D_{a}^{-}$.

\paragraph{Properties of $%
%TCIMACRO{\TeXButton{R}{\slR}}%
%BeginExpansion
\slR
%EndExpansion
_{2}(B)$ and $%
%TCIMACRO{\TeXButton{R}{\slR}}%
%BeginExpansion
\slR
%EndExpansion
_{1}(b)$}

On a Riemann or Lorentz \textit{MCGSS} $(U_{o},%
%TCIMACRO{\TeXButton{itg}{\itg}}%
%BeginExpansion
\itg
%EndExpansion
,\lambda)$ we use simplified notations, the Riemann $(2,2)$- extensor field is
denoted$\ B\mapsto%
%TCIMACRO{\TeXButton{R}{\slR}}%
%BeginExpansion
\slR
%EndExpansion
_{2}(B)$ and the Ricci $(1,1)$-extensor is denoted $b\mapsto%
%TCIMACRO{\TeXButton{R}{\slR}}%
%BeginExpansion
\slR
%EndExpansion
_{1}(b)$. Those objects now have three additional properties besides the ones
those objects have on a general \textit{MCGSS }$(U_{o},%
%TCIMACRO{\TeXButton{itg}{\itg}}%
%BeginExpansion
\itg
%EndExpansion
,\gamma)$. We have

\textbf{(i)} $%
%TCIMACRO{\TeXButton{R}{\slR}}%
%BeginExpansion
\slR
%EndExpansion
_{2}(B)$ \emph{is adjoint symmetric}, i.e.,
\begin{equation}%
%TCIMACRO{\TeXButton{R}{\slR}}%
%BeginExpansion
\slR
%EndExpansion
_{2}(B)=%
%TCIMACRO{\TeXButton{R}{\slR}}%
%BeginExpansion
\slR
%EndExpansion
_{2}^{\dagger}(B). \label{3.43}%
\end{equation}

\textbf{(ii)} $%
%TCIMACRO{\TeXButton{R}{\slR}}%
%BeginExpansion
\slR
%EndExpansion
_{1}(b)$ \emph{is adjoint symmetric}, i.e.,
\begin{equation}%
%TCIMACRO{\TeXButton{R}{\slR}}%
%BeginExpansion
\slR
%EndExpansion
_{1}(b)=%
%TCIMACRO{\TeXButton{R}{\slR}}%
%BeginExpansion
\slR
%EndExpansion
_{1}^{\dagger}(b). \label{3.44}%
\end{equation}

\textbf{(iii)}$.$The\emph{ matrix element }for the Ricci extensor field is
given by the notable formula
\begin{equation}%
%TCIMACRO{\TeXButton{R}{\slR}}%
%BeginExpansion
\slR
%EndExpansion
_{1}(b)\cdot c=\partial_{a}\cdot\rho(a,b,c), \label{3.45}%
\end{equation}
i.e., the \textit{divergent of the curvature operator with respect to the
first variable.}

\subsubsection{Levi-Civita Differential Operators}

On the Riemann or Lorentz \textit{GSS} $(U_{o},%
%TCIMACRO{\TeXButton{itg}{\itg}}%
%BeginExpansion
\itg
%EndExpansion
,\lambda)$ we can introduce three differential operators : the gradient $%
%TCIMACRO{\TeXButton{D}{\slD}}%
%BeginExpansion
\slD
%EndExpansion
\underset{%
%TCIMACRO{\TeXButton{sig}{\sitg}}%
%BeginExpansion
\sitg
%EndExpansion
}{}$, the divergent $%
%TCIMACRO{\TeXButton{D}{\slD}}%
%BeginExpansion
\slD
%EndExpansion
\underset{%
%TCIMACRO{\TeXButton{sig}{\sitg}}%
%BeginExpansion
\sitg
%EndExpansion
}{\lrcorner}$ and the \emph{rotational} $%
%TCIMACRO{\TeXButton{D}{\slD}}%
%BeginExpansion
\slD
%EndExpansion
\wedge$ acting on smooth multiform fields.

\begin{itemize}
\item The \emph{gradient of a smooth multiform field} $X$ $\in\sec%
%TCIMACRO{\dbigwedge }%
%BeginExpansion
{\displaystyle\bigwedge}
%EndExpansion
T^{\ast}U$ is defined by
\begin{equation}%
%TCIMACRO{\TeXButton{D}{\slD}}%
%BeginExpansion
\slD
%EndExpansion
\underset{%
%TCIMACRO{\TeXButton{sig}{\sitg}}%
%BeginExpansion
\sitg
%EndExpansion
}{}X=\partial_{a}\underset{%
%TCIMACRO{\TeXButton{sig}{\sitg}}%
%BeginExpansion
\sitg
%EndExpansion
}{}(D_{a}^{-}X), \label{3.46}%
\end{equation}
i.e., $%
%TCIMACRO{\TeXButton{D}{\slD}}%
%BeginExpansion
\slD
%EndExpansion
\underset{%
%TCIMACRO{\TeXButton{sig}{\sitg}}%
%BeginExpansion
\sitg
%EndExpansion
}{}X=\varepsilon^{\mu}\underset{%
%TCIMACRO{\TeXButton{sig}{\sitg}}%
%BeginExpansion
\sitg
%EndExpansion
}{}(D_{\varepsilon_{\mu}}^{-}X)=\varepsilon_{\mu}\underset{%
%TCIMACRO{\TeXButton{sig}{\sitg}}%
%BeginExpansion
\sitg
%EndExpansion
}{}(D_{\varepsilon^{\mu}}^{-}X),$ where $(\{\varepsilon^{\mu}\},\{\varepsilon
_{\mu}\})$ is a pair of arbitrary reciprocal basis defined on $U_{o}$.

\item The\textbf{ }$\emph{divergent}$\textit{ }\emph{of a smooth multiform
field} $X\in\sec%
%TCIMACRO{\dbigwedge }%
%BeginExpansion
{\displaystyle\bigwedge}
%EndExpansion
T^{\ast}U$ is defined by
\begin{equation}%
%TCIMACRO{\TeXButton{D}{\slD}}%
%BeginExpansion
\slD
%EndExpansion
\underset{%
%TCIMACRO{\TeXButton{sig}{\sitg}}%
%BeginExpansion
\sitg
%EndExpansion
}{\lrcorner}X=\partial_{a}\underset{%
%TCIMACRO{\TeXButton{sig}{\sitg}}%
%BeginExpansion
\sitg
%EndExpansion
}{\lrcorner}(D_{a}^{-}X). \label{3.47}%
\end{equation}

\item The $\emph{rotacional}$ \emph{of a smooth multiform field} $X\in\sec%
%TCIMACRO{\dbigwedge }%
%BeginExpansion
{\displaystyle\bigwedge}
%EndExpansion
T^{\ast}U$ is defined by
\begin{equation}%
%TCIMACRO{\TeXButton{D}{\slD}}%
%BeginExpansion
\slD
%EndExpansion
\wedge X=\partial_{a}\wedge(D_{a}^{-}X). \label{3.48}%
\end{equation}
\medskip
\end{itemize}

The main properties of those operators are:

\textbf{(i)} For any smooth multiform field $X\in\sec%
%TCIMACRO{\dbigwedge }%
%BeginExpansion
{\displaystyle\bigwedge}
%EndExpansion
T^{\ast}U,$
\begin{equation}%
%TCIMACRO{\TeXButton{D}{\slD}}%
%BeginExpansion
\slD
%EndExpansion
\underset{%
%TCIMACRO{\TeXButton{sig}{\sitg}}%
%BeginExpansion
\sitg
%EndExpansion
}{}X=%
%TCIMACRO{\TeXButton{D}{\slD}}%
%BeginExpansion
\slD
%EndExpansion
\underset{%
%TCIMACRO{\TeXButton{sig}{\sitg}}%
%BeginExpansion
\sitg
%EndExpansion
}{\lrcorner}X+%
%TCIMACRO{\TeXButton{D}{\slD}}%
%BeginExpansion
\slD
%EndExpansion
\wedge X, \label{3.49}%
\end{equation}
i.e., $%
%TCIMACRO{\TeXButton{D}{\slD}}%
%BeginExpansion
\slD
%EndExpansion
\underset{%
%TCIMACRO{\TeXButton{sig}{\sitg}}%
%BeginExpansion
\sitg
%EndExpansion
}{}=%
%TCIMACRO{\TeXButton{D}{\slD}}%
%BeginExpansion
\slD
%EndExpansion
\underset{%
%TCIMACRO{\TeXButton{sig}{\sitg}}%
%BeginExpansion
\sitg
%EndExpansion
}{\lrcorner}+%
%TCIMACRO{\TeXButton{D}{\slD}}%
%BeginExpansion
\slD
%EndExpansion
\wedge.$

\textbf{(ii)} For any smooth multiform field $X\in\sec%
%TCIMACRO{\dbigwedge }%
%BeginExpansion
{\displaystyle\bigwedge}
%EndExpansion
T^{\ast}U$
\begin{align}%
%TCIMACRO{\TeXButton{D}{\slD}}%
%BeginExpansion
\slD
%EndExpansion
\underset{%
%TCIMACRO{\TeXButton{sig}{\sitg}}%
%BeginExpansion
\sitg
%EndExpansion
}{\lrcorner}X  &  =\frac{1}{\sqrt{\left\vert \det[%
%TCIMACRO{\TeXButton{itg}{\itg}}%
%BeginExpansion
\itg
%EndExpansion
]\right\vert }}\underline{g}(\partial\lrcorner\sqrt{\left\vert \det[%
%TCIMACRO{\TeXButton{itg}{\itg}}%
%BeginExpansion
\itg
%EndExpansion
]\right\vert }\underline{g}^{-1}(X)),\label{3.50}\\%
%TCIMACRO{\TeXButton{D}{\slD}}%
%BeginExpansion
\slD
%EndExpansion
\wedge X  &  =\partial\wedge X. \label{3.51}%
\end{align}

\subsection{Deformation of \textit{MCGSS }Structures}

\subsubsection{Enter the Plastic Distortion Field $%
%TCIMACRO{\TeXButton{h}{\slh}}%
%BeginExpansion
\slh
%EndExpansion
$}

\paragraph{Minkowski Metric on $U_{o}$}

Let $M$ be a smooth manifold with $\dim M=$ $4.$ Consider the canonical spaces
$\mathbf{U}$ and $U$ defined by local coordinates $(\mathcal{U},\phi)_{o}$,
and the canonical dual bases $\{\mathbf{b}_{\mu}\}$ and $\{%
%TCIMACRO{\TeXButton{beta}{\mbox{\boldmath{$\beta$}}}}%
%BeginExpansion
\mbox{\boldmath{$\beta$}}%
%EndExpansion
^{\mu}\}$ for $\mathbf{U}$ and $U$, $%
%TCIMACRO{\TeXButton{beta}{\mbox{\boldmath{$\beta$}}}}%
%BeginExpansion
\mbox{\boldmath{$\beta$}}%
%EndExpansion
^{\mu}(\mathbf{b}_{\nu})=\delta_{\nu}^{\mu}$. Moreover, we denote the
reciprocal basis of the basis $\{%
%TCIMACRO{\TeXButton{beta}{\mbox{\boldmath{$\beta$}}}}%
%BeginExpansion
\mbox{\boldmath{$\beta$}}%
%EndExpansion
^{\mu}\}$ by $\{\beta_{\mu}\}$, with $%
%TCIMACRO{\TeXButton{beta}{\mbox{\boldmath{$\beta$}}}}%
%BeginExpansion
\mbox{\boldmath{$\beta$}}%
%EndExpansion
^{\mu}\cdot\beta_{\nu}=\delta_{\nu}^{\mu}$. Note that due to the
identifications that define the canonical space, we can also write that $%
%TCIMACRO{\TeXButton{beta}{\mbox{\boldmath{$\beta$}}}}%
%BeginExpansion
\mbox{\boldmath{$\beta$}}%
%EndExpansion
^{\mu},\beta_{\nu}\in\sec%
%TCIMACRO{\dbigwedge \nolimits^{1}}%
%BeginExpansion
{\displaystyle\bigwedge\nolimits^{1}}
%EndExpansion
T^{\ast}U$ for all $x\in U_{0}$.

Recalling Section 2.8.4 we introduce the smooth extensor field on $U_{0}$, say
$\eta\in\sec(1,1)$-$extU$ defined for any $a\in\sec%
%TCIMACRO{\dbigwedge \nolimits^{1}}%
%BeginExpansion
{\displaystyle\bigwedge\nolimits^{1}}
%EndExpansion
T^{\ast}U$ by $a\mapsto\eta(a)\in\sec%
%TCIMACRO{\dbigwedge \nolimits^{1}}%
%BeginExpansion
{\displaystyle\bigwedge\nolimits^{1}}
%EndExpansion
T^{\ast}U$, such that%
\begin{equation}
\eta(a)=%
%TCIMACRO{\TeXButton{beta}{\mbox{\boldmath{$\beta$}}}}%
%BeginExpansion
\mbox{\boldmath{$\beta$}}%
%EndExpansion
^{0}a%
%TCIMACRO{\TeXButton{beta}{\mbox{\boldmath{$\beta$}}}}%
%BeginExpansion
\mbox{\boldmath{$\beta$}}%
%EndExpansion
^{0},
\end{equation}
The field $\eta$ is a well defined metric extensor field on $U_{0}$ (i.e.,
$\eta$ is symmetric and non degenerated) and is called \textit{Minkowski
metric extensor}\footnote{We eventually call $\eta$ the Minkowski metric when
no confusion arises.
\par
{}}

The main properties of $\eta$ are:

\textbf{(i)} $\eta$ is a constant $(1,1)$-extensor field, i.e., $a\cdot
\partial\eta=0$.

\textbf{(ii)} The 1-forms of the canonical basis $\{%
%TCIMACRO{\TeXButton{beta}{\mbox{\boldmath{$\beta$}}}}%
%BeginExpansion
\mbox{\boldmath{$\beta$}}%
%EndExpansion
^{\mu}\}$ are $\eta$ orthonormal (i.e., $\eta(%
%TCIMACRO{\TeXButton{beta}{\mbox{\boldmath{$\beta$}}}}%
%BeginExpansion
\mbox{\boldmath{$\beta$}}%
%EndExpansion
^{\mu})\cdot%
%TCIMACRO{\TeXButton{beta}{\mbox{\boldmath{$\beta$}}}}%
%BeginExpansion
\mbox{\boldmath{$\beta$}}%
%EndExpansion
^{\nu}=\delta^{\mu\nu}$); $%
%TCIMACRO{\TeXButton{beta}{\mbox{\boldmath{$\beta$}}}}%
%BeginExpansion
\mbox{\boldmath{$\beta$}}%
%EndExpansion
^{0}$ is an eigenvector with eigenvalue $1,$ i.e., $\eta(%
%TCIMACRO{\TeXButton{beta}{\mbox{\boldmath{$\beta$}}}}%
%BeginExpansion
\mbox{\boldmath{$\beta$}}%
%EndExpansion
^{0})=%
%TCIMACRO{\TeXButton{beta}{\mbox{\boldmath{$\beta$}}}}%
%BeginExpansion
\mbox{\boldmath{$\beta$}}%
%EndExpansion
^{0},$ and $%
%TCIMACRO{\TeXButton{beta}{\mbox{\boldmath{$\beta$}}}}%
%BeginExpansion
\mbox{\boldmath{$\beta$}}%
%EndExpansion
^{k}$ $(k=1,2,3)$ are eigenvectors with eigenvalues $-1,$ i.e., $\eta(%
%TCIMACRO{\TeXButton{beta}{\mbox{\boldmath{$\beta$}}}}%
%BeginExpansion
\mbox{\boldmath{$\beta$}}%
%EndExpansion
^{k})=-%
%TCIMACRO{\TeXButton{beta}{\mbox{\boldmath{$\beta$}}}}%
%BeginExpansion
\mbox{\boldmath{$\beta$}}%
%EndExpansion
^{k}.$

Then, $\eta$ has signature\emph{ }$(1,3).$

\textbf{(iii)} $\mathrm{tr}[\eta]=-2$ and $\det[\eta]=-1.$

\textbf{(iv)} The extended of $\eta$ has the same generator of $\eta,$ i.e.,
for any $X\in\sec%
%TCIMACRO{\dbigwedge \nolimits^{1}}%
%BeginExpansion
{\displaystyle\bigwedge\nolimits^{1}}
%EndExpansion
T^{\ast}U$, $\underline{\eta}(X)=%
%TCIMACRO{\TeXButton{beta}{\mbox{\boldmath{$\beta$}}}}%
%BeginExpansion
\mbox{\boldmath{$\beta$}}%
%EndExpansion
^{0}X%
%TCIMACRO{\TeXButton{beta}{\mbox{\boldmath{$\beta$}}}}%
%BeginExpansion
\mbox{\boldmath{$\beta$}}%
%EndExpansion
^{0}$.

\textbf{(v)} $\eta$ is orthogonal canonical, i.e., $\eta=\eta^{\ast}$ (recall
that $\eta^{\ast}=(\eta^{\dagger})^{-1}=(\eta^{-1})^{\dagger}$). Then,
$\eta^{2}=i_{d}$.

\subsubsection{Lorentzian Metric}

Given a metric tensor $%
%TCIMACRO{\TeXButton{slg}{\slg}}%
%BeginExpansion
\slg
%EndExpansion
\in\sec T_{2}^{0}M$ of signature $(1,3)$, if the $(1,1)$-extensor field
associated to $%
%TCIMACRO{\TeXButton{slg}{\slg}}%
%BeginExpansion
\slg
%EndExpansion
$ is $\mathbf{g}$ and its representative on $U_{0}$ is $%
%TCIMACRO{\TeXButton{itg}{\itg}}%
%BeginExpansion
\itg
%EndExpansion
$, then $%
%TCIMACRO{\TeXButton{itg}{\itg}}%
%BeginExpansion
\itg
%EndExpansion
$ also has the same signature as the Minkowski metric extensor, i.e.,
\textrm{signature}$(%
%TCIMACRO{\TeXButton{itg}{\itg}}%
%BeginExpansion
\itg
%EndExpansion
)=$ \textrm{signature}$(\eta)=(1,3)$. In what follows $\mathbf{g}$ (and its
representative $%
%TCIMACRO{\TeXButton{itg}{\itg}}%
%BeginExpansion
\itg
%EndExpansion
$) will be called a \textit{Lorentzian metric extensor field}, or when non
confusion arises, simply Lorentzian metric.

Recalling Section 2.8.4 \ we know that there exists a (non unique)
$(1,1)$-extensor field $%
%TCIMACRO{\TeXButton{h}{\slh}}%
%BeginExpansion
\slh
%EndExpansion
$ given by:
\begin{equation}%
%TCIMACRO{\TeXButton{h}{\slh}}%
%BeginExpansion
\slh
%EndExpansion
(a)=\overset{3}{\underset{\mu=0}{\sum}}\sqrt{\left\vert \lambda^{\mu
}\right\vert }(a\cdot v^{\mu})%
%TCIMACRO{\TeXButton{beta}{\mbox{\boldmath{$\beta$}}}}%
%BeginExpansion
\mbox{\boldmath{$\beta$}}%
%EndExpansion
^{\mu},
\end{equation}
where the $\lambda^{\mu}\in\sec%
%TCIMACRO{\dbigwedge \nolimits^{0}}%
%BeginExpansion
{\displaystyle\bigwedge\nolimits^{0}}
%EndExpansion
T^{\ast}U$ are the eigenvalue fields of $%
%TCIMACRO{\TeXButton{itg}{\itg}}%
%BeginExpansion
\itg
%EndExpansion
$) and $v^{\mu}\in\sec%
%TCIMACRO{\dbigwedge \nolimits^{1}}%
%BeginExpansion
{\displaystyle\bigwedge\nolimits^{1}}
%EndExpansion
T^{\ast}U$ are the associated eigenvector\ fields of $%
%TCIMACRO{\TeXButton{itg}{\itg}}%
%BeginExpansion
\itg
%EndExpansion
$ such that%
\begin{equation}%
%TCIMACRO{\TeXButton{itg}{\itg}}%
%BeginExpansion
\itg
%EndExpansion
=%
%TCIMACRO{\TeXButton{h}{\slh}}%
%BeginExpansion
\slh
%EndExpansion
^{\dagger}\eta%
%TCIMACRO{\TeXButton{h}{\slh} }%
%BeginExpansion
\slh
%EndExpansion
\label{4.1bis}%
\end{equation}
The 1-form $\{v^{\mu}\}$ defines an\emph{ }\textit{euclidean orthonormal basis
}for $V$, i.e., $v^{\mu}\cdot v^{\nu}=\delta^{\mu\nu}$. As we know $%
%TCIMACRO{\TeXButton{h}{\slh}}%
%BeginExpansion
\slh
%EndExpansion
$ is not unique and is defined modulo a local Lorentz transformation, i.e., $%
%TCIMACRO{\TeXButton{h}{\slh}}%
%BeginExpansion
\slh
%EndExpansion
$ and $%
%TCIMACRO{\TeXButton{h}{\slh}}%
%BeginExpansion
\slh
%EndExpansion
^{\prime}=\Lambda%
%TCIMACRO{\TeXButton{h}{\slh}}%
%BeginExpansion
\slh
%EndExpansion
$ determine the same \ if $\Lambda\in\sec(1,1)$-$extU$ is a local Lorentz
transformation, i.e., for any $a,b\in\sec%
%TCIMACRO{\dbigwedge \nolimits^{1}}%
%BeginExpansion
{\displaystyle\bigwedge\nolimits^{1}}
%EndExpansion
T^{\ast}U$, $\Lambda(a)\underset{\eta}{\cdot}\Lambda(b)=a\underset{\eta
}{\cdot}b$. In what follows we call $%
%TCIMACRO{\TeXButton{h}{\slh}}%
%BeginExpansion
\slh
%EndExpansion
$ a \textit{plastic gauge distortion field} on $U_{o}$

\subsubsection{On Elastic and Plastic Deformations}

The wording gauge is of course, well justified, since $%
%TCIMACRO{\TeXButton{h}{\slh}}%
%BeginExpansion
\slh
%EndExpansion
$ can only be determined modulus a local Lorentz transformation. The wording
distortion is also well justified in view of Eq.(\ref{4.1bis}), $%
%TCIMACRO{\TeXButton{h}{\slh}}%
%BeginExpansion
\slh
%EndExpansion
$ distorts $\eta$ into $%
%TCIMACRO{\TeXButton{itg}{\itg}}%
%BeginExpansion
\itg
%EndExpansion
$. The wording plastic is justified as follows. In the theory of deformations
and defects on continuum media \cite{Zorawski} \ we introduce two different
kinds of deformations for any medium that lives in a manifold $M$ carrying a
metric field, say $\overset{\circ}{%
%TCIMACRO{\TeXButton{g}{\slg}}%
%BeginExpansion
\slg
%EndExpansion
}$. Those deformations are:\ 

\textbf{(i)} an \textit{elastic} one where the deformation of the medium is
described by diffeomorphism \texttt{h}$:M\rightarrow M$ which induces a
\textit{deformed} metric given by the pullback metric\footnote{Take notice
that here $\mathtt{h}^{\ast}$ denotes the pulback mapping.} $%
%TCIMACRO{\TeXButton{g}{\slg}}%
%BeginExpansion
\slg
%EndExpansion
=\mathtt{h}^{\ast}$ $\overset{\circ}{%
%TCIMACRO{\TeXButton{g}{\slg}}%
%BeginExpansion
\slg
%EndExpansion
}$ on $M$ and which is used in the definition of the \textit{Cauchy-Green
tensor} \cite{frankel}.

\textbf{(ii)}: a \textit{plastic }one where the deformation of the medium is
such that the new effective metric $%
%TCIMACRO{\TeXButton{g}{\slg}}%
%BeginExpansion
\slg
%EndExpansion
$ on $M$ defining distance measurements \textit{cannot }be described by the
pullback of any diffeomorphism as in the elastic case.

In case \textbf{(i)} we can show the pullback $D^{\prime}=\mathtt{h}^{\ast}D$
of the Levi-Civita connection $D$ of $\overset{\circ}{%
%TCIMACRO{\TeXButton{g}{\slg}}%
%BeginExpansion
\slg
%EndExpansion
}$, has the same torsion and Riemann curvature tensors then $D$, i.e., they
are null\footnote{This result, which can be easily proven is shown, e.g, in
Remark 250 in \cite{rodcap2007}. Also, we take the opportunity to call the
reader's attention that in \cite{notterod} it is presented a theory for the
gravitational field where the field $%
%TCIMACRO{\TeXButton{g}{\slg}}%
%BeginExpansion
\slg
%EndExpansion
$ has been identified as a result of an elastic deformation. Such
identification is of course, incorrect. The arXiv version of the paper
contains a corretion.}, whereas in case \textbf{(ii)} the medium can
conveniently be described by new effective connection $\nabla$ that is $%
%TCIMACRO{\TeXButton{g}{\slg}}%
%BeginExpansion
\slg
%EndExpansion
$ compatible ($\nabla%
%TCIMACRO{\TeXButton{g}{\slg}}%
%BeginExpansion
\slg
%EndExpansion
=0$) but that has in general non null torsion and Riemann curvature tensors.

Now, although there exists a distortion field $%
%TCIMACRO{\TeXButton{h}{\slh}}%
%BeginExpansion
\slh
%EndExpansion
$ corresponding to the pullback $\mathtt{h}^{\ast}$ of any diffeomorphism
\cite{fmr074}, there are distortion fields $%
%TCIMACRO{\TeXButton{h}{\slh}}%
%BeginExpansion
\slh
%EndExpansion
$ to which there corresponds \textit{no} diffeomorphism. Thus the wording
\textit{plastic} is justified, even more because we are going to see below
that $%
%TCIMACRO{\TeXButton{h}{\slh}}%
%BeginExpansion
\slh
%EndExpansion
$ also deforms in a precise \textit{sense} the Levi-Civita connection of $%
%TCIMACRO{\TeXButton{g}{\slg}}%
%BeginExpansion
\slg
%EndExpansion
$ and thus produces a connection on $M$ that is not compatible with
$\overset{\circ}{%
%TCIMACRO{\TeXButton{g}{\slg}}%
%BeginExpansion
\slg
%EndExpansion
}$. And in this sense we can also say that $%
%TCIMACRO{\TeXButton{h}{\slh}}%
%BeginExpansion
\slh
%EndExpansion
$ distorts the parallelism structure defined by the Levi-Civita connection of
$\overset{\circ}{%
%TCIMACRO{\TeXButton{g}{\slg}}%
%BeginExpansion
\slg
%EndExpansion
}$.

\subsubsection{Construction of a Lorentzian Metric Field on $U_{o}$}

\textbf{Proposition 4.1}\emph{\ \cite{fmr073} }Let\emph{ }$%
%TCIMACRO{\TeXButton{h}{\slh}}%
%BeginExpansion
\slh
%EndExpansion
$ be a $(1,1)$-extensor field on $U_{0},$ defined by
\begin{equation}%
%TCIMACRO{\TeXButton{h}{\slh}}%
%BeginExpansion
\slh
%EndExpansion
(a)=\overset{3}{\underset{\mu=0}{\sum}}\rho^{\mu}\Lambda(a)\cdot\beta^{\mu
}\beta^{\mu}, \label{4.3}%
\end{equation}
where $\rho^{\mu}\in\sec%
%TCIMACRO{\dbigwedge }%
%BeginExpansion
{\displaystyle\bigwedge}
%EndExpansion
T^{\ast}U$ are positive scalar fields on $U_{o}$ (i.e., $\rho^{\mu}(x)>0,$ for
$x\in U_{0}$) and $\Lambda$ is an orthogonal $\ (1,1)$-extensor field on
$U_{0}$ (i.e., $\Lambda=\Lambda^{\diamond}$; then the $(1,1)$-extensor field $%
%TCIMACRO{\TeXButton{itg}{\itg}}%
%BeginExpansion
\itg
%EndExpansion
$ on $U_{o}$ given by
\begin{equation}%
%TCIMACRO{\TeXButton{itg}{\itg}}%
%BeginExpansion
\itg
%EndExpansion
=%
%TCIMACRO{\TeXButton{h}{\slh}}%
%BeginExpansion
\slh
%EndExpansion
^{\dagger}\eta%
%TCIMACRO{\TeXButton{h}{\slh}}%
%BeginExpansion
\slh
%EndExpansion
, \label{4.4}%
\end{equation}
is a Lorentzian metric extensor field (i.e., $\mathrm{signa}$\textrm{ture}$(%
%TCIMACRO{\TeXButton{itg}{\itg}}%
%BeginExpansion
\itg
%EndExpansion
)=(1,3)$).

The non null scalar fields $(\rho^{0})^{2}$ and $-(\rho^{1})^{2},-(\rho
^{2})^{2},-(\rho^{3})^{2}$ are the eigenvalue fields of $%
%TCIMACRO{\TeXButton{itg}{\itg}}%
%BeginExpansion
\itg
%EndExpansion
,$ and the non null $1$-form fields $\Lambda^{\dagger}(%
%TCIMACRO{\TeXButton{beta}{\mbox{\boldmath{$\beta$}}}}%
%BeginExpansion
\mbox{\boldmath{$\beta$}}%
%EndExpansion
^{0})$ and $\Lambda^{\dagger}(%
%TCIMACRO{\TeXButton{beta}{\mbox{\boldmath{$\beta$}}}}%
%BeginExpansion
\mbox{\boldmath{$\beta$}}%
%EndExpansion
^{1}),%
%TCIMACRO{\TeXButton{beta}{\mbox{\boldmath{$\beta$}}}}%
%BeginExpansion
\mbox{\boldmath{$\beta$}}%
%EndExpansion
^{\dagger}(\beta^{2}),%
%TCIMACRO{\TeXButton{beta}{\mbox{\boldmath{$\beta$}}}}%
%BeginExpansion
\mbox{\boldmath{$\beta$}}%
%EndExpansion
^{\dagger}(\beta^{3})$ are the associated eigenvector fields of $%
%TCIMACRO{\TeXButton{itg}{\itg}}%
%BeginExpansion
\itg
%EndExpansion
$.

\subsection{Deformation of a Minkowski-Cartan $(U_{o},\eta,\varkappa)$
\textit{MCGSS} into a Lorentz-Cartan \textit{MCGSS }$(U_{o},%
%TCIMACRO{\TeXButton{itg}{\itg}}%
%BeginExpansion
\itg
%EndExpansion
,\gamma)$}

Let $M$ be $4$-dimensional (with enough structure to be part of a Lorentzian
spacetime structure as defined in Section 1). Let $(\mathcal{U},\phi)_{0}$ a
chart for $\mathcal{U\subset}M$\ containing $o\in\mathcal{U}$ and let
$U_{o}\subset U$ be defined as previously. A \textit{MCGSS} $(U_{o}%
,\eta,\varkappa)$ where $\eta$ is the Minkowski metric extensor on $U_{o}$ and
$\varkappa$ is a connection extensor field on $U_{o}$ compatible with $\eta$
is said a\emph{ Minkowski-Cartan \textit{MCGSS.}}

The metric compatibility may be written $\overset{\varkappa\eta}{\mathcal{D}%
}_{a}^{-}\eta=0,$ where $\overset{\varkappa\eta}{\mathcal{D}}_{a}^{-}$ is one
of the covariant derivative operators associated to $\varkappa$. Besides going
on a crucial observation is needed.\medskip

\textbf{Remark} \textbf{4.3 }Note that we did not impose that the manifold $M$
carries a Minkowski metric field $%
%TCIMACRO{\TeXButton{eta}{\mbox{\boldmath{$\eta$}}}}%
%BeginExpansion
\mbox{\boldmath{$\eta$}}%
%EndExpansion
$ as \textit{given} by Eq.(\ref{0.66a}) for if that was the case we would have
$M\simeq\mathbb{R}^{4}$, which is not being considered at this moment.\medskip

Now, let us recall that the Theorem 4.1 (Eq.(\ref{3.28})) implies the
existence of a\emph{ }$\varkappa\eta$\emph{-gauge rotation field}, say
$\overset{\varkappa\eta}{\Omega}$ such
\begin{equation}
\varkappa_{a}(b)=\overset{\varkappa\eta}{\Omega}(a)\underset{\eta}{\times
}b=\overset{\varkappa\eta}{\Omega}(a)\times\eta(b). \label{4.5}%
\end{equation}

Also, according to Eq.(\ref{3.7a}), the covariant derivative
$\overset{\varkappa\eta}{\mathcal{D}}_{a}$ when acting on a smooth $1$-form
field is given by
\begin{equation}
\overset{\varkappa\eta}{\mathcal{D}}_{a}b=a\cdot\partial b+\overset{\varkappa
\eta}{\Omega}(a)\underset{\eta}{\times}b. \label{4.6}%
\end{equation}
\medskip

\textbf{Remark 4.4 }Keep in mind that in a general Minkowski-Cartan
$\ $\textit{GSS }$(U_{o},\eta,\varkappa)$, $\overset{\varkappa\eta
}{\mathcal{D}}_{a}$ is not the representative of the Levi-Civita connection of
$%
%TCIMACRO{\TeXButton{eta}{\mbox{\boldmath{$\eta$}}}}%
%BeginExpansion
\mbox{\boldmath{$\eta$}}%
%EndExpansion
$ and indeed, in general possess non null torsion and Riemann curvature tensors.

We recall that a \textit{MCGSS} $(M,%
%TCIMACRO{\TeXButton{g}{\slg}}%
%BeginExpansion
\slg
%EndExpansion
,\mathbf{\nabla})$ where $M$ is a $4$-dimensional \ manifold as in the
previous subsection, $%
%TCIMACRO{\TeXButton{g}{\slg}}%
%BeginExpansion
\slg
%EndExpansion
$ is a Lorentzian metric $\mathbf{\nabla}$ is an arbitrary metric compatible
connection on $M$, i.e., $\mathbf{\nabla}%
%TCIMACRO{\TeXButton{g}{\slg}}%
%BeginExpansion
\slg
%EndExpansion
=0$, is said to be a \emph{ Lorentz-Cartan }\textit{MCGSS}. Let $U_{o},$ $%
%TCIMACRO{\TeXButton{itg}{\itg}}%
%BeginExpansion
\itg
%EndExpansion
$, and $\gamma$ as defined previously. We recall that we agreed in calling
also $(U_{o},%
%TCIMACRO{\TeXButton{itg}{\itg}}%
%BeginExpansion
\itg
%EndExpansion
,\gamma)$ a Lorentz-Cartan \textit{MCGSS}.

The compatibility between $%
%TCIMACRO{\TeXButton{itg}{\itg}}%
%BeginExpansion
\itg
%EndExpansion
$ and $\gamma$ may be written as $\overset{\gamma%
%TCIMACRO{\TeXButton{sig}{\sitg}}%
%BeginExpansion
\sitg
%EndExpansion
}{\nabla_{a}^{-}}%
%TCIMACRO{\TeXButton{itg}{\itg}}%
%BeginExpansion
\itg
%EndExpansion
=0$, where $\overset{\gamma%
%TCIMACRO{\TeXButton{sig}{\sitg}}%
%BeginExpansion
\sitg
%EndExpansion
}{\nabla_{a}^{-}}$ is one of the covariant derivative operators associated to
$\gamma$.

The Theorem 4.1 implies in this case the existence of a $\gamma%
%TCIMACRO{\TeXButton{itg}{\itg}}%
%BeginExpansion
\itg
%EndExpansion
$-gauge rotation field $\overset{\gamma%
%TCIMACRO{\TeXButton{sig}{\sitg}}%
%BeginExpansion
\sitg
%EndExpansion
}{\omega}(a)$ such that
\begin{equation}
\gamma_{a}(b)=\frac{1}{2}%
%TCIMACRO{\TeXButton{itg}{\itg}}%
%BeginExpansion
\itg
%EndExpansion
^{-1}(a\cdot\partial%
%TCIMACRO{\TeXButton{itg}{\itg}}%
%BeginExpansion
\itg
%EndExpansion
)(b)+\overset{\gamma%
%TCIMACRO{\TeXButton{sig}{\sitg}}%
%BeginExpansion
\sitg
%EndExpansion
}{\omega}(a)\underset{%
%TCIMACRO{\TeXButton{sig}{\sitg}}%
%BeginExpansion
\sitg
%EndExpansion
}{\times}b=\frac{1}{2}%
%TCIMACRO{\TeXButton{itg}{\itg}}%
%BeginExpansion
\itg
%EndExpansion
^{-1}(a\cdot\partial%
%TCIMACRO{\TeXButton{itg}{\itg}}%
%BeginExpansion
\itg
%EndExpansion
)(b)+\overset{\gamma%
%TCIMACRO{\TeXButton{sig}{\sitg}}%
%BeginExpansion
\sitg
%EndExpansion
}{\omega}(a)\times%
%TCIMACRO{\TeXButton{itg}{\itg}}%
%BeginExpansion
\itg
%EndExpansion
(b). \label{4.7}%
\end{equation}

According to Eq.(\ref{3.7a}), the covariant derivative $\overset{\gamma%
%TCIMACRO{\TeXButton{sig}{\sitg}}%
%BeginExpansion
\sitg
%EndExpansion
}{\nabla_{a}},$ when acting on a smooth 1-form field is given by
\begin{equation}
\overset{\gamma%
%TCIMACRO{\TeXButton{sig}{\sitg}}%
%BeginExpansion
\sitg
%EndExpansion
}{\nabla_{a}}b=a\cdot\partial b+\frac{1}{2}%
%TCIMACRO{\TeXButton{itg}{\itg}}%
%BeginExpansion
\itg
%EndExpansion
^{-1}(a\cdot\partial%
%TCIMACRO{\TeXButton{itg}{\itg}}%
%BeginExpansion
\itg
%EndExpansion
)(b)+\overset{\gamma%
%TCIMACRO{\TeXButton{sig}{\sitg}}%
%BeginExpansion
\sitg
%EndExpansion
}{\omega}(a)\underset{%
%TCIMACRO{\TeXButton{sig}{\sitg}}%
%BeginExpansion
\sitg
%EndExpansion
}{\times}b. \label{4.8}%
\end{equation}

\subsubsection{$%
%TCIMACRO{\TeXButton{h}{\slh}}%
%BeginExpansion
\slh
%EndExpansion
$-Distortions of Covariant Derivatives}

\textbf{Theorem 4.2}. The operators $\overset{\gamma%
%TCIMACRO{\TeXButton{sig}{\sitg}}%
%BeginExpansion
\sitg
%EndExpansion
}{\nabla_{a}}$\ and $\overset{\gamma%
%TCIMACRO{\TeXButton{sig}{\sitg}}%
%BeginExpansion
\sitg
%EndExpansion
}{\nabla_{a}^{-}}$ of the Lorentz-Cartan \textit{GSS }are related with the
operators $\overset{\varkappa\eta}{\mathcal{D}}_{a}$\ and $\overset{\varkappa
\eta}{\mathcal{D}}_{a}^{-}$ of the Minkowski-Cartan \textit{GSS} by
\begin{align}
\overset{\gamma%
%TCIMACRO{\TeXButton{sig}{\sitg}}%
%BeginExpansion
\sitg
%EndExpansion
}{\nabla_{a}}b  &  =%
%TCIMACRO{\TeXButton{h}{\slh}}%
%BeginExpansion
\slh
%EndExpansion
^{-1}(\overset{\varkappa\eta}{\mathcal{D}}_{a}%
%TCIMACRO{\TeXButton{h}{\slh}}%
%BeginExpansion
\slh
%EndExpansion
(b)),\label{4.9a}\\
\overset{\gamma%
%TCIMACRO{\TeXButton{sig}{\sitg}}%
%BeginExpansion
\sitg
%EndExpansion
}{\nabla_{a}^{-}}b  &  =%
%TCIMACRO{\TeXButton{h}{\slh}}%
%BeginExpansion
\slh
%EndExpansion
^{\dagger}(\overset{\varkappa\eta}{\mathcal{D}}_{a}^{-}%
%TCIMACRO{\TeXButton{h}{\slh}}%
%BeginExpansion
\slh
%EndExpansion
^{\clubsuit}(b)), \label{4.9b}%
\end{align}
where $%
%TCIMACRO{\TeXButton{h}{\slh}}%
%BeginExpansion
\slh
%EndExpansion
$ is the\textit{ plastic distortion field} introduce above such that $%
%TCIMACRO{\TeXButton{itg}{\itg}}%
%BeginExpansion
\itg
%EndExpansion
=%
%TCIMACRO{\TeXButton{h}{\slh}}%
%BeginExpansion
\slh
%EndExpansion
^{\dagger}\eta%
%TCIMACRO{\TeXButton{h}{\slh}}%
%BeginExpansion
\slh
%EndExpansion
.$

In order to prove the Theorem 4.3 we shall need the results of the following
two lemmas.\medskip

\textbf{Lemma 4.1} Let $(\overset{1}{\mathcal{\nabla}}_{a},$
$\overset{1}{\nabla_{a}^{-}})$ be any pair of associated covariant derivative
operators i.e., $\overset{1}{\mathcal{\nabla}}_{a}$ and $\overset{1}{\nabla
_{a}^{-}}$ satisfy Eq.(\ref{3.12}),%
\begin{equation}
a\cdot\partial(b\cdot c)=(\overset{1}{\nabla}_{a}b)\cdot c+b\cdot
\overset{1}{\nabla_{a}^{-}}c.
\end{equation}
Then the pair of covariant derivative operators $(\overset{2}{\nabla},$
$\overset{2}{\nabla_{a}^{-}})$, defined by
\begin{align}
\overset{2}{\nabla}_{a}b  &  =%
%TCIMACRO{\TeXButton{h}{\slh}}%
%BeginExpansion
\slh
%EndExpansion
^{-1}(\overset{1}{\nabla}_{a}%
%TCIMACRO{\TeXButton{h}{\slh}}%
%BeginExpansion
\slh
%EndExpansion
(b))\label{4.9i}\\
\overset{2}{\nabla_{a}^{-}}b  &  =%
%TCIMACRO{\TeXButton{h}{\slh}}%
%BeginExpansion
\slh
%EndExpansion
^{\dagger}(\overset{1}{\nabla_{a}^{-}}%
%TCIMACRO{\TeXButton{h}{\slh}}%
%BeginExpansion
\slh
%EndExpansion
^{\clubsuit}(b)), \label{4.9ii}%
\end{align}
where $%
%TCIMACRO{\TeXButton{h}{\slh}}%
%BeginExpansion
\slh
%EndExpansion
$ is a smooth invertible $(1,1)$-extensor field is also a pair of
associated\ covariant derivative operators, i.e., $\overset{2}{\nabla}_{a}$
and $\overset{2}{\nabla_{a}^{-}}$ satisfy Eq.(\ref{3.12}),%
\begin{equation}
a\cdot\partial(b\cdot c)=(\overset{2}{\nabla}_{a}b)\cdot c+b\cdot
\overset{2}{\nabla_{a}^{-}}c).
\end{equation}

Given that $\overset{1}{\nabla}_{a}$ and $\overset{1}{\nabla_{a}^{-}}$ are
well defined covariant derivative operators they satisfy, Eqs.(\ref{3.8}),
(\ref{3.88}) and (\ref{3.888}). Then a simple algebraic manipulation shows
that $\overset{2}{\nabla}_{a}$ and $\overset{2}{\nabla_{a}^{-}}$ also satisfy
those same equations. Thus, $\overset{2}{\nabla}_{a}$ and $\overset{2}{\nabla
_{a}^{-}}$ are also well defined covariant derivative operators.

On the other side, using Eq.(\ref{3.12}) we have
\[
\overset{1}{\nabla}_{a}%
%TCIMACRO{\TeXButton{h}{\slh}}%
%BeginExpansion
\slh
%EndExpansion
(b)\cdot%
%TCIMACRO{\TeXButton{h}{\slh}}%
%BeginExpansion
\slh
%EndExpansion
^{\clubsuit}(c)+%
%TCIMACRO{\TeXButton{h}{\slh}}%
%BeginExpansion
\slh
%EndExpansion
(b)\cdot\overset{1}{\nabla_{a}^{-}}%
%TCIMACRO{\TeXButton{h}{\slh}}%
%BeginExpansion
\slh
%EndExpansion
^{\clubsuit}(c)=a\cdot\partial(%
%TCIMACRO{\TeXButton{h}{\slh}}%
%BeginExpansion
\slh
%EndExpansion
(b)\cdot%
%TCIMACRO{\TeXButton{h}{\slh}}%
%BeginExpansion
\slh
%EndExpansion
^{\clubsuit}(c)),
\]
and after some algebraic manipulations,using Eqs.(\ref{4.9i}) and
(\ref{4.9ii}) we get
\begin{align*}%
%TCIMACRO{\TeXButton{h}{\slh}}%
%BeginExpansion
\slh
%EndExpansion
^{-1}(\overset{1}{\nabla}_{a}%
%TCIMACRO{\TeXButton{h}{\slh}}%
%BeginExpansion
\slh
%EndExpansion
(b))\cdot c+b\cdot%
%TCIMACRO{\TeXButton{h}{\slh}}%
%BeginExpansion
\slh
%EndExpansion
^{\dagger}(\overset{1}{\nabla_{a}^{-}}%
%TCIMACRO{\TeXButton{h}{\slh}}%
%BeginExpansion
\slh
%EndExpansion
^{\clubsuit}(c))  &  =a\cdot\partial(%
%TCIMACRO{\TeXButton{h}{\slh}}%
%BeginExpansion
\slh
%EndExpansion
^{-1}%
%TCIMACRO{\TeXButton{h}{\slh}}%
%BeginExpansion
\slh
%EndExpansion
(b)\cdot c)\\
(\overset{2}{\nabla}_{a}b)\cdot c+b\cdot\overset{2}{\nabla_{a}^{-}}c  &
=a\cdot\partial(b\cdot c).
\end{align*}

We see then that $\overset{2}{\nabla}_{a}$ and $\overset{2}{\nabla_{a}^{-}}$
also satisfy Eq.(\ref{3.12}), i.e., $(\overset{2}{\nabla}_{a},$
$\overset{2}{\nabla_{a}^{-}}$ $)$ is a pair of associated covariant derivative
operators.%
%TCIMACRO{\TeXButton{End Proof}{\endproof}}%
%BeginExpansion
\endproof
%EndExpansion
\medskip

\textbf{Lemma 4.2 } Let $%
%TCIMACRO{\TeXButton{h}{\slh}}%
%BeginExpansion
\slh
%EndExpansion
$ be a plastic distortion field such that $%
%TCIMACRO{\TeXButton{itg}{\itg}}%
%BeginExpansion
\itg
%EndExpansion
=%
%TCIMACRO{\TeXButton{h}{\slh}}%
%BeginExpansion
\slh
%EndExpansion
^{\dagger}\eta%
%TCIMACRO{\TeXButton{h}{\slh}}%
%BeginExpansion
\slh
%EndExpansion
,$ and let $\overset{1}{\nabla_{a}^{-}}$ and $\overset{2}{\nabla_{a}^{-}}$ be
a pair of \ associated covariant derivative operators acting on the module of
the smooth $(p,q)$-extensor fields. Then
\begin{equation}
\overset{2}{\nabla_{a}^{-}}%
%TCIMACRO{\TeXButton{itg}{\itg}}%
%BeginExpansion
\itg
%EndExpansion
=%
%TCIMACRO{\TeXButton{h}{\slh}}%
%BeginExpansion
\slh
%EndExpansion
^{\dagger}(\overset{1}{\nabla_{a}^{-}}\eta)%
%TCIMACRO{\TeXButton{h}{\slh}}%
%BeginExpansion
\slh
%EndExpansion
. \label{4.9iii}%
\end{equation}

Indeed, according to the Eq.(\ref{3.13b}),utilizing Eqs.(\ref{4.9i}) and
(\ref{4.9i}) and also taking into account the Theorem 4.1, we have
\begin{align*}
(\overset{2}{\nabla_{a}^{-}}%
%TCIMACRO{\TeXButton{itg}{\itg}}%
%BeginExpansion
\itg
%EndExpansion
)(b)  &  =\overset{2}{\nabla_{a}^{-}}%
%TCIMACRO{\TeXButton{itg}{\itg}}%
%BeginExpansion
\itg
%EndExpansion
(b)-%
%TCIMACRO{\TeXButton{itg}{\itg}}%
%BeginExpansion
\itg
%EndExpansion
(\overset{2}{\nabla}_{a}b)=%
%TCIMACRO{\TeXButton{h}{\slh}}%
%BeginExpansion
\slh
%EndExpansion
^{\dagger}\overset{1}{\nabla_{a}^{-}}%
%TCIMACRO{\TeXButton{h}{\slh}}%
%BeginExpansion
\slh
%EndExpansion
^{\clubsuit}%
%TCIMACRO{\TeXButton{h}{\slh}}%
%BeginExpansion
\slh
%EndExpansion
^{\dagger}\eta%
%TCIMACRO{\TeXButton{h}{\slh}}%
%BeginExpansion
\slh
%EndExpansion
(b)-%
%TCIMACRO{\TeXButton{h}{\slh}}%
%BeginExpansion
\slh
%EndExpansion
^{\dagger}\eta%
%TCIMACRO{\TeXButton{h}{\slh}}%
%BeginExpansion
\slh
%EndExpansion%
%TCIMACRO{\TeXButton{h}{\slh}}%
%BeginExpansion
\slh
%EndExpansion
^{-1}\overset{1}{\nabla}_{a}%
%TCIMACRO{\TeXButton{h}{\slh}}%
%BeginExpansion
\slh
%EndExpansion
(b)\\
&  =%
%TCIMACRO{\TeXButton{h}{\slh}}%
%BeginExpansion
\slh
%EndExpansion
^{\dagger}(\overset{1}{\nabla_{a}^{-}}\eta%
%TCIMACRO{\TeXButton{h}{\slh}}%
%BeginExpansion
\slh
%EndExpansion
(b))-\eta\overset{1}{\nabla}_{a}%
%TCIMACRO{\TeXButton{h}{\slh}}%
%BeginExpansion
\slh
%EndExpansion
(b))=%
%TCIMACRO{\TeXButton{h}{\slh}}%
%BeginExpansion
\slh
%EndExpansion
^{\dagger}(\overset{1}{\nabla_{a}^{-}}\eta)%
%TCIMACRO{\TeXButton{h}{\slh}}%
%BeginExpansion
\slh
%EndExpansion
(b),
\end{align*}
i.e., $\overset{2}{\nabla_{a}^{-}}%
%TCIMACRO{\TeXButton{itg}{\itg}}%
%BeginExpansion
\itg
%EndExpansion
=%
%TCIMACRO{\TeXButton{h}{\slh}}%
%BeginExpansion
\slh
%EndExpansion
^{\dagger}(\overset{1}{\nabla_{a}^{-}}\eta)%
%TCIMACRO{\TeXButton{h}{\slh}}%
%BeginExpansion
\slh
%EndExpansion
.$%
%TCIMACRO{\TeXButton{End Proof}{\endproof}}%
%BeginExpansion
\endproof
%EndExpansion

Eq.(\ref{4.9iii}) shows that if $\overset{1}{\nabla_{a}^{-}}$ is $\eta
$-compatible (i.e., $\overset{1}{\nabla_{a}^{-}}\eta=0$) iff
$\overset{2}{\nabla_{a}^{-}}$ is $%
%TCIMACRO{\TeXButton{itg}{\itg}}%
%BeginExpansion
\itg
%EndExpansion
$-compatible (i.e., $\overset{2}{\nabla_{a}^{-}}%
%TCIMACRO{\TeXButton{itg}{\itg}}%
%BeginExpansion
\itg
%EndExpansion
=0$).

Or in other worlds, if $\overset{1}{\nabla}_{a}$ and $\overset{1}{\nabla
_{a}^{-}}$ are covariant derivative operators on the Minkowski-Cartan
\textit{MCGSS} iff $\overset{2}{\nabla}_{a}$ and $\overset{2}{\nabla_{a}^{-}}$
are covariant derivative operators on the Lorentz-Cartan \textit{MCGSS}%
.\medskip

\textbf{Proof of Theorem 4.3 } The theorem follows at once from Lemmas 4.1.
and 4.2 using the identifications:
\[
\overset{\varkappa\eta}{\mathcal{D}}_{a}\equiv\overset{1}{\nabla}%
_{a},\emph{\ }\overset{\varkappa\eta}{\mathcal{D}}_{a}^{-}\equiv
\overset{1}{\nabla_{a}^{-}}and\overset{\gamma%
%TCIMACRO{\TeXButton{sig}{\sitg}}%
%BeginExpansion
\sitg
%EndExpansion
}{\nabla_{a}}\ \equiv\overset{2}{\nabla}_{a},\overset{\gamma%
%TCIMACRO{\TeXButton{sig}{\sitg}}%
%BeginExpansion
\sitg
%EndExpansion
}{\nabla_{a}^{-}}\equiv\overset{2}{\nabla_{a}^{-}}.%
%TCIMACRO{\TeXButton{End Proof}{\endproof}}%
%BeginExpansion
\endproof
%EndExpansion
\]

\subsubsection{Existence of a $2$-Extensor Coupling Field Between the
Minkowski-Cartan \textit{MCGSS} and Lorentz-Cartan \textit{MCGSS}}

\textbf{Proposition 4.3 \ }There exists a smooth $2$-extensor field on $U_{0}%
$, $f\in\sec(%
%TCIMACRO{\dbigwedge \nolimits^{1}}%
%BeginExpansion
{\displaystyle\bigwedge\nolimits^{1}}
%EndExpansion
T^{\ast}U\times\sec%
%TCIMACRO{\dbigwedge \nolimits^{1}}%
%BeginExpansion
{\displaystyle\bigwedge\nolimits^{1}}
%EndExpansion
T^{\ast}U)\rightarrow\sec%
%TCIMACRO{\dbigwedge \nolimits^{1}}%
%BeginExpansion
{\displaystyle\bigwedge\nolimits^{1}}
%EndExpansion
T^{\ast}U$, $(a,b)\mapsto f_{a}(b),$ which is $\eta$-adjoint antisymmetric
\footnote{The $%
%TCIMACRO{\TeXButton{itg}{\itg}}%
%BeginExpansion
\itg
%EndExpansion
$-adjoint (or metric adjoint) of a $(1,1)$-extensor $t$ is the $(1,1)$%
-extensor $\ t^{\dagger(%
%TCIMACRO{\TeXButton{sig}{\sitg}}%
%BeginExpansion
\sitg
%EndExpansion
)}\equiv%
%TCIMACRO{\TeXButton{itg}{\itg}}%
%BeginExpansion
\itg
%EndExpansion
\circ t^{\dagger}\circ%
%TCIMACRO{\TeXButton{itg}{\itg}}%
%BeginExpansion
\itg
%EndExpansion
^{-1}$.} (i.e., $f_{a}=-f_{a}^{\dagger(\eta)}$) such that
\begin{equation}
\underline{%
%TCIMACRO{\TeXButton{h}{\slh}}%
%BeginExpansion
\slh
%EndExpansion
}\overset{\gamma%
%TCIMACRO{\TeXButton{sig}{\sitg}}%
%BeginExpansion
\sitg
%EndExpansion
}{\omega}(a)=\overset{\varkappa\eta}{\Omega}(a)+\frac{1}{2}\overset{\eta
}{\mathrm{bif}}[f_{a}]. \label{4.10}%
\end{equation}
Such a $2$-extensor field is given by the formula
\begin{equation}
f_{a}(b)=(a\cdot\partial%
%TCIMACRO{\TeXButton{h}{\slh}}%
%BeginExpansion
\slh
%EndExpansion
)%
%TCIMACRO{\TeXButton{h}{\slh}}%
%BeginExpansion
\slh
%EndExpansion
^{-1}(b)-\frac{1}{2}\eta%
%TCIMACRO{\TeXButton{h}{\slh}}%
%BeginExpansion
\slh
%EndExpansion
^{\clubsuit}(a\cdot\partial%
%TCIMACRO{\TeXButton{itg}{\itg}}%
%BeginExpansion
\itg
%EndExpansion
)%
%TCIMACRO{\TeXButton{h}{\slh}}%
%BeginExpansion
\slh
%EndExpansion
^{-1}(b). \label{4.0a}%
\end{equation}
\medskip

The proof is a simple calculation.

\subsubsection{The Gauge Riemann and Ricci Fields}

Let $(a,b,c)\mapsto\overset{\varkappa\eta}{\rho}(a,b,c)$ and $(a,b,c)\mapsto
\overset{\gamma%
%TCIMACRO{\TeXButton{sig}{\sitg}}%
%BeginExpansion
\sitg
%EndExpansion
}{\rho}(a,b,c)$ be the curvature operators of \ the \textit{MCGSS}'s
$(U_{o},\eta,\varkappa)$ and $(U_{o},%
%TCIMACRO{\TeXButton{itg}{\itg}}%
%BeginExpansion
\itg
%EndExpansion
,\gamma)$ as previously defined and $(a,b,c,w)\mapsto\overset{\varkappa
\eta}{\mathbf{R}}_{3}(a,b,c,w)$ and $(a,b,c,w)\mapsto\overset{\gamma%
%TCIMACRO{\TeXButton{sig}{\sitg}}%
%BeginExpansion
\sitg
%EndExpansion
}{\mathbf{R}}_{3}(a,b,c,w)$ the corresponding Riemann $4$-extensor fields.

Let $B\mapsto\overset{\varkappa\eta}{\mathbf{R}_{2}}(B)$ and $B\mapsto
\overset{\gamma%
%TCIMACRO{\TeXButton{sig}{\sitg}}%
%BeginExpansion
\sitg
%EndExpansion
}{\mathbf{R}_{2}}(B)$ be the Riemann $2$-extensor fields of ($M,\eta
,\varkappa)$ and $(M,%
%TCIMACRO{\TeXButton{itg}{\itg}}%
%BeginExpansion
\itg
%EndExpansion
,\gamma).$

We related $\overset{\varkappa\eta}{\rho}(a,b,c)$ with $\overset{\gamma%
%TCIMACRO{\TeXButton{sig}{\sitg}}%
%BeginExpansion
\sitg
%EndExpansion
}{\rho}(a,b,c)$, utilizing Eq.(\ref{4.9a}) Indeed,
\begin{align}
\overset{\gamma%
%TCIMACRO{\TeXButton{sig}{\sitg}}%
%BeginExpansion
\sitg
%EndExpansion
}{\rho}(a,b,c)  &  =[\nabla_{a},\nabla_{b}]c-\nabla_{\lbrack a,b]}c\nonumber\\
&  =%
%TCIMACRO{\TeXButton{h}{\slh}}%
%BeginExpansion
\slh
%EndExpansion
^{-1}([\mathcal{D}_{a},\mathcal{D}_{b}]%
%TCIMACRO{\TeXButton{h}{\slh}}%
%BeginExpansion
\slh
%EndExpansion
(c)-\mathcal{D}_{[a,b]}%
%TCIMACRO{\TeXButton{h}{\slh}}%
%BeginExpansion
\slh
%EndExpansion
(c))\nonumber\\
\overset{\gamma%
%TCIMACRO{\TeXButton{sig}{\sitg}}%
%BeginExpansion
\sitg
%EndExpansion
}{\rho}(a,b,c)  &  =%
%TCIMACRO{\TeXButton{h}{\slh}}%
%BeginExpansion
\slh
%EndExpansion
^{-1}(\overset{\varkappa}{\rho}(a,b,%
%TCIMACRO{\TeXButton{h}{\slh}}%
%BeginExpansion
\slh
%EndExpansion
(c))). \label{4.11}%
\end{align}

We relate $\overset{\varkappa\eta}{\mathbf{R}}_{3}(a,b,c,w)$with
$\overset{\gamma%
%TCIMACRO{\TeXButton{sig}{\sitg}}%
%BeginExpansion
\sitg
%EndExpansion
}{\mathbf{R}}_{3}(a,b,c,w)$, utilizing Eq.(\ref{4.11}). Indeed,
\begin{align}
\overset{\gamma%
%TCIMACRO{\TeXButton{sig}{\sitg}}%
%BeginExpansion
\sitg
%EndExpansion
}{\mathbf{R}_{3}}(a,b,c,w)  &  =-\overset{\gamma%
%TCIMACRO{\TeXButton{sig}{\sitg}}%
%BeginExpansion
\sitg
%EndExpansion
}{\rho}(a,b,c)\cdot%
%TCIMACRO{\TeXButton{itg}{\itg}}%
%BeginExpansion
\itg
%EndExpansion
(w)\nonumber\\
&  =-%
%TCIMACRO{\TeXButton{h}{\slh}}%
%BeginExpansion
\slh
%EndExpansion
^{-1}(\overset{\varkappa\eta}{\rho}(a,b,%
%TCIMACRO{\TeXButton{h}{\slh}}%
%BeginExpansion
\slh
%EndExpansion
(c)))\cdot%
%TCIMACRO{\TeXButton{h}{\slh}}%
%BeginExpansion
\slh
%EndExpansion
^{\dagger}\eta%
%TCIMACRO{\TeXButton{h}{\slh}}%
%BeginExpansion
\slh
%EndExpansion
(w)\nonumber\\
&  =-\overset{\varkappa\eta}{\rho}(a,b,%
%TCIMACRO{\TeXButton{h}{\slh}}%
%BeginExpansion
\slh
%EndExpansion
(c))\cdot\eta%
%TCIMACRO{\TeXButton{h}{\slh}}%
%BeginExpansion
\slh
%EndExpansion
(w)\nonumber\\
\overset{\gamma%
%TCIMACRO{\TeXButton{sig}{\sitg}}%
%BeginExpansion
\sitg
%EndExpansion
}{\mathbf{R}_{3}}(a,b,c,w)  &  =\overset{\varkappa\eta}{\mathbf{R}_{3}}(a,b,%
%TCIMACRO{\TeXButton{h}{\slh}}%
%BeginExpansion
\slh
%EndExpansion
(c),%
%TCIMACRO{\TeXButton{h}{\slh}}%
%BeginExpansion
\slh
%EndExpansion
(w)). \label{4.12}%
\end{align}

We related $\overset{\varkappa\eta}{%
%TCIMACRO{\TeXButton{R}{\slR}}%
%BeginExpansion
\slR
%EndExpansion
_{2}}(B)$ with $\overset{\gamma%
%TCIMACRO{\TeXButton{sig}{\sitg}}%
%BeginExpansion
\sitg
%EndExpansion
}{%
%TCIMACRO{\TeXButton{R}{\slR}}%
%BeginExpansion
\slR
%EndExpansion
_{2}}(B),$ utilizing the factorization of\emph{ }$\overset{\gamma%
%TCIMACRO{\TeXButton{sig}{\sitg}}%
%BeginExpansion
\sitg
%EndExpansion
}{\mathbf{R}}_{3}$, and the Eqs.(\ref{3.33}) and .(\ref{4.12}). Indeed,
\begin{align*}
\overset{\gamma%
%TCIMACRO{\TeXButton{sig}{\sitg}}%
%BeginExpansion
\sitg
%EndExpansion
}{%
%TCIMACRO{\TeXButton{R}{\slR}}%
%BeginExpansion
\slR
%EndExpansion
_{2}}(a\wedge b)\cdot(c\wedge w)  &  =\overset{\gamma%
%TCIMACRO{\TeXButton{sig}{\sitg}}%
%BeginExpansion
\sitg
%EndExpansion
}{\mathbf{R}}_{3}(a,b,c,w)\\
&  =\overset{\varkappa\eta}{\mathbf{R}}_{3}(a,b,%
%TCIMACRO{\TeXButton{h}{\slh}}%
%BeginExpansion
\slh
%EndExpansion
(c),%
%TCIMACRO{\TeXButton{h}{\slh}}%
%BeginExpansion
\slh
%EndExpansion
(w))\\
&  =\overset{\varkappa\eta}{%
%TCIMACRO{\TeXButton{R}{\slR}}%
%BeginExpansion
\slR
%EndExpansion
_{2}}(a\wedge b)\cdot(%
%TCIMACRO{\TeXButton{h}{\slh}}%
%BeginExpansion
\slh
%EndExpansion
(c)\wedge%
%TCIMACRO{\TeXButton{h}{\slh}}%
%BeginExpansion
\slh
%EndExpansion
(w))\\
&  =\underline{h}^{\dagger}\overset{\varkappa\eta}{%
%TCIMACRO{\TeXButton{R}{\slR}}%
%BeginExpansion
\slR
%EndExpansion
_{2}}(a\wedge b)\cdot(c\wedge w),
\end{align*}
which implies
\[
\overset{\gamma%
%TCIMACRO{\TeXButton{sig}{\sitg}}%
%BeginExpansion
\sitg
%EndExpansion
}{%
%TCIMACRO{\TeXButton{R}{\slR}}%
%BeginExpansion
\slR
%EndExpansion
_{2}}(a\wedge b)=\underline{%
%TCIMACRO{\TeXButton{h}{\slh}}%
%BeginExpansion
\slh
%EndExpansion
}^{\dagger}\overset{\varkappa\eta}{%
%TCIMACRO{\TeXButton{R}{\slR}}%
%BeginExpansion
\slR
%EndExpansion
_{2}}(a\wedge b),
\]
i.e.,
\begin{equation}
\overset{\gamma%
%TCIMACRO{\TeXButton{sig}{\sitg}}%
%BeginExpansion
\sitg
%EndExpansion
}{%
%TCIMACRO{\TeXButton{R}{\slR}}%
%BeginExpansion
\slR
%EndExpansion
}(B)=\underline{%
%TCIMACRO{\TeXButton{h}{\slh}}%
%BeginExpansion
\slh
%EndExpansion
}^{\dagger}\overset{\varkappa\eta}{%
%TCIMACRO{\TeXButton{R}{\slR}}%
%BeginExpansion
\slR
%EndExpansion
_{2}}(B). \label{4.13}%
\end{equation}

We get now $\overset{\varkappa\eta}{\rho}(a,b,c)$ and $\overset{\varkappa
\eta}{%
%TCIMACRO{\TeXButton{R}{\slR}}%
%BeginExpansion
\slR
%EndExpansion
_{2}}(a\wedge b)$ in terms of the $\varkappa\eta$-gauge rotation field
$\overset{\varkappa\eta}{\Omega}$ of the Minkowski-Cartan structure; For
$\overset{\varkappa}{\rho}(a,b,c)$ we have:
\begin{align}
\overset{\varkappa\eta}{\rho}(a,b,c)  &  =[\overset{\varkappa}{\mathcal{D}%
}_{a},\overset{\varkappa}{\mathcal{D}}_{a}]c-\overset{\varkappa}{\mathcal{D}%
}_{[a,b]}c\nonumber\\
&  =(a\cdot\partial\overset{\varkappa\eta}{\Omega}(b)-b\cdot\partial
\overset{\varkappa\eta}{\Omega}(a)-\overset{\varkappa\eta}{\Omega
}([a,b])+\overset{\varkappa\eta}{\Omega}(a)\underset{\eta}{\times
}\overset{\varkappa\eta}{\Omega}(b))\underset{\eta}{\times}c\nonumber\\
\overset{\varkappa\eta}{\rho}(a,b,c)  &  =((a\cdot\partial\overset{\varkappa
\eta}{\Omega})(b)-(b\cdot\partial\overset{\varkappa\eta}{\Omega}%
)(a)+\overset{\varkappa\eta}{\Omega}(a)\underset{\eta}{\times}%
\overset{\varkappa\eta}{\Omega}(b))\underset{\eta}{\times}c \label{4.14}%
\end{align}

For $\overset{\varkappa\eta}{%
%TCIMACRO{\TeXButton{R}{\slR}}%
%BeginExpansion
\slR
%EndExpansion
_{2}}(a\wedge b),$ utilizing the factorization theorem for \emph{
}$\overset{\gamma%
%TCIMACRO{\TeXButton{sig}{\sitg}}%
%BeginExpansion
\sitg
%EndExpansion
}{\mathbf{R}}_{3}$, (Eq.(\ref{3.33})), the Eq.(\ref{4.14}) and the multiform
identities $B\underset{%
%TCIMACRO{\TeXButton{sig}{\sitg}}%
%BeginExpansion
\sitg
%EndExpansion
}{\times}b=B\underset{%
%TCIMACRO{\TeXButton{sig}{\sitg}}%
%BeginExpansion
\sitg
%EndExpansion
}{\llcorner}b$ and $(B\underset{%
%TCIMACRO{\TeXButton{sig}{\sitg}}%
%BeginExpansion
\sitg
%EndExpansion
}{\llcorner}b)\underset{%
%TCIMACRO{\TeXButton{sig}{\sitg}}%
%BeginExpansion
\sitg
%EndExpansion
}{\cdot}a=B\underset{%
%TCIMACRO{\TeXButton{sig}{\sitg}}%
%BeginExpansion
\sitg
%EndExpansion
}{\cdot}(a\wedge b)$, with $B\in\sec%
%TCIMACRO{\dbigwedge \nolimits^{2}}%
%BeginExpansion
{\displaystyle\bigwedge\nolimits^{2}}
%EndExpansion
T^{\ast}U$ and $a,b\in\sec%
%TCIMACRO{\dbigwedge \nolimits^{1}}%
%BeginExpansion
{\displaystyle\bigwedge\nolimits^{1}}
%EndExpansion
T^{\ast}U$, we have:
\begin{align*}
\overset{\varkappa\eta}{%
%TCIMACRO{\TeXButton{R}{\slR}}%
%BeginExpansion
\slR
%EndExpansion
_{2}}(a\wedge b)\cdot(c\wedge w)  &  =\overset{\varkappa\eta}{\mathbf{R}_{3}%
}(a,b,c,w)\\
&  =-\overset{\varkappa\eta}{\rho}(a,b,c)\underset{\eta}{\cdot}w\\
&  =-((a\cdot\partial\overset{\varkappa\eta}{\Omega}(b)-b\cdot\partial
\overset{\varkappa\eta}{\Omega}(a)-\overset{\varkappa\eta}{\Omega
}([a,b])+\overset{\varkappa\eta}{\Omega}(a)\underset{\eta}{\times
}\overset{\varkappa\eta}{\Omega}(b))\underset{\eta}{\times}c)\underset{\eta
}{\cdot}w\\
&  =(a\cdot\partial\overset{\varkappa\eta}{\Omega}(b)-b\cdot\partial
\overset{\varkappa}{\Omega}(a)-\overset{\varkappa\eta}{\Omega}%
([a,b])+\overset{\varkappa}{\Omega}(a)\underset{\eta}{\times}%
\overset{\varkappa\eta}{\Omega}(b))\cdot\underline{\eta}(c\wedge w).
\end{align*}
Relabeling $c\rightarrow\eta(c)$ and $w\rightarrow\eta(w)$, and recalling that
$\eta^{2}=i_{d}$, we have
\[
\overset{\varkappa\eta}{%
%TCIMACRO{\TeXButton{R}{\slR}}%
%BeginExpansion
\slR
%EndExpansion
_{2}}(a\wedge b)\cdot\underline{\eta}(c\wedge w)=(a\cdot\partial
\overset{\varkappa\eta}{\Omega}(b)-b\cdot\partial\overset{\varkappa
\eta}{\Omega}(a)-\overset{\varkappa\eta}{\Omega}([a,b])+\overset{\varkappa
\eta}{\Omega}(a)\underset{\eta}{\times}\overset{\varkappa\eta}{\Omega
}(b))\cdot(c\wedge w),
\]
which implies that
\begin{equation}
\underline{\eta}\overset{\varkappa\eta}{%
%TCIMACRO{\TeXButton{R}{\slR}}%
%BeginExpansion
\slR
%EndExpansion
_{2}}(a\wedge b)=a\cdot\partial\overset{\varkappa\eta}{\Omega}(b)-b\cdot
\partial\overset{\varkappa\eta}{\Omega}(a)-\overset{\varkappa\eta}{\Omega
}([a,b])+\overset{\varkappa\eta}{\Omega}(a)\underset{\eta}{\times
}\overset{\varkappa}{\Omega}(b). \label{4.15}%
\end{equation}

\subsubsection{Gauge Extensor Fields Associated to a Lorentz-Cartan
\textit{MCGSS} $(U_{o},%
%TCIMACRO{\TeXButton{itg}{\itg}}%
%BeginExpansion
\itg
%EndExpansion
,\gamma)$}

We introduce now three smooth \textit{gauge} extensor fields on $U_{o}$
obtained from the Minkowski-Cartan \textit{MCGSS }$(U_{o},\eta,\varkappa)$ and
which encodes all the information encoded in the Lorentz-Cartan.\textit{MCGSS}
$(U_{o},%
%TCIMACRO{\TeXButton{itg}{\itg}}%
%BeginExpansion
\itg
%EndExpansion
,\gamma)$ where .$%
%TCIMACRO{\TeXButton{itg}{\itg}}%
%BeginExpansion
\itg
%EndExpansion
=%
%TCIMACRO{\TeXButton{h}{\slh}}%
%BeginExpansion
\slh
%EndExpansion
^{\dagger}\eta%
%TCIMACRO{\TeXButton{h}{\slh}}%
%BeginExpansion
\slh
%EndExpansion
$. First we define

\textbf{(i)} The $(2,2)$-extensor field, $B\mapsto\overset{\gamma%
%TCIMACRO{\TeXButton{sh}{\sslh}}%
%BeginExpansion
\sslh
%EndExpansion
}{\mathcal{R}}_{2}(B)$, given by
\begin{equation}
\overset{\gamma%
%TCIMACRO{\TeXButton{sh}{\sslh}}%
%BeginExpansion
\sslh
%EndExpansion
}{\mathcal{R}}_{2}(B)=\underline{\eta}\overset{\varkappa\eta}{%
%TCIMACRO{\TeXButton{R}{\slR}}%
%BeginExpansion
\slR
%EndExpansion
_{2}}(B), \label{4.16}%
\end{equation}
which is called the \emph{Riemann gauge field for }$(U_{o},%
%TCIMACRO{\TeXButton{itg}{\itg}}%
%BeginExpansion
\itg
%EndExpansion
,\gamma)$, and the reason for that name will be clear in a while.\emph{ }

\textbf{(ii)} The $(1,1)$-extensor field, $b\mapsto\overset{\gamma%
%TCIMACRO{\TeXButton{sh}{\sslh}}%
%BeginExpansion
\sslh
%EndExpansion
}{\mathcal{R}}_{1}(b)$, given by
\begin{equation}
\overset{\gamma%
%TCIMACRO{\TeXButton{sh}{\sslh}}%
%BeginExpansion
\sslh
%EndExpansion
}{\mathcal{R}}_{1}(b)=%
%TCIMACRO{\TeXButton{h}{\slh}}%
%BeginExpansion
\slh
%EndExpansion
^{\clubsuit}(\partial_{a})\lrcorner\overset{\gamma%
%TCIMACRO{\TeXButton{sh}{\sslh}}%
%BeginExpansion
\sslh
%EndExpansion
}{\mathcal{R}}_{2}(a\wedge b), \label{4.17}%
\end{equation}
called the Ricci\emph{ gauge field for }$(U_{o},%
%TCIMACRO{\TeXButton{itg}{\itg}}%
%BeginExpansion
\itg
%EndExpansion
,\gamma)$.

\textbf{(iii)} The scalar field $\overset{\gamma%
%TCIMACRO{\TeXButton{sh}{\sslh}}%
%BeginExpansion
\sslh
%EndExpansion
}{\mathcal{R}}$, given by by
\begin{equation}
\overset{\gamma%
%TCIMACRO{\TeXButton{sh}{\sslh}}%
%BeginExpansion
\sslh
%EndExpansion
}{\mathcal{R}}=%
%TCIMACRO{\TeXButton{h}{\slh}}%
%BeginExpansion
\slh
%EndExpansion
^{\clubsuit}(\partial_{b})\cdot\overset{\gamma%
%TCIMACRO{\TeXButton{sh}{\sslh}}%
%BeginExpansion
\sslh
%EndExpansion
}{\mathcal{R}}_{1}(b)=\underline{%
%TCIMACRO{\TeXButton{h}{\slh}}%
%BeginExpansion
\slh
%EndExpansion
}^{\ast}(\partial_{a}\wedge\partial_{b})\cdot\overset{\gamma%
%TCIMACRO{\TeXButton{sh}{\sslh}}%
%BeginExpansion
\sslh
%EndExpansion
}{\mathcal{R}}_{2}(a\wedge b), \label{4.18}%
\end{equation}
called the \textit{gauge Ricci scalar field} \emph{for }$(U_{o},%
%TCIMACRO{\TeXButton{itg}{\itg}}%
%BeginExpansion
\itg
%EndExpansion
,\gamma)$.

\textbf{(iv)} The $(1,1)$-extensor field, $a\mapsto\overset{\gamma%
%TCIMACRO{\TeXButton{sh}{\sslh}}%
%BeginExpansion
\sslh
%EndExpansion
}{\mathcal{G}}(a)$, given by
\begin{equation}
\overset{\gamma%
%TCIMACRO{\TeXButton{sh}{\sslh}}%
%BeginExpansion
\sslh
%EndExpansion
}{\mathcal{G}}(a)=\overset{\gamma%
%TCIMACRO{\TeXButton{sh}{\sslh}}%
%BeginExpansion
\sslh
%EndExpansion
}{\mathcal{R}}_{1}(a)-\frac{1}{2}%
%TCIMACRO{\TeXButton{h}{\slh}}%
%BeginExpansion
\slh
%EndExpansion
(a)\overset{\gamma%
%TCIMACRO{\TeXButton{sh}{\sslh}}%
%BeginExpansion
\sslh
%EndExpansion
}{\mathcal{R}}, \label{4.19}%
\end{equation}
called \emph{the Einstein gauge field for }$(U_{o},%
%TCIMACRO{\TeXButton{itg}{\itg}}%
%BeginExpansion
\itg
%EndExpansion
,\gamma)$.\medskip

\textbf{Proposition 4.5} The Riemann gauge field has the fundamental property
\begin{equation}
\overset{\gamma%
%TCIMACRO{\TeXButton{sh}{\sslh}}%
%BeginExpansion
\sslh
%EndExpansion
}{\mathcal{R}}_{2}(a\wedge b)=a\cdot\partial\overset{\varkappa\eta}{\Omega
}(b)-b\cdot\partial\overset{\varkappa\eta}{\Omega}(a)-\overset{\varkappa
\eta}{\Omega}([a,b])+\overset{\varkappa\eta}{\Omega}(a)\underset{\eta}{\times
}\overset{\varkappa\eta}{\Omega}(b) \label{4.20}%
\end{equation}

\textbf{Proof: }Eq.(\ref{4.20}) is an immediate consequence of Eq.(\ref{4.16})
and Eq.(\ref{4.15}).
%TCIMACRO{\TeXButton{End Proof}{\endproof}}%
%BeginExpansion
\endproof
%EndExpansion
\medskip

\textbf{Proposition 4.6 }The Ricci scalar field $\overset{\gamma%
%TCIMACRO{\TeXButton{sig}{\sitg}}%
%BeginExpansion
\sitg
%EndExpansion
}{R}$ (associated with $(U_{o},%
%TCIMACRO{\TeXButton{itg}{\itg}}%
%BeginExpansion
\itg
%EndExpansion
,\gamma)$) is equal to $\overset{\gamma%
%TCIMACRO{\TeXButton{sh}{\sslh}}%
%BeginExpansion
\sslh
%EndExpansion
}{\mathcal{R}},$ i.e.,
\begin{equation}
\overset{\gamma%
%TCIMACRO{\TeXButton{sig}{\sitg}}%
%BeginExpansion
\sitg
%EndExpansion
}{R}=\overset{\gamma%
%TCIMACRO{\TeXButton{sh}{\sslh}}%
%BeginExpansion
\sslh
%EndExpansion
}{\mathcal{R}} \label{4.21}%
\end{equation}

\textbf{Proof: }By a simple algebraic manipulation of Eq.(\ref{3.35}),
utilizing the Theorem 4.1, the Eq.(\ref{4.13}) and Eq.(\ref{4.14}), we get
that
\begin{align*}
\overset{\gamma%
%TCIMACRO{\TeXButton{sig}{\sitg}}%
%BeginExpansion
\sitg
%EndExpansion
}{R}  &  =\underline{g}^{-1}(\partial_{a}\wedge\partial_{b})\cdot
\overset{\gamma%
%TCIMACRO{\TeXButton{sig}{\sitg}}%
%BeginExpansion
\sitg
%EndExpansion
}{%
%TCIMACRO{\TeXButton{R}{\slR}}%
%BeginExpansion
\slR
%EndExpansion
_{2}}(a\wedge b)=\underline{%
%TCIMACRO{\TeXButton{h}{\slh}}%
%BeginExpansion
\slh
%EndExpansion
}^{-1}\underline{\eta}\underline{%
%TCIMACRO{\TeXButton{h}{\slh}}%
%BeginExpansion
\slh
%EndExpansion
}^{\ast}(\partial_{a}\wedge\partial_{b})\cdot\underline{%
%TCIMACRO{\TeXButton{h}{\slh}}%
%BeginExpansion
\slh
%EndExpansion
}^{\dagger}\overset{\varkappa\eta}{%
%TCIMACRO{\TeXButton{R}{\slR}}%
%BeginExpansion
\slR
%EndExpansion
_{2}}(a\wedge b)\\
&  =\underline{%
%TCIMACRO{\TeXButton{h}{\slh}}%
%BeginExpansion
\slh
%EndExpansion
}^{\ast}(\partial_{a}\wedge\partial_{b})\cdot\underline{\eta}%
\overset{\varkappa\eta}{%
%TCIMACRO{\TeXButton{R}{\slR}}%
%BeginExpansion
\slR
%EndExpansion
_{2}}(a\wedge b)=\underline{%
%TCIMACRO{\TeXButton{h}{\slh}}%
%BeginExpansion
\slh
%EndExpansion
}^{\ast}(\partial_{a}\wedge\partial_{b})\cdot\overset{\gamma%
%TCIMACRO{\TeXButton{sh}{\sslh}}%
%BeginExpansion
\sslh
%EndExpansion
}{\mathcal{R}}_{2}(a\wedge b),
\end{align*}
i.e., by Eq.(\ref{4.18}), $\overset{\gamma%
%TCIMACRO{\TeXButton{sig}{\sitg}}%
%BeginExpansion
\sitg
%EndExpansion
}{R}=\overset{\gamma%
%TCIMACRO{\TeXButton{sh}{\sslh}}%
%BeginExpansion
\sslh
%EndExpansion
}{\mathcal{R}}.%
%TCIMACRO{\TeXButton{End Proof}{\endproof}}%
%BeginExpansion
\endproof
%EndExpansion
\medskip$

\textbf{Proposition 4.7} The Ricci $(1,1)$-extensor field, $b\mapsto
\overset{\gamma%
%TCIMACRO{\TeXButton{sig}{\sitg}}%
%BeginExpansion
\sitg
%EndExpansion
}{%
%TCIMACRO{\TeXButton{R}{\slR}}%
%BeginExpansion
\slR
%EndExpansion
_{1}}(b)$, and Einstein $(1,1)$-extensor field, $a\mapsto\overset{\gamma%
%TCIMACRO{\TeXButton{sig}{\sitg}}%
%BeginExpansion
\sitg
%EndExpansion
}{%
%TCIMACRO{\TeXButton{G}{\slG}}%
%BeginExpansion
\slG
%EndExpansion
}(a),$ associated to the Lorentz-Cartan \textit{MCGSS }$(U_{o},%
%TCIMACRO{\TeXButton{itg}{\itg}}%
%BeginExpansion
\itg
%EndExpansion
,\gamma)$ are related to the gauge fields $b\mapsto\overset{\gamma%
%TCIMACRO{\TeXButton{sh}{\sslh}}%
%BeginExpansion
\sslh
%EndExpansion
}{\mathcal{R}}_{1}(b)$ and $a\mapsto\overset{\gamma%
%TCIMACRO{\TeXButton{sh}{\sslh}}%
%BeginExpansion
\sslh
%EndExpansion
}{\mathcal{G}}(a)$ by
\begin{align}
\overset{\gamma%
%TCIMACRO{\TeXButton{sig}{\sitg}}%
%BeginExpansion
\sitg
%EndExpansion
}{%
%TCIMACRO{\TeXButton{R}{\slR}}%
%BeginExpansion
\slR
%EndExpansion
_{1}}(b)  &  =%
%TCIMACRO{\TeXButton{h}{\slh}}%
%BeginExpansion
\slh
%EndExpansion
^{\dagger}\eta\overset{\gamma%
%TCIMACRO{\TeXButton{sh}{\sslh}}%
%BeginExpansion
\sslh
%EndExpansion
}{\mathcal{R}}_{1}(b),\label{4.22}\\
\overset{\gamma%
%TCIMACRO{\TeXButton{sig}{\sitg}}%
%BeginExpansion
\sitg
%EndExpansion
}{%
%TCIMACRO{\TeXButton{G}{\slG}}%
%BeginExpansion
\slG
%EndExpansion
}(a)  &  =%
%TCIMACRO{\TeXButton{h}{\slh}}%
%BeginExpansion
\slh
%EndExpansion
^{\dagger}\eta\overset{\gamma%
%TCIMACRO{\TeXButton{sh}{\sslh}}%
%BeginExpansion
\sslh
%EndExpansion
}{\mathcal{G}}(a). \label{4.23}%
\end{align}

\textbf{Proof:\ }By simple algebraic manipulation of Eq.(\ref{3.35}),
utilizing Eq.(\ref{4.13}), Eq. (\ref{4.16}), Eq.(\ref{1.40}) and the Theorem
4.1, we get that
\begin{align*}
\overset{\gamma%
%TCIMACRO{\TeXButton{sig}{\sitg}}%
%BeginExpansion
\sitg
%EndExpansion
}{%
%TCIMACRO{\TeXButton{R}{\slR}}%
%BeginExpansion
\slR
%EndExpansion
_{2}}(b)  &  =%
%TCIMACRO{\TeXButton{itg}{\itg}}%
%BeginExpansion
\itg
%EndExpansion
^{-1}(\partial_{a})\lrcorner\overset{\gamma%
%TCIMACRO{\TeXButton{sig}{\sitg}}%
%BeginExpansion
\sitg
%EndExpansion
}{%
%TCIMACRO{\TeXButton{R}{\slR}}%
%BeginExpansion
\slR
%EndExpansion
_{2}}(a\wedge b)=%
%TCIMACRO{\TeXButton{itg}{\itg}}%
%BeginExpansion
\itg
%EndExpansion
^{-1}(\partial_{a})\lrcorner\underline{%
%TCIMACRO{\TeXButton{h}{\slh}}%
%BeginExpansion
\slh
%EndExpansion
}^{\dagger}\overset{\varkappa\eta}{%
%TCIMACRO{\TeXButton{R}{\slR}}%
%BeginExpansion
\slR
%EndExpansion
_{2}}(a\wedge b)\\
&  =%
%TCIMACRO{\TeXButton{itg}{\itg}}%
%BeginExpansion
\itg
%EndExpansion
^{-1}(\partial_{a})\lrcorner\underline{%
%TCIMACRO{\TeXButton{h}{\slh}}%
%BeginExpansion
\slh
%EndExpansion
}^{\dagger}\underline{\eta}\overset{\gamma%
%TCIMACRO{\TeXButton{sh}{\sslh}}%
%BeginExpansion
\sslh
%EndExpansion
}{\mathcal{R}}_{2}(a\wedge b)=%
%TCIMACRO{\TeXButton{h}{\slh}}%
%BeginExpansion
\slh
%EndExpansion
^{\dagger}\eta(\eta%
%TCIMACRO{\TeXButton{h}{\slh}}%
%BeginExpansion
\slh
%EndExpansion%
%TCIMACRO{\TeXButton{h}{\slh}}%
%BeginExpansion
\slh
%EndExpansion
^{-1}\eta%
%TCIMACRO{\TeXButton{h}{\slh}}%
%BeginExpansion
\slh
%EndExpansion
^{\clubsuit}(\partial_{a})\lrcorner\overset{\gamma%
%TCIMACRO{\TeXButton{sh}{\sslh}}%
%BeginExpansion
\sslh
%EndExpansion
}{\mathcal{R}}_{2}(a\wedge b))\\
&  =%
%TCIMACRO{\TeXButton{h}{\slh}}%
%BeginExpansion
\slh
%EndExpansion
^{\dagger}\eta(%
%TCIMACRO{\TeXButton{h}{\slh}}%
%BeginExpansion
\slh
%EndExpansion
^{\clubsuit}(\partial_{a})\lrcorner\overset{\gamma%
%TCIMACRO{\TeXButton{sh}{\sslh}}%
%BeginExpansion
\sslh
%EndExpansion
}{\mathcal{R}}_{2}(a\wedge b)),
\end{align*}
and recalling Eq.(\ref{4.17}). we have that $\overset{\gamma}{%
%TCIMACRO{\TeXButton{R}{\slR}}%
%BeginExpansion
\slR
%EndExpansion
_{1}}(b)=%
%TCIMACRO{\TeXButton{h}{\slh}}%
%BeginExpansion
\slh
%EndExpansion
^{\dagger}\eta\overset{\gamma%
%TCIMACRO{\TeXButton{sh}{\sslh}}%
%BeginExpansion
\sslh
%EndExpansion
}{\mathcal{R}}_{1}(b).$

Also, putting Eqs.(\ref{4.22}) and (\ref{4.21}) in Eq.(\ref{3.36}) and
utilizing \ the Theorem 4.1, we get
\begin{align*}
\overset{\gamma%
%TCIMACRO{\TeXButton{sig}{\sitg}}%
%BeginExpansion
\sitg
%EndExpansion
}{%
%TCIMACRO{\TeXButton{G}{\slG}}%
%BeginExpansion
\slG
%EndExpansion
}(a)  &  =\overset{\gamma%
%TCIMACRO{\TeXButton{sig}{\sitg}}%
%BeginExpansion
\sitg
%EndExpansion
}{%
%TCIMACRO{\TeXButton{R}{\slR}}%
%BeginExpansion
\slR
%EndExpansion
_{1}}(a)-\frac{1}{2}%
%TCIMACRO{\TeXButton{itg}{\itg}}%
%BeginExpansion
\itg
%EndExpansion
(a)\overset{\gamma%
%TCIMACRO{\TeXButton{sig}{\sitg}}%
%BeginExpansion
\sitg
%EndExpansion
}{R}=%
%TCIMACRO{\TeXButton{h}{\slh}}%
%BeginExpansion
\slh
%EndExpansion
^{\dagger}\eta\overset{\gamma%
%TCIMACRO{\TeXButton{sh}{\sslh}}%
%BeginExpansion
\sslh
%EndExpansion
}{\mathcal{R}}_{1}(a)-\frac{1}{2}%
%TCIMACRO{\TeXButton{itg}{\itg}}%
%BeginExpansion
\itg
%EndExpansion
(a)\overset{\gamma%
%TCIMACRO{\TeXButton{sh}{\sslh}}%
%BeginExpansion
\sslh
%EndExpansion
}{\mathcal{R}}\\
&  =%
%TCIMACRO{\TeXButton{h}{\slh}}%
%BeginExpansion
\slh
%EndExpansion
^{\dagger}\eta\overset{\gamma%
%TCIMACRO{\TeXButton{sh}{\sslh}}%
%BeginExpansion
\sslh
%EndExpansion
}{\mathcal{R}}_{1}(a)-\frac{1}{2}%
%TCIMACRO{\TeXButton{h}{\slh}}%
%BeginExpansion
\slh
%EndExpansion
^{\dagger}\eta%
%TCIMACRO{\TeXButton{h}{\slh}}%
%BeginExpansion
\slh
%EndExpansion
(a)\overset{\gamma%
%TCIMACRO{\TeXButton{sh}{\sslh}}%
%BeginExpansion
\sslh
%EndExpansion
}{\mathcal{R}}=%
%TCIMACRO{\TeXButton{h}{\slh}}%
%BeginExpansion
\slh
%EndExpansion
^{\dagger}\eta(\overset{\gamma%
%TCIMACRO{\TeXButton{sh}{\sslh}}%
%BeginExpansion
\sslh
%EndExpansion
}{\mathcal{R}}(a)-\frac{1}{2}%
%TCIMACRO{\TeXButton{h}{\slh}}%
%BeginExpansion
\slh
%EndExpansion
(a)\overset{\gamma%
%TCIMACRO{\TeXButton{sh}{\sslh}}%
%BeginExpansion
\sslh
%EndExpansion
}{\mathcal{R}}),
\end{align*}
and recalling Eq.(\ref{4.19}) it follows that $\overset{\gamma%
%TCIMACRO{\TeXButton{sig}{\sitg}}%
%BeginExpansion
\sitg
%EndExpansion
}{%
%TCIMACRO{\TeXButton{G}{\slG}}%
%BeginExpansion
\slG
%EndExpansion
}(a)=%
%TCIMACRO{\TeXButton{h}{\slh}}%
%BeginExpansion
\slh
%EndExpansion
^{\dagger}\eta\overset{\gamma%
%TCIMACRO{\TeXButton{sh}{\sslh}}%
%BeginExpansion
\sslh
%EndExpansion
}{\mathcal{G}}(a).$%
%TCIMACRO{\TeXButton{End Proof}{\endproof}}%
%BeginExpansion
\endproof
%EndExpansion
\medskip

The last results may be interpreted by saying that the plastic distortion
field $%
%TCIMACRO{\TeXButton{h}{\slh}}%
%BeginExpansion
\slh
%EndExpansion
$ living on the Minkowski \textit{MCGSS }$(U_{o},\eta,\mathring{\varkappa}%
)$\ deforms its natural parallel transport rule thus generating a
Minkowski-Cartan \textit{MCGSS }$(U_{o},\eta,\varkappa)$\footnote{The way in
which $\mathcal{\mathring{D}}_{a}$ and $\mathcal{D}_{a}$ are related is a
particular case of the way that two different connections defined on a
manifold $M$ are related, and are briefly recalled in Appendix D.}.
Equivalently, we may say that the field $%
%TCIMACRO{\TeXButton{h}{\slh}}%
%BeginExpansion
\slh
%EndExpansion
$ generates an effective Lorentzian metric (extensor) $%
%TCIMACRO{\TeXButton{itg}{\itg}}%
%BeginExpansion
\itg
%EndExpansion
$ and that \ the Lorentz \textit{MCGSS }$(U_{o},%
%TCIMACRO{\TeXButton{itg}{\itg}}%
%BeginExpansion
\itg
%EndExpansion
,\gamma)$ is gauge equivalent to the Minkowski-Cartan \textit{MCGSS }%
$(U_{o},\eta,\varkappa)$. This is the case because the Riemann gauge field
$\overset{\gamma%
%TCIMACRO{\TeXButton{sh}{\sslh}}%
%BeginExpansion
\sslh
%EndExpansion
}{\mathcal{R}}(B)$ \emph{encodes all the information contained in the Ricci}
$(1,1)$-extensor field $\overset{\gamma%
%TCIMACRO{\TeXButton{sig}{\sitg}}%
%BeginExpansion
\sitg
%EndExpansion
}{%
%TCIMACRO{\TeXButton{R}{\slR}}%
%BeginExpansion
\slR
%EndExpansion
_{1}}(a)$, in the Ricci scalar field $\overset{\gamma%
%TCIMACRO{\TeXButton{sig}{\sitg}}%
%BeginExpansion
\sitg
%EndExpansion
}{R}$ and in the Einstein $(1,1)$-extensor field $\overset{\gamma%
%TCIMACRO{\TeXButton{sig}{\sitg}}%
%BeginExpansion
\sitg
%EndExpansion
}{%
%TCIMACRO{\TeXButton{G}{\slG}}%
%BeginExpansion
\slG
%EndExpansion
}(a)$ of the Lorentz-Cartan \textit{MCGSS. }Moreover, we have the\textit{
}following proposition.\medskip

\textbf{Proposition\ 4.8} The Ricci scalar field $\overset{\gamma%
%TCIMACRO{\TeXButton{sh}{\sslh}}%
%BeginExpansion
\sslh
%EndExpansion
}{\mathcal{R}}$ ($=\overset{\gamma%
%TCIMACRO{\TeXButton{sig}{\sitg}}%
%BeginExpansion
\sitg
%EndExpansion
}{R}$) is a \ \textit{scalar }$%
%TCIMACRO{\TeXButton{h}{\slh}}%
%BeginExpansion
\slh
%EndExpansion
^{\clubsuit}$\textit{-divergent} of the Riemann gauge field gauge de Riemann
$\overset{\gamma%
%TCIMACRO{\TeXButton{sh}{\sslh}}%
%BeginExpansion
\sslh
%EndExpansion
}{\mathcal{R}}_{2}(B)$, i.e.,
\begin{align}
\overset{\gamma%
%TCIMACRO{\TeXButton{sh}{\sslh}}%
%BeginExpansion
\sslh
%EndExpansion
}{\mathcal{R}}  &  =\underline{%
%TCIMACRO{\TeXButton{h}{\slh}}%
%BeginExpansion
\slh
%EndExpansion
}^{\ast}(\partial_{a}\wedge\partial_{b})\cdot\overset{\gamma%
%TCIMACRO{\TeXButton{sh}{\sslh}}%
%BeginExpansion
\sslh
%EndExpansion
}{\mathcal{R}}_{2}(a\wedge b)\nonumber\\
&  =\underline{%
%TCIMACRO{\TeXButton{h}{\slh}}%
%BeginExpansion
\slh
%EndExpansion
}^{\ast}(\partial_{a}\wedge\partial_{b})\cdot(a\cdot\partial\overset{\varkappa
\eta}{\Omega}(b)-b\cdot\partial\overset{\varkappa\eta}{\Omega}%
(a)-\overset{\varkappa\eta}{\Omega}([a,b])+\overset{\varkappa\eta}{\Omega
}(a)\underset{\eta}{\times}\overset{\varkappa\eta}{\Omega}(b))\nonumber\\
\overset{\gamma%
%TCIMACRO{\TeXButton{sh}{\sslh}}%
%BeginExpansion
\sslh
%EndExpansion
}{\mathcal{R}}  &  =\underline{%
%TCIMACRO{\TeXButton{h}{\slh}}%
%BeginExpansion
\slh
%EndExpansion
}^{\ast}(\partial_{a}\wedge\partial_{b})\cdot((a\cdot\partial
\overset{\varkappa\eta}{\Omega})(b)-(b\cdot\partial\overset{\varkappa
\eta}{\Omega})(a)+\overset{\varkappa\eta}{\Omega}(a)\underset{\eta}{\times
}\overset{\varkappa\eta}{\Omega}(b)). \label{4.24}%
\end{align}

The proof of Proposition 4.8 follows trivially from the previous
results.\medskip

\textbf{Remark 4.5 }Proposition 4.8 shows that $\overset{\gamma%
%TCIMACRO{\TeXButton{sh}{\sslh}}%
%BeginExpansion
\sslh
%EndExpansion
}{\mathcal{R}}$ may be interpreted as a \textit{scalar functional }of the
plastic gauge distortion field $%
%TCIMACRO{\TeXButton{h}{\slh}}%
%BeginExpansion
\slh
%EndExpansion
$ an the $\varkappa\eta$-gauge rotation field $\overset{\varkappa\eta}{\Omega
}$ (and their directional derivatives $\cdot\partial\overset{\varkappa
\eta}{\Omega}$). However to get a simple theory for the gravitational field we
need an additional result which is valid for a Lorentz \textit{MCGSS }$(U_{o},%
%TCIMACRO{\TeXButton{itg}{\itg}}%
%BeginExpansion
\itg
%EndExpansion
,\lambda)$

\subsubsection{Lorentz \textit{MCGSS} as $%
%TCIMACRO{\TeXButton{h}{\slh}}%
%BeginExpansion
\slh
%EndExpansion
$-Deformation of a Particular Minkowski-Cartan \textit{MCGSS }}

Let $(M,%
%TCIMACRO{\TeXButton{g}{\slg}}%
%BeginExpansion
\slg
%EndExpansion
,D,\tau_{%
%TCIMACRO{\TeXButton{g}{\sslg}}%
%BeginExpansion
\sslg
%EndExpansion
},\uparrow)$ be a Lorentzian spacetime structure as defined in the beginning
of Section 1. We already called the triple $(M,%
%TCIMACRO{\TeXButton{g}{\slg}}%
%BeginExpansion
\slg
%EndExpansion
,D)$ Lorentz \textit{MCGSS}. As in the previous sections let .$\mathcal{U}%
\subset M$, $U$ be the canonical vector space and $U_{o}\subset U$ the
representative of the points of $\mathcal{U}$. Moreover let $%
%TCIMACRO{\TeXButton{itg}{\itg}}%
%BeginExpansion
\itg
%EndExpansion
$ be the metric extensor on $U_{o}$ (which represents $%
%TCIMACRO{\TeXButton{g}{\slg}}%
%BeginExpansion
\slg
%EndExpansion
$) and $\lambda$ the $\mathit{\nabla}$ \textit{connection }$2$%
\textit{-extensor} on $U_{o}$ representing the de Levi-Civita connection $D$.
Under those conditions we also say that $(U_{o},%
%TCIMACRO{\TeXButton{itg}{\itg}}%
%BeginExpansion
\itg
%EndExpansion
,\lambda)$ is a Lorentz \textit{MCGSS.}

Observe that the Lorentz $(U_{o},%
%TCIMACRO{\TeXButton{itg}{\itg}}%
%BeginExpansion
\itg
%EndExpansion
,\lambda)$\ \textit{MCGSS}\ is a particular Lorentz-Cartan \textit{MCGSS}
$(U_{o},%
%TCIMACRO{\TeXButton{itg}{\itg}}%
%BeginExpansion
\itg
%EndExpansion
,\gamma)$where the connection $2$-extensor field is $\gamma=\lambda$ (the
Levi-Civita connection $2$-extensor field).

Let $D_{a}$\ be the covariant\ derivative operator defined by $(U_{o},%
%TCIMACRO{\TeXButton{itg}{\itg}}%
%BeginExpansion
\itg
%EndExpansion
,\lambda)$ and let $a,b\in\sec%
%TCIMACRO{\dbigwedge \nolimits^{1}}%
%BeginExpansion
{\displaystyle\bigwedge\nolimits^{1}}
%EndExpansion
T^{\ast}U$. Under those conditions we shall use the following
\textit{notation} (recall Eq.(\ref{4.8}))%
\begin{equation}
D_{a}b=a\cdot\partial b+\frac{1}{2}%
%TCIMACRO{\TeXButton{itg}{\itg}}%
%BeginExpansion
\itg
%EndExpansion
^{-1}(a\cdot\partial%
%TCIMACRO{\TeXButton{itg}{\itg}}%
%BeginExpansion
\itg
%EndExpansion
)(b)+\overset{}{\omega}(a)\times%
%TCIMACRO{\TeXButton{itg}{\itg}}%
%BeginExpansion
\itg
%EndExpansion
(b), \label{4.250}%
\end{equation}
where the $%
%TCIMACRO{\TeXButton{itg}{\itg}}%
%BeginExpansion
\itg
%EndExpansion
$-gauge rotation field $\overset{}{\omega}$ of the $(U_{o},%
%TCIMACRO{\TeXButton{itg}{\itg}}%
%BeginExpansion
\itg
%EndExpansion
,\lambda)$ GSS is given by the formula \cite{fmr073}
\begin{equation}
\overset{}{\omega}(a)=-\frac{1}{2}\underline{%
%TCIMACRO{\TeXButton{itg}{\itg}}%
%BeginExpansion
\itg
%EndExpansion
}^{-1}(\partial_{b}\wedge\partial_{c})A(a,b,c), \label{4.26}%
\end{equation}
where
\begin{equation}
A(a,b,c)\equiv\dfrac{1}{2}a\cdot(b\cdot\partial%
%TCIMACRO{\TeXButton{itg}{\itg}}%
%BeginExpansion
\itg
%EndExpansion
(c)-c\cdot\partial%
%TCIMACRO{\TeXButton{itg}{\itg}}%
%BeginExpansion
\itg
%EndExpansion
(b)-%
%TCIMACRO{\TeXButton{itg}{\itg}}%
%BeginExpansion
\itg
%EndExpansion
([b,c]))=\dfrac{1}{2}a\cdot((b\cdot\partial%
%TCIMACRO{\TeXButton{itg}{\itg}}%
%BeginExpansion
\itg
%EndExpansion
)(c)-(c\cdot\partial%
%TCIMACRO{\TeXButton{itg}{\itg}}%
%BeginExpansion
\itg
%EndExpansion
)(b)), \label{4.266}%
\end{equation}
is a scalar $3$-extensor field (obviously antisymmetric with respect to the
second and third variables).

If $%
%TCIMACRO{\TeXButton{itg}{\itg}}%
%BeginExpansion
\itg
%EndExpansion
=%
%TCIMACRO{\TeXButton{h}{\slh}}%
%BeginExpansion
\slh
%EndExpansion
^{\dagger}\eta%
%TCIMACRO{\TeXButton{h}{\slh}}%
%BeginExpansion
\slh
%EndExpansion
$ then according to the Theorem 4.3$\ $\ we may interpret the Lorentz
\textit{MCGSS} $(U_{o},%
%TCIMACRO{\TeXButton{itg}{\itg}}%
%BeginExpansion
\itg
%EndExpansion
,\lambda)$ as a deformation of a \textit{particular} Minkowski-Cartan
\textit{MCGSS} which will be denoted by $(U_{o},\eta,\mu)$. Indeed, in this
case we have the following result:\medskip

\textbf{Proposition 4.9} \cite{fmr073} If the covariant derivative operator
defined by$(U_{o},\eta,\mu)$ is denoted $\overset{}{\mathcal{D}}_{a}$then%
\begin{equation}
D_{a}b=%
%TCIMACRO{\TeXButton{h}{\slh}}%
%BeginExpansion
\slh
%EndExpansion
^{-1}(\overset{}{\mathcal{D}}_{a}%
%TCIMACRO{\TeXButton{h}{\slh}}%
%BeginExpansion
\slh
%EndExpansion
(b)), \label{4.266a}%
\end{equation}
where
\begin{equation}
\overset{}{\mathcal{D}}_{a}b=a\cdot\partial b+\text{ }\overset{}{\Omega
}(a)\underset{\eta}{\times}b \label{4.25a}%
\end{equation}
and the $\eta$-gauge rotation operator $\overset{}{\Omega}$ of the
Minkowski-Cartan $(U_{o},\eta,\mu)$ is given by
\begin{equation}
\overset{}{\Omega}(a)=-\frac{1}{2}\underline{\eta}^{-1}(\partial_{b}%
\wedge\partial_{c})[a,%
%TCIMACRO{\TeXButton{h}{\slh}}%
%BeginExpansion
\slh
%EndExpansion
^{-1}(b),%
%TCIMACRO{\TeXButton{h}{\slh}}%
%BeginExpansion
\slh
%EndExpansion
^{-1}(c)], \label{4.25}%
\end{equation}
where $[x,y,z]$ \ defined by Eq.(\ref{3.2}) is the first Christoffel operator
associated to the Lorentz metric extensor $%
%TCIMACRO{\TeXButton{itg}{\itg}}%
%BeginExpansion
\itg
%EndExpansion
=%
%TCIMACRO{\TeXButton{h}{\slh}}%
%BeginExpansion
\slh
%EndExpansion
^{\dagger}\eta%
%TCIMACRO{\TeXButton{h}{\slh}}%
%BeginExpansion
\slh
%EndExpansion
$.\medskip

\textbf{Remark 4.6 }\ It is absolutely clear form the above formulas that the
biform field $\overset{}{\Omega}(a)$ may be interpreted as a \textit{biform
functional }of the plastic distortion gauge field $%
%TCIMACRO{\TeXButton{h}{\slh}}%
%BeginExpansion
\slh
%EndExpansion
$ of its first directional derivatives $\cdot\partial%
%TCIMACRO{\TeXButton{h}{\slh}}%
%BeginExpansion
\slh
%EndExpansion
$ and of its second order directional derivatives $\cdot\partial\cdot\partial%
%TCIMACRO{\TeXButton{h}{\slh}}%
%BeginExpansion
\slh
%EndExpansion
$. It is this result that summed to the one recalled in Remark 4.5 that permit
us to formulate a gravitational theory on Minkowski spacetime where this field
is the plastic distortion field $%
%TCIMACRO{\TeXButton{h}{\slh}}%
%BeginExpansion
\slh
%EndExpansion
$. This is indeed the case because if $R$ \ is the curvature scalar in
Einstein's \ \textit{GRT }(the same as $R$ in the Lorentz \textit{GSS }%
$(U_{o},%
%TCIMACRO{\TeXButton{itg}{\itg}}%
%BeginExpansion
\itg
%EndExpansion
,\lambda)$), then according to Eq.(\ref{4.21}) (where now $\gamma=\lambda$),
$R\equiv\overset{\lambda%
%TCIMACRO{\TeXButton{sig}{\sitg}}%
%BeginExpansion
\sitg
%EndExpansion
}{R}=\overset{\lambda%
%TCIMACRO{\TeXButton{sh}{\sslh}}%
%BeginExpansion
\sslh
%EndExpansion
}{\mathcal{R}}$ is a \textit{scalar functional} of $%
%TCIMACRO{\TeXButton{h}{\slh}}%
%BeginExpansion
\slh
%EndExpansion
^{\clubsuit}$, $\cdot\partial%
%TCIMACRO{\TeXButton{h}{\slh}}%
%BeginExpansion
\slh
%EndExpansion
^{\clubsuit}$ and $\cdot\partial\cdot\partial%
%TCIMACRO{\TeXButton{h}{\slh}}%
%BeginExpansion
\slh
%EndExpansion
^{\clubsuit}$.

\section{Gravitation as Plastic Distortion of the Lorentz Vacuum}

Remark 4.6 clearly reveals the way that the dynamics of the plastic gauge
deformations $%
%TCIMACRO{\TeXButton{h}{\slh}}%
%BeginExpansion
\slh
%EndExpansion
^{\clubsuit}$ of the Lorentz vacuum must be described in a world that unless
experimental facts (and none exists until now, for the best of our knowledge)
demonstrate the contrary must be described by an event manifold $M\simeq
\mathbb{R}^{4}$. We elaborate this point as follows.

Let $(M\simeq\mathbb{R}^{4},%
%TCIMACRO{\TeXButton{eta}{\mbox{\boldmath{$\eta$}}}}%
%BeginExpansion
\mbox{\boldmath{$\eta$}}%
%EndExpansion
,D,\tau_{\eta},\uparrow)$ be the structure representing Minkowski spacetime as
defined in Section 1. In our theory we suppose that all physical fields and/or
particles live and interact in the arena defined by Minkowski spacetime. Now,
let $\{M\simeq\mathbb{R}^{4},%
%TCIMACRO{\TeXButton{g}{\slg}}%
%BeginExpansion
\slg
%EndExpansion
,D,\tau_{%
%TCIMACRO{\TeXButton{g}{\\slg}}%
%BeginExpansion
\\slg%
%EndExpansion
},\uparrow\}$ be a Lorentzian manifold that as we know represent a particular
gravitational field in Einstein's \textit{GRT. }Let moreover $\mathtt{\eta
}\mathbf{,}\mathtt{g}$ $\in\sec T_{0}^{2}M$ be the metric tensors on the
cotangent bundle such that in, the global chart with global coordinates
\footnote{Called coordinates in the Einstein-Lorentz-Poincar\'{e} gauge.}
$\{\mathtt{x}^{\alpha}\}$ associated to a chart \ $(M,\phi)_{o}$.where%
\begin{equation}%
%TCIMACRO{\TeXButton{eta}{\mbox{\boldmath{$\eta$}}}}%
%BeginExpansion
\mbox{\boldmath{$\eta$}}%
%EndExpansion
=\eta_{\alpha\beta}d\mathtt{x}^{\alpha}\otimes d\mathtt{x}^{\beta},\text{
\ \ }%
%TCIMACRO{\TeXButton{g}{\slg}}%
%BeginExpansion
\slg
%EndExpansion
=g_{\alpha\beta}d\mathtt{x}^{\alpha}\otimes d\mathtt{x}^{\beta}%
\end{equation}
we have%
\begin{equation}
\mathtt{\eta}=\eta^{\alpha\beta}\frac{\partial}{\partial\mathtt{x}^{\alpha}%
}\otimes\frac{\partial}{\partial\mathtt{x}^{\beta}}\text{, \ \ \ \ \ }%
\mathtt{g}=g^{\alpha\beta}\frac{\partial}{\partial\mathtt{x}^{\alpha}}%
\otimes\frac{\partial}{\partial\mathtt{x}^{\beta}}%
\end{equation}
with $\eta_{\alpha\beta}\eta^{\beta\nu}=\delta_{\alpha}^{\nu}$.and
$g_{\alpha\beta}g^{\beta\nu}=\delta_{\alpha}^{\nu}$ Let also $\eta$ be the
extensor field corresponding to $\mathtt{\eta},$ i.e.,
\begin{equation}
\mathtt{\eta}\mathbf{(}d\mathtt{x}^{\alpha},d\mathtt{x}^{\beta}\mathbf{)=}%
\eta(d\mathtt{x}^{\alpha})\cdot d\mathtt{x}^{\beta} \label{g2}%
\end{equation}
where in Eq.(\ref{g2})\ the symbol $\cdot$ denotes the canonical scalar
product on $M\simeq$ $\mathbb{R}^{4}$, i.e., $d\mathtt{x}^{\alpha}\cdot
d\mathtt{x}^{\beta}=\delta^{\alpha\beta}.$

Since $M\simeq\mathbb{R}^{4}$, utilizing the global coordinates $\{\mathtt{x}%
^{\alpha}\}$ we have immediately that $M\simeq U_{o}\simeq U$ (where $U_{o}$
and $U$ are as previously introduced) and the equivalence classes of the
$d\mathtt{x}^{\alpha}$ are\ $%
%TCIMACRO{\TeXButton{vt}{\mbox{\boldmath{$\vartheta$}}}}%
%BeginExpansion
\mbox{\boldmath{$\vartheta$}}%
%EndExpansion
^{\alpha}:=[d\mathtt{x}^{\alpha}]$, which $\{%
%TCIMACRO{\TeXButton{vt}{\mbox{\boldmath{$\vartheta$}}}}%
%BeginExpansion
\mbox{\boldmath{$\vartheta$}}%
%EndExpansion
^{\alpha}\}$ defining a basis\footnote{The reciprocal basis of\ $\{%
%TCIMACRO{\TeXButton{vt}{\mbox{\boldmath{$\vartheta$}}}}%
%BeginExpansion
\mbox{\boldmath{$\vartheta$}}%
%EndExpansion
^{\alpha}\}$ of \ $U$ is denoted by $\{\vartheta_{\mathbf{\alpha}}\}$. The
basis \ $\{%
%TCIMACRO{\TeXButton{vt}{\mbox{\boldmath{$\vartheta$}}}}%
%BeginExpansion
\mbox{\boldmath{$\vartheta$}}%
%EndExpansion
_{\mathbf{\alpha}}\}$ of $U$ is defined by the condition \ $%
%TCIMACRO{\TeXButton{vt}{\mbox{\boldmath{$\vartheta$}}}}%
%BeginExpansion
\mbox{\boldmath{$\vartheta$}}%
%EndExpansion
^{\alpha}\underset{\eta^{-1}}{\cdot}%
%TCIMACRO{\TeXButton{vt}{\mbox{\boldmath{$\vartheta$}}}}%
%BeginExpansion
\mbox{\boldmath{$\vartheta$}}%
%EndExpansion
_{\mathbf{\beta}}=%
%TCIMACRO{\TeXButton{vt}{\mbox{\boldmath{$\vartheta$}}}}%
%BeginExpansion
\mbox{\boldmath{$\vartheta$}}%
%EndExpansion
^{\alpha}\underset{\eta}{\cdot}%
%TCIMACRO{\TeXButton{vt}{\mbox{\boldmath{$\vartheta$}}}}%
%BeginExpansion
\mbox{\boldmath{$\vartheta$}}%
%EndExpansion
_{\mathbf{\beta}}=\delta_{\beta}^{\alpha}$. It may be called the $\eta
$-reciprocal basis of $\{%
%TCIMACRO{\TeXButton{vt}{\mbox{\boldmath{$\vartheta$}}}}%
%BeginExpansion
\mbox{\boldmath{$\vartheta$}}%
%EndExpansion
^{\alpha}\}$.} for $U$.

Its reciprocal basis will be denoted $\{\vartheta_{\mathbf{\alpha}}\}$In this
case given a smooth invertible extensor field $\mathbf{h:}\sec%
%TCIMACRO{\dbigwedge \nolimits^{1}}%
%BeginExpansion
{\displaystyle\bigwedge\nolimits^{1}}
%EndExpansion
TM\rightarrow\sec%
%TCIMACRO{\dbigwedge \nolimits^{1}}%
%BeginExpansion
{\displaystyle\bigwedge\nolimits^{1}}
%EndExpansion
TM$ its representative on $U$ is the smooth extensor field $%
%TCIMACRO{\TeXButton{h}{\slh}}%
%BeginExpansion
\slh
%EndExpansion
:\sec%
%TCIMACRO{\dbigwedge \nolimits^{1}}%
%BeginExpansion
{\displaystyle\bigwedge\nolimits^{1}}
%EndExpansion
T^{\ast}U\rightarrow\sec%
%TCIMACRO{\dbigwedge \nolimits^{1}}%
%BeginExpansion
{\displaystyle\bigwedge\nolimits^{1}}
%EndExpansion
T^{\ast}U$ . Recall moreover that the metric tensor $%
%TCIMACRO{\TeXButton{g}{\slg}}%
%BeginExpansion
\slg
%EndExpansion
$ is represented on $U$ by the extensor field $%
%TCIMACRO{\TeXButton{itg}{\itg}}%
%BeginExpansion
\itg
%EndExpansion
:\sec%
%TCIMACRO{\dbigwedge \nolimits^{1}}%
%BeginExpansion
{\displaystyle\bigwedge\nolimits^{1}}
%EndExpansion
T^{\ast}U\rightarrow\sec%
%TCIMACRO{\dbigwedge \nolimits^{1}}%
%BeginExpansion
{\displaystyle\bigwedge\nolimits^{1}}
%EndExpansion
T^{\ast}U$,%
\begin{equation}%
%TCIMACRO{\TeXButton{itg}{\itg}}%
%BeginExpansion
\itg
%EndExpansion
=%
%TCIMACRO{\TeXButton{h}{\slh}}%
%BeginExpansion
\slh
%EndExpansion
^{\dagger}\eta%
%TCIMACRO{\TeXButton{h}{\slh}}%
%BeginExpansion
\slh
%EndExpansion
, \label{def g_}%
\end{equation}
whereas the representative of \texttt{g }is%
\begin{equation}%
%TCIMACRO{\TeXButton{itg}{\itg}}%
%BeginExpansion
\itg
%EndExpansion
^{-1}=%
%TCIMACRO{\TeXButton{h}{\slh}}%
%BeginExpansion
\slh
%EndExpansion
^{-1}\eta%
%TCIMACRO{\TeXButton{h}{\slh}}%
%BeginExpansion
\slh
%EndExpansion
^{-1\dagger} \label{def g1}%
\end{equation}
with (for $\mathbf{x\in}U$ and $x\in M,$ $\mathbf{x}=\phi(x)$)%
\begin{equation}
\left.
%TCIMACRO{\TeXButton{itg}{\itg}}%
%BeginExpansion
\itg
%EndExpansion
^{-1}\right\vert _{\mathbf{x}}(%
%TCIMACRO{\TeXButton{vt}{\mbox{\boldmath{$\vartheta$}}}}%
%BeginExpansion
\mbox{\boldmath{$\vartheta$}}%
%EndExpansion
^{\alpha})\cdot%
%TCIMACRO{\TeXButton{vt}{\mbox{\boldmath{$\vartheta$}}}}%
%BeginExpansion
\mbox{\boldmath{$\vartheta$}}%
%EndExpansion
^{\beta}:=\left.  \mathtt{g}(d\mathtt{x}^{\alpha},d\mathtt{x}^{\beta
})\right\vert _{x},
\end{equation}
In what follows we present a gravitational theory where this field which
\textit{lives} in Minkowski spacetime is represented by smooth invertible
extensor field $%
%TCIMACRO{\TeXButton{h}{\slh}}%
%BeginExpansion
\slh
%EndExpansion
:\sec%
%TCIMACRO{\dbigwedge \nolimits^{1}}%
%BeginExpansion
{\displaystyle\bigwedge\nolimits^{1}}
%EndExpansion
T^{\ast}U\rightarrow\sec%
%TCIMACRO{\dbigwedge \nolimits^{1}}%
%BeginExpansion
{\displaystyle\bigwedge\nolimits^{1}}
%EndExpansion
T^{\ast}U$ describing (where it is \textit{not} the identity) a plastic
deformation of the Lorentz vacuum, putting it in a\ `excited' state which may
be described by Minkowski-Cartan \textit{MCGSS }$(M,\eta,\varkappa$) which as
we already know is gauge equivalent to the Lorentz \textit{MCGSS }$(M,%
%TCIMACRO{\TeXButton{itg}{\itg}}%
%BeginExpansion
\itg
%EndExpansion
,\gamma)$. We will give the Lagrangian encoding the dynamics of $%
%TCIMACRO{\TeXButton{h}{\slh}}%
%BeginExpansion
\slh
%EndExpansion
$ and its interaction with the matter fields and will derive its equation of motion.

\subsection{Lagrangian for the Free $%
%TCIMACRO{\TeXButton{h}{\slh}}%
%BeginExpansion
\slh
%EndExpansion
^{\clubsuit}$ Field}

Under the conditions established above we consider that the Lagrangian for the
gravitational field $%
%TCIMACRO{\TeXButton{h}{\slh}}%
%BeginExpansion
\slh
%EndExpansion
:\sec%
%TCIMACRO{\dbigwedge \nolimits^{1}}%
%BeginExpansion
{\displaystyle\bigwedge\nolimits^{1}}
%EndExpansion
T^{\ast}U\rightarrow\sec%
%TCIMACRO{\dbigwedge \nolimits^{1}}%
%BeginExpansion
{\displaystyle\bigwedge\nolimits^{1}}
%EndExpansion
T^{\ast}U$ is encoded in the following Lagrangian, which using notations
introduced in the previous section is written as
\begin{equation}
\lbrack%
%TCIMACRO{\TeXButton{h}{\slh}}%
%BeginExpansion
\slh
%EndExpansion
^{\clubsuit},\cdot\partial%
%TCIMACRO{\TeXButton{h}{\slh}}%
%BeginExpansion
\slh
%EndExpansion
^{\clubsuit},\cdot\partial\cdot\partial%
%TCIMACRO{\TeXButton{h}{\slh}}%
%BeginExpansion
\slh
%EndExpansion
^{\clubsuit}]\mapsto\mathfrak{L}_{EH}[%
%TCIMACRO{\TeXButton{h}{\slh}}%
%BeginExpansion
\slh
%EndExpansion
^{\clubsuit},\cdot\partial%
%TCIMACRO{\TeXButton{h}{\slh}}%
%BeginExpansion
\slh
%EndExpansion
^{\clubsuit},\cdot\partial\cdot\partial%
%TCIMACRO{\TeXButton{h}{\slh}}%
%BeginExpansion
\slh
%EndExpansion
^{\clubsuit}]:=\frac{1}{2}\mathcal{R}\det[%
%TCIMACRO{\TeXButton{h}{\slh}}%
%BeginExpansion
\slh
%EndExpansion
], \label{4.27}%
\end{equation}
where recalling Eq.(\ref{4.24}) and Eq.(\ref{4.25})) we see that
$\mathcal{R}\equiv\overset{\lambda%
%TCIMACRO{\TeXButton{sh}{\sslh}}%
%BeginExpansion
\sslh
%EndExpansion
}{\mathcal{R}}$ may indeed be expressed as a scalar functional field of
extensor variables $%
%TCIMACRO{\TeXButton{h}{\slh}}%
%BeginExpansion
\slh
%EndExpansion
^{\clubsuit},$ $\cdot\partial%
%TCIMACRO{\TeXButton{h}{\slh}}%
%BeginExpansion
\slh
%EndExpansion
^{\clubsuit}$ and $\cdot\partial\cdot\partial%
%TCIMACRO{\TeXButton{h}{\slh}}%
%BeginExpansion
\slh
%EndExpansion
^{\clubsuit}$, i.e.,
\begin{align}
\mathcal{R}  &  =\underline{%
%TCIMACRO{\TeXButton{h}{\slh}}%
%BeginExpansion
\slh
%EndExpansion
}^{\ast}(\partial_{a}\wedge\partial_{b})\cdot\mathcal{R}_{2}(a\wedge
b)\nonumber\\
&  =\underline{%
%TCIMACRO{\TeXButton{h}{\slh}}%
%BeginExpansion
\slh
%EndExpansion
}^{\ast}(\partial_{a}\wedge\partial_{b})\cdot(a\cdot\partial(\Omega
b)-b\cdot\partial\Omega(a)-\Omega([a,b])+\Omega(a)\underset{\eta}{\times
}\Omega(b)), \label{4.28}%
\end{align}
where the biform field $\Omega(a)$ given by Eq.(\ref{4.26}) is also expressed
as a functional for the extensor fields $%
%TCIMACRO{\TeXButton{h}{\slh}}%
%BeginExpansion
\slh
%EndExpansion
^{\clubsuit},$ $\cdot\partial%
%TCIMACRO{\TeXButton{h}{\slh}}%
%BeginExpansion
\slh
%EndExpansion
^{\clubsuit}$ and $\cdot\partial\cdot\partial%
%TCIMACRO{\TeXButton{h}{\slh}}%
%BeginExpansion
\slh
%EndExpansion
^{\clubsuit}$.

To continue we must express also $\det[%
%TCIMACRO{\TeXButton{h}{\slh}}%
%BeginExpansion
\slh
%EndExpansion
]$ as a scalar functional of $%
%TCIMACRO{\TeXButton{h}{\slh}}%
%BeginExpansion
\slh
%EndExpansion
^{\clubsuit}$, i.e.,
\begin{equation}
\det[%
%TCIMACRO{\TeXButton{h}{\slh}}%
%BeginExpansion
\slh
%EndExpansion
]=\frac{1}{\det[%
%TCIMACRO{\TeXButton{h}{\slh}}%
%BeginExpansion
\slh
%EndExpansion
^{\clubsuit}]}. \label{4.29}%
\end{equation}

The action functional for the field $%
%TCIMACRO{\TeXButton{h}{\slh}}%
%BeginExpansion
\slh
%EndExpansion
^{\clubsuit}$ on $U$ is
\begin{align}
\mathcal{A}  &  =\int_{U}\mathfrak{L}_{eh}[%
%TCIMACRO{\TeXButton{h}{\slh}}%
%BeginExpansion
\slh
%EndExpansion
^{\clubsuit},\cdot\partial%
%TCIMACRO{\TeXButton{h}{\slh}}%
%BeginExpansion
\slh
%EndExpansion
^{\clubsuit},\cdot\partial\cdot\partial%
%TCIMACRO{\TeXButton{h}{\slh}}%
%BeginExpansion
\slh
%EndExpansion
^{\clubsuit}]\text{ }\tau\nonumber\\
&  =\frac{1}{2}\int_{U}\mathcal{R}\det[%
%TCIMACRO{\TeXButton{h}{\slh}}%
%BeginExpansion
\slh
%EndExpansion
]\text{ }\tau, \label{4.30}%
\end{align}
where $\tau$ is an arbitrary euclidean volume element, e.g., we can take
$\tau=d\mathtt{x}^{0}\wedge d\mathtt{x}^{1}\wedge d\mathtt{x}^{2}\wedge
d\mathtt{x}^{3}$

Let $%
%TCIMACRO{\TeXButton{delta}{\mbox{\boldmath{$\delta$}}}}%
%BeginExpansion
\mbox{\boldmath{$\delta$}}%
%EndExpansion
_{%
%TCIMACRO{\TeXButton{sh}{\sslh}}%
%BeginExpansion
\sslh
%EndExpansion
^{\clubsuit}}^{w}$ be the variational operator with respect to $%
%TCIMACRO{\TeXButton{h}{\slh}}%
%BeginExpansion
\slh
%EndExpansion
^{\clubsuit}$\ in the direction of the smooth $(1,1)$-extensor field $w,$ such
that $\left.  w\right\vert _{\partial U}=0$ and $\left.  \cdot\partial
w\right\vert _{\partial U}=0$.

\subsubsection{Equations of Motion for $%
%TCIMACRO{\TeXButton{h}{\slh}}%
%BeginExpansion
\slh
%EndExpansion
^{\clubsuit}$}

As usual in the Lagrangian formalism we suppose that the dynamics of $%
%TCIMACRO{\TeXButton{h}{\slh}}%
%BeginExpansion
\slh
%EndExpansion
^{\clubsuit}$ is given by the principle of stationary action,
\begin{equation}
\int_{U}%
%TCIMACRO{\TeXButton{delta}{\mbox{\boldmath{$\delta$}}}}%
%BeginExpansion
\mbox{\boldmath{$\delta$}}%
%EndExpansion
_{%
%TCIMACRO{\TeXButton{sh}{\sslh}}%
%BeginExpansion
\sslh
%EndExpansion
^{\clubsuit}}^{w}\mathcal{L}_{eh}[%
%TCIMACRO{\TeXButton{h}{\slh}}%
%BeginExpansion
\slh
%EndExpansion
^{\clubsuit},\cdot\partial%
%TCIMACRO{\TeXButton{h}{\slh}}%
%BeginExpansion
\slh
%EndExpansion
^{\clubsuit},\cdot\partial\cdot\partial%
%TCIMACRO{\TeXButton{h}{\slh}}%
%BeginExpansion
\slh
%EndExpansion
^{\clubsuit}]\tau=\int_{U}%
%TCIMACRO{\TeXButton{delta}{\mbox{\boldmath{$\delta$}}}}%
%BeginExpansion
\mbox{\boldmath{$\delta$}}%
%EndExpansion
_{%
%TCIMACRO{\TeXButton{sh}{\sslh}}%
%BeginExpansion
\sslh
%EndExpansion
^{\clubsuit}}^{w}(\mathcal{R}\det[%
%TCIMACRO{\TeXButton{h}{\slh}}%
%BeginExpansion
\slh
%EndExpansion
])\text{ }\tau=0, \label{4.31a}%
\end{equation}
which implies in a functional differential equation (Euler-Lagrange equation)
for $%
%TCIMACRO{\TeXButton{h}{\slh}}%
%BeginExpansion
\slh
%EndExpansion
^{\clubsuit}$ that we now derive.

As is well known the solution of that problem implies in the use
of\ variational formulas, the Gauss-Stokes theorem (for star shape regions),
i.e.,
\begin{equation}
\int_{U}\partial\cdot a\text{ }\tau=\oint_{\partial U}%
%TCIMACRO{\TeXButton{vt}{\mbox{\boldmath{$\vartheta$}}}}%
%BeginExpansion
\mbox{\boldmath{$\vartheta$}}%
%EndExpansion
^{\mu}\cdot a\text{ }\tau_{\mu}, \label{4.43}%
\end{equation}
where $\tau_{\mu}$ are $3$-form fields on the boundary $\partial U$ of $U$,
$a$ is a smooth 1-form field and,a fundamental lemma of integration theory,
which says that if $\int_{U}A\cdot X$ $\tau=0$ for all multiform field $A$
then the $X=0$.

Before starting the calculations we recall from Section 3.8 that any
variational operator $%
%TCIMACRO{\TeXButton{delta}{\mbox{\boldmath{$\delta$}}}}%
%BeginExpansion
\mbox{\boldmath{$\delta$}}%
%EndExpansion
_{t}^{w}$ satisfies the following rules. Let $\Phi,\Psi$ \ be arbitrary
multiform functionals of the extensor field $t$. Moreover, let $\digamma$be a
scalar functional of the extensor field $t$ with $a$ a smooth 1-form field and
$\varphi$ an arbitrary function. Then:
\begin{subequations}
\begin{align}%
%TCIMACRO{\TeXButton{delta}{\mbox{\boldmath{$\delta$}}}}%
%BeginExpansion
\mbox{\boldmath{$\delta$}}%
%EndExpansion
_{t}^{w}(\Phi\lbrack t]\ast\Psi\lbrack t])  &  =(%
%TCIMACRO{\TeXButton{delta}{\mbox{\boldmath{$\delta$}}}}%
%BeginExpansion
\mbox{\boldmath{$\delta$}}%
%EndExpansion
_{t}^{w}\Phi\lbrack t])\ast\Psi\lbrack t]+\Phi\lbrack t]\ast(%
%TCIMACRO{\TeXButton{delta}{\mbox{\boldmath{$\delta$}}}}%
%BeginExpansion
\mbox{\boldmath{$\delta$}}%
%EndExpansion
_{t}^{w}\Psi\lbrack t]),\label{LRa}\\%
%TCIMACRO{\TeXButton{delta}{\mbox{\boldmath{$\delta$}}}}%
%BeginExpansion
\mbox{\boldmath{$\delta$}}%
%EndExpansion
_{t}^{w}t(a)  &  =w(a),\label{LRb}\\%
%TCIMACRO{\TeXButton{delta}{\mbox{\boldmath{$\delta$}}}}%
%BeginExpansion
\mbox{\boldmath{$\delta$}}%
%EndExpansion
_{t}^{w}(a\cdot\partial\Phi\lbrack t])  &  =a\cdot\partial(%
%TCIMACRO{\TeXButton{delta}{\mbox{\boldmath{$\delta$}}}}%
%BeginExpansion
\mbox{\boldmath{$\delta$}}%
%EndExpansion
_{t}^{w}\Phi\lbrack t]),\label{LRc}\\%
%TCIMACRO{\TeXButton{delta}{\mbox{\boldmath{$\delta$}}}}%
%BeginExpansion
\mbox{\boldmath{$\delta$}}%
%EndExpansion
_{t}^{w}\varphi(\digamma\lbrack t])  &  =\varphi^{\prime}(\digamma\lbrack t])(%
%TCIMACRO{\TeXButton{delta}{\mbox{\boldmath{$\delta$}}}}%
%BeginExpansion
\mbox{\boldmath{$\delta$}}%
%EndExpansion
_{t}^{w}\digamma\lbrack t]), \label{LRd}%
\end{align}
where in Eq.(\ref{LRa}) $\ast$ denotes here (as previously agreed) any one of
the multiform products.

Utilizing Eq.(\ref{LRa}) in Eq.(\ref{4.27}), and taking into account
Eq.(\ref{4.28}), we get
\end{subequations}
\begin{align}%
%TCIMACRO{\TeXButton{delta}{\mbox{\boldmath{$\delta$}}}}%
%BeginExpansion
\mbox{\boldmath{$\delta$}}%
%EndExpansion
_{%
%TCIMACRO{\TeXButton{sh}{\sslh}}%
%BeginExpansion
\sslh
%EndExpansion
^{\clubsuit}}^{w}(\underline{%
%TCIMACRO{\TeXButton{h}{\slh}}%
%BeginExpansion
\slh
%EndExpansion
}^{\clubsuit}(\partial_{a}\wedge\partial_{b})\cdot\mathcal{R}_{2}(a\wedge
b)\det[%
%TCIMACRO{\TeXButton{h}{\slh}}%
%BeginExpansion
\slh
%EndExpansion
])  &  =(%
%TCIMACRO{\TeXButton{delta}{\mbox{\boldmath{$\delta$}}}}%
%BeginExpansion
\mbox{\boldmath{$\delta$}}%
%EndExpansion
_{%
%TCIMACRO{\TeXButton{sh}{\sslh}}%
%BeginExpansion
\sslh
%EndExpansion
^{\clubsuit}}^{w}\underline{%
%TCIMACRO{\TeXButton{h}{\slh}}%
%BeginExpansion
\slh
%EndExpansion
}^{\clubsuit}(\partial_{a}\wedge\partial_{b}))\cdot\mathcal{R}_{2}(a\wedge
b)\det[%
%TCIMACRO{\TeXButton{h}{\slh}}%
%BeginExpansion
\slh
%EndExpansion
]\nonumber\\
&  +\underline{%
%TCIMACRO{\TeXButton{h}{\slh}}%
%BeginExpansion
\slh
%EndExpansion
}^{\clubsuit}(\partial_{a}\wedge\partial_{b})\cdot(%
%TCIMACRO{\TeXButton{delta}{\mbox{\boldmath{$\delta$}}}}%
%BeginExpansion
\mbox{\boldmath{$\delta$}}%
%EndExpansion
_{%
%TCIMACRO{\TeXButton{h}{\slh}}%
%BeginExpansion
\slh
%EndExpansion
^{\clubsuit}}^{w}\mathcal{R}_{2}(a\wedge b))\det[%
%TCIMACRO{\TeXButton{h}{\slh}}%
%BeginExpansion
\slh
%EndExpansion
]\nonumber\\
&  +\underline{%
%TCIMACRO{\TeXButton{h}{\slh}}%
%BeginExpansion
\slh
%EndExpansion
}^{\clubsuit}(\partial_{a}\wedge\partial_{b})\cdot\mathcal{R}_{2}(a\wedge b)(%
%TCIMACRO{\TeXButton{delta}{\mbox{\boldmath{$\delta$}}}}%
%BeginExpansion
\mbox{\boldmath{$\delta$}}%
%EndExpansion
_{%
%TCIMACRO{\TeXButton{sh}{\sslh}}%
%BeginExpansion
\sslh
%EndExpansion
^{\clubsuit}}^{w}\det[%
%TCIMACRO{\TeXButton{h}{\slh}}%
%BeginExpansion
\slh
%EndExpansion
]). \label{4.43i}%
\end{align}

We next obtain a functional identity for the scalar product in the first term
of Eq.(\ref{4.43i}). We have,
\begin{align*}
&  (%
%TCIMACRO{\TeXButton{delta}{\mbox{\boldmath{$\delta$}}}}%
%BeginExpansion
\mbox{\boldmath{$\delta$}}%
%EndExpansion
_{%
%TCIMACRO{\TeXButton{sh}{\sslh}}%
%BeginExpansion
\sslh
%EndExpansion
^{\clubsuit}}^{w}\underline{%
%TCIMACRO{\TeXButton{h}{\slh}}%
%BeginExpansion
\slh
%EndExpansion
}^{\clubsuit}(\partial_{a}\wedge\partial_{b}))\cdot\mathcal{R}_{2}(a\wedge
b)\\
&  =(%
%TCIMACRO{\TeXButton{delta}{\mbox{\boldmath{$\delta$}}}}%
%BeginExpansion
\mbox{\boldmath{$\delta$}}%
%EndExpansion
_{%
%TCIMACRO{\TeXButton{h}{\slh}}%
%BeginExpansion
\slh
%EndExpansion
^{\clubsuit}}^{w}%
%TCIMACRO{\TeXButton{h}{\slh}}%
%BeginExpansion
\slh
%EndExpansion
^{\clubsuit}(\partial_{a})\wedge%
%TCIMACRO{\TeXButton{h}{\slh}}%
%BeginExpansion
\slh
%EndExpansion
^{\clubsuit}(\partial_{b})+%
%TCIMACRO{\TeXButton{h}{\slh}}%
%BeginExpansion
\slh
%EndExpansion
^{\clubsuit}(\partial_{a})\wedge%
%TCIMACRO{\TeXButton{delta}{\mbox{\boldmath{$\delta$}}}}%
%BeginExpansion
\mbox{\boldmath{$\delta$}}%
%EndExpansion
_{%
%TCIMACRO{\TeXButton{sh}{\sslh}}%
%BeginExpansion
\sslh
%EndExpansion
^{\clubsuit}}^{w}%
%TCIMACRO{\TeXButton{h}{\slh}}%
%BeginExpansion
\slh
%EndExpansion
^{\clubsuit}(\partial_{b}))\cdot\mathcal{R}_{2}(a\wedge b)\\
&  =(w(\partial_{a})\wedge%
%TCIMACRO{\TeXButton{h}{\slh}}%
%BeginExpansion
\slh
%EndExpansion
^{\clubsuit}(\partial_{b})+%
%TCIMACRO{\TeXButton{h}{\slh}}%
%BeginExpansion
\slh
%EndExpansion
^{\clubsuit}(\partial_{a})\wedge w(\partial_{b}))\cdot\mathcal{R}_{2}(a\wedge
b)\\
&  =(w(\partial_{a})\wedge%
%TCIMACRO{\TeXButton{h}{\slh}}%
%BeginExpansion
\slh
%EndExpansion
^{\clubsuit}(\partial_{b}))\cdot\mathcal{R}_{2}(a\wedge b)+(%
%TCIMACRO{\TeXButton{h}{\slh}}%
%BeginExpansion
\slh
%EndExpansion
^{\clubsuit}(\partial_{a})\wedge w(\partial_{b}))\cdot\mathcal{R}_{2}(a\wedge
b)\\
&  =(%
%TCIMACRO{\TeXButton{h}{\slh}}%
%BeginExpansion
\slh
%EndExpansion
^{\clubsuit}(\partial_{b})\wedge w(\partial_{a}))\cdot\mathcal{R}_{2}(b\wedge
a)+(%
%TCIMACRO{\TeXButton{h}{\slh}}%
%BeginExpansion
\slh
%EndExpansion
^{\clubsuit}(\partial_{b})\wedge w(\partial_{a}))\cdot\mathcal{R}_{2}(b\wedge
a)\\
&  =2(%
%TCIMACRO{\TeXButton{h}{\slh}}%
%BeginExpansion
\slh
%EndExpansion
^{\clubsuit}(\partial_{b})\wedge w(\partial_{a}))\cdot\mathcal{R}_{2}(b\wedge
a)\\
&  =2w(\partial_{a})\cdot(%
%TCIMACRO{\TeXButton{h}{\slh}}%
%BeginExpansion
\slh
%EndExpansion
^{\clubsuit}(\partial_{b})\lrcorner\mathcal{R}_{2}(b\wedge a)),
\end{align*}
i.e.,%
\begin{equation}
(%
%TCIMACRO{\TeXButton{delta}{\mbox{\boldmath{$\delta$}}}}%
%BeginExpansion
\mbox{\boldmath{$\delta$}}%
%EndExpansion
_{%
%TCIMACRO{\TeXButton{sh}{\sslh}}%
%BeginExpansion
\sslh
%EndExpansion
^{\clubsuit}}^{w}\underline{%
%TCIMACRO{\TeXButton{h}{\slh}}%
%BeginExpansion
\slh
%EndExpansion
}^{\clubsuit}(\partial_{a}\wedge\partial_{b}))\cdot\mathcal{R}_{2}(a\wedge
b)=2w(\partial_{a})\cdot\mathcal{R}_{1}(a), \label{4.43ii}%
\end{equation}
where we utilized Eq.(\ref{LRa}) in the form $%
%TCIMACRO{\TeXButton{delta}{\mbox{\boldmath{$\delta$}}}}%
%BeginExpansion
\mbox{\boldmath{$\delta$}}%
%EndExpansion
_{t}^{w}(\mathbf{\Phi}[t]\wedge\mathbf{\Psi}[t])=(%
%TCIMACRO{\TeXButton{delta}{\mbox{\boldmath{$\delta$}}}}%
%BeginExpansion
\mbox{\boldmath{$\delta$}}%
%EndExpansion
_{t}^{w}\mathbf{\Phi}[t])\wedge\mathbf{\Psi}[t]+\mathbf{\Phi}[t])\wedge(%
%TCIMACRO{\TeXButton{delta}{\mbox{\boldmath{$\delta$}}}}%
%BeginExpansion
\mbox{\boldmath{$\delta$}}%
%EndExpansion
_{t}^{w}\mathcal{\Psi}[t])$, Eq.(\ref{LRb}) and Eq.(\ref{4.17}) which defines
the Ricci gauge field.

We next obtain a second functional identity for the scalar product in the
second term\footnote{Keep in mind that that $\{\vartheta_{\mu}\}$ is the
reciprocal basis of $\{%
%TCIMACRO{\TeXButton{vt}{\mbox{\boldmath{$\vartheta$}}}}%
%BeginExpansion
\mbox{\boldmath{$\vartheta$}}%
%EndExpansion
^{\mu}\}$, i.e., $%
%TCIMACRO{\TeXButton{vt}{\mbox{\boldmath{$\vartheta$}}}}%
%BeginExpansion
\mbox{\boldmath{$\vartheta$}}%
%EndExpansion
^{\mu}\cdot\vartheta_{\nu}=\delta_{\nu}^{\mu}$.} of Eq.(\ref{4.43i}),%
\begin{align}
\underline{%
%TCIMACRO{\TeXButton{h}{\slh}}%
%BeginExpansion
\slh
%EndExpansion
}^{\clubsuit}(\partial_{a}\wedge\partial_{b})\cdot%
%TCIMACRO{\TeXButton{delta}{\mbox{\boldmath{$\delta$}}}}%
%BeginExpansion
\mbox{\boldmath{$\delta$}}%
%EndExpansion
_{%
%TCIMACRO{\TeXButton{sh}{\sslh}}%
%BeginExpansion
\sslh
%EndExpansion
^{\clubsuit}}^{w}\mathcal{R}_{2}(a\wedge b)  &  =\underline{%
%TCIMACRO{\TeXButton{h}{\slh}}%
%BeginExpansion
\slh
%EndExpansion
}^{\clubsuit}(%
%TCIMACRO{\TeXButton{vt}{\mbox{\boldmath{$\vartheta$}}}}%
%BeginExpansion
\mbox{\boldmath{$\vartheta$}}%
%EndExpansion
^{\mu}\wedge%
%TCIMACRO{\TeXButton{vt}{\mbox{\boldmath{$\vartheta$}}}}%
%BeginExpansion
\mbox{\boldmath{$\vartheta$}}%
%EndExpansion
^{\nu})\cdot%
%TCIMACRO{\TeXButton{delta}{\mbox{\boldmath{$\delta$}}}}%
%BeginExpansion
\mbox{\boldmath{$\delta$}}%
%EndExpansion
_{%
%TCIMACRO{\TeXButton{sh}{\sslh}}%
%BeginExpansion
\sslh
%EndExpansion
^{\clubsuit}}^{w}\mathcal{R}_{2}(\vartheta_{\mu}\wedge\vartheta_{\nu
})\nonumber\\
&  =\underline{%
%TCIMACRO{\TeXButton{h}{\slh}}%
%BeginExpansion
\slh
%EndExpansion
}^{\clubsuit}(%
%TCIMACRO{\TeXButton{vt}{\mbox{\boldmath{$\vartheta$}}}}%
%BeginExpansion
\mbox{\boldmath{$\vartheta$}}%
%EndExpansion
^{\mu}\wedge%
%TCIMACRO{\TeXButton{vt}{\mbox{\boldmath{$\vartheta$}}}}%
%BeginExpansion
\mbox{\boldmath{$\vartheta$}}%
%EndExpansion
^{\nu})\cdot%
%TCIMACRO{\TeXButton{delta}{\mbox{\boldmath{$\delta$}}}}%
%BeginExpansion
\mbox{\boldmath{$\delta$}}%
%EndExpansion
_{%
%TCIMACRO{\TeXButton{sh}{\sslh}}%
%BeginExpansion
\sslh
%EndExpansion
^{\clubsuit}}^{w}(\vartheta_{\mu}\cdot\partial\Omega(\vartheta_{\nu
})-\vartheta_{\nu}\cdot\partial\Omega(\vartheta_{\mu})\nonumber\\
&  +\Omega(\vartheta_{\mu})\underset{\eta}{\times}\Omega(\vartheta_{\nu
}))\nonumber\\
&  =\underline{%
%TCIMACRO{\TeXButton{h}{\slh}}%
%BeginExpansion
\slh
%EndExpansion
}^{\clubsuit}(%
%TCIMACRO{\TeXButton{vt}{\mbox{\boldmath{$\vartheta$}}}}%
%BeginExpansion
\mbox{\boldmath{$\vartheta$}}%
%EndExpansion
^{\mu}\wedge%
%TCIMACRO{\TeXButton{vt}{\mbox{\boldmath{$\vartheta$}}}}%
%BeginExpansion
\mbox{\boldmath{$\vartheta$}}%
%EndExpansion
^{\nu})\cdot(\vartheta_{\mu}\cdot\partial%
%TCIMACRO{\TeXButton{delta}{\mbox{\boldmath{$\delta$}}}}%
%BeginExpansion
\mbox{\boldmath{$\delta$}}%
%EndExpansion
_{%
%TCIMACRO{\TeXButton{sh}{\sslh}}%
%BeginExpansion
\sslh
%EndExpansion
^{\clubsuit}}^{w}\Omega(\vartheta_{\nu})-\vartheta_{\nu}\cdot\partial%
%TCIMACRO{\TeXButton{delta}{\mbox{\boldmath{$\delta$}}}}%
%BeginExpansion
\mbox{\boldmath{$\delta$}}%
%EndExpansion
_{%
%TCIMACRO{\TeXButton{sh}{\sslh}}%
%BeginExpansion
\sslh
%EndExpansion
^{\clubsuit}}^{w}\Omega(\vartheta_{\mu})\nonumber\\
&  +\delta_{%
%TCIMACRO{\TeXButton{h}{\slh}}%
%BeginExpansion
\slh
%EndExpansion
^{\clubsuit}}^{w}\Omega(\vartheta_{\mu})\underset{\eta}{\times}\Omega
(\vartheta_{\nu})+\Omega(\vartheta_{\mu})\underset{\eta}{\times}%
%TCIMACRO{\TeXButton{delta}{\mbox{\boldmath{$\delta$}}}}%
%BeginExpansion
\mbox{\boldmath{$\delta$}}%
%EndExpansion
_{%
%TCIMACRO{\TeXButton{h}{\slh}}%
%BeginExpansion
\slh
%EndExpansion
^{\clubsuit}}^{w}\Omega(\vartheta_{\nu}))\nonumber\\
&  =\underline{%
%TCIMACRO{\TeXButton{h}{\slh}}%
%BeginExpansion
\slh
%EndExpansion
}^{\clubsuit}(%
%TCIMACRO{\TeXButton{vt}{\mbox{\boldmath{$\vartheta$}}}}%
%BeginExpansion
\mbox{\boldmath{$\vartheta$}}%
%EndExpansion
^{\mu}\wedge%
%TCIMACRO{\TeXButton{vt}{\mbox{\boldmath{$\vartheta$}}}}%
%BeginExpansion
\mbox{\boldmath{$\vartheta$}}%
%EndExpansion
^{\nu})\cdot(\mathcal{D}_{\vartheta_{\mu}}%
%TCIMACRO{\TeXButton{delta}{\mbox{\boldmath{$\delta$}}}}%
%BeginExpansion
\mbox{\boldmath{$\delta$}}%
%EndExpansion
_{%
%TCIMACRO{\TeXButton{h}{\slh}}%
%BeginExpansion
\slh
%EndExpansion
^{\clubsuit}}^{w}\Omega(\vartheta_{\nu})-\mathcal{D}_{\vartheta_{\nu}}%
%TCIMACRO{\TeXButton{delta}{\mbox{\boldmath{$\delta$}}}}%
%BeginExpansion
\mbox{\boldmath{$\delta$}}%
%EndExpansion
_{%
%TCIMACRO{\TeXButton{h}{\slh}}%
%BeginExpansion
\slh
%EndExpansion
^{\clubsuit}}^{w}\Omega(\vartheta_{\mu}))\nonumber\\
&  =2\underline{%
%TCIMACRO{\TeXButton{h}{\slh}}%
%BeginExpansion
\slh
%EndExpansion
}^{\clubsuit}(%
%TCIMACRO{\TeXButton{vt}{\mbox{\boldmath{$\vartheta$}}}}%
%BeginExpansion
\mbox{\boldmath{$\vartheta$}}%
%EndExpansion
^{\mu}\wedge%
%TCIMACRO{\TeXButton{vt}{\mbox{\boldmath{$\vartheta$}}}}%
%BeginExpansion
\mbox{\boldmath{$\vartheta$}}%
%EndExpansion
^{\nu})\cdot\mathcal{D}_{\vartheta_{\mu}}\delta_{%
%TCIMACRO{\TeXButton{sh}{\sslh}}%
%BeginExpansion
\sslh
%EndExpansion
^{\clubsuit}}^{w}\Omega(\vartheta_{\nu})\nonumber\\
&  =2%
%TCIMACRO{\TeXButton{h}{\slh}}%
%BeginExpansion
\slh
%EndExpansion
^{\clubsuit}(%
%TCIMACRO{\TeXButton{vt}{\mbox{\boldmath{$\vartheta$}}}}%
%BeginExpansion
\mbox{\boldmath{$\vartheta$}}%
%EndExpansion
^{\nu})\cdot(%
%TCIMACRO{\TeXButton{h}{\slh}}%
%BeginExpansion
\slh
%EndExpansion
^{\clubsuit}(%
%TCIMACRO{\TeXButton{vt}{\mbox{\boldmath{$\vartheta$}}}}%
%BeginExpansion
\mbox{\boldmath{$\vartheta$}}%
%EndExpansion
^{\mu})\lrcorner\mathcal{D}_{\vartheta_{\mu}}\delta_{%
%TCIMACRO{\TeXButton{sh}{\sslh}}%
%BeginExpansion
\sslh
%EndExpansion
^{\clubsuit}}^{w}\Omega(\vartheta_{\nu}))\nonumber\\
\underline{%
%TCIMACRO{\TeXButton{h}{\slh}}%
%BeginExpansion
\slh
%EndExpansion
}^{\clubsuit}(\partial_{a}\wedge\partial_{b})\cdot%
%TCIMACRO{\TeXButton{delta}{\mbox{\boldmath{$\delta$}}}}%
%BeginExpansion
\mbox{\boldmath{$\delta$}}%
%EndExpansion
_{%
%TCIMACRO{\TeXButton{sh}{\sslh}}%
%BeginExpansion
\sslh
%EndExpansion
^{\clubsuit}}^{w}\mathcal{R}_{2}(a\wedge b)  &  =2%
%TCIMACRO{\TeXButton{h}{\slh}}%
%BeginExpansion
\slh
%EndExpansion
^{\clubsuit}(\vartheta^{\nu})\cdot%
%TCIMACRO{\TeXButton{D}{\slD}}%
%BeginExpansion
\slD
%EndExpansion
\lrcorner%
%TCIMACRO{\TeXButton{delta}{\mbox{\boldmath{$\delta$}}}}%
%BeginExpansion
\mbox{\boldmath{$\delta$}}%
%EndExpansion
_{%
%TCIMACRO{\TeXButton{sh}{\sslh}}%
%BeginExpansion
\sslh
%EndExpansion
^{\clubsuit}}^{w}\Omega(\vartheta_{\nu}). \label{4.43iii}%
\end{align}
On those equations we utilized essentially Eq.(\ref{LRc}) once more the
Eq.(\ref{LRa}), this time in the form, $%
%TCIMACRO{\TeXButton{delta}{\mbox{\boldmath{$\delta$}}}}%
%BeginExpansion
\mbox{\boldmath{$\delta$}}%
%EndExpansion
_{t}^{w}(\mathbf{\Phi}[t]\underset{\eta}{\times}\mathbf{\Psi}[t])=(%
%TCIMACRO{\TeXButton{delta}{\mbox{\boldmath{$\delta$}}}}%
%BeginExpansion
\mbox{\boldmath{$\delta$}}%
%EndExpansion
_{t}^{w}\mathbf{\Phi}[t])\underset{\eta}{\times}\mathbf{\Psi}[t]+\mathbf{\Phi
}[t])\underset{\eta}{\times}(%
%TCIMACRO{\TeXButton{delta}{\mbox{\boldmath{$\delta$}}}}%
%BeginExpansion
\mbox{\boldmath{$\delta$}}%
%EndExpansion
_{t}^{w}\mathbf{\Psi}[t])$, and the definition of the gauge divergent
$\overset{}{%
%TCIMACRO{\TeXButton{D}{\slD}}%
%BeginExpansion
\slD
%EndExpansion
}\lrcorner X=%
%TCIMACRO{\TeXButton{h}{\slh}}%
%BeginExpansion
\slh
%EndExpansion
^{\clubsuit}(\partial_{a})\lrcorner\overset{}{\mathcal{D}_{a}}X$, where
$\overset{}{\mathcal{D}_{a}}$ is the covariant derivative operator, which
applied on a smooth multiform field $X$ gives
\begin{equation}
\mathcal{D}_{a}X:=a\cdot\partial X+\Omega(a)\underset{\eta}{\times}X.
\end{equation}

We may also write the right hand side of Eq.(\ref{4.43}) which is a scalar
product as the scalar divergent of a smooth $1$-form field, utilizing the
following identity (which is easily obtained if we take into account
Eq.(\ref{3.50}),
\begin{equation}%
%TCIMACRO{\TeXButton{D}{\slD}}%
%BeginExpansion
\slD
%EndExpansion
\lrcorner X=\dfrac{1}{\det[%
%TCIMACRO{\TeXButton{h}{\slh}}%
%BeginExpansion
\slh
%EndExpansion
]}\underline{%
%TCIMACRO{\TeXButton{h}{\slh}}%
%BeginExpansion
\slh
%EndExpansion
}(\partial\lrcorner\det[%
%TCIMACRO{\TeXButton{h}{\slh}}%
%BeginExpansion
\slh
%EndExpansion
]\underline{%
%TCIMACRO{\TeXButton{h}{\slh}}%
%BeginExpansion
\slh
%EndExpansion
}^{-1}(X)).
\end{equation}
We then have
\begin{align*}%
%TCIMACRO{\TeXButton{h}{\slh}}%
%BeginExpansion
\slh
%EndExpansion
^{\clubsuit}(%
%TCIMACRO{\TeXButton{vt}{\mbox{\boldmath{$\vartheta$}}}}%
%BeginExpansion
\mbox{\boldmath{$\vartheta$}}%
%EndExpansion
^{\nu})\cdot%
%TCIMACRO{\TeXButton{D}{\slD}}%
%BeginExpansion
\slD
%EndExpansion
\lrcorner%
%TCIMACRO{\TeXButton{delta}{\mbox{\boldmath{$\delta$}}}}%
%BeginExpansion
\mbox{\boldmath{$\delta$}}%
%EndExpansion
_{%
%TCIMACRO{\TeXButton{sh}{\sslh}}%
%BeginExpansion
\sslh
%EndExpansion
^{\clubsuit}}^{w}\Omega(\vartheta_{\nu})  &  =%
%TCIMACRO{\TeXButton{h}{\slh}}%
%BeginExpansion
\slh
%EndExpansion
^{\clubsuit}(%
%TCIMACRO{\TeXButton{vt}{\mbox{\boldmath{$\vartheta$}}}}%
%BeginExpansion
\mbox{\boldmath{$\vartheta$}}%
%EndExpansion
^{\nu})\cdot\dfrac{1}{\det[%
%TCIMACRO{\TeXButton{h}{\slh}}%
%BeginExpansion
\slh
%EndExpansion
]}%
%TCIMACRO{\TeXButton{h}{\slh}}%
%BeginExpansion
\slh
%EndExpansion
(\partial\lrcorner\det[%
%TCIMACRO{\TeXButton{h}{\slh}}%
%BeginExpansion
\slh
%EndExpansion
]\underline{%
%TCIMACRO{\TeXButton{h}{\slh}}%
%BeginExpansion
\slh
%EndExpansion
}^{-1}(%
%TCIMACRO{\TeXButton{delta}{\mbox{\boldmath{$\delta$}}}}%
%BeginExpansion
\mbox{\boldmath{$\delta$}}%
%EndExpansion
_{%
%TCIMACRO{\TeXButton{sh}{\sslh}}%
%BeginExpansion
\sslh
%EndExpansion
^{\clubsuit}}^{w}\Omega(\vartheta_{\nu})))\\
&  =\dfrac{1}{\det[%
%TCIMACRO{\TeXButton{h}{\slh}}%
%BeginExpansion
\slh
%EndExpansion
]}%
%TCIMACRO{\TeXButton{vt}{\mbox{\boldmath{$\vartheta$}}}}%
%BeginExpansion
\mbox{\boldmath{$\vartheta$}}%
%EndExpansion
^{\nu}\cdot\partial\lrcorner(\det[%
%TCIMACRO{\TeXButton{h}{\slh}}%
%BeginExpansion
\slh
%EndExpansion
]\underline{%
%TCIMACRO{\TeXButton{h}{\slh}}%
%BeginExpansion
\slh
%EndExpansion
}^{-1}(%
%TCIMACRO{\TeXButton{delta}{\mbox{\boldmath{$\delta$}}}}%
%BeginExpansion
\mbox{\boldmath{$\delta$}}%
%EndExpansion
_{%
%TCIMACRO{\TeXButton{sh}{\sslh}}%
%BeginExpansion
\sslh
%EndExpansion
^{\clubsuit}}^{w}\Omega(\vartheta_{\nu}))\\
&  =\dfrac{1}{\det[%
%TCIMACRO{\TeXButton{h}{\slh}}%
%BeginExpansion
\slh
%EndExpansion
]}%
%TCIMACRO{\TeXButton{vt}{\mbox{\boldmath{$\vartheta$}}}}%
%BeginExpansion
\mbox{\boldmath{$\vartheta$}}%
%EndExpansion
^{\nu}\cdot(%
%TCIMACRO{\TeXButton{vt}{\mbox{\boldmath{$\vartheta$}}}}%
%BeginExpansion
\mbox{\boldmath{$\vartheta$}}%
%EndExpansion
^{\mu}\lrcorner\vartheta_{\mu}\cdot\partial(\det[%
%TCIMACRO{\TeXButton{h}{\slh}}%
%BeginExpansion
\slh
%EndExpansion
]\underline{%
%TCIMACRO{\TeXButton{h}{\slh}}%
%BeginExpansion
\slh
%EndExpansion
}^{-1}(%
%TCIMACRO{\TeXButton{delta}{\mbox{\boldmath{$\delta$}}}}%
%BeginExpansion
\mbox{\boldmath{$\delta$}}%
%EndExpansion
_{%
%TCIMACRO{\TeXButton{sh}{\sslh}}%
%BeginExpansion
\sslh
%EndExpansion
^{\clubsuit}}^{w}\Omega(\vartheta_{\nu})))
\end{align*}%
\begin{align*}
&  =\dfrac{1}{\det[%
%TCIMACRO{\TeXButton{h}{\slh}}%
%BeginExpansion
\slh
%EndExpansion
]}(%
%TCIMACRO{\TeXButton{vt}{\mbox{\boldmath{$\vartheta$}}}}%
%BeginExpansion
\mbox{\boldmath{$\vartheta$}}%
%EndExpansion
^{\mu}\wedge%
%TCIMACRO{\TeXButton{vt}{\mbox{\boldmath{$\vartheta$}}}}%
%BeginExpansion
\mbox{\boldmath{$\vartheta$}}%
%EndExpansion
^{\nu})\cdot\vartheta_{\mu}\cdot\partial(\det[%
%TCIMACRO{\TeXButton{h}{\slh}}%
%BeginExpansion
\slh
%EndExpansion
]\underline{%
%TCIMACRO{\TeXButton{h}{\slh}}%
%BeginExpansion
\slh
%EndExpansion
}^{-1}(%
%TCIMACRO{\TeXButton{delta}{\mbox{\boldmath{$\delta$}}}}%
%BeginExpansion
\mbox{\boldmath{$\delta$}}%
%EndExpansion
_{%
%TCIMACRO{\TeXButton{sh}{\sslh}}%
%BeginExpansion
\sslh
%EndExpansion
^{\clubsuit}}^{w}\Omega(\vartheta_{\nu}))\\
&  =-\dfrac{1}{\det[%
%TCIMACRO{\TeXButton{h}{\slh}}%
%BeginExpansion
\slh
%EndExpansion
]}(%
%TCIMACRO{\TeXButton{vt}{\mbox{\boldmath{$\vartheta$}}}}%
%BeginExpansion
\mbox{\boldmath{$\vartheta$}}%
%EndExpansion
^{\mu}\wedge%
%TCIMACRO{\TeXButton{vt}{\mbox{\boldmath{$\vartheta$}}}}%
%BeginExpansion
\mbox{\boldmath{$\vartheta$}}%
%EndExpansion
^{\nu})\cdot\vartheta_{\mu}\cdot\partial(\det[%
%TCIMACRO{\TeXButton{h}{\slh}}%
%BeginExpansion
\slh
%EndExpansion
]\underline{%
%TCIMACRO{\TeXButton{h}{\slh}}%
%BeginExpansion
\slh
%EndExpansion
}^{-1}(%
%TCIMACRO{\TeXButton{delta}{\mbox{\boldmath{$\delta$}}}}%
%BeginExpansion
\mbox{\boldmath{$\delta$}}%
%EndExpansion
_{%
%TCIMACRO{\TeXButton{sh}{\sslh}}%
%BeginExpansion
\sslh
%EndExpansion
^{\clubsuit}}^{w}\Omega(\vartheta_{\nu}))\\
&  =-\dfrac{1}{\det[%
%TCIMACRO{\TeXButton{h}{\slh}}%
%BeginExpansion
\slh
%EndExpansion
]}%
%TCIMACRO{\TeXButton{vt}{\mbox{\boldmath{$\vartheta$}}}}%
%BeginExpansion
\mbox{\boldmath{$\vartheta$}}%
%EndExpansion
^{\mu}\cdot(%
%TCIMACRO{\TeXButton{vt}{\mbox{\boldmath{$\vartheta$}}}}%
%BeginExpansion
\mbox{\boldmath{$\vartheta$}}%
%EndExpansion
^{\nu}\lrcorner\vartheta_{\mu}\cdot\partial(\det[%
%TCIMACRO{\TeXButton{h}{\slh}}%
%BeginExpansion
\slh
%EndExpansion
]\underline{%
%TCIMACRO{\TeXButton{h}{\slh}}%
%BeginExpansion
\slh
%EndExpansion
}^{-1}(%
%TCIMACRO{\TeXButton{delta}{\mbox{\boldmath{$\delta$}}}}%
%BeginExpansion
\mbox{\boldmath{$\delta$}}%
%EndExpansion
_{%
%TCIMACRO{\TeXButton{sh}{\sslh}}%
%BeginExpansion
\sslh
%EndExpansion
^{\clubsuit}}^{w}\Omega(\vartheta_{\nu})))\\
&  =-\dfrac{1}{\det[%
%TCIMACRO{\TeXButton{h}{\slh}}%
%BeginExpansion
\slh
%EndExpansion
]}%
%TCIMACRO{\TeXButton{vt}{\mbox{\boldmath{$\vartheta$}}}}%
%BeginExpansion
\mbox{\boldmath{$\vartheta$}}%
%EndExpansion
^{\mu}\cdot\vartheta_{\mu}\cdot\partial(%
%TCIMACRO{\TeXButton{vt}{\mbox{\boldmath{$\vartheta$}}}}%
%BeginExpansion
\mbox{\boldmath{$\vartheta$}}%
%EndExpansion
^{\nu}\lrcorner\det[%
%TCIMACRO{\TeXButton{h}{\slh}}%
%BeginExpansion
\slh
%EndExpansion
]\underline{%
%TCIMACRO{\TeXButton{h}{\slh}}%
%BeginExpansion
\slh
%EndExpansion
}^{-1}(%
%TCIMACRO{\TeXButton{delta}{\mbox{\boldmath{$\delta$}}}}%
%BeginExpansion
\mbox{\boldmath{$\delta$}}%
%EndExpansion
_{%
%TCIMACRO{\TeXButton{sh}{\sslh}}%
%BeginExpansion
\sslh
%EndExpansion
^{\clubsuit}}^{w}\Omega(\vartheta_{\nu})))\\
&  =-\dfrac{1}{\det[%
%TCIMACRO{\TeXButton{h}{\slh}}%
%BeginExpansion
\slh
%EndExpansion
]}\partial\cdot(\det[%
%TCIMACRO{\TeXButton{h}{\slh}}%
%BeginExpansion
\slh
%EndExpansion
]%
%TCIMACRO{\TeXButton{vt}{\mbox{\boldmath{$\vartheta$}}}}%
%BeginExpansion
\mbox{\boldmath{$\vartheta$}}%
%EndExpansion
^{\nu}\lrcorner\underline{%
%TCIMACRO{\TeXButton{h}{\slh}}%
%BeginExpansion
\slh
%EndExpansion
}^{-1}(%
%TCIMACRO{\TeXButton{delta}{\mbox{\boldmath{$\delta$}}}}%
%BeginExpansion
\mbox{\boldmath{$\delta$}}%
%EndExpansion
_{%
%TCIMACRO{\TeXButton{sh}{\sslh}}%
%BeginExpansion
\sslh
%EndExpansion
^{\clubsuit}}^{w}\Omega(\vartheta_{\nu}))).
\end{align*}
i.e.,
\begin{equation}%
%TCIMACRO{\TeXButton{h}{\slh}}%
%BeginExpansion
\slh
%EndExpansion
^{\clubsuit}(%
%TCIMACRO{\TeXButton{vt}{\mbox{\boldmath{$\vartheta$}}}}%
%BeginExpansion
\mbox{\boldmath{$\vartheta$}}%
%EndExpansion
^{\nu})\cdot%
%TCIMACRO{\TeXButton{D}{\slD}}%
%BeginExpansion
\slD
%EndExpansion
\lrcorner%
%TCIMACRO{\TeXButton{delta}{\mbox{\boldmath{$\delta$}}}}%
%BeginExpansion
\mbox{\boldmath{$\delta$}}%
%EndExpansion
_{%
%TCIMACRO{\TeXButton{sh}{\sslh}}%
%BeginExpansion
\sslh
%EndExpansion
^{\clubsuit}}^{w}\Omega(\vartheta_{\nu})=-\dfrac{1}{\det[%
%TCIMACRO{\TeXButton{h}{\slh}}%
%BeginExpansion
\slh
%EndExpansion
]}\partial\cdot(\det[%
%TCIMACRO{\TeXButton{h}{\slh}}%
%BeginExpansion
\slh
%EndExpansion
]\partial_{a}\lrcorner\underline{%
%TCIMACRO{\TeXButton{h}{\slh}}%
%BeginExpansion
\slh
%EndExpansion
}^{-1}(%
%TCIMACRO{\TeXButton{delta}{\mbox{\boldmath{$\delta$}}}}%
%BeginExpansion
\mbox{\boldmath{$\delta$}}%
%EndExpansion
_{%
%TCIMACRO{\TeXButton{sh}{\sslh}}%
%BeginExpansion
\sslh
%EndExpansion
^{\clubsuit}}^{w}\Omega(a))). \label{4.43iv}%
\end{equation}

Now, putting Eq.(\ref{4.43iv}) in Eq.(\ref{4.43iii}) we get an analogous of
the well known \emph{ Palatini identity, }i.e.,\emph{ }
\begin{equation}
\underline{%
%TCIMACRO{\TeXButton{h}{\slh}}%
%BeginExpansion
\slh
%EndExpansion
}^{\clubsuit}(\partial_{a}\wedge\partial_{b})\cdot%
%TCIMACRO{\TeXButton{delta}{\mbox{\boldmath{$\delta$}}}}%
%BeginExpansion
\mbox{\boldmath{$\delta$}}%
%EndExpansion
_{%
%TCIMACRO{\TeXButton{sh}{\sslh}}%
%BeginExpansion
\sslh
%EndExpansion
^{\clubsuit}}^{w}\mathcal{R}_{2}(a\wedge b)=-2\dfrac{1}{\det[%
%TCIMACRO{\TeXButton{h}{\slh}}%
%BeginExpansion
\slh
%EndExpansion
]}\partial\cdot(\det[%
%TCIMACRO{\TeXButton{h}{\slh}}%
%BeginExpansion
\slh
%EndExpansion
]\partial_{a}\lrcorner\underline{%
%TCIMACRO{\TeXButton{h}{\slh}}%
%BeginExpansion
\slh
%EndExpansion
}^{-1}(%
%TCIMACRO{\TeXButton{delta}{\mbox{\boldmath{$\delta$}}}}%
%BeginExpansion
\mbox{\boldmath{$\delta$}}%
%EndExpansion
_{%
%TCIMACRO{\TeXButton{sh}{\sslh}}%
%BeginExpansion
\sslh
%EndExpansion
^{\clubsuit}}^{w}\Omega(a))). \label{4.43v}%
\end{equation}

Next we calculate the variation of $\det[%
%TCIMACRO{\TeXButton{h}{\slh}}%
%BeginExpansion
\slh
%EndExpansion
]$ with respect to $%
%TCIMACRO{\TeXButton{h}{\slh}}%
%BeginExpansion
\slh
%EndExpansion
^{\clubsuit}$ in the direction of $w$. In order to do so we utilize the chain
rule given by Eq.(\ref{LRd}) and the variational formula (recall
Eq.(\ref{vdet}))%
\begin{equation}%
%TCIMACRO{\TeXButton{delta}{\mbox{\boldmath{$\delta$}}}}%
%BeginExpansion
\mbox{\boldmath{$\delta$}}%
%EndExpansion
_{t}^{w}\det[t]=w(\partial_{a})\cdot t^{\clubsuit}(a)\det[t],
\end{equation}
which permit us to write%
\begin{align}%
%TCIMACRO{\TeXButton{delta}{\mbox{\boldmath{$\delta$}}}}%
%BeginExpansion
\mbox{\boldmath{$\delta$}}%
%EndExpansion
_{%
%TCIMACRO{\TeXButton{sh}{\sslh}}%
%BeginExpansion
\sslh
%EndExpansion
^{\clubsuit}}^{w}\det[%
%TCIMACRO{\TeXButton{h}{\slh}}%
%BeginExpansion
\slh
%EndExpansion
]  &  =%
%TCIMACRO{\TeXButton{delta}{\mbox{\boldmath{$\delta$}}}}%
%BeginExpansion
\mbox{\boldmath{$\delta$}}%
%EndExpansion
_{%
%TCIMACRO{\TeXButton{sh}{\sslh}}%
%BeginExpansion
\sslh
%EndExpansion
^{\clubsuit}}^{w}\dfrac{1}{\det[%
%TCIMACRO{\TeXButton{h}{\slh}}%
%BeginExpansion
\slh
%EndExpansion
^{\clubsuit}]}\nonumber\\
&  =-\dfrac{1}{(\det[%
%TCIMACRO{\TeXButton{h}{\slh}}%
%BeginExpansion
\slh
%EndExpansion
^{\clubsuit}])^{2}}%
%TCIMACRO{\TeXButton{delta}{\mbox{\boldmath{$\delta$}}}}%
%BeginExpansion
\mbox{\boldmath{$\delta$}}%
%EndExpansion
_{%
%TCIMACRO{\TeXButton{sh}{\sslh}}%
%BeginExpansion
\sslh
%EndExpansion
^{\clubsuit}}^{w}\det[%
%TCIMACRO{\TeXButton{h}{\slh}}%
%BeginExpansion
\slh
%EndExpansion
^{\clubsuit}]\nonumber\\
&  =-\dfrac{1}{(\det[%
%TCIMACRO{\TeXButton{h}{\slh}}%
%BeginExpansion
\slh
%EndExpansion
^{\clubsuit}])^{2}}w(\partial_{a})\cdot%
%TCIMACRO{\TeXButton{h}{\slh}}%
%BeginExpansion
\slh
%EndExpansion
(a)\det[%
%TCIMACRO{\TeXButton{h}{\slh}}%
%BeginExpansion
\slh
%EndExpansion
^{\clubsuit}]\nonumber\\
&  =-\dfrac{1}{\det[%
%TCIMACRO{\TeXButton{h}{\slh}}%
%BeginExpansion
\slh
%EndExpansion
^{\clubsuit}]}w(\partial_{a})\cdot%
%TCIMACRO{\TeXButton{h}{\slh}}%
%BeginExpansion
\slh
%EndExpansion
(a)\nonumber\\%
%TCIMACRO{\TeXButton{delta}{\mbox{\boldmath{$\delta$}}}}%
%BeginExpansion
\mbox{\boldmath{$\delta$}}%
%EndExpansion
_{%
%TCIMACRO{\TeXButton{sh}{\sslh}}%
%BeginExpansion
\sslh
%EndExpansion
^{\clubsuit}}^{w}\det[%
%TCIMACRO{\TeXButton{h}{\slh}}%
%BeginExpansion
\slh
%EndExpansion
]  &  =-w(\partial_{a})\cdot%
%TCIMACRO{\TeXButton{h}{\slh}}%
%BeginExpansion
\slh
%EndExpansion
(a)\det[%
%TCIMACRO{\TeXButton{h}{\slh}}%
%BeginExpansion
\slh
%EndExpansion
]. \label{4.43vi}%
\end{align}

Finally, utilizing Eqs.(\ref{4.43ii}), (\ref{4.43v}) and (\ref{4.43vi}) in
Eq.(\ref{4.43i}) we get the variation with respect to $%
%TCIMACRO{\TeXButton{h}{\slh}}%
%BeginExpansion
\slh
%EndExpansion
^{\clubsuit}$ in the direction of $w$ of the Lagrangian given by
Eq.(\ref{4.27}),
\begin{align}%
%TCIMACRO{\TeXButton{delta}{\mbox{\boldmath{$\delta$}}}}%
%BeginExpansion
\mbox{\boldmath{$\delta$}}%
%EndExpansion
_{%
%TCIMACRO{\TeXButton{sh}{\sslh}}%
%BeginExpansion
\sslh
%EndExpansion
^{\clubsuit}}^{w}(\mathcal{R}\det[%
%TCIMACRO{\TeXButton{h}{\slh}}%
%BeginExpansion
\slh
%EndExpansion
])  &  =2w(\partial_{a})\cdot\mathcal{R}_{1}(a)\det[%
%TCIMACRO{\TeXButton{h}{\slh}}%
%BeginExpansion
\slh
%EndExpansion
]\nonumber\\
&  -2\dfrac{1}{\det[%
%TCIMACRO{\TeXButton{h}{\slh}}%
%BeginExpansion
\slh
%EndExpansion
]}\partial\cdot(\det[%
%TCIMACRO{\TeXButton{h}{\slh}}%
%BeginExpansion
\slh
%EndExpansion
]\partial_{a}\lrcorner\underline{%
%TCIMACRO{\TeXButton{h}{\slh}}%
%BeginExpansion
\slh
%EndExpansion
}^{-1}(%
%TCIMACRO{\TeXButton{delta}{\mbox{\boldmath{$\delta$}}}}%
%BeginExpansion
\mbox{\boldmath{$\delta$}}%
%EndExpansion
_{%
%TCIMACRO{\TeXButton{sh}{\sslh}}%
%BeginExpansion
\sslh
%EndExpansion
^{\clubsuit}}^{w}\Omega(a)))\det[%
%TCIMACRO{\TeXButton{h}{\slh}}%
%BeginExpansion
\slh
%EndExpansion
]\nonumber\\
&  -\underline{%
%TCIMACRO{\TeXButton{h}{\slh}}%
%BeginExpansion
\slh
%EndExpansion
}^{\clubsuit}(\partial_{a}\wedge\partial_{b})\cdot\mathcal{R}_{2}(a\wedge
b)(w(\partial_{a})\cdot%
%TCIMACRO{\TeXButton{h}{\slh}}%
%BeginExpansion
\slh
%EndExpansion
(a)\det[%
%TCIMACRO{\TeXButton{h}{\slh}}%
%BeginExpansion
\slh
%EndExpansion
])\nonumber\\
&  =2w(\partial_{a})\cdot(\mathcal{R}_{1}(a)-\frac{1}{2}%
%TCIMACRO{\TeXButton{h}{\slh}}%
%BeginExpansion
\slh
%EndExpansion
(a)\mathcal{R})\det[%
%TCIMACRO{\TeXButton{h}{\slh}}%
%BeginExpansion
\slh
%EndExpansion
]-2\partial\cdot(\det[%
%TCIMACRO{\TeXButton{h}{\slh}}%
%BeginExpansion
\slh
%EndExpansion
]\partial_{a}\lrcorner\underline{%
%TCIMACRO{\TeXButton{h}{\slh}}%
%BeginExpansion
\slh
%EndExpansion
}^{-1}(%
%TCIMACRO{\TeXButton{delta}{\mbox{\boldmath{$\delta$}}}}%
%BeginExpansion
\mbox{\boldmath{$\delta$}}%
%EndExpansion
_{%
%TCIMACRO{\TeXButton{h}{\slh}}%
%BeginExpansion
\slh
%EndExpansion
^{\clubsuit}}^{w}\Omega(a)))\nonumber\\%
%TCIMACRO{\TeXButton{delta}{\mbox{\boldmath{$\delta$}}}}%
%BeginExpansion
\mbox{\boldmath{$\delta$}}%
%EndExpansion
_{%
%TCIMACRO{\TeXButton{sh}{\sslh}}%
%BeginExpansion
\sslh
%EndExpansion
^{\clubsuit}}^{w}(\mathcal{R}\det[%
%TCIMACRO{\TeXButton{h}{\slh}}%
%BeginExpansion
\slh
%EndExpansion
])  &  =2w(\partial_{a})\cdot\mathcal{G}(a)\det[%
%TCIMACRO{\TeXButton{h}{\slh}}%
%BeginExpansion
\slh
%EndExpansion
]-2\partial\cdot(\det[%
%TCIMACRO{\TeXButton{h}{\slh}}%
%BeginExpansion
\slh
%EndExpansion
]\partial_{a}\lrcorner\underline{%
%TCIMACRO{\TeXButton{h}{\slh}}%
%BeginExpansion
\slh
%EndExpansion
}^{-1}(%
%TCIMACRO{\TeXButton{delta}{\mbox{\boldmath{$\delta$}}}}%
%BeginExpansion
\mbox{\boldmath{$\delta$}}%
%EndExpansion
_{%
%TCIMACRO{\TeXButton{sh}{\sslh}}%
%BeginExpansion
\sslh
%EndExpansion
^{\clubsuit}}^{w}\Omega(a))). \label{4.34}%
\end{align}
In the last step we recalled the definition (Eq.(\ref{4.19})) of the Einstein
gauge field .$\mathcal{G}$ for the particular case of the Lorentz
\textit{MCGSS.}

Tanking into account Eq.(\ref{4.34}), the contour problem for the dynamic
variable $%
%TCIMACRO{\TeXButton{h}{\slh}}%
%BeginExpansion
\slh
%EndExpansion
^{\clubsuit}$ becomes then
\begin{equation}
\int_{U}w(\partial_{a})\cdot\mathcal{G}(a)\det[%
%TCIMACRO{\TeXButton{h}{\slh}}%
%BeginExpansion
\slh
%EndExpansion
]\text{ }\tau-\int_{U}\partial\cdot(\det[%
%TCIMACRO{\TeXButton{h}{\slh}}%
%BeginExpansion
\slh
%EndExpansion
]\partial_{a}\lrcorner\underline{%
%TCIMACRO{\TeXButton{h}{\slh}}%
%BeginExpansion
\slh
%EndExpansion
}^{-1}(%
%TCIMACRO{\TeXButton{delta}{\mbox{\boldmath{$\delta$}}}}%
%BeginExpansion
\mbox{\boldmath{$\delta$}}%
%EndExpansion
_{%
%TCIMACRO{\TeXButton{sh}{\sslh}}%
%BeginExpansion
\sslh
%EndExpansion
^{\clubsuit}}^{w}\Omega(a)))\text{ }\tau=0, \label{4.35}%
\end{equation}
for all $w$ satisfying the boundary conditions $\left.  w\right\vert
_{\partial U}=0$ and $\left.  a\cdot\partial w\right\vert _{\partial U}=0.$

Utilizing then the Gauss-Stokes theorem with the above boundary conditions the
second term of Eq.(\ref{4.35}) may be integrated and gives
\begin{equation}
\int_{U}\partial\cdot(\det[%
%TCIMACRO{\TeXButton{h}{\slh}}%
%BeginExpansion
\slh
%EndExpansion
]\partial_{a}\lrcorner\underline{%
%TCIMACRO{\TeXButton{h}{\slh}}%
%BeginExpansion
\slh
%EndExpansion
}^{-1}(%
%TCIMACRO{\TeXButton{delta}{\mbox{\boldmath{$\delta$}}}}%
%BeginExpansion
\mbox{\boldmath{$\delta$}}%
%EndExpansion
_{%
%TCIMACRO{\TeXButton{sh}{\sslh}}%
%BeginExpansion
\sslh
%EndExpansion
^{\clubsuit}}^{w}\Omega(a)))\text{ }\tau=\oint_{\partial U}\det[%
%TCIMACRO{\TeXButton{h}{\slh}}%
%BeginExpansion
\slh
%EndExpansion
]%
%TCIMACRO{\TeXButton{vt}{\mbox{\boldmath{$\vartheta$}}}}%
%BeginExpansion
\mbox{\boldmath{$\vartheta$}}%
%EndExpansion
^{\mu}\cdot(\partial_{a}\lrcorner\underline{%
%TCIMACRO{\TeXButton{h}{\slh}}%
%BeginExpansion
\slh
%EndExpansion
}^{-1}(%
%TCIMACRO{\TeXButton{delta}{\mbox{\boldmath{$\delta$}}}}%
%BeginExpansion
\mbox{\boldmath{$\delta$}}%
%EndExpansion
_{%
%TCIMACRO{\TeXButton{sh}{\sslh}}%
%BeginExpansion
\sslh
%EndExpansion
^{\clubsuit}}^{w}\Omega(a)))\text{ }\tau_{\mu}=0, \label{4.36}%
\end{equation}
since $%
%TCIMACRO{\TeXButton{delta}{\mbox{\boldmath{$\delta$}}}}%
%BeginExpansion
\mbox{\boldmath{$\delta$}}%
%EndExpansion
_{%
%TCIMACRO{\TeXButton{h}{\slh}}%
%BeginExpansion
\slh
%EndExpansion
^{\clubsuit}}^{w}\Omega(a)=0$ under the boundary conditions $\left.
w\right\vert _{\partial U}=0$ and $\left.  a\cdot\partial w\right\vert
_{\partial U}=0$.

Thus, utilizing Eq.(\ref{4.36}) in Eq.(\ref{4.35}) it follows that
\begin{equation}
\int_{U}w(\partial_{a})\cdot\mathcal{G}(a)\det[%
%TCIMACRO{\TeXButton{h}{\slh}}%
%BeginExpansion
\slh
%EndExpansion
]\text{ }\tau=0, \label{4.37}%
\end{equation}
for all $w$, and since $w$ \ is arbitrary a fundamental lemma of integration
theory yields
\begin{equation}
\mathcal{G}(a)\det[%
%TCIMACRO{\TeXButton{h}{\slh}}%
%BeginExpansion
\slh
%EndExpansion
]=0, \label{4.38}%
\end{equation}
i.e.,
\begin{equation}
\mathcal{G}(a)=\mathcal{R}_{1}(a)-\frac{1}{2}%
%TCIMACRO{\TeXButton{h}{\slh}}%
%BeginExpansion
\slh
%EndExpansion
(a)\mathcal{R}=0, \label{4.39}%
\end{equation}
which is the field equation for the distortion gauge field\emph{ }$%
%TCIMACRO{\TeXButton{h}{\slh}}%
%BeginExpansion
\slh
%EndExpansion
.$

If we recall the relation between the Einstein $(1,1)$-extensor field,
$a\mapsto%
%TCIMACRO{\TeXButton{G}{\slG}}%
%BeginExpansion
\slG
%EndExpansion
(a)$ and the Einstein gauge field, $a\mapsto\mathcal{G}(a),$ (i.e., $%
%TCIMACRO{\TeXButton{G}{\slG}}%
%BeginExpansion
\slG
%EndExpansion
(a)=%
%TCIMACRO{\TeXButton{h}{\slh}}%
%BeginExpansion
\slh
%EndExpansion
^{\dagger}\eta\mathcal{G}(a)$), we get%
\begin{equation}%
%TCIMACRO{\TeXButton{G}{\slG}}%
%BeginExpansion
\slG
%EndExpansion
(a)=0, \label{4.40}%
\end{equation}
which we recognize as equivalent to Einstein equation for $%
%TCIMACRO{\TeXButton{itg}{\itg}}%
%BeginExpansion
\itg
%EndExpansion
$.

\subsubsection{Lagrangian for the Gravitational Field Plus Matter Field
Including a Cosmological Constant Term}

Since cosmological data seems to suggest an expansion of the universe (when it
is described in terms of the Lorentzian spacetime model) we postulate here
that the total Lagrangian describing the interaction of the gravitational
field with matter is\footnote{Note that we are using geometrical units. The
matter Lagrangian density is normaly written as $\mathfrak{L}_{m}%
=-\kappa\mathfrak{L}_{m}^{\prime}$, where $\kappa=-8\pi G$, where $G$ is
Newton's gravitational constant.}%
\begin{equation}
\mathfrak{L=L}_{eh}+\lambda\det[%
%TCIMACRO{\TeXButton{h}{\slh}}%
%BeginExpansion
\slh
%EndExpansion
]+\mathfrak{L}_{m}\det[%
%TCIMACRO{\TeXButton{h}{\slh}}%
%BeginExpansion
\slh
%EndExpansion
], \label{4t1}%
\end{equation}
where $\lambda$ is the cosmological constant and where $\mathfrak{L}_{m}$ is
the matter Lagrangian. The equations of motion for $%
%TCIMACRO{\TeXButton{h}{\slh}}%
%BeginExpansion
\slh
%EndExpansion
$ are obtained from the variational principle,%
\begin{equation}
\int_{U}%
%TCIMACRO{\TeXButton{delta}{\mbox{\boldmath{$\delta$}}}}%
%BeginExpansion
\mbox{\boldmath{$\delta$}}%
%EndExpansion
_{%
%TCIMACRO{\TeXButton{sh}{\sslh}}%
%BeginExpansion
\sslh
%EndExpansion
^{\clubsuit}}^{w}(\frac{1}{2}\mathcal{R+\lambda+}\mathfrak{L}_{m})\det[%
%TCIMACRO{\TeXButton{h}{\slh}}%
%BeginExpansion
\slh
%EndExpansion
]\text{ }\tau=0. \label{4.T2}%
\end{equation}
We then get defining conveniently the energy-momentum extensor of matter
$T(a)$ by
\begin{equation}
(w(\partial_{a})\cdot\eta%
%TCIMACRO{\TeXButton{h}{\slh}}%
%BeginExpansion
\slh
%EndExpansion
^{\clubsuit}T(a))\det[%
%TCIMACRO{\TeXButton{h}{\slh}}%
%BeginExpansion
\slh
%EndExpansion
]:=%
%TCIMACRO{\TeXButton{delta}{\mbox{\boldmath{$\delta$}}}}%
%BeginExpansion
\mbox{\boldmath{$\delta$}}%
%EndExpansion
_{%
%TCIMACRO{\TeXButton{sh}{\sslh}}%
%BeginExpansion
\sslh
%EndExpansion
^{\clubsuit}}^{w}(\mathfrak{L}_{m}\det[%
%TCIMACRO{\TeXButton{h}{\slh}}%
%BeginExpansion
\slh
%EndExpansion
]), \label{4t3}%
\end{equation}
that%
\begin{equation}
\overset{}{\mathcal{R}}_{1}(a)-\frac{1}{2}%
%TCIMACRO{\TeXButton{h}{\slh}}%
%BeginExpansion
\slh
%EndExpansion
(a)\overset{}{\mathcal{R}}-\lambda%
%TCIMACRO{\TeXButton{h}{\slh}}%
%BeginExpansion
\slh
%EndExpansion
(a)=-\eta%
%TCIMACRO{\TeXButton{h}{\slh}}%
%BeginExpansion
\slh
%EndExpansion
^{\clubsuit}T(a). \label{4.t4}%
\end{equation}
Multiplying Eq.(\ref{4.t4}) on both sides by $%
%TCIMACRO{\TeXButton{h}{\slh}}%
%BeginExpansion
\slh
%EndExpansion
^{\dagger}\eta$ we get:%
\begin{equation}%
%TCIMACRO{\TeXButton{G}{\slG}}%
%BeginExpansion
\slG
%EndExpansion
(a)-\lambda%
%TCIMACRO{\TeXButton{itg}{\itg}}%
%BeginExpansion
\itg
%EndExpansion
(a)=-T(a). \label{4.t5}%
\end{equation}

\section{Formulation of the Gravitational Theory in Terms of the Potentials
$\mathfrak{g}^{\mathbf{\alpha}}=%
%TCIMACRO{\TeXButton{h}{\slh}}%
%BeginExpansion
\slh
%EndExpansion
^{\dagger}(%
%TCIMACRO{\TeXButton{vt}{\mbox{\boldmath{$\vartheta$}}}}%
%BeginExpansion
\mbox{\boldmath{$\vartheta$}}%
%EndExpansion
^{\mathbf{\alpha}})$}

Let $\{\mathtt{x}^{\mu}\}$ be global coordinates in the Einstein-Lorentz
Poincar\'{e} gauge for $M\simeq\mathbb{R}^{4}$ and write as above
\begin{equation}%
%TCIMACRO{\TeXButton{vt}{\mbox{\boldmath{$\vartheta$}}}}%
%BeginExpansion
\mbox{\boldmath{$\vartheta$}}%
%EndExpansion
^{\mathbf{\mu}}=d\mathtt{x}^{\mu}\text{.} \label{g1}%
\end{equation}
Then
\begin{equation}%
%TCIMACRO{\TeXButton{eta}{\mbox{\boldmath{$\eta$}}}}%
%BeginExpansion
\mbox{\boldmath{$\eta$}}%
%EndExpansion
=\eta_{\mathbf{\alpha\beta}}%
%TCIMACRO{\TeXButton{vt}{\mbox{\boldmath{$\vartheta$}}}}%
%BeginExpansion
\mbox{\boldmath{$\vartheta$}}%
%EndExpansion
^{\mathbf{\alpha}}\otimes%
%TCIMACRO{\TeXButton{vt}{\mbox{\boldmath{$\vartheta$}}}}%
%BeginExpansion
\mbox{\boldmath{$\vartheta$}}%
%EndExpansion
^{\mathbf{\beta}}. \label{g1'}%
\end{equation}

Define next the \textit{gravitational potentials}\footnote{Take notice that in
the calculations of this section we used the $%
%TCIMACRO{\TeXButton{itg}{\itg}}%
%BeginExpansion
\itg
%EndExpansion
$-reciprocal basis $\{\mathfrak{g}_{\alpha}\}$ of the basis $\{\mathfrak{g}%
^{\mathbf{\mu}}\}$, i.e., $\mathfrak{g}^{\mathbf{\mu}}\underset{%
%TCIMACRO{\TeXButton{sig}{\sitg}}%
%BeginExpansion
\sitg
%EndExpansion
^{{\tiny -1}}}{\cdot}\mathfrak{g}_{\mathbf{\alpha}}=\delta_{\mathbf{\alpha}%
}^{\mathbf{\mu}}$, i.e., $\mathfrak{g}_{\mathbf{\alpha}}=%
%TCIMACRO{\TeXButton{h}{\slh}}%
%BeginExpansion
\slh
%EndExpansion
^{\dagger}\eta(\vartheta_{\mathbf{\alpha}})=\eta_{\mathbf{\alpha\beta}%
}\mathfrak{g}^{\mathbf{\beta}}$.}%
\begin{equation}
\mathfrak{g}^{\mathbf{\mu}}=%
%TCIMACRO{\TeXButton{h}{\slh}}%
%BeginExpansion
\slh
%EndExpansion
^{\dagger}(%
%TCIMACRO{\TeXButton{vt}{\mbox{\boldmath{$\vartheta$}}}}%
%BeginExpansion
\mbox{\boldmath{$\vartheta$}}%
%EndExpansion
^{\mathbf{\mu}})\in\sec%
%TCIMACRO{\dbigwedge \nolimits^{1}}%
%BeginExpansion
{\displaystyle\bigwedge\nolimits^{1}}
%EndExpansion
T^{\ast}U. \label{g1a}%
\end{equation}
We immediately have that
\begin{align}
\mathfrak{g}^{\mathbf{\alpha}}\underset{%
%TCIMACRO{\TeXButton{sig}{\sitg}}%
%BeginExpansion
\sitg
%EndExpansion
^{{\tiny -1}}}{\cdot}\mathfrak{g}^{\mathbf{\beta}}  &  =%
%TCIMACRO{\TeXButton{h}{\slh}}%
%BeginExpansion
\slh
%EndExpansion
^{-1}\eta%
%TCIMACRO{\TeXButton{h}{\slh}}%
%BeginExpansion
\slh
%EndExpansion
^{\clubsuit}\mathfrak{g}^{\mathbf{\alpha}}\cdot\mathfrak{g}^{\mathbf{\beta}%
}=\eta%
%TCIMACRO{\TeXButton{h}{\slh}}%
%BeginExpansion
\slh
%EndExpansion
^{\clubsuit}\mathfrak{g}^{\mathbf{\alpha}}\cdot%
%TCIMACRO{\TeXButton{h}{\slh}}%
%BeginExpansion
\slh
%EndExpansion
^{\clubsuit}\mathfrak{g}^{\mathbf{\beta}}\\
&  =%
%TCIMACRO{\TeXButton{h}{\slh}}%
%BeginExpansion
\slh
%EndExpansion
^{\clubsuit}\mathfrak{g}^{\mathbf{\alpha}}\underset{\eta^{-1}}{\cdot}%
%TCIMACRO{\TeXButton{h}{\slh}}%
%BeginExpansion
\slh
%EndExpansion
^{\clubsuit}\mathfrak{g}^{\mathbf{\beta}}=\nonumber\\
&  =%
%TCIMACRO{\TeXButton{vt}{\mbox{\boldmath{$\vartheta$}}}}%
%BeginExpansion
\mbox{\boldmath{$\vartheta$}}%
%EndExpansion
^{\mathbf{\alpha}}\underset{\eta^{-1}}{\cdot}%
%TCIMACRO{\TeXButton{vt}{\mbox{\boldmath{$\vartheta$}}}}%
%BeginExpansion
\mbox{\boldmath{$\vartheta$}}%
%EndExpansion
^{\mathbf{\beta}}=%
%TCIMACRO{\TeXButton{vt}{\mbox{\boldmath{$\vartheta$}}}}%
%BeginExpansion
\mbox{\boldmath{$\vartheta$}}%
%EndExpansion
^{\mathbf{\alpha}}\underset{\eta}{\cdot}%
%TCIMACRO{\TeXButton{vt}{\mbox{\boldmath{$\vartheta$}}}}%
%BeginExpansion
\mbox{\boldmath{$\vartheta$}}%
%EndExpansion
^{\mathbf{\beta}}=\eta^{\mathbf{\alpha\beta}}, \label{1d}%
\end{align}
i.e., the $\mathfrak{g}^{\mathbf{\mu}}$ are $%
%TCIMACRO{\TeXButton{itg}{\itg}}%
%BeginExpansion
\itg
%EndExpansion
^{-1}$-orthonormal, which means that $\{\mathfrak{g}^{\mathbf{\mu}}\}$ is a
section of the $%
%TCIMACRO{\TeXButton{itg}{\itg}}%
%BeginExpansion
\itg
%EndExpansion
^{-1}$-orthonormal coframe bundle of ($M\simeq\mathbb{R}^{4},%
%TCIMACRO{\TeXButton{g}{\slg}}%
%BeginExpansion
\slg
%EndExpansion
)$. Recalling now Eq.(\ref{4.21}) which says that $R\equiv\overset{%
%TCIMACRO{\TeXButton{sig}{\sitg}}%
%BeginExpansion
\sitg
%EndExpansion
}{R}=\mathcal{R}$ we can write the gravitational Lagrangian (Eq.(\ref{4.27}))
as
\begin{equation}
\overset{}{\mathfrak{L}_{eh}}:=\frac{1}{2}\overset{}{R}\det[%
%TCIMACRO{\TeXButton{h}{\slh}}%
%BeginExpansion
\slh
%EndExpansion
], \label{g3}%
\end{equation}
and the Lagrangian density\ $\mathcal{L}_{eh}=\overset{}{\mathfrak{L}_{eh}%
}\tau$ is then%
\begin{align}
\mathcal{L}_{eh}  &  =\frac{1}{2}\{%
%TCIMACRO{\TeXButton{h}{\slh}}%
%BeginExpansion
\slh
%EndExpansion
^{\clubsuit}(\partial_{a}\wedge\partial_{b})\cdot\mathcal{R}_{2}%
\mathcal{(}a\wedge b)\}\det%
%TCIMACRO{\TeXButton{h}{\slh}}%
%BeginExpansion
\slh
%EndExpansion
\tau\nonumber\\
&  =\frac{1}{2}\{%
%TCIMACRO{\TeXButton{h}{\slh}}%
%BeginExpansion
\slh
%EndExpansion
^{\clubsuit}(\partial_{a}\wedge\partial_{b})\cdot\eta\overset{\varkappa
\eta}{\mathbf{%
%TCIMACRO{\TeXButton{R}{\slR}}%
%BeginExpansion
\slR
%EndExpansion
}}_{2}\mathcal{(}a\wedge b)\}\tau_{%
%TCIMACRO{\TeXButton{sig}{\sitg}}%
%BeginExpansion
\sitg
%EndExpansion
}, \label{g4}%
\end{align}
where we have used that $\tau_{%
%TCIMACRO{\TeXButton{sig}{\sitg}}%
%BeginExpansion
\sitg
%EndExpansion
}=\det%
%TCIMACRO{\TeXButton{h}{\slh}}%
%BeginExpansion
\slh
%EndExpansion
\tau$ and \ Eq.(\ref{4.16}) which says that $\mathcal{R}_{2}\mathcal{(}a\wedge
b)=\eta\overset{\mu\eta}{\mathbf{%
%TCIMACRO{\TeXButton{R}{\slR}}%
%BeginExpansion
\slR
%EndExpansion
}}_{2}\mathcal{(}a\wedge b)$. Now, we write the $1$-form fields $a$,$b$ and
the multiform derivatives $\partial_{a},\partial_{b}$ as \
\begin{align}
a  &  =a^{\mathbf{\kappa}}\mathfrak{g}_{\mathbf{\kappa}}\text{, }%
b=b^{\mathbf{\kappa}}\mathfrak{g}_{\mathbf{\kappa}},\nonumber\\
\partial_{a}  &  =\mathfrak{g}^{\mathbf{\kappa}}\frac{\partial}{\partial
a^{\mathbf{\kappa}}}\text{, }\partial_{b}=\mathfrak{g}^{\mathbf{\kappa}}%
\frac{\partial}{\partial b^{\mathbf{\kappa}}}. \label{g5}%
\end{align}

Then, using Eq.(\ref{g5}) in Eq.(\ref{g4}) and taking into account
Eq.(\ref{4.15}) which says that $\overset{%
%TCIMACRO{\TeXButton{sig}{\sitg}}%
%BeginExpansion
\sitg
%EndExpansion
}{%
%TCIMACRO{\TeXButton{R}{\slR}}%
%BeginExpansion
\slR
%EndExpansion
}(B)\equiv\overset{\lambda%
%TCIMACRO{\TeXButton{sig}{\sitg}}%
%BeginExpansion
\sitg
%EndExpansion
}{%
%TCIMACRO{\TeXButton{R}{\slR}}%
%BeginExpansion
\slR
%EndExpansion
}(B)=\underline{%
%TCIMACRO{\TeXButton{h}{\slh}}%
%BeginExpansion
\slh
%EndExpansion
}^{\dagger}\overset{\varkappa\eta}{%
%TCIMACRO{\TeXButton{R}{\slR}}%
%BeginExpansion
\slR
%EndExpansion
_{2}}(B)$ we get%
\begin{align}
2\mathcal{L}_{eh}  &  =\{%
%TCIMACRO{\TeXButton{h}{\slh}}%
%BeginExpansion
\slh
%EndExpansion
^{\clubsuit}(\mathfrak{g}^{\mathbf{\kappa}}\wedge\mathfrak{g}^{\mathbf{\iota}%
})\cdot\eta\overset{\mu\eta}{\mathbf{%
%TCIMACRO{\TeXButton{R}{\slR}}%
%BeginExpansion
\slR
%EndExpansion
}}_{2}\mathcal{(}\mathfrak{g}_{\mathbf{\kappa}}\wedge\mathfrak{g}%
_{\mathbf{\iota}})\}\tau_{%
%TCIMACRO{\TeXButton{sig}{\sitg}}%
%BeginExpansion
\sitg
%EndExpansion
}\nonumber\\
&  =\{%
%TCIMACRO{\TeXButton{h}{\slh}}%
%BeginExpansion
\slh
%EndExpansion
^{\clubsuit}(\mathfrak{g}^{\mathbf{\kappa}}\wedge\mathfrak{g}^{\mathbf{\iota}%
})\cdot\eta%
%TCIMACRO{\TeXButton{h}{\slh}}%
%BeginExpansion
\slh
%EndExpansion
^{\clubsuit}%
%TCIMACRO{\TeXButton{h}{\slh}}%
%BeginExpansion
\slh
%EndExpansion
^{\dagger}\overset{\mu\eta}{\mathbf{%
%TCIMACRO{\TeXButton{R}{\slR}}%
%BeginExpansion
\slR
%EndExpansion
}_{2}}\mathcal{(}\mathfrak{g}_{\mathbf{\kappa}}\wedge\mathfrak{g}%
_{\mathbf{\iota}})\}\tau_{%
%TCIMACRO{\TeXButton{sig}{\sitg}}%
%BeginExpansion
\sitg
%EndExpansion
}\nonumber\\
&  =\{%
%TCIMACRO{\TeXButton{h}{\slh}}%
%BeginExpansion
\slh
%EndExpansion
^{-1}\eta%
%TCIMACRO{\TeXButton{h}{\slh}}%
%BeginExpansion
\slh
%EndExpansion
^{\clubsuit}(\mathfrak{g}^{\mathbf{\kappa}}\wedge\mathfrak{g}^{\mathbf{\iota}%
})\cdot%
%TCIMACRO{\TeXButton{h}{\slh}}%
%BeginExpansion
\slh
%EndExpansion
_{2}^{\dagger}\overset{\mu\eta}{\mathbf{%
%TCIMACRO{\TeXButton{R}{\slR}}%
%BeginExpansion
\slR
%EndExpansion
}_{2}}\mathcal{(}\mathfrak{g}_{\mathbf{\kappa}}\wedge\mathfrak{g}%
_{\mathbf{\iota l}})\}\nonumber\\
&  =\{%
%TCIMACRO{\TeXButton{h}{\slh}}%
%BeginExpansion
\slh
%EndExpansion
^{-1}\eta%
%TCIMACRO{\TeXButton{h}{\slh}}%
%BeginExpansion
\slh
%EndExpansion
^{\clubsuit}(\mathfrak{g}^{\mathbf{\kappa}}\wedge\mathfrak{g}^{\mathbf{\iota}%
})\cdot\overset{%
%TCIMACRO{\TeXButton{sig}{\sitg}}%
%BeginExpansion
\sitg
%EndExpansion
}{\mathbf{%
%TCIMACRO{\TeXButton{R}{\slR}}%
%BeginExpansion
\slR
%EndExpansion
}_{2}}\mathcal{(}\mathfrak{g}_{\mathbf{\kappa}}\wedge\mathfrak{g}%
_{\mathbf{\iota}})\}\tau_{%
%TCIMACRO{\TeXButton{sig}{\sitg}}%
%BeginExpansion
\sitg
%EndExpansion
}\nonumber\\
&  \{(\mathfrak{g}^{\mathbf{\kappa}}\wedge\mathfrak{g}^{\mathbf{\iota}%
})\underset{%
%TCIMACRO{\TeXButton{sig}{\sitg}}%
%BeginExpansion
\sitg
%EndExpansion
^{-1}}{\cdot}\overset{%
%TCIMACRO{\TeXButton{sig}{\sitg}}%
%BeginExpansion
\sitg
%EndExpansion
}{\mathbf{%
%TCIMACRO{\TeXButton{R}{\slR}}%
%BeginExpansion
\slR
%EndExpansion
}_{2}}\mathcal{(}\mathfrak{g}_{\mathbf{\kappa}}\wedge\mathfrak{g}%
_{\mathbf{\iota}})\}\tau_{%
%TCIMACRO{\TeXButton{sig}{\sitg}}%
%BeginExpansion
\sitg
%EndExpansion
} \label{g6}%
\end{align}
We now recall that from Eq.(\ref{3.33}) it is
\begin{equation}
\overset{%
%TCIMACRO{\TeXButton{sig}{\sitg}}%
%BeginExpansion
\sitg
%EndExpansion
}{\mathbf{%
%TCIMACRO{\TeXButton{R}{\slR}}%
%BeginExpansion
\slR
%EndExpansion
}_{2}}\mathcal{(}\mathfrak{g}_{\mathbf{\kappa}}\wedge\mathfrak{g}%
_{\mathbf{\iota}})\equiv\mathcal{R}_{\mathbf{\kappa\iota}} \label{g7}%
\end{equation}
where $\mathcal{R}_{\mathbf{\kappa\iota}}:U\rightarrow%
%TCIMACRO{\dbigwedge \nolimits^{2}}%
%BeginExpansion
{\displaystyle\bigwedge\nolimits^{2}}
%EndExpansion
U$ \ are the representatives of the \textit{Cartan curvature 2-form fields.}
Then recalling the definition of the Hodge star operator (Eq.(\ref{4.4c})) we
can write
\begin{align}
\mathcal{L}_{eh}  &  =\frac{1}{2}\{(\mathfrak{g}^{\mathbf{\kappa}}%
\wedge\mathfrak{g}^{\mathbf{\iota}})\underset{%
%TCIMACRO{\TeXButton{sig}{\sitg}}%
%BeginExpansion
\sitg
%EndExpansion
^{-1}}{\cdot}\mathcal{R}_{\mathbf{\kappa\iota}}\}\tau_{%
%TCIMACRO{\TeXButton{sig}{\sitg}}%
%BeginExpansion
\sitg
%EndExpansion
}\nonumber\\
&  =\frac{1}{2}(\mathfrak{g}^{\mathbf{\kappa}}\wedge\mathfrak{g}%
^{\mathbf{\iota}})\wedge\underset{%
%TCIMACRO{\TeXButton{sig}{\sitg}}%
%BeginExpansion
\sitg
%EndExpansion
}{\star}\mathcal{R}_{\mathbf{\kappa\iota}} \label{g8}%
\end{align}

Now, $\mathcal{L}_{eh}$ may be written (see, e.g., \cite{rodcap2007}) once we
realize that the connection $1$-forms can be written as%
\begin{equation}
\omega^{\mathbf{\gamma\delta}}=\frac{1}{2}\left[  \mathfrak{g}^{\mathbf{\delta
}}\underset{%
%TCIMACRO{\TeXButton{sig}{\sitg}}%
%BeginExpansion
\sitg
%EndExpansion
}{\lrcorner}d\mathfrak{g}^{\mathbf{\gamma}}-\mathfrak{g}^{\mathbf{\gamma}%
}\underset{%
%TCIMACRO{\TeXButton{sig}{\sitg}}%
%BeginExpansion
\sitg
%EndExpansion
}{\lrcorner}d\mathfrak{g}^{\mathbf{\delta}}+\mathfrak{g}^{\mathbf{\gamma}%
}\underset{%
%TCIMACRO{\TeXButton{sig}{\sitg}}%
%BeginExpansion
\sitg
%EndExpansion
}{\lrcorner}\left(  \mathfrak{g}^{\mathbf{\delta}}\underset{%
%TCIMACRO{\TeXButton{sig}{\sitg}}%
%BeginExpansion
\sitg
%EndExpansion
}{\lrcorner}d\mathfrak{g}_{\mathbf{\alpha}}\right)  \mathfrak{g}%
^{\mathbf{\alpha}}\right]  \label{7.10.18a}%
\end{equation}
as%
\begin{equation}
\mathcal{L}_{eh}=-\frac{1}{2}d\mathfrak{g}^{\mathbf{\alpha}}\wedge\underset{%
%TCIMACRO{\TeXButton{sig}{\sitg}}%
%BeginExpansion
\sitg
%EndExpansion
}{\star}d\mathfrak{g}_{\mathbf{\alpha}}+\frac{1}{2}\underset{%
%TCIMACRO{\TeXButton{sig}{\sitg}}%
%BeginExpansion
\sitg
%EndExpansion
}{\delta}\mathfrak{g}^{\mathbf{\alpha}}\wedge\underset{%
%TCIMACRO{\TeXButton{sig}{\sitg}}%
%BeginExpansion
\sitg
%EndExpansion
}{\star}\underset{%
%TCIMACRO{\TeXButton{sig}{\sitg}}%
%BeginExpansion
\sitg
%EndExpansion
}{\delta}\mathfrak{g}_{\mathbf{\alpha}}+\frac{1}{4}d\mathfrak{g}%
^{\mathbf{\alpha}}\wedge\mathfrak{g}_{\mathbf{\alpha}}\wedge\underset{%
%TCIMACRO{\TeXButton{sig}{\sitg}}%
%BeginExpansion
\sitg
%EndExpansion
}{\star}(d\mathfrak{g}^{\mathbf{\beta}}\wedge\mathfrak{g}_{\mathbf{\beta}%
})-d(\mathfrak{g}^{\mathbf{\alpha}}\wedge\underset{%
%TCIMACRO{\TeXButton{sig}{\sitg}}%
%BeginExpansion
\sitg
%EndExpansion
}{\star}d\mathfrak{g}_{\mathbf{\beta}}) \label{g9}%
\end{equation}
where $d$ is the differential operator and $\underset{%
%TCIMACRO{\TeXButton{sig}{\sitg}}%
%BeginExpansion
\sitg
%EndExpansion
}{\delta}$ is the Hodge coderivative operator given by a $r$-form field
$A_{r}$ by (see e.g., \cite{rodcap2007})
\begin{equation}
\underset{%
%TCIMACRO{\TeXButton{sig}{\sitg}}%
%BeginExpansion
\sitg
%EndExpansion
}{\delta}A_{r}=(-1)^{r}\underset{%
%TCIMACRO{\TeXButton{sig}{\sitg}}%
%BeginExpansion
\sitg
%EndExpansion
}{\star}^{-1}d\underset{%
%TCIMACRO{\TeXButton{sig}{\sitg}}%
%BeginExpansion
\sitg
%EndExpansion
}{\star}A_{r}. \label{hodgecod}%
\end{equation}
The proof of Eq.(\ref{g9}) is in Appendix D. Writing%
\begin{equation}
\mathcal{L}_{g}=-\frac{1}{2}d\mathfrak{g}^{\mathbf{\alpha}}\wedge\underset{%
%TCIMACRO{\TeXButton{sig}{\sitg}}%
%BeginExpansion
\sitg
%EndExpansion
}{\star}d\mathfrak{g}_{\mathbf{\alpha}}+\frac{1}{2}\underset{%
%TCIMACRO{\TeXButton{sig}{\sitg}}%
%BeginExpansion
\sitg
%EndExpansion
}{\delta}\mathfrak{g}^{\mathbf{\alpha}}\wedge\underset{%
%TCIMACRO{\TeXButton{sig}{\sitg}}%
%BeginExpansion
\sitg
%EndExpansion
}{\star}\underset{%
%TCIMACRO{\TeXButton{sig}{\sitg}}%
%BeginExpansion
\sitg
%EndExpansion
}{\delta}\mathfrak{g}_{\mathbf{\alpha}}+\frac{1}{4}d\mathfrak{g}%
^{\mathbf{\alpha}}\wedge\mathfrak{g}_{\mathbf{\alpha}}\wedge\underset{%
%TCIMACRO{\TeXButton{sig}{\sitg}}%
%BeginExpansion
\sitg
%EndExpansion
}{\star}(d\mathfrak{g}^{\mathbf{\beta}}\wedge\mathfrak{g}_{\mathbf{\beta}})
\label{g10}%
\end{equation}
and taking notice of a notable identity \cite{rodquin}%

\begin{equation}
-\frac{1}{2}d\mathfrak{g}^{\mathbf{a}}\wedge\underset{%
%TCIMACRO{\TeXButton{sig}{\sitg}}%
%BeginExpansion
\sitg
%EndExpansion
}{\star}d\mathfrak{g}_{\mathbf{a}}+\frac{1}{2}\underset{%
%TCIMACRO{\TeXButton{sig}{\sitg}}%
%BeginExpansion
\sitg
%EndExpansion
}{\delta}\mathfrak{g}^{\mathbf{a}}\wedge\underset{%
%TCIMACRO{\TeXButton{sig}{\sitg}}%
%BeginExpansion
\sitg
%EndExpansion
}{\star}\underset{%
%TCIMACRO{\TeXButton{sig}{\sitg}}%
%BeginExpansion
\sitg
%EndExpansion
}{\delta}\mathfrak{g}_{\mathbf{a}}=-\frac{1}{2}(d\mathfrak{g}^{\mathbf{a}%
}\wedge\mathfrak{g}^{\mathbf{b}})\wedge\underset{%
%TCIMACRO{\TeXButton{sig}{\sitg}}%
%BeginExpansion
\sitg
%EndExpansion
}{\star}(d\mathfrak{g}_{\mathbf{b}}\wedge\mathfrak{g}_{\mathbf{a}}),
\end{equation}
we can write Eq.(\ref{g10}) as
\begin{equation}
\mathcal{L}_{g}=-\frac{1}{2}(d\mathfrak{g}^{\mathbf{a}}\wedge\mathfrak{g}%
^{\mathbf{b}})\wedge\underset{%
%TCIMACRO{\TeXButton{sig}{\sitg}}%
%BeginExpansion
\sitg
%EndExpansion
}{\star}(d\mathfrak{g}_{\mathbf{b}}\wedge\mathfrak{g}_{\mathbf{a}})+\frac
{1}{4}d\mathfrak{g}^{\mathbf{\alpha}}\wedge\mathfrak{g}_{\mathbf{\alpha}%
}\wedge\underset{%
%TCIMACRO{\TeXButton{sig}{\sitg}}%
%BeginExpansion
\sitg
%EndExpansion
}{\star}(d\mathfrak{g}^{\mathbf{\beta}}\wedge\mathfrak{g}_{\mathbf{\beta}})
\label{g10'}%
\end{equation}
Using Eq.(\ref{4.4a}) and Eq.(\ref{4.4ee}) we can write:
\begin{align}
(d\mathfrak{g}^{\mathbf{\alpha}}\wedge\mathfrak{g}^{\mathbf{\beta}}%
)\wedge\underset{%
%TCIMACRO{\TeXButton{sig}{\sitg}}%
%BeginExpansion
\sitg
%EndExpansion
}{\star}(d\mathfrak{g}_{\mathbf{\alpha}}\wedge\mathfrak{g}_{\mathbf{\beta}})
&  =(d\mathfrak{g}^{\mathbf{\alpha}}\wedge\mathfrak{g}^{\mathbf{\beta}}%
)\wedge\underline{%
%TCIMACRO{\TeXButton{h}{\slh}}%
%BeginExpansion
\slh
%EndExpansion
}^{\dagger}\underset{\eta}{\star}\underline{%
%TCIMACRO{\TeXButton{h}{\slh}}%
%BeginExpansion
\slh
%EndExpansion
}^{\clubsuit}(d\mathfrak{g}_{\mathbf{\alpha}}\wedge\mathfrak{g}_{\mathbf{\beta
}})\nonumber\\
&  =\underline{%
%TCIMACRO{\TeXButton{h}{\slh}}%
%BeginExpansion
\slh
%EndExpansion
}^{\dagger}[\underline{%
%TCIMACRO{\TeXButton{h}{\slh}}%
%BeginExpansion
\slh
%EndExpansion
}^{\clubsuit}d\mathfrak{g}^{\mathbf{\alpha}}\wedge%
%TCIMACRO{\TeXButton{h}{\slh}}%
%BeginExpansion
\slh
%EndExpansion
^{\clubsuit}\mathfrak{g}^{\mathbf{\beta}}\wedge\underset{\eta}{\star
}\underline{%
%TCIMACRO{\TeXButton{h}{\slh}}%
%BeginExpansion
\slh
%EndExpansion
}^{\clubsuit}d\mathfrak{g}_{\mathbf{\alpha}}\wedge%
%TCIMACRO{\TeXButton{h}{\slh}}%
%BeginExpansion
\slh
%EndExpansion
^{\clubsuit}\mathfrak{g}_{\mathbf{\beta}}].\\
& \\
(d\mathfrak{g}^{\mathbf{\alpha}}\wedge\mathfrak{g}_{\mathbf{\alpha}}%
)\wedge\underset{%
%TCIMACRO{\TeXButton{sig}{\sitg}}%
%BeginExpansion
\sitg
%EndExpansion
}{\star}(d\mathfrak{g}^{\mathbf{\beta}}\wedge\mathfrak{g}_{\mathbf{\beta}})
&  =\underline{%
%TCIMACRO{\TeXButton{h}{\slh}}%
%BeginExpansion
\slh
%EndExpansion
}^{\dagger}[\underline{%
%TCIMACRO{\TeXButton{h}{\slh}}%
%BeginExpansion
\slh
%EndExpansion
}^{\clubsuit}d\mathfrak{g}^{\mathbf{\alpha}}\wedge%
%TCIMACRO{\TeXButton{h}{\slh}}%
%BeginExpansion
\slh
%EndExpansion
^{\clubsuit}\mathfrak{g}_{\mathbf{\alpha}}\wedge\underset{\eta}{\star
}\underline{%
%TCIMACRO{\TeXButton{h}{\slh}}%
%BeginExpansion
\slh
%EndExpansion
}^{\clubsuit}d\mathfrak{g}^{\mathbf{\beta}}\wedge%
%TCIMACRO{\TeXButton{h}{\slh}}%
%BeginExpansion
\slh
%EndExpansion
^{\clubsuit}\mathfrak{g}_{\mathbf{\beta}})].
\end{align}
and write $\mathcal{L}_{h}$ $=\underline{%
%TCIMACRO{\TeXButton{h}{\slh}}%
%BeginExpansion
\slh
%EndExpansion
}^{\clubsuit}\mathcal{L}_{g}$ as%
\begin{equation}
\mathcal{L}_{h}=\left(  -\frac{1}{2}\underline{%
%TCIMACRO{\TeXButton{h}{\slh}}%
%BeginExpansion
\slh
%EndExpansion
}^{\clubsuit}(\mathfrak{g}^{\mathbf{\alpha}}\wedge\mathfrak{g}^{\mathbf{\beta
}})\underset{\eta^{-1}}{\cdot}\underline{%
%TCIMACRO{\TeXButton{h}{\slh}}%
%BeginExpansion
\slh
%EndExpansion
}^{\clubsuit}(d\mathfrak{g}_{\mathbf{\alpha}}\wedge\mathfrak{g}_{\mathbf{\beta
}})+\frac{1}{4}\underline{%
%TCIMACRO{\TeXButton{h}{\slh}}%
%BeginExpansion
\slh
%EndExpansion
}^{\clubsuit}(d\mathfrak{g}^{\mathbf{\alpha}}\wedge\mathfrak{g}%
_{\mathbf{\alpha}})\underset{\eta^{-1}}{\cdot}\underline{%
%TCIMACRO{\TeXButton{h}{\slh}}%
%BeginExpansion
\slh
%EndExpansion
}^{\clubsuit}(d\mathfrak{g}^{\mathbf{\beta}}\wedge\mathfrak{g}_{\mathbf{\beta
}})\right)  \tau_{\eta} \label{10.47}%
\end{equation}
\medskip\medskip\medskip

\textbf{Remark 6.1} A careful inspection of the Lagrangian density given by
Eq.(\ref{g10}) reveals its remarkable structure. Indeed, the first term is of
the Yang-Mills kind, the second term may be called a gauge fixing term, since,
e.g., $\underset{%
%TCIMACRO{\TeXButton{sig}{\sitg}}%
%BeginExpansion
\sitg
%EndExpansion
}{\delta}\mathfrak{g}^{\mathbf{\alpha}}=0$ is analogous of the Lorenz gauge in
electromagnetic theory. Finally the third term is an auto-interaction term,
describing the coupling of the \textit{vorticities} $(d\mathfrak{g}%
^{\mathbf{\alpha}}\wedge\mathfrak{g}_{\mathbf{\alpha}}$) of the fields
$\mathfrak{g}^{\mathbf{\alpha}}$.One way of saying that is: the distortion
field $%
%TCIMACRO{\TeXButton{h}{\slh}}%
%BeginExpansion
\slh
%EndExpansion
$ puts the Lorentz vacuum medium in motion. That motion is described by
$\mathfrak{g}^{\mathbf{\alpha}}=$ $%
%TCIMACRO{\TeXButton{h}{\slh}}%
%BeginExpansion
\slh
%EndExpansion
^{\dagger}(%
%TCIMACRO{\TeXButton{vt}{\mbox{\boldmath{$\vartheta$}}}}%
%BeginExpansion
\mbox{\boldmath{$\vartheta$}}%
%EndExpansion
^{\mathbf{\alpha}})$, saying that the deformed cosmic lattice $%
%TCIMACRO{\TeXButton{h}{\slh}}%
%BeginExpansion
\slh
%EndExpansion
^{\dagger}(%
%TCIMACRO{\TeXButton{vt}{\mbox{\boldmath{$\vartheta$}}}}%
%BeginExpansion
\mbox{\boldmath{$\vartheta$}}%
%EndExpansion
^{\mathbf{\alpha}})$ is in motion relative to the ground state cosmic lattice
which is defined (modulo a global Lorentz transformation) by $\{%
%TCIMACRO{\TeXButton{vt}{\mbox{\boldmath{$\vartheta$}}}}%
%BeginExpansion
\mbox{\boldmath{$\vartheta$}}%
%EndExpansion
^{\mathbf{\alpha}}\}$.\medskip

\textbf{Remark 6.2 }In the gravitational theory with the Lagrangian density
given by Eq.(\ref{g10}) no mention of any connection in world manifold
appears. But if some one wants to give a geometrical model interpretation for
such a theory,of course, the simple one is to say that the gravitational field
generates a teleparallel geometry with a \ metric compatible connection
defined on the world manifold by\footnote{The $\{\mathfrak{e}_{\mathbf{\alpha
}}\}$ is the dual basis of $\{\mathfrak{g}^{\mathbf{\alpha}}\}$.}
$\nabla_{\mathfrak{e}_{\alpha}}^{-}\mathfrak{g}^{\mathbf{\beta}}=0,$
$\nabla^{-}%
%TCIMACRO{\TeXButton{g}{\slg}}%
%BeginExpansion
\slg
%EndExpansion
=0$. For that connection the Riemann curvature tensor is null the whereas the
torsion $2$-form fields are given by $\Theta^{\mathbf{\alpha}}=d\mathfrak{g}%
^{\mathbf{\alpha}}$ (i.e., the gravitational field is torsion in this model).
Only on a second thought someone would think to introduce a Levi-Civita
connection $D$ on the world manifold such that \ $D_{\mathfrak{e}_{\alpha}%
}^{-}\mathfrak{g}^{\mathbf{\beta}}=-L_{\mathbf{\alpha\gamma}}^{\mathbf{\beta}%
}\mathfrak{g}^{\gamma}$, $D^{-}%
%TCIMACRO{\TeXButton{g}{\slg}}%
%BeginExpansion
\slg
%EndExpansion
=0$, for which the torsion 2-forms $\Theta^{\mathbf{\alpha}}=d\mathfrak{g}%
^{\mathbf{\alpha}}+\omega_{\mathbf{\beta}}^{\mathbf{\alpha}}\wedge
\mathfrak{g}^{\mathbf{\beta}}=0$ and the curvature 2-forms $\mathcal{R}%
_{\mathbf{\beta}}^{\mathbf{\alpha}}=d\omega_{\mathbf{\beta}}^{\mathbf{\alpha}%
}+\omega_{\mathbf{\beta}}^{\mathbf{\alpha}}\wedge\omega_{\mathbf{\beta}%
}^{\mathbf{\alpha}}\neq0$. However, since history did not follow that path, it
was the geometry associated to the Levi-Civita connection that was first
discovered by Einstein and Grossmann. More on this issue in \cite{nororo2009}%
.\medskip

\textbf{Remark 6.3 }Eq.(\ref{10.47}) express $\mathcal{L}_{h}$ as a functional
of $%
%TCIMACRO{\TeXButton{h}{\slh}}%
%BeginExpansion
\slh
%EndExpansion
^{\clubsuit}$ and uses only objects belonging to the Minkowski spacetime
structure. However to perform the variation of $\mathcal{L}_{h}$ to determine
the equations of motion for the extensor field $%
%TCIMACRO{\TeXButton{h}{\slh}}%
%BeginExpansion
\slh
%EndExpansion
^{\clubsuit}$ is an almost impracticable task, which may leads one to
appreciate the tricks used in Section 5 to derive the equation of motion for
that extensor field $%
%TCIMACRO{\TeXButton{h}{\slh}}%
%BeginExpansion
\slh
%EndExpansion
$. The derivation of the equations of motion directly for the potential fields
$\mathfrak{g}^{\mathbf{a}}$ is a more easy \ (but still involved) task and is
given in Appendix E.

\subsection{The Total Lagrangian for the Massive Gravitational Field Plus the
Matter Fields}

In this section as we study the interaction of the gravitational potentials
$\mathfrak{g}^{\mathbf{a}}$ with the matter fields described by a Lagrangian
density $\mathcal{L}_{m}$. Moreover, as before in Section 5.1.2 we add a term
corresponding to the Lorentz vacuum energy density, the one given in terms of
the cosmological constant, but which we prefer \textit{here} to write as a
`graviton mass' term, i.e., we have:
\begin{equation}
\mathcal{L=L}_{eh}+\frac{1}{2}m^{2}\mathfrak{g}_{\mathbf{\alpha}}%
\wedge\underset{%
%TCIMACRO{\TeXButton{sig}{\sitg}}%
%BeginExpansion
\sitg
%EndExpansion
}{\star}\mathfrak{g}^{\mathbf{\alpha}}+\mathcal{L}_{m} \label{n10.49}%
\end{equation}
or equivalently, since $\mathcal{L}_{eh}$ and $\mathcal{L}_{g}$ differs from
an exact differential.
\begin{align*}
\mathcal{L}  &  \mathcal{=L}_{g}+\frac{1}{2}m^{2}\mathfrak{g}_{\mathbf{\alpha
}}\wedge\underset{%
%TCIMACRO{\TeXButton{sig}{\sitg}}%
%BeginExpansion
\sitg
%EndExpansion
}{\star}\mathfrak{g}^{\mathbf{\alpha}}+\mathcal{L}_{m}\\
&  =\mathcal{L}_{g}^{\prime}+\mathcal{L}_{m}%
\end{align*}

The resulting equations of motion (whose derivation for completeness are given
in the Appendix E) may be written putting%
\begin{equation}
\star t^{\mathbf{\alpha}}=\frac{\partial\mathcal{L}_{g}^{\prime}}%
{\partial\mathfrak{g}_{\mathbf{\alpha}}}\text{, }\star\mathfrak{t}%
^{\mathbf{\alpha}}=\frac{\partial\mathcal{L}_{g}}{\partial\mathfrak{g}%
_{\mathbf{\alpha}}}\text{, }\underset{%
%TCIMACRO{\TeXButton{sig}{\sitg}}%
%BeginExpansion
\sitg
%EndExpansion
}{\star}\mathcal{S}^{\mathbf{\alpha}}=\frac{\partial\mathcal{L}_{g}^{\prime}%
}{\partial d\mathfrak{g}_{\mathbf{\alpha}}}\text{, }-\underset{%
%TCIMACRO{\TeXButton{sig}{\sitg}}%
%BeginExpansion
\sitg
%EndExpansion
}{\star}\mathcal{T}^{\mathbf{\alpha}}=\underset{%
%TCIMACRO{\TeXButton{sig}{\sitg}}%
%BeginExpansion
\sitg
%EndExpansion
}{\star}T^{\mathbf{\alpha}}:=-\frac{\partial\mathcal{L}_{m}}{\partial
\mathfrak{g}_{\mathbf{\alpha}}}. \label{prel}%
\end{equation}

\begin{equation}%
\begin{tabular}
[c]{|l|}\hline
$-\underset{%
%TCIMACRO{\TeXButton{sig}{\sitg}}%
%BeginExpansion
\sitg
%EndExpansion
}{\star}G^{\mathbf{\alpha}}=d\underset{%
%TCIMACRO{\TeXButton{sig}{\sitg}}%
%BeginExpansion
\sitg
%EndExpansion
}{\star}\mathcal{S}^{\mathbf{\alpha}}+\underset{%
%TCIMACRO{\TeXButton{sig}{\sitg}}%
%BeginExpansion
\sitg
%EndExpansion
}{\star}\mathfrak{t}^{\mathbf{\alpha}}=-\underset{%
%TCIMACRO{\TeXButton{sig}{\sitg}}%
%BeginExpansion
\sitg
%EndExpansion
}{\star}\mathcal{T}^{\mathbf{\alpha}},$\\\hline
\end{tabular}
\ \label{gem}%
\end{equation}
where $\underset{%
%TCIMACRO{\TeXButton{sig}{\sitg}}%
%BeginExpansion
\sitg
%EndExpansion
}{\star}G^{\mathbf{\alpha}}\in\sec%
%TCIMACRO{\dbigwedge \nolimits^{3}}%
%BeginExpansion
{\displaystyle\bigwedge\nolimits^{3}}
%EndExpansion
T^{\ast}U$ are the Einstein 3-form fields while $\underset{%
%TCIMACRO{\TeXButton{sig}{\sitg}}%
%BeginExpansion
\sitg
%EndExpansion
}{\star}\mathfrak{t}_{\mathbf{\ }}^{\mathbf{\kappa}}\in\sec%
%TCIMACRO{\dbigwedge \nolimits^{3}}%
%BeginExpansion
{\displaystyle\bigwedge\nolimits^{3}}
%EndExpansion
T^{\ast}U$ and $\underset{%
%TCIMACRO{\TeXButton{sig}{\sitg}}%
%BeginExpansion
\sitg
%EndExpansion
}{\star}\mathcal{S}^{\mathbf{\alpha}}\in\sec%
%TCIMACRO{\dbigwedge \nolimits^{2}}%
%BeginExpansion
{\displaystyle\bigwedge\nolimits^{2}}
%EndExpansion
T^{\ast}U$ are given by (recall Eq.(\ref{special}) and Eq.(\ref{sp'}) of
Appendix E):%

\begin{equation}
\underset{%
%TCIMACRO{\TeXButton{sig}{\sitg}}%
%BeginExpansion
\sitg
%EndExpansion
}{\star}\mathfrak{t}_{\mathbf{\ }}^{\mathbf{\kappa}}=\frac{\partial
\mathcal{L}_{g}^{\prime}}{\partial\mathfrak{g}_{\mathbf{\kappa}}}%
=\frac{\partial\mathcal{L}_{g}}{\partial\mathfrak{g}_{\mathbf{\kappa}}}%
+m^{2}\mathfrak{g}^{\mathbf{\kappa}}=\mathfrak{g}_{\mathbf{\kappa}}\underset{%
%TCIMACRO{\TeXButton{sig}{\sitg}}%
%BeginExpansion
\sitg
%EndExpansion
}{\lrcorner}\mathcal{L}_{g}-(\mathfrak{g}_{\mathbf{\kappa}}\underset{%
%TCIMACRO{\TeXButton{sig}{\sitg}}%
%BeginExpansion
\sitg
%EndExpansion
}{\lrcorner}d\mathfrak{g}_{\mathbf{\alpha}})\wedge\underset{%
%TCIMACRO{\TeXButton{sig}{\sitg}}%
%BeginExpansion
\sitg
%EndExpansion
}{\star}\mathcal{S}^{\mathbf{\alpha}}+m^{2}\mathfrak{g}^{\mathbf{\kappa}}
\label{7.10.16}%
\end{equation}

\begin{equation}
\underset{%
%TCIMACRO{\TeXButton{sig}{\sitg}}%
%BeginExpansion
\sitg
%EndExpansion
}{\star}\mathcal{S}^{\mathbf{\kappa}}=\frac{\partial\mathcal{L}_{g}}{\partial
d\mathfrak{g}_{\mathbf{\kappa}}}=-\mathfrak{g}^{\mathbf{\alpha}}%
\wedge\underset{%
%TCIMACRO{\TeXButton{sig}{\sitg}}%
%BeginExpansion
\sitg
%EndExpansion
}{\star}(d\mathfrak{g}_{\mathbf{\alpha}}\wedge\mathfrak{g}_{\mathbf{\kappa}%
})+\frac{1}{2}\mathfrak{g}_{\mathbf{\kappa}}\wedge\underset{%
%TCIMACRO{\TeXButton{sig}{\sitg}}%
%BeginExpansion
\sitg
%EndExpansion
}{\star}(d\mathfrak{g}^{\mathbf{\alpha}}\wedge\mathfrak{g}_{\mathbf{\alpha}}).
\label{7.10.17}%
\end{equation}

We write moreover
\begin{align}
\underset{%
%TCIMACRO{\TeXButton{sig}{\sitg}}%
%BeginExpansion
\sitg
%EndExpansion
}{\star}\mathcal{S}^{\mathbf{\kappa}}  &  =-\underset{%
%TCIMACRO{\TeXButton{sig}{\sitg}}%
%BeginExpansion
\sitg
%EndExpansion
}{\star}d\mathfrak{g}^{\mathbf{\kappa}}+\underset{%
%TCIMACRO{\TeXButton{sig}{\sitg}}%
%BeginExpansion
\sitg
%EndExpansion
}{\star}\mathfrak{K}^{\mathbf{\kappa}},\label{SS}\\
\underset{%
%TCIMACRO{\TeXButton{sig}{\sitg}}%
%BeginExpansion
\sitg
%EndExpansion
}{\star}\mathfrak{K}^{\mathbf{\kappa}}  &  =-(\mathfrak{g}^{\mathbf{\kappa}%
}\underset{%
%TCIMACRO{\TeXButton{sig}{\sitg}}%
%BeginExpansion
\sitg
%EndExpansion
}{\lrcorner}\underset{%
%TCIMACRO{\TeXButton{sig}{\sitg}}%
%BeginExpansion
\sitg
%EndExpansion
}{\star}\mathfrak{g}^{\mathbf{\alpha}})\wedge\underset{%
%TCIMACRO{\TeXButton{sig}{\sitg}}%
%BeginExpansion
\sitg
%EndExpansion
}{\star}d\underset{%
%TCIMACRO{\TeXButton{sig}{\sitg}}%
%BeginExpansion
\sitg
%EndExpansion
}{\star}\mathfrak{g}_{\mathbf{\alpha}}+\frac{1}{2}\mathfrak{g}^{\mathbf{\kappa
}}\wedge\underset{%
%TCIMACRO{\TeXButton{sig}{\sitg}}%
%BeginExpansion
\sitg
%EndExpansion
}{\star}(d\mathfrak{g}^{\mathbf{\alpha}}\wedge\mathfrak{g}_{\mathbf{\alpha}})
\end{align}
and insert this result in Eq.(\ref{gem}) obtaining:%
\begin{equation}
-d\underset{%
%TCIMACRO{\TeXButton{sig}{\sitg}}%
%BeginExpansion
\sitg
%EndExpansion
}{\star}d\mathfrak{g}^{\mathbf{\kappa}}+m^{2}\underset{%
%TCIMACRO{\TeXButton{sig}{\sitg}}%
%BeginExpansion
\sitg
%EndExpansion
}{\star}\mathfrak{g}^{\mathbf{\kappa}}=-\underset{%
%TCIMACRO{\TeXButton{sig}{\sitg}}%
%BeginExpansion
\sitg
%EndExpansion
}{\star}\left(  t^{\mathbf{\kappa}}+\mathcal{T}^{\mathbf{\kappa}}+\underset{%
%TCIMACRO{\TeXButton{sig}{\sitg}}%
%BeginExpansion
\sitg
%EndExpansion
}{\star}^{-1}d\underset{%
%TCIMACRO{\TeXButton{sig}{\sitg}}%
%BeginExpansion
\sitg
%EndExpansion
}{\star}\mathfrak{K}^{\mathbf{\kappa}}\right)  \label{SSS}%
\end{equation}
Applying apply the operator $\underset{%
%TCIMACRO{\TeXButton{sig}{\sitg}}%
%BeginExpansion
\sitg
%EndExpansion
}{\star}^{-1}$ to both sides of that equation, we get%
\begin{equation}
-\underset{%
%TCIMACRO{\TeXButton{sig}{\sitg}}%
%BeginExpansion
\sitg
%EndExpansion
}{\delta}d\mathfrak{g}^{\mathbf{\kappa}}+m^{2}\mathfrak{g}^{\mathbf{\kappa}%
}=-\left(  t^{\mathbf{\kappa}}+\mathcal{T}^{\mathbf{\kappa}}+\underset{%
%TCIMACRO{\TeXButton{sig}{\sitg}}%
%BeginExpansion
\sitg
%EndExpansion
}{\delta}\mathfrak{K}^{\mathbf{\kappa}}\right)  \label{SD}%
\end{equation}

Calling
\[
\mathfrak{F}^{\mathbf{\kappa}}=d\mathfrak{g}^{\mathbf{\kappa}}%
\]
we have he following \textit{Maxwell like} equations for the gravitational
fields $\mathfrak{F}^{\mathbf{\kappa}}$:%

\begin{subequations}
\label{mlg}%
\begin{align}
d\mathfrak{F}^{\mathbf{\kappa}}  &  =0,\label{a}\\
\underset{%
%TCIMACRO{\TeXButton{sig}{\sitg}}%
%BeginExpansion
\sitg
%EndExpansion
}{\delta}\mathfrak{F}^{\mathbf{\kappa}}  &  =\left(  t^{\mathbf{\kappa}%
}+\mathcal{T}^{\mathbf{\kappa}}+\underset{%
%TCIMACRO{\TeXButton{sig}{\sitg}}%
%BeginExpansion
\sitg
%EndExpansion
}{\delta}\mathfrak{K}^{\mathbf{\kappa}}+m^{2}\mathfrak{g}^{\mathbf{\kappa}%
}\right)  . \label{b}%
\end{align}
However, take notice that the system is non linear in the gravitational
potentials $\mathfrak{g}^{\mathbf{\kappa}}$.

We now may define the total energy-momentum of the system consisting of
the\ matter plus the gravitational field either as or $\mathbf{P=}%
P_{\mathbf{\kappa}}\vartheta^{\mathbf{\kappa}}$%
\end{subequations}
\begin{equation}
P_{\mathbf{\kappa}}=-%
%TCIMACRO{\dint \nolimits_{U}}%
%BeginExpansion
{\displaystyle\int\nolimits_{U}}
%EndExpansion
\underset{%
%TCIMACRO{\TeXButton{sig}{\sitg}}%
%BeginExpansion
\sitg
%EndExpansion
}{\star}(t_{\kappa}+\mathcal{T}_{\mathbf{\kappa}}+\underset{%
%TCIMACRO{\TeXButton{sig}{\sitg}}%
%BeginExpansion
\sitg
%EndExpansion
}{\star}^{-1}d\underset{%
%TCIMACRO{\TeXButton{sig}{\sitg}}%
%BeginExpansion
\sitg
%EndExpansion
}{\star}\mathfrak{K}^{\mathbf{\kappa}}+m^{2}\mathfrak{g}_{\mathbf{\kappa}})=%
%TCIMACRO{\dint \nolimits_{\partial U}}%
%BeginExpansion
{\displaystyle\int\nolimits_{\partial U}}
%EndExpansion
\underset{%
%TCIMACRO{\TeXButton{sig}{\sitg}}%
%BeginExpansion
\sitg
%EndExpansion
}{\star}d\mathfrak{g}_{\mathbf{\kappa}}. \label{em2}%
\end{equation}
or $\mathbf{P}^{\prime}\mathbf{=}P_{\mathbf{\kappa}}^{\prime}\vartheta
^{\mathbf{\kappa}}$%
\begin{equation}
P_{\mathbf{\kappa}}^{\prime}=-%
%TCIMACRO{\dint \nolimits_{U}}%
%BeginExpansion
{\displaystyle\int\nolimits_{U}}
%EndExpansion
\underset{%
%TCIMACRO{\TeXButton{sig}{\sitg}}%
%BeginExpansion
\sitg
%EndExpansion
}{\star}(t_{\kappa}+\mathcal{T}_{\mathbf{\kappa}}+m^{2}\mathfrak{g}%
_{\mathbf{\kappa}})=%
%TCIMACRO{\dint \nolimits_{\partial U}}%
%BeginExpansion
{\displaystyle\int\nolimits_{\partial U}}
%EndExpansion
\underset{%
%TCIMACRO{\TeXButton{sig}{\sitg}}%
%BeginExpansion
\sitg
%EndExpansion
}{\star}\mathcal{S}_{\mathbf{\kappa}}, \label{em1}%
\end{equation}

\subsubsection{Energy-Momentum Conservation Law}

The Maxwell like formulation (Eqs. (\ref{mlg})) or Eq.(\ref{gem}) immediately
imply tow distinct energy-momentum conservation laws in our theory (i.e., both
the $P_{\mathbf{\kappa}}$ as well as the $P_{\mathbf{\kappa}}^{\prime}$ are
conserved), but in order to avoid any misunderstanding some remarks are
necessary.\medskip

\textbf{Remark 6.4 }The values of $P_{\mathbf{\kappa}}$ (Eq.(\ref{em2})) as
well as $P_{\mathbf{\kappa}}^{\prime}$ (Eq.(\ref{em1})) are, of course,
independent of the coordinate system used to calculate them. This is also the
case in \textit{GRT} where identical equations hold for models with global $%
%TCIMACRO{\TeXButton{g}{\slg}}%
%BeginExpansion
\slg
%EndExpansion
$-orthonormal cotetrad fields\textit{.} However, in \textit{GRT} expressions
like $\mathbf{P=}P_{\mathbf{\kappa}}\vartheta^{\mathbf{\kappa}}$ or
$\mathbf{P}^{\prime}\mathbf{=}P_{\mathbf{\kappa}}^{\prime}\vartheta
^{\mathbf{\kappa}}$ have meaning only for asymptotically flat spacetimes.
Also, it is obvious that in our theory as well in \textit{GRT} both
$P_{\mathbf{\kappa}}$ or $P_{\mathbf{\kappa}}^{\prime}$ depend on the choice
of the cotetrad field $\{\mathfrak{g}^{\alpha}\}$. The difference is that in
our theory the $\{\mathfrak{g}^{\alpha}\}$, the gravitational potentials
(defined by the extensor field $%
%TCIMACRO{\TeXButton{h}{\slh}}%
%BeginExpansion
\slh
%EndExpansion
$ modulus an arbitrary local Lorentz rotation $\Lambda$ which is hidden in the
definition of $%
%TCIMACRO{\TeXButton{itg}{\itg}}%
%BeginExpansion
\itg
%EndExpansion
=%
%TCIMACRO{\TeXButton{h}{\slh}}%
%BeginExpansion
\slh
%EndExpansion
^{\dagger}\eta%
%TCIMACRO{\TeXButton{h}{\slh}}%
%BeginExpansion
\slh
%EndExpansion
=%
%TCIMACRO{\TeXButton{h}{\slh}}%
%BeginExpansion
\slh
%EndExpansion
^{\dagger}\Lambda^{\dagger}\eta\Lambda%
%TCIMACRO{\TeXButton{h}{\slh}}%
%BeginExpansion
\slh
%EndExpansion
)$ are by chance a section of the $%
%TCIMACRO{\TeXButton{g}{\slg}}%
%BeginExpansion
\slg
%EndExpansion
$-orthonormal frame bundle of an effective Lorentzian spacetime (with the
$\underset{%
%TCIMACRO{\TeXButton{sig}{\sitg}}%
%BeginExpansion
\sitg
%EndExpansion
}{\star}t_{\kappa}$ being thus gauge dependent objects) whereas in GRT the
$\{\mathfrak{g}^{\alpha}\}$ do not have any interpretation different from the
one of being a section of the Lorentzian spacetime representing a particular
gravitational field. Moreover in \textit{GRT} we can write equations similar
to Eq.(\ref{7.10.16}) also for coordinate basis of the linear frame bundle of
the Lorentzian spacetime, thus obtaining expressions different
\ `energy-mometum pseudo-tensors', for which the analogous integral to
Eq.(\ref{em1}) depends on the coordinate system used for their calculations
\cite{boro,notterod2009}.\medskip

\textbf{Remark 6.5 }The condition of possessing a global $%
%TCIMACRO{\TeXButton{g}{\slg}}%
%BeginExpansion
\slg
%EndExpansion
$-orthonormal tetrad field is, of course, not satisfied by a general
Lorentzian structure\footnote{Such a condition, as well known \cite{geroch,
rodcap2007} is a necessary one for existence of spinor fields} $(M,%
%TCIMACRO{\TeXButton{g}{\slg}}%
%BeginExpansion
\slg
%EndExpansion
)$ and if we want that \textit{GRT} admits a total energy-momentum
conservation law for the system consisting of the matter and gravitational
fields, such a condition must be added as an extra condition for the spacetime
structure, a fact that we already mention in Section 1.\medskip

\textbf{Remark 6.6 }Another very well known fact concerning the
$t^{\mathbf{\alpha}}$ is that their components are not symmetric. Then, for
example in \cite{thirringa}\ \ it is claimed that in order to obtain a angular
momentum conservation law it is necessary to use superpotentials different
from the $\mathcal{S}^{\alpha}$ which produce a symmetric $t^{\mathbf{\alpha}%
}$. The choice made in \cite{thirringa} is then to use the so-called
Landau-Lifschitz pseudo-energy momentum tensor \cite{landau}, which as well
known is a symmetric object\footnote{For more details on those issues see
\cite{notterod2009}.}. However, we will show next that with the original non
symmetric $t^{\mathbf{\alpha}}$ we can get a total angular momentum
conservation law which includes the spin density of the gravitational field.

\subsubsection{Angular Momentum Conservation Law}

Consider the global chart for $\mathbb{R}^{4}$ with coordinates in the
Einstein-Lorentz-Poincar\'{e} gauge $\{\mathtt{x}^{\mu}\}$ and as previously
put \ $%
%TCIMACRO{\TeXButton{vt}{\mbox{\boldmath{$\vartheta$}}}}%
%BeginExpansion
\mbox{\boldmath{$\vartheta$}}%
%EndExpansion
^{\mathbf{\mu}}=d\mathtt{x}^{\mu}$. Calling
\begin{equation}
\underset{%
%TCIMACRO{\TeXButton{sig}{\sitg}}%
%BeginExpansion
\sitg
%EndExpansion
}{\star}\mathbf{T}^{\mathbf{\alpha}}:=\underset{%
%TCIMACRO{\TeXButton{sig}{\sitg}}%
%BeginExpansion
\sitg
%EndExpansion
}{\star}(t^{\kappa}+\mathfrak{T}^{\kappa}++m^{2}\mathfrak{g}^{\kappa})
\label{am1}%
\end{equation}
the (non symmetric) \textit{total} energy momentum 3-forms of matter plus the
gravitational field, we define the \textit{chart dependent} \ density of
\textit{total orbital angular momentum} of the matter as the $3$-form fields%

\begin{equation}
\underset{%
%TCIMACRO{\TeXButton{sig}{\sitg}}%
%BeginExpansion
\sitg
%EndExpansion
}{\star}\mathbf{L}_{m}^{\mathbf{\alpha\beta}}:=\underline{%
%TCIMACRO{\TeXButton{h}{\slh}}%
%BeginExpansion
\slh
%EndExpansion
}^{\dagger}\{%
%TCIMACRO{\TeXButton{h}{\slh}}%
%BeginExpansion
\slh
%EndExpansion
^{\clubsuit}(\mathtt{x}^{\mathbf{\alpha}})\wedge\underset{\eta}{\star}%
%TCIMACRO{\TeXButton{h}{\slh}}%
%BeginExpansion
\slh
%EndExpansion
^{\clubsuit}\mathbf{T}^{\mathbf{\beta}}-%
%TCIMACRO{\TeXButton{h}{\slh}}%
%BeginExpansion
\slh
%EndExpansion
^{\clubsuit}(\mathtt{x}^{\mathbf{\beta}})\wedge\underset{\eta}{\star}%
%TCIMACRO{\TeXButton{h}{\slh}}%
%BeginExpansion
\slh
%EndExpansion
^{\clubsuit}\mathbf{T}^{\mathbf{\alpha}}\}, \label{am1'}%
\end{equation}
which can be written as%

\begin{equation}
\underset{%
%TCIMACRO{\TeXButton{sig}{\sitg}}%
%BeginExpansion
\sitg
%EndExpansion
}{\star}\mathbf{L}_{m}^{\mathbf{\alpha\beta}}=\mathtt{x}^{\mathbf{\alpha}%
}\underset{%
%TCIMACRO{\TeXButton{sig}{\sitg}}%
%BeginExpansion
\sitg
%EndExpansion
}{\star}\mathbf{T}^{\mathbf{\beta}}-\mathtt{x}^{\mathbf{\beta}}\underset{%
%TCIMACRO{\TeXButton{sig}{\sitg}}%
%BeginExpansion
\sitg
%EndExpansion
}{\star}\mathbf{T}^{\mathbf{\alpha}}. \label{am2}%
\end{equation}
Of course,
\begin{equation}
d\underset{%
%TCIMACRO{\TeXButton{sig}{\sitg}}%
%BeginExpansion
\sitg
%EndExpansion
}{\star}\mathbf{L}_{m}^{\mathbf{\alpha\beta}}=%
%TCIMACRO{\TeXButton{vt}{\mbox{\boldmath{$\vartheta$}}}}%
%BeginExpansion
\mbox{\boldmath{$\vartheta$}}%
%EndExpansion
^{\mathbf{\alpha}}\wedge\underset{%
%TCIMACRO{\TeXButton{sig}{\sitg}}%
%BeginExpansion
\sitg
%EndExpansion
}{\star}\mathbf{T}^{\mathbf{\beta}}-%
%TCIMACRO{\TeXButton{vt}{\mbox{\boldmath{$\vartheta$}}}}%
%BeginExpansion
\mbox{\boldmath{$\vartheta$}}%
%EndExpansion
^{\mathbf{\beta}}\wedge\underset{%
%TCIMACRO{\TeXButton{sig}{\sitg}}%
%BeginExpansion
\sitg
%EndExpansion
}{\star}\mathbf{T}^{\mathbf{\alpha}}\neq0. \label{am3}%
\end{equation}

However, let us define the density ($3$-forms) of \textit{orbital }angular
momentum of the gravitational field by%
\begin{equation}
\underset{%
%TCIMACRO{\TeXButton{sig}{\sitg}}%
%BeginExpansion
\sitg
%EndExpansion
}{\star}\mathbf{L}_{g}^{\mathbf{\alpha\beta}}:=\underline{%
%TCIMACRO{\TeXButton{h}{\slh}}%
%BeginExpansion
\slh
%EndExpansion
}^{\dagger}\{\mathfrak{g}^{\mathbf{\alpha}}\wedge\underset{\eta}{\star}%
%TCIMACRO{\TeXButton{h}{\slh}}%
%BeginExpansion
\slh
%EndExpansion
^{\clubsuit}\mathfrak{F}^{\mathbf{\beta}}-\mathfrak{g}^{\mathbf{\beta}}%
\wedge\underset{\eta}{\star}%
%TCIMACRO{\TeXButton{h}{\slh}}%
%BeginExpansion
\slh
%EndExpansion
^{\clubsuit}\mathfrak{F}^{\mathbf{\alpha}}\}, \label{am3'}%
\end{equation}
which can be written as%
\begin{equation}
\underset{%
%TCIMACRO{\TeXButton{sig}{\sitg}}%
%BeginExpansion
\sitg
%EndExpansion
}{\star}\mathbf{L}_{g}^{\mathbf{\alpha\beta}}=%
%TCIMACRO{\TeXButton{vt}{\mbox{\boldmath{$\vartheta$}}}}%
%BeginExpansion
\mbox{\boldmath{$\vartheta$}}%
%EndExpansion
^{\mathbf{\alpha}}\wedge\underset{%
%TCIMACRO{\TeXButton{sig}{\sitg}}%
%BeginExpansion
\sitg
%EndExpansion
}{\star}\mathcal{S}^{\beta}-%
%TCIMACRO{\TeXButton{vt}{\mbox{\boldmath{$\vartheta$}}}}%
%BeginExpansion
\mbox{\boldmath{$\vartheta$}}%
%EndExpansion
^{\mathbf{\beta}}\wedge\underset{%
%TCIMACRO{\TeXButton{sig}{\sitg}}%
%BeginExpansion
\sitg
%EndExpansion
}{\star}\mathcal{S}^{\mathbf{\alpha}}, \label{am4}%
\end{equation}
and the density of \textit{total orbital angular momentum} of the matter\ plus
the gravitational field by%
\begin{equation}
\underset{%
%TCIMACRO{\TeXButton{sig}{\sitg}}%
%BeginExpansion
\sitg
%EndExpansion
}{\star}\mathbf{L}_{t}^{\mathbf{\alpha\beta}}=\underset{%
%TCIMACRO{\TeXButton{sig}{\sitg}}%
%BeginExpansion
\sitg
%EndExpansion
}{\star}\mathbf{L}_{m}^{\mathbf{\alpha\beta}}+\underset{%
%TCIMACRO{\TeXButton{sig}{\sitg}}%
%BeginExpansion
\sitg
%EndExpansion
}{\star}\mathbf{L}_{g}^{\mathbf{\alpha\beta}}. \label{am5}%
\end{equation}
Then we immediately get that%
\begin{align}
d\underset{%
%TCIMACRO{\TeXButton{sig}{\sitg}}%
%BeginExpansion
\sitg
%EndExpansion
}{\star}\mathbf{L}_{t}^{\mathbf{\alpha\beta}}  &  =%
%TCIMACRO{\TeXButton{vt}{\mbox{\boldmath{$\vartheta$}}}}%
%BeginExpansion
\mbox{\boldmath{$\vartheta$}}%
%EndExpansion
^{\mathbf{\alpha}}\wedge\underset{%
%TCIMACRO{\TeXButton{sig}{\sitg}}%
%BeginExpansion
\sitg
%EndExpansion
}{\star}\mathbf{T}^{\mathbf{\beta}}-%
%TCIMACRO{\TeXButton{vt}{\mbox{\boldmath{$\vartheta$}}}}%
%BeginExpansion
\mbox{\boldmath{$\vartheta$}}%
%EndExpansion
^{\mathbf{\beta}}\wedge\underset{%
%TCIMACRO{\TeXButton{sig}{\sitg}}%
%BeginExpansion
\sitg
%EndExpansion
}{\star}\mathbf{T}^{\mathbf{\alpha}}-%
%TCIMACRO{\TeXButton{vt}{\mbox{\boldmath{$\vartheta$}}}}%
%BeginExpansion
\mbox{\boldmath{$\vartheta$}}%
%EndExpansion
^{\mathbf{\alpha}}\wedge d\underset{%
%TCIMACRO{\TeXButton{sig}{\sitg}}%
%BeginExpansion
\sitg
%EndExpansion
}{\star}\mathcal{S}^{\beta}+%
%TCIMACRO{\TeXButton{vt}{\mbox{\boldmath{$\vartheta$}}}}%
%BeginExpansion
\mbox{\boldmath{$\vartheta$}}%
%EndExpansion
^{\mathbf{\beta}}\wedge d\underset{%
%TCIMACRO{\TeXButton{sig}{\sitg}}%
%BeginExpansion
\sitg
%EndExpansion
}{\star}\mathcal{S}^{\alpha}\nonumber\\
&  =%
%TCIMACRO{\TeXButton{vt}{\mbox{\boldmath{$\vartheta$}}}}%
%BeginExpansion
\mbox{\boldmath{$\vartheta$}}%
%EndExpansion
^{\mathbf{\alpha}}\wedge\underset{%
%TCIMACRO{\TeXButton{sig}{\sitg}}%
%BeginExpansion
\sitg
%EndExpansion
}{\star}\mathbf{T}^{\mathbf{\beta}}-%
%TCIMACRO{\TeXButton{vt}{\mbox{\boldmath{$\vartheta$}}}}%
%BeginExpansion
\mbox{\boldmath{$\vartheta$}}%
%EndExpansion
^{\mathbf{\beta}}\wedge\underset{%
%TCIMACRO{\TeXButton{sig}{\sitg}}%
%BeginExpansion
\sitg
%EndExpansion
}{\star}\mathbf{T}^{\mathbf{\alpha}}-%
%TCIMACRO{\TeXButton{vt}{\mbox{\boldmath{$\vartheta$}}}}%
%BeginExpansion
\mbox{\boldmath{$\vartheta$}}%
%EndExpansion
^{\mathbf{\alpha}}\wedge\underset{%
%TCIMACRO{\TeXButton{sig}{\sitg}}%
%BeginExpansion
\sitg
%EndExpansion
}{\star}\mathbf{T}^{\mathbf{\beta}}+%
%TCIMACRO{\TeXButton{vt}{\mbox{\boldmath{$\vartheta$}}}}%
%BeginExpansion
\mbox{\boldmath{$\vartheta$}}%
%EndExpansion
^{\mathbf{\beta}}\wedge\underset{%
%TCIMACRO{\TeXButton{sig}{\sitg}}%
%BeginExpansion
\sitg
%EndExpansion
}{\star}\mathbf{T}^{\mathbf{\alpha}}\nonumber\\
&  =0. \label{am6}%
\end{align}
However, Eq.(\ref{am6}) does not contains yet the spin angular momentum of
matter and the spin angular momentum of the gravitational field and thus
cannot be considered as a satisfactory equation for the conservation of total
angular momentum. To go on, we proceed as follows. First we write the total
Lagrangian density for the gravitational field as
\begin{equation}
\mathcal{L=L}_{eh}+\mathcal{L}_{m}, \label{am7}%
\end{equation}
where $\mathcal{L}_{eh}$ is defined by Eq.(\ref{g8}), i.e.,
\begin{equation}
\mathcal{L}_{eh}=\frac{1}{2}\mathcal{R}_{\mathbf{\kappa\iota}}\wedge\underset{%
%TCIMACRO{\TeXButton{sig}{\sitg}}%
%BeginExpansion
\sitg
%EndExpansion
}{\star}(\mathfrak{g}^{\mathbf{\kappa}}\wedge\mathfrak{g}^{\mathbf{\iota}}),
\label{am8}%
\end{equation}
but where now we suppose that $\mathcal{L}_{eh}$ is a functional of the
gravitational potentials $\mathfrak{g}^{\mathbf{\kappa}}$ and the connection
$1$-form fields $\omega_{\mathbf{\kappa\iota}}$ appearing in $\mathcal{R}%
_{\mathbf{\kappa\iota}}$ which we take as independent variables to start.
Moreover,\ we suppose that $\mathcal{L}_{m}$ depends on the field variables
$\phi^{A}$ (which in general are indexed form fields) and their exterior
covariant derivatives, i.e., it is a functional of the kind $\mathcal{L}%
_{m}=\mathcal{L}_{m}(\phi^{A},d\phi^{A},\omega_{\mathbf{\kappa\iota}})$.

If we make the variation of the action $%
%TCIMACRO{\dint \nolimits_{U}}%
%BeginExpansion
{\displaystyle\int\nolimits_{U}}
%EndExpansion
(\mathcal{L}_{eh}+\mathcal{L}_{m})$ with respect to $\omega_{\mathbf{\kappa
\iota}}$ we get%
\begin{align}
&
%TCIMACRO{\TeXButton{delta}{\mbox{\boldmath{$\delta$}}}}%
%BeginExpansion
\mbox{\boldmath{$\delta$}}%
%EndExpansion%
%TCIMACRO{\dint _{U}}%
%BeginExpansion
{\displaystyle\int_{U}}
%EndExpansion
(\mathcal{L}_{eh}+\mathcal{L}_{m})\nonumber\\
&  =%
%TCIMACRO{\dint _{U}}%
%BeginExpansion
{\displaystyle\int_{U}}
%EndExpansion
(%
%TCIMACRO{\TeXButton{delta}{\mbox{\boldmath{$\delta$}}}}%
%BeginExpansion
\mbox{\boldmath{$\delta$}}%
%EndExpansion
\omega_{\mathbf{\kappa\iota}}\wedge\frac{\partial\mathcal{L}_{eh}}%
{\partial\omega_{\mathbf{\kappa\iota}}}+%
%TCIMACRO{\TeXButton{delta}{\mbox{\boldmath{$\delta$}}}}%
%BeginExpansion
\mbox{\boldmath{$\delta$}}%
%EndExpansion
d\omega_{\mathbf{\kappa\iota}}\wedge\frac{\partial\mathcal{L}_{eh}}{\partial
d\omega_{\mathbf{\kappa\iota}}}+%
%TCIMACRO{\TeXButton{delta}{\mbox{\boldmath{$\delta$}}}}%
%BeginExpansion
\mbox{\boldmath{$\delta$}}%
%EndExpansion
\omega_{\mathbf{\kappa\iota}}\wedge\frac{\partial\mathcal{L}_{m}}%
{\partial\omega_{\mathbf{\kappa\iota}}})\nonumber\\
&  =%
%TCIMACRO{\dint _{U}}%
%BeginExpansion
{\displaystyle\int_{U}}
%EndExpansion
(%
%TCIMACRO{\TeXButton{delta}{\mbox{\boldmath{$\delta$}}}}%
%BeginExpansion
\mbox{\boldmath{$\delta$}}%
%EndExpansion
\omega_{\mathbf{\kappa\iota}}\wedge\left[  \frac{\partial\mathcal{L}_{eh}%
}{\partial\omega_{\mathbf{\kappa\iota}}}+d\left(  \frac{\partial
\mathcal{L}_{eh}}{\partial d\omega_{\mathbf{\kappa\iota}}}\right)
+\frac{\partial\mathcal{L}_{m}}{\partial\omega_{\mathbf{\kappa\iota}}}\right]
+d\left(
%TCIMACRO{\TeXButton{delta}{\mbox{\boldmath{$\delta$}}}}%
%BeginExpansion
\mbox{\boldmath{$\delta$}}%
%EndExpansion
\omega_{\mathbf{\kappa\iota}}\wedge\frac{\partial\mathcal{L}_{eh}}{\partial
d\omega_{\mathbf{\kappa\iota}}}\right)  , \label{am9}%
\end{align}
from where we obtain from the principle of stationary action and with the
usual hypothesis that the variations vanishes on the boundary $\partial U$ of
$U$ that
\begin{equation}
\frac{\partial\mathcal{L}_{eh}}{\partial\omega_{\mathbf{\kappa\iota}}%
}+d\left(  \frac{\partial\mathcal{L}_{eh}}{\partial d\omega_{\mathbf{\kappa
\iota}}}\right)  +\frac{\partial\mathcal{L}_{m}}{\partial\omega
_{\mathbf{\kappa\iota}}}=0 \label{am10}%
\end{equation}
In metric affine theories the spin \cite{benn} (density) angular momentum of
matter is defined by
\begin{equation}
\underset{%
%TCIMACRO{\TeXButton{sig}{\sitg}}%
%BeginExpansion
\sitg
%EndExpansion
}{\star}\mathbf{S}_{m}^{\mathbf{\kappa\iota}}=\frac{\partial\mathcal{L}_{m}%
}{\partial\omega_{\mathbf{\kappa\iota}}}. \label{am11}%
\end{equation}
and we will accept it as a good one. Now,
\begin{align}
\frac{\partial\mathcal{L}_{eh}}{\partial\omega_{\mathbf{\kappa\iota}}}  &
=\frac{1}{2}\left(  \omega_{\mathbf{\alpha}}^{\mathbf{\kappa}}\wedge\underset{%
%TCIMACRO{\TeXButton{sig}{\sitg}}%
%BeginExpansion
\sitg
%EndExpansion
}{\star}(\mathfrak{g}^{\mathbf{\alpha}}\wedge\mathfrak{g}^{\mathbf{\iota}%
})+\omega_{\mathbf{\alpha}}^{\mathbf{\iota}}\wedge\underset{%
%TCIMACRO{\TeXButton{sig}{\sitg}}%
%BeginExpansion
\sitg
%EndExpansion
}{\star}(\mathfrak{g}^{\mathbf{\kappa}}\wedge\mathfrak{g}^{\mathbf{\alpha}%
})\right)  ,\nonumber\\
d\left(  \frac{\partial\mathcal{L}_{eh}}{\partial d\omega_{\mathbf{\kappa
\iota}}}\right)   &  =\frac{1}{2}\underset{%
%TCIMACRO{\TeXButton{sig}{\sitg}}%
%BeginExpansion
\sitg
%EndExpansion
}{\star}(\mathfrak{g}^{\mathbf{\kappa}}\wedge\mathfrak{g}^{\mathbf{l}}),
\label{am12}%
\end{align}
Now, calling
\begin{equation}
\underset{%
%TCIMACRO{\TeXButton{sig}{\sitg}}%
%BeginExpansion
\sitg
%EndExpansion
}{\star}\mathbf{S}_{g}^{\mathbf{\kappa\iota}}=\omega_{\mathbf{\alpha}%
}^{\mathbf{\kappa}}\wedge\underset{%
%TCIMACRO{\TeXButton{sig}{\sitg}}%
%BeginExpansion
\sitg
%EndExpansion
}{\star}(\mathfrak{g}^{\mathbf{\alpha}}\wedge\mathfrak{g}^{\mathbf{\iota}%
})+\omega_{\mathbf{\alpha}}^{\mathbf{\iota}}\wedge\underset{%
%TCIMACRO{\TeXButton{sig}{\sitg}}%
%BeginExpansion
\sitg
%EndExpansion
}{\star}(\mathfrak{g}^{\mathbf{\kappa}}\wedge\mathfrak{g}^{\mathbf{\alpha}})
\label{am13}%
\end{equation}
the spin (density) angular momentum of the gravitational field, we get
\begin{equation}
d\underset{%
%TCIMACRO{\TeXButton{sig}{\sitg}}%
%BeginExpansion
\sitg
%EndExpansion
}{\star}(\mathfrak{g}^{\mathbf{\kappa}}\wedge\mathfrak{g}^{\mathbf{l}%
})=-2\underset{%
%TCIMACRO{\TeXButton{sig}{\sitg}}%
%BeginExpansion
\sitg
%EndExpansion
}{\star}\left(  \mathbf{S}_{m}^{\mathbf{\kappa\iota}}+\mathbf{S}%
_{g}^{\mathbf{\kappa\iota}}\right)  , \label{am14}%
\end{equation}
and thus\footnote{An equations anlogous to this one has been found originally
by Bramsom \cite{bram}.}
\begin{equation}
d\underset{%
%TCIMACRO{\TeXButton{sig}{\sitg}}%
%BeginExpansion
\sitg
%EndExpansion
}{\star}\left(  \mathbf{S}_{m}^{\mathbf{\kappa\iota}}+\mathbf{S}%
_{g}^{\mathbf{\kappa\iota}}\right)  =0. \label{am15}%
\end{equation}
We now define the total angular momentum of the matter field plus the
gravitational field as%
\begin{equation}
\underset{%
%TCIMACRO{\TeXButton{sig}{\sitg}}%
%BeginExpansion
\sitg
%EndExpansion
}{\star}\mathbf{J}^{\mathbf{\alpha\beta}}=\underset{%
%TCIMACRO{\TeXButton{sig}{\sitg}}%
%BeginExpansion
\sitg
%EndExpansion
}{\star}(\mathbf{L}_{m}^{\mathbf{\alpha\beta}}+\mathbf{S}_{m}^{\mathbf{\kappa
\iota}})+\underset{%
%TCIMACRO{\TeXButton{sig}{\sitg}}%
%BeginExpansion
\sitg
%EndExpansion
}{\star}(\mathbf{L}_{g}^{\mathbf{\alpha\beta}}++\mathbf{S}_{g}^{\mathbf{\kappa
\iota}}), \label{am16}%
\end{equation}
and due Eq.(\ref{am6}) and Eq.(\ref{am15}) we have%
\begin{equation}
d\underset{%
%TCIMACRO{\TeXButton{sig}{\sitg}}%
%BeginExpansion
\sitg
%EndExpansion
}{\star}\mathbf{J}^{\mathbf{\alpha\beta}}=0. \label{am17}%
\end{equation}
Eq.(\ref{am17}) express the law of conservation of total angular momentum in
our theory.

\subsubsection{Wave Equation for the $\mathfrak{g}^{\mathbf{\kappa}}$}

Wave equation for the gravitational potential $\mathfrak{g}^{\mathbf{\kappa}}$
can be written in the \textit{effective} Lorentzian spacetime structure $(M,%
%TCIMACRO{\TeXButton{g}{\slg}}%
%BeginExpansion
\slg
%EndExpansion
,D,\tau_{%
%TCIMACRO{\TeXButton{sg}{\sslg}}%
%BeginExpansion
\sslg
%EndExpansion
},\uparrow)$ already introduced. To get the wave equation we add the term
$d\underset{%
%TCIMACRO{\TeXButton{sig}{\sitg}}%
%BeginExpansion
\sitg
%EndExpansion
}{\delta}\mathfrak{g}^{\kappa}$ to both members Eq.(\ref{SD}) getting
\begin{equation}
-\underset{%
%TCIMACRO{\TeXButton{sig}{\sitg}}%
%BeginExpansion
\sitg
%EndExpansion
}{\delta}d\mathfrak{g}^{\kappa}-d\underset{%
%TCIMACRO{\TeXButton{sig}{\sitg}}%
%BeginExpansion
\sitg
%EndExpansion
}{\delta}\mathfrak{g}^{\kappa}+m^{2}\mathfrak{g}^{\kappa}=-\left(
\mathfrak{t}^{\kappa}+\mathfrak{T}^{\kappa}+\underset{%
%TCIMACRO{\TeXButton{sig}{\sitg}}%
%BeginExpansion
\sitg
%EndExpansion
}{\delta}\mathfrak{K}^{\kappa}+d\underset{%
%TCIMACRO{\TeXButton{sig}{\sitg}}%
%BeginExpansion
\sitg
%EndExpansion
}{\delta}\mathfrak{g}^{\kappa}\right)  . \label{cl2}%
\end{equation}
We now recall the definition of the Hodge D'Alembertian, which in the Clifford
bundle formalism is the square of the Dirac operator (${%
%TCIMACRO{\TeXButton{dirac}{\mbox{\boldmath$\partial$}}}%
%BeginExpansion
\mbox{\boldmath$\partial$}%
%EndExpansion
}:=\mathfrak{g}^{\kappa}\underset{%
%TCIMACRO{\TeXButton{sig}{\sitg}}%
%BeginExpansion
\sitg
%EndExpansion
}{}D_{\mathfrak{g}_{\mathbf{\kappa}}}^{-}=d-\underset{%
%TCIMACRO{\TeXButton{sig}{\sitg}}%
%BeginExpansion
\sitg
%EndExpansion
}{\delta}$) acting on \ multiform fields \cite{rodcap2007}, i.e., we can
write
\begin{equation}
\mathring{\lozenge}\mathfrak{g}^{\kappa}:={%
%TCIMACRO{\TeXButton{dirac}{\mbox{\boldmath$\partial$}}}%
%BeginExpansion
\mbox{\boldmath$\partial$}%
%EndExpansion
}^{2}\mathfrak{g}^{\kappa}=(-\underset{%
%TCIMACRO{\TeXButton{sig}{\sitg}}%
%BeginExpansion
\sitg
%EndExpansion
}{\delta}d-d\underset{%
%TCIMACRO{\TeXButton{sig}{\sitg}}%
%BeginExpansion
\sitg
%EndExpansion
}{\delta})\mathfrak{g}^{\kappa}. \label{cl3}%
\end{equation}
We recall moreover the following nontrivial decomposition \cite{rodcap2007} of
${%
%TCIMACRO{\TeXButton{dirac}{\mbox{\boldmath$\partial$}}}%
%BeginExpansion
\mbox{\boldmath$\partial$}%
%EndExpansion
}^{2}$,%
\begin{equation}
{%
%TCIMACRO{\TeXButton{dirac}{\mbox{\boldmath$\partial$}}}%
%BeginExpansion
\mbox{\boldmath$\partial$}%
%EndExpansion
}\,^{2}\mathfrak{g}^{\kappa}={%
%TCIMACRO{\TeXButton{dirac}{\mbox{\boldmath$\partial$}}}%
%BeginExpansion
\mbox{\boldmath$\partial$}%
%EndExpansion
}\underset{%
%TCIMACRO{\TeXButton{sig}{\sitg}}%
%BeginExpansion
\sitg
%EndExpansion
}{\cdot}{%
%TCIMACRO{\TeXButton{dirac}{\mbox{\boldmath$\partial$}}}%
%BeginExpansion
\mbox{\boldmath$\partial$}%
%EndExpansion
}\mathfrak{g}^{\kappa}+{%
%TCIMACRO{\TeXButton{dirac}{\mbox{\boldmath$\partial$}}}%
%BeginExpansion
\mbox{\boldmath$\partial$}%
%EndExpansion
}\wedge{%
%TCIMACRO{\TeXButton{dirac}{\mbox{\boldmath$\partial$}}}%
%BeginExpansion
\mbox{\boldmath$\partial$}%
%EndExpansion
}\mathfrak{g}^{\kappa}, \label{cl4}%
\end{equation}
where $\square=%
%TCIMACRO{\TeXButton{sd}{\bpartial}}%
%BeginExpansion
\bpartial
%EndExpansion
\underset{%
%TCIMACRO{\TeXButton{sig}{\sitg}}%
%BeginExpansion
\sitg
%EndExpansion
}{\cdot}%
%TCIMACRO{\TeXButton{sd}{\bpartial}}%
%BeginExpansion
\bpartial
%EndExpansion
$ is the \textit{covariant D'Alembertian} and $%
%TCIMACRO{\TeXButton{sd}{\bpartial}}%
%BeginExpansion
\bpartial
%EndExpansion
\wedge%
%TCIMACRO{\TeXButton{sd}{\bpartial}}%
%BeginExpansion
\bpartial
%EndExpansion
$ is the \textit{Ricci operator associated to the Levi-Civita connection of }$%
%TCIMACRO{\TeXButton{itg}{\itg}}%
%BeginExpansion
\itg
%EndExpansion
$. If
\begin{equation}
D_{\mathfrak{g}_{\mathbf{\alpha}}}^{-}\mathfrak{g}^{\beta}=-L_{\mathbf{\alpha
\rho}}^{\mathbf{\beta}}\mathfrak{g}_{\mathbf{\rho}},
\end{equation}
then%
\begin{equation}%
\begin{tabular}
[c]{c}%
(a)\\
(b)
\end{tabular}%
\begin{array}
[c]{ccl}%
{%
%TCIMACRO{\TeXButton{dirac}{\mbox{\boldmath$\partial$}}}%
%BeginExpansion
\mbox{\boldmath$\partial$}%
%EndExpansion
}\underset{%
%TCIMACRO{\TeXButton{sig}{\sitg}}%
%BeginExpansion
\sitg
%EndExpansion
}{{\cdot}}{%
%TCIMACRO{\TeXButton{dirac}{\mbox{\boldmath$\partial$}}}%
%BeginExpansion
\mbox{\boldmath$\partial$}%
%EndExpansion
} & = & \eta^{\mathbf{\alpha\beta}}(D_{\mathfrak{g}_{\mathbf{\alpha}}}%
^{-}D_{\mathfrak{g}_{\beta}}^{-}-L_{\mathbf{\alpha\beta}}^{\mathbf{\rho}%
}D_{\mathfrak{g}_{\rho}}^{-})\\
{%
%TCIMACRO{\TeXButton{dirac}{\mbox{\boldmath$\partial$}}}%
%BeginExpansion
\mbox{\boldmath$\partial$}%
%EndExpansion
}\wedge{%
%TCIMACRO{\TeXButton{dirac}{\mbox{\boldmath$\partial$}}}%
%BeginExpansion
\mbox{\boldmath$\partial$}%
%EndExpansion
} & = & \mathfrak{g}^{\alpha}\wedge\mathfrak{g}^{\beta}(D_{\mathfrak{g}%
_{\mathbf{\alpha}}}^{-}D_{\mathfrak{g}_{\mathbf{\beta}}}^{-}-L_{\mathbf{\alpha
\beta}}^{\mathbf{\rho}}D_{\mathfrak{g}_{\mathbf{\rho}}}^{-}),
\end{array}
\label{1792}%
\end{equation}
and an easily computation shows that%
\begin{equation}
{%
%TCIMACRO{\TeXButton{dirac}{\mbox{\boldmath$\partial$}}}%
%BeginExpansion
\mbox{\boldmath$\partial$}%
%EndExpansion
}\wedge{%
%TCIMACRO{\TeXButton{dirac}{\mbox{\boldmath$\partial$}}}%
%BeginExpansion
\mbox{\boldmath$\partial$}%
%EndExpansion
}\mathfrak{g}^{\kappa}{=}\mathcal{R}^{\mathbf{\kappa}},
\end{equation}
where $\mathcal{R}^{\mathbf{\kappa}}:U\rightarrow%
%TCIMACRO{\dbigwedge \nolimits^{1}}%
%BeginExpansion
{\displaystyle\bigwedge\nolimits^{1}}
%EndExpansion
U$ are the Ricci 1-form fields, given by:%
\begin{equation}
\mathcal{R}^{\mathbf{\kappa}}=R_{\mathbf{\iota}}^{\mathbf{\kappa}}%
\mathfrak{g}^{\iota}, \label{r1f}%
\end{equation}
where $R_{\mathbf{\iota}}^{\mathbf{\kappa}}$ are the components of the Ricci
tensor in the basis defined by $\{\mathfrak{g}^{\mathbf{\alpha}}\}$. Then we
have the following wave equations for the gravitational potentials:%
\begin{equation}
\square\mathfrak{g}^{\kappa}+m^{2}\mathfrak{g}^{\kappa}=-(\mathfrak{T}%
^{\kappa}+\mathfrak{t}^{\kappa}+\underset{%
%TCIMACRO{\TeXButton{sig}{\sitg}}%
%BeginExpansion
\sitg
%EndExpansion
}{\delta}\mathfrak{K}^{\kappa}+d\underset{%
%TCIMACRO{\TeXButton{sig}{\sitg}}%
%BeginExpansion
\sitg
%EndExpansion
}{\delta}\mathfrak{g}^{\kappa}). \label{weg}%
\end{equation}

\section{Hamiltonian Formalism}

\subsection{The Hamiltonian $3$-form Density $\mathcal{H}$}

We start with the Lagrangian density\footnote{A rigorous formulation of the
Lagrangian and Hamiltonian formalism in field theory needs at least the
introduction of the concepts of jet bundles and the Legendre bundles. Such a
theory is finely presented, e.g., in the excellent texts \cite{gimasa,kiwlo}.
Here we give a formulation of the theory without mentioning those concepts,
but which is very similar to the standard one used in physicists books on
field theory, and which for the best of our knowledge has ben first used in
\cite{tw78} \ and developed with maestry in \cite{wallner2} (see also
\cite{ko}). Also, we recall that the Hamiltonian formalism for \textit{GRT} as
usualy presented on some textbooks has been introduced in \cite{adm}, where
the concept of \textit{ADM} energy first appeared. See Section 7.3.} given by
Eq.(\ref{g10}), i.e.,%
\begin{equation}
\mathcal{L}_{g}(\mathfrak{g}^{\mathbf{\alpha}},d\mathfrak{g}^{\mathbf{\alpha}%
})=-\frac{1}{2}d\mathfrak{g}^{\mathbf{\alpha}}\wedge\underset{%
%TCIMACRO{\TeXButton{sig}{\sitg}}%
%BeginExpansion
\sitg
%EndExpansion
}{\star}d\mathfrak{g}_{\mathbf{\alpha}}+\frac{1}{2}\underset{%
%TCIMACRO{\TeXButton{sig}{\sitg}}%
%BeginExpansion
\sitg
%EndExpansion
}{\delta}\mathfrak{g}^{\mathbf{\alpha}}\wedge\underset{%
%TCIMACRO{\TeXButton{sig}{\sitg}}%
%BeginExpansion
\sitg
%EndExpansion
}{\star}\underset{%
%TCIMACRO{\TeXButton{sig}{\sitg}}%
%BeginExpansion
\sitg
%EndExpansion
}{\delta}\mathfrak{g}_{\mathbf{\alpha}}+\frac{1}{4}d\mathfrak{g}%
^{\mathbf{\alpha}}\wedge\mathfrak{g}_{\mathbf{\alpha}}\wedge\underset{%
%TCIMACRO{\TeXButton{sig}{\sitg}}%
%BeginExpansion
\sitg
%EndExpansion
}{\star}(d\mathfrak{g}^{\mathbf{\beta}}\wedge\mathfrak{g}_{\mathbf{\beta}}),
\label{g10bis}%
\end{equation}
and as usual in the Lagrangian formalism we define the conjugate momenta
$\mathfrak{p}_{\mathbf{\alpha}}\in\sec%
%TCIMACRO{\dbigwedge \nolimits^{3}}%
%BeginExpansion
{\displaystyle\bigwedge\nolimits^{3}}
%EndExpansion
T^{\ast}M\hookrightarrow\mathcal{C\ell(}\mathtt{g},M\mathcal{)}$ to the
potentials $\mathfrak{g}^{\mathbf{\alpha}}$ by:%
\begin{equation}
\mathfrak{p}_{\mathbf{\alpha}}=\frac{\partial\mathcal{L}_{g}}{\partial
d\mathfrak{g}^{\mathbf{\alpha}}}, \label{h2}%
\end{equation}
and before proceeding we recall that according to Eq.(\ref{7.10.17}), it is:%
\begin{equation}
\mathfrak{p}_{\mathbf{\alpha}}\equiv\underset{%
%TCIMACRO{\TeXButton{sig}{\sitg}}%
%BeginExpansion
\sitg
%EndExpansion
}{\star}\mathcal{S}_{\mathbf{\alpha}}. \label{h2'}%
\end{equation}
Next, supposing that we can solve Eq.(\ref{h2}) for the $d\mathfrak{g}%
^{\mathbf{\alpha}}$ as functions of the $\mathfrak{p}_{\mathbf{\alpha}}$, we
introduce a \textit{Legendre} transformation with respect to the fields
$d\mathfrak{g}^{\mathbf{\alpha}}$ by
\begin{equation}
\mathbf{L}:(\mathfrak{g}^{\mathbf{\alpha}},\mathfrak{p}_{\mathbf{\alpha}%
})\mapsto\mathbf{L}(\mathfrak{g}^{\mathbf{\alpha}},\mathfrak{p}%
_{\mathbf{\alpha}})=d\mathfrak{g}^{\mathbf{\alpha}}\wedge p_{\mathbf{\alpha}%
}-\mathcal{L}_{g}(\mathfrak{g}^{\mathbf{\alpha}},d\mathfrak{g}^{\mathbf{\alpha
}}(\mathfrak{p}_{\mathbf{\alpha}})) \label{h3}%
\end{equation}

We write in what follows%
\begin{equation}
\mathfrak{L}_{g}(\mathfrak{g}^{\mathbf{\alpha}},\mathfrak{p}_{\alpha
}):=\mathcal{L}_{g}(\mathfrak{g}^{\mathbf{\alpha}},d\mathfrak{g}%
^{\mathbf{\alpha}}(\mathfrak{p}_{\alpha})) \label{defl}%
\end{equation}

We observe that
\begin{align}%
%TCIMACRO{\TeXButton{delta}{\mbox{\boldmath{$\delta$}}}}%
%BeginExpansion
\mbox{\boldmath{$\delta$}}%
%EndExpansion
\mathcal{L}_{g}(\mathfrak{g}^{\mathbf{\alpha}},d\mathfrak{g}^{\mathbf{\alpha}%
})  &  =d(%
%TCIMACRO{\TeXButton{delta}{\mbox{\boldmath{$\delta$}}}}%
%BeginExpansion
\mbox{\boldmath{$\delta$}}%
%EndExpansion
\mathfrak{g}^{\mathbf{\alpha}}\wedge\frac{\partial\mathcal{L}_{g}%
(\mathfrak{g}^{\mathbf{\alpha}},d\mathfrak{g}^{\mathbf{\alpha}})}{\partial
d\mathfrak{g}^{\mathbf{\alpha}}})+%
%TCIMACRO{\TeXButton{delta}{\mbox{\boldmath{$\delta$}}}}%
%BeginExpansion
\mbox{\boldmath{$\delta$}}%
%EndExpansion
\mathfrak{g}^{\mathbf{\alpha}}\wedge\left[  \frac{\partial\mathcal{L}%
_{g}(\mathfrak{g}^{\mathbf{\alpha}},d\mathfrak{g}^{\mathbf{\alpha}})}%
{\partial\mathfrak{g}^{\mathbf{\alpha}}}-d(\frac{\partial\mathcal{L}%
_{g}(\mathfrak{g}^{\mathbf{\alpha}},d\mathfrak{g}^{\mathbf{\alpha}})}{\partial
d\mathfrak{g}^{\mathbf{\alpha}}})\right] \nonumber\\
&  =d(%
%TCIMACRO{\TeXButton{delta}{\mbox{\boldmath{$\delta$}}}}%
%BeginExpansion
\mbox{\boldmath{$\delta$}}%
%EndExpansion
\mathfrak{g}^{\mathbf{\alpha}}\wedge\mathfrak{p}_{\alpha})+%
%TCIMACRO{\TeXButton{delta}{\mbox{\boldmath{$\delta$}}}}%
%BeginExpansion
\mbox{\boldmath{$\delta$}}%
%EndExpansion
\mathfrak{g}^{\mathbf{\alpha}}\wedge\frac{%
%TCIMACRO{\TeXButton{delta}{\mbox{\boldmath{$\delta$}}}}%
%BeginExpansion
\mbox{\boldmath{$\delta$}}%
%EndExpansion
\mathcal{L}_{g}(\mathfrak{g}^{\mathbf{\alpha}},d\mathfrak{g}^{\mathbf{\alpha}%
})}{%
%TCIMACRO{\TeXButton{delta}{\mbox{\boldmath{$\delta$}}}}%
%BeginExpansion
\mbox{\boldmath{$\delta$}}%
%EndExpansion
\mathfrak{g}^{\mathbf{\alpha}}}. \label{h4}%
\end{align}
Also from Eq.(\ref{h3}) we immediately have:
\begin{align}%
%TCIMACRO{\TeXButton{delta}{\mbox{\boldmath{$\delta$}}}}%
%BeginExpansion
\mbox{\boldmath{$\delta$}}%
%EndExpansion
\mathfrak{L}_{g}(\mathfrak{g}^{\mathbf{\alpha}},\mathfrak{p}_{\mathbf{\alpha}%
})  &  =%
%TCIMACRO{\TeXButton{delta}{\mbox{\boldmath{$\delta$}}}}%
%BeginExpansion
\mbox{\boldmath{$\delta$}}%
%EndExpansion
(d\mathfrak{g}^{\mathbf{\alpha}}\wedge\mathfrak{p}_{\mathbf{\alpha}})-%
%TCIMACRO{\TeXButton{delta}{\mbox{\boldmath{$\delta$}}}}%
%BeginExpansion
\mbox{\boldmath{$\delta$}}%
%EndExpansion
\mathfrak{g}^{\mathbf{\alpha}}\wedge\frac{\partial\mathbf{L}}{\partial
\mathfrak{g}^{\mathbf{\alpha}}}-%
%TCIMACRO{\TeXButton{delta}{\mbox{\boldmath{$\delta$}}}}%
%BeginExpansion
\mbox{\boldmath{$\delta$}}%
%EndExpansion
\mathfrak{p}_{\mathbf{\alpha}}\wedge\frac{\partial\mathbf{L}}{\partial
\mathfrak{p}_{\mathbf{\alpha}}}\nonumber\\
&  =d(%
%TCIMACRO{\TeXButton{delta}{\mbox{\boldmath{$\delta$}}}}%
%BeginExpansion
\mbox{\boldmath{$\delta$}}%
%EndExpansion
\mathfrak{g}^{\mathbf{\alpha}}\wedge\mathfrak{p}_{\mathbf{\alpha}})+%
%TCIMACRO{\TeXButton{delta}{\mbox{\boldmath{$\delta$}}}}%
%BeginExpansion
\mbox{\boldmath{$\delta$}}%
%EndExpansion
\mathfrak{g}^{\mathbf{\alpha}}\left(  -d\mathfrak{p}_{\mathbf{\alpha}}%
-\frac{\partial\mathbf{L}}{\partial\mathfrak{g}^{\mathbf{\alpha}}}\right)  +%
%TCIMACRO{\TeXButton{delta}{\mbox{\boldmath{$\delta$}}}}%
%BeginExpansion
\mbox{\boldmath{$\delta$}}%
%EndExpansion
(d\mathfrak{g}^{\mathbf{\alpha}}-\frac{\partial\mathbf{L}}{\partial
\mathfrak{p}_{\mathbf{\alpha}}})\wedge\mathfrak{p}_{\mathbf{\alpha}%
}\nonumber\\
&  =d(%
%TCIMACRO{\TeXButton{delta}{\mbox{\boldmath{$\delta$}}}}%
%BeginExpansion
\mbox{\boldmath{$\delta$}}%
%EndExpansion
\mathfrak{g}^{\mathbf{\alpha}}\wedge\mathfrak{p}_{\mathbf{\alpha}})+%
%TCIMACRO{\TeXButton{delta}{\mbox{\boldmath{$\delta$}}}}%
%BeginExpansion
\mbox{\boldmath{$\delta$}}%
%EndExpansion
\mathfrak{g}^{\mathbf{\alpha}}\wedge\left(  \frac{%
%TCIMACRO{\TeXButton{delta}{\mbox{\boldmath{$\delta$}}}}%
%BeginExpansion
\mbox{\boldmath{$\delta$}}%
%EndExpansion
\mathfrak{L}_{g}(\mathfrak{g}^{\mathbf{\alpha}},\mathfrak{p}_{\mathbf{\alpha}%
})}{%
%TCIMACRO{\TeXButton{delta}{\mbox{\boldmath{$\delta$}}}}%
%BeginExpansion
\mbox{\boldmath{$\delta$}}%
%EndExpansion
\mathfrak{g}^{\mathbf{\alpha}}}\right)  +\left(  \frac{%
%TCIMACRO{\TeXButton{delta}{\mbox{\boldmath{$\delta$}}}}%
%BeginExpansion
\mbox{\boldmath{$\delta$}}%
%EndExpansion
\mathfrak{L}_{g}(\mathfrak{g}^{\mathbf{\alpha}},\mathfrak{p}_{\mathbf{\alpha}%
})}{%
%TCIMACRO{\TeXButton{delta}{\mbox{\boldmath{$\delta$}}}}%
%BeginExpansion
\mbox{\boldmath{$\delta$}}%
%EndExpansion
\mathfrak{p}_{\mathbf{\alpha}}}\right)  \wedge%
%TCIMACRO{\TeXButton{delta}{\mbox{\boldmath{$\delta$}}}}%
%BeginExpansion
\mbox{\boldmath{$\delta$}}%
%EndExpansion
\mathfrak{p}_{\mathbf{\alpha}}, \label{h5}%
\end{align}
where we put:%
\begin{align}
\frac{%
%TCIMACRO{\TeXButton{delta}{\mbox{\boldmath{$\delta$}}}}%
%BeginExpansion
\mbox{\boldmath{$\delta$}}%
%EndExpansion
\mathfrak{L}_{g}(\mathfrak{g}^{\mathbf{\alpha}},\mathfrak{p}_{\mathbf{\alpha}%
})}{%
%TCIMACRO{\TeXButton{delta}{\mbox{\boldmath{$\delta$}}}}%
%BeginExpansion
\mbox{\boldmath{$\delta$}}%
%EndExpansion
\mathfrak{g}^{\mathbf{\alpha}}}  &  :=-d\mathfrak{p}_{\mathbf{\alpha}}%
-\frac{\partial\mathbf{L}}{\partial\mathfrak{g}^{\mathbf{\alpha}}},\nonumber\\
\frac{%
%TCIMACRO{\TeXButton{delta}{\mbox{\boldmath{$\delta$}}}}%
%BeginExpansion
\mbox{\boldmath{$\delta$}}%
%EndExpansion
\mathfrak{L}_{g}(\mathfrak{g}^{\mathbf{\alpha}},\mathfrak{p}_{\mathbf{\alpha}%
})}{%
%TCIMACRO{\TeXButton{delta}{\mbox{\boldmath{$\delta$}}}}%
%BeginExpansion
\mbox{\boldmath{$\delta$}}%
%EndExpansion
\mathfrak{p}_{\mathbf{\alpha}}}  &  :=d\mathfrak{g}^{\mathbf{\alpha}}%
-\frac{\partial\mathbf{L}}{\partial\mathfrak{p}_{\mathbf{\alpha}}}. \label{h6}%
\end{align}
From Eqs. (\ref{h4}) and (\ref{h5}) we have
\begin{equation}%
%TCIMACRO{\TeXButton{delta}{\mbox{\boldmath{$\delta$}}}}%
%BeginExpansion
\mbox{\boldmath{$\delta$}}%
%EndExpansion
\mathfrak{g}^{\mathbf{\alpha}}\wedge\frac{%
%TCIMACRO{\TeXButton{delta}{\mbox{\boldmath{$\delta$}}}}%
%BeginExpansion
\mbox{\boldmath{$\delta$}}%
%EndExpansion
\mathcal{L}_{g}(\mathfrak{g}^{\mathbf{\alpha}},d\mathfrak{g}^{\mathbf{\alpha}%
})}{%
%TCIMACRO{\TeXButton{delta}{\mbox{\boldmath{$\delta$}}}}%
%BeginExpansion
\mbox{\boldmath{$\delta$}}%
%EndExpansion
\mathfrak{g}^{\mathbf{\alpha}}}=%
%TCIMACRO{\TeXButton{delta}{\mbox{\boldmath{$\delta$}}}}%
%BeginExpansion
\mbox{\boldmath{$\delta$}}%
%EndExpansion
\mathfrak{g}^{\mathbf{\alpha}}\wedge\left(  \frac{%
%TCIMACRO{\TeXButton{delta}{\mbox{\boldmath{$\delta$}}}}%
%BeginExpansion
\mbox{\boldmath{$\delta$}}%
%EndExpansion
\mathfrak{L}_{g}(\mathfrak{g}^{\mathbf{\alpha}},\mathfrak{p}_{\mathbf{\alpha}%
})}{%
%TCIMACRO{\TeXButton{delta}{\mbox{\boldmath{$\delta$}}}}%
%BeginExpansion
\mbox{\boldmath{$\delta$}}%
%EndExpansion
\mathfrak{g}^{\mathbf{\alpha}}}\right)  +\left(  \frac{%
%TCIMACRO{\TeXButton{delta}{\mbox{\boldmath{$\delta$}}}}%
%BeginExpansion
\mbox{\boldmath{$\delta$}}%
%EndExpansion
\mathfrak{L}_{g}(\mathfrak{g}^{\mathbf{\alpha}},\mathfrak{p}_{\mathbf{\alpha}%
})}{%
%TCIMACRO{\TeXButton{delta}{\mbox{\boldmath{$\delta$}}}}%
%BeginExpansion
\mbox{\boldmath{$\delta$}}%
%EndExpansion
\mathfrak{p}_{\mathbf{\alpha}}}\right)  \wedge%
%TCIMACRO{\TeXButton{delta}{\mbox{\boldmath{$\delta$}}}}%
%BeginExpansion
\mbox{\boldmath{$\delta$}}%
%EndExpansion
\mathfrak{p}_{\mathbf{\alpha}}. \label{h6'}%
\end{equation}

To define the Hamiltonian form we need to choice a time for our manifold, and
we choose this time to be given by the flow of an arbitrary timelike vector
field $\mathbf{Z}\in\sec TM$ such that $%
%TCIMACRO{\TeXButton{g}{\slg}}%
%BeginExpansion
\slg
%EndExpansion
(\mathbf{Z,Z)}=1$. Moreover we define $Z=%
%TCIMACRO{\TeXButton{g}{\slg}}%
%BeginExpansion
\slg
%EndExpansion
(\mathbf{Z,)}\in\sec%
%TCIMACRO{\dbigwedge \nolimits^{1}}%
%BeginExpansion
{\displaystyle\bigwedge\nolimits^{1}}
%EndExpansion
T^{\ast}M\hookrightarrow\mathcal{C\ell(}\mathtt{g},M\mathcal{)}$. With this
choice the variation $%
%TCIMACRO{\TeXButton{delta}{\mbox{\boldmath{$\delta$}}}}%
%BeginExpansion
\mbox{\boldmath{$\delta$}}%
%EndExpansion
$ is generated by the Lie derivative $\pounds _{\mathbf{Z}}$. Using Cartan's
`magical formula' (see, e.g., \cite{rodcap2007}) we have
\begin{equation}%
%TCIMACRO{\TeXButton{delta}{\mbox{\boldmath{$\delta$}}}}%
%BeginExpansion
\mbox{\boldmath{$\delta$}}%
%EndExpansion
\mathfrak{L}_{g}=\pounds _{\mathbf{Z}}\mathfrak{L}_{g}=d(Z\lrcorner
\mathfrak{L}_{g})+Z\lrcorner d\mathfrak{L}_{g}=d(Z\lrcorner\mathfrak{L}_{g}).
\label{h7}%
\end{equation}
Then, using Eq.(\ref{h5}) we can write
\begin{equation}
d(Z\lrcorner\mathfrak{L}_{g})=d(\pounds _{\mathbf{Z}}\mathfrak{g}%
^{\mathbf{\alpha}}\wedge\mathfrak{p}_{\mathbf{\alpha}})+%
%TCIMACRO{\tciLaplace}%
%BeginExpansion
\mathcal{L}%
%EndExpansion
_{\mathbf{Z}}\mathfrak{g}^{\mathbf{\alpha}}\wedge\frac{%
%TCIMACRO{\TeXButton{delta}{\mbox{\boldmath{$\delta$}}}}%
%BeginExpansion
\mbox{\boldmath{$\delta$}}%
%EndExpansion
\mathfrak{L}_{g}}{%
%TCIMACRO{\TeXButton{delta}{\mbox{\boldmath{$\delta$}}}}%
%BeginExpansion
\mbox{\boldmath{$\delta$}}%
%EndExpansion
\mathfrak{g}^{\mathbf{\alpha}}}+%
%TCIMACRO{\tciLaplace}%
%BeginExpansion
\mathcal{L}%
%EndExpansion
_{\mathbf{Z}}\mathfrak{p}_{\mathbf{\alpha}}\wedge\left(  \frac{%
%TCIMACRO{\TeXButton{delta}{\mbox{\boldmath{$\delta$}}}}%
%BeginExpansion
\mbox{\boldmath{$\delta$}}%
%EndExpansion
\mathfrak{L}}{%
%TCIMACRO{\TeXButton{delta}{\mbox{\boldmath{$\delta$}}}}%
%BeginExpansion
\mbox{\boldmath{$\delta$}}%
%EndExpansion
\mathfrak{p}_{\mathbf{\alpha}}}\right)  \label{h8}%
\end{equation}
and taking into account Eq.(\ref{h6'}) we get%
\begin{equation}
d(\pounds _{\mathbf{Z}}\mathfrak{g}^{\mathbf{\alpha}}\wedge\mathfrak{p}%
_{\mathbf{\alpha}}-Z\lrcorner\mathfrak{L}_{g})=\pounds _{\mathbf{Z}%
}\mathfrak{g}^{\mathbf{\alpha}}\wedge\frac{%
%TCIMACRO{\TeXButton{delta}{\mbox{\boldmath{$\delta$}}}}%
%BeginExpansion
\mbox{\boldmath{$\delta$}}%
%EndExpansion
\mathcal{L}_{g}}{%
%TCIMACRO{\TeXButton{delta}{\mbox{\boldmath{$\delta$}}}}%
%BeginExpansion
\mbox{\boldmath{$\delta$}}%
%EndExpansion
\mathfrak{g}^{\mathbf{\alpha}}}. \label{h9}%
\end{equation}
Now, we define the \textit{Hamiltonian} $3$-form by%
\begin{equation}
\mathcal{H(\mathfrak{g}^{\mathbf{\alpha}}},\mathcal{\mathfrak{p}%
_{\mathbf{\alpha}})}:\mathcal{=}\pounds _{\mathbf{Z}}\mathfrak{g}%
^{\mathbf{\alpha}}\wedge\mathfrak{p}_{\mathbf{\alpha}}-Z\lrcorner
\mathfrak{L}_{g}. \label{h10}%
\end{equation}
We immediately have in view of Eq.(\ref{h9}) that when the field equations for
the \textit{free }gravitational field are satisfied, i.e., when the
Euler-Lagrange functional is null, $\frac{%
%TCIMACRO{\TeXButton{delta}{\mbox{\boldmath{$\delta$}}}}%
%BeginExpansion
\mbox{\boldmath{$\delta$}}%
%EndExpansion
\mathcal{L}_{g}}{%
%TCIMACRO{\TeXButton{delta}{\mbox{\boldmath{$\delta$}}}}%
%BeginExpansion
\mbox{\boldmath{$\delta$}}%
%EndExpansion
\mathfrak{g}^{\mathbf{\alpha}}}=0$ that
\begin{equation}
d\mathcal{H}=0. \label{h11}%
\end{equation}
Thus $\mathcal{H}$ is a conserved Noether current. We next write
\begin{equation}
\mathcal{H}=Z^{\mathbf{\alpha}}\mathcal{H}_{\mathbf{\alpha}}+dB \label{h12}%
\end{equation}
and find some equivalent expressions for $\mathcal{H}_{\mathbf{\alpha}}$. To
start, \ we rewrite Eq.(\ref{h10})\ taking into account Eq.(\ref{h2}),
Cartan's magical formula and some Clifford algebra identities as
\begin{align}
\mathcal{H}  &  =-Z\lrcorner\mathfrak{L}_{g}+\pounds _{\mathbf{Z}}%
\mathfrak{g}^{\mathbf{\alpha}}\wedge\mathfrak{p}_{\mathbf{\alpha}}\nonumber\\
&  =-Z\lrcorner\mathfrak{L}_{g}+d(Z\lrcorner\mathfrak{g}^{\mathbf{\alpha}%
})\wedge\mathfrak{p}_{\mathbf{\alpha}}+(Z\lrcorner d\mathfrak{g}%
^{\mathbf{\alpha}})\wedge\mathfrak{p}_{\mathbf{\alpha}}\nonumber\\
&  -Z\lrcorner\mathfrak{L}_{g}+d\left(  (Z\lrcorner\mathfrak{g}%
^{\mathbf{\alpha}})\wedge\mathfrak{p}_{\mathbf{\alpha}}\right)  -(Z\lrcorner
\mathfrak{g}^{\mathbf{\alpha}})\wedge d\mathfrak{p}_{\mathbf{\alpha}%
}+(Z\lrcorner d\mathfrak{g}^{\mathbf{\alpha}})\wedge\mathfrak{p}%
_{\mathbf{\alpha}}\nonumber\\
&  =Z^{\mathbf{\alpha}}\left(  \mathfrak{g}_{\mathbf{\alpha}}\lrcorner
\mathfrak{L}_{g}+(\mathfrak{g}_{\mathbf{\alpha}}\lrcorner d\mathfrak{g}%
^{\mathbf{\kappa}})\wedge\mathfrak{p}_{\mathbf{\kappa}}-d\mathfrak{p}%
_{\mathbf{\alpha}}\right)  +d\left(  Z^{\mathbf{\alpha}}\mathfrak{p}%
_{\mathbf{\alpha}}\right)  , \label{h13}%
\end{align}
from where we get%
\begin{align}
&
\begin{tabular}
[c]{|l|}\hline
$\mathcal{H}_{\mathbf{\alpha}}=\mathfrak{g}_{\mathbf{\alpha}}\lrcorner
\mathfrak{L}_{g}+(\mathfrak{g}_{\mathbf{\alpha}}\lrcorner d\mathfrak{g}%
^{\mathbf{\kappa}})\wedge\mathfrak{p}_{\mathbf{\kappa}}-d\mathfrak{p}%
_{\mathbf{\alpha}},$\\\hline
\end{tabular}
\nonumber\\
&
\begin{tabular}
[c]{|l|}\hline
$B=Z^{\mathbf{\alpha}}\mathfrak{p}_{\mathbf{\alpha}}.$\\\hline
\end{tabular}
\ \ \ \ \label{h14}%
\end{align}
\ 

Next we utilize the definition of the Legendre transform (Eq.(\ref{h3})), and
some Clifford algebra identities to write Eq.(\ref{h10}) as :%
\begin{align}
\mathcal{H}  &  \mathcal{=-}Z\lrcorner\mathfrak{L}_{g}+d(Z\lrcorner
\mathfrak{g}^{\mathbf{\alpha}})\wedge\mathfrak{p}_{\mathbf{\alpha}%
}+(Z\lrcorner d\mathfrak{g}^{\mathbf{\alpha}})\wedge\mathfrak{p}%
_{\mathbf{\alpha}}\nonumber\\
&  =Z\lrcorner\mathbf{L-}Z\lrcorner(\mathfrak{g}^{\mathbf{\kappa}}%
\wedge\mathfrak{p}_{\mathbf{\kappa}})+d(Z\lrcorner\mathfrak{g}^{\mathbf{\kappa
}})\wedge\mathfrak{p}_{\mathbf{\kappa}}+(Z\lrcorner d\mathfrak{g}%
^{\mathbf{\kappa}})\wedge\mathfrak{p}_{\mathbf{\kappa}}\nonumber\\
&  =Z^{\mathbf{\alpha}}\left(  \mathbf{L}_{\mathbf{\alpha}}+(\mathfrak{g}%
_{\mathbf{\alpha}}\lrcorner\mathfrak{p}_{\mathbf{\kappa}})\wedge
d\mathfrak{g}^{\mathbf{\kappa}}-(\mathfrak{g}_{\mathbf{\alpha}}\lrcorner
\mathfrak{g}^{\mathbf{\kappa}})d\mathfrak{p}_{\mathbf{\kappa}}\right)
+d\left(  Z^{\mathbf{\alpha}}\mathfrak{p}_{\mathbf{\alpha}}\right)  ,
\label{h15}%
\end{align}
where we put $\mathbf{L}_{\mathbf{\alpha}}=\mathfrak{g}_{\mathbf{\alpha}%
}\lrcorner\mathbf{L}$. Thus we also have
\begin{align}
&
\begin{tabular}
[c]{|l|}\hline
$\mathcal{H}_{\mathbf{\alpha}}=\mathbf{L}_{\mathbf{\alpha}}+(\mathfrak{g}%
_{\mathbf{\alpha}}\lrcorner\mathfrak{p}_{\mathbf{\kappa}})\wedge
d\mathfrak{g}^{\mathbf{\kappa}}-d\mathfrak{p}_{\mathbf{\alpha}},$\\\hline
\end{tabular}
\nonumber\\
&
\begin{tabular}
[c]{|l|}\hline
$B=Z^{\mathbf{\alpha}}\mathfrak{p}_{\mathbf{\alpha}}.$\\\hline
\end{tabular}
\ \ \label{h16}%
\end{align}

Now, recalling that Eq.(\ref{h9}) we can write%
\begin{equation}
d(Z^{\mathbf{\alpha}}\mathcal{H}_{\mathbf{\alpha}}+dB)+\pounds _{\mathbf{Z}%
}\mathfrak{g}^{\mathbf{\alpha}}\wedge\frac{%
%TCIMACRO{\TeXButton{delta}{\mbox{\boldmath{$\delta$}}}}%
%BeginExpansion
\mbox{\boldmath{$\delta$}}%
%EndExpansion
\mathcal{L}_{g}}{%
%TCIMACRO{\TeXButton{delta}{\mbox{\boldmath{$\delta$}}}}%
%BeginExpansion
\mbox{\boldmath{$\delta$}}%
%EndExpansion
\mathfrak{g}^{\mathbf{\alpha}}}=0. \label{h17}%
\end{equation}
Then,%

\begin{align}
&  dZ^{\mathbf{\alpha}}\wedge\mathcal{H}_{\mathbf{\alpha}}+Z^{\mathbf{\alpha}%
}d\mathcal{H}_{\mathbf{\alpha}}+d(Z\lrcorner\mathfrak{g}^{\mathbf{\alpha}%
})\wedge\frac{%
%TCIMACRO{\TeXButton{delta}{\mbox{\boldmath{$\delta$}}}}%
%BeginExpansion
\mbox{\boldmath{$\delta$}}%
%EndExpansion
\mathcal{L}_{g}}{%
%TCIMACRO{\TeXButton{delta}{\mbox{\boldmath{$\delta$}}}}%
%BeginExpansion
\mbox{\boldmath{$\delta$}}%
%EndExpansion
\mathfrak{g}^{\mathbf{\alpha}}}+(Z\lrcorner d\mathfrak{g}^{\mathbf{\alpha}%
})\wedge\frac{%
%TCIMACRO{\TeXButton{delta}{\mbox{\boldmath{$\delta$}}}}%
%BeginExpansion
\mbox{\boldmath{$\delta$}}%
%EndExpansion
\mathcal{L}_{g}}{%
%TCIMACRO{\TeXButton{delta}{\mbox{\boldmath{$\delta$}}}}%
%BeginExpansion
\mbox{\boldmath{$\delta$}}%
%EndExpansion
\mathfrak{g}^{\mathbf{\alpha}}}\nonumber\\
&  =dZ^{\mathbf{\alpha}}\wedge\left(  \mathcal{H}_{\mathbf{\alpha}}+\frac{%
%TCIMACRO{\TeXButton{delta}{\mbox{\boldmath{$\delta$}}}}%
%BeginExpansion
\mbox{\boldmath{$\delta$}}%
%EndExpansion
\mathcal{L}_{g}}{%
%TCIMACRO{\TeXButton{delta}{\mbox{\boldmath{$\delta$}}}}%
%BeginExpansion
\mbox{\boldmath{$\delta$}}%
%EndExpansion
\mathfrak{g}^{\mathbf{\alpha}}}\right)  +Z^{\mathbf{\alpha}}\left(
d\mathcal{H}_{\mathbf{\alpha}}+d(\mathfrak{g}_{\mathbf{\alpha}}\lrcorner
\mathfrak{g}^{\mathbf{\kappa}})\wedge\frac{%
%TCIMACRO{\TeXButton{delta}{\mbox{\boldmath{$\delta$}}}}%
%BeginExpansion
\mbox{\boldmath{$\delta$}}%
%EndExpansion
\mathcal{L}_{g}}{%
%TCIMACRO{\TeXButton{delta}{\mbox{\boldmath{$\delta$}}}}%
%BeginExpansion
\mbox{\boldmath{$\delta$}}%
%EndExpansion
\mathfrak{g}^{\mathbf{\kappa}}}+(\mathfrak{g}_{\mathbf{\alpha}}\lrcorner
d\mathfrak{g}^{\mathbf{\kappa}})\wedge\frac{%
%TCIMACRO{\TeXButton{delta}{\mbox{\boldmath{$\delta$}}}}%
%BeginExpansion
\mbox{\boldmath{$\delta$}}%
%EndExpansion
\mathcal{L}_{g}}{%
%TCIMACRO{\TeXButton{delta}{\mbox{\boldmath{$\delta$}}}}%
%BeginExpansion
\mbox{\boldmath{$\delta$}}%
%EndExpansion
\mathfrak{g}^{\mathbf{\kappa}}}\right)  =0, \label{h18}%
\end{align}
which means that
\begin{equation}
\mathcal{H}_{\mathbf{\alpha}}=-\frac{%
%TCIMACRO{\TeXButton{delta}{\mbox{\boldmath{$\delta$}}}}%
%BeginExpansion
\mbox{\boldmath{$\delta$}}%
%EndExpansion
\mathcal{L}_{g}}{%
%TCIMACRO{\TeXButton{delta}{\mbox{\boldmath{$\delta$}}}}%
%BeginExpansion
\mbox{\boldmath{$\delta$}}%
%EndExpansion
\mathfrak{g}^{\mathbf{\alpha}}} \label{h19}%
\end{equation}
and%
\begin{equation}
d\mathcal{H}_{\mathbf{\alpha}}+d(\mathfrak{g}_{\mathbf{\alpha}}\lrcorner
\mathfrak{g}^{\mathbf{\kappa}})\wedge\frac{%
%TCIMACRO{\TeXButton{delta}{\mbox{\boldmath{$\delta$}}}}%
%BeginExpansion
\mbox{\boldmath{$\delta$}}%
%EndExpansion
\mathcal{L}_{g}}{%
%TCIMACRO{\TeXButton{delta}{\mbox{\boldmath{$\delta$}}}}%
%BeginExpansion
\mbox{\boldmath{$\delta$}}%
%EndExpansion
\mathfrak{g}^{\mathbf{\kappa}}}+(\mathfrak{g}_{\mathbf{\alpha}}\lrcorner
d\mathfrak{g}^{\mathbf{\kappa}})\wedge\frac{%
%TCIMACRO{\TeXButton{delta}{\mbox{\boldmath{$\delta$}}}}%
%BeginExpansion
\mbox{\boldmath{$\delta$}}%
%EndExpansion
\mathcal{L}_{g}}{%
%TCIMACRO{\TeXButton{delta}{\mbox{\boldmath{$\delta$}}}}%
%BeginExpansion
\mbox{\boldmath{$\delta$}}%
%EndExpansion
\mathfrak{g}^{\mathbf{\kappa}}}=0. \label{h20}%
\end{equation}
From Eq.(\ref{h20}) we have
\begin{align}
d\mathcal{H}_{\mathbf{\alpha}}  &  =-d(\mathfrak{g}_{\mathbf{\alpha}}%
\lrcorner\mathfrak{g}^{\mathbf{\kappa}})\wedge\frac{%
%TCIMACRO{\TeXButton{delta}{\mbox{\boldmath{$\delta$}}}}%
%BeginExpansion
\mbox{\boldmath{$\delta$}}%
%EndExpansion
\mathcal{L}_{g}}{%
%TCIMACRO{\TeXButton{delta}{\mbox{\boldmath{$\delta$}}}}%
%BeginExpansion
\mbox{\boldmath{$\delta$}}%
%EndExpansion
\mathfrak{g}^{\mathbf{\kappa}}}-(\mathfrak{g}_{\mathbf{\alpha}}\lrcorner
d\mathfrak{g}^{\mathbf{\kappa}})\wedge\frac{%
%TCIMACRO{\TeXButton{delta}{\mbox{\boldmath{$\delta$}}}}%
%BeginExpansion
\mbox{\boldmath{$\delta$}}%
%EndExpansion
\mathcal{L}_{g}}{%
%TCIMACRO{\TeXButton{delta}{\mbox{\boldmath{$\delta$}}}}%
%BeginExpansion
\mbox{\boldmath{$\delta$}}%
%EndExpansion
\mathfrak{g}^{\mathbf{\kappa}}}\nonumber\\
&  =-(\mathfrak{g}_{\mathbf{\alpha}}\lrcorner d\mathfrak{g}^{\mathbf{\kappa}%
})\wedge\frac{%
%TCIMACRO{\TeXButton{delta}{\mbox{\boldmath{$\delta$}}}}%
%BeginExpansion
\mbox{\boldmath{$\delta$}}%
%EndExpansion
\mathcal{L}_{g}}{%
%TCIMACRO{\TeXButton{delta}{\mbox{\boldmath{$\delta$}}}}%
%BeginExpansion
\mbox{\boldmath{$\delta$}}%
%EndExpansion
\mathfrak{g}^{\mathbf{\kappa}}}-d\left[  (\mathfrak{g}_{\mathbf{\alpha}%
}\lrcorner\mathfrak{g}^{\mathbf{\kappa}})\wedge\frac{%
%TCIMACRO{\TeXButton{delta}{\mbox{\boldmath{$\delta$}}}}%
%BeginExpansion
\mbox{\boldmath{$\delta$}}%
%EndExpansion
\mathcal{L}_{g}}{%
%TCIMACRO{\TeXButton{delta}{\mbox{\boldmath{$\delta$}}}}%
%BeginExpansion
\mbox{\boldmath{$\delta$}}%
%EndExpansion
\mathfrak{g}^{\mathbf{\kappa}}}\right]  +(\mathfrak{g}_{\mathbf{\alpha}%
}\lrcorner\mathfrak{g}^{\mathbf{\kappa}})\wedge d\left(  \frac{%
%TCIMACRO{\TeXButton{delta}{\mbox{\boldmath{$\delta$}}}}%
%BeginExpansion
\mbox{\boldmath{$\delta$}}%
%EndExpansion
\mathcal{L}_{g}}{%
%TCIMACRO{\TeXButton{delta}{\mbox{\boldmath{$\delta$}}}}%
%BeginExpansion
\mbox{\boldmath{$\delta$}}%
%EndExpansion
\mathfrak{g}^{\mathbf{\kappa}}}\right) \nonumber\\
&  =-(\mathfrak{g}_{\mathbf{\alpha}}\lrcorner d\mathfrak{g}^{\mathbf{\kappa}%
})\wedge\frac{%
%TCIMACRO{\TeXButton{delta}{\mbox{\boldmath{$\delta$}}}}%
%BeginExpansion
\mbox{\boldmath{$\delta$}}%
%EndExpansion
\mathcal{L}_{g}}{%
%TCIMACRO{\TeXButton{delta}{\mbox{\boldmath{$\delta$}}}}%
%BeginExpansion
\mbox{\boldmath{$\delta$}}%
%EndExpansion
\mathfrak{g}^{\mathbf{\kappa}}}-d\left[  \frac{%
%TCIMACRO{\TeXButton{delta}{\mbox{\boldmath{$\delta$}}}}%
%BeginExpansion
\mbox{\boldmath{$\delta$}}%
%EndExpansion
\mathcal{L}_{g}}{%
%TCIMACRO{\TeXButton{delta}{\mbox{\boldmath{$\delta$}}}}%
%BeginExpansion
\mbox{\boldmath{$\delta$}}%
%EndExpansion
\mathfrak{g}^{\mathbf{\alpha}}}\right]  +d\left(  \frac{%
%TCIMACRO{\TeXButton{delta}{\mbox{\boldmath{$\delta$}}}}%
%BeginExpansion
\mbox{\boldmath{$\delta$}}%
%EndExpansion
\mathcal{L}_{g}}{%
%TCIMACRO{\TeXButton{delta}{\mbox{\boldmath{$\delta$}}}}%
%BeginExpansion
\mbox{\boldmath{$\delta$}}%
%EndExpansion
\mathfrak{g}^{\mathbf{\alpha}}}\right) \nonumber\\
&  =-(\mathfrak{g}_{\mathbf{\alpha}}\lrcorner d\mathfrak{g}^{\mathbf{\kappa}%
})\wedge\frac{%
%TCIMACRO{\TeXButton{delta}{\mbox{\boldmath{$\delta$}}}}%
%BeginExpansion
\mbox{\boldmath{$\delta$}}%
%EndExpansion
\mathcal{L}_{g}}{%
%TCIMACRO{\TeXButton{delta}{\mbox{\boldmath{$\delta$}}}}%
%BeginExpansion
\mbox{\boldmath{$\delta$}}%
%EndExpansion
\mathfrak{g}^{\mathbf{\kappa}}}+d\mathcal{H}_{\mathbf{\alpha}}+d\left(  \frac{%
%TCIMACRO{\TeXButton{delta}{\mbox{\boldmath{$\delta$}}}}%
%BeginExpansion
\mbox{\boldmath{$\delta$}}%
%EndExpansion
\mathcal{L}_{g}}{%
%TCIMACRO{\TeXButton{delta}{\mbox{\boldmath{$\delta$}}}}%
%BeginExpansion
\mbox{\boldmath{$\delta$}}%
%EndExpansion
\mathfrak{g}^{\mathbf{\alpha}}}\right)  , \label{h21}%
\end{align}
from where it follows the identity:%
\begin{equation}
(\mathfrak{g}_{\mathbf{\alpha}}\lrcorner d\mathfrak{g}^{\mathbf{\kappa}%
})\wedge\frac{%
%TCIMACRO{\TeXButton{delta}{\mbox{\boldmath{$\delta$}}}}%
%BeginExpansion
\mbox{\boldmath{$\delta$}}%
%EndExpansion
\mathcal{L}_{g}}{%
%TCIMACRO{\TeXButton{delta}{\mbox{\boldmath{$\delta$}}}}%
%BeginExpansion
\mbox{\boldmath{$\delta$}}%
%EndExpansion
\mathfrak{g}^{\mathbf{\kappa}}}=d\left(  \frac{%
%TCIMACRO{\TeXButton{delta}{\mbox{\boldmath{$\delta$}}}}%
%BeginExpansion
\mbox{\boldmath{$\delta$}}%
%EndExpansion
\mathcal{L}_{g}}{%
%TCIMACRO{\TeXButton{delta}{\mbox{\boldmath{$\delta$}}}}%
%BeginExpansion
\mbox{\boldmath{$\delta$}}%
%EndExpansion
\mathfrak{g}^{\mathbf{\alpha}}}\right)  . \label{h22}%
\end{equation}

We now return to Eq.(\ref{h17}) and recalling Eq.(\ref{h6'}) we can write:%
\begin{equation}
d(Z^{\mathbf{\alpha}}\mathcal{H}_{\mathbf{\alpha}}+dB)+%
%TCIMACRO{\TeXButton{delta}{\mbox{\boldmath{$\delta$}}}}%
%BeginExpansion
\mbox{\boldmath{$\delta$}}%
%EndExpansion
\mathfrak{g}^{\mathbf{\alpha}}\wedge\left(  \frac{%
%TCIMACRO{\TeXButton{delta}{\mbox{\boldmath{$\delta$}}}}%
%BeginExpansion
\mbox{\boldmath{$\delta$}}%
%EndExpansion
\mathfrak{L}_{g}}{%
%TCIMACRO{\TeXButton{delta}{\mbox{\boldmath{$\delta$}}}}%
%BeginExpansion
\mbox{\boldmath{$\delta$}}%
%EndExpansion
\mathfrak{g}^{\mathbf{\alpha}}}\right)  +\left(  \frac{%
%TCIMACRO{\TeXButton{delta}{\mbox{\boldmath{$\delta$}}}}%
%BeginExpansion
\mbox{\boldmath{$\delta$}}%
%EndExpansion
\mathfrak{L}_{g}}{%
%TCIMACRO{\TeXButton{delta}{\mbox{\boldmath{$\delta$}}}}%
%BeginExpansion
\mbox{\boldmath{$\delta$}}%
%EndExpansion
\mathfrak{p}_{\mathbf{\alpha}}}\right)  \wedge%
%TCIMACRO{\TeXButton{delta}{\mbox{\boldmath{$\delta$}}}}%
%BeginExpansion
\mbox{\boldmath{$\delta$}}%
%EndExpansion
\mathfrak{p}_{\mathbf{\alpha}}=0, \label{h23}%
\end{equation}
from where we get after some algebra%

\begin{align}
&
\begin{tabular}
[c]{|l|}\hline
$\mathcal{H}_{\mathbf{\alpha}}(\mathfrak{g}^{\mathbf{\alpha}},\mathfrak{p}%
_{\mathbf{\alpha}})=\frac{%
%TCIMACRO{\TeXButton{delta}{\mbox{\boldmath{$\delta$}}}}%
%BeginExpansion
\mbox{\boldmath{$\delta$}}%
%EndExpansion
\mathfrak{L}_{g}}{%
%TCIMACRO{\TeXButton{delta}{\mbox{\boldmath{$\delta$}}}}%
%BeginExpansion
\mbox{\boldmath{$\delta$}}%
%EndExpansion
\mathfrak{g}^{\mathbf{\alpha}}}-(\mathfrak{g}_{\mathbf{\alpha}}\lrcorner
\wedge\mathfrak{p}_{\mathbf{\kappa}})\wedge\frac{%
%TCIMACRO{\TeXButton{delta}{\mbox{\boldmath{$\delta$}}}}%
%BeginExpansion
\mbox{\boldmath{$\delta$}}%
%EndExpansion
\mathfrak{L}_{g}}{%
%TCIMACRO{\TeXButton{delta}{\mbox{\boldmath{$\delta$}}}}%
%BeginExpansion
\mbox{\boldmath{$\delta$}}%
%EndExpansion
\mathfrak{p}_{\mathbf{\kappa}}},$\\\hline
\end{tabular}
\nonumber\\
&
\begin{tabular}
[c]{|l|}\hline
$B=Z^{\mathbf{\alpha}}\mathfrak{p}_{\mathbf{\alpha}}.$\\\hline
\end{tabular}
\ \label{h24}%
\end{align}

\subsection{The Quasi Local Energy}

Now, let us investigate the meaning of the boundary term in Eq.(\ref{h12}).
Consider an arbitrary spacelike hypersuface $\sigma$. Then we define%
\begin{align}
\mathbf{H}  &  =%
%TCIMACRO{\dint \nolimits_{\sigma}}%
%BeginExpansion
{\displaystyle\int\nolimits_{\sigma}}
%EndExpansion
(Z^{\alpha}\mathcal{H}_{\alpha}+dB)\nonumber\\
&  =%
%TCIMACRO{\dint \nolimits_{\sigma}}%
%BeginExpansion
{\displaystyle\int\nolimits_{\sigma}}
%EndExpansion
Z^{\alpha}\mathcal{H}_{\alpha}+%
%TCIMACRO{\dint \nolimits_{\partial\sigma}}%
%BeginExpansion
{\displaystyle\int\nolimits_{\partial\sigma}}
%EndExpansion
B. \label{h25}%
\end{align}
If we take into account Eq.(\ref{h19}) we see that the first term in
Eq.(\ref{h19}) is null when the field equations (for the free gravitational
field) are satisfied and we are thus left with
\begin{equation}
\mathbf{E}=%
%TCIMACRO{\dint \nolimits_{\partial\sigma}}%
%BeginExpansion
{\displaystyle\int\nolimits_{\partial\sigma}}
%EndExpansion
B, \label{h26}%
\end{equation}
which is called the quasi local energy. \footnote{For an up to date review on
the concept of quasilocal energy in \textit{GRT}, its success and drawbacks
see \cite{szabados}.}

Before proceeding, recall that if $\{\mathfrak{e}_{\mathbf{\alpha}}\}$ is the
dual basis of $\{\mathfrak{g}^{\mathbf{\alpha}}\}$ we have
\begin{equation}
\mathfrak{g}^{\mathbf{0}}(\mathfrak{e}_{\mathbf{i}})=0,\text{ }\mathbf{i=}%
1,2,3, \label{h127}%
\end{equation}
and at spatial infinity $\mathfrak{e}_{\mathbf{0}}\rightarrow\partial
/\partial\mathrm{x}^{0}$ Then, if we take $\mathbf{Z=}$ $\mathfrak{e}%
_{\mathbf{0}}$ we can even choose $\sigma$ such that at spatial
\textit{infinity }$\mathfrak{e}_{\mathbf{0}}$ is orthogonal to the
hypersurface $\sigma$,(and the $\mathfrak{e}_{\mathbf{i}},\mathbf{i=}1,2,3$
are tangent to $\sigma$), we get recalling that in our theory it is
$\mathfrak{p}_{\mathbf{\alpha}}=\underset{%
%TCIMACRO{\TeXButton{sig}{\sitg}}%
%BeginExpansion
\sitg
%EndExpansion
}{\star}\mathcal{S}_{\mathbf{\alpha}}$ (Eq.(\ref{7.10.17})) that
\begin{equation}
\mathbf{E}=%
%TCIMACRO{\dint \nolimits_{\partial\sigma}}%
%BeginExpansion
{\displaystyle\int\nolimits_{\partial\sigma}}
%EndExpansion
\underset{%
%TCIMACRO{\TeXButton{sig}{\sitg}}%
%BeginExpansion
\sitg
%EndExpansion
}{\star}\mathcal{S}_{0}, \label{h128}%
\end{equation}
which we recognize as being the same conserved quantity as defined by
$P_{0}^{\prime}$ in Eq.(\ref{em2}), which is the total energy contained in the
gravitational plus matter fields for a solution of Einstein's equation such
that the source term goes to zero at spatial infinity

\subsection{Hamilton's Equations}

Before we explore in details the meaning of this\ quasi local energy, let us
remind that the role of an Hamiltonian density is to derive Hamilton's
equation. We have
\begin{equation}%
%TCIMACRO{\TeXButton{delta}{\mbox{\boldmath{$\delta$}}}}%
%BeginExpansion
\mbox{\boldmath{$\delta$}}%
%EndExpansion
\ \mathbf{H=}%
%TCIMACRO{\dint \nolimits_{\sigma}}%
%BeginExpansion
{\displaystyle\int\nolimits_{\sigma}}
%EndExpansion%
%TCIMACRO{\TeXButton{delta}{\mbox{\boldmath{$\delta$}}}}%
%BeginExpansion
\mbox{\boldmath{$\delta$}}%
%EndExpansion
\mathcal{H=}%
%TCIMACRO{\dint \nolimits_{\sigma}}%
%BeginExpansion
{\displaystyle\int\nolimits_{\sigma}}
%EndExpansion%
%TCIMACRO{\TeXButton{delta}{\mbox{\boldmath{$\delta$}}}}%
%BeginExpansion
\mbox{\boldmath{$\delta$}}%
%EndExpansion
(Z^{\alpha}\mathcal{H}_{\alpha}+dB) \label{h27}%
\end{equation}
where $%
%TCIMACRO{\TeXButton{delta}{\mbox{\boldmath{$\delta$}}}}%
%BeginExpansion
\mbox{\boldmath{$\delta$}}%
%EndExpansion
$ is an \textit{arbitrary} variation not generated by the flow of the vector
field $\mathbf{Z}$. To perform the variation we recall that we obtained above
three different (but equivalent) expressions for $\mathcal{H}_{\mathbf{\alpha
}}$, namely Eq.(\ref{h14}), Eq.(\ref{h16}) and Eq.(\ref{h26}).

Taking into account that $%
%TCIMACRO{\TeXButton{delta}{\mbox{\boldmath{$\delta$}}}}%
%BeginExpansion
\mbox{\boldmath{$\delta$}}%
%EndExpansion
\mathcal{L}_{g}(\mathfrak{g}^{\mathbf{\alpha}},d\mathfrak{g}^{\mathbf{\alpha}%
})=%
%TCIMACRO{\TeXButton{delta}{\mbox{\boldmath{$\delta$}}}}%
%BeginExpansion
\mbox{\boldmath{$\delta$}}%
%EndExpansion
\mathfrak{L}_{g}(\mathfrak{g}^{\mathbf{\alpha}},\mathfrak{p}_{\mathbf{\alpha}%
})$, we have using Eq.(\ref{h4}) and Cartan's magical formula
\begin{align}%
%TCIMACRO{\TeXButton{delta}{\mbox{\boldmath{$\delta$}}}}%
%BeginExpansion
\mbox{\boldmath{$\delta$}}%
%EndExpansion
\mathcal{H}  &  \mathcal{=}%
%TCIMACRO{\TeXButton{delta}{\mbox{\boldmath{$\delta$}}}}%
%BeginExpansion
\mbox{\boldmath{$\delta$}}%
%EndExpansion
\left(  \pounds _{\mathbf{Z}}\mathfrak{g}^{\mathbf{\alpha}}\wedge
\mathfrak{p}_{\mathbf{\alpha}}-Z\lrcorner\mathfrak{L}_{g}\right) \nonumber\\
&  =-Z\lrcorner\left(  d(%
%TCIMACRO{\TeXButton{delta}{\mbox{\boldmath{$\delta$}}}}%
%BeginExpansion
\mbox{\boldmath{$\delta$}}%
%EndExpansion
\mathfrak{g}^{\mathbf{\alpha}}\wedge\mathfrak{p}_{\alpha})+%
%TCIMACRO{\TeXButton{delta}{\mbox{\boldmath{$\delta$}}}}%
%BeginExpansion
\mbox{\boldmath{$\delta$}}%
%EndExpansion
\mathfrak{g}^{\mathbf{\alpha}}\wedge\frac{%
%TCIMACRO{\TeXButton{delta}{\mbox{\boldmath{$\delta$}}}}%
%BeginExpansion
\mbox{\boldmath{$\delta$}}%
%EndExpansion
\mathcal{L}_{g}}{%
%TCIMACRO{\TeXButton{delta}{\mbox{\boldmath{$\delta$}}}}%
%BeginExpansion
\mbox{\boldmath{$\delta$}}%
%EndExpansion
\mathfrak{g}^{\mathbf{\alpha}}}\right)  +\pounds _{\mathbf{Z}}%
%TCIMACRO{\TeXButton{delta}{\mbox{\boldmath{$\delta$}}}}%
%BeginExpansion
\mbox{\boldmath{$\delta$}}%
%EndExpansion
\mathfrak{g}^{\mathbf{\alpha}}\wedge\mathfrak{p}_{\mathbf{\alpha}%
}+\pounds _{\mathbf{Z}}\mathfrak{g}^{\mathbf{\alpha}}\wedge%
%TCIMACRO{\TeXButton{delta}{\mbox{\boldmath{$\delta$}}}}%
%BeginExpansion
\mbox{\boldmath{$\delta$}}%
%EndExpansion
\mathfrak{p}_{\mathbf{\alpha}}\nonumber\\
&  =-%
%TCIMACRO{\TeXButton{delta}{\mbox{\boldmath{$\delta$}}}}%
%BeginExpansion
\mbox{\boldmath{$\delta$}}%
%EndExpansion
\mathfrak{g}^{\mathbf{\alpha}}\wedge\pounds _{\mathbf{Z}}\mathfrak{p}%
_{\mathbf{\alpha}}+%
%TCIMACRO{\TeXButton{delta}{\mbox{\boldmath{$\delta$}}}}%
%BeginExpansion
\mbox{\boldmath{$\delta$}}%
%EndExpansion
\mathfrak{p}_{\mathbf{\alpha}}\wedge\pounds _{\mathbf{Z}}\mathfrak{g}%
^{\mathbf{\alpha}}-Z\lrcorner(%
%TCIMACRO{\TeXButton{delta}{\mbox{\boldmath{$\delta$}}}}%
%BeginExpansion
\mbox{\boldmath{$\delta$}}%
%EndExpansion
\mathfrak{g}^{\mathbf{\alpha}}\wedge\frac{%
%TCIMACRO{\TeXButton{delta}{\mbox{\boldmath{$\delta$}}}}%
%BeginExpansion
\mbox{\boldmath{$\delta$}}%
%EndExpansion
\mathcal{L}_{g}}{%
%TCIMACRO{\TeXButton{delta}{\mbox{\boldmath{$\delta$}}}}%
%BeginExpansion
\mbox{\boldmath{$\delta$}}%
%EndExpansion
\mathfrak{g}^{\mathbf{\alpha}}})+d[Z\lrcorner(%
%TCIMACRO{\TeXButton{delta}{\mbox{\boldmath{$\delta$}}}}%
%BeginExpansion
\mbox{\boldmath{$\delta$}}%
%EndExpansion
\mathfrak{g}^{\mathbf{\alpha}}\wedge\mathfrak{p}_{\mathbf{\alpha}})].
\label{h28}%
\end{align}
This last equation can also be written using Eq.(\ref{h6'}) as
\begin{align}%
%TCIMACRO{\TeXButton{delta}{\mbox{\boldmath{$\delta$}}}}%
%BeginExpansion
\mbox{\boldmath{$\delta$}}%
%EndExpansion
\mathcal{H}  &  \mathcal{=}-%
%TCIMACRO{\TeXButton{delta}{\mbox{\boldmath{$\delta$}}}}%
%BeginExpansion
\mbox{\boldmath{$\delta$}}%
%EndExpansion
\mathfrak{g}^{\mathbf{\alpha}}\wedge\pounds _{\mathbf{Z}}\mathfrak{p}%
_{\mathbf{\alpha}}+%
%TCIMACRO{\TeXButton{delta}{\mbox{\boldmath{$\delta$}}}}%
%BeginExpansion
\mbox{\boldmath{$\delta$}}%
%EndExpansion
\mathfrak{p}_{\mathbf{\alpha}}\wedge\pounds _{\mathbf{Z}}\mathfrak{g}%
^{\mathbf{\alpha}}\nonumber\\
&  -Z\lrcorner(%
%TCIMACRO{\TeXButton{delta}{\mbox{\boldmath{$\delta$}}}}%
%BeginExpansion
\mbox{\boldmath{$\delta$}}%
%EndExpansion
\mathfrak{g}^{\mathbf{\alpha}}\wedge\frac{%
%TCIMACRO{\TeXButton{delta}{\mbox{\boldmath{$\delta$}}}}%
%BeginExpansion
\mbox{\boldmath{$\delta$}}%
%EndExpansion
\mathfrak{L}_{g}}{%
%TCIMACRO{\TeXButton{delta}{\mbox{\boldmath{$\delta$}}}}%
%BeginExpansion
\mbox{\boldmath{$\delta$}}%
%EndExpansion
\mathfrak{g}^{\mathbf{\alpha}}}+%
%TCIMACRO{\TeXButton{delta}{\mbox{\boldmath{$\delta$}}}}%
%BeginExpansion
\mbox{\boldmath{$\delta$}}%
%EndExpansion
\mathfrak{p}_{\mathbf{\alpha}}\wedge\frac{%
%TCIMACRO{\TeXButton{delta}{\mbox{\boldmath{$\delta$}}}}%
%BeginExpansion
\mbox{\boldmath{$\delta$}}%
%EndExpansion
\mathfrak{L}_{g}}{%
%TCIMACRO{\TeXButton{delta}{\mbox{\boldmath{$\delta$}}}}%
%BeginExpansion
\mbox{\boldmath{$\delta$}}%
%EndExpansion
\mathfrak{p}_{\mathbf{\alpha}}})+d[Z\lrcorner(%
%TCIMACRO{\TeXButton{delta}{\mbox{\boldmath{$\delta$}}}}%
%BeginExpansion
\mbox{\boldmath{$\delta$}}%
%EndExpansion
\mathfrak{g}^{\mathbf{\alpha}}\wedge\mathfrak{p}_{\mathbf{\alpha}})],
\label{h29}%
\end{align}
and a simple calculation shows that this is also the form $%
%TCIMACRO{\TeXButton{delta}{\mbox{\boldmath{$\delta$}}}}%
%BeginExpansion
\mbox{\boldmath{$\delta$}}%
%EndExpansion
\mathcal{H}$ that we get varying Eq.(\ref{h20}). From the above results it
follows that
\begin{align}%
%TCIMACRO{\TeXButton{delta}{\mbox{\boldmath{$\delta$}}}}%
%BeginExpansion
\mbox{\boldmath{$\delta$}}%
%EndExpansion
\mathbf{H}  &  \mathbf{=}%
%TCIMACRO{\dint \nolimits_{\sigma}}%
%BeginExpansion
{\displaystyle\int\nolimits_{\sigma}}
%EndExpansion
(-%
%TCIMACRO{\TeXButton{delta}{\mbox{\boldmath{$\delta$}}}}%
%BeginExpansion
\mbox{\boldmath{$\delta$}}%
%EndExpansion
\mathfrak{g}^{\mathbf{\alpha}}\wedge\pounds _{\mathbf{Z}}\mathfrak{p}%
_{\mathbf{\alpha}}+%
%TCIMACRO{\TeXButton{delta}{\mbox{\boldmath{$\delta$}}}}%
%BeginExpansion
\mbox{\boldmath{$\delta$}}%
%EndExpansion
\mathfrak{p}_{\mathbf{\alpha}}\wedge\pounds _{\mathbf{Z}}\mathfrak{g}%
^{\mathbf{\alpha}})\nonumber\\
&
%TCIMACRO{\dint \nolimits_{\sigma}}%
%BeginExpansion
{\displaystyle\int\nolimits_{\sigma}}
%EndExpansion
-Z\lrcorner(%
%TCIMACRO{\TeXButton{delta}{\mbox{\boldmath{$\delta$}}}}%
%BeginExpansion
\mbox{\boldmath{$\delta$}}%
%EndExpansion
\mathfrak{g}^{\mathbf{\alpha}}\wedge\frac{%
%TCIMACRO{\TeXButton{delta}{\mbox{\boldmath{$\delta$}}}}%
%BeginExpansion
\mbox{\boldmath{$\delta$}}%
%EndExpansion
\mathfrak{L}_{g}}{%
%TCIMACRO{\TeXButton{delta}{\mbox{\boldmath{$\delta$}}}}%
%BeginExpansion
\mbox{\boldmath{$\delta$}}%
%EndExpansion
\mathfrak{g}^{\mathbf{\alpha}}}+%
%TCIMACRO{\TeXButton{delta}{\mbox{\boldmath{$\delta$}}}}%
%BeginExpansion
\mbox{\boldmath{$\delta$}}%
%EndExpansion
\mathfrak{p}_{\mathbf{\alpha}}\wedge\frac{%
%TCIMACRO{\TeXButton{delta}{\mbox{\boldmath{$\delta$}}}}%
%BeginExpansion
\mbox{\boldmath{$\delta$}}%
%EndExpansion
\mathfrak{L}_{g}}{%
%TCIMACRO{\TeXButton{delta}{\mbox{\boldmath{$\delta$}}}}%
%BeginExpansion
\mbox{\boldmath{$\delta$}}%
%EndExpansion
\mathfrak{p}_{\mathbf{\alpha}}})\nonumber\\
&  +%
%TCIMACRO{\dint \nolimits_{\partial\sigma}}%
%BeginExpansion
{\displaystyle\int\nolimits_{\partial\sigma}}
%EndExpansion
Z\lrcorner(%
%TCIMACRO{\TeXButton{delta}{\mbox{\boldmath{$\delta$}}}}%
%BeginExpansion
\mbox{\boldmath{$\delta$}}%
%EndExpansion
\mathfrak{g}^{\mathbf{\alpha}}\wedge\mathfrak{p}_{\mathbf{\alpha}}).
\label{h29''}%
\end{align}

Thus, only if the field equations are satisfied (i.e., $\frac{%
%TCIMACRO{\TeXButton{delta}{\mbox{\boldmath{$\delta$}}}}%
%BeginExpansion
\mbox{\boldmath{$\delta$}}%
%EndExpansion
\mathcal{L}_{g}}{%
%TCIMACRO{\TeXButton{delta}{\mbox{\boldmath{$\delta$}}}}%
%BeginExpansion
\mbox{\boldmath{$\delta$}}%
%EndExpansion
\mathfrak{g}^{\mathbf{\alpha}}}=0$ or $\frac{%
%TCIMACRO{\TeXButton{delta}{\mbox{\boldmath{$\delta$}}}}%
%BeginExpansion
\mbox{\boldmath{$\delta$}}%
%EndExpansion
\mathfrak{L}_{g}}{%
%TCIMACRO{\TeXButton{delta}{\mbox{\boldmath{$\delta$}}}}%
%BeginExpansion
\mbox{\boldmath{$\delta$}}%
%EndExpansion
\mathfrak{g}^{\mathbf{\alpha}}}=0$, $\frac{%
%TCIMACRO{\TeXButton{delta}{\mbox{\boldmath{$\delta$}}}}%
%BeginExpansion
\mbox{\boldmath{$\delta$}}%
%EndExpansion
\mathfrak{L}_{g}}{%
%TCIMACRO{\TeXButton{delta}{\mbox{\boldmath{$\delta$}}}}%
%BeginExpansion
\mbox{\boldmath{$\delta$}}%
%EndExpansion
\mathfrak{p}_{\mathbf{\alpha}}}=0$) and
\begin{equation}%
%TCIMACRO{\dint \nolimits_{\partial\sigma}}%
%BeginExpansion
{\displaystyle\int\nolimits_{\partial\sigma}}
%EndExpansion
Z\lrcorner(%
%TCIMACRO{\TeXButton{delta}{\mbox{\boldmath{$\delta$}}}}%
%BeginExpansion
\mbox{\boldmath{$\delta$}}%
%EndExpansion
\mathfrak{g}^{\mathbf{\alpha}}\wedge\mathfrak{p}_{\mathbf{\alpha}})=0,
\label{h29'}%
\end{equation}
\ for arbitrary variations $%
%TCIMACRO{\TeXButton{delta}{\mbox{\boldmath{$\delta$}}}}%
%BeginExpansion
\mbox{\boldmath{$\delta$}}%
%EndExpansion
\mathfrak{g}^{\mathbf{\alpha}}$ not necessarily vanishing at the boundary
\cite{rete} $\partial\sigma$, it follows that the variation $%
%TCIMACRO{\TeXButton{delta}{\mbox{\boldmath{$\delta$}}}}%
%BeginExpansion
\mbox{\boldmath{$\delta$}}%
%EndExpansion
\mathbf{H}$ gives Hamilton's equations :%
\begin{equation}
\pounds _{\mathbf{Z}}\mathfrak{p}_{\mathbf{\alpha}}=-\frac{%
%TCIMACRO{\TeXButton{delta}{\mbox{\boldmath{$\delta$}}}}%
%BeginExpansion
\mbox{\boldmath{$\delta$}}%
%EndExpansion
\mathcal{H}}{%
%TCIMACRO{\TeXButton{delta}{\mbox{\boldmath{$\delta$}}}}%
%BeginExpansion
\mbox{\boldmath{$\delta$}}%
%EndExpansion
\mathfrak{g}^{\mathbf{\alpha}}},\text{ }\pounds _{\mathbf{Z}}\mathfrak{g}%
^{\mathbf{\alpha}}=\frac{%
%TCIMACRO{\TeXButton{delta}{\mbox{\boldmath{$\delta$}}}}%
%BeginExpansion
\mbox{\boldmath{$\delta$}}%
%EndExpansion
\mathcal{H}}{%
%TCIMACRO{\TeXButton{delta}{\mbox{\boldmath{$\delta$}}}}%
%BeginExpansion
\mbox{\boldmath{$\delta$}}%
%EndExpansion
\mathfrak{p}_{\mathbf{\alpha}}}. \label{h30}%
\end{equation}
Now the term $d[Z\lrcorner(%
%TCIMACRO{\TeXButton{delta}{\mbox{\boldmath{$\delta$}}}}%
%BeginExpansion
\mbox{\boldmath{$\delta$}}%
%EndExpansion
\mathfrak{g}^{\mathbf{\alpha}}\wedge\mathfrak{p}_{\mathbf{\alpha}})]$ comes as
part of the variation of the term $dB$. Indeed, since
\begin{align}
B  &  =Z^{\mathbf{\alpha}}\mathfrak{p}_{\mathbf{\alpha}}=(Z\lrcorner
\mathfrak{g}^{\mathbf{\alpha}})\wedge\mathfrak{p}_{\mathbf{\alpha}}\nonumber\\
&  =Z\lrcorner(\mathfrak{g}^{\mathbf{\alpha}}\wedge\mathfrak{p}%
_{\mathbf{\alpha}})+\mathfrak{g}^{\mathbf{\alpha}}\wedge(Z\lrcorner
\mathfrak{p}_{\mathbf{\alpha}}), \label{h32}%
\end{align}
it is%
\begin{equation}
dB=d(Z\lrcorner(\mathfrak{g}^{\mathbf{\alpha}}\wedge\mathfrak{p}%
_{\mathbf{\alpha}}))+d(\mathfrak{g}^{\mathbf{\alpha}}\wedge(Z\lrcorner
\mathfrak{p}_{\mathbf{\alpha}})), \label{h33}%
\end{equation}
and%

\begin{align}%
%TCIMACRO{\TeXButton{delta}{\mbox{\boldmath{$\delta$}}}}%
%BeginExpansion
\mbox{\boldmath{$\delta$}}%
%EndExpansion
dB  &  =d(Z\lrcorner(%
%TCIMACRO{\TeXButton{delta}{\mbox{\boldmath{$\delta$}}}}%
%BeginExpansion
\mbox{\boldmath{$\delta$}}%
%EndExpansion
\mathfrak{g}^{\mathbf{\alpha}}\wedge\mathfrak{p}_{\mathbf{\alpha}%
}))+d(Z\lrcorner(\mathfrak{g}^{\mathbf{\alpha}}\wedge%
%TCIMACRO{\TeXButton{delta}{\mbox{\boldmath{$\delta$}}}}%
%BeginExpansion
\mbox{\boldmath{$\delta$}}%
%EndExpansion
\mathfrak{p}_{\mathbf{\alpha}}))\nonumber\\
&  +d(%
%TCIMACRO{\TeXButton{delta}{\mbox{\boldmath{$\delta$}}}}%
%BeginExpansion
\mbox{\boldmath{$\delta$}}%
%EndExpansion
\mathfrak{g}^{\mathbf{\alpha}}\wedge(Z\lrcorner\mathfrak{p}_{\mathbf{\alpha}%
}))+d(\mathfrak{g}^{\mathbf{\alpha}}\wedge(Z\lrcorner%
%TCIMACRO{\TeXButton{delta}{\mbox{\boldmath{$\delta$}}}}%
%BeginExpansion
\mbox{\boldmath{$\delta$}}%
%EndExpansion
\mathfrak{p}_{\mathbf{\alpha}}))\nonumber\\
&  =d(Z\lrcorner(%
%TCIMACRO{\TeXButton{delta}{\mbox{\boldmath{$\delta$}}}}%
%BeginExpansion
\mbox{\boldmath{$\delta$}}%
%EndExpansion
\mathfrak{g}^{\mathbf{\alpha}}\wedge\mathfrak{p}_{\mathbf{\alpha}%
}))+d((Z\lrcorner(\mathfrak{g}^{\mathbf{\alpha}})\wedge%
%TCIMACRO{\TeXButton{delta}{\mbox{\boldmath{$\delta$}}}}%
%BeginExpansion
\mbox{\boldmath{$\delta$}}%
%EndExpansion
\mathfrak{p}_{\mathbf{\alpha}})-\mathfrak{g}^{\mathbf{\alpha}}\wedge
(Z\lrcorner%
%TCIMACRO{\TeXButton{delta}{\mbox{\boldmath{$\delta$}}}}%
%BeginExpansion
\mbox{\boldmath{$\delta$}}%
%EndExpansion
\mathfrak{p}_{\mathbf{\alpha}})+\mathfrak{g}^{\mathbf{\alpha}}\wedge
(Z\lrcorner%
%TCIMACRO{\TeXButton{delta}{\mbox{\boldmath{$\delta$}}}}%
%BeginExpansion
\mbox{\boldmath{$\delta$}}%
%EndExpansion
\mathfrak{p}_{\mathbf{\alpha}}))\nonumber\\
&  +d(%
%TCIMACRO{\TeXButton{delta}{\mbox{\boldmath{$\delta$}}}}%
%BeginExpansion
\mbox{\boldmath{$\delta$}}%
%EndExpansion
\mathfrak{g}^{\mathbf{\alpha}}\wedge(Z\lrcorner\mathfrak{p}_{\mathbf{\alpha}%
}))\nonumber\\
&  =d(Z\lrcorner(%
%TCIMACRO{\TeXButton{delta}{\mbox{\boldmath{$\delta$}}}}%
%BeginExpansion
\mbox{\boldmath{$\delta$}}%
%EndExpansion
\mathfrak{g}^{\mathbf{\alpha}}\wedge\mathfrak{p}_{\mathbf{\alpha}%
}))+d((Z\lrcorner(\mathfrak{g}^{\mathbf{\alpha}})\wedge%
%TCIMACRO{\TeXButton{delta}{\mbox{\boldmath{$\delta$}}}}%
%BeginExpansion
\mbox{\boldmath{$\delta$}}%
%EndExpansion
\mathfrak{p}_{\mathbf{\alpha}})+d(Z\lrcorner(%
%TCIMACRO{\TeXButton{delta}{\mbox{\boldmath{$\delta$}}}}%
%BeginExpansion
\mbox{\boldmath{$\delta$}}%
%EndExpansion
\mathfrak{g}^{\mathbf{\alpha}}\wedge\mathfrak{p}_{\mathbf{\alpha}}))
\label{h34}%
\end{align}

This result is important since it shows that if Eq.(\ref{h29'}) is not
satisfied we must modify in an appropriate way the boundary term $B$ in
Eq.(\ref{h12}) in order for that condition to hold. This was exactly the idea
originally used by Nester and collaborators \cite{ccn,chennes,meng} in their
proposal for the use of the quasi local energy concept. Here we analyze what
happens in our theory when we choose, e.g., $Z=\mathfrak{g}^{\mathbf{0}}$. So,
we examine the integrand in Eq.(\ref{h29'}). We have:%
\begin{equation}
Z\lrcorner(%
%TCIMACRO{\TeXButton{delta}{\mbox{\boldmath{$\delta$}}}}%
%BeginExpansion
\mbox{\boldmath{$\delta$}}%
%EndExpansion
\mathfrak{g}^{\mathbf{\alpha}}\wedge\mathfrak{p}_{\mathbf{\alpha}%
})=(Z\lrcorner%
%TCIMACRO{\TeXButton{delta}{\mbox{\boldmath{$\delta$}}}}%
%BeginExpansion
\mbox{\boldmath{$\delta$}}%
%EndExpansion
\mathfrak{g}^{\mathbf{\alpha}})\wedge\mathfrak{p}_{\mathbf{\alpha}}-%
%TCIMACRO{\TeXButton{delta}{\mbox{\boldmath{$\delta$}}}}%
%BeginExpansion
\mbox{\boldmath{$\delta$}}%
%EndExpansion
\mathfrak{g}^{\mathbf{\alpha}}\wedge\left(  Z\lrcorner\mathfrak{p}%
_{\mathbf{\alpha}}\right)  . \label{hh1}%
\end{equation}
and with $Z=\mathfrak{g}^{\mathbf{0}}$ it is
\begin{equation}
(Z\lrcorner%
%TCIMACRO{\TeXButton{delta}{\mbox{\boldmath{$\delta$}}}}%
%BeginExpansion
\mbox{\boldmath{$\delta$}}%
%EndExpansion
\mathfrak{g}^{\mathbf{\alpha}})\wedge\mathfrak{p}_{\mathbf{\alpha}}=%
%TCIMACRO{\TeXButton{delta}{\mbox{\boldmath{$\delta$}}}}%
%BeginExpansion
\mbox{\boldmath{$\delta$}}%
%EndExpansion
[(Z\lrcorner\mathfrak{g}^{\mathbf{\alpha}})\wedge\mathfrak{p}_{\mathbf{\alpha
}}]-(%
%TCIMACRO{\TeXButton{delta}{\mbox{\boldmath{$\delta$}}}}%
%BeginExpansion
\mbox{\boldmath{$\delta$}}%
%EndExpansion
Z\lrcorner\mathfrak{g}^{\mathbf{\alpha}})\wedge\mathfrak{p}_{\mathbf{\alpha}%
}-(Z\lrcorner\mathfrak{g}^{\mathbf{\alpha}})\wedge%
%TCIMACRO{\TeXButton{delta}{\mbox{\boldmath{$\delta$}}}}%
%BeginExpansion
\mbox{\boldmath{$\delta$}}%
%EndExpansion
\mathfrak{p}_{\mathbf{\alpha}} \label{hh2}%
\end{equation}
So, if $Z=\mathfrak{g}^{\mathbf{0}}$.is maintained \textit{fixed} on
$\partial\sigma$ (for we used this hypothesis in deriving Eq.(\ref{h34}) we
get%
\begin{equation}
Z\lrcorner(%
%TCIMACRO{\TeXButton{delta}{\mbox{\boldmath{$\delta$}}}}%
%BeginExpansion
\mbox{\boldmath{$\delta$}}%
%EndExpansion
\mathfrak{g}^{\mathbf{\alpha}}\wedge\mathfrak{p}_{\mathbf{\alpha}})=%
%TCIMACRO{\TeXButton{delta}{\mbox{\boldmath{$\delta$}}}}%
%BeginExpansion
\mbox{\boldmath{$\delta$}}%
%EndExpansion
\mathfrak{p}_{0}-%
%TCIMACRO{\TeXButton{delta}{\mbox{\boldmath{$\delta$}}}}%
%BeginExpansion
\mbox{\boldmath{$\delta$}}%
%EndExpansion
\mathfrak{p}_{0}=0, \label{hh3}%
\end{equation}
and thus the boundary term in $%
%TCIMACRO{\TeXButton{delta}{\mbox{\boldmath{$\delta$}}}}%
%BeginExpansion
\mbox{\boldmath{$\delta$}}%
%EndExpansion
\mathbf{H}$ null, thus implying the validity of Hamilton's equations of
motion.\medskip

\textbf{Remark 7.1 }If we choose $Z\neq\mathfrak{g}^{\mathbf{0}}$ we will have
in general that $Z\lrcorner(%
%TCIMACRO{\TeXButton{delta}{\mbox{\boldmath{$\delta$}}}}%
%BeginExpansion
\mbox{\boldmath{$\delta$}}%
%EndExpansion
\mathfrak{g}^{\mathbf{\alpha}}\wedge\mathfrak{p}_{\mathbf{\alpha}})\neq0$ and
in this case a modification of the boundary term will indeed be necessary in
order to get Hamilton's equations of motion. According to the ideas first
presented (for the best of our knowledge) on the \cite{kiwlo} concerning a
symplectic framework for field theories, Chen and Nester presented the
following prescription for modification of the $B(Z)=(Z\lrcorner
\mathfrak{g}^{\mathbf{\alpha}})\wedge\mathfrak{p}_{\mathbf{\alpha}%
}=Z^{\mathbf{\alpha}}\wedge\mathfrak{p}_{\mathbf{\alpha}}$ term: introduce a
reference configuration $(\mathfrak{\mathring{g}}^{\mathbf{\alpha}%
},\mathfrak{\mathring{p}}_{\mathbf{\alpha}})$ and adjust $B(Z)$ to one of the
two covariant symplectic forms,%
\begin{equation}
B_{\mathfrak{g}^{\mathbf{\alpha}}}(Z)=(Z\lrcorner\mathfrak{g}^{\mathbf{\alpha
}})\wedge\triangle\mathfrak{p}_{\mathbf{\alpha}}+\triangle\mathfrak{g}%
^{\mathbf{\alpha}}\wedge(Z\lrcorner\mathfrak{\mathring{p}}_{\mathbf{\alpha}%
}),\label{h35}%
\end{equation}
or%
\begin{equation}
B_{\mathfrak{p}_{\mathbf{\alpha}}}(Z)=(Z\lrcorner\mathfrak{\mathring{g}%
}^{\mathbf{\alpha}})\wedge\triangle\mathfrak{p}_{\mathbf{\alpha}}%
+\triangle\mathfrak{g}^{\mathbf{\alpha}}\wedge(Z\lrcorner\mathfrak{p}%
_{\mathbf{\alpha}}),\label{h36}%
\end{equation}
where $\triangle\mathfrak{g}^{\mathbf{\alpha}}=\mathfrak{g}^{\mathbf{\alpha}%
}-\mathfrak{\mathring{g}}^{\mathbf{\alpha}}$ and $\triangle\mathfrak{p}%
_{\mathbf{\alpha}}=\mathfrak{p}_{\mathbf{\alpha}}-\mathfrak{\mathring{p}%
}_{\mathbf{\alpha}}$. Next introduce the improved Hamiltonian $3$-forms:%
\begin{equation}
\mathcal{H}_{\mathfrak{g}^{\mathbf{\alpha}}}(Z\mathcal{)}=Z^{\mu}%
\mathcal{H}_{\mu}+dB_{\mathfrak{g}^{\mathbf{\alpha}}}(Z),\label{h37}%
\end{equation}
and%
\begin{equation}
\mathcal{H}_{\mathfrak{p}_{\mathbf{\alpha}}}(Z\mathcal{)}=Z^{\mu}%
\mathcal{H}_{\mu}+dB_{\mathfrak{p}_{\mathbf{\alpha}}}(Z).\label{h38}%
\end{equation}
For those improved Hamiltonians we immediately get for arbitrary variations
(not generated by the flow of $\mathbf{Z}$)
\begin{align}%
%TCIMACRO{\TeXButton{delta}{\mbox{\boldmath{$\delta$}}}}%
%BeginExpansion
\mbox{\boldmath{$\delta$}}%
%EndExpansion
\mathcal{H}_{\mathfrak{g}^{\mathbf{\alpha}}}(Z\mathcal{)} &  \mathcal{=}-%
%TCIMACRO{\TeXButton{delta}{\mbox{\boldmath{$\delta$}}}}%
%BeginExpansion
\mbox{\boldmath{$\delta$}}%
%EndExpansion
\mathfrak{g}^{\mathbf{\alpha}}\wedge\pounds _{\mathbf{Z}}\mathfrak{p}%
_{\mathbf{\alpha}}+%
%TCIMACRO{\TeXButton{delta}{\mbox{\boldmath{$\delta$}}}}%
%BeginExpansion
\mbox{\boldmath{$\delta$}}%
%EndExpansion
\mathfrak{p}_{\mathbf{\alpha}}\wedge\pounds _{\mathbf{Z}}\mathfrak{g}%
^{\mathbf{\alpha}}\nonumber\\
&  -Z\lrcorner(%
%TCIMACRO{\TeXButton{delta}{\mbox{\boldmath{$\delta$}}}}%
%BeginExpansion
\mbox{\boldmath{$\delta$}}%
%EndExpansion
\mathfrak{g}^{\mathbf{\alpha}}\wedge\frac{%
%TCIMACRO{\TeXButton{delta}{\mbox{\boldmath{$\delta$}}}}%
%BeginExpansion
\mbox{\boldmath{$\delta$}}%
%EndExpansion
\mathfrak{L}_{g}}{%
%TCIMACRO{\TeXButton{delta}{\mbox{\boldmath{$\delta$}}}}%
%BeginExpansion
\mbox{\boldmath{$\delta$}}%
%EndExpansion
\mathfrak{g}^{\mathbf{\alpha}}}+%
%TCIMACRO{\TeXButton{delta}{\mbox{\boldmath{$\delta$}}}}%
%BeginExpansion
\mbox{\boldmath{$\delta$}}%
%EndExpansion
\mathfrak{p}_{\mathbf{\alpha}}\wedge\frac{%
%TCIMACRO{\TeXButton{delta}{\mbox{\boldmath{$\delta$}}}}%
%BeginExpansion
\mbox{\boldmath{$\delta$}}%
%EndExpansion
\mathfrak{L}_{g}}{%
%TCIMACRO{\TeXButton{delta}{\mbox{\boldmath{$\delta$}}}}%
%BeginExpansion
\mbox{\boldmath{$\delta$}}%
%EndExpansion
\mathfrak{p}_{\mathbf{\alpha}}})+d[Z\lrcorner(%
%TCIMACRO{\TeXButton{delta}{\mbox{\boldmath{$\delta$}}}}%
%BeginExpansion
\mbox{\boldmath{$\delta$}}%
%EndExpansion
\mathfrak{g}^{\mathbf{\alpha}}\wedge\triangle\mathfrak{p}_{\mathbf{\alpha}%
})],\label{h39}%
\end{align}
and%

\begin{align}%
%TCIMACRO{\TeXButton{delta}{\mbox{\boldmath{$\delta$}}}}%
%BeginExpansion
\mbox{\boldmath{$\delta$}}%
%EndExpansion
\mathcal{H}_{\mathfrak{p}_{\mathbf{\alpha}}}(Z\mathcal{)}  &  \mathcal{=}-%
%TCIMACRO{\TeXButton{delta}{\mbox{\boldmath{$\delta$}}}}%
%BeginExpansion
\mbox{\boldmath{$\delta$}}%
%EndExpansion
\mathfrak{g}^{\mathbf{\alpha}}\wedge\pounds _{\mathbf{Z}}\mathfrak{p}%
_{\mathbf{\alpha}}+%
%TCIMACRO{\TeXButton{delta}{\mbox{\boldmath{$\delta$}}}}%
%BeginExpansion
\mbox{\boldmath{$\delta$}}%
%EndExpansion
\mathfrak{p}_{\mathbf{\alpha}}\wedge\pounds _{\mathbf{Z}}\mathfrak{g}%
^{\mathbf{\alpha}}\nonumber\\
&  -Z\lrcorner(%
%TCIMACRO{\TeXButton{delta}{\mbox{\boldmath{$\delta$}}}}%
%BeginExpansion
\mbox{\boldmath{$\delta$}}%
%EndExpansion
\mathfrak{g}^{\mathbf{\alpha}}\wedge\frac{%
%TCIMACRO{\TeXButton{delta}{\mbox{\boldmath{$\delta$}}}}%
%BeginExpansion
\mbox{\boldmath{$\delta$}}%
%EndExpansion
\mathfrak{L}_{g}}{%
%TCIMACRO{\TeXButton{delta}{\mbox{\boldmath{$\delta$}}}}%
%BeginExpansion
\mbox{\boldmath{$\delta$}}%
%EndExpansion
\mathfrak{g}^{\mathbf{\alpha}}}+%
%TCIMACRO{\TeXButton{delta}{\mbox{\boldmath{$\delta$}}}}%
%BeginExpansion
\mbox{\boldmath{$\delta$}}%
%EndExpansion
\mathfrak{p}_{\mathbf{\alpha}}\wedge\frac{%
%TCIMACRO{\TeXButton{delta}{\mbox{\boldmath{$\delta$}}}}%
%BeginExpansion
\mbox{\boldmath{$\delta$}}%
%EndExpansion
\mathfrak{L}_{g}}{%
%TCIMACRO{\TeXButton{delta}{\mbox{\boldmath{$\delta$}}}}%
%BeginExpansion
\mbox{\boldmath{$\delta$}}%
%EndExpansion
\mathfrak{p}_{\mathbf{\alpha}}})-d[Z\lrcorner(\triangle\mathfrak{g}%
^{\mathbf{\alpha}}\wedge%
%TCIMACRO{\TeXButton{delta}{\mbox{\boldmath{$\delta$}}}}%
%BeginExpansion
\mbox{\boldmath{$\delta$}}%
%EndExpansion
\mathfrak{p}_{\mathbf{\alpha}})]. \label{h40}%
\end{align}

The variations $%
%TCIMACRO{\TeXButton{delta}{\mbox{\boldmath{$\delta$}}}}%
%BeginExpansion
\mbox{\boldmath{$\delta$}}%
%EndExpansion
\mathcal{H}_{\mathfrak{g}^{\mathbf{\alpha}}}(Z\mathcal{)}$ and $%
%TCIMACRO{\TeXButton{delta}{\mbox{\boldmath{$\delta$}}}}%
%BeginExpansion
\mbox{\boldmath{$\delta$}}%
%EndExpansion
\mathcal{H}_{\mathfrak{p}_{\mathbf{\alpha}}}(Z\mathcal{)}$ includes then,
besides the field equations the following boundary terms,
\begin{subequations}
\label{h41}%
\begin{align}
dC_{\mathfrak{g}^{\mathbf{\alpha}}} &  :=d[Z\lrcorner(%
%TCIMACRO{\TeXButton{delta}{\mbox{\boldmath{$\delta$}}}}%
%BeginExpansion
\mbox{\boldmath{$\delta$}}%
%EndExpansion
\mathfrak{g}^{\mathbf{\alpha}}\wedge\triangle\mathfrak{p}_{\mathbf{\alpha}%
})],\label{h41a}\\
-dC_{\mathfrak{p}_{\mathbf{\alpha}}} &  :=-d[Z\lrcorner(\triangle
\mathfrak{g}^{\mathbf{\alpha}}\wedge%
%TCIMACRO{\TeXButton{delta}{\mbox{\boldmath{$\delta$}}}}%
%BeginExpansion
\mbox{\boldmath{$\delta$}}%
%EndExpansion
\mathfrak{p}_{\mathbf{\alpha}})]\label{h41b}%
\end{align}
which are the projections on the boundary of covariant symplectic structures
$dC_{\mathfrak{g}^{\mathbf{\alpha}}}$ and $-dC_{\mathfrak{p}_{\mathbf{\alpha}%
}}$, which simply reflect according the general methodology introduced in
\cite{kiwlo} the choice of the control mode.

Now, we can take arbitrary variations on the boundary such that $%
%TCIMACRO{\TeXButton{delta}{\mbox{\boldmath{$\delta$}}}}%
%BeginExpansion
\mbox{\boldmath{$\delta$}}%
%EndExpansion
\mathfrak{g}^{\mathbf{\alpha}}=o(1/r)$ and $%
%TCIMACRO{\TeXButton{delta}{\mbox{\boldmath{$\delta$}}}}%
%BeginExpansion
\mbox{\boldmath{$\delta$}}%
%EndExpansion
\mathfrak{p}_{\mathbf{\alpha}}=o(1/r^{2})$ which give $%
%TCIMACRO{\dint \nolimits_{\partial\sigma}}%
%BeginExpansion
{\displaystyle\int\nolimits_{\partial\sigma}}
%EndExpansion
B_{\mathfrak{g}^{\mathbf{\alpha}}}(Z)=0$ and $%
%TCIMACRO{\dint \nolimits_{\partial\sigma}}%
%BeginExpansion
{\displaystyle\int\nolimits_{\partial\sigma}}
%EndExpansion
B_{\mathfrak{p}_{\mathbf{\alpha}}}(Z)=0$, warranting the validity of
Hamilton's equations. Moreover, we have
\end{subequations}
\begin{equation}
B_{\mathfrak{g}^{\mathbf{\alpha}}}(Z)-B_{\mathfrak{p}_{\mathbf{\alpha}}%
}(Z)=Z\lrcorner(\triangle\mathfrak{g}^{\mathbf{\alpha}}\wedge\triangle
\mathfrak{p}_{\mathbf{\alpha}}), \label{h42}%
\end{equation}
and $\triangle\mathfrak{g}^{\mathbf{\alpha}}\wedge\triangle\mathfrak{p}%
_{\mathbf{\alpha}}$ is null at spatial infinity if we $\mathfrak{\mathring{g}%
}^{\mathbf{\alpha}}$ and $\mathfrak{\mathring{p}}_{\mathbf{\alpha}}$ are the
references concerning the Lorentz vacuum (Minkowski spacetime). Thus we can
use $B_{\mathfrak{g}^{\mathbf{\alpha}}}(Z)$ or $B_{\mathfrak{p}%
_{\mathbf{\alpha}}}(Z)$ as good boundary terms, but each one gives a different
concept of energy, which may be useful in applications.

\subsection{The \textit{ADM} Energy.}

The relation of the previous sections with the original \textit{ADM} formalism
\cite{adm} can be seen as follows. Instead of choosing an orthonormal vector
field $\mathbf{Z}$, start with a global timelike vector field $\mathbf{n}%
\in\sec TM$ such that \ $n=%
%TCIMACRO{\TeXButton{slg}{\slg}}%
%BeginExpansion
\slg
%EndExpansion
(\mathbf{N,})=N^{2}dt\in\sec%
%TCIMACRO{\dbigwedge \nolimits^{1}}%
%BeginExpansion
{\displaystyle\bigwedge\nolimits^{1}}
%EndExpansion
T^{\ast}M\hookrightarrow\mathcal{C\ell(}M,\mathtt{g})$, with
$N:\mathbb{R\supset I\rightarrow R}$, a positive function called the lapse
function of $M$. Then $n\wedge dn=0$ and according to Frobenius theorem $n$
induces a foliation of $M$, i.e., topologically it is $M=\mathbb{I\times
}\sigma_{t},$ where $\sigma_{t}$ is a spacelike hypersurface with normal given
by $\mathbf{n}$. Now, we can decompose any $A\in\sec%
%TCIMACRO{\dbigwedge \nolimits^{p}}%
%BeginExpansion
{\displaystyle\bigwedge\nolimits^{p}}
%EndExpansion
T^{\ast}M\hookrightarrow\mathcal{C\ell(}M,\mathtt{g})$ into\ a tangent
component $\underline{A}$ to $\sigma_{t}$ and an orthogonal component
$\ ^{\perp}A$ to $\sigma_{t}$ by \cite{helewa,wallner2}:%
\begin{equation}
A=\underline{A}+\text{ }^{\perp}A\text{ }dt\wedge A_{\perp} \label{h43}%
\end{equation}
where%
\begin{align}
\underline{A}  &  =n\lrcorner(dt\wedge A)\text{, }^{\perp}A=dt\wedge A_{\perp
}\label{h44}\\
A_{\perp}\text{ }  &  =n\lrcorner A.
\end{align}
Introduce also the parallel component \underline{$d$} of the differential
operator $d$ by:%
\begin{equation}
\underline{d}A=n\lrcorner(dt\wedge dA) \label{h45}%
\end{equation}
from where it follows (taking into account Cartan's magical formula) that
\begin{equation}
dA=dt\wedge(\pounds _{\mathbf{n}}\underline{A}-\underline{d}A_{\perp
})+\underline{dA}. \label{h46}%
\end{equation}
Call \
\begin{align}%
%TCIMACRO{\TeXButton{h}{\slh}}%
%BeginExpansion
\slh
%EndExpansion
&  =-%
%TCIMACRO{\TeXButton{g}{slg}}%
%BeginExpansion
slg%
%EndExpansion
+n\otimes n\nonumber\\
&  =\underline{\mathfrak{g}^{i}}\otimes\underline{\mathfrak{g}}_{i},
\label{h46bis}%
\end{align}
the first fundamental form on $\sigma_{t}$. and next introduce the Hodge dual
operator associated\footnote{Recall that $\sigma$ has an internal orientation
compatible with the orientation of $M$.} to $%
%TCIMACRO{\TeXButton{h}{\slh}}%
%BeginExpansion
\slh
%EndExpansion
$ \ acting on the (horizontal forms) forms $\underline{A}$ by
\begin{equation}
\underset{%
%TCIMACRO{\TeXButton{sh}{\sslh}}%
%BeginExpansion
\sslh
%EndExpansion
}{\star}\underline{A}=\underset{%
%TCIMACRO{\TeXButton{sg}{\sslg}}%
%BeginExpansion
\sslg
%EndExpansion
}{\star}(\frac{n}{N}\wedge\underline{A}). \label{H47}%
\end{equation}
At this point we come back to the Lagrangian density Eq.(\ref{h10}) and
proceeding like in Sections 7.1 and 7.2, but now leaving $%
%TCIMACRO{\TeXButton{delta}{\mbox{\boldmath{$\delta$}}}}%
%BeginExpansion
\mbox{\boldmath{$\delta$}}%
%EndExpansion
n^{\mathbf{\alpha}}$ to be non null, we arrive to the following Hamiltonian
density
\begin{equation}
\mathcal{H(\underline{\mathfrak{g}}}^{i},\underline{\mathfrak{p}_{i}%
}\mathcal{)}=\pounds _{\mathbf{n}}\underline{\mathfrak{g}}^{i}\wedge\underset{%
%TCIMACRO{\TeXButton{sh}{\sslh}}%
%BeginExpansion
\sslh
%EndExpansion
}{\star}\underline{\mathfrak{p}_{i}}-\mathcal{K}_{g}, \label{h50bis}%
\end{equation}
where%
\begin{equation}
\mathfrak{g}^{i}\mathfrak{-}\underline{\mathfrak{g}}^{i}=dt\wedge
(n\lrcorner\mathfrak{g}^{i})=n^{i}dt, \label{h52}%
\end{equation}
and where $K_{g}$ depends on $(n^{i},\underline{d}n^{i}%
,\underline{\mathfrak{g}}^{i},\underline{d\mathfrak{g}}^{i}%
,\pounds _{\mathbf{n}}\underline{\mathfrak{g}}^{i})$. As in Section 7.1 we can
show (after some tedious but straightforward algebra\footnote{Details may be
found in \cite{wallner2}.}) that $\mathcal{H(\underline{\mathfrak{g}}}%
^{i},\underline{\mathfrak{p}_{i}}\mathcal{)}$ can be put into the form%
\begin{equation}
\mathcal{H=}n^{i}\mathcal{H}_{i}+\underline{d}B^{\prime}, \label{h51}%
\end{equation}
with
\begin{equation}
\mathcal{H}_{i}=\frac{%
%TCIMACRO{\TeXButton{delta}{\mbox{\boldmath{$\delta$}}}}%
%BeginExpansion
\mbox{\boldmath{$\delta$}}%
%EndExpansion
\mathcal{L}_{g}}{%
%TCIMACRO{\TeXButton{delta}{\mbox{\boldmath{$\delta$}}}}%
%BeginExpansion
\mbox{\boldmath{$\delta$}}%
%EndExpansion
\mathfrak{g}^{i}}=\frac{%
%TCIMACRO{\TeXButton{delta}{\mbox{\boldmath{$\delta$}}}}%
%BeginExpansion
\mbox{\boldmath{$\delta$}}%
%EndExpansion
\mathcal{K}_{g}}{%
%TCIMACRO{\TeXButton{delta}{\mbox{\boldmath{$\delta$}}}}%
%BeginExpansion
\mbox{\boldmath{$\delta$}}%
%EndExpansion
n^{i}} \label{h54}%
\end{equation}
and
\begin{equation}
B^{\prime}=-N\underline{\mathfrak{g}}_{i}\wedge\underset{%
%TCIMACRO{\TeXButton{sh}{\sslh}}%
%BeginExpansion
\sslh
%EndExpansion
}{\star}\underline{d}\text{ }\underline{\mathfrak{g}^{i}} \label{h55}%
\end{equation}
Then, on shell, i.e., when the field equations are satisfied we get
\begin{equation}
\mathbf{E}^{\prime}\mathbf{=-}%
%TCIMACRO{\dint \nolimits_{\partial\sigma_{t}}}%
%BeginExpansion
{\displaystyle\int\nolimits_{\partial\sigma_{t}}}
%EndExpansion
N\underline{\mathfrak{g}}i\wedge\underset{%
%TCIMACRO{\TeXButton{sh}{\sslh}}%
%BeginExpansion
\sslh
%EndExpansion
}{\star}\underline{d}\text{ }\underline{\mathfrak{g}^{i}} \label{h56}%
\end{equation}
which is exactly the \textit{ADM} energy, as can be seem if we take into
account that taking $\partial\sigma_{t}$ as an two-sphere at the infinity we
have (using coordinates in the ELP gauge) $\underline{\mathfrak{g}_{{}}%
}i=h_{ij}\underline{d}\mathrm{x}^{j}$ and $h_{ij},N\rightarrow1$. Then%
\begin{equation}
\underline{\mathfrak{g}_{i}}\wedge\underset{%
%TCIMACRO{\TeXButton{sh}{\sslh}}%
%BeginExpansion
\sslh
%EndExpansion
}{\star}\underline{d\mathfrak{g}^{i}}=h^{ij}(\frac{\partial h_{ij}}%
{\partial\mathrm{x}^{k}}-\frac{\partial h_{ik}}{\partial\mathrm{x}^{j}%
})\underset{%
%TCIMACRO{\TeXButton{sh}{\sslh}}%
%BeginExpansion
\sslh
%EndExpansion
}{\star}\underline{\mathfrak{g}^{k}} \label{h57}%
\end{equation}
and under the above conditions we have the ADM formula
\begin{equation}
\mathbf{E}^{\prime}\mathbf{=}%
%TCIMACRO{\dint \nolimits_{\partial\sigma_{t}}}%
%BeginExpansion
{\displaystyle\int\nolimits_{\partial\sigma_{t}}}
%EndExpansion
\left(  \frac{\partial h_{ik}}{\partial\mathrm{x}^{i}}-\frac{\partial h_{ii}%
}{\partial\mathrm{x}^{k}}\right)  \underset{%
%TCIMACRO{\TeXButton{sh}{\sslh}}%
%BeginExpansion
\sslh
%EndExpansion
}{\star}\underline{\mathfrak{g}^{k}}., \label{h58}%
\end{equation}
which, as is well known \footnote{As shown by Wallner (\cite{wallner2}) the
formalism used in this section permits to give an alternative (and very
simple) proof of the positive mass theorem, different from the one in
\cite{witten}.}, is positive definite. If we choose $n=\mathfrak{g}^{0}$\ it
may happen, of course, that $\mathfrak{g}^{0}\wedge d\mathfrak{g}^{0}\neq0$
and thus it does not determine a spacelike hypersurface $\sigma_{t}$. However
all \ algebraic calculations above\ up to Eq.(\ref{h55}) are valid (and of
course, $\mathfrak{g}^{\mathbf{i}}=\underline{\mathfrak{g}}^{\mathbf{i}})$.
So, if we take a spacelike hypersurface $\sigma$ such that at spatial infinity
the $\mathfrak{e}_{\mathbf{i}}$ ($\mathfrak{g}^{\mathbf{i}}(\mathfrak{e}%
_{\mathbf{j}})=\delta_{\mathbf{j}}^{\mathbf{i}}$) are tangent to $\sigma$, and
$\mathfrak{e}_{\mathbf{0}}\rightarrow\partial/\partial t$ is orthonormal to
$\sigma$, then we have $\mathbf{E}^{\prime}=\mathbf{E}$ since in this case
$-N\underline{\mathfrak{g}}_{i}\wedge\underset{%
%TCIMACRO{\TeXButton{sm}{\sslm}}%
%BeginExpansion
\sslm
%EndExpansion
}{\star}\underline{d}\underline{\mathfrak{g}}^{i}\rightarrow-\mathfrak{g}%
_{i}\wedge\underset{%
%TCIMACRO{\TeXButton{sg}{\sslg}}%
%BeginExpansion
\sslg
%EndExpansion
}{\star}(\mathfrak{g}^{0}\wedge d\mathfrak{g}^{i})$ which recalling Eq.(6.27)
is the asymptotic value of $\underset{%
%TCIMACRO{\TeXButton{sg}{\sslg}}%
%BeginExpansion
\sslg
%EndExpansion
}{\star}\mathcal{S}^{0}$ (taking into account that at spatial infinity
$d\mathfrak{g}^{0}=0$)

\section{Conclusions}

In this paper we present a theory of the gravitational field where this field
is mathematically represented a $(1,1)$-extensor field $%
%TCIMACRO{\TeXButton{h}{\slh}}%
%BeginExpansion
\slh
%EndExpansion
$ living in Minkowski spacetime $(M,%
%TCIMACRO{\TeXButton{eta}{\mbox{\boldmath{$\eta$}}}}%
%BeginExpansion
\mbox{\boldmath{$\eta$}}%
%EndExpansion
,D,\tau_{\mathbf{\eta}},\uparrow)$ and\ describing a plastic distortion of a
medium that we called the Lorentz vacuum. We showed (Remark 6.2) that the
presence of a nontrivial distortion field can be described by distinct
effective geometrical structures $(M=\mathbb{R}^{4},%
%TCIMACRO{\TeXButton{itg}{\itg}}%
%BeginExpansion
\itg
%EndExpansion
=%
%TCIMACRO{\TeXButton{h}{\slh}}%
%BeginExpansion
\slh
%EndExpansion
\eta%
%TCIMACRO{\TeXButton{h}{\slh}}%
%BeginExpansion
\slh
%EndExpansion
^{\dagger},\nabla,\tau_{%
%TCIMACRO{\TeXButton{sg}{\sslg}}%
%BeginExpansion
\sslg
%EndExpansion
})$ which may described a Lorentzian spacetime, a teleparallel spacetime or
even a more general Riemann-Cartan spacetime \footnote{In that latter case we
can get equations of motion similar to Einstein equations but where the matter
tensor appears as a combination of terms proportional to the torsion tensor,
that as well known in continuum theories describe some special deffects in the
medium. This possibility will be discussed with more details in another
publication.}. We postulated a Lagrangian density for the distortion field in
interaction with matter, and developed with details the mathematical theory
(the multiform and extensor calculus) necessary to obtain the equations of
motion of the theory from a variational principle. Moreover, we showed that
when our theory is written in terms of the global potentials $\mathfrak{g}%
^{\mathbf{\alpha}}=%
%TCIMACRO{\TeXButton{h}{\slh}}%
%BeginExpansion
\slh
%EndExpansion
(%
%TCIMACRO{\TeXButton{vt}{\mbox{\boldmath{$\vartheta$}}}}%
%BeginExpansion
\mbox{\boldmath{$\vartheta$}}%
%EndExpansion
^{\mathbf{\alpha}})\in\sec%
%TCIMACRO{\dbigwedge \nolimits^{1}}%
%BeginExpansion
{\displaystyle\bigwedge\nolimits^{1}}
%EndExpansion
T^{\ast}M$ obtained by deformation of the (continuum) cosmic lattice
describing the Lorentz vacuum in its ground state the Lagrangian density
contains a term of the Yang-Mills type plus a gauge fixing term and plus a
term describing the interaction of the\ `vorticities' of the potentials
$\mathfrak{g}^{\mathbf{\alpha}}$ and no connection appears in it. We proved
that in our theory (in contrast with \textit{GRT}) there are trustful
conservation laws of energy-momentum and angular momentum. We showed moreover
how the results obtained for the energy-momentum conservation from the
Lagrangian formalism appears directly in the Hamiltonian formalism of the
theory and how the energy obtained through the Hamiltonian formalism is also
related to the concept of \textit{ADM} energy used in \textit{GRT. }We also
discussed in Appendix F why a gauge theory of gravitation described in and
apparently similar to ours do not work. To end we would like to observe that
eventually some readers will not fell happy for seeing us to introduce an
universal medium filling all spacetime, because it eventually looks like the
revival of ether of the $XIX^{th}$ century. To those people we ask to study
the arguments for the existence of such a medium in, e.g.,
\cite{laughlin,schmelzer,schmelzer1,schmelzer2,volovik} and also analyze
references \cite{unzicker0,unzicker1} which show that if we interpret
particles as defects in the medium\footnote{Those topological defects
introduce obstructions for the Pfaff topology generated by the $\mathfrak{g}%
^{\mathbf{\alpha}}$ \cite{kiehn,kiehn1}, a subject that we shall discuss in
another publication.}, then some characteristic relativistic and quantum field
theory properties of those objects emerge in a natural way and also there is
hope to also describe the electromagnetic field as some kind of distortion of
the medium. We will come back to those issues in a forthcoming paper. We hope
to have motivated the reader to the idea that eventually is time for
deconstruct the idea that gravitation must be necessarily associated with the
geometry of spacetime and look for its real nature.

\begin{acknowledgement}
The authors are grateful to Dr. E. A. Notte-Cuello, Dr. A. M. Moya, Dr. R. da
Rocha and particularly to Dr. Q. A. G. Souza for many useful discussions on
the subject of this paper.
\end{acknowledgement}

\appendix

\section{May a Torus with Null Riemann Curvature Live on $\mathbb{E}^{3}$?}

We can also give a flat Riemann curvature tensor for\footnote{Recall that
$\mathcal{T}_{1}$ and $\mathcal{T}_{2}$ have been defined in Section 1.1.}
$\mathcal{T}_{3}\simeq S^{1}\times S^{1}\simeq\mathcal{T}\simeq\mathcal{T}%
_{2}$. First parametrize $\mathcal{T}_{3}$ by $(x^{1},x^{2})$, $x^{1},x^{2}%
\in\mathbb{R}$, such that its coordinates in $\mathbb{R}^{3}$ are $%
%TCIMACRO{\TeXButton{x}{\slx}}%
%BeginExpansion
\slx
%EndExpansion
(x^{1},x^{2})=%
%TCIMACRO{\TeXButton{x}{\slx}}%
%BeginExpansion
\slx
%EndExpansion
(x^{1}+2\pi,x^{2}+2\pi)$ and%
\begin{equation}%
%TCIMACRO{\TeXButton{x}{\slx}}%
%BeginExpansion
\slx
%EndExpansion
(x^{1},x^{2})=(h(x^{1})\cos x^{2},h(x^{1})\sin x^{2},l(x^{1}), \label{0.3a}%
\end{equation}
with%
\begin{equation}
h(x^{1})=R+r\cos x^{1}\text{, }l(x^{1})=b\sin x^{1}, \label{0.3b}%
\end{equation}
where \textit{here} $R$ and $r$ are real positive constants and $R>r$ (See
Figure 1). Now, $T\mathcal{T}_{3}$(the tangent bundle of $\mathcal{T}_{3}$)
has a global coordinate basis $\{\partial/\partial x^{1},\partial/\partial
x^{2}\}$and the induced metric on $\mathcal{T}_{3}$ is easily found as%
\begin{equation}%
%TCIMACRO{\TeXButton{g}{\slg}}%
%BeginExpansion
\slg
%EndExpansion
=r^{2}dx^{1}\otimes dx^{1}+(R+r\cos x^{1})^{2}dx^{2}\otimes dx^{2}
\label{0.3c'}%
\end{equation}

Now, let us introduced a connection $\nabla$ on $\mathcal{T}_{3}$ such that
\begin{equation}
\nabla_{\partial/\partial x^{i}}\partial/\partial x^{j}=0 \label{0.3d}%
\end{equation}

With respect to this connection it is immediate to verify that its torsion and
curvature tensors are null, but the \emph{nonmetricity} of the connection is
non null. Indeed,
\begin{equation}
\nabla_{\partial/\partial x^{i}}%
%TCIMACRO{\TeXButton{g}{\slg}}%
%BeginExpansion
\slg
%EndExpansion
=-2(R+r\cos x^{1})\sin x^{1}. \label{0.3e}%
\end{equation}
So, in this case we have a particular Riemann-Cartan-Weyl \textit{GSS}
$(\mathcal{T}_{3},%
%TCIMACRO{\TeXButton{g}{\slg}}%
%BeginExpansion
\slg
%EndExpansion
,\nabla)$ with $\mathfrak{A}\neq0,$ $\Theta=0,$ $\mathfrak{R}=0$.

We can alternatively define the parallelism rule on $\mathcal{T}_{3}$ by
introducing this time a metric compatible connection $\nabla^{\prime}$ as
follows. First we introduce the global orthonormal basis for $T\mathcal{T}%
_{3}$, $\{\mathbf{e}_{i}\}$ with%
\begin{equation}
\mathbf{e}_{\mathbf{1}}=\frac{1}{r}\frac{\partial}{\partial x^{1}}\text{,
}\mathbf{e}_{\mathbf{2}}=\frac{1}{(R+r\cos x^{1})}\frac{\partial}{\partial
x^{2}}, \label{0.3f}%
\end{equation}
and then postulate that
\begin{equation}
\nabla_{\mathbf{e}_{i}}^{\prime}\mathbf{e}_{\mathbf{j}}=0. \label{0.3g}%
\end{equation}

Since
\begin{equation}
\lbrack\mathbf{e}_{\mathbf{1}},\mathbf{e}_{\mathbf{2}}]=\frac{r\sin x^{1}%
}{r(R+r\cos x^{1})}\mathbf{e}_{\mathbf{2}} \label{0.3h}%
\end{equation}
we immediately get that the torsion operator of $\nabla^{\prime}$ is%
\begin{equation}
\overset{\triangledown^{\prime}}{\tau}(\mathbf{e}_{\mathbf{1}},\mathbf{e}%
_{\mathbf{2}})=[\mathbf{e}_{\mathbf{1}},\mathbf{e}_{\mathbf{2}}]=\frac{R\sin
x^{1}}{r(R+r\cos x^{1})}\mathbf{e}_{\mathbf{2}}, \label{0.3i}%
\end{equation}
and the non null components of the torsion tensor are $T_{\mathbf{12}%
}^{\mathbf{2}}$ and $T_{\mathbf{21}}^{\mathbf{2}}=-T_{\mathbf{12}}%
^{\mathbf{2}},$
\begin{align}
T_{\mathbf{12}}^{\mathbf{2}}  &  =\text{ }\overset{\triangledown^{\prime
}}{\Theta}(\theta^{\mathbf{2}},%
%TCIMACRO{\TeXButton{e}{\sle}}%
%BeginExpansion
\sle
%EndExpansion
_{\mathbf{1}},%
%TCIMACRO{\TeXButton{e}{\sle}}%
%BeginExpansion
\sle
%EndExpansion
_{\mathbf{2}})=-\theta^{\mathbf{2}}(\mathbf{\tau}(%
%TCIMACRO{\TeXButton{e}{\sle}}%
%BeginExpansion
\sle
%EndExpansion
_{\mathbf{1}},%
%TCIMACRO{\TeXButton{e}{\sle}}%
%BeginExpansion
\sle
%EndExpansion
_{\mathbf{2}}))\label{0.3ii}\\
&  =-\theta^{\mathbf{2}}\left(  -[%
%TCIMACRO{\TeXButton{e}{\sle}}%
%BeginExpansion
\sle
%EndExpansion
_{\mathbf{1}},%
%TCIMACRO{\TeXButton{e}{\sle}}%
%BeginExpansion
\sle
%EndExpansion
_{\mathbf{2}}]\right)  =-\frac{R\sin x^{1}}{r(R+r\cos x^{1})}. \label{0.3iii}%
\end{align}

Also it is trivial to see that the curvature operator is null and that
$\nabla^{\prime}%
%TCIMACRO{\TeXButton{g}{\slg}}%
%BeginExpansion
\slg
%EndExpansion
=0$ and in his case we have a particular Riemann-Cartan \textit{MCGSS
}$(\mathcal{T}_{3},%
%TCIMACRO{\TeXButton{g}{\slg}}%
%BeginExpansion
\slg
%EndExpansion
,\nabla^{\prime})$, with $\Theta\neq0$ and $\mathfrak{R}=0$.

%

%TCIMACRO{\FRAME{dtbphFUX}{4.2099in}{1.7097in}{0pt}{\Qcb{Some Possible
%\QTR{it}{GSS }for the Torus $\QTR{cal}{T}_{3}$ living in $\QTR{Bbb}{E}^{3}$
%where the grid defines the parallelism}}{\Qlb{KXXNS601}}{Figure}%
%{\special{ language "Scientific Word";  type "GRAPHIC";  display "USEDEF";
%valid_file "T";  width 4.2099in;  height 1.7097in;  depth 0pt;
%original-width 15.0503in;  original-height 5.9378in;  cropleft "0";
%tempfilename 'LN8XY900.wmf';tempfile-properties "XPR";}} }%
%BeginExpansion
\begin{center}
\fbox{\includegraphics[
natheight=5.937800in,
natwidth=15.050300in,
height=1.7097in,
width=4.2099in
]%
{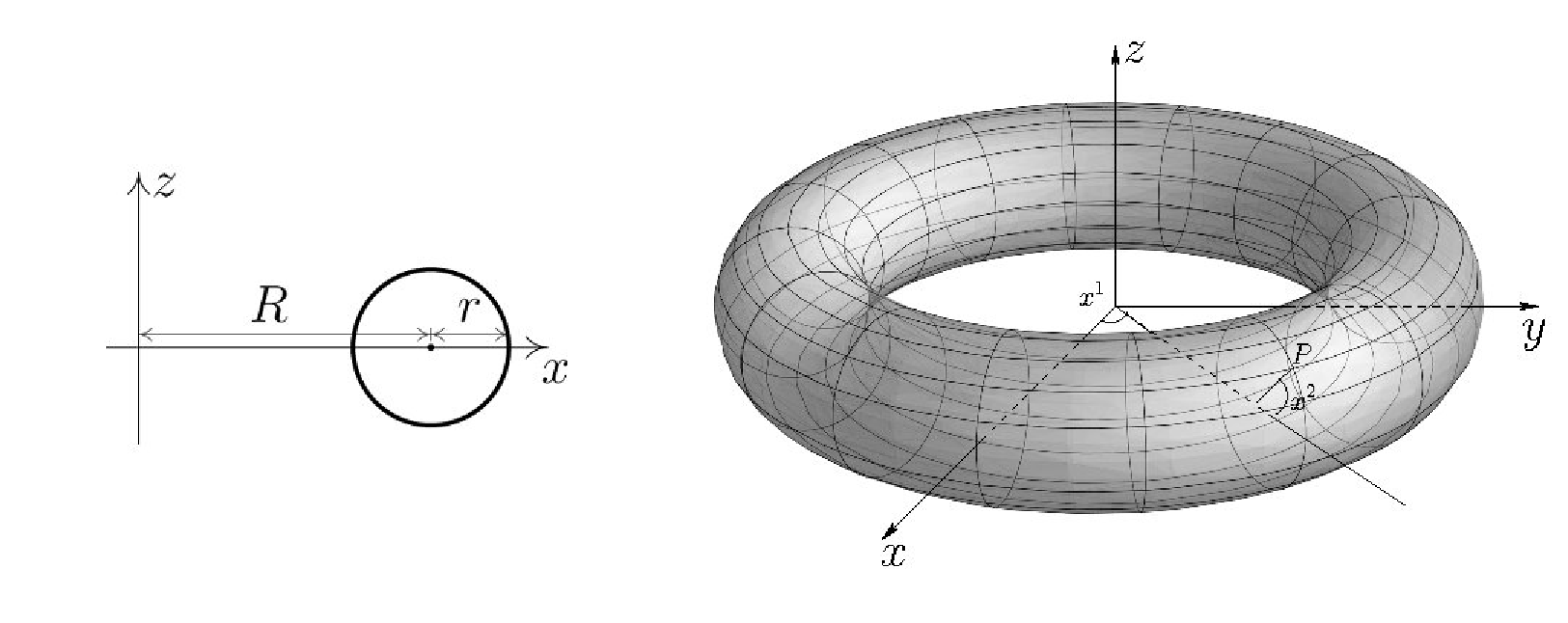}%
}\\
Some Possible \textit{GSS }for the Torus $\mathcal{T}_{3}$ living in
$\mathbb{E}^{3}$ where the grid defines the parallelism
\label{KXXNS601}%
\end{center}
%EndExpansion

\section{The Levi-Civita and the Nunes Connection on the Punctured Sphere}

First consider $S^{2}$, an sphere of radius $\mathfrak{R}=1$ embedded in
$\mathbb{R}^{3}$. Let $(x^{1},x^{2})=(\vartheta,\varphi)$ $0<\vartheta<\pi$,
$0<\varphi<2\pi$, be the standard spherical coordinates of $S^{2}$, which
covers all the open set $U$ which is $S^{2}$ with the \textit{exclusion} of a
semi-circle uniting the \textit{north} and south \textit{poles}.

Introduce the \textit{coordinate bases}\textbf{ }
\begin{equation}
\{{\mbox{\boldmath$\partial$}}_{\mu}\},\{\theta^{\mu}=dx^{\mu}\}
\end{equation}
for $T^{\ast}U$ and $T^{\ast}U$. Next introduce the \textit{orthonormal
bases}\textbf{ }$\{%
%TCIMACRO{\TeXButton{e}{\sle}}%
%BeginExpansion
\sle
%EndExpansion
_{\mathbf{a}}\},\{\theta^{\mathbf{a}}\}$ for $T^{\ast}U$ and $T^{\ast}U$ with%
\begin{subequations}
\begin{align}%
%TCIMACRO{\TeXButton{e}{\sle}}%
%BeginExpansion
\sle
%EndExpansion
_{\mathbf{1}}  &  ={\mbox{\boldmath$\partial$}}_{1}\text{, }%
%TCIMACRO{\TeXButton{e}{\sle}}%
%BeginExpansion
\sle
%EndExpansion
_{\mathbf{2}}=\frac{1}{\sin x^{1}}{\mbox{\boldmath$\partial$}}_{2}%
,\label{ba}\\
\theta^{\mathbf{1}}  &  =dx^{1}\text{, }\theta^{\mathbf{2}}=\sin x^{1}dx^{2}.
\end{align}
Then,%
\end{subequations}
\begin{align}
\lbrack%
%TCIMACRO{\TeXButton{e}{\sle}}%
%BeginExpansion
\sle
%EndExpansion
_{\mathbf{i}},%
%TCIMACRO{\TeXButton{e}{\sle}}%
%BeginExpansion
\sle
%EndExpansion
_{\mathbf{j}}]  &  =c_{\mathbf{ij}}^{\mathbf{k}}%
%TCIMACRO{\TeXButton{e}{\sle}}%
%BeginExpansion
\sle
%EndExpansion
_{\mathbf{k}},\\
c_{\mathbf{12}}^{\mathbf{2}}  &  =-c_{\mathbf{21}}^{\mathbf{2}}=-\cot
x^{\mathbf{1}}.\nonumber
\end{align}
Moreover the metric $%
%TCIMACRO{\TeXButton{slg}{\slg}}%
%BeginExpansion
\slg
%EndExpansion
$\texttt{ }$\in\sec T_{2}^{0}S^{2}$ inherited form the ambient Euclidean
metric is:
\begin{align}%
%TCIMACRO{\TeXButton{g}{\slg}}%
%BeginExpansion
\slg
%EndExpansion
&  =dx^{1}\otimes dx^{1}+\sin^{2}x^{1}dx^{2}\otimes dx^{2}\nonumber\\
&  =\theta^{\mathbf{1}}\otimes\theta^{\mathbf{1}}+\theta^{\mathbf{2}}%
\otimes\theta^{\mathbf{2}}.
\end{align}

The Levi-Civita connection $D$ of $%
%TCIMACRO{\TeXButton{g}{\slg}}%
%BeginExpansion
\slg
%EndExpansion
$ has the following non null connections coefficients $\Gamma_{\mu\nu}^{\rho}$
in the coordinate basis (just introduced):%
\begin{align}
D_{\mbox{\tiny\boldmath$\partial$}_{\mu}}{\mbox{\boldmath$\partial$}}_{\nu}
&  =\Gamma_{\mu\nu}^{\rho}{\mbox{\boldmath$\partial$}}_{\rho},\nonumber\\
\text{ }\Gamma_{21}^{2}  &  =\Gamma_{\theta\varphi}^{\varphi}=\Gamma_{12}%
^{2}=\Gamma_{\varphi\theta}^{\varphi}=\cot\vartheta\text{, }\Gamma_{22}%
^{1}=\Gamma_{\varphi\varphi}^{\vartheta}=-\cos\vartheta\sin\vartheta.
\label{10x}%
\end{align}
Also, in the basis $\{%
%TCIMACRO{\TeXButton{e}{\sle}}%
%BeginExpansion
\sle
%EndExpansion
_{\mathbf{a}}\}$, $D_{%
%TCIMACRO{\TeXButton{e}{\sle}}%
%BeginExpansion
\sle
%EndExpansion
_{\mathbf{i}}}%
%TCIMACRO{\TeXButton{e}{\sle}}%
%BeginExpansion
\sle
%EndExpansion
_{\mathbf{j}}=\omega_{\mathbf{ij}}^{\mathbf{k}}%
%TCIMACRO{\TeXButton{e}{\sle}}%
%BeginExpansion
\sle
%EndExpansion
_{\mathbf{k}}$ and the non null coefficients are:
\begin{equation}
\omega_{\mathbf{21}}^{\mathbf{2}}=\cot\vartheta\text{, }\omega_{\mathbf{22}%
}^{\mathbf{1}}=-\cot\vartheta. \label{crist}%
\end{equation}

The torsion \ and the (Riemann) curvature tensors of $D$ are%

\begin{equation}
\Theta(\theta^{\mathbf{k}},%
%TCIMACRO{\TeXButton{e}{\sle}}%
%BeginExpansion
\sle
%EndExpansion
_{\mathbf{i}},%
%TCIMACRO{\TeXButton{e}{\sle}}%
%BeginExpansion
\sle
%EndExpansion
_{\mathbf{j}})=\theta^{\mathbf{k}}(\mathbf{\tau}(%
%TCIMACRO{\TeXButton{e}{\sle}}%
%BeginExpansion
\sle
%EndExpansion
_{\mathbf{i}},%
%TCIMACRO{\TeXButton{e}{\sle}}%
%BeginExpansion
\sle
%EndExpansion
_{\mathbf{j}}))=\theta^{\mathbf{k}}\left(  D_{%
%TCIMACRO{\TeXButton{e}{\sle}}%
%BeginExpansion
\sle
%EndExpansion
_{\mathbf{j}}}%
%TCIMACRO{\TeXButton{e}{\sle}}%
%BeginExpansion
\sle
%EndExpansion
_{\mathbf{i}}-D_{%
%TCIMACRO{\TeXButton{e}{\sle}}%
%BeginExpansion
\sle
%EndExpansion
_{\mathbf{i}}}%
%TCIMACRO{\TeXButton{e}{\sle}}%
%BeginExpansion
\sle
%EndExpansion
_{\mathbf{j}}-[%
%TCIMACRO{\TeXButton{e}{\sle}}%
%BeginExpansion
\sle
%EndExpansion
_{\mathbf{i}},%
%TCIMACRO{\TeXButton{e}{\sle}}%
%BeginExpansion
\sle
%EndExpansion
_{\mathbf{j}}]\right)  , \label{tolevi}%
\end{equation}%
\begin{equation}
\mathbf{R(}%
%TCIMACRO{\TeXButton{e}{\sle}}%
%BeginExpansion
\sle
%EndExpansion
_{\mathbf{k}},\theta^{\mathbf{a}},%
%TCIMACRO{\TeXButton{e}{\sle}}%
%BeginExpansion
\sle
%EndExpansion
_{\mathbf{i}},%
%TCIMACRO{\TeXButton{e}{\sle}}%
%BeginExpansion
\sle
%EndExpansion
_{\mathbf{j}}\mathbf{)}=\theta^{\mathbf{a}}\left(  \left[  D_{%
%TCIMACRO{\TeXButton{e}{\sle}}%
%BeginExpansion
\sle
%EndExpansion
_{\mathbf{i}}}D_{%
%TCIMACRO{\TeXButton{e}{\sle}}%
%BeginExpansion
\sle
%EndExpansion
_{\mathbf{j}}}-D_{%
%TCIMACRO{\TeXButton{e}{\sle}}%
%BeginExpansion
\sle
%EndExpansion
_{\mathbf{j}}}D_{%
%TCIMACRO{\TeXButton{e}{\sle}}%
%BeginExpansion
\sle
%EndExpansion
_{\mathbf{i}}}-D_{[%
%TCIMACRO{\TeXButton{e}{\sle}}%
%BeginExpansion
\sle
%EndExpansion
_{\mathbf{i}},\mathtt{\ }%
%TCIMACRO{\TeXButton{e}{\sle}}%
%BeginExpansion
\sle
%EndExpansion
_{\mathbf{j}}]}\right]
%TCIMACRO{\TeXButton{e}{\sle}}%
%BeginExpansion
\sle
%EndExpansion
_{\mathbf{k}}\right)  , \label{RIEMLEVI}%
\end{equation}
which results in\ $\mathcal{T}=0$ and that the non null components of
$\mathbf{R}$ are $R_{\mathbf{1\;21}}^{\;\mathbf{1}}=-R_{\mathbf{1\;12}%
}^{\;\mathbf{1}}=R_{\mathbf{1\;12}}^{\;\mathbf{2}}=-R_{\mathbf{1\;12}%
}^{\;\mathbf{2}}=-1$.

Since the Riemann curvature tensor is non null the parallel transport of a
given vector depends on the path to be followed. We say that a vector (say
$\mathbf{v}_{0}$) is parallel transported along a generic path
$\mathbb{R\supset}I\mapsto\gamma(s)\in\mathbb{R}^{3}$ (say, from $A=\gamma(0)$
to $B=\gamma(1)$) with tangent vector $\gamma_{\ast}(s)$ (at $\gamma(s)$) if
it determines a vector field $\mathbf{V}$ along $\gamma$ satisfying
\begin{equation}
D_{\gamma_{\ast}}\mathbf{V}=0, \label{lctransp}%
\end{equation}
and such that $\mathbf{V}(\gamma(0))=\mathbf{v}_{0}$. When the path is a
geodesic\footnote{We recall that a geodesic of $D$ also determines the minimal
distance (as given by the metric $%
%TCIMACRO{\TeXButton{g}{\slg}}%
%BeginExpansion
\slg
%EndExpansion
$) between any two points on $S^{2}$.} of the connection $D$, i.e.,a curve
\ $\mathbb{R\supset}I\mapsto c(s)\in\mathbb{R}^{3}$ with tangent vector
$c_{\ast}(s)$ (at $c(s)$) satisfying%
\begin{equation}
D_{c_{\ast}}c_{\ast}=0, \label{geodlv}%
\end{equation}
the parallel transported vector along a $c$ forms a constant angle with
$c_{\ast}$. Indeed, from Eq.(\ref{lctransp}) it is $\gamma_{\ast}\underset{%
%TCIMACRO{\TeXButton{sg}{\sslg}}%
%BeginExpansion
\sslg
%EndExpansion
}{\cdot}D_{\gamma_{\ast}}\mathbf{V}=0$. Then taking into account
Eq.(\ref{geodlv}) it follows that
\[
D_{\gamma_{\ast}}(\gamma_{\ast}\underset{%
%TCIMACRO{\TeXButton{sg}{\sslg}}%
%BeginExpansion
\sslg
%EndExpansion
}{\cdot}\mathbf{V})=0.
\]
i.e., $\gamma_{\ast}\underset{%
%TCIMACRO{\TeXButton{sg}{\sslg}}%
%BeginExpansion
\sslg
%EndExpansion
}{\cdot}\mathbf{V}=$ \emph{constant}$.$This is clearly illustrated in Figure 1
(adapted from \cite{berg92}).

%

%TCIMACRO{\FRAME{fhFUX}{2.2355in}{2.1395in}{0pt}{\Qcb{Levi-Civita and Nunes
%transport of a vector $\QTR{bf}{v}_{0}$ starting at $p$ through the paths
%$psr$ and $pqr$. Levi-Civita transport through psr leads to $\QTR{bf}{v}_{1}$
%whereas Nunes transport leads to $\QTR{bf}{v}_{2}$. Along $pqr$ both
%Levi-Civita and Nunes transport agree and leads to $\QTR{bf}{v}_{2}.$}}%
%{}{Figure}{\special{ language "Scientific Word";  type "GRAPHIC";
%display "USEDEF";  valid_file "T";  width 2.2355in;  height 2.1395in;
%depth 0pt;  original-width 3.1548in;  original-height 2.9966in;
%cropleft "0";  croptop "1";  cropright "1";  cropbottom "0";
%tempfilename 'LN8XY901.wmf';tempfile-properties "XPR";}} }%
%BeginExpansion
\begin{figure}[h]%
\centering
\fbox{\includegraphics[
natheight=2.996600in,
natwidth=3.154800in,
height=2.1395in,
width=2.2355in
]%
{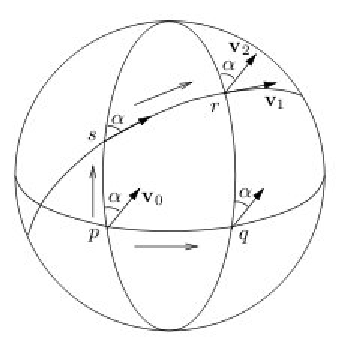}%
}\caption{Levi-Civita and Nunes transport of a vector $\mathbf{v}_{0}$
starting at $p$ through the paths $psr$ and $pqr$. Levi-Civita transport
through psr leads to $\mathbf{v}_{1}$ whereas Nunes transport leads to
$\mathbf{v}_{2}$. Along $pqr$ both Levi-Civita and Nunes transport agree and
leads to $\mathbf{v}_{2}.$}%
\end{figure}
%EndExpansion

Consider next the manifold $\mathring{S}^{2}$ $=\{S^{2}\backslash
$\textrm{north pole + south pole}$\}\subset\mathbb{R}^{3}$, which is our
sphere of radius $\mathfrak{R}=1$ but this time \textit{excluding} the north
and south poles. Let again \texttt{ }$%
%TCIMACRO{\TeXButton{slg}{\slg}}%
%BeginExpansion
\slg
%EndExpansion
\in\sec T_{2}^{0}\mathring{S}^{2}$ be the metric field on $\mathring{S}^{2}$
inherited from the ambient space $\mathbb{R}^{3}$ and introduce on
$\mathring{S}^{2}$ the Nunes (or navigator) connection $\mathbf{\nabla}$
defined by the following parallel transport rule: a vector at an arbitrary
point of $\mathring{S}^{2}$ is parallel transported along a curve $\gamma$, if
it determines a vector field on $\gamma$ such that at any point of $\gamma$
the angle between the \ transported vector and the vector tangent to the
latitude line passing through that point is constant during the transport.
This is clearly illustrated in Figure 3. and to distinguish the Nunes
transport from the Levi-Civita transport we ask also for the reader to study
with attention the caption of Figure 1.%

%TCIMACRO{\FRAME{dtbphFUX}{4.3076in}{2.4889in}{0pt}{\Qcb{Characterization of
%the Nunes connection.}}{}{Figure}{\special{ language "Scientific Word";
%type "GRAPHIC";  display "USEDEF";  valid_file "T";  width 4.3076in;
%height 2.4889in;  depth 0pt;  original-width 6.8787in;
%original-height 3.9401in;  cropleft "0";  croptop "1";  cropright "1";
%cropbottom "0";  tempfilename 'LN8XY902.wmf';tempfile-properties "XPR";}} }%
%BeginExpansion
\begin{center}
\fbox{\includegraphics[
natheight=3.940100in,
natwidth=6.878700in,
height=2.4889in,
width=4.3076in
]%
{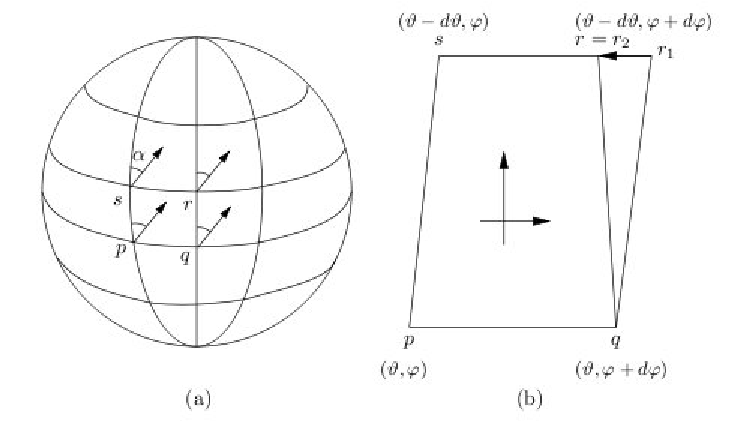}%
}\\
Characterization of the Nunes connection.
\end{center}
%EndExpansion

We recall that\emph{ }from the calculation of the Riemann tensor $\mathbf{R}$
it follows that the structures $(\mathring{S}^{2},%
%TCIMACRO{\TeXButton{g}{\slg}}%
%BeginExpansion
\slg
%EndExpansion
,D,\tau_{%
%TCIMACRO{\TeXButton{sg}{\sslg}}%
%BeginExpansion
\sslg
%EndExpansion
})$ and also $(S^{2},%
%TCIMACRO{\TeXButton{g}{\slg}}%
%BeginExpansion
\slg
%EndExpansion
,D,\tau_{%
%TCIMACRO{\TeXButton{sg}{\sslg}}%
%BeginExpansion
\sslg
%EndExpansion
})$ are Riemann spaces of \textit{constant curvature}. We now show that the
structure $(\mathring{S}^{2},%
%TCIMACRO{\TeXButton{g}{\slg}}%
%BeginExpansion
\slg
%EndExpansion
,\mathbf{\nabla},\tau_{%
%TCIMACRO{\TeXButton{sg}{\sslg}}%
%BeginExpansion
\sslg
%EndExpansion
})$ is a \textit{teleparallel space\footnote{As recalled in Section 1, a
teleparallel manifold $M$ is characterized by the existence of global vector
fields which is a basis for $T_{x}M$ for any $x\in M$. The reason for
considering $\mathring{S}^{2}$ for introducing the Nunes connection is that as
well known (see, e.g., \cite{docarmo}) $S^{2}$ does not admit a continuous
vector field that is nonnull at on points of it. }}, with zero Riemann
curvature tensor, but non zero torsion tensor.

Indeed, from Figure 2 it is clear that (a) if a vector is transported along
the \textit{infinitesimal} quadrilateral $pqrs$ composed of latitudes and
longitudes, first starting from $p$ along $pqr$ and then starting from $p$
along $psr$ the parallel transported vectors that result in both cases will
coincide (study also the caption of Figure 2.

Now, the vector fields $%
%TCIMACRO{\TeXButton{e}{\sle}}%
%BeginExpansion
\sle
%EndExpansion
_{\mathbf{1}}$ and $%
%TCIMACRO{\TeXButton{e}{\sle}}%
%BeginExpansion
\sle
%EndExpansion
_{\mathbf{2}}$ in Eq.(\ref{ba}) define a basis for each point $p$ of
$T_{p}\mathring{S}^{2}$ and $\mathbf{\nabla}$ is clearly characterized by:
\begin{equation}
\mathbf{\nabla}_{%
%TCIMACRO{\TeXButton{e}{\sle}}%
%BeginExpansion
\sle
%EndExpansion
_{\mathbf{j}}}%
%TCIMACRO{\TeXButton{e}{\sle}}%
%BeginExpansion
\sle
%EndExpansion
_{\mathbf{i}}=0. \label{6x}%
\end{equation}

The components of curvature operator are:%

\begin{equation}
\overset{\triangledown}{\mathbf{R}}\mathbf{(}%
%TCIMACRO{\TeXButton{e}{\sle}}%
%BeginExpansion
\sle
%EndExpansion
_{\mathbf{k}},\theta^{\mathbf{a}},%
%TCIMACRO{\TeXButton{e}{\sle}}%
%BeginExpansion
\sle
%EndExpansion
_{\mathbf{i}},%
%TCIMACRO{\TeXButton{e}{\sle}}%
%BeginExpansion
\sle
%EndExpansion
_{\mathbf{j}}\mathbf{)}=\theta^{\mathbf{a}}\left(  \left[  \mathbf{\nabla}_{%
%TCIMACRO{\TeXButton{e}{\sle}}%
%BeginExpansion
\sle
%EndExpansion
_{\mathbf{i}}}\mathbf{\nabla}_{%
%TCIMACRO{\TeXButton{e}{\sle}}%
%BeginExpansion
\sle
%EndExpansion
_{\mathbf{j}}}-\mathbf{\nabla}_{%
%TCIMACRO{\TeXButton{e}{\sle}}%
%BeginExpansion
\sle
%EndExpansion
_{\mathbf{j}}}\mathbf{\nabla}_{%
%TCIMACRO{\TeXButton{e}{\sle}}%
%BeginExpansion
\sle
%EndExpansion
_{\mathbf{i}}}-\mathbf{\nabla}_{[%
%TCIMACRO{\TeXButton{e}{\sle}}%
%BeginExpansion
\sle
%EndExpansion
_{\mathbf{i}},%
%TCIMACRO{\TeXButton{e}{\sle}}%
%BeginExpansion
\sle
%EndExpansion
_{\mathbf{j}}]}\right]
%TCIMACRO{\TeXButton{e}{\sle}}%
%BeginExpansion
\sle
%EndExpansion
_{\mathbf{k}}\right)  =0,
\end{equation}
and the torsion operator is
\begin{align}
\overset{\triangledown}{\mathbf{\tau}}(%
%TCIMACRO{\TeXButton{e}{\sle}}%
%BeginExpansion
\sle
%EndExpansion
_{\mathbf{i}},%
%TCIMACRO{\TeXButton{e}{\sle}}%
%BeginExpansion
\sle
%EndExpansion
_{\mathbf{j}})  &  =\mathbf{\nabla}_{%
%TCIMACRO{\TeXButton{e}{\sle}}%
%BeginExpansion
\sle
%EndExpansion
_{\mathbf{j}}}%
%TCIMACRO{\TeXButton{e}{\sle}}%
%BeginExpansion
\sle
%EndExpansion
_{\mathbf{i}}-\mathbf{\nabla}_{%
%TCIMACRO{\TeXButton{e}{\sle}}%
%BeginExpansion
\sle
%EndExpansion
_{\mathbf{i}}}%
%TCIMACRO{\TeXButton{e}{\sle}}%
%BeginExpansion
\sle
%EndExpansion
_{\mathbf{j}}-[%
%TCIMACRO{\TeXButton{e}{\sle}}%
%BeginExpansion
\sle
%EndExpansion
_{\mathbf{i}},%
%TCIMACRO{\TeXButton{e}{\sle}}%
%BeginExpansion
\sle
%EndExpansion
_{\mathbf{j}}]\nonumber\\
&  =[%
%TCIMACRO{\TeXButton{e}{\sle}}%
%BeginExpansion
\sle
%EndExpansion
_{\mathbf{i}},%
%TCIMACRO{\TeXButton{e}{\sle}}%
%BeginExpansion
\sle
%EndExpansion
_{\mathbf{j}}],
\end{align}
which gives for the components of the torsion tensor, $T_{\mathbf{12}%
}^{\mathbf{2}}=-T_{\mathbf{12}}^{\mathbf{2}}=\cot\vartheta$ . It follows that
$\mathring{S}^{2}$ considered as part of the structure $(\mathring{S}^{2},%
%TCIMACRO{\TeXButton{g}{\slg}}%
%BeginExpansion
\slg
%EndExpansion
,\mathbf{\nabla},\tau_{%
%TCIMACRO{\TeXButton{sg}{\sslg}}%
%BeginExpansion
\sslg
%EndExpansion
})$ is\textit{ flat }(but has torsion)!

If you still need more details to grasp this last result, consider Figure 2
(b) which shows the standard parametrization of the points $p,q,r,s$ in terms
of the spherical coordinates introduced above. According to the geometrical
meaning of torsion, its value at a given point is determined by calculating
the difference between the (infinitesimal)\footnote{This wording, of course,
means that those vectors are identified as elements of the appropriate tangent
spaces.} vectors $pr_{1}$and $pr_{2}$. If the vector $pq$ is transported along
$ps$ one get (recalling that $\mathfrak{R}=1\mathfrak{)}$ the vector
$\mathbf{v}=sr_{1}$ such that $\left\vert
%TCIMACRO{\TeXButton{slg}{\slg}}%
%BeginExpansion
\slg
%EndExpansion
(\mathbf{v},\mathbf{v})\right\vert ^{\frac{1}{2}}=\sin\vartheta\triangle
\varphi$. On the other hand, if the vector $ps$ is transported along $pq$ one
get the vector $qr_{2}=qr$. Let $\mathbf{w}=sr$. Then,%

\begin{equation}
\left\vert
%TCIMACRO{\TeXButton{g}{\slg}}%
%BeginExpansion
\slg
%EndExpansion
(\mathbf{w},\mathbf{w})\right\vert =\sin(\vartheta-\triangle\vartheta
)\triangle\varphi\simeq\sin\vartheta\triangle\varphi-\cos\vartheta
\triangle\vartheta\triangle\varphi,
\end{equation}
Also,%

\begin{equation}
\mathbf{u}=r_{1}r_{2}=-u(\frac{1}{\sin\vartheta}\frac{\partial}{\partial
\varphi})\text{, }u=\left\vert
%TCIMACRO{\TeXButton{g}{\slg}}%
%BeginExpansion
\slg
%EndExpansion
(\mathbf{u},\mathbf{u})\right\vert =\cos\vartheta\triangle\vartheta
\triangle\varphi.
\end{equation}
Then, the connection $\mathbf{\nabla}$ of the structure $(\mathring{S}^{2},%
%TCIMACRO{\TeXButton{g}{\slg}}%
%BeginExpansion
\slg
%EndExpansion
,\mathbf{\nabla},\tau_{%
%TCIMACRO{\TeXButton{sg}{\sslg}}%
%BeginExpansion
\sslg
%EndExpansion
})$ has a non null torsion tensor $\mathcal{\bar{T}}$. \ Indeed, the component
of $\mathbf{u}=r_{1}r_{2}$ in the direction $\partial/\partial\varphi$ is
precisely $\bar{T}_{\vartheta\varphi}^{\varphi}\triangle\vartheta
\triangle\varphi$. So, one get (recalling that $\mathbf{\nabla}_{\partial
j}\partial_{i}=\Gamma_{ji}^{k}\partial_{k}$)
\begin{equation}
\bar{T}_{\vartheta\varphi}^{\varphi}=\left(  \Gamma_{\vartheta\varphi
}^{\varphi}-\Gamma_{\varphi\vartheta}^{\varphi}\right)  =-\cot\theta.
\end{equation}
To end this Appendix it is worth to show that $\mathbf{\nabla}$ is metrical
compatible, i.e., that $\mathbf{\nabla}%
%TCIMACRO{\TeXButton{g}{\slg}}%
%BeginExpansion
\slg
%EndExpansion
=0$. Indeed, we have:%
\begin{align}
0  &  =\text{ }\mathbf{\nabla}_{%
%TCIMACRO{\TeXButton{e}{\sle}}%
%BeginExpansion
\sle
%EndExpansion
_{\mathbf{c}}}%
%TCIMACRO{\TeXButton{g}{\slg}}%
%BeginExpansion
\slg
%EndExpansion
(%
%TCIMACRO{\TeXButton{e}{\sle}}%
%BeginExpansion
\sle
%EndExpansion
_{\mathbf{i}},%
%TCIMACRO{\TeXButton{e}{\sle}}%
%BeginExpansion
\sle
%EndExpansion
_{\mathbf{j}})=(\mathbf{\nabla}_{%
%TCIMACRO{\TeXButton{e}{\sle}}%
%BeginExpansion
\sle
%EndExpansion
_{\mathbf{c}}}%
%TCIMACRO{\TeXButton{g}{\slg}}%
%BeginExpansion
\slg
%EndExpansion
)(%
%TCIMACRO{\TeXButton{e}{\sle}}%
%BeginExpansion
\sle
%EndExpansion
_{\mathbf{i}},%
%TCIMACRO{\TeXButton{e}{\sle}}%
%BeginExpansion
\sle
%EndExpansion
_{\mathbf{j}})+\mathtt{\ }%
%TCIMACRO{\TeXButton{g}{\slg}}%
%BeginExpansion
\slg
%EndExpansion
(\mathbf{\nabla}_{%
%TCIMACRO{\TeXButton{e}{\sle}}%
%BeginExpansion
\sle
%EndExpansion
_{\mathbf{c}}}%
%TCIMACRO{\TeXButton{e}{\sle}}%
%BeginExpansion
\sle
%EndExpansion
_{\mathbf{i}},%
%TCIMACRO{\TeXButton{e}{\sle}}%
%BeginExpansion
\sle
%EndExpansion
_{\mathbf{j}})+\mathtt{\ }%
%TCIMACRO{\TeXButton{g}{\slg}}%
%BeginExpansion
\slg
%EndExpansion
(%
%TCIMACRO{\TeXButton{e}{\sle}}%
%BeginExpansion
\sle
%EndExpansion
_{\mathbf{i}},\mathbf{\nabla}_{%
%TCIMACRO{\TeXButton{e}{\sle}}%
%BeginExpansion
\sle
%EndExpansion
_{\mathbf{c}}}%
%TCIMACRO{\TeXButton{e}{\sle}}%
%BeginExpansion
\sle
%EndExpansion
_{\mathbf{j}})\nonumber\\
&  =(\mathbf{\nabla}_{%
%TCIMACRO{\TeXButton{e}{\sle}}%
%BeginExpansion
\sle
%EndExpansion
_{\mathbf{c}}}%
%TCIMACRO{\TeXButton{g}{\slg}}%
%BeginExpansion
\slg
%EndExpansion
)(%
%TCIMACRO{\TeXButton{e}{\sle}}%
%BeginExpansion
\sle
%EndExpansion
_{\mathbf{i}},%
%TCIMACRO{\TeXButton{e}{\sle}}%
%BeginExpansion
\sle
%EndExpansion
_{\mathbf{j}}).
\end{align}

\section{Gravitational Theory for Independent $%
%TCIMACRO{\TeXButton{h}{\slh}}%
%BeginExpansion
\slh
%EndExpansion
$ and $\Omega$ Fields on Minkowski Spacetime}

In this section we present for completeness a theory where the dynamics of the
gravitational field is such that it must be described by two independent
fields, namely the plastic distortion field $%
%TCIMACRO{\TeXButton{h}{\slh}}%
%BeginExpansion
\slh
%EndExpansion
$ (more precisely $%
%TCIMACRO{\TeXButton{h}{\slh}}%
%BeginExpansion
\slh
%EndExpansion
^{\clubsuit}$) and for a $\varkappa\eta$-gauge rotation field $\Omega
\equiv\overset{\varkappa\eta}{\Omega}$ defined on the canonical space $U$.
Those two independent fields distorts the Lorentz vacuum creating an effective
\ Minkowski-Cartan \textit{MCGSS} $(U,\eta,\varkappa)$ or equivalently a
Lorentz-Cartan \textit{MCGSS} $(U,%
%TCIMACRO{\TeXButton{itg}{\itg}}%
%BeginExpansion
\itg
%EndExpansion
=%
%TCIMACRO{\TeXButton{h}{\slh}}%
%BeginExpansion
\slh
%EndExpansion
^{\dagger}\eta%
%TCIMACRO{\TeXButton{h}{\slh}}%
%BeginExpansion
\slh
%EndExpansion
,\gamma)$ which which is viewed as gauge equivalent to $(U,\eta,\varkappa)$ as
already discussed in the main text. \ From the mathematical point of view our
theory is then described a Lagrangian where no the fields $%
%TCIMACRO{\TeXButton{h}{\slh}}%
%BeginExpansion
\slh
%EndExpansion
$ and $\Omega$ are the dynamic variables to be varied in the action
functional. The Lagrangian is postulated to be the scalar functional
\begin{equation}
\lbrack%
%TCIMACRO{\TeXButton{h}{\slh}}%
%BeginExpansion
\slh
%EndExpansion
^{\clubsuit},\Omega,\cdot\partial\Omega]\mapsto\mathcal{L}[%
%TCIMACRO{\TeXButton{h}{\slh}}%
%BeginExpansion
\slh
%EndExpansion
^{\clubsuit},\Omega,\cdot\partial\Omega]=\mathcal{R}\det[%
%TCIMACRO{\TeXButton{h}{\slh}}%
%BeginExpansion
\slh
%EndExpansion
], \label{4.41}%
\end{equation}
where $\mathcal{R}(=\overset{\gamma%
%TCIMACRO{\TeXButton{sh}{\sslh}}%
%BeginExpansion
\sslh
%EndExpansion
}{\mathcal{R}})$ \ (given by Eq.(\ref{4.18}) ) is expressed as a functional of
the fields $%
%TCIMACRO{\TeXButton{h}{\slh}}%
%BeginExpansion
\slh
%EndExpansion
^{\clubsuit},$ $\Omega$ and $\cdot\partial\Omega$, i.e.,
\begin{align}
\mathcal{R}  &  =\underline{%
%TCIMACRO{\TeXButton{h}{\slh}}%
%BeginExpansion
\slh
%EndExpansion
}^{\clubsuit}(\partial_{a}\wedge\partial_{b})\cdot\mathcal{R}_{2}(a\wedge
b)\nonumber\\
&  =\underline{%
%TCIMACRO{\TeXButton{h}{\slh}}%
%BeginExpansion
\slh
%EndExpansion
}^{\clubsuit}(\partial_{a}\wedge\partial_{b})\cdot(a\cdot\partial
\Omega(b)-b\cdot\partial\Omega(a)-\Omega([a,b])+\Omega(a)\underset{\eta
}{\times}\Omega(b)), \label{4.42}%
\end{align}
and $\det[%
%TCIMACRO{\TeXButton{h}{\slh}}%
%BeginExpansion
\slh
%EndExpansion
]$, as in Section 5.1 is expressed as a functional of $%
%TCIMACRO{\TeXButton{h}{\slh}}%
%BeginExpansion
\slh
%EndExpansion
^{\clubsuit},$ i.e., $\det[%
%TCIMACRO{\TeXButton{h}{\slh}}%
%BeginExpansion
\slh
%EndExpansion
]=(\det[%
%TCIMACRO{\TeXButton{h}{\slh}}%
%BeginExpansion
\slh
%EndExpansion
^{\clubsuit}])^{-1}$.

The Lagrangian formalism for extensor fields supposes the existence of an
action functional
\begin{equation}
\mathcal{A}=\int_{U}\mathcal{L}[%
%TCIMACRO{\TeXButton{h}{\slh}}%
%BeginExpansion
\slh
%EndExpansion
^{\clubsuit},\Omega,\cdot\partial\Omega]\text{ }\tau=\int_{U}\mathcal{R}\det[%
%TCIMACRO{\TeXButton{h}{\slh}}%
%BeginExpansion
\slh
%EndExpansion
]\text{ }\tau\label{4.44}%
\end{equation}
which according to the principle of stationary action gives for any one of the
dynamic variables a contour problem.

For the dynamic variable $%
%TCIMACRO{\TeXButton{h}{\slh}}%
%BeginExpansion
\slh
%EndExpansion
^{\clubsuit}$ it is
\begin{equation}
\int_{U}%
%TCIMACRO{\TeXButton{delta}{\mbox{\boldmath{$\delta$}}}}%
%BeginExpansion
\mbox{\boldmath{$\delta$}}%
%EndExpansion
_{%
%TCIMACRO{\TeXButton{sh}{\sslh}}%
%BeginExpansion
\sslh
%EndExpansion
^{\clubsuit}}^{w}\mathcal{L}[%
%TCIMACRO{\TeXButton{h}{\slh}}%
%BeginExpansion
\slh
%EndExpansion
^{\clubsuit},\Omega,\cdot\partial\Omega]\text{ }\tau=0, \label{4.45}%
\end{equation}
for any smooth $(1,1)$-extensor field $w$ such that $\left.  w\right\vert
_{\partial U}=0$.

For the dynamic variable $\Omega$ it is
\begin{equation}
\int_{U}%
%TCIMACRO{\TeXButton{delta}{\mbox{\boldmath{$\delta$}}}}%
%BeginExpansion
\mbox{\boldmath{$\delta$}}%
%EndExpansion
_{\Omega}^{B}\mathcal{L}[%
%TCIMACRO{\TeXButton{h}{\slh}}%
%BeginExpansion
\slh
%EndExpansion
^{\clubsuit},\Omega,\cdot\partial\Omega]\text{ }\tau=0, \label{4.46}%
\end{equation}
for any smooth $(1,2)$-extensor field $B,$ such that $\left.  B\right\vert
_{\partial U}=0$.

We now show that Eqs.(\ref{4.45}) and (\ref{4.46}) give to coupled
differential equations for the fields $%
%TCIMACRO{\TeXButton{h}{\slh}}%
%BeginExpansion
\slh
%EndExpansion
^{\clubsuit}$ and $\Omega$.

Of course, the solution of the contour problems given by Eqs.(\ref{4.45}) and
(\ref{4.46}) is obtained following the traditional path, i.e., by using the
appropriated variational formulas, the Gauss-Stokes and a fundamental lemma of
integration theory (already mentioned in Section 5.1).

First we calculate the variation in Eq.(\ref{4.45}). We have:
\begin{align}%
%TCIMACRO{\TeXButton{delta}{\mbox{\boldmath{$\delta$}}}}%
%BeginExpansion
\mbox{\boldmath{$\delta$}}%
%EndExpansion
_{%
%TCIMACRO{\TeXButton{sh}{\sslh}}%
%BeginExpansion
\sslh
%EndExpansion
^{\clubsuit}}^{w}(\underline{%
%TCIMACRO{\TeXButton{h}{\slh}}%
%BeginExpansion
\slh
%EndExpansion
}^{\clubsuit}(\partial_{a}\wedge\partial_{b})\cdot\mathcal{R}_{2}(a\wedge
b)\det[%
%TCIMACRO{\TeXButton{h}{\slh}}%
%BeginExpansion
\slh
%EndExpansion
])  &  =(%
%TCIMACRO{\TeXButton{delta}{\mbox{\boldmath{$\delta$}}}}%
%BeginExpansion
\mbox{\boldmath{$\delta$}}%
%EndExpansion
_{%
%TCIMACRO{\TeXButton{sh}{\sslh}}%
%BeginExpansion
\sslh
%EndExpansion
^{\clubsuit}}^{w}\underline{%
%TCIMACRO{\TeXButton{h}{\slh}}%
%BeginExpansion
\slh
%EndExpansion
}^{\clubsuit}(\partial_{a}\wedge\partial_{b}))\cdot\mathcal{R}_{2}(a\wedge
b)\det[%
%TCIMACRO{\TeXButton{h}{\slh}}%
%BeginExpansion
\slh
%EndExpansion
]\nonumber\\
&  +\underline{%
%TCIMACRO{\TeXButton{h}{\slh}}%
%BeginExpansion
\slh
%EndExpansion
}^{\clubsuit}(\partial_{a}\wedge\partial_{b})\cdot\mathcal{R}_{2}(a\wedge
b)(\delta_{%
%TCIMACRO{\TeXButton{h}{\slh}}%
%BeginExpansion
\slh
%EndExpansion
^{\clubsuit}}^{w}\det[%
%TCIMACRO{\TeXButton{h}{\slh}}%
%BeginExpansion
\slh
%EndExpansion
]). \label{4.46i}%
\end{align}

For the scalar product in the first member of Eq.(\ref{4.46i}), an algebraic
manipulation analogous to the one already utilized for deriving
Eq.(\ref{4.43ii}) gives
\begin{equation}
(%
%TCIMACRO{\TeXButton{delta}{\mbox{\boldmath{$\delta$}}}}%
%BeginExpansion
\mbox{\boldmath{$\delta$}}%
%EndExpansion
_{%
%TCIMACRO{\TeXButton{sh}{\sslh}}%
%BeginExpansion
\sslh
%EndExpansion
^{\clubsuit}}^{w}\underline{%
%TCIMACRO{\TeXButton{h}{\slh}}%
%BeginExpansion
\slh
%EndExpansion
}^{\clubsuit}(\partial_{a}\wedge\partial_{b}))\cdot\mathcal{R}_{2}(a\wedge
b)=2w(\partial_{a})\cdot\mathcal{R}_{1}(a), \label{4.46ii}%
\end{equation}
where $\mathcal{R}(a)$ is the Ricci \ gauge field (in the Lorentz-Cartan
\textit{MCGSS}).

The calculation of $%
%TCIMACRO{\TeXButton{delta}{\mbox{\boldmath{$\delta$}}}}%
%BeginExpansion
\mbox{\boldmath{$\delta$}}%
%EndExpansion
_{%
%TCIMACRO{\TeXButton{sh}{\sslh}}%
%BeginExpansion
\sslh
%EndExpansion
^{\clubsuit}}^{w}\det[%
%TCIMACRO{\TeXButton{h}{\slh}}%
%BeginExpansion
\slh
%EndExpansion
]$ has already been obtained (recall Eq.(\ref{4.43vi})). So, substituting
Eqs.(\ref{4.46ii}) and (\ref{4.43vi}) in Eq.(\ref{4.46i}) and recalling the
definition of $\mathcal{G\equiv}\overset{\gamma%
%TCIMACRO{\TeXButton{sh}{\sslh}}%
%BeginExpansion
\sslh
%EndExpansion
}{\mathcal{G}}$ (Eq.(\ref{4.19})) we get that
\begin{align}%
%TCIMACRO{\TeXButton{delta}{\mbox{\boldmath{$\delta$}}}}%
%BeginExpansion
\mbox{\boldmath{$\delta$}}%
%EndExpansion
_{%
%TCIMACRO{\TeXButton{sh}{\sslh}}%
%BeginExpansion
\sslh
%EndExpansion
^{\clubsuit}}^{w}(\mathcal{R}\det[%
%TCIMACRO{\TeXButton{h}{\slh}}%
%BeginExpansion
\slh
%EndExpansion
])  &  =2w(\partial_{a})\cdot\mathcal{R}_{1}(a)\det[%
%TCIMACRO{\TeXButton{h}{\slh}}%
%BeginExpansion
\slh
%EndExpansion
]-\mathcal{R}w(\partial_{a})\cdot%
%TCIMACRO{\TeXButton{h}{\slh}}%
%BeginExpansion
\slh
%EndExpansion
(a)\det[%
%TCIMACRO{\TeXButton{h}{\slh}}%
%BeginExpansion
\slh
%EndExpansion
]\nonumber\\
&  =2w(\partial_{a})\cdot(\mathcal{R}_{1}(a)-\frac{1}{2}%
%TCIMACRO{\TeXButton{h}{\slh}}%
%BeginExpansion
\slh
%EndExpansion
(a)\mathcal{R})\det[%
%TCIMACRO{\TeXButton{h}{\slh}}%
%BeginExpansion
\slh
%EndExpansion
]\nonumber\\%
%TCIMACRO{\TeXButton{delta}{\mbox{\boldmath{$\delta$}}}}%
%BeginExpansion
\mbox{\boldmath{$\delta$}}%
%EndExpansion
_{%
%TCIMACRO{\TeXButton{sh}{\sslh}}%
%BeginExpansion
\sslh
%EndExpansion
^{\clubsuit}}^{w}(\mathcal{R}\det[%
%TCIMACRO{\TeXButton{h}{\slh}}%
%BeginExpansion
\slh
%EndExpansion
])  &  =2w(\partial_{a})\cdot\mathcal{G}(a)\det[%
%TCIMACRO{\TeXButton{h}{\slh}}%
%BeginExpansion
\slh
%EndExpansion
]. \label{4.47}%
\end{align}

Substituting Eq.(\ref{4.47}) in Eq.(\ref{4.45}), the contour problem for the
dynamic variable $%
%TCIMACRO{\TeXButton{h}{\slh}}%
%BeginExpansion
\slh
%EndExpansion
^{\clubsuit}$ becomes
\begin{equation}
\int_{\mathcal{U}}w(\partial_{a})\cdot\mathcal{G}(a)\det[%
%TCIMACRO{\TeXButton{h}{\slh}}%
%BeginExpansion
\slh
%EndExpansion
]\text{ }\tau=0, \label{4.48}%
\end{equation}
for any $w$, and since $w$ is arbitrary, recalling a fundamental lemma of
integration theory, we get
\begin{equation}
\mathcal{G}(a)=0. \label{4.49}%
\end{equation}

This field equation looks like the one already obtained for $%
%TCIMACRO{\TeXButton{h}{\slh}}%
%BeginExpansion
\slh
%EndExpansion
$ in Section 6.1., but here it is not a differential equation for the field $%
%TCIMACRO{\TeXButton{h}{\slh}}%
%BeginExpansion
\slh
%EndExpansion
$. To continue recall that Eq.(\ref{4.49}) is equivalent to $\mathcal{R}%
(a)=0.$

Indeed, taking the scalar $%
%TCIMACRO{\TeXButton{h}{\slh}}%
%BeginExpansion
\slh
%EndExpansion
^{\clubsuit}$-divergent of $\mathcal{R}_{1}(a)$ we have
\begin{align*}
\mathcal{G}(a)=0  &  \Rightarrow\mathcal{R}_{1}(a)-\frac{1}{2}%
%TCIMACRO{\TeXButton{h}{\slh}}%
%BeginExpansion
\slh
%EndExpansion
(a)\mathcal{R}=0\\
&  \Rightarrow%
%TCIMACRO{\TeXButton{h}{\slh}}%
%BeginExpansion
\slh
%EndExpansion
^{\clubsuit}(\partial_{a})\cdot\mathcal{R}_{1}(a)-\frac{1}{2}%
%TCIMACRO{\TeXButton{h}{\slh}}%
%BeginExpansion
\slh
%EndExpansion
^{\clubsuit}(\partial_{a})\cdot%
%TCIMACRO{\TeXButton{h}{\slh}}%
%BeginExpansion
\slh
%EndExpansion
(a)\mathcal{R}=0\\
&  \Rightarrow\mathcal{R}-\frac{1}{2}4\mathcal{R}=0\\
&  \Rightarrow\mathcal{R}=0,
\end{align*}
which implies
\begin{equation}
\mathcal{R}_{1}(a)=0. \label{4.49a}%
\end{equation}

Also it quite clear that Eq.(\ref{4.49a}) implies Eq.(\ref{4.49}).

Let us express now Eq.(\ref{4.49a}) in terms of the \textit{canonical }%
$1$-forms of\emph{ }$%
%TCIMACRO{\TeXButton{h}{\slh}}%
%BeginExpansion
\slh
%EndExpansion
$ (i.e., $%
%TCIMACRO{\TeXButton{h}{\slh}}%
%BeginExpansion
\slh
%EndExpansion
(%
%TCIMACRO{\TeXButton{vt}{\mbox{\boldmath{$\vartheta$}}}}%
%BeginExpansion
\mbox{\boldmath{$\vartheta$}}%
%EndExpansion
^{0}),\ldots,%
%TCIMACRO{\TeXButton{h}{\slh}}%
%BeginExpansion
\slh
%EndExpansion
(%
%TCIMACRO{\TeXButton{vt}{\mbox{\boldmath{$\vartheta$}}}}%
%BeginExpansion
\mbox{\boldmath{$\vartheta$}}%
%EndExpansion
^{3})$) and the canonical \emph{biforms }of $\Omega$ (i.e., $\Omega
(\vartheta_{0}),\ldots,\Omega(\vartheta_{3})$),
\begin{align*}
\mathcal{R}_{1}(a)=0  &  \Leftrightarrow\mathcal{R}_{1}(\vartheta_{\nu})=0\\
&  \Leftrightarrow%
%TCIMACRO{\TeXButton{h}{\slh}}%
%BeginExpansion
\slh
%EndExpansion
^{\clubsuit}(%
%TCIMACRO{\TeXButton{vt}{\mbox{\boldmath{$\vartheta$}}}}%
%BeginExpansion
\mbox{\boldmath{$\vartheta$}}%
%EndExpansion
^{\mu})\lrcorner\mathcal{R}_{2}(\vartheta_{\mu}\wedge\vartheta_{\nu})=0,
\end{align*}
i.e.,
\begin{equation}%
%TCIMACRO{\TeXButton{h}{\slh}}%
%BeginExpansion
\slh
%EndExpansion
^{\clubsuit}(%
%TCIMACRO{\TeXButton{vt}{\mbox{\boldmath{$\vartheta$}}}}%
%BeginExpansion
\mbox{\boldmath{$\vartheta$}}%
%EndExpansion
^{\mu})\lrcorner(\vartheta_{\mu}\cdot\partial\Omega(\vartheta_{\nu}%
)-\vartheta_{\nu}\cdot\partial\Omega(\vartheta_{\mu})+\Omega(\vartheta_{\mu
})\underset{\eta}{\times}\Omega(\vartheta_{\nu}))=0. \label{4.49b}%
\end{equation}

Eq.(\ref{4.49b}) resumes a set of four coupled multiform equations which are
algebraic on the canonical $\emph{1}$-form of $%
%TCIMACRO{\TeXButton{h}{\slh}}%
%BeginExpansion
\slh
%EndExpansion
$ and differential on the canonical biforms of $\Omega.$

Let us now calculate the variation in Eq.(\ref{4.46}). We have%
\begin{equation}%
%TCIMACRO{\TeXButton{delta}{\mbox{\boldmath{$\delta$}}}}%
%BeginExpansion
\mbox{\boldmath{$\delta$}}%
%EndExpansion
_{\Omega}^{B}(\underline{%
%TCIMACRO{\TeXButton{h}{\slh}}%
%BeginExpansion
\slh
%EndExpansion
}^{\clubsuit}(\partial_{a}\wedge\partial_{b})\cdot\mathcal{R}_{2}(a\wedge
b)\det[%
%TCIMACRO{\TeXButton{h}{\slh}}%
%BeginExpansion
\slh
%EndExpansion
])=\underline{%
%TCIMACRO{\TeXButton{h}{\slh}}%
%BeginExpansion
\slh
%EndExpansion
}^{\clubsuit}(\partial_{a}\wedge\partial_{b})\cdot(%
%TCIMACRO{\TeXButton{delta}{\mbox{\boldmath{$\delta$}}}}%
%BeginExpansion
\mbox{\boldmath{$\delta$}}%
%EndExpansion
_{\Omega}^{B}\mathcal{R}_{2}(a\wedge b))\det[%
%TCIMACRO{\TeXButton{h}{\slh}}%
%BeginExpansion
\slh
%EndExpansion
]. \label{4.46iv}%
\end{equation}

For the scalar product on the right side of Eq.(\ref{4.46iv}), utilizing the
commutation property involving $%
%TCIMACRO{\TeXButton{delta}{\mbox{\boldmath{$\delta$}}}}%
%BeginExpansion
\mbox{\boldmath{$\delta$}}%
%EndExpansion
_{\Omega}^{B}$ and $a\cdot\partial$, the property $%
%TCIMACRO{\TeXButton{delta}{\mbox{\boldmath{$\delta$}}}}%
%BeginExpansion
\mbox{\boldmath{$\delta$}}%
%EndExpansion
_{\Omega}^{B}\Omega(a)=B(a)$ and the definition of the covariant derivative
operator $\mathcal{D}_{a}\equiv$ $\overset{\mathcal{\varkappa\eta
}}{\mathcal{D}}_{a}$ , i.e., $\mathcal{D}_{a}X=a\cdot\partial X+\Omega
(a)\underset{\eta}{\times}X,$ we have
\begin{align}
\underline{%
%TCIMACRO{\TeXButton{h}{\slh}}%
%BeginExpansion
\slh
%EndExpansion
}^{\clubsuit}(\partial_{a}\wedge\partial_{b})\cdot(%
%TCIMACRO{\TeXButton{delta}{\mbox{\boldmath{$\delta$}}}}%
%BeginExpansion
\mbox{\boldmath{$\delta$}}%
%EndExpansion
_{\Omega}^{B}\mathcal{R}_{2}(a\wedge b))  &  =\underline{%
%TCIMACRO{\TeXButton{h}{\slh}}%
%BeginExpansion
\slh
%EndExpansion
}^{\clubsuit}(%
%TCIMACRO{\TeXButton{vt}{\mbox{\boldmath{$\vartheta$}}}}%
%BeginExpansion
\mbox{\boldmath{$\vartheta$}}%
%EndExpansion
^{\mu}\wedge%
%TCIMACRO{\TeXButton{vt}{\mbox{\boldmath{$\vartheta$}}}}%
%BeginExpansion
\mbox{\boldmath{$\vartheta$}}%
%EndExpansion
^{\nu})\cdot(%
%TCIMACRO{\TeXButton{delta}{\mbox{\boldmath{$\delta$}}}}%
%BeginExpansion
\mbox{\boldmath{$\delta$}}%
%EndExpansion
_{\Omega}^{B}\mathcal{R}_{2}(\vartheta_{\mu}\wedge\vartheta_{\nu}))\nonumber\\
&  =\underline{%
%TCIMACRO{\TeXButton{h}{\slh}}%
%BeginExpansion
\slh
%EndExpansion
}^{\clubsuit}(%
%TCIMACRO{\TeXButton{vt}{\mbox{\boldmath{$\vartheta$}}}}%
%BeginExpansion
\mbox{\boldmath{$\vartheta$}}%
%EndExpansion
^{\mu}\wedge%
%TCIMACRO{\TeXButton{vt}{\mbox{\boldmath{$\vartheta$}}}}%
%BeginExpansion
\mbox{\boldmath{$\vartheta$}}%
%EndExpansion
^{\nu})\cdot(\vartheta_{\mu}\cdot\partial B(\vartheta_{\nu})-\vartheta_{\nu
}\cdot\partial B(\vartheta_{\mu})+\nonumber\\
&  B(\vartheta_{\mu})\underset{\eta^{-1}}{\times}\Omega(\vartheta_{\nu
})+\Omega(\vartheta_{\mu})\underset{\eta^{-1}}{\times}B(\vartheta_{\nu
}))\nonumber\\
&  =2\underline{%
%TCIMACRO{\TeXButton{h}{\slh}}%
%BeginExpansion
\slh
%EndExpansion
}^{\clubsuit}(%
%TCIMACRO{\TeXButton{vt}{\mbox{\boldmath{$\vartheta$}}}}%
%BeginExpansion
\mbox{\boldmath{$\vartheta$}}%
%EndExpansion
^{\mu}\wedge%
%TCIMACRO{\TeXButton{vt}{\mbox{\boldmath{$\vartheta$}}}}%
%BeginExpansion
\mbox{\boldmath{$\vartheta$}}%
%EndExpansion
^{\nu})\cdot\mathcal{D}_{\vartheta_{\mu}}B(\vartheta_{\nu}). \label{4.46v}%
\end{align}

After some algebraic manipulations the right side of Eq.(\ref{4.46v}) may be
written as
\begin{align*}
\underline{%
%TCIMACRO{\TeXButton{h}{\slh}}%
%BeginExpansion
\slh
%EndExpansion
}^{\clubsuit}(%
%TCIMACRO{\TeXButton{vt}{\mbox{\boldmath{$\vartheta$}}}}%
%BeginExpansion
\mbox{\boldmath{$\vartheta$}}%
%EndExpansion
^{\mu}\wedge%
%TCIMACRO{\TeXButton{vt}{\mbox{\boldmath{$\vartheta$}}}}%
%BeginExpansion
\mbox{\boldmath{$\vartheta$}}%
%EndExpansion
^{\nu})\cdot\mathcal{D}_{\vartheta_{\mu}}B(\vartheta_{\nu})  &  =\underline{%
%TCIMACRO{\TeXButton{h}{\slh}}%
%BeginExpansion
\slh
%EndExpansion
}^{\clubsuit}(%
%TCIMACRO{\TeXButton{vt}{\mbox{\boldmath{$\vartheta$}}}}%
%BeginExpansion
\mbox{\boldmath{$\vartheta$}}%
%EndExpansion
^{\mu}\wedge%
%TCIMACRO{\TeXButton{vt}{\mbox{\boldmath{$\vartheta$}}}}%
%BeginExpansion
\mbox{\boldmath{$\vartheta$}}%
%EndExpansion
^{\nu})\cdot\vartheta_{\mu}\cdot\partial B(\vartheta_{\nu})+\vartheta_{\mu
}\cdot\partial\underline{%
%TCIMACRO{\TeXButton{h}{\slh}}%
%BeginExpansion
\slh
%EndExpansion
}^{\clubsuit}(%
%TCIMACRO{\TeXButton{vt}{\mbox{\boldmath{$\vartheta$}}}}%
%BeginExpansion
\mbox{\boldmath{$\vartheta$}}%
%EndExpansion
^{\mu}\wedge%
%TCIMACRO{\TeXButton{vt}{\mbox{\boldmath{$\vartheta$}}}}%
%BeginExpansion
\mbox{\boldmath{$\vartheta$}}%
%EndExpansion
^{\nu})\cdot B(\vartheta_{\nu})\\
&  +\underline{%
%TCIMACRO{\TeXButton{h}{\slh}}%
%BeginExpansion
\slh
%EndExpansion
}^{\clubsuit}(%
%TCIMACRO{\TeXButton{vt}{\mbox{\boldmath{$\vartheta$}}}}%
%BeginExpansion
\mbox{\boldmath{$\vartheta$}}%
%EndExpansion
^{\mu}\wedge%
%TCIMACRO{\TeXButton{vt}{\mbox{\boldmath{$\vartheta$}}}}%
%BeginExpansion
\mbox{\boldmath{$\vartheta$}}%
%EndExpansion
^{\nu})\cdot(\Omega(\vartheta_{\mu})\underset{\eta^{-1}}{\times}%
B(\vartheta_{\nu}))-\vartheta_{\mu}\cdot\partial\underline{%
%TCIMACRO{\TeXButton{h}{\slh}}%
%BeginExpansion
\slh
%EndExpansion
}^{\clubsuit}(%
%TCIMACRO{\TeXButton{vt}{\mbox{\boldmath{$\vartheta$}}}}%
%BeginExpansion
\mbox{\boldmath{$\vartheta$}}%
%EndExpansion
^{\mu}\wedge%
%TCIMACRO{\TeXButton{vt}{\mbox{\boldmath{$\vartheta$}}}}%
%BeginExpansion
\mbox{\boldmath{$\vartheta$}}%
%EndExpansion
^{\nu})\cdot B(\vartheta_{\nu})\\
&  =\vartheta_{\mu}\cdot\partial(\underline{%
%TCIMACRO{\TeXButton{h}{\slh}}%
%BeginExpansion
\slh
%EndExpansion
}^{\clubsuit}(%
%TCIMACRO{\TeXButton{vt}{\mbox{\boldmath{$\vartheta$}}}}%
%BeginExpansion
\mbox{\boldmath{$\vartheta$}}%
%EndExpansion
^{\mu}\wedge%
%TCIMACRO{\TeXButton{vt}{\mbox{\boldmath{$\vartheta$}}}}%
%BeginExpansion
\mbox{\boldmath{$\vartheta$}}%
%EndExpansion
^{\nu})\cdot B(\vartheta_{\nu}))-(\vartheta_{\mu}\cdot\partial\underline{%
%TCIMACRO{\TeXButton{h}{\slh}}%
%BeginExpansion
\slh
%EndExpansion
}^{\clubsuit}(%
%TCIMACRO{\TeXButton{vt}{\mbox{\boldmath{$\vartheta$}}}}%
%BeginExpansion
\mbox{\boldmath{$\vartheta$}}%
%EndExpansion
^{\mu}\wedge%
%TCIMACRO{\TeXButton{vt}{\mbox{\boldmath{$\vartheta$}}}}%
%BeginExpansion
\mbox{\boldmath{$\vartheta$}}%
%EndExpansion
^{\nu})\\
&  +\underline{\eta}(\Omega(\vartheta_{\mu})\underset{\eta}{\times
}\underline{\eta}\underline{%
%TCIMACRO{\TeXButton{h}{\slh}}%
%BeginExpansion
\slh
%EndExpansion
}^{\clubsuit}(%
%TCIMACRO{\TeXButton{vt}{\mbox{\boldmath{$\vartheta$}}}}%
%BeginExpansion
\mbox{\boldmath{$\vartheta$}}%
%EndExpansion
^{\mu}\wedge%
%TCIMACRO{\TeXButton{vt}{\mbox{\boldmath{$\vartheta$}}}}%
%BeginExpansion
\mbox{\boldmath{$\vartheta$}}%
%EndExpansion
^{\nu})))\cdot B(\vartheta_{\nu})\\
&  =\vartheta_{\mu}\cdot\partial(\underline{%
%TCIMACRO{\TeXButton{h}{\slh}}%
%BeginExpansion
\slh
%EndExpansion
}^{\clubsuit}(%
%TCIMACRO{\TeXButton{vt}{\mbox{\boldmath{$\vartheta$}}}}%
%BeginExpansion
\mbox{\boldmath{$\vartheta$}}%
%EndExpansion
^{\mu}\wedge%
%TCIMACRO{\TeXButton{vt}{\mbox{\boldmath{$\vartheta$}}}}%
%BeginExpansion
\mbox{\boldmath{$\vartheta$}}%
%EndExpansion
^{\nu})\cdot B(\vartheta_{\nu}))-\underline{\eta}(\vartheta_{\mu}\cdot
\partial\eta\underline{%
%TCIMACRO{\TeXButton{h}{\slh}}%
%BeginExpansion
\slh
%EndExpansion
}^{\clubsuit}(\vartheta^{\mu}\wedge\vartheta^{\nu})\\
&  +\Omega(\vartheta_{\mu})\underset{\eta}{\times}\underline{\eta}\underline{%
%TCIMACRO{\TeXButton{h}{\slh}}%
%BeginExpansion
\slh
%EndExpansion
}^{\clubsuit}(%
%TCIMACRO{\TeXButton{vt}{\mbox{\boldmath{$\vartheta$}}}}%
%BeginExpansion
\mbox{\boldmath{$\vartheta$}}%
%EndExpansion
^{\mu}\wedge%
%TCIMACRO{\TeXButton{vt}{\mbox{\boldmath{$\vartheta$}}}}%
%BeginExpansion
\mbox{\boldmath{$\vartheta$}}%
%EndExpansion
^{\nu}))\cdot B(\vartheta_{\nu}),
\end{align*}
i.e.,
\begin{align}
\underline{%
%TCIMACRO{\TeXButton{h}{\slh}}%
%BeginExpansion
\slh
%EndExpansion
}^{\clubsuit}(%
%TCIMACRO{\TeXButton{vt}{\mbox{\boldmath{$\vartheta$}}}}%
%BeginExpansion
\mbox{\boldmath{$\vartheta$}}%
%EndExpansion
^{\mu}\wedge%
%TCIMACRO{\TeXButton{vt}{\mbox{\boldmath{$\vartheta$}}}}%
%BeginExpansion
\mbox{\boldmath{$\vartheta$}}%
%EndExpansion
^{\nu})\cdot\mathcal{D}_{\vartheta_{\mu}}B(\vartheta_{\nu})  &  =\vartheta
_{\mu}\cdot\partial(\underline{%
%TCIMACRO{\TeXButton{h}{\slh}}%
%BeginExpansion
\slh
%EndExpansion
}^{\clubsuit}(%
%TCIMACRO{\TeXButton{vt}{\mbox{\boldmath{$\vartheta$}}}}%
%BeginExpansion
\mbox{\boldmath{$\vartheta$}}%
%EndExpansion
^{\mu}\wedge%
%TCIMACRO{\TeXButton{vt}{\mbox{\boldmath{$\vartheta$}}}}%
%BeginExpansion
\mbox{\boldmath{$\vartheta$}}%
%EndExpansion
^{\nu})\cdot B(\vartheta_{\nu}))\nonumber\\
&  -\underline{\eta}\mathcal{D}_{\vartheta_{\mu}}\underline{\eta}\underline{%
%TCIMACRO{\TeXButton{h}{\slh}}%
%BeginExpansion
\slh
%EndExpansion
}^{\clubsuit}(%
%TCIMACRO{\TeXButton{vt}{\mbox{\boldmath{$\vartheta$}}}}%
%BeginExpansion
\mbox{\boldmath{$\vartheta$}}%
%EndExpansion
^{\mu}\wedge%
%TCIMACRO{\TeXButton{vt}{\mbox{\boldmath{$\vartheta$}}}}%
%BeginExpansion
\mbox{\boldmath{$\vartheta$}}%
%EndExpansion
^{\nu})\cdot B(\vartheta_{\nu}). \label{4.46vi}%
\end{align}

Also, the first member on the right side of Eq.(\ref{4.46vi}) may be written
as a scalar divergent of a 1-form field, i.e.,
\begin{equation}
\vartheta_{\mu}\cdot\partial(\underline{%
%TCIMACRO{\TeXButton{h}{\slh}}%
%BeginExpansion
\slh
%EndExpansion
}^{\clubsuit}(%
%TCIMACRO{\TeXButton{vt}{\mbox{\boldmath{$\vartheta$}}}}%
%BeginExpansion
\mbox{\boldmath{$\vartheta$}}%
%EndExpansion
^{\mu}\wedge%
%TCIMACRO{\TeXButton{vt}{\mbox{\boldmath{$\vartheta$}}}}%
%BeginExpansion
\mbox{\boldmath{$\vartheta$}}%
%EndExpansion
^{\nu})\cdot B(\vartheta_{\nu}))=\partial\cdot(\vartheta_{\mu}\underline{%
%TCIMACRO{\TeXButton{h}{\slh}}%
%BeginExpansion
\slh
%EndExpansion
}^{\clubsuit}(%
%TCIMACRO{\TeXButton{vt}{\mbox{\boldmath{$\vartheta$}}}}%
%BeginExpansion
\mbox{\boldmath{$\vartheta$}}%
%EndExpansion
^{\mu}\wedge%
%TCIMACRO{\TeXButton{vt}{\mbox{\boldmath{$\vartheta$}}}}%
%BeginExpansion
\mbox{\boldmath{$\vartheta$}}%
%EndExpansion
^{\nu})\cdot B(\vartheta_{\nu})), \label{4.46vii}%
\end{equation}
where we used the fact that $\partial\cdot(\vartheta_{\mu}\phi)=\vartheta
_{\mu}\cdot\partial\phi,$ where $\phi$ is a scalar field.

After putting Eq.(\ref{4.46vii}) in Eq.(\ref{4.46vi}) and using the result
obtained in Eq.(\ref{4.46v}), the Eq.(\ref{4.46iv}) gives finally the
variational formula
\begin{align}%
%TCIMACRO{\TeXButton{delta}{\mbox{\boldmath{$\delta$}}}}%
%BeginExpansion
\mbox{\boldmath{$\delta$}}%
%EndExpansion
_{\Omega}^{B}(\underline{%
%TCIMACRO{\TeXButton{h}{\slh}}%
%BeginExpansion
\slh
%EndExpansion
}^{\clubsuit}(\partial_{a}\wedge\partial_{b})\cdot\mathcal{R}_{2}(a\wedge
b)\det[%
%TCIMACRO{\TeXButton{h}{\slh}}%
%BeginExpansion
\slh
%EndExpansion
])  &  =-2(\underline{\eta}\mathcal{D}_{\vartheta_{\mu}}\underline{\eta
}\underline{%
%TCIMACRO{\TeXButton{h}{\slh}}%
%BeginExpansion
\slh
%EndExpansion
}^{\clubsuit}(%
%TCIMACRO{\TeXButton{vt}{\mbox{\boldmath{$\vartheta$}}}}%
%BeginExpansion
\mbox{\boldmath{$\vartheta$}}%
%EndExpansion
^{\mu}\wedge%
%TCIMACRO{\TeXButton{vt}{\mbox{\boldmath{$\vartheta$}}}}%
%BeginExpansion
\mbox{\boldmath{$\vartheta$}}%
%EndExpansion
^{\nu})\cdot B(\vartheta_{\nu})\nonumber\\
&  -\partial\cdot(\vartheta_{\mu}\underline{%
%TCIMACRO{\TeXButton{h}{\slh}}%
%BeginExpansion
\slh
%EndExpansion
}^{\clubsuit}(%
%TCIMACRO{\TeXButton{vt}{\mbox{\boldmath{$\vartheta$}}}}%
%BeginExpansion
\mbox{\boldmath{$\vartheta$}}%
%EndExpansion
^{\mu}\wedge%
%TCIMACRO{\TeXButton{vt}{\mbox{\boldmath{$\vartheta$}}}}%
%BeginExpansion
\mbox{\boldmath{$\vartheta$}}%
%EndExpansion
^{\nu})\cdot B(\vartheta_{\nu})))\det[%
%TCIMACRO{\TeXButton{h}{\slh}}%
%BeginExpansion
\slh
%EndExpansion
]. \label{4.50}%
\end{align}

Putting Eq.(\ref{4.50}) in Eq.(\ref{4.46}), the contour problem for the
dynamic variable $\Omega$ becomes
\begin{align}
&  \int_{U}\underline{\eta}\mathcal{D}_{\vartheta_{\mu}}\underline{\eta
}\underline{%
%TCIMACRO{\TeXButton{h}{\slh}}%
%BeginExpansion
\slh
%EndExpansion
}^{\clubsuit}(%
%TCIMACRO{\TeXButton{vt}{\mbox{\boldmath{$\vartheta$}}}}%
%BeginExpansion
\mbox{\boldmath{$\vartheta$}}%
%EndExpansion
^{\mu}\wedge%
%TCIMACRO{\TeXButton{vt}{\mbox{\boldmath{$\vartheta$}}}}%
%BeginExpansion
\mbox{\boldmath{$\vartheta$}}%
%EndExpansion
^{\nu})\cdot B(\vartheta_{\nu})\det[%
%TCIMACRO{\TeXButton{h}{\slh}}%
%BeginExpansion
\slh
%EndExpansion
]\text{ }\tau\nonumber\\
&  -\int_{U}\partial\cdot(\vartheta_{\mu}\underline{%
%TCIMACRO{\TeXButton{h}{\slh}}%
%BeginExpansion
\slh
%EndExpansion
}^{\clubsuit}(%
%TCIMACRO{\TeXButton{vt}{\mbox{\boldmath{$\vartheta$}}}}%
%BeginExpansion
\mbox{\boldmath{$\vartheta$}}%
%EndExpansion
^{\mu}\wedge%
%TCIMACRO{\TeXButton{vt}{\mbox{\boldmath{$\vartheta$}}}}%
%BeginExpansion
\mbox{\boldmath{$\vartheta$}}%
%EndExpansion
^{\nu})\cdot B(\vartheta_{\nu}))\det[%
%TCIMACRO{\TeXButton{h}{\slh}}%
%BeginExpansion
\slh
%EndExpansion
]\text{ }\tau=0, \label{4.51}%
\end{align}
for all $B$ \ such $\left.  B\right\vert _{\partial U}=0$. Now, the second
member of Eq.(\ref{4.51}) may be easily transformed in such a way that a
\ scalar divergent of a 1-form field appears. Indeed, if we recall that and
$\partial\cdot(a\phi)=a\cdot\partial\phi+(\partial\cdot a)\phi$, where $a$ is
a 1-form field and $\phi$ is a scalar field we have
\begin{align}
\int_{U}\partial\cdot(\vartheta_{\mu}\underline{%
%TCIMACRO{\TeXButton{h}{\slh}}%
%BeginExpansion
\slh
%EndExpansion
}^{\clubsuit}(%
%TCIMACRO{\TeXButton{vt}{\mbox{\boldmath{$\vartheta$}}}}%
%BeginExpansion
\mbox{\boldmath{$\vartheta$}}%
%EndExpansion
^{\mu}\wedge%
%TCIMACRO{\TeXButton{vt}{\mbox{\boldmath{$\vartheta$}}}}%
%BeginExpansion
\mbox{\boldmath{$\vartheta$}}%
%EndExpansion
^{\nu})\cdot B(\vartheta_{\nu}))\det[%
%TCIMACRO{\TeXButton{h}{\slh}}%
%BeginExpansion
\slh
%EndExpansion
]\text{ }\tau &  =\int_{U}\partial\cdot(\vartheta_{\mu}\underline{%
%TCIMACRO{\TeXButton{h}{\slh}}%
%BeginExpansion
\slh
%EndExpansion
}^{\clubsuit}(%
%TCIMACRO{\TeXButton{vt}{\mbox{\boldmath{$\vartheta$}}}}%
%BeginExpansion
\mbox{\boldmath{$\vartheta$}}%
%EndExpansion
^{\mu}\wedge%
%TCIMACRO{\TeXButton{vt}{\mbox{\boldmath{$\vartheta$}}}}%
%BeginExpansion
\mbox{\boldmath{$\vartheta$}}%
%EndExpansion
^{\nu})\cdot B(\vartheta_{\nu}))\det[%
%TCIMACRO{\TeXButton{h}{\slh}}%
%BeginExpansion
\slh
%EndExpansion
])\text{ }\tau\nonumber\\
&  -\int_{U}\underline{%
%TCIMACRO{\TeXButton{h}{\slh}}%
%BeginExpansion
\slh
%EndExpansion
}^{\clubsuit}(%
%TCIMACRO{\TeXButton{vt}{\mbox{\boldmath{$\vartheta$}}}}%
%BeginExpansion
\mbox{\boldmath{$\vartheta$}}%
%EndExpansion
^{\mu}\wedge%
%TCIMACRO{\TeXButton{vt}{\mbox{\boldmath{$\vartheta$}}}}%
%BeginExpansion
\mbox{\boldmath{$\vartheta$}}%
%EndExpansion
^{\nu})\cdot B(\vartheta_{\nu})\vartheta_{\mu}\cdot\partial\det[%
%TCIMACRO{\TeXButton{h}{\slh}}%
%BeginExpansion
\slh
%EndExpansion
]\text{ }\tau. \label{4.46viii}%
\end{align}

Putting Eq.(\ref{4.46viii}) in Eq.(\ref{4.51}) we \ get
\begin{align}
\int_{U}(\underline{\eta}\mathcal{D}_{\vartheta_{\mu}}\underline{\eta
}\underline{%
%TCIMACRO{\TeXButton{h}{\slh}}%
%BeginExpansion
\slh
%EndExpansion
}^{\clubsuit}(%
%TCIMACRO{\TeXButton{vt}{\mbox{\boldmath{$\vartheta$}}}}%
%BeginExpansion
\mbox{\boldmath{$\vartheta$}}%
%EndExpansion
^{\mu}\wedge%
%TCIMACRO{\TeXButton{vt}{\mbox{\boldmath{$\vartheta$}}}}%
%BeginExpansion
\mbox{\boldmath{$\vartheta$}}%
%EndExpansion
^{\nu})\det[%
%TCIMACRO{\TeXButton{h}{\slh}}%
%BeginExpansion
\slh
%EndExpansion
]+\underline{%
%TCIMACRO{\TeXButton{h}{\slh}}%
%BeginExpansion
\slh
%EndExpansion
}^{\clubsuit}(%
%TCIMACRO{\TeXButton{vt}{\mbox{\boldmath{$\vartheta$}}}}%
%BeginExpansion
\mbox{\boldmath{$\vartheta$}}%
%EndExpansion
^{\mu}\wedge%
%TCIMACRO{\TeXButton{vt}{\mbox{\boldmath{$\vartheta$}}}}%
%BeginExpansion
\mbox{\boldmath{$\vartheta$}}%
%EndExpansion
^{\nu})\vartheta_{\mu}\cdot\partial\det[%
%TCIMACRO{\TeXButton{h}{\slh}}%
%BeginExpansion
\slh
%EndExpansion
])\cdot B(\vartheta_{\nu})\text{ }\tau & \nonumber\\
-\int_{U}\partial\cdot(\vartheta_{\mu}\underline{%
%TCIMACRO{\TeXButton{h}{\slh}}%
%BeginExpansion
\slh
%EndExpansion
}^{\clubsuit}(%
%TCIMACRO{\TeXButton{vt}{\mbox{\boldmath{$\vartheta$}}}}%
%BeginExpansion
\mbox{\boldmath{$\vartheta$}}%
%EndExpansion
^{\mu}\wedge%
%TCIMACRO{\TeXButton{vt}{\mbox{\boldmath{$\vartheta$}}}}%
%BeginExpansion
\mbox{\boldmath{$\vartheta$}}%
%EndExpansion
^{\nu})\cdot B(\vartheta_{\nu}))\det[%
%TCIMACRO{\TeXButton{h}{\slh}}%
%BeginExpansion
\slh
%EndExpansion
])\text{ }\tau &  =0, \label{4.53}%
\end{align}
for all $B$ $\left.  B\right\vert _{\partial U}=0$. Utilizing next the
Gauss-Stokes theorem with the boundary condition $\left.  B\right\vert
_{\partial U}=0$, the second term of Eq.(\ref{4.53}) may be integrated and
gives
\begin{align}
\int_{U}\partial\cdot(\vartheta_{\mu}\underline{%
%TCIMACRO{\TeXButton{h}{\slh}}%
%BeginExpansion
\slh
%EndExpansion
}^{\clubsuit}(%
%TCIMACRO{\TeXButton{vt}{\mbox{\boldmath{$\vartheta$}}}}%
%BeginExpansion
\mbox{\boldmath{$\vartheta$}}%
%EndExpansion
^{\mu}\wedge%
%TCIMACRO{\TeXButton{vt}{\mbox{\boldmath{$\vartheta$}}}}%
%BeginExpansion
\mbox{\boldmath{$\vartheta$}}%
%EndExpansion
^{\nu})\cdot B(%
%TCIMACRO{\TeXButton{vt}{\mbox{\boldmath{$\vartheta$}}}}%
%BeginExpansion
\mbox{\boldmath{$\vartheta$}}%
%EndExpansion
_{\nu}))\det[%
%TCIMACRO{\TeXButton{h}{\slh}}%
%BeginExpansion
\slh
%EndExpansion
])\text{ }\tau &  =\oint_{\partial U}%
%TCIMACRO{\TeXButton{vt}{\mbox{\boldmath{$\vartheta$}}}}%
%BeginExpansion
\mbox{\boldmath{$\vartheta$}}%
%EndExpansion
^{\rho}\cdot\vartheta_{\mu}\underline{%
%TCIMACRO{\TeXButton{h}{\slh}}%
%BeginExpansion
\slh
%EndExpansion
}^{\clubsuit}(%
%TCIMACRO{\TeXButton{vt}{\mbox{\boldmath{$\vartheta$}}}}%
%BeginExpansion
\mbox{\boldmath{$\vartheta$}}%
%EndExpansion
^{\mu}\wedge%
%TCIMACRO{\TeXButton{vt}{\mbox{\boldmath{$\vartheta$}}}}%
%BeginExpansion
\mbox{\boldmath{$\vartheta$}}%
%EndExpansion
^{\nu})\cdot B(\vartheta_{\nu})\det[%
%TCIMACRO{\TeXButton{h}{\slh}}%
%BeginExpansion
\slh
%EndExpansion
]\text{ }\tau_{\rho}\nonumber\\
&  =\oint_{\partial U}\underline{%
%TCIMACRO{\TeXButton{h}{\slh}}%
%BeginExpansion
\slh
%EndExpansion
}^{\clubsuit}(%
%TCIMACRO{\TeXButton{vt}{\mbox{\boldmath{$\vartheta$}}}}%
%BeginExpansion
\mbox{\boldmath{$\vartheta$}}%
%EndExpansion
^{\mu}\wedge%
%TCIMACRO{\TeXButton{vt}{\mbox{\boldmath{$\vartheta$}}}}%
%BeginExpansion
\mbox{\boldmath{$\vartheta$}}%
%EndExpansion
^{\nu})\cdot B(\vartheta_{\nu})\det[%
%TCIMACRO{\TeXButton{h}{\slh}}%
%BeginExpansion
\slh
%EndExpansion
]\text{ }\tau_{\mu}=0. \label{4.53a}%
\end{align}

So, putting Eq.(\ref{4.53a}) in Eq.(\ref{4.53}) gives
\begin{equation}
\int_{U}(\underline{\eta}\mathcal{D}_{\vartheta_{\mu}}\underline{\eta
}\underline{%
%TCIMACRO{\TeXButton{h}{\slh}}%
%BeginExpansion
\slh
%EndExpansion
}^{\clubsuit}(%
%TCIMACRO{\TeXButton{vt}{\mbox{\boldmath{$\vartheta$}}}}%
%BeginExpansion
\mbox{\boldmath{$\vartheta$}}%
%EndExpansion
^{\mu}\wedge%
%TCIMACRO{\TeXButton{vt}{\mbox{\boldmath{$\vartheta$}}}}%
%BeginExpansion
\mbox{\boldmath{$\vartheta$}}%
%EndExpansion
^{\nu})\det[%
%TCIMACRO{\TeXButton{h}{\slh}}%
%BeginExpansion
\slh
%EndExpansion
]+\underline{%
%TCIMACRO{\TeXButton{h}{\slh}}%
%BeginExpansion
\slh
%EndExpansion
}^{\clubsuit}(%
%TCIMACRO{\TeXButton{vt}{\mbox{\boldmath{$\vartheta$}}}}%
%BeginExpansion
\mbox{\boldmath{$\vartheta$}}%
%EndExpansion
^{\mu}\wedge%
%TCIMACRO{\TeXButton{vt}{\mbox{\boldmath{$\vartheta$}}}}%
%BeginExpansion
\mbox{\boldmath{$\vartheta$}}%
%EndExpansion
^{\nu})\vartheta_{\mu}\cdot\partial\det[%
%TCIMACRO{\TeXButton{h}{\slh}}%
%BeginExpansion
\slh
%EndExpansion
])\cdot B(\vartheta_{\nu})\text{ }\tau=0, \label{4.54}%
\end{equation}
for all $B$ such $\left.  B\right\vert _{\partial U}=0$, and since $B$ is
arbitrary a fundamental lemma of integration theory finally yields
\[
\underline{\eta}\mathcal{D}_{\vartheta_{\mu}}\underline{\eta}\underline{%
%TCIMACRO{\TeXButton{h}{\slh}}%
%BeginExpansion
\slh
%EndExpansion
}^{\clubsuit}(%
%TCIMACRO{\TeXButton{vt}{\mbox{\boldmath{$\vartheta$}}}}%
%BeginExpansion
\mbox{\boldmath{$\vartheta$}}%
%EndExpansion
^{\mu}\wedge%
%TCIMACRO{\TeXButton{vt}{\mbox{\boldmath{$\vartheta$}}}}%
%BeginExpansion
\mbox{\boldmath{$\vartheta$}}%
%EndExpansion
^{\nu})\det[%
%TCIMACRO{\TeXButton{h}{\slh}}%
%BeginExpansion
\slh
%EndExpansion
]+\underline{%
%TCIMACRO{\TeXButton{h}{\slh}}%
%BeginExpansion
\slh
%EndExpansion
}^{\clubsuit}(%
%TCIMACRO{\TeXButton{vt}{\mbox{\boldmath{$\vartheta$}}}}%
%BeginExpansion
\mbox{\boldmath{$\vartheta$}}%
%EndExpansion
^{\mu}\wedge%
%TCIMACRO{\TeXButton{vt}{\mbox{\boldmath{$\vartheta$}}}}%
%BeginExpansion
\mbox{\boldmath{$\vartheta$}}%
%EndExpansion
^{\nu})\vartheta_{\mu}\cdot\partial\det[%
%TCIMACRO{\TeXButton{h}{\slh}}%
%BeginExpansion
\slh
%EndExpansion
]=0,
\]
i.e.,
\begin{equation}
\mathcal{D}_{\vartheta_{\mu}}\underline{\eta}\underline{%
%TCIMACRO{\TeXButton{h}{\slh}}%
%BeginExpansion
\slh
%EndExpansion
}^{\clubsuit}(%
%TCIMACRO{\TeXButton{vt}{\mbox{\boldmath{$\vartheta$}}}}%
%BeginExpansion
\mbox{\boldmath{$\vartheta$}}%
%EndExpansion
^{\mu}\wedge%
%TCIMACRO{\TeXButton{vt}{\mbox{\boldmath{$\vartheta$}}}}%
%BeginExpansion
\mbox{\boldmath{$\vartheta$}}%
%EndExpansion
^{\nu})=-(\vartheta_{\mu}\cdot\partial\ln\left\vert \det[%
%TCIMACRO{\TeXButton{h}{\slh}}%
%BeginExpansion
\slh
%EndExpansion
]\right\vert )\underline{\eta}\underline{%
%TCIMACRO{\TeXButton{h}{\slh}}%
%BeginExpansion
\slh
%EndExpansion
}^{\ast}(%
%TCIMACRO{\TeXButton{vt}{\mbox{\boldmath{$\vartheta$}}}}%
%BeginExpansion
\mbox{\boldmath{$\vartheta$}}%
%EndExpansion
^{\mu}\wedge%
%TCIMACRO{\TeXButton{vt}{\mbox{\boldmath{$\vartheta$}}}}%
%BeginExpansion
\mbox{\boldmath{$\vartheta$}}%
%EndExpansion
^{\nu}), \label{4.55}%
\end{equation}
or yet,
\begin{equation}
(\vartheta_{\mu}\cdot\partial\ln\left\vert \det[%
%TCIMACRO{\TeXButton{h}{\slh}}%
%BeginExpansion
\slh
%EndExpansion
]\right\vert )\underline{\eta}\underline{%
%TCIMACRO{\TeXButton{h}{\slh}}%
%BeginExpansion
\slh
%EndExpansion
}^{\clubsuit}(%
%TCIMACRO{\TeXButton{vt}{\mbox{\boldmath{$\vartheta$}}}}%
%BeginExpansion
\mbox{\boldmath{$\vartheta$}}%
%EndExpansion
^{\mu}\wedge%
%TCIMACRO{\TeXButton{vt}{\mbox{\boldmath{$\vartheta$}}}}%
%BeginExpansion
\mbox{\boldmath{$\vartheta$}}%
%EndExpansion
^{\nu})+\vartheta_{\mu}\cdot\partial\underline{\eta}\underline{%
%TCIMACRO{\TeXButton{h}{\slh}}%
%BeginExpansion
\slh
%EndExpansion
}^{\clubsuit}(%
%TCIMACRO{\TeXButton{vt}{\mbox{\boldmath{$\vartheta$}}}}%
%BeginExpansion
\mbox{\boldmath{$\vartheta$}}%
%EndExpansion
^{\mu}\wedge%
%TCIMACRO{\TeXButton{vt}{\mbox{\boldmath{$\vartheta$}}}}%
%BeginExpansion
\mbox{\boldmath{$\vartheta$}}%
%EndExpansion
^{\nu})+\Omega(\vartheta_{\mu})\underset{\eta}{\times}\underline{\eta
}\underline{%
%TCIMACRO{\TeXButton{h}{\slh}}%
%BeginExpansion
\slh
%EndExpansion
}^{\clubsuit}(%
%TCIMACRO{\TeXButton{vt}{\mbox{\boldmath{$\vartheta$}}}}%
%BeginExpansion
\mbox{\boldmath{$\vartheta$}}%
%EndExpansion
^{\mu}\wedge%
%TCIMACRO{\TeXButton{vt}{\mbox{\boldmath{$\vartheta$}}}}%
%BeginExpansion
\mbox{\boldmath{$\vartheta$}}%
%EndExpansion
^{\nu})=0 \label{4.56}%
\end{equation}

This is the field equation for the $\varkappa\eta$-gauge rotation field
$\Omega$. It resumes a set of four coupled multiform equations, differential
in the canonical 1-forms of $%
%TCIMACRO{\TeXButton{h}{\slh}}%
%BeginExpansion
\slh
%EndExpansion
^{\clubsuit}$ (i.e., $%
%TCIMACRO{\TeXButton{h}{\slh}}%
%BeginExpansion
\slh
%EndExpansion
^{\clubsuit}(%
%TCIMACRO{\TeXButton{vt}{\mbox{\boldmath{$\vartheta$}}}}%
%BeginExpansion
\mbox{\boldmath{$\vartheta$}}%
%EndExpansion
^{0}),\ldots,%
%TCIMACRO{\TeXButton{h}{\slh}}%
%BeginExpansion
\slh
%EndExpansion
^{\clubsuit}(%
%TCIMACRO{\TeXButton{vt}{\mbox{\boldmath{$\vartheta$}}}}%
%BeginExpansion
\mbox{\boldmath{$\vartheta$}}%
%EndExpansion
^{3})$) \ and algebraic in the canonical biforms of $\Omega$ (i.e.,
$\Omega(\vartheta_{0}),\ldots,\Omega(\vartheta_{3})$).

We end this Appendix, writing once more the Euler-Lagrange equations for the
fields $%
%TCIMACRO{\TeXButton{h}{\slh}}%
%BeginExpansion
\slh
%EndExpansion
$ and $\Omega,$ in terms of the canonical $1$-form and biform fields.
\begin{align*}%
%TCIMACRO{\TeXButton{h}{\slh}}%
%BeginExpansion
\slh
%EndExpansion
^{\clubsuit}(%
%TCIMACRO{\TeXButton{vt}{\mbox{\boldmath{$\vartheta$}}}}%
%BeginExpansion
\mbox{\boldmath{$\vartheta$}}%
%EndExpansion
^{\mu})\lrcorner(\vartheta_{\mu}\cdot\partial\Omega(\vartheta_{\nu}%
)-\vartheta_{\nu}\cdot\partial\Omega(\vartheta_{\mu})+\Omega(\vartheta_{\mu
})\underset{\eta}{\times}\Omega(\vartheta_{\nu}))  &  =0\\
(\vartheta_{\mu}\cdot\partial\ln\left\vert \det[%
%TCIMACRO{\TeXButton{h}{\slh}}%
%BeginExpansion
\slh
%EndExpansion
]\right\vert )\underline{\eta}\underline{%
%TCIMACRO{\TeXButton{h}{\slh}}%
%BeginExpansion
\slh
%EndExpansion
}^{\clubsuit}(%
%TCIMACRO{\TeXButton{vt}{\mbox{\boldmath{$\vartheta$}}}}%
%BeginExpansion
\mbox{\boldmath{$\vartheta$}}%
%EndExpansion
^{\mu}\wedge%
%TCIMACRO{\TeXButton{vt}{\mbox{\boldmath{$\vartheta$}}}}%
%BeginExpansion
\mbox{\boldmath{$\vartheta$}}%
%EndExpansion
^{\nu})+\vartheta_{\mu}\cdot\partial\underline{\eta}\underline{%
%TCIMACRO{\TeXButton{h}{\slh}}%
%BeginExpansion
\slh
%EndExpansion
}^{\clubsuit}(%
%TCIMACRO{\TeXButton{vt}{\mbox{\boldmath{$\vartheta$}}}}%
%BeginExpansion
\mbox{\boldmath{$\vartheta$}}%
%EndExpansion
^{\mu}\wedge%
%TCIMACRO{\TeXButton{vt}{\mbox{\boldmath{$\vartheta$}}}}%
%BeginExpansion
\mbox{\boldmath{$\vartheta$}}%
%EndExpansion
^{\nu})+\Omega(\vartheta_{\mu})\underset{\eta}{\times}\underline{\eta
}\underline{%
%TCIMACRO{\TeXButton{h}{\slh}}%
%BeginExpansion
\slh
%EndExpansion
}^{\clubsuit}(%
%TCIMACRO{\TeXButton{vt}{\mbox{\boldmath{$\vartheta$}}}}%
%BeginExpansion
\mbox{\boldmath{$\vartheta$}}%
%EndExpansion
^{\mu}\wedge%
%TCIMACRO{\TeXButton{vt}{\mbox{\boldmath{$\vartheta$}}}}%
%BeginExpansion
\mbox{\boldmath{$\vartheta$}}%
%EndExpansion
^{\nu})  &  =0.
\end{align*}

\section{Proof of Eq.(\ref{g9})}

\textbf{Proposition\cite{rodcap2007} }The Einstein-Hilbert Lagrangian density
$L_{eh}$ can be written as
\begin{equation}
\mathcal{L}_{eh}=-d\left(  \mathfrak{g}^{\mathbf{\alpha}}\wedge\underset{%
%TCIMACRO{\TeXButton{sig}{\sitg}}%
%BeginExpansion
\sitg
%EndExpansion
}{\star}d\mathfrak{g}_{\mathbf{\alpha}}\right)  +\mathcal{L}_{g},
\label{10.20}%
\end{equation}
where%
\begin{equation}
\mathcal{L}_{g}=-\frac{1}{2}d\mathfrak{g}^{\mathbf{\alpha}}\wedge\underset{%
%TCIMACRO{\TeXButton{sig}{\sitg}}%
%BeginExpansion
\sitg
%EndExpansion
}{\star}d\mathfrak{g}_{\mathbf{\alpha}}+\frac{1}{2}\underset{%
%TCIMACRO{\TeXButton{sig}{\sitg}}%
%BeginExpansion
\sitg
%EndExpansion
}{\delta}\mathfrak{g}^{\mathbf{\alpha}}\wedge\underset{%
%TCIMACRO{\TeXButton{sig}{\sitg}}%
%BeginExpansion
\sitg
%EndExpansion
}{\star}\underset{%
%TCIMACRO{\TeXButton{sig}{\sitg}}%
%BeginExpansion
\sitg
%EndExpansion
}{\delta}\mathfrak{g}_{\mathbf{\alpha}}+\frac{1}{4}d\mathfrak{g}%
^{\mathbf{\alpha}}\wedge\mathfrak{g}_{\mathbf{\alpha}}\wedge\underset{%
%TCIMACRO{\TeXButton{sig}{\sitg}}%
%BeginExpansion
\sitg
%EndExpansion
}{\star}(d\mathfrak{g}^{\mathbf{\alpha}}\wedge\mathfrak{g}_{\mathbf{\alpha}})
\label{10.20a}%
\end{equation}
is the first order Lagrangian density (first introduced by Einstein).

\begin{proof}
We will establish that $\mathcal{L}_{g}$ may be written as%
\begin{equation}
\mathcal{L}_{g}=-\frac{1}{2}\tau_{\mathbf{%
%TCIMACRO{\TeXButton{sig}{\sitg}}%
%BeginExpansion
\sitg
%EndExpansion
}}\mathfrak{g}^{\mathbf{\gamma}}\underset{\mathbf{%
%TCIMACRO{\TeXButton{sig}{\sitg}}%
%BeginExpansion
\sitg
%EndExpansion
}^{-1}}{\lrcorner}\mathfrak{g}^{\mathbf{\beta}}\underset{\mathbf{%
%TCIMACRO{\TeXButton{sig}{\sitg}}%
%BeginExpansion
\sitg
%EndExpansion
}^{-1}}{\lrcorner}\left(  \omega_{\mathbf{\alpha\beta}}\wedge\omega
_{\mathbf{\gamma}}^{\mathbf{\alpha}}\right)  ,
\end{equation}
which is nothing more than the \textit{intrinsic} form of the Einstein first
order Lagrangian in the gauge defined by the basis $\{\mathfrak{g}%
^{\mathbf{\alpha}}\}$. To see this first recall Cartan's structure equations
for the Lorentzian structure $(M\simeq\mathbb{R}^{4},%
%TCIMACRO{\TeXButton{g}{\slg}}%
%BeginExpansion
\slg
%EndExpansion
,D)$. These are
\begin{subequations}
\label{c}%
\begin{align}
d\mathfrak{g}^{\mathbf{\alpha}}  &  =-\omega_{\mathbf{\beta}}^{\mathbf{\alpha
}}\wedge\mathfrak{g}^{\mathbf{\beta}},\label{c1}\\
\mathcal{R}_{\mathbf{\beta}}^{\mathbf{\alpha}}  &  =d\omega_{\mathbf{\beta}%
}^{\mathbf{\alpha}}+\omega_{\mathbf{\gamma}}^{\mathbf{\alpha}}\wedge
\omega_{\mathbf{\beta}}^{\mathbf{\gamma}}, \label{c2}%
\end{align}
\ where $\omega_{\mathbf{b}}^{\mathbf{a}}$ are the so called connection
$1$-form fields, which on the basis $\{\mathfrak{g}^{\mathbf{a}}\}$ is written
as%
\end{subequations}
\begin{equation}
\omega_{\mathbf{\beta}}^{\mathbf{\alpha}}=L_{\mathbf{\beta\gamma}%
}^{\mathbf{\alpha}}\mathfrak{g}^{\mathbf{\gamma}}. \label{c3}%
\end{equation}
Now, using one of the identities in Eq.(\ref{1.71}) we easily get
\begin{equation}
\mathfrak{g}^{\mathbf{\gamma}}\underset{\mathbf{%
%TCIMACRO{\TeXButton{sig}{\sitg}}%
%BeginExpansion
\sitg
%EndExpansion
}^{-1}}{\lrcorner}\mathfrak{g}^{\mathbf{\beta}}\underset{\mathbf{%
%TCIMACRO{\TeXButton{sig}{\sitg}}%
%BeginExpansion
\sitg
%EndExpansion
}^{-1}}{\lrcorner}\left(  \omega_{\mathbf{\alpha\beta}}\wedge\omega
_{\mathbf{\gamma}}^{\mathbf{\alpha}}\right)  =\eta^{\mathbf{\beta\kappa}%
}\left(  L_{\mathbf{\kappa\gamma}}^{\mathbf{\delta}}L_{\mathbf{\delta\beta}%
}^{\mathbf{\gamma}}-L_{\mathbf{\delta\gamma}}^{\mathbf{\delta}}%
L_{\mathbf{\kappa\beta}}^{\mathbf{\gamma}}\right)  . \label{10.22}%
\end{equation}
Next we recall that Eq.(\ref{c1}) can easily be solved for $\omega
_{\mathbf{b}}^{\mathbf{a}}$ in terms of the $d\mathfrak{g}^{\mathbf{a}}$'s and
the $\mathfrak{g}^{\mathbf{b}}$'s. We get%
\begin{equation}
\omega^{\mathbf{\gamma\delta}}=\frac{1}{2}\left[  \mathfrak{g}^{\mathbf{\delta
}}\underset{\mathtt{\ }\mathbf{%
%TCIMACRO{\TeXButton{sig}{\sitg}}%
%BeginExpansion
\sitg
%EndExpansion
}^{-1}}{\lrcorner}d\mathfrak{g}^{\mathbf{\gamma}}-\mathfrak{g}^{\mathbf{\gamma
}}\underset{\mathtt{\ }\mathbf{%
%TCIMACRO{\TeXButton{sig}{\sitg}}%
%BeginExpansion
\sitg
%EndExpansion
}^{-1}}{\lrcorner}d\mathfrak{g}^{\mathbf{\delta}}+\mathfrak{g}^{\mathbf{\gamma
}}\underset{\mathtt{\ }\mathbf{%
%TCIMACRO{\TeXButton{sig}{\sitg}}%
%BeginExpansion
\sitg
%EndExpansion
}^{-1}}{\lrcorner}\left(  \mathfrak{g}^{\mathbf{\delta}}\underset{\mathtt{\ }%
\mathbf{%
%TCIMACRO{\TeXButton{sig}{\sitg}}%
%BeginExpansion
\sitg
%EndExpansion
}^{-1}}{\lrcorner}d\mathfrak{g}_{\mathbf{\alpha}}\right)  \mathfrak{g}%
^{\mathbf{\alpha}}\right]  \label{w}%
\end{equation}

We are now prepared to prove that $\mathcal{L}_{eh}$ can be written as in
Eq.(\ref{10.20}) we start using Cartan's second structure equation
(Eq.(\ref{c2})) to write Eq.(\ref{g8}) as:%
\begin{align}
\mathcal{L}_{eh}  &  =\frac{1}{2}d\omega_{\mathbf{\alpha\beta}}\wedge
\underset{\mathbf{%
%TCIMACRO{\TeXButton{sig}{\sitg}}%
%BeginExpansion
\sitg
%EndExpansion
}}{\star}(\mathfrak{g}^{\mathbf{\alpha}}\wedge\mathfrak{g}^{\mathbf{\beta}%
})+\frac{1}{2}\omega_{\mathbf{\alpha\gamma}}\wedge\omega_{\mathbf{\beta}%
}^{\mathbf{\gamma}}\wedge\underset{\mathbf{%
%TCIMACRO{\TeXButton{sig}{\sitg}}%
%BeginExpansion
\sitg
%EndExpansion
}}{\star}(\mathfrak{g}^{\mathbf{\alpha}}\wedge\mathfrak{g}^{\mathbf{\beta}%
})\nonumber\\
&  =\frac{1}{2}d[\omega_{\mathbf{\alpha\beta}}\wedge\underset{\mathbf{%
%TCIMACRO{\TeXButton{sig}{\sitg}}%
%BeginExpansion
\sitg
%EndExpansion
}}{\star}(\mathfrak{g}^{\mathbf{\alpha}}\wedge\mathfrak{g}^{\mathbf{\beta}%
})]+\frac{1}{2}\omega_{\mathbf{\alpha\beta}}\wedge\underset{\mathbf{%
%TCIMACRO{\TeXButton{sig}{\sitg}}%
%BeginExpansion
\sitg
%EndExpansion
}}{\star}d(\mathfrak{g}^{\mathbf{\alpha}}\wedge\mathfrak{g}^{\mathbf{\beta}%
})+\frac{1}{2}\omega_{\mathbf{\alpha\gamma}}\wedge\omega_{\mathbf{\beta}%
}^{\mathbf{\gamma}}\wedge\underset{\mathbf{%
%TCIMACRO{\TeXButton{sig}{\sitg}}%
%BeginExpansion
\sitg
%EndExpansion
}}{\star}(\mathfrak{g}^{\mathbf{\alpha}}\wedge\mathfrak{g}^{\mathbf{\beta}%
})\nonumber\\
&  =\frac{1}{2}d[\omega_{\mathbf{\alpha\beta}}\wedge\underset{\mathbf{%
%TCIMACRO{\TeXButton{sig}{\sitg}}%
%BeginExpansion
\sitg
%EndExpansion
}}{\star}(\mathfrak{g}^{\mathbf{\alpha}}\wedge\mathfrak{g}^{\mathbf{\beta}%
})]-\frac{1}{2}\omega_{\mathbf{\alpha\beta}}\wedge\omega_{\mathbf{\gamma}%
}^{\mathbf{\alpha}}\wedge\underset{\mathbf{%
%TCIMACRO{\TeXButton{sig}{\sitg}}%
%BeginExpansion
\sitg
%EndExpansion
}}{\star}(\mathfrak{g}^{\mathbf{\gamma}}\wedge\mathfrak{g}^{\mathbf{\beta}}).
\label{10.23}%
\end{align}
Next using again some of the identities in Eq.(\ref{440new}) we get (recall
that $\omega^{\mathbf{\gamma\delta}}=-\omega^{\mathbf{\delta\gamma}}$)
\begin{align}
(\mathfrak{g}^{\mathbf{\gamma}}\wedge\mathfrak{g}^{\mathbf{\delta}}%
)\wedge\underset{\mathbf{%
%TCIMACRO{\TeXButton{sig}{\sitg}}%
%BeginExpansion
\sitg
%EndExpansion
}}{\star}\omega_{\mathbf{\gamma\delta}}  &  =-\underset{\mathbf{%
%TCIMACRO{\TeXButton{sig}{\sitg}}%
%BeginExpansion
\sitg
%EndExpansion
}}{\star}[\omega^{\mathbf{\gamma\delta}}\underset{\mathbf{%
%TCIMACRO{\TeXButton{sig}{\sitg}}%
%BeginExpansion
\sitg
%EndExpansion
}}{\lrcorner}(\mathfrak{g}_{\mathbf{\gamma}}\wedge\mathfrak{g}_{\mathbf{\delta
}})]\nonumber\\
&  =\text{ }\underset{\mathbf{%
%TCIMACRO{\TeXButton{sig}{\sitg}}%
%BeginExpansion
\sitg
%EndExpansion
}}{\star}[(\omega^{\mathbf{\gamma\delta}}\underset{\mathbf{%
%TCIMACRO{\TeXButton{sig}{\sitg}}%
%BeginExpansion
\sitg
%EndExpansion
}}{\cdot}\mathfrak{g}_{\mathbf{\delta}})\theta_{\mathbf{\gamma}}%
-(\omega^{\mathbf{\gamma\delta}}\underset{\mathbf{%
%TCIMACRO{\TeXButton{sig}{\sitg}}%
%BeginExpansion
\sitg
%EndExpansion
}}{\cdot}\mathfrak{g}_{\mathbf{\gamma}})\mathfrak{g}_{\mathbf{\delta}%
}]=2\underset{\mathbf{%
%TCIMACRO{\TeXButton{sig}{\sitg}}%
%BeginExpansion
\sitg
%EndExpansion
}}{\star}[(\omega^{\mathbf{\gamma\delta}}\underset{\mathbf{%
%TCIMACRO{\TeXButton{sig}{\sitg}}%
%BeginExpansion
\sitg
%EndExpansion
}}{\cdot}\mathfrak{g}_{\mathbf{\delta}})\mathfrak{g}_{\mathbf{\gamma}}],
\label{10.24}%
\end{align}
and from Cartan's first structure equation (Eq.(\ref{c1})) we have%

\begin{align}
\mathfrak{g}^{\mathbf{\alpha}}\wedge\underset{\mathbf{%
%TCIMACRO{\TeXButton{sig}{\sitg}}%
%BeginExpansion
\sitg
%EndExpansion
}}{\star}d\mathfrak{g}_{\mathbf{\alpha}}  &  =\mathfrak{g}^{\mathbf{\alpha}%
}\wedge\underset{\mathbf{%
%TCIMACRO{\TeXButton{sig}{\sitg}}%
%BeginExpansion
\sitg
%EndExpansion
}}{\star}(\omega_{\mathbf{\beta\alpha}}\wedge\mathfrak{g}^{\mathbf{\beta}%
})=-\underset{\mathbf{%
%TCIMACRO{\TeXButton{sig}{\sitg}}%
%BeginExpansion
\sitg
%EndExpansion
}}{\star}[\mathfrak{g}_{\mathbf{\alpha}}\underset{\mathbf{%
%TCIMACRO{\TeXButton{sig}{\sitg}}%
%BeginExpansion
\sitg
%EndExpansion
}}{\lrcorner}(\omega^{\mathbf{\beta\alpha}}\wedge\mathfrak{g}_{\mathbf{\beta}%
})]\nonumber\\
&  =-\underset{\mathbf{%
%TCIMACRO{\TeXButton{sig}{\sitg}}%
%BeginExpansion
\sitg
%EndExpansion
}}{\star}[\mathfrak{g}_{\mathbf{\alpha}}\underset{\mathbf{%
%TCIMACRO{\TeXButton{sig}{\sitg}}%
%BeginExpansion
\sitg
%EndExpansion
}}{\cdot}\omega^{\mathbf{\beta\alpha}})\mathfrak{g}_{\mathbf{\beta}%
}]=-\underset{\mathbf{%
%TCIMACRO{\TeXButton{sig}{\sitg}}%
%BeginExpansion
\sitg
%EndExpansion
}}{\star}[(\mathfrak{g}^{\mathbf{\alpha}}\underset{\mathbf{%
%TCIMACRO{\TeXButton{sig}{\sitg}}%
%BeginExpansion
\sitg
%EndExpansion
}}{\cdot}\omega_{\mathbf{\beta\alpha}})\mathfrak{g}^{\mathbf{\beta}}],
\label{10.25}%
\end{align}
from where it follows that%
\begin{equation}
\frac{1}{2}d[(\mathfrak{g}^{\mathbf{\gamma}}\wedge\mathfrak{g}^{\mathbf{\delta
}})\wedge\underset{\mathbf{%
%TCIMACRO{\TeXButton{sig}{\sitg}}%
%BeginExpansion
\sitg
%EndExpansion
}}{\star}\omega_{\mathbf{\gamma\delta}}]=-d(\mathfrak{g}^{\mathbf{\alpha}%
}\wedge\underset{\mathbf{%
%TCIMACRO{\TeXButton{sig}{\sitg}}%
%BeginExpansion
\sitg
%EndExpansion
}}{\star}d\mathfrak{g}_{\mathbf{\alpha}}). \label{10.26}%
\end{equation}
On the other hand the second term in the last line of Eq.(\ref{10.23}) can be
written as%
\begin{align*}
&  \frac{1}{2}\omega_{\mathbf{\alpha\beta}}\wedge\omega_{\mathbf{\gamma}%
}^{\mathbf{\alpha}}\wedge\underset{\mathbf{%
%TCIMACRO{\TeXButton{sig}{\sitg}}%
%BeginExpansion
\sitg
%EndExpansion
}}{\star}(\mathfrak{g}^{\mathbf{\gamma}}\wedge\mathfrak{g}^{\mathbf{\beta}})\\
&  =-\frac{1}{2}\underset{g}{\star}[(\mathfrak{g}^{\mathbf{\beta}%
}\underset{\mathbf{%
%TCIMACRO{\TeXButton{sig}{\sitg}}%
%BeginExpansion
\sitg
%EndExpansion
}}{\cdot}\omega_{\mathbf{\alpha\beta}})(\mathfrak{g}^{\mathbf{\gamma}%
}\underset{\mathbf{%
%TCIMACRO{\TeXButton{sig}{\sitg}}%
%BeginExpansion
\sitg
%EndExpansion
}}{\cdot}\omega_{\mathbf{\gamma}}^{\mathbf{\alpha}})-(\mathfrak{g}%
^{\mathbf{\beta}}\underset{\mathbf{%
%TCIMACRO{\TeXButton{sig}{\sitg}}%
%BeginExpansion
\sitg
%EndExpansion
}}{\cdot}\omega_{\mathbf{\gamma}}^{\mathbf{\alpha}})(\mathfrak{g}%
^{\mathbf{\gamma}}\underset{\mathbf{%
%TCIMACRO{\TeXButton{sig}{\sitg}}%
%BeginExpansion
\sitg
%EndExpansion
}}{\cdot}\omega_{\mathbf{\alpha\beta}})].
\end{align*}
Now,
\begin{align*}
&  (\mathfrak{g}^{\mathbf{\beta}}\underset{\mathbf{%
%TCIMACRO{\TeXButton{sig}{\sitg}}%
%BeginExpansion
\sitg
%EndExpansion
}}{\cdot}\omega_{\mathbf{\alpha\beta}})(\mathfrak{g}^{\mathbf{\gamma}%
}\underset{\mathbf{%
%TCIMACRO{\TeXButton{sig}{\sitg}}%
%BeginExpansion
\sitg
%EndExpansion
}}{\cdot}\omega_{\mathbf{\gamma}}^{\mathbf{\alpha}})\\
&  =\omega_{\mathbf{\alpha\beta}}\underset{\mathbf{%
%TCIMACRO{\TeXButton{sig}{\sitg}}%
%BeginExpansion
\sitg
%EndExpansion
}}{\cdot}[(\mathfrak{g}^{\mathbf{\gamma}}\underset{\mathbf{%
%TCIMACRO{\TeXButton{sig}{\sitg}}%
%BeginExpansion
\sitg
%EndExpansion
}}{\cdot}\omega_{\mathbf{\gamma}}^{\mathbf{\alpha}})\mathfrak{g}%
^{\mathbf{\beta}}]\\
&  =\omega_{\mathbf{\alpha\beta}}\underset{\mathbf{%
%TCIMACRO{\TeXButton{sig}{\sitg}}%
%BeginExpansion
\sitg
%EndExpansion
}}{\cdot}[\mathfrak{g}^{\mathbf{\gamma}}\underset{\mathbf{%
%TCIMACRO{\TeXButton{sig}{\sitg}}%
%BeginExpansion
\sitg
%EndExpansion
}}{\lrcorner}(\omega_{\mathbf{\gamma}}^{\mathbf{\alpha}}\wedge\mathfrak{g}%
^{\mathbf{\beta}})+\omega^{\mathbf{\alpha\beta}}]\\
&  =(\omega_{\mathbf{\alpha\beta}}\wedge\mathfrak{g}^{\mathbf{\gamma}%
})\underset{\mathbf{%
%TCIMACRO{\TeXButton{sig}{\sitg}}%
%BeginExpansion
\sitg
%EndExpansion
}}{\lrcorner}(\omega_{\mathbf{\gamma}}^{\mathbf{\alpha}}\wedge\mathfrak{g}%
^{\mathbf{\beta}})+\omega_{\mathbf{\gamma\delta}}\underset{\mathbf{%
%TCIMACRO{\TeXButton{sig}{\sitg}}%
%BeginExpansion
\sitg
%EndExpansion
}}{\cdot}\omega^{\mathbf{\gamma\delta}}]
\end{align*}
and taking into account that $d\mathfrak{g}^{\mathbf{\alpha}}=-\omega
_{\mathbf{\beta}}^{\mathbf{\alpha}}\wedge\mathfrak{g}^{\mathbf{\beta}}$,
$\underset{\mathbf{%
%TCIMACRO{\TeXButton{sig}{\sitg}}%
%BeginExpansion
\sitg
%EndExpansion
}}{\star}^{-1}d\underset{\mathbf{%
%TCIMACRO{\TeXButton{sig}{\sitg}}%
%BeginExpansion
\sitg
%EndExpansion
}}{\star}\mathfrak{g}^{\mathbf{\alpha}}=-\omega_{\mathbf{\beta}}%
^{\mathbf{\alpha}}\underset{\mathbf{%
%TCIMACRO{\TeXButton{sig}{\sitg}}%
%BeginExpansion
\sitg
%EndExpansion
}}{\cdot}\mathfrak{g}^{\mathbf{\beta}}$ and that $\underset{\mathbf{%
%TCIMACRO{\TeXButton{sig}{\sitg}}%
%BeginExpansion
\sitg
%EndExpansion
}}{\delta}\mathfrak{g}_{\mathbf{\alpha}}=-\underset{\mathbf{%
%TCIMACRO{\TeXButton{sig}{\sitg}}%
%BeginExpansion
\sitg
%EndExpansion
}}{\star}^{-1}d\underset{\mathbf{%
%TCIMACRO{\TeXButton{sig}{\sitg}}%
%BeginExpansion
\sitg
%EndExpansion
}}{\star}\mathfrak{g}_{\mathbf{\alpha}}$ we have%

\begin{equation}
\frac{1}{2}\omega_{\mathbf{\alpha\beta}}\wedge\omega_{\mathbf{\gamma}%
}^{\mathbf{\alpha}}\wedge\underset{\mathbf{%
%TCIMACRO{\TeXButton{sig}{\sitg}}%
%BeginExpansion
\sitg
%EndExpansion
}}{\star}(\mathfrak{g}^{\mathbf{\gamma}}\wedge\mathfrak{g}^{\mathbf{\beta}%
})=\frac{1}{2}[-d\mathfrak{g}^{\mathbf{\alpha}}\wedge\underset{\mathbf{%
%TCIMACRO{\TeXButton{sig}{\sitg}}%
%BeginExpansion
\sitg
%EndExpansion
}}{\star}d\mathfrak{g}_{\mathbf{\alpha}}-\text{ }\underset{\mathbf{%
%TCIMACRO{\TeXButton{sig}{\sitg}}%
%BeginExpansion
\sitg
%EndExpansion
}}{\delta}\mathfrak{g}^{\mathbf{\alpha}}\wedge\underset{\mathbf{%
%TCIMACRO{\TeXButton{sig}{\sitg}}%
%BeginExpansion
\sitg
%EndExpansion
}}{\star}\underset{\mathbf{%
%TCIMACRO{\TeXButton{sig}{\sitg}}%
%BeginExpansion
\sitg
%EndExpansion
}}{\delta}\mathfrak{g}_{\mathbf{\alpha}}+\omega_{\mathbf{\gamma\delta}}%
\wedge\underset{\mathbf{%
%TCIMACRO{\TeXButton{sig}{\sitg}}%
%BeginExpansion
\sitg
%EndExpansion
}}{\star}\omega^{\mathbf{\gamma\delta}}] \label{10.27}%
\end{equation}
Next, using Eq.(\ref{w}) the last term in the last equation can after some
algebra be written as%
\[
\frac{1}{2}\omega_{\mathbf{\gamma\delta}}\wedge\underset{\mathbf{%
%TCIMACRO{\TeXButton{sig}{\sitg}}%
%BeginExpansion
\sitg
%EndExpansion
}}{\star}\omega^{\mathbf{\gamma\delta}}=d\mathfrak{g}^{\mathbf{\alpha}}%
\wedge\underset{\mathbf{%
%TCIMACRO{\TeXButton{sig}{\sitg}}%
%BeginExpansion
\sitg
%EndExpansion
}}{\star}d\mathfrak{g}_{\mathbf{\alpha}}-\frac{1}{4}\left(  d\mathfrak{g}%
^{\mathbf{\alpha}}\wedge\mathfrak{g}_{\mathbf{\alpha}}\right)  \wedge
\underset{\mathtt{\mathbf{g}}}{\star}\left(  d\mathfrak{g}^{\mathbf{\alpha}%
}\wedge\mathfrak{g}_{\mathbf{\alpha}}\right)
\]
and we finally get%
\begin{equation}
\mathcal{L}_{g}=-\frac{1}{2}d\mathfrak{g}^{\mathbf{\alpha}}\wedge
\underset{\mathbf{%
%TCIMACRO{\TeXButton{sig}{\sitg}}%
%BeginExpansion
\sitg
%EndExpansion
}}{\star}d\mathfrak{g}_{\mathbf{\alpha}}+\frac{1}{2}\underset{\mathbf{%
%TCIMACRO{\TeXButton{sig}{\sitg}}%
%BeginExpansion
\sitg
%EndExpansion
}}{\delta}\mathfrak{g}^{\mathbf{\alpha}}\wedge\underset{\mathbf{%
%TCIMACRO{\TeXButton{sig}{\sitg}}%
%BeginExpansion
\sitg
%EndExpansion
}}{\star}\underset{\mathbf{%
%TCIMACRO{\TeXButton{sig}{\sitg}}%
%BeginExpansion
\sitg
%EndExpansion
}}{\delta}\mathfrak{g}_{\mathbf{\alpha}}+\frac{1}{4}\left(  d\mathfrak{g}%
^{\mathbf{\alpha}}\wedge\mathfrak{g}_{\mathbf{\alpha}}\right)  \wedge
\underset{\mathbf{%
%TCIMACRO{\TeXButton{sig}{\sitg}}%
%BeginExpansion
\sitg
%EndExpansion
}}{\star}\left(  d\mathfrak{g}^{\mathbf{\beta}}\wedge\mathfrak{g}%
_{\mathbf{\beta}}\right)  , \label{10.28}%
\end{equation}
and the proposition is proved.
\end{proof}

\section{Derivation of the Field Equations from the Einstein-Hilbert
Lagrangian Density}

Let $X$ $\in\sec%
%TCIMACRO{\dbigwedge \nolimits^{p}}%
%BeginExpansion
{\displaystyle\bigwedge\nolimits^{p}}
%EndExpansion
T^{\ast}U$. \textit{A multiform functional} $F$ of $X$ (not depending
\textit{explicitly} on $x\in%
%TCIMACRO{\dbigwedge \nolimits^{1}}%
%BeginExpansion
{\displaystyle\bigwedge\nolimits^{1}}
%EndExpansion
U$) is a mapping%
\[
F:\sec%
%TCIMACRO{\dbigwedge \nolimits^{p}}%
%BeginExpansion
{\displaystyle\bigwedge\nolimits^{p}}
%EndExpansion
T^{\ast}U\rightarrow\sec%
%TCIMACRO{\dbigwedge \nolimits^{r}}%
%BeginExpansion
{\displaystyle\bigwedge\nolimits^{r}}
%EndExpansion
T^{\ast}U
\]
As, in Section 3, when no confusion arises we use a sloppy notation and denote
the image $F(X)\in\sec%
%TCIMACRO{\dbigwedge \nolimits^{r}}%
%BeginExpansion
{\displaystyle\bigwedge\nolimits^{r}}
%EndExpansion
T^{\ast}U$ simply by $F$. Eventually we also denote a functional $F$ by
$F(X)$. Which object we are talking about is always obvious from the context
of the equations where they appear.

Let $w:=$ $%
%TCIMACRO{\TeXButton{bdelta}{\mbox{\boldmath{$\delta$}}}}%
%BeginExpansion
\mbox{\boldmath{$\delta$}}%
%EndExpansion
X\in\sec%
%TCIMACRO{\dbigwedge \nolimits^{p}}%
%BeginExpansion
{\displaystyle\bigwedge\nolimits^{p}}
%EndExpansion
T^{\ast}U$. As in Exercise 3.6. we write the \textit{variation} of $F$ in the
direction of $%
%TCIMACRO{\TeXButton{bdelta}{\mbox{\boldmath{$\delta$}}}}%
%BeginExpansion
\mbox{\boldmath{$\delta$}}%
%EndExpansion
X$ as the functional $%
%TCIMACRO{\TeXButton{bdelta}{\mbox{\boldmath{$\delta$}}}}%
%BeginExpansion
\mbox{\boldmath{$\delta$}}%
%EndExpansion
F:%
%TCIMACRO{\dbigwedge \nolimits^{p}}%
%BeginExpansion
{\displaystyle\bigwedge\nolimits^{p}}
%EndExpansion
U\rightarrow%
%TCIMACRO{\dbigwedge \nolimits^{r}}%
%BeginExpansion
{\displaystyle\bigwedge\nolimits^{r}}
%EndExpansion
U$ given by%
\begin{equation}%
%TCIMACRO{\TeXButton{bdelta}{\mbox{\boldmath{$\delta$}}}}%
%BeginExpansion
\mbox{\boldmath{$\delta$}}%
%EndExpansion
F=\lim_{\lambda\rightarrow0}\frac{F(X+\lambda%
%TCIMACRO{\TeXButton{bdelta}{\mbox{\boldmath{$\delta$}}}}%
%BeginExpansion
\mbox{\boldmath{$\delta$}}%
%EndExpansion
X)-F(X)}{\lambda}. \label{ad2}%
\end{equation}
Recall from Remark 3.1 that the\textit{ algebraic derivative} of $F$ relative
to $X$, $\frac{\partial F}{\partial X}$ is such that:%
\begin{equation}%
%TCIMACRO{\TeXButton{bdelta}{\mbox{\boldmath{$\delta$}}}}%
%BeginExpansion
\mbox{\boldmath{$\delta$}}%
%EndExpansion
F=%
%TCIMACRO{\TeXButton{bdelta}{\mbox{\boldmath{$\delta$}}}}%
%BeginExpansion
\mbox{\boldmath{$\delta$}}%
%EndExpansion
X\wedge\frac{\partial F}{\partial X}. \label{ad1}%
\end{equation}
Moreover, given the functionals $F:%
%TCIMACRO{\dbigwedge \nolimits^{p}}%
%BeginExpansion
{\displaystyle\bigwedge\nolimits^{p}}
%EndExpansion
U\rightarrow%
%TCIMACRO{\dbigwedge \nolimits^{p}}%
%BeginExpansion
{\displaystyle\bigwedge\nolimits^{p}}
%EndExpansion
U\rightarrow$ and $G:%
%TCIMACRO{\dbigwedge \nolimits^{p}}%
%BeginExpansion
{\displaystyle\bigwedge\nolimits^{p}}
%EndExpansion
U\rightarrow%
%TCIMACRO{\dbigwedge \nolimits^{s}}%
%BeginExpansion
{\displaystyle\bigwedge\nolimits^{s}}
%EndExpansion
U$ the variation $%
%TCIMACRO{\TeXButton{bdelta}{\mbox{\boldmath{$\delta$}}}}%
%BeginExpansion
\mbox{\boldmath{$\delta$}}%
%EndExpansion
$ satisfies
\begin{equation}%
%TCIMACRO{\TeXButton{bdelta}{\mbox{\boldmath{$\delta$}}}}%
%BeginExpansion
\mbox{\boldmath{$\delta$}}%
%EndExpansion
(F\wedge G)=%
%TCIMACRO{\TeXButton{bdelta}{\mbox{\boldmath{$\delta$}}}}%
%BeginExpansion
\mbox{\boldmath{$\delta$}}%
%EndExpansion
F\wedge G+F\wedge%
%TCIMACRO{\TeXButton{bdelta}{\mbox{\boldmath{$\delta$}}}}%
%BeginExpansion
\mbox{\boldmath{$\delta$}}%
%EndExpansion
G, \label{ad3}%
\end{equation}
and the algebraic derivative (as it is trivial to verify) satisfies
\begin{equation}
\frac{\partial}{\partial X}(F\wedge G)=\frac{\partial F}{\partial X}\wedge
G+(-1)^{rp}F\wedge\frac{\partial G}{\partial X}. \label{ad4}%
\end{equation}

An important property of $%
%TCIMACRO{\TeXButton{bdelta}{\mbox{\boldmath{$\delta$}}}}%
%BeginExpansion
\mbox{\boldmath{$\delta$}}%
%EndExpansion
$ is that it commutes with the exterior derivative operator $d$, i.e., for any
given functional $F$%
\begin{equation}
d%
%TCIMACRO{\TeXButton{bdelta}{\mbox{\boldmath{$\delta$}}}}%
%BeginExpansion
\mbox{\boldmath{$\delta$}}%
%EndExpansion
F=%
%TCIMACRO{\TeXButton{bdelta}{\mbox{\boldmath{$\delta$}}}}%
%BeginExpansion
\mbox{\boldmath{$\delta$}}%
%EndExpansion
dF. \label{ad5}%
\end{equation}

In general we may have functionals depending on several different multiform
forms fields, say, $F:\sec(%
%TCIMACRO{\dbigwedge \nolimits^{p}}%
%BeginExpansion
{\displaystyle\bigwedge\nolimits^{p}}
%EndExpansion
T^{\ast}U\times%
%TCIMACRO{\dbigwedge \nolimits^{q}}%
%BeginExpansion
{\displaystyle\bigwedge\nolimits^{q}}
%EndExpansion
T^{\ast}U)\rightarrow\sec%
%TCIMACRO{\dbigwedge \nolimits^{r}}%
%BeginExpansion
{\displaystyle\bigwedge\nolimits^{r}}
%EndExpansion
T^{\ast}U$, with $(X,Y)\mapsto F(X,Y)\in\sec%
%TCIMACRO{\dbigwedge \nolimits^{p}}%
%BeginExpansion
{\displaystyle\bigwedge\nolimits^{p}}
%EndExpansion
T^{\ast}U$. In this case we have:%
\begin{equation}%
%TCIMACRO{\TeXButton{bdelta}{\mbox{\boldmath{$\delta$}}}}%
%BeginExpansion
\mbox{\boldmath{$\delta$}}%
%EndExpansion
F=%
%TCIMACRO{\TeXButton{bdelta}{\mbox{\boldmath{$\delta$}}}}%
%BeginExpansion
\mbox{\boldmath{$\delta$}}%
%EndExpansion
X\wedge\frac{\partial F}{\partial X}+%
%TCIMACRO{\TeXButton{bdelta}{\mbox{\boldmath{$\delta$}}}}%
%BeginExpansion
\mbox{\boldmath{$\delta$}}%
%EndExpansion
Y\wedge\frac{\partial F}{\partial Y}. \label{ad6}%
\end{equation}
An important case is the one where the functional $F$ is such that
$F(X,dX):\sec(%
%TCIMACRO{\dbigwedge \nolimits^{p}}%
%BeginExpansion
{\displaystyle\bigwedge\nolimits^{p}}
%EndExpansion
T^{\ast}U\times%
%TCIMACRO{\dbigwedge \nolimits^{p+1}}%
%BeginExpansion
{\displaystyle\bigwedge\nolimits^{p+1}}
%EndExpansion
T^{\ast}U)\rightarrow\sec%
%TCIMACRO{\dbigwedge \nolimits^{4}}%
%BeginExpansion
{\displaystyle\bigwedge\nolimits^{4}}
%EndExpansion
T^{\ast}U$. We then can write supposing that the variation $%
%TCIMACRO{\TeXButton{bdelta}{\mbox{\boldmath{$\delta$}}}}%
%BeginExpansion
\mbox{\boldmath{$\delta$}}%
%EndExpansion
X$ is chosen to be null in the boundary $\partial U^{^{\prime}}$, $U^{\prime
}\subset U$ (or that $\left.  \frac{\partial F}{\partial dX}\right\vert
_{\partial U^{\prime}}=0$) and taking into account Stokes theorem,%
\begin{align}%
%TCIMACRO{\TeXButton{bdelta}{\mbox{\boldmath{$\delta$}}}}%
%BeginExpansion
\mbox{\boldmath{$\delta$}}%
%EndExpansion%
%TCIMACRO{\dint \nolimits_{U^{\prime}}}%
%BeginExpansion
{\displaystyle\int\nolimits_{U^{\prime}}}
%EndExpansion
F  &  =%
%TCIMACRO{\dint \nolimits_{U^{\prime}}}%
%BeginExpansion
{\displaystyle\int\nolimits_{U^{\prime}}}
%EndExpansion%
%TCIMACRO{\TeXButton{bdelta}{\mbox{\boldmath{$\delta$}}}}%
%BeginExpansion
\mbox{\boldmath{$\delta$}}%
%EndExpansion
F=%
%TCIMACRO{\dint \nolimits_{U^{\prime}}}%
%BeginExpansion
{\displaystyle\int\nolimits_{U^{\prime}}}
%EndExpansion%
%TCIMACRO{\TeXButton{bdelta}{\mbox{\boldmath{$\delta$}}}}%
%BeginExpansion
\mbox{\boldmath{$\delta$}}%
%EndExpansion
X\wedge\frac{\partial F}{\partial X}+%
%TCIMACRO{\TeXButton{bdelta}{\mbox{\boldmath{$\delta$}}}}%
%BeginExpansion
\mbox{\boldmath{$\delta$}}%
%EndExpansion
dX\wedge\frac{\partial F}{\partial dX}\nonumber\\
&  =%
%TCIMACRO{\dint \nolimits_{U^{\prime}}}%
%BeginExpansion
{\displaystyle\int\nolimits_{U^{\prime}}}
%EndExpansion%
%TCIMACRO{\TeXButton{bdelta}{\mbox{\boldmath{$\delta$}}}}%
%BeginExpansion
\mbox{\boldmath{$\delta$}}%
%EndExpansion
X\wedge\left[  \frac{\partial F}{\partial X}-(-1)^{p}d\left(  \frac{\partial
F}{\partial dX}\right)  \right]  +d\left(
%TCIMACRO{\TeXButton{bdelta}{\mbox{\boldmath{$\delta$}}}}%
%BeginExpansion
\mbox{\boldmath{$\delta$}}%
%EndExpansion
X\wedge\frac{\partial F}{\partial dX}\right) \nonumber\\
&  =%
%TCIMACRO{\dint \nolimits_{U^{\prime}}}%
%BeginExpansion
{\displaystyle\int\nolimits_{U^{\prime}}}
%EndExpansion%
%TCIMACRO{\TeXButton{bdelta}{\mbox{\boldmath{$\delta$}}}}%
%BeginExpansion
\mbox{\boldmath{$\delta$}}%
%EndExpansion
X\wedge\left[  \frac{\partial F}{\partial X}-(-1)^{p}d\left(  \frac{\partial
F}{\partial dX}\right)  \right]  +%
%TCIMACRO{\dint \nolimits_{\partial U^{\prime}}}%
%BeginExpansion
{\displaystyle\int\nolimits_{\partial U^{\prime}}}
%EndExpansion%
%TCIMACRO{\TeXButton{bdelta}{\mbox{\boldmath{$\delta$}}}}%
%BeginExpansion
\mbox{\boldmath{$\delta$}}%
%EndExpansion
X\wedge\frac{\partial F}{\partial dX}\nonumber\\
&  =%
%TCIMACRO{\dint \nolimits_{U^{\prime}}}%
%BeginExpansion
{\displaystyle\int\nolimits_{U^{\prime}}}
%EndExpansion%
%TCIMACRO{\TeXButton{bdelta}{\mbox{\boldmath{$\delta$}}}}%
%BeginExpansion
\mbox{\boldmath{$\delta$}}%
%EndExpansion
X\wedge\frac{%
%TCIMACRO{\TeXButton{bdelta}{\mbox{\boldmath{$\delta$}}}}%
%BeginExpansion
\mbox{\boldmath{$\delta$}}%
%EndExpansion
F}{%
%TCIMACRO{\TeXButton{bdelta}{\mbox{\boldmath{$\delta$}}}}%
%BeginExpansion
\mbox{\boldmath{$\delta$}}%
%EndExpansion
X}, \label{ad7}%
\end{align}
where $\frac{%
%TCIMACRO{\TeXButton{bdelta}{\mbox{\boldmath{$\delta$}}}}%
%BeginExpansion
\mbox{\boldmath{$\delta$}}%
%EndExpansion
}{%
%TCIMACRO{\TeXButton{bdelta}{\mbox{\boldmath{$\delta$}}}}%
%BeginExpansion
\mbox{\boldmath{$\delta$}}%
%EndExpansion
X}F(X,dX)$ $:\sec(%
%TCIMACRO{\dbigwedge \nolimits^{p}}%
%BeginExpansion
{\displaystyle\bigwedge\nolimits^{p}}
%EndExpansion
T^{\ast}U\times%
%TCIMACRO{\dbigwedge \nolimits^{p+1}}%
%BeginExpansion
{\displaystyle\bigwedge\nolimits^{p+1}}
%EndExpansion
$ $T^{\ast}U)\rightarrow%
%TCIMACRO{\dbigwedge \nolimits^{4-p}}%
%BeginExpansion
{\displaystyle\bigwedge\nolimits^{4-p}}
%EndExpansion
T^{\ast}U$ is called the \textit{functional derivative} of $F$ and we have:%
\begin{equation}
\frac{%
%TCIMACRO{\TeXButton{bdelta}{\mbox{\boldmath{$\delta$}}}}%
%BeginExpansion
\mbox{\boldmath{$\delta$}}%
%EndExpansion
F}{%
%TCIMACRO{\TeXButton{bdelta}{\mbox{\boldmath{$\delta$}}}}%
%BeginExpansion
\mbox{\boldmath{$\delta$}}%
%EndExpansion
X}=\frac{\partial F}{\partial X}-(-1)^{p}d\left(  \frac{\partial F}{\partial
dX}\right)  . \label{ad8}%
\end{equation}
When $F=\mathcal{L}$ is a Lagrangian density in field theory $\frac{%
%TCIMACRO{\TeXButton{bdelta}{\mbox{\boldmath{$\delta$}}}}%
%BeginExpansion
\mbox{\boldmath{$\delta$}}%
%EndExpansion
\mathcal{L}}{%
%TCIMACRO{\TeXButton{bdelta}{\mbox{\boldmath{$\delta$}}}}%
%BeginExpansion
\mbox{\boldmath{$\delta$}}%
%EndExpansion
X}$ is the \textit{Euler-Lagrange functional}.

We now obtain the variation of the Einstein-Hilbert Lagrangian density
$\mathcal{L}_{eh}$ given by Eq.(\ref{g8}), i.e.,%
\[
\mathcal{L}_{eh}=\frac{1}{2}(\mathfrak{g}^{\mathbf{\kappa}}\wedge
\mathfrak{g}^{\mathbf{\iota}})\wedge\underset{%
%TCIMACRO{\TeXButton{sig}{\sitg}}%
%BeginExpansion
\sitg
%EndExpansion
}{\star}\mathcal{R}_{\mathbf{\kappa\iota}}=\frac{1}{2}\mathcal{R}%
_{\mathbf{\kappa\iota}}\wedge\underset{%
%TCIMACRO{\TeXButton{sig}{\sitg}}%
%BeginExpansion
\sitg
%EndExpansion
}{\star}(\mathfrak{g}^{\mathbf{\kappa}}\wedge\mathfrak{g}^{\mathbf{\iota}}),
\]
by varying the $\omega_{\mathbf{\gamma\kappa}}$ and the $\mathfrak{g}%
^{\mathbf{\kappa}}$ independently.

We have\footnote{We use only constrained variations of the $\mathfrak{g}%
^{\mathbf{\alpha}}$ that do no change the metric field $%
%TCIMACRO{\TeXButton{itg}{\itg}}%
%BeginExpansion
\itg
%EndExpansion
$.}%

\begin{align}%
%TCIMACRO{\TeXButton{bdelta}{\mbox{\boldmath{$\delta$}}}}%
%BeginExpansion
\mbox{\boldmath{$\delta$}}%
%EndExpansion
\mathcal{L}_{eh}  &  =\frac{1}{2}%
%TCIMACRO{\TeXButton{bdelta}{\mbox{\boldmath{$\delta$}}}}%
%BeginExpansion
\mbox{\boldmath{$\delta$}}%
%EndExpansion
[\mathcal{R}_{\mathbf{\gamma\delta}}\wedge\underset{%
%TCIMACRO{\TeXButton{sig}{\sitg}}%
%BeginExpansion
\sitg
%EndExpansion
}{\star}(\mathfrak{g}^{\mathbf{\gamma}}\wedge\mathfrak{g}^{\mathbf{\delta}%
})]\nonumber\\
&  =\frac{1}{2}%
%TCIMACRO{\TeXButton{bdelta}{\mbox{\boldmath{$\delta$}}}}%
%BeginExpansion
\mbox{\boldmath{$\delta$}}%
%EndExpansion
\mathcal{R}_{\mathbf{\gamma\delta}}\wedge\underset{%
%TCIMACRO{\TeXButton{sig}{\sitg}}%
%BeginExpansion
\sitg
%EndExpansion
}{\star}(\mathfrak{g}^{\mathbf{\gamma}}\wedge\mathfrak{g}^{\mathbf{\delta}%
})+\frac{1}{2}\mathcal{R}_{\mathbf{\gamma\delta}}\wedge%
%TCIMACRO{\TeXButton{bdelta}{\mbox{\boldmath{$\delta$}}}}%
%BeginExpansion
\mbox{\boldmath{$\delta$}}%
%EndExpansion
\underset{%
%TCIMACRO{\TeXButton{sig}{\sitg}}%
%BeginExpansion
\sitg
%EndExpansion
}{\star}(\mathfrak{g}^{\mathbf{\gamma}}\wedge\mathfrak{g}^{\mathbf{\delta}}).
\label{v5}%
\end{align}
From Cartan's second structure equation we can write
\begin{align}
&
%TCIMACRO{\TeXButton{bdelta}{\mbox{\boldmath{$\delta$}}}}%
%BeginExpansion
\mbox{\boldmath{$\delta$}}%
%EndExpansion
\mathcal{R}_{\mathbf{\gamma\delta}}\wedge\underset{%
%TCIMACRO{\TeXButton{sig}{\sitg}}%
%BeginExpansion
\sitg
%EndExpansion
}{\star}(\mathfrak{g}^{\mathbf{\gamma}}\wedge\mathfrak{g}^{\mathbf{\delta}%
})\nonumber\\
&  =%
%TCIMACRO{\TeXButton{bdelta}{\mbox{\boldmath{$\delta$}}}}%
%BeginExpansion
\mbox{\boldmath{$\delta$}}%
%EndExpansion
d\omega_{\mathbf{\gamma\delta}}\wedge\underset{%
%TCIMACRO{\TeXButton{sig}{\sitg}}%
%BeginExpansion
\sitg
%EndExpansion
}{\star}(\mathfrak{g}^{\mathbf{\gamma}}\wedge\mathfrak{g}^{\mathbf{\delta}})+%
%TCIMACRO{\TeXButton{bdelta}{\mbox{\boldmath{$\delta$}}}}%
%BeginExpansion
\mbox{\boldmath{$\delta$}}%
%EndExpansion
\omega_{\mathbf{\gamma\kappa}}\wedge\omega_{\mathbf{\delta}}^{\mathbf{\kappa}%
}\wedge\star(\mathfrak{g}^{\mathbf{\gamma}}\wedge\mathfrak{g}^{\mathbf{\delta
}})+\omega_{\mathbf{\gamma\kappa}}\wedge%
%TCIMACRO{\TeXButton{bdelta}{\mbox{\boldmath{$\delta$}}}}%
%BeginExpansion
\mbox{\boldmath{$\delta$}}%
%EndExpansion
\omega_{\mathbf{\delta}}^{\mathbf{\kappa}}\wedge\underset{%
%TCIMACRO{\TeXButton{sig}{\sitg}}%
%BeginExpansion
\sitg
%EndExpansion
}{\star}(\mathfrak{g}^{\mathbf{\gamma}}\wedge\mathfrak{g}^{\mathbf{\delta}%
})\nonumber\\
&  =%
%TCIMACRO{\TeXButton{bdelta}{\mbox{\boldmath{$\delta$}}}}%
%BeginExpansion
\mbox{\boldmath{$\delta$}}%
%EndExpansion
d\omega_{\mathbf{\gamma\delta}}\wedge\underset{%
%TCIMACRO{\TeXButton{sig}{\sitg}}%
%BeginExpansion
\sitg
%EndExpansion
}{\star}(\mathfrak{g}^{\mathbf{\gamma}}\wedge\mathfrak{g}^{\mathbf{\delta}%
})\label{v6}\\
&  =d[%
%TCIMACRO{\TeXButton{bdelta}{\mbox{\boldmath{$\delta$}}}}%
%BeginExpansion
\mbox{\boldmath{$\delta$}}%
%EndExpansion
\omega_{\mathbf{\gamma\delta}}\wedge\underset{%
%TCIMACRO{\TeXButton{sig}{\sitg}}%
%BeginExpansion
\sitg
%EndExpansion
}{\star}(\mathfrak{g}^{\mathbf{\gamma}}\wedge\mathfrak{g}^{\mathbf{\delta}})]-%
%TCIMACRO{\TeXButton{bdelta}{\mbox{\boldmath{$\delta$}}}}%
%BeginExpansion
\mbox{\boldmath{$\delta$}}%
%EndExpansion
\omega_{\mathbf{\gamma\delta}}\wedge d[\underset{%
%TCIMACRO{\TeXButton{sig}{\sitg}}%
%BeginExpansion
\sitg
%EndExpansion
}{\star}(\mathfrak{g}^{\mathbf{\gamma}}\wedge\mathfrak{g}^{\mathbf{\delta}%
})].\nonumber\\
&  =d[%
%TCIMACRO{\TeXButton{bdelta}{\mbox{\boldmath{$\delta$}}}}%
%BeginExpansion
\mbox{\boldmath{$\delta$}}%
%EndExpansion
\omega_{\mathbf{\gamma\delta}}\wedge\underset{%
%TCIMACRO{\TeXButton{sig}{\sitg}}%
%BeginExpansion
\sitg
%EndExpansion
}{\star}(\mathfrak{g}^{\mathbf{\gamma}}\wedge\mathfrak{g}^{\mathbf{\delta}})]-%
%TCIMACRO{\TeXButton{bdelta}{\mbox{\boldmath{$\delta$}}}}%
%BeginExpansion
\mbox{\boldmath{$\delta$}}%
%EndExpansion
\omega_{\mathbf{\gamma\delta}}\wedge\lbrack-\omega_{\mathbf{\kappa}%
}^{\mathbf{\alpha}}\wedge\underset{%
%TCIMACRO{\TeXButton{sig}{\sitg}}%
%BeginExpansion
\sitg
%EndExpansion
}{\star}(\mathfrak{g}^{\mathbf{\kappa}}\wedge\mathfrak{g}^{\mathbf{\kappa}%
})-\omega_{\mathbf{\kappa}}^{\mathbf{\delta}}\wedge\underset{%
%TCIMACRO{\TeXButton{sig}{\sitg}}%
%BeginExpansion
\sitg
%EndExpansion
}{\star}(\mathfrak{g}^{\mathbf{\gamma}}\wedge\mathfrak{g}^{\mathbf{\kappa}%
})]\nonumber\\
&  =d[%
%TCIMACRO{\TeXButton{bdelta}{\mbox{\boldmath{$\delta$}}}}%
%BeginExpansion
\mbox{\boldmath{$\delta$}}%
%EndExpansion
\omega_{\mathbf{\gamma\delta}}\wedge\underset{%
%TCIMACRO{\TeXButton{sig}{\sitg}}%
%BeginExpansion
\sitg
%EndExpansion
}{\star}(\mathfrak{g}^{\mathbf{\gamma}}\wedge\mathfrak{g}^{\mathbf{\delta}%
})].\nonumber
\end{align}
Moreover, using the definition of algebraic derivative (Eq.(\ref{ad2})) we
have immediately
\begin{equation}%
%TCIMACRO{\TeXButton{bdelta}{\mbox{\boldmath{$\delta$}}}}%
%BeginExpansion
\mbox{\boldmath{$\delta$}}%
%EndExpansion
\underset{%
%TCIMACRO{\TeXButton{sig}{\sitg}}%
%BeginExpansion
\sitg
%EndExpansion
}{\star}(\mathfrak{g}^{\mathbf{\gamma}}\wedge\mathfrak{g}^{\mathbf{\delta}}):=%
%TCIMACRO{\TeXButton{bdelta}{\mbox{\boldmath{$\delta$}}}}%
%BeginExpansion
\mbox{\boldmath{$\delta$}}%
%EndExpansion
\mathfrak{g}^{\mathbf{\kappa}}\wedge\frac{\partial\lbrack\underset{%
%TCIMACRO{\TeXButton{sig}{\sitg}}%
%BeginExpansion
\sitg
%EndExpansion
}{\star}(\mathfrak{g}^{\mathbf{\gamma}}\wedge\mathfrak{g}^{\mathbf{\delta}}%
)]}{\partial\mathfrak{g}^{\mathbf{\kappa}}} \label{v7}%
\end{equation}

Now recalling Eq.(\ref{fhd}) of Section 4 we can write
\begin{align}%
%TCIMACRO{\TeXButton{bdelta}{\mbox{\boldmath{$\delta$}}}}%
%BeginExpansion
\mbox{\boldmath{$\delta$}}%
%EndExpansion
\underset{%
%TCIMACRO{\TeXButton{sig}{\sitg}}%
%BeginExpansion
\sitg
%EndExpansion
}{\star}(\mathfrak{g}^{\mathbf{\gamma}}\wedge\mathfrak{g}^{\mathbf{\delta}})
&  =%
%TCIMACRO{\TeXButton{bdelta}{\mbox{\boldmath{$\delta$}}}}%
%BeginExpansion
\mbox{\boldmath{$\delta$}}%
%EndExpansion
(\frac{1}{2}\eta^{\mathbf{\gamma\kappa}}\eta^{\mathbf{\delta\iota}}%
\epsilon_{\mathbf{\kappa\iota\mu\nu}}\mathfrak{g}^{\mathbf{\mu}}%
\wedge\mathfrak{g}^{\mathbf{\nu}})\nonumber\\
&  =%
%TCIMACRO{\TeXButton{bdelta}{\mbox{\boldmath{$\delta$}}}}%
%BeginExpansion
\mbox{\boldmath{$\delta$}}%
%EndExpansion
\mathfrak{g}^{\mathbf{\mu}}\wedge(\eta^{\mathbf{\gamma\kappa}}\eta
^{\mathbf{\delta\iota}}\epsilon_{\mathbf{\kappa\iota\mu\nu}}\mathfrak{g}%
^{\mathbf{\nu}}), \label{v8}%
\end{align}
from where we get%
\begin{equation}
\frac{\partial\underset{%
%TCIMACRO{\TeXButton{sig}{\sitg}}%
%BeginExpansion
\sitg
%EndExpansion
}{\star}(\mathfrak{g}^{\mathbf{\gamma}}\wedge\mathfrak{g}^{\mathbf{\delta}}%
)}{\partial\mathfrak{g}^{\mathbf{m}}}=\eta^{\mathbf{\gamma\kappa}}%
\eta^{\mathbf{\delta\iota}}\epsilon_{\mathbf{\kappa\iota\mu\nu}}%
\mathfrak{g}^{\mathbf{\nu}}. \label{v9}%
\end{equation}
On the other hand we have
\begin{align}
&  \mathfrak{g}_{\mathbf{\mu}}\lrcorner\underset{%
%TCIMACRO{\TeXButton{sig}{\sitg}}%
%BeginExpansion
\sitg
%EndExpansion
}{\star}(\mathfrak{g}^{\mathbf{\gamma}}\wedge\mathfrak{g}^{\mathbf{\delta}%
})=\mathfrak{g}_{\mathbf{\mu}}\underset{%
%TCIMACRO{\TeXButton{sig}{\sitg}}%
%BeginExpansion
\sitg
%EndExpansion
}{\lrcorner}(\frac{1}{2}\eta^{\mathbf{\gamma\kappa}}\eta^{\mathbf{\delta\iota
}}\epsilon_{\mathbf{\kappa\iota\rho\sigma}}\mathfrak{g}^{\mathbf{\rho}}%
\wedge\mathfrak{g}^{\mathbf{\sigma}})\nonumber\\
&  =\eta^{\mathbf{\gamma\kappa}}\eta^{\mathbf{\delta\iota}}\epsilon
_{\mathbf{\kappa\iota\mu\nu}}\mathfrak{g}^{\mathbf{\nu}}. \label{v10}%
\end{align}
Moreover, using the fourth formula in Eq.(\ref{440new}), we can write
\begin{align}
\frac{\partial\lbrack\underset{%
%TCIMACRO{\TeXButton{sig}{\sitg}}%
%BeginExpansion
\sitg
%EndExpansion
}{\star}(\mathfrak{g}^{\mathbf{\gamma}}\wedge\mathfrak{g}^{\mathbf{\delta}}%
)]}{\partial\mathfrak{g}^{\mathbf{\kappa}}}  &  =\mathfrak{g}_{\mathbf{\kappa
}}\underset{%
%TCIMACRO{\TeXButton{sig}{\sitg}}%
%BeginExpansion
\sitg
%EndExpansion
}{\lrcorner}\underset{%
%TCIMACRO{\TeXButton{sig}{\sitg}}%
%BeginExpansion
\sitg
%EndExpansion
}{\star}(\mathfrak{g}^{\mathbf{\gamma}}\wedge\mathfrak{g}^{\mathbf{\delta}%
})\nonumber\\
&  =\underset{%
%TCIMACRO{\TeXButton{sig}{\sitg}}%
%BeginExpansion
\sitg
%EndExpansion
}{\star}[\mathfrak{g}_{\mathbf{\kappa}}\wedge(\mathfrak{g}^{\mathbf{\gamma}%
}\wedge\mathfrak{g}^{\mathbf{\delta}})]=\underset{%
%TCIMACRO{\TeXButton{sig}{\sitg}}%
%BeginExpansion
\sitg
%EndExpansion
}{\star}(\mathfrak{g}^{\mathbf{\gamma}}\wedge\mathfrak{g}^{\mathbf{\delta}%
}\wedge\mathfrak{g}_{\mathbf{\kappa}}). \label{v11}%
\end{align}
Finally,%
\begin{equation}%
%TCIMACRO{\TeXButton{bdelta}{\mbox{\boldmath{$\delta$}}}}%
%BeginExpansion
\mbox{\boldmath{$\delta$}}%
%EndExpansion
\underset{%
%TCIMACRO{\TeXButton{sig}{\sitg}}%
%BeginExpansion
\sitg
%EndExpansion
}{\star}(\mathfrak{g}^{\mathbf{\gamma}}\wedge\mathfrak{g}^{\mathbf{\delta}})=%
%TCIMACRO{\TeXButton{bdelta}{\mbox{\boldmath{$\delta$}}}}%
%BeginExpansion
\mbox{\boldmath{$\delta$}}%
%EndExpansion
\mathfrak{g}^{\mathbf{\kappa}}\wedge\underset{%
%TCIMACRO{\TeXButton{sig}{\sitg}}%
%BeginExpansion
\sitg
%EndExpansion
}{\star}(\mathfrak{g}^{\mathbf{\gamma}}\wedge\mathfrak{g}^{\mathbf{\delta}%
}\wedge\mathfrak{g}_{\mathbf{\kappa}}). \label{v12}%
\end{equation}
Then using Eq.(\ref{v6}) and Eq.(\ref{v12}) in Eq.(\ref{v5}) we get%
\begin{equation}%
%TCIMACRO{\TeXButton{bdelta}{\mbox{\boldmath{$\delta$}}}}%
%BeginExpansion
\mbox{\boldmath{$\delta$}}%
%EndExpansion
\mathcal{L}_{eh}=\frac{1}{2}d[%
%TCIMACRO{\TeXButton{bdelta}{\mbox{\boldmath{$\delta$}}}}%
%BeginExpansion
\mbox{\boldmath{$\delta$}}%
%EndExpansion
\omega_{\mathbf{\gamma\delta}}\wedge\underset{%
%TCIMACRO{\TeXButton{sig}{\sitg}}%
%BeginExpansion
\sitg
%EndExpansion
}{\star}(\mathfrak{g}^{\mathbf{\gamma}}\wedge\mathfrak{g}^{\mathbf{\delta}})]+%
%TCIMACRO{\TeXButton{bdelta}{\mbox{\boldmath{$\delta$}}}}%
%BeginExpansion
\mbox{\boldmath{$\delta$}}%
%EndExpansion
\mathfrak{g}^{\mathbf{\kappa}}\wedge\lbrack\frac{1}{2}\mathcal{R}%
_{\mathbf{\alpha\beta}}\wedge\underset{%
%TCIMACRO{\TeXButton{sig}{\sitg}}%
%BeginExpansion
\sitg
%EndExpansion
}{\star}(\mathfrak{g}^{\mathbf{\alpha}}\wedge\mathfrak{g}^{\mathbf{\beta}%
}\wedge\mathfrak{g}_{\mathbf{\kappa}})]. \label{v13}%
\end{equation}

Now, the curvature $2$-form fields are given by $\mathcal{R}_{\alpha\beta
}=\frac{1}{2}R_{\mathbf{\alpha\beta\gamma\kappa}}\mathfrak{g}^{\mathbf{\gamma
}}\wedge\mathfrak{g}^{\mathbf{\kappa}}$ where $R_{\mathbf{\alpha\beta
\gamma\kappa}}$ are the components of the Riemann tensor and then%
\begin{align}
\frac{1}{2}\mathcal{R}_{\mathbf{\alpha\beta}}\wedge\underset{%
%TCIMACRO{\TeXButton{sig}{\sitg}}%
%BeginExpansion
\sitg
%EndExpansion
}{\star}(\mathfrak{g}^{\mathbf{\alpha}}\wedge\mathfrak{g}^{\mathbf{\beta}%
}\wedge\mathfrak{g}_{\mathbf{\mu}})  &  =-\frac{1}{2}\underset{%
%TCIMACRO{\TeXButton{sig}{\sitg}}%
%BeginExpansion
\sitg
%EndExpansion
}{\star}[\mathcal{R}_{\mathbf{\alpha\beta}}\underset{%
%TCIMACRO{\TeXButton{sig}{\sitg}}%
%BeginExpansion
\sitg
%EndExpansion
}{\lrcorner}(\mathfrak{g}^{\mathbf{\alpha}}\wedge\mathfrak{g}^{\mathbf{\beta}%
}\wedge\mathfrak{g}_{\mathbf{\mu}})]\nonumber\\
&  =-\frac{1}{4}R_{\mathbf{\alpha\beta\gamma\kappa}}\underset{%
%TCIMACRO{\TeXButton{sig}{\sitg}}%
%BeginExpansion
\sitg
%EndExpansion
}{\star}[(\mathfrak{g}^{\mathbf{\gamma}}\wedge\mathfrak{g}^{\mathbf{\kappa}%
})\underset{%
%TCIMACRO{\TeXButton{sig}{\sitg}}%
%BeginExpansion
\sitg
%EndExpansion
}{\lrcorner}(\mathfrak{g}^{\mathbf{\alpha}}\wedge\mathfrak{g}^{\mathbf{\beta}%
}\wedge\mathfrak{g}_{\mathbf{\mu}})]\nonumber\\
&  =-\star(\mathcal{R}_{\mathbf{\mu}}-\frac{1}{2}R\mathfrak{g}_{\mathbf{\mu}%
})=:-\star\mathcal{G}_{\mathbf{\mu}}, \label{v14}%
\end{align}
where the $\mathcal{R}_{\mathbf{\mu}}$ are the Ricci 1-form fields and
$\mathcal{G}_{\mathbf{\mu}}$ are the Einstein 1-form fields. So, finally, we
have and so we can write
\begin{equation}
\int%
%TCIMACRO{\TeXButton{bdelta}{\mbox{\boldmath{$\delta$}}}}%
%BeginExpansion
\mbox{\boldmath{$\delta$}}%
%EndExpansion
(\mathcal{L}_{eh}+\mathcal{L}_{m})=\int%
%TCIMACRO{\TeXButton{bdelta}{\mbox{\boldmath{$\delta$}}}}%
%BeginExpansion
\mbox{\boldmath{$\delta$}}%
%EndExpansion
\mathfrak{g}^{\mathbf{\alpha}}\wedge(-\star\mathcal{G}_{\mathbf{\alpha}}%
+\frac{\partial\mathfrak{L}_{m}}{\partial\mathfrak{g}^{\mathbf{\alpha}}})=0.
\label{v15}%
\end{equation}

To have the equations of motion in the form of Eq.(\ref{gem}) (here with
$m=0$), i.e.,%

\begin{equation}
d\underset{%
%TCIMACRO{\TeXButton{sig}{\sitg}}%
%BeginExpansion
\sitg
%EndExpansion
}{\star}\mathcal{S}^{\mathbf{\gamma}}+\underset{%
%TCIMACRO{\TeXButton{sig}{\sitg}}%
%BeginExpansion
\sitg
%EndExpansion
}{\star}t^{\gamma}=-\underset{%
%TCIMACRO{\TeXButton{sig}{\sitg}}%
%BeginExpansion
\sitg
%EndExpansion
}{\star}\mathcal{T}^{\gamma}%
\end{equation}
it remains to prove that \
\begin{equation}
d\underset{%
%TCIMACRO{\TeXButton{sig}{\sitg}}%
%BeginExpansion
\sitg
%EndExpansion
}{\star}\mathcal{S}^{\mathbf{\gamma}}+\underset{%
%TCIMACRO{\TeXButton{sig}{\sitg}}%
%BeginExpansion
\sitg
%EndExpansion
}{\star}t^{\mathbf{\gamma}}=-\star\mathcal{G}^{\mathbf{\gamma}}. \label{hs}%
\end{equation}

In order to do that we recall that $\mathfrak{L}_{eh}$ may be written (recall
Eq.(\ref{g9}) and Eq.(\ref{g10'})) as%
\begin{align}
\mathcal{L}_{eh}  &  =\mathcal{L}_{g}-d(\mathfrak{g}^{\mathbf{\alpha}}%
\wedge\underset{%
%TCIMACRO{\TeXButton{sig}{\sitg}}%
%BeginExpansion
\sitg
%EndExpansion
}{\star}d\mathfrak{g}_{\mathbf{\beta}})\nonumber\\
&  =-\frac{1}{2}d\mathfrak{g}^{\mathbf{\alpha}}\wedge\underset{%
%TCIMACRO{\TeXButton{sig}{\sitg}}%
%BeginExpansion
\sitg
%EndExpansion
}{\star}d\mathfrak{g}_{\mathbf{\alpha}}+\frac{1}{2}\underset{%
%TCIMACRO{\TeXButton{sig}{\sitg}}%
%BeginExpansion
\sitg
%EndExpansion
}{\delta}\mathfrak{g}^{\mathbf{\alpha}}\wedge\underset{%
%TCIMACRO{\TeXButton{sig}{\sitg}}%
%BeginExpansion
\sitg
%EndExpansion
}{\star}\underset{%
%TCIMACRO{\TeXButton{sig}{\sitg}}%
%BeginExpansion
\sitg
%EndExpansion
}{\delta}\mathfrak{g}_{\mathbf{\alpha}}+\frac{1}{4}d\mathfrak{g}%
^{\mathbf{\alpha}}\wedge\mathfrak{g}_{\mathbf{\alpha}}\wedge\underset{%
%TCIMACRO{\TeXButton{sig}{\sitg}}%
%BeginExpansion
\sitg
%EndExpansion
}{\star}(d\mathfrak{g}^{\mathbf{\beta}}\wedge\mathfrak{g}_{\mathbf{\beta}%
})-d(\mathfrak{g}^{\mathbf{\alpha}}\wedge\underset{%
%TCIMACRO{\TeXButton{sig}{\sitg}}%
%BeginExpansion
\sitg
%EndExpansion
}{\star}d\mathfrak{g}_{\mathbf{\beta}})\nonumber\\
&  =-\frac{1}{2}d\mathfrak{g}^{\mathbf{\alpha}}\wedge\mathfrak{g}_{\beta
}\wedge\underset{%
%TCIMACRO{\TeXButton{sig}{\sitg}}%
%BeginExpansion
\sitg
%EndExpansion
}{\star}(d\mathfrak{g}^{\mathbf{\beta}}\wedge\mathfrak{g}_{\mathbf{\alpha}%
})+\frac{1}{4}d\mathfrak{g}^{\mathbf{\alpha}}\wedge\mathfrak{g}%
_{\mathbf{\alpha}}\wedge\underset{%
%TCIMACRO{\TeXButton{sig}{\sitg}}%
%BeginExpansion
\sitg
%EndExpansion
}{\star}(d\mathfrak{g}^{\mathbf{\beta}}\wedge\mathfrak{g}_{\mathbf{\beta}%
})-d(\mathfrak{g}^{\mathbf{\alpha}}\wedge\underset{%
%TCIMACRO{\TeXButton{sig}{\sitg}}%
%BeginExpansion
\sitg
%EndExpansion
}{\star}d\mathfrak{g}_{\mathbf{\beta}})\nonumber\\
&  =\frac{1}{2}d\mathfrak{g}_{\mathbf{\alpha}}\wedge\underset{%
%TCIMACRO{\TeXButton{sig}{\sitg}}%
%BeginExpansion
\sitg
%EndExpansion
}{\star}\mathcal{S}^{\mathbf{\alpha}}-d(\mathfrak{g}^{\mathbf{\alpha}}%
\wedge\underset{%
%TCIMACRO{\TeXButton{sig}{\sitg}}%
%BeginExpansion
\sitg
%EndExpansion
}{\star}d\mathfrak{g}_{\mathbf{\beta}}) \label{a1}%
\end{align}

Then%
\begin{align}%
%TCIMACRO{\TeXButton{bdelta}{\mbox{\boldmath{$\delta$}}}}%
%BeginExpansion
\mbox{\boldmath{$\delta$}}%
%EndExpansion
\mathcal{L}_{g}  &  =%
%TCIMACRO{\TeXButton{bdelta}{\mbox{\boldmath{$\delta$}}}}%
%BeginExpansion
\mbox{\boldmath{$\delta$}}%
%EndExpansion
\mathfrak{g}^{\mathbf{\alpha}}\wedge\frac{\partial\mathcal{L}_{g}}%
{\partial\mathfrak{g}^{\mathbf{\alpha}}}+%
%TCIMACRO{\TeXButton{bdelta}{\mbox{\boldmath{$\delta$}}}}%
%BeginExpansion
\mbox{\boldmath{$\delta$}}%
%EndExpansion
d\mathfrak{g}^{\mathbf{\alpha}}\wedge\frac{\partial\mathcal{L}_{g}}{\partial
d\mathfrak{g}^{\mathbf{\alpha}}}\nonumber\\
&
%TCIMACRO{\TeXButton{bdelta}{\mbox{\boldmath{$\delta$}}}}%
%BeginExpansion
\mbox{\boldmath{$\delta$}}%
%EndExpansion
\mathfrak{g}^{\mathbf{\alpha}}\wedge\left[  \frac{\partial\mathcal{L}_{g}%
}{\partial\mathfrak{g}^{\mathbf{\alpha}}}+d\left(  \frac{\partial
\mathcal{L}_{g}}{\partial d\mathfrak{g}^{\mathbf{\alpha}}}\right)  \right]
+d\left(
%TCIMACRO{\TeXButton{bdelta}{\mbox{\boldmath{$\delta$}}}}%
%BeginExpansion
\mbox{\boldmath{$\delta$}}%
%EndExpansion
\mathfrak{g}^{\mathbf{\alpha}}\wedge\frac{\partial\mathcal{L}_{g}}{\partial
d\mathfrak{g}^{\mathbf{\alpha}}}\right) \nonumber\\
&  =%
%TCIMACRO{\TeXButton{bdelta}{\mbox{\boldmath{$\delta$}}}}%
%BeginExpansion
\mbox{\boldmath{$\delta$}}%
%EndExpansion
\mathfrak{g}^{\mathbf{\alpha}}\wedge\left(  \underset{%
%TCIMACRO{\TeXButton{sig}{\sitg}}%
%BeginExpansion
\sitg
%EndExpansion
}{\star}t_{\mathbf{\alpha}}+d\underset{%
%TCIMACRO{\TeXButton{sig}{\sitg}}%
%BeginExpansion
\sitg
%EndExpansion
}{\star}\mathcal{S}_{\mathbf{\alpha}}\right)  +d(%
%TCIMACRO{\TeXButton{bdelta}{\mbox{\boldmath{$\delta$}}}}%
%BeginExpansion
\mbox{\boldmath{$\delta$}}%
%EndExpansion
\mathfrak{g}^{\mathbf{\alpha}}\wedge\underset{%
%TCIMACRO{\TeXButton{sig}{\sitg}}%
%BeginExpansion
\sitg
%EndExpansion
}{\star}\mathcal{S}_{\mathbf{\alpha}}), \label{a2}%
\end{align}
with
\begin{equation}
\underset{%
%TCIMACRO{\TeXButton{sig}{\sitg}}%
%BeginExpansion
\sitg
%EndExpansion
}{\star}t_{\mathbf{\alpha}}=\frac{\partial\mathcal{L}_{g}}{\partial
\mathfrak{g}^{\mathbf{\alpha}}},\text{ \ }\underset{%
%TCIMACRO{\TeXButton{sig}{\sitg}}%
%BeginExpansion
\sitg
%EndExpansion
}{\star}\mathcal{S}_{\mathbf{\alpha}}:=\frac{\partial\mathcal{L}_{g}}{\partial
d\mathfrak{g}^{\mathbf{\alpha}}} \label{a3}%
\end{equation}

To calculate $\frac{\partial\mathfrak{L}_{g}}{\partial\mathfrak{g}%
^{\mathbf{\alpha}}}$ we first recall from Eq.(\ref{1.21}) that we can write
\begin{equation}
\mathfrak{g}_{\mathbf{\kappa}}\underset{%
%TCIMACRO{\TeXButton{sig}{\sitg}}%
%BeginExpansion
\sitg
%EndExpansion
}{\lrcorner}(\frac{1}{2}d\mathfrak{g}_{\mathbf{\alpha}}\wedge\underset{%
%TCIMACRO{\TeXButton{sig}{\sitg}}%
%BeginExpansion
\sitg
%EndExpansion
}{\star}\mathcal{S}^{\mathbf{\alpha}})=\frac{1}{2}(\mathfrak{g}%
_{\mathbf{\kappa}}\underset{%
%TCIMACRO{\TeXButton{sig}{\sitg}}%
%BeginExpansion
\sitg
%EndExpansion
}{\lrcorner}d\mathfrak{g}_{\mathbf{\alpha}})\wedge\underset{%
%TCIMACRO{\TeXButton{sig}{\sitg}}%
%BeginExpansion
\sitg
%EndExpansion
}{\star}\mathcal{S}^{\mathbf{\alpha}}+\frac{1}{2}d\mathfrak{g}_{\mathbf{\alpha
}}\wedge(\mathfrak{g}_{\mathbf{\kappa}}\underset{%
%TCIMACRO{\TeXButton{sig}{\sitg}}%
%BeginExpansion
\sitg
%EndExpansion
}{\lrcorner}\underset{%
%TCIMACRO{\TeXButton{sig}{\sitg}}%
%BeginExpansion
\sitg
%EndExpansion
}{\star}\mathcal{S}^{\mathbf{\alpha}}) \label{a4}%
\end{equation}
and thus since from Eq.(\ref{a1}) it is $\mathfrak{L}_{g}=\frac{1}%
{2}d\mathfrak{g}_{\mathbf{\alpha}}\wedge\underset{%
%TCIMACRO{\TeXButton{sig}{\sitg}}%
%BeginExpansion
\sitg
%EndExpansion
}{\star}\mathcal{S}^{\mathbf{\alpha}}$, we have
\begin{equation}
\mathfrak{g}_{\mathbf{\kappa}}\underset{%
%TCIMACRO{\TeXButton{sig}{\sitg}}%
%BeginExpansion
\sitg
%EndExpansion
}{\lrcorner}\mathfrak{L}_{g}-(\mathfrak{g}_{\mathbf{\kappa}}\underset{%
%TCIMACRO{\TeXButton{sig}{\sitg}}%
%BeginExpansion
\sitg
%EndExpansion
}{\lrcorner}d\mathfrak{g}_{\mathbf{\alpha}})\wedge\underset{%
%TCIMACRO{\TeXButton{sig}{\sitg}}%
%BeginExpansion
\sitg
%EndExpansion
}{\star}\mathcal{S}^{\mathbf{\alpha}}=\frac{1}{2}\left[  d\mathfrak{g}%
_{\mathbf{\alpha}}\wedge(\mathfrak{g}_{\mathbf{\kappa}}\underset{%
%TCIMACRO{\TeXButton{sig}{\sitg}}%
%BeginExpansion
\sitg
%EndExpansion
}{\lrcorner}\underset{%
%TCIMACRO{\TeXButton{sig}{\sitg}}%
%BeginExpansion
\sitg
%EndExpansion
}{\star}\mathcal{S}^{\mathbf{\alpha}})-(\mathfrak{g}_{\mathbf{\kappa}%
}\underset{%
%TCIMACRO{\TeXButton{sig}{\sitg}}%
%BeginExpansion
\sitg
%EndExpansion
}{\lrcorner}d\mathfrak{g}_{\mathbf{\alpha}})\wedge\underset{%
%TCIMACRO{\TeXButton{sig}{\sitg}}%
%BeginExpansion
\sitg
%EndExpansion
}{\star}\mathcal{S}^{\mathbf{\alpha}}\right]  \label{a5}%
\end{equation}

Next we recall that
\begin{align}%
%TCIMACRO{\TeXButton{bdelta}{\mbox{\boldmath{$\delta$}}}}%
%BeginExpansion
\mbox{\boldmath{$\delta$}}%
%EndExpansion
\mathcal{S}^{\mathbf{\alpha}}  &  =\frac{1}{2}\mathcal{S}_{\kappa\iota
}^{\mathbf{\alpha}}%
%TCIMACRO{\TeXButton{bdelta}{\mbox{\boldmath{$\delta$}}}}%
%BeginExpansion
\mbox{\boldmath{$\delta$}}%
%EndExpansion
(\mathfrak{g}^{\mathbf{\kappa}}\wedge\mathfrak{g}^{\iota})=%
%TCIMACRO{\TeXButton{bdelta}{\mbox{\boldmath{$\delta$}}}}%
%BeginExpansion
\mbox{\boldmath{$\delta$}}%
%EndExpansion
\mathfrak{g}^{\mathbf{\kappa}}\wedge(\mathfrak{g}_{\mathbf{\kappa}}\underset{%
%TCIMACRO{\TeXButton{sig}{\sitg}}%
%BeginExpansion
\sitg
%EndExpansion
}{\lrcorner}\mathcal{S}^{\mathbf{\alpha}}),\nonumber\\%
%TCIMACRO{\TeXButton{bdelta}{\mbox{\boldmath{$\delta$}}}}%
%BeginExpansion
\mbox{\boldmath{$\delta$}}%
%EndExpansion
\underset{%
%TCIMACRO{\TeXButton{sig}{\sitg}}%
%BeginExpansion
\sitg
%EndExpansion
}{\star}\mathcal{S}^{\mathbf{\alpha}}  &  =%
%TCIMACRO{\TeXButton{bdelta}{\mbox{\boldmath{$\delta$}}}}%
%BeginExpansion
\mbox{\boldmath{$\delta$}}%
%EndExpansion
\mathfrak{g}^{\mathbf{\kappa}}\wedge(\mathfrak{g}_{\mathbf{\kappa}}\underset{%
%TCIMACRO{\TeXButton{sig}{\sitg}}%
%BeginExpansion
\sitg
%EndExpansion
}{\lrcorner}\underset{%
%TCIMACRO{\TeXButton{sig}{\sitg}}%
%BeginExpansion
\sitg
%EndExpansion
}{\star}\mathcal{S}^{\mathbf{\alpha}}) \label{a66}%
\end{align}
and so we have
\begin{align}
\left[
%TCIMACRO{\TeXButton{bdelta}{\mbox{\boldmath{$\delta$}}}}%
%BeginExpansion
\mbox{\boldmath{$\delta$}}%
%EndExpansion
,\underset{%
%TCIMACRO{\TeXButton{sig}{\sitg}}%
%BeginExpansion
\sitg
%EndExpansion
}{\star}\right]  \mathcal{S}^{\mathbf{\alpha}}  &  =%
%TCIMACRO{\TeXButton{bdelta}{\mbox{\boldmath{$\delta$}}}}%
%BeginExpansion
\mbox{\boldmath{$\delta$}}%
%EndExpansion
\underset{%
%TCIMACRO{\TeXButton{sig}{\sitg}}%
%BeginExpansion
\sitg
%EndExpansion
}{\star}\mathcal{S}^{\mathbf{\alpha}}-\underset{%
%TCIMACRO{\TeXButton{sig}{\sitg}}%
%BeginExpansion
\sitg
%EndExpansion
}{\star}%
%TCIMACRO{\TeXButton{bdelta}{\mbox{\boldmath{$\delta$}}}}%
%BeginExpansion
\mbox{\boldmath{$\delta$}}%
%EndExpansion
\mathcal{S}^{\mathbf{\alpha}}\nonumber\\
&  =%
%TCIMACRO{\TeXButton{bdelta}{\mbox{\boldmath{$\delta$}}}}%
%BeginExpansion
\mbox{\boldmath{$\delta$}}%
%EndExpansion
\mathfrak{g}^{\mathbf{\kappa}}\wedge(\mathfrak{g}_{\mathbf{\kappa}}\underset{%
%TCIMACRO{\TeXButton{sig}{\sitg}}%
%BeginExpansion
\sitg
%EndExpansion
}{\lrcorner}\mathcal{S}^{\mathbf{\alpha}})-%
%TCIMACRO{\TeXButton{bdelta}{\mbox{\boldmath{$\delta$}}}}%
%BeginExpansion
\mbox{\boldmath{$\delta$}}%
%EndExpansion
\mathfrak{g}^{\mathbf{\kappa}}\wedge(\mathfrak{g}_{\mathbf{\kappa}}\underset{%
%TCIMACRO{\TeXButton{sig}{\sitg}}%
%BeginExpansion
\sitg
%EndExpansion
}{\lrcorner}\underset{%
%TCIMACRO{\TeXButton{sig}{\sitg}}%
%BeginExpansion
\sitg
%EndExpansion
}{\star}\mathcal{S}^{\mathbf{\alpha}}). \label{a7}%
\end{align}
Multiplying Eq.(\ref{a7}) on the left by $d\mathfrak{g}_{\mathbf{\alpha}%
}\wedge$ and moreover adding $%
%TCIMACRO{\TeXButton{bdelta}{\mbox{\boldmath{$\delta$}}}}%
%BeginExpansion
\mbox{\boldmath{$\delta$}}%
%EndExpansion
d\mathfrak{g}^{\mathbf{\kappa}}\wedge\underset{%
%TCIMACRO{\TeXButton{sig}{\sitg}}%
%BeginExpansion
\sitg
%EndExpansion
}{\star}\mathcal{S}^{\mathbf{\alpha}}$ to both sides we get
\begin{align*}%
%TCIMACRO{\TeXButton{bdelta}{\mbox{\boldmath{$\delta$}}}}%
%BeginExpansion
\mbox{\boldmath{$\delta$}}%
%EndExpansion
(\frac{1}{2}\mathfrak{g}^{\mathbf{\kappa}}\wedge\underset{%
%TCIMACRO{\TeXButton{sig}{\sitg}}%
%BeginExpansion
\sitg
%EndExpansion
}{\star}\mathcal{S}^{\mathbf{\alpha}})  &  =%
%TCIMACRO{\TeXButton{bdelta}{\mbox{\boldmath{$\delta$}}}}%
%BeginExpansion
\mbox{\boldmath{$\delta$}}%
%EndExpansion
d\mathfrak{g}^{\mathbf{\kappa}}\wedge\frac{1}{2}\underset{%
%TCIMACRO{\TeXButton{sig}{\sitg}}%
%BeginExpansion
\sitg
%EndExpansion
}{\star}\mathcal{S}^{\mathbf{\alpha}}+\frac{1}{2}d\mathfrak{g}^{\mathbf{\kappa
}}\wedge\underset{%
%TCIMACRO{\TeXButton{sig}{\sitg}}%
%BeginExpansion
\sitg
%EndExpansion
}{\star}%
%TCIMACRO{\TeXButton{bdelta}{\mbox{\boldmath{$\delta$}}}}%
%BeginExpansion
\mbox{\boldmath{$\delta$}}%
%EndExpansion
\mathcal{S}^{\mathbf{\alpha}}\\
&  +%
%TCIMACRO{\TeXButton{bdelta}{\mbox{\boldmath{$\delta$}}}}%
%BeginExpansion
\mbox{\boldmath{$\delta$}}%
%EndExpansion
\mathfrak{g}^{\mathbf{\kappa}}\wedge\frac{1}{2}\left(  d\mathfrak{g}%
_{\mathbf{\alpha}}\wedge(\mathfrak{g}_{\mathbf{\kappa}}\underset{%
%TCIMACRO{\TeXButton{sig}{\sitg}}%
%BeginExpansion
\sitg
%EndExpansion
}{\lrcorner}\underset{%
%TCIMACRO{\TeXButton{sig}{\sitg}}%
%BeginExpansion
\sitg
%EndExpansion
}{\star}\mathcal{S}^{\mathbf{\alpha}})-(\mathfrak{g}_{\mathbf{\kappa}%
}\underset{%
%TCIMACRO{\TeXButton{sig}{\sitg}}%
%BeginExpansion
\sitg
%EndExpansion
}{\lrcorner}d\mathfrak{g}_{\mathbf{\alpha}})\wedge\underset{%
%TCIMACRO{\TeXButton{sig}{\sitg}}%
%BeginExpansion
\sitg
%EndExpansion
}{\star}\mathcal{S}^{\mathbf{\alpha}}\right)
\end{align*}
or%
\begin{align}%
%TCIMACRO{\TeXButton{bdelta}{\mbox{\boldmath{$\delta$}}}}%
%BeginExpansion
\mbox{\boldmath{$\delta$}}%
%EndExpansion
\mathcal{L}_{g}  &  =%
%TCIMACRO{\TeXButton{bdelta}{\mbox{\boldmath{$\delta$}}}}%
%BeginExpansion
\mbox{\boldmath{$\delta$}}%
%EndExpansion
d\mathfrak{g}^{\mathbf{\kappa}}\wedge\frac{1}{2}\underset{%
%TCIMACRO{\TeXButton{sig}{\sitg}}%
%BeginExpansion
\sitg
%EndExpansion
}{\star}\mathcal{S}^{\mathbf{\alpha}}+\frac{1}{2}d\mathfrak{g}^{\mathbf{\kappa
}}\wedge\underset{%
%TCIMACRO{\TeXButton{sig}{\sitg}}%
%BeginExpansion
\sitg
%EndExpansion
}{\star}%
%TCIMACRO{\TeXButton{bdelta}{\mbox{\boldmath{$\delta$}}}}%
%BeginExpansion
\mbox{\boldmath{$\delta$}}%
%EndExpansion
\mathcal{S}^{\mathbf{\alpha}}\nonumber\\
&  +%
%TCIMACRO{\TeXButton{bdelta}{\mbox{\boldmath{$\delta$}}}}%
%BeginExpansion
\mbox{\boldmath{$\delta$}}%
%EndExpansion
\mathfrak{g}^{\mathbf{\kappa}}\wedge\frac{1}{2}\left(  d\mathfrak{g}%
_{\mathbf{\alpha}}\wedge(\mathfrak{g}_{\mathbf{\kappa}}\underset{%
%TCIMACRO{\TeXButton{sig}{\sitg}}%
%BeginExpansion
\sitg
%EndExpansion
}{\lrcorner}\underset{%
%TCIMACRO{\TeXButton{sig}{\sitg}}%
%BeginExpansion
\sitg
%EndExpansion
}{\star}\mathcal{S}^{\mathbf{\alpha}})-(\mathfrak{g}_{\mathbf{\kappa}%
}\underset{%
%TCIMACRO{\TeXButton{sig}{\sitg}}%
%BeginExpansion
\sitg
%EndExpansion
}{\lrcorner}d\mathfrak{g}_{\mathbf{\alpha}})\wedge\underset{%
%TCIMACRO{\TeXButton{sig}{\sitg}}%
%BeginExpansion
\sitg
%EndExpansion
}{\star}\mathcal{S}^{\mathbf{\alpha}}\right)  . \label{a8}%
\end{align}
Thus comparing the coefficient of $%
%TCIMACRO{\TeXButton{bdelta}{\mbox{\boldmath{$\delta$}}}}%
%BeginExpansion
\mbox{\boldmath{$\delta$}}%
%EndExpansion
\mathfrak{g}^{\mathbf{\kappa}}$ in Eq.(\ref{a8}) with the one appearing at the
first line in Eq.(\ref{a2}) and taking into account the identity given by
Eq.(\ref{a5}) we get
\begin{equation}%
\begin{tabular}
[c]{l}%
$\underset{%
%TCIMACRO{\TeXButton{sig}{\sitg}}%
%BeginExpansion
\sitg
%EndExpansion
}{\star}t_{\mathbf{\alpha}}=\frac{\partial\mathcal{L}_{g}}{\partial
\mathfrak{g}^{\mathbf{\alpha}}}=\mathfrak{g}_{\mathbf{\alpha}}\underset{%
%TCIMACRO{\TeXButton{sig}{\sitg}}%
%BeginExpansion
\sitg
%EndExpansion
}{\lrcorner}\mathfrak{L}_{g}-(\mathfrak{g}_{\mathbf{\alpha}}\underset{%
%TCIMACRO{\TeXButton{sig}{\sitg}}%
%BeginExpansion
\sitg
%EndExpansion
}{\lrcorner}d\mathfrak{g}^{\mathbf{\kappa}})\wedge\underset{%
%TCIMACRO{\TeXButton{sig}{\sitg}}%
%BeginExpansion
\sitg
%EndExpansion
}{\star}\mathcal{S}_{\mathbf{\kappa}}.$%
\end{tabular}
\ \ \label{special}%
\end{equation}
Also take into account that
\begin{equation}%
\begin{tabular}
[c]{l}%
$\underset{%
%TCIMACRO{\TeXButton{sig}{\sitg}}%
%BeginExpansion
\sitg
%EndExpansion
}{\star}\mathcal{S}_{\mathbf{\alpha}}=\frac{\partial\mathcal{L}_{g}}{\partial
d\mathfrak{g}^{\mathbf{\alpha}}}=-\mathfrak{g}_{\mathbf{\kappa}}%
\wedge\underset{%
%TCIMACRO{\TeXButton{sig}{\sitg}}%
%BeginExpansion
\sitg
%EndExpansion
}{\star}(d\mathfrak{g}^{\mathbf{\kappa}}\wedge\mathfrak{g}_{\mathbf{\alpha}%
})+\frac{1}{2}\mathfrak{g}_{\mathbf{\alpha}}\wedge\underset{%
%TCIMACRO{\TeXButton{sig}{\sitg}}%
%BeginExpansion
\sitg
%EndExpansion
}{\star}(d\mathfrak{g}^{\mathbf{\kappa}}\wedge\mathfrak{g}_{\mathbf{\kappa}%
}),$%
\end{tabular}
\ \ \label{sp'}%
\end{equation}
or%
\begin{equation}
\underset{%
%TCIMACRO{\TeXButton{sig}{\sitg}}%
%BeginExpansion
\sitg
%EndExpansion
}{\star}\mathcal{S}^{\mathbf{\kappa}}=-\underset{%
%TCIMACRO{\TeXButton{sig}{\sitg}}%
%BeginExpansion
\sitg
%EndExpansion
}{\star}d\mathfrak{g}^{\mathbf{\kappa}}-(\mathfrak{g}^{\mathbf{\kappa}%
}\underset{%
%TCIMACRO{\TeXButton{sig}{\sitg}}%
%BeginExpansion
\sitg
%EndExpansion
}{\lrcorner}\underset{%
%TCIMACRO{\TeXButton{sig}{\sitg}}%
%BeginExpansion
\sitg
%EndExpansion
}{\star}\mathfrak{g}^{\mathbf{\alpha}})\wedge\underset{%
%TCIMACRO{\TeXButton{sig}{\sitg}}%
%BeginExpansion
\sitg
%EndExpansion
}{\star}d\underset{%
%TCIMACRO{\TeXButton{sig}{\sitg}}%
%BeginExpansion
\sitg
%EndExpansion
}{\star}\mathfrak{g}_{\mathbf{\alpha}}+\frac{1}{2}\mathfrak{g}^{\mathbf{\kappa
}}\wedge\underset{%
%TCIMACRO{\TeXButton{sig}{\sitg}}%
%BeginExpansion
\sitg
%EndExpansion
}{\star}(d\mathfrak{g}^{\mathbf{\alpha}}\wedge\mathfrak{g}_{\mathbf{\alpha}}).
\label{sp''}%
\end{equation}
Finally comparing the coefficient of $%
%TCIMACRO{\TeXButton{bdelta}{\mbox{\boldmath{$\delta$}}}}%
%BeginExpansion
\mbox{\boldmath{$\delta$}}%
%EndExpansion
\mathfrak{g}^{\mathbf{\kappa}}$in Eq.(\ref{a8}) with the one appearing on
Eq.(\ref{v13}) we get the proof of Eq.(\ref{hs}).\medskip

\textbf{ Remark F1 }Before proceeding recall that in our theory the $\star
t_{\mathbf{\ }}^{\mathbf{c}}$ are $3$-form fields defined directly as
functions of the gravitational potentials $\mathfrak{g}^{\mathbf{\alpha}}$ and
the fields $d\mathfrak{g}^{\mathbf{\alpha}}$ (Eqs.(\ref{7.10.16}) and
(\ref{7.10.17})), where the $\{\mathfrak{g}^{\mathbf{\alpha}}\}$ define by
chance a basis for the $%
%TCIMACRO{\TeXButton{ig}{\itg}}%
%BeginExpansion
\itg
%EndExpansion
$-orthonormal bundle. However, from Remark 6.4 we know the cotetrad fields
$\{\mathfrak{g}^{\alpha}\}$ defining the gravitational potentials \ are
associated to the extensor field $%
%TCIMACRO{\TeXButton{h}{\slh}}%
%BeginExpansion
\slh
%EndExpansion
$ modulus an arbitrary local Lorentz rotation $\Lambda$ which is hidden in the
definition of $%
%TCIMACRO{\TeXButton{itg}{\itg}}%
%BeginExpansion
\itg
%EndExpansion
=%
%TCIMACRO{\TeXButton{h}{\slh}}%
%BeginExpansion
\slh
%EndExpansion
^{\dagger}\eta%
%TCIMACRO{\TeXButton{h}{\slh}}%
%BeginExpansion
\slh
%EndExpansion
=%
%TCIMACRO{\TeXButton{h}{\slh}}%
%BeginExpansion
\slh
%EndExpansion
^{\dagger}\Lambda^{\dagger}\eta\Lambda%
%TCIMACRO{\TeXButton{h}{\slh}}%
%BeginExpansion
\slh
%EndExpansion
$. Thus the energy-momentum 3-forms $\underset{%
%TCIMACRO{\TeXButton{sig}{\sitg}}%
%BeginExpansion
\sitg
%EndExpansion
}{\star}t^{\mathbf{\gamma}}$ are gauge dependent. However, once a gauge is
chosen the $\star t_{\mathbf{\ }}^{\mathbf{\gamma}}$ are, of course,
independent on the basis (coordinate or otherwise) where they are expressed,
i.e., they are legitimate tensor fields. However, in \textit{GRT }(where the
$\underset{%
%TCIMACRO{\TeXButton{sig}{\sitg}}%
%BeginExpansion
\sitg
%EndExpansion
}{\star}\mathcal{S}^{\gamma}:U\rightarrow\bigwedge\nolimits^{2}U$ are called
\textit{superpotentials}\footnote{Those objects are called in \cite{szabados},
the Sparling forms \cite{sparling}. However, they was already used much
earlier by Thirring and Wallner in \cite{tw78}.}) and the $\star
t_{\mathbf{\ }}^{\mathbf{\gamma}}$ are called the \textit{gravitational
energy-momentum pseudo }$3$\textit{-forms}) this is not the case The reason
has to do with the name \textit{pseudo }$3$\textit{- forms}. for\textit{
}$\star t_{\mathbf{\ }}^{\mathbf{\gamma}}$. Indeed, in \textit{GRT} those
objects are directly associated to the connection $1$-forms $\omega
_{\mathbf{\alpha\beta}}$ associated to a particular section of the $%
%TCIMACRO{\TeXButton{g}{\slg}}%
%BeginExpansion
\slg
%EndExpansion
$-orthonormal bundle and thus besides being gauge dependent, their components
in any basis do not define a true tensor field.

Taking into account Remark F1 we provide yet alternative expressions for
$\underset{%
%TCIMACRO{\TeXButton{sig}{\sitg}}%
%BeginExpansion
\sitg
%EndExpansion
}{\star}t_{\mathbf{\ }}^{\mathbf{\gamma}}$ and $\underset{%
%TCIMACRO{\TeXButton{sig}{\sitg}}%
%BeginExpansion
\sitg
%EndExpansion
}{\star}\mathcal{S}^{\mathbf{\gamma}}$ given by Eqs.(\ref{special}) and
(\ref{sp'}) as
\begin{align}
\underset{%
%TCIMACRO{\TeXButton{sig}{\sitg}}%
%BeginExpansion
\sitg
%EndExpansion
}{\star}t_{\mathbf{\ }}^{\mathbf{\gamma}}  &  =-\frac{1}{2}\omega
_{\mathbf{\alpha\beta}}\wedge\lbrack\omega_{\mathbf{\delta}}^{\mathbf{\gamma}%
}\wedge\underset{%
%TCIMACRO{\TeXButton{sig}{\sitg}}%
%BeginExpansion
\sitg
%EndExpansion
}{\star}(\mathfrak{g}^{\mathbf{\alpha}}\wedge\mathfrak{g}^{\mathbf{\beta}%
}\wedge\mathfrak{g}^{\mathbf{\delta}})+\omega_{\mathbf{\delta}}^{\mathbf{\beta
}}\wedge\underset{%
%TCIMACRO{\TeXButton{sig}{\sitg}}%
%BeginExpansion
\sitg
%EndExpansion
}{\star}(\mathfrak{g}^{\mathbf{\alpha}}\wedge\mathfrak{g}^{\mathbf{\beta}%
}\wedge\mathfrak{g}^{\mathbf{\gamma}})],\nonumber\\
\underset{%
%TCIMACRO{\TeXButton{sig}{\sitg}}%
%BeginExpansion
\sitg
%EndExpansion
}{\star}\mathcal{S}^{\mathbf{\gamma}}  &  =\frac{1}{2}\omega_{\mathbf{\alpha
\beta}}\wedge\underset{%
%TCIMACRO{\TeXButton{sig}{\sitg}}%
%BeginExpansion
\sitg
%EndExpansion
}{\star}(\mathfrak{g}^{\mathbf{\alpha}}\wedge\mathfrak{g}^{\mathbf{\beta}%
}\wedge\mathfrak{g}^{\gamma}), \label{19a}%
\end{align}
where we use Eq.(\ref{w}) for packing part of complicated formulas involving
$\mathfrak{g}^{\mathbf{\alpha}}$ and $d\mathfrak{g}^{\mathbf{\alpha}}$ in the
form of $\omega_{\mathbf{\alpha\beta}}$.\ With Eqs.(\ref{19a}) we can provide
a simple proof of Eq.(\ref{hs}). Indeed, let us compute$\ -2\underset{%
%TCIMACRO{\TeXButton{sig}{\sitg}}%
%BeginExpansion
\sitg
%EndExpansion
}{\star}\mathcal{G}^{\mathbf{\delta}}$ using Eq.(\ref{v14}) and Cartan's
second structure equation.%
\begin{align*}
-2\underset{%
%TCIMACRO{\TeXButton{sig}{\sitg}}%
%BeginExpansion
\sitg
%EndExpansion
}{\star}\mathcal{G}^{\mathbf{\delta}}  &  =d%
%TCIMACRO{\TeXButton{omega}{\mbox{\boldmath{$\omega$}}}}%
%BeginExpansion
\mbox{\boldmath{$\omega$}}%
%EndExpansion
_{\mathbf{\alpha\beta}}\wedge\underset{%
%TCIMACRO{\TeXButton{sig}{\sitg}}%
%BeginExpansion
\sitg
%EndExpansion
}{\star}(\mathfrak{g}^{\mathbf{\alpha}}\wedge\mathfrak{g}^{\mathbf{\beta}%
}\wedge\mathfrak{g}^{\mathbf{\delta}})+%
%TCIMACRO{\TeXButton{omega}{\mbox{\boldmath{$\omega$}}}}%
%BeginExpansion
\mbox{\boldmath{$\omega$}}%
%EndExpansion
_{\mathbf{\alpha\gamma}}\wedge%
%TCIMACRO{\TeXButton{omega}{\mbox{\boldmath{$\omega$}}}}%
%BeginExpansion
\mbox{\boldmath{$\omega$}}%
%EndExpansion
_{\mathbf{\beta}}^{\mathbf{\gamma}}\wedge\underset{%
%TCIMACRO{\TeXButton{sig}{\sitg}}%
%BeginExpansion
\sitg
%EndExpansion
}{\star}(\mathfrak{g}^{\mathbf{\alpha}}\wedge\mathfrak{g}^{\mathbf{\beta}%
}\wedge\mathfrak{g}^{\mathbf{\delta}})\\
&  =d[%
%TCIMACRO{\TeXButton{omega}{\mbox{\boldmath{$\omega$}}}}%
%BeginExpansion
\mbox{\boldmath{$\omega$}}%
%EndExpansion
_{\mathbf{\alpha\beta}}\wedge\underset{%
%TCIMACRO{\TeXButton{sig}{\sitg}}%
%BeginExpansion
\sitg
%EndExpansion
}{\star}(\mathfrak{g}^{\mathbf{\alpha}}\wedge\mathfrak{g}^{\mathbf{\beta}%
}\wedge\mathfrak{g}^{\mathbf{\delta}})]+%
%TCIMACRO{\TeXButton{omega}{\mbox{\boldmath{$\omega$}}}}%
%BeginExpansion
\mbox{\boldmath{$\omega$}}%
%EndExpansion
_{\mathbf{\alpha\beta}}\wedge d\underset{%
%TCIMACRO{\TeXButton{sig}{\sitg}}%
%BeginExpansion
\sitg
%EndExpansion
}{\star}(\mathfrak{g}^{\mathbf{\alpha}}\wedge\mathfrak{g}^{\mathbf{\beta}%
}\wedge\mathfrak{g}^{\mathbf{\delta}})\\
&  +%
%TCIMACRO{\TeXButton{omega}{\mbox{\boldmath{$\omega$}}}}%
%BeginExpansion
\mbox{\boldmath{$\omega$}}%
%EndExpansion
_{\mathbf{\alpha\gamma}}\wedge%
%TCIMACRO{\TeXButton{omega}{\mbox{\boldmath{$\omega$}}}}%
%BeginExpansion
\mbox{\boldmath{$\omega$}}%
%EndExpansion
_{\mathbf{\beta}}^{\mathbf{\gamma}}\wedge\underset{%
%TCIMACRO{\TeXButton{sig}{\sitg}}%
%BeginExpansion
\sitg
%EndExpansion
}{\star}(\mathfrak{g}^{\mathbf{\alpha}}\wedge\mathfrak{g}^{\mathbf{\beta}%
}\wedge\mathfrak{g}^{\mathbf{\delta}})
\end{align*}

\begin{align}
&  =d[%
%TCIMACRO{\TeXButton{omega}{\mbox{\boldmath{$\omega$}}}}%
%BeginExpansion
\mbox{\boldmath{$\omega$}}%
%EndExpansion
_{\mathbf{\alpha\beta}}\wedge\underset{%
%TCIMACRO{\TeXButton{sig}{\sitg}}%
%BeginExpansion
\sitg
%EndExpansion
}{\star}(\mathfrak{g}^{\mathbf{\alpha}}\wedge\mathfrak{g}^{\mathbf{\beta}%
}\wedge\mathfrak{g}^{\mathbf{\delta}})]-%
%TCIMACRO{\TeXButton{omega}{\mbox{\boldmath{$\omega$}}}}%
%BeginExpansion
\mbox{\boldmath{$\omega$}}%
%EndExpansion
_{\mathbf{\alpha\beta}}\wedge%
%TCIMACRO{\TeXButton{omega}{\mbox{\boldmath{$\omega$}}}}%
%BeginExpansion
\mbox{\boldmath{$\omega$}}%
%EndExpansion
_{\mathbf{\rho}}^{\mathbf{\alpha}}\wedge\underset{%
%TCIMACRO{\TeXButton{sig}{\sitg}}%
%BeginExpansion
\sitg
%EndExpansion
}{\star}(\mathfrak{g}^{\mathbf{\rho}}\wedge\mathfrak{g}^{\mathbf{\beta}}%
\wedge\mathfrak{g}^{\mathbf{\delta}})\nonumber\\
&  -%
%TCIMACRO{\TeXButton{omega}{\mbox{\boldmath{$\omega$}}}}%
%BeginExpansion
\mbox{\boldmath{$\omega$}}%
%EndExpansion
_{\mathbf{\alpha\beta}}\wedge%
%TCIMACRO{\TeXButton{omega}{\mbox{\boldmath{$\omega$}}}}%
%BeginExpansion
\mbox{\boldmath{$\omega$}}%
%EndExpansion
_{\mathbf{\rho}}^{\mathbf{\beta}}\wedge\underset{%
%TCIMACRO{\TeXButton{sig}{\sitg}}%
%BeginExpansion
\sitg
%EndExpansion
}{\star}(\mathfrak{g}^{\mathbf{\alpha}}\wedge\mathfrak{g}^{\mathbf{\rho}%
}\wedge\mathfrak{g}^{\mathbf{\delta}})-%
%TCIMACRO{\TeXButton{omega}{\mbox{\boldmath{$\omega$}}}}%
%BeginExpansion
\mbox{\boldmath{$\omega$}}%
%EndExpansion
_{\mathbf{\alpha\beta}}\wedge%
%TCIMACRO{\TeXButton{omega}{\mbox{\boldmath{$\omega$}}}}%
%BeginExpansion
\mbox{\boldmath{$\omega$}}%
%EndExpansion
_{\mathbf{\rho}}^{\mathbf{\delta}}\wedge\underset{%
%TCIMACRO{\TeXButton{sig}{\sitg}}%
%BeginExpansion
\sitg
%EndExpansion
}{\star}(\mathfrak{g}^{\mathbf{\alpha}}\wedge\mathfrak{g}^{\mathbf{\beta}%
}\wedge\mathfrak{g}^{\mathbf{\rho}})\nonumber\\
&  +%
%TCIMACRO{\TeXButton{omega}{\mbox{\boldmath{$\omega$}}}}%
%BeginExpansion
\mbox{\boldmath{$\omega$}}%
%EndExpansion
_{\mathbf{\alpha\gamma}}\wedge%
%TCIMACRO{\TeXButton{omega}{\mbox{\boldmath{$\omega$}}}}%
%BeginExpansion
\mbox{\boldmath{$\omega$}}%
%EndExpansion
_{\mathbf{\beta}}^{\mathbf{\gamma}}\wedge\underset{%
%TCIMACRO{\TeXButton{sig}{\sitg}}%
%BeginExpansion
\sitg
%EndExpansion
}{\star}(\mathfrak{g}^{\mathbf{\alpha}}\wedge\mathfrak{g}^{\mathbf{\beta}%
}\wedge\mathfrak{g}^{\mathbf{\delta}})\nonumber\\
&  =d[%
%TCIMACRO{\TeXButton{omega}{\mbox{\boldmath{$\omega$}}}}%
%BeginExpansion
\mbox{\boldmath{$\omega$}}%
%EndExpansion
_{\mathbf{\alpha\beta}}\wedge\underset{%
%TCIMACRO{\TeXButton{sig}{\sitg}}%
%BeginExpansion
\sitg
%EndExpansion
}{\star}(\mathfrak{g}^{\mathbf{\alpha}}\wedge\mathfrak{g}^{\mathbf{\beta}%
}\wedge\mathfrak{g}^{\mathbf{\delta}})]-%
%TCIMACRO{\TeXButton{omega}{\mbox{\boldmath{$\omega$}}}}%
%BeginExpansion
\mbox{\boldmath{$\omega$}}%
%EndExpansion
_{\mathbf{\alpha\beta}}\wedge\lbrack%
%TCIMACRO{\TeXButton{omega}{\mbox{\boldmath{$\omega$}}}}%
%BeginExpansion
\mbox{\boldmath{$\omega$}}%
%EndExpansion
_{\mathbf{\rho}}^{\mathbf{\delta}}\wedge\underset{%
%TCIMACRO{\TeXButton{sig}{\sitg}}%
%BeginExpansion
\sitg
%EndExpansion
}{\star}(\mathfrak{g}^{\mathbf{\alpha}}\wedge\mathfrak{g}^{\mathbf{\beta}%
}\wedge\mathfrak{g}^{\mathbf{\rho}})\nonumber\\
&  +%
%TCIMACRO{\TeXButton{omega}{\mbox{\boldmath{$\omega$}}}}%
%BeginExpansion
\mbox{\boldmath{$\omega$}}%
%EndExpansion
_{\mathbf{\rho}}^{\mathbf{\beta}}\wedge\underset{%
%TCIMACRO{\TeXButton{sig}{\sitg}}%
%BeginExpansion
\sitg
%EndExpansion
}{\star}(\mathfrak{g}^{\mathbf{\alpha}}\wedge\mathfrak{g}^{\mathbf{\rho}%
}\wedge\mathfrak{g}^{\mathbf{\delta}})]\nonumber\\
&  =2(d\underset{%
%TCIMACRO{\TeXButton{sig}{\sitg}}%
%BeginExpansion
\sitg
%EndExpansion
}{\star}\mathcal{S}^{\mathbf{\delta}}+\underset{%
%TCIMACRO{\TeXButton{sig}{\sitg}}%
%BeginExpansion
\sitg
%EndExpansion
}{\star}t^{\mathbf{\delta}}). \label{7.10.20}%
\end{align}

\section{A Comment on the Gauge Theory of Gravitation of References
\cite{dola} and \cite{hestenes2005}}

In footnote 1 we recalled that in \textit{GRT} a gravitational field is
defined by an equivalence class of pentuples, where $(M,%
%TCIMACRO{\TeXButton{g}{\slg}}%
%BeginExpansion
\slg
%EndExpansion
,D,\tau_{%
%TCIMACRO{\TeXButton{g}{\sslg}}%
%BeginExpansion
\sslg
%EndExpansion
},\uparrow)$ \ and $(M^{\prime},%
%TCIMACRO{\TeXButton{g}{\slg}}%
%BeginExpansion
\slg
%EndExpansion
^{\prime},D^{\prime},\tau_{%
%TCIMACRO{\TeXButton{g}{\sslg}}%
%BeginExpansion
\sslg
%EndExpansion
}^{\prime},\uparrow^{\prime})$ are said equivalent if there is a
diffeomorphism $f:M\rightarrow M^{\prime}$, such that $%
%TCIMACRO{\TeXButton{g}{\slg}}%
%BeginExpansion
\slg
%EndExpansion
^{\prime}=f^{\ast}%
%TCIMACRO{\TeXButton{g}{\slg}}%
%BeginExpansion
\slg
%EndExpansion
$, $D^{\prime}=f^{\ast}D$, $\tau_{%
%TCIMACRO{\TeXButton{g}{\sslg}}%
%BeginExpansion
\sslg
%EndExpansion
}^{\prime}=f^{\ast}\tau_{%
%TCIMACRO{\TeXButton{g}{\sslg}}%
%BeginExpansion
\sslg
%EndExpansion
},\uparrow^{\prime}=f^{\ast}\uparrow,$ (where $f^{\ast}$ here denotes the
pullback mapping). For more details, see, e.g., \cite{sawu,
rodrosa,rodcap2007}. Moreover, in \textit{GRT} when one is studying the
coupling of the matter fields, represented, say by some sections of the
exterior algebra bundle\footnote{If the exterior algebra bundle $%
%TCIMACRO{\dbigwedge }%
%BeginExpansion
{\displaystyle\bigwedge}
%EndExpansion
T^{\ast}M$ is viewed as embedded in the Clifford algebra bundle
$\mathcal{C\ell(}M,\mathtt{g})$ ($%
%TCIMACRO{\dbigwedge }%
%BeginExpansion
{\displaystyle\bigwedge}
%EndExpansion
T^{\ast}M\hookrightarrow\mathcal{C\ell(}M,\mathtt{g})$) then even spinor
fields can be represented in that formalism as some appropriate equivalence
classes of sums of even non homegeneous differential forms. Details may be
found in \cite{rodcap2007}.} $\psi_{1,...,}\psi_{n}$, which satisfy certain
coupled differential equations (\textit{the equations of motion}\footnote{The
equations of motion, for any particular problem satisfy appropriate initial
and boundary conditions.}), two models \ $\{(M,%
%TCIMACRO{\TeXButton{g}{\slg}}%
%BeginExpansion
\slg
%EndExpansion
,D,\tau_{%
%TCIMACRO{\TeXButton{g}{\sslg}}%
%BeginExpansion
\sslg
%EndExpansion
},\uparrow),\psi_{1,...,}\psi_{n}\}$ \ and $(M^{\prime},%
%TCIMACRO{\TeXButton{g}{\slg}}%
%BeginExpansion
\slg
%EndExpansion
^{\prime},D^{\prime},\tau_{%
%TCIMACRO{\TeXButton{g}{\sslg}}%
%BeginExpansion
\sslg
%EndExpansion
}^{\prime},\uparrow^{\prime}),\psi_{1,...,}^{\prime}\psi_{n}^{\prime}\}$ are
said to be dynamically equivalent iff there is a diffeomorphism $\mathtt{h}%
:M\rightarrow M^{\prime}$, such that $(M,%
%TCIMACRO{\TeXButton{g}{\slg}}%
%BeginExpansion
\slg
%EndExpansion
,D,\tau_{%
%TCIMACRO{\TeXButton{g}{\sslg}}%
%BeginExpansion
\sslg
%EndExpansion
},\uparrow)$ \ and $(M^{\prime},%
%TCIMACRO{\TeXButton{g}{\slg}}%
%BeginExpansion
\slg
%EndExpansion
^{\prime},D^{\prime},\tau_{%
%TCIMACRO{\TeXButton{g}{\sslg}}%
%BeginExpansion
\sslg
%EndExpansion
}^{\prime},\uparrow^{\prime})$ and moreover $\psi_{i}^{\prime}=f^{\ast}%
\psi_{i}$, $i=1,...,n$.

This kind of equivalence is not a particularity of \textit{GRT}, it is an
obvious mathematical requirement that any theory formulated with tensor fields
living in an arbitrary manifold must satisfy\footnote{Some people now call
this property background independence, although \ we do not think that o be a
very appropriate name. Others call this property general
covariance.\ Eventually it is most aprropriate to simply say that the theory
is invariant under diffeomorphisms. Moreover, we recall that there is a
mathematical theorem saying that the \textit{action functional} for the matter
fields \ represented by sections of the exterior algebra bundle (plus the
gravitational field)\ is invariant under the action of one-parameter group of
diffeormorphisms\ for Lagrangians that vanish on the boundary of the region
where the action functional is defined. Details, e.g., in \cite{rodcap2007}.}.
This fact is sometimes confused with the fact that for particular theory, say
$\mathfrak{T}$ it may happen that there are diffeomorphisms $f:M\rightarrow M$
such that for a given model $\{(M,%
%TCIMACRO{\TeXButton{g}{\slg}}%
%BeginExpansion
\slg
%EndExpansion
,D,\tau_{%
%TCIMACRO{\TeXButton{g}{\sslg}}%
%BeginExpansion
\sslg
%EndExpansion
},\uparrow),\psi_{1,...,}\psi_{n}\}$ it is $%
%TCIMACRO{\TeXButton{g}{\slg}}%
%BeginExpansion
\slg
%EndExpansion
f^{\ast}%
%TCIMACRO{\TeXButton{g}{\slg}}%
%BeginExpansion
\slg
%EndExpansion
$, $D^{\prime}=f^{\ast}D$, $\tau_{%
%TCIMACRO{\TeXButton{g}{\sslg}}%
%BeginExpansion
\sslg
%EndExpansion
}^{\prime}=f^{\ast}\tau_{%
%TCIMACRO{\TeXButton{g}{\sslg}}%
%BeginExpansion
\sslg
%EndExpansion
},\uparrow^{\prime}=f^{\ast}\uparrow$, but $\psi_{i}^{\prime}\neq f^{\ast}%
\psi_{i}$. When this happen the set of diffeomorphisms satisfying that
property defines a group, called a symmetry group of $\mathfrak{T}$, and of
course, knowing the possible symmetry groups of a given theory facilitates the
finding of solutions for the equations of motion \footnote{See some examples,
e.g., in Chapter 5 of \cite{rodcap2007}.}.

Having saying that, we now analyze first the motivations for the `gauge theory
of the gravitational field' in Minkowski spacetime of references \cite{dola}
and \cite{hestenes2005} formulated on a vector manifold $U$ \footnote{The
vector manifod $U$ is as defined in Chapter 5, but with the elements of $U$
being vectors instead of forms, and the elements of $%
%TCIMACRO{\dbigwedge }%
%BeginExpansion
{\displaystyle\bigwedge}
%EndExpansion
TU$ being multivector fields instead of multiform fields.}. Those authors
taking advantage of the obvious identification of the tangent spaces of $U$
write that for a diffeomorphism $f:U\rightarrow U$, $x\mapsto f(x)$ we have
some particularly useful representations for the pullback mapping ($f^{\ast},$
$f_{x^{\prime}}^{\ast}:T_{x^{\prime}}^{\ast}U\rightarrow T_{x}^{\ast}U$ and
pushforward mapping $\mathtt{(}f_{\ast},$ $f_{\ast_{x}}:T_{x}U\rightarrow
T_{x^{\prime}}U)$. Indeed, for representing the pushforward mapping
\cite{dola} and \cite{hestenes2005} introduce the operator $\mathtt{f}$ such
that%
\begin{equation}%
\begin{array}
[c]{ccccc}%
\mathtt{f} & : & \sec%
%TCIMACRO{\dbigwedge \nolimits^{1}}%
%BeginExpansion
{\displaystyle\bigwedge\nolimits^{1}}
%EndExpansion
TU & \rightarrow & \sec%
%TCIMACRO{\dbigwedge \nolimits^{1}}%
%BeginExpansion
{\displaystyle\bigwedge\nolimits^{1}}
%EndExpansion
TU,\\
&  & a & \mapsto & \mathtt{f}(a)=a\underset{\eta}{\cdot}\partial_{x}f
\end{array}
\label{l0}%
\end{equation}
where%
\begin{equation}
\partial_{x}=\gamma^{\mu}\frac{\partial}{\partial\mathtt{x}^{\mu}},
\label{l01}%
\end{equation}
with \{$\mathtt{x}^{\mu}$\} coordinates for $U$ in the
Einstein-Lorentz-Poincar\'{e} gauge and $\{\gamma^{\mu}\}$ the dual basis of
the basis $\{\vartheta_{\mu}\}$, as introduced in Chapter 5\footnote{The
reciprocal basis of $\{\gamma^{\mu}\}$ is the coordinate basis $\{\gamma_{\mu
}\}$ such that $\gamma_{\mu}\underset{\eta}{\cdot}\gamma^{\nu}=\delta_{\mu
}^{\nu}$.}. For representing the pullback mapping \cite{dola} and
\cite{hestenes2005} introduce the operator $\mathtt{f}^{\dagger}$ such
that\footnote{Recall that $\underline{\mathtt{f}}^{\dagger}$ is the extended
of \texttt{f.}}%
\begin{align}
\underline{\mathtt{f}}^{\dagger}  &  :\sec%
%TCIMACRO{\dbigwedge }%
%BeginExpansion
{\displaystyle\bigwedge}
%EndExpansion
T_{x}U\rightarrow\sec%
%TCIMACRO{\dbigwedge }%
%BeginExpansion
{\displaystyle\bigwedge}
%EndExpansion
T_{x}U,\nonumber\\
\psi_{i}(x)  &  \mapsto\psi_{i}^{\prime}(x)=(\underline{\mathtt{f}}^{\dagger
}\psi_{i})(x)=\psi_{i}(f(x)) \label{l1}%
\end{align}
Next those authors argue that the description of physics $\psi_{i}$ or
$\psi_{i}^{^{\prime}}$ is not \textit{covariant. }To sane this \ `deficiency'
they propose to introduce a gauge field $%
%TCIMACRO{\TeXButton{h}{\slh}}%
%BeginExpansion
\slh
%EndExpansion
:\sec%
%TCIMACRO{\dbigwedge \nolimits^{1}}%
%BeginExpansion
{\displaystyle\bigwedge\nolimits^{1}}
%EndExpansion
TU\rightarrow\sec%
%TCIMACRO{\dbigwedge \nolimits^{1}}%
%BeginExpansion
{\displaystyle\bigwedge\nolimits^{1}}
%EndExpansion
TU$ such that,
\begin{align}
\psi_{i}(x)  &  =%
%TCIMACRO{\TeXButton{h}{\slh}}%
%BeginExpansion
\slh
%EndExpansion
^{\dagger}[\Psi_{i}(x)],\nonumber\\
\psi_{i}^{\prime}(x)  &  =%
%TCIMACRO{\TeXButton{h}{\slh}}%
%BeginExpansion
\slh
%EndExpansion
^{\dagger\prime}[\Psi_{i}(x)], \label{l2}%
\end{align}
where $\Psi_{i}(x)$ is said to be the covariant description of $\psi_{i}(x)$.
This implies in the following transformation law for $%
%TCIMACRO{\TeXButton{h}{\slh}}%
%BeginExpansion
\slh
%EndExpansion
$ under a diffeomorphism $f:U\rightarrow U$,%
\begin{equation}%
%TCIMACRO{\TeXButton{h}{\slh}}%
%BeginExpansion
\slh
%EndExpansion
^{\dagger\prime}=\underline{\mathtt{f}}^{\dagger}%
%TCIMACRO{\TeXButton{h}{\slh}}%
%BeginExpansion
\slh
%EndExpansion
^{\dagger}. \label{l3}%
\end{equation}
From this point, using heuristic arguments, those authors say that this
permits the introduction of a new metric extensor field on $U$, that they
write as%
\begin{equation}%
%TCIMACRO{\TeXButton{itg}{\itg}}%
%BeginExpansion
\itg
%EndExpansion
=%
%TCIMACRO{\TeXButton{h}{\slh}}%
%BeginExpansion
\slh
%EndExpansion
^{\dagger}%
%TCIMACRO{\TeXButton{h}{\slh}}%
%BeginExpansion
\slh
%EndExpansion
, \label{l4}%
\end{equation}
because differently from what we did in this paper they choose as basic
algebra for performing calculations $\mathcal{C}\ell(U,\eta$), instead of
$\mathcal{C}\ell(U,\cdot$). Take notice also that in \cite{dola,hestenes2005}
it is $%
%TCIMACRO{\TeXButton{h}{\slh}}%
%BeginExpansion
\slh
%EndExpansion
=h^{-1}$.

The reasoning behind Eqs.(\ref{l2}) is that the field $\Psi_{i}\in\sec%
%TCIMACRO{\dbigwedge }%
%BeginExpansion
{\displaystyle\bigwedge}
%EndExpansion
TU$ has been selected as a representative of the class of the
diffeomorphically equivalent fields $\{\underline{\mathtt{f}}^{\dagger}%
\psi_{i}\}$, for $f\in$\textrm{Diff}$U$, the diffeomorphism group of $U$. This
is a very important point to take in mind. Indeed, in \cite{dola} and
\cite{hestenes2005} \ in a region $V\subset U$ the gravitational potentials
(called \textit{displacement gauge invariant frame }$\{g^{\mu}\}$) are
introduced as follows.\ First, take arbitrary coordinates $\{x^{\mu}\}$
covering $V$ and introduce the coordinate vector fields%
\begin{equation}
e_{\mu}=\frac{\partial x}{\partial x^{\mu}}=\partial_{\mu}x, \label{l5}%
\end{equation}
where the position vector $x=\mathtt{x}^{\mu}(x^{\alpha})\gamma_{\alpha}$ and
the reciprocal basis of $\{e_{\mu}\}$ is the basis $\{e^{\mu}\}$ such
that\footnote{Recall that $\partial_{x}=\gamma^{\mu}\frac{\partial}%
{\partial\mathtt{x}^{\mu}}=e^{\mu}\frac{\partial}{\partial x^{\mu}}$, with
$e^{\mu}=\gamma^{\alpha}\frac{\partial x^{\mu}}{\partial\mathtt{x}^{\alpha}}%
$.}
\begin{equation}
e^{\mu}=\partial_{x}x^{\mu}. \label{l6}%
\end{equation}

Then,%
\begin{equation}
g_{\mu}=%
%TCIMACRO{\TeXButton{h}{\slh}}%
%BeginExpansion
\slh
%EndExpansion
^{\dagger}(e_{\mu})=h_{\mu}^{\alpha}e_{\alpha}, \label{l7}%
\end{equation}%
\begin{equation}
g^{\mu}=%
%TCIMACRO{\TeXButton{h}{\slh}}%
%BeginExpansion
\slh
%EndExpansion
^{-1}(e^{\mu})=(%
%TCIMACRO{\TeXButton{h}{\slh}}%
%BeginExpansion
\slh
%EndExpansion
^{-1}\partial_{x})x^{\mu} \label{l8}%
\end{equation}
and
\begin{equation}%
%TCIMACRO{\TeXButton{h}{\slh}}%
%BeginExpansion
\slh
%EndExpansion
^{\dagger}(e_{\mu})\underset{\eta}{\cdot}%
%TCIMACRO{\TeXButton{h}{\slh}}%
%BeginExpansion
\slh
%EndExpansion
^{\dagger}(e_{\nu})=%
%TCIMACRO{\TeXButton{itg}{\itg}}%
%BeginExpansion
\itg
%EndExpansion
(e_{\mu})\underset{\eta}{\cdot}e_{\nu}=g_{\mu\nu}, \label{l9}%
\end{equation}%
\begin{equation}%
%TCIMACRO{\TeXButton{h}{\slh}}%
%BeginExpansion
\slh
%EndExpansion
^{-1}(e^{\mu})\underset{\eta}{\cdot}%
%TCIMACRO{\TeXButton{h}{\slh}}%
%BeginExpansion
\slh
%EndExpansion
^{-1}(e^{\nu})=%
%TCIMACRO{\TeXButton{itg}{\itg}}%
%BeginExpansion
\itg
%EndExpansion
^{-1}(e^{\mu})\underset{\eta}{\cdot}e^{\nu}=g^{\mu\nu}. \label{l10}%
\end{equation}

Form its definition the commutator $[e_{\mu},e_{\nu}]=e_{\mu}\underset{\eta
}{\cdot}\partial_{x}e_{\nu}-e_{\nu}\underset{\eta}{\cdot}\partial_{x}e_{\mu
}=0$, but in general%
\begin{align}
\lbrack g_{\mu},g_{\nu}]  &  =[%
%TCIMACRO{\TeXButton{h}{\slh}}%
%BeginExpansion
\slh
%EndExpansion
^{\dagger}e_{\mu},%
%TCIMACRO{\TeXButton{h}{\slh}}%
%BeginExpansion
\slh
%EndExpansion
^{\dagger}e_{\nu}]:=(%
%TCIMACRO{\TeXButton{h}{\slh}}%
%BeginExpansion
\slh
%EndExpansion
^{\dagger}e_{\mu}\underset{\eta}{\cdot}\partial_{x})%
%TCIMACRO{\TeXButton{h}{\slh}}%
%BeginExpansion
\slh
%EndExpansion
^{\dagger}e_{\nu}-(%
%TCIMACRO{\TeXButton{h}{\slh}}%
%BeginExpansion
\slh
%EndExpansion
^{\dagger}e_{\nu}\underset{\eta}{\cdot}\partial_{x})%
%TCIMACRO{\TeXButton{h}{\slh}}%
%BeginExpansion
\slh
%EndExpansion
^{\dagger}e_{\mu}\nonumber\\
&  =(\partial_{\mu}h_{v}^{\alpha}-\partial_{\nu}h_{\mu}^{\alpha})e_{\alpha
}=c_{\mu\nu}^{\alpha}e_{\alpha}\neq0, \label{l11}%
\end{align}
i.e., $\{g_{\mu}\}$ is not a coordinate basis, because the structure
coefficients $c_{\mu\nu}^{\alpha}$ of the basis $g_{\mu}$ are non
null.\medskip

\textbf{Remark F1}. Before proceeding, we recall that at page 477 of
\cite{dola} it is stated that the main property that must be required from a
nontrivial field strength $%
%TCIMACRO{\TeXButton{h}{\slh}}%
%BeginExpansion
\slh
%EndExpansion
$ is that we should not have $%
%TCIMACRO{\TeXButton{h}{\slh}}%
%BeginExpansion
\slh
%EndExpansion
^{\dagger}(a)=\mathtt{f}^{\dagger}(a)$ where \texttt{f }is obtained from a
diffeomorphism $f:U\rightarrow U$. This implies that $\partial_{x}\wedge%
%TCIMACRO{\TeXButton{h}{\slh}}%
%BeginExpansion
\slh
%EndExpansion
^{\dagger}(e_{\mu})\neq0$, i.e., $(\partial_{\mu}h_{v}^{\alpha}-\partial_{\nu
}h_{\mu}^{\alpha})e_{\alpha}\neq0$.%

%TCIMACRO{\U{b4}}%
%BeginExpansion
\'{}%
%EndExpansion
The next step in \cite{dola} towards the formulation of their gravitational
theory is the introduction of what those authors call a \ `gauge covariant
derivative' that `converts $\partial_{\mu}$ into a covariant derivative'. If
$M=M^{\alpha}e_{\alpha}\in\sec%
%TCIMACRO{\dbigwedge }%
%BeginExpansion
{\displaystyle\bigwedge}
%EndExpansion
TU$ they define%

\begin{equation}%
\begin{tabular}
[c]{|c|}\hline
$\mathcal{D}_{\mu}\mathcal{M}=\partial_{\mu}\mathcal{M}+\Omega_{\mu}%
\times\mathcal{M},$\\\hline
\end{tabular}
\ \ \ \ \ \ \label{s error}%
\end{equation}
with $\Omega:$ sec$%
%TCIMACRO{\dbigwedge \nolimits^{1}}%
%BeginExpansion
{\displaystyle\bigwedge\nolimits^{1}}
%EndExpansion
TU\rightarrow\sec%
%TCIMACRO{\dbigwedge \nolimits^{2}}%
%BeginExpansion
{\displaystyle\bigwedge\nolimits^{2}}
%EndExpansion
TU$, $\Omega_{\mu}=\Omega(e_{\mu})$ and%
\begin{equation}
\mathcal{M}:\mathcal{=}\boldsymbol{h}^{\dagger}(M)=M^{\alpha}g_{\alpha
},\label{s error1}%
\end{equation}
where take notice in Eq.(\ref{s error1}) $M^{\alpha}$ are the components of
$M$ in the coordinate basis $\{e_{\mu}\}$.\medskip

Before going on we must observe that the correct notation for $\mathcal{D}%
_{\mu}$ in order to avoid confusion\footnote{In \cite{hestenes2005}
Eq.(\ref{l12}) is written $\mathcal{D}_{\mu}\mathcal{M}=\partial_{\mu
}\mathcal{M}+\omega(g_{\mu})\times\mathcal{M}$. We must immediately conclude
that $\omega(g_{\mu})=\omega(%
%TCIMACRO{\TeXButton{h}{\slh}}%
%BeginExpansion
\slh
%EndExpansion
^{\dagger}(e_{\mu}))=\Omega(e_{\mu})$, i.e., $\omega=%
%TCIMACRO{\TeXButton{h}{\slh}}%
%BeginExpansion
\slh
%EndExpansion
^{\clubsuit}\Omega$. Moreover from the notation used in \cite{hestenes2005} we
should not infer that $\mathcal{D}_{\mu}$ is representing $\mathcal{D}%
_{g_{\mu}}$, because if that was the case then the defining equation for
$\mathcal{D}_{g_{\mu}}$ should read $\mathcal{D}_{g_{\mu}}\mathcal{M}=h_{\mu
}^{\alpha}\partial_{\mu}\mathcal{M}+\Omega(g_{\mu})\times\mathcal{M}$.} must
\ be $\mathcal{D}_{e_{\mu}}$ and we shall use it in what follows, thus to
start, we write%
\begin{equation}
\mathcal{D}_{e_{\mu}}\mathcal{M}=\partial_{\mu}\mathcal{M}+\Omega_{\mu}%
\times\mathcal{M}. \label{L12}%
\end{equation}
\textbf{Remark F2} \ In \cite{dola,hestenes2005} there is \emph{no} clear
specification of the relation of the gauge covariant derivative $\mathcal{D}%
_{e_{\mu}}$ with the standard Levi-Civita covariant derivative $D_{e_{\mu}}$of
$%
%TCIMACRO{\TeXButton{g}{\slg}}%
%BeginExpansion
\slg
%EndExpansion
$. So, when we first analyzed the contents of \cite{dola,hestenes2005} long
ago we necessarily had to proceed as follows to evaluate $\mathcal{D}_{e_{\mu
}}%
%TCIMACRO{\TeXButton{eta}{\mbox{\boldmath{$\eta$}}}}%
%BeginExpansion
\mbox{\boldmath{$\eta$}}%
%EndExpansion
$ and $\mathcal{D}_{e_{\mu}}%
%TCIMACRO{\TeXButton{g}{\slg}}%
%BeginExpansion
\slg
%EndExpansion
.$

We introduce the basis $\{\mathfrak{\theta}^{\mu}\}$ of $T^{\ast}U$ dual of
the basis $\{e_{\mu}\}$ of $TU$ and also the basis $\{\zeta^{\mu}\}$ of
$T^{\ast}U$ dual of the basis $\{g_{\mu}\}$ of $TU$ and write like in
\cite{dola,hestenes2005}:%
\begin{equation}
\mathcal{D}_{e_{\mu}}g_{\nu}=L_{\mu\nu}^{\alpha}g_{\alpha},\text{
\ \ }\mathcal{D}_{e_{\mu}}g^{\alpha}=-L_{\mu\nu}^{\alpha}g^{\nu}%
,\text{\ \ \ }\mathcal{D}_{e_{\mu}}\zeta^{\alpha}=-L_{\mu\nu}^{\alpha}%
\zeta^{\nu}\text{\ } \label{l13}%
\end{equation}
We also write
\begin{equation}
\mathcal{D}_{e_{\mu}}e_{\nu}=\mathbf{L}_{\mu\nu}^{\alpha}g_{\alpha},\text{
\ \ }\mathcal{D}_{e_{\mu}}e^{\alpha}=-\mathbf{L}_{\mu\nu}^{\alpha}e^{\nu
},\text{ \ \ }\mathcal{D}_{e_{\mu}}\theta^{\alpha}=-\mathbf{L}_{\mu\nu
}^{\alpha}\theta^{\nu}. \label{l14}%
\end{equation}
Note now that whereas the indices $\mu$ and $\nu$ in $\mathbf{L}_{\mu\nu
}^{\alpha}$ refers to the same basis\footnote{They are holonomic, since
$\{e_{\mu}\}$ is a coordinate basis.}, namely $\{e_{\mu}\}$ the indices $\mu$
and $\nu$ in $L_{\mu\nu}^{\alpha}$ are hybrid\footnote{The indice $\mu$ is
holonomic and the indice $\nu$ is non holonomic.}, i.e., the first ($\mu$)
corresponds to the basis $\{e_{\mu}\}$ whereas the second ($\nu$) corresponds
to the basis $\{g_{\nu}\}$. This observation is very important, for indeed, if
we write $%
%TCIMACRO{\TeXButton{eta}{\mbox{\boldmath{$\eta$}}}}%
%BeginExpansion
\mbox{\boldmath{$\eta$}}%
%EndExpansion
$ and $%
%TCIMACRO{\TeXButton{g}{\slg}}%
%BeginExpansion
\slg
%EndExpansion
$ the metric tensors which correspond to the metric extensors $\eta^{-1}$ and
$%
%TCIMACRO{\TeXButton{itg}{\itg}}%
%BeginExpansion
\itg
%EndExpansion
^{-1}$, we have immediately from $g_{\mu}\underset{\eta}{\cdot}g_{\nu}%
=g_{\mu\nu}$ and $e_{\mu}\underset{%
%TCIMACRO{\TeXButton{sig}{\sitg}}%
%BeginExpansion
\sitg
%EndExpansion
}{\cdot}e_{\nu}=g_{\mu\nu}$ that
\begin{equation}%
%TCIMACRO{\TeXButton{eta}{\mbox{\boldmath{$\eta$}}}}%
%BeginExpansion
\mbox{\boldmath{$\eta$}}%
%EndExpansion
=g_{\mu\nu}\zeta^{\mu}\otimes\zeta^{\nu}\text{, \ \ \ }%
%TCIMACRO{\TeXButton{g}{\slg}}%
%BeginExpansion
\slg
%EndExpansion
=g_{\mu\nu}\theta^{\mu}\otimes\theta^{\nu}\text{.} \label{l15}%
\end{equation}

We then get%
\begin{align}
\mathcal{D}_{e_{\mu}}%
%TCIMACRO{\TeXButton{eta}{\mbox{\boldmath{$\eta$}}}}%
%BeginExpansion
\mbox{\boldmath{$\eta$}}%
%EndExpansion
&  =\mathcal{D}_{e_{\mu}}(g_{\alpha\beta}\zeta^{\alpha}\otimes\zeta^{\beta
})\nonumber\\
&  =(\partial_{\mu}g_{\alpha\beta}-g_{\rho\beta}L_{\mu\alpha}^{\rho}%
-g_{\alpha\rho}L_{\mu\beta}^{\rho})\zeta^{\alpha}\otimes\zeta^{\beta},
\label{12}%
\end{align}
i.e., the connection $D$ will be metric compatible with $%
%TCIMACRO{\TeXButton{eta}{\mbox{\boldmath{$\eta$}}}}%
%BeginExpansion
\mbox{\boldmath{$\eta$}}%
%EndExpansion
$ iff%
\begin{equation}
\partial_{\mu}g_{\alpha\beta}-g_{\rho\beta}L_{\mu\alpha}^{\rho}-g_{\alpha\rho
}L_{\mu\beta}^{\rho}=0. \label{l16}%
\end{equation}

Now, in \cite{dola,hestenes2005} Eq.(\ref{l16}) is trivially
true\footnote{Indeed, we have for one side that $\mathcal{D}_{e_{\mu}%
}(g_{\alpha}\underset{\eta}{\cdot}g_{\beta})=\partial_{\mu}g_{\alpha\beta}$
and on the other side $\mathcal{D}_{e_{\mu}}(g_{\alpha}\underset{\eta}{\cdot
}g_{\beta})=\mathcal{D}_{e_{\mu}}g_{\alpha}\underset{\eta}{\cdot}g_{\beta
}+g_{\alpha}\underset{\eta}{\cdot}\mathcal{D}_{e_{\mu}}g_{\beta}=g_{\rho\beta
}L_{\mu\alpha}^{\rho}+g_{\alpha\rho}L_{\mu\beta}^{\rho}.$}, and is interpreted
by those authors as meaning that $\mathcal{D}_{e_{\mu}}$ is compatible with $%
%TCIMACRO{\TeXButton{g}{\slg}}%
%BeginExpansion
\slg
%EndExpansion
$. Without a clear specification of the relation of the gauge covariant
derivative $\mathcal{D}_{e_{\mu}}$ with the Levi-Civita covariant derivative
$D_{e_{\mu}}$of $%
%TCIMACRO{\TeXButton{g}{\slg}}%
%BeginExpansion
\slg
%EndExpansion
$ this claim is not justified and leads readers of \cite{dola,hestenes2005} at
the conclusion that authors made a confusion with the hybrid indices
(holonomic and non holonomic) in $L_{\mu\alpha}^{\rho}$, something briefly
mentioned in \cite{femoro}. And, indeed, a trivial calculation shows
immediately that%

\begin{equation}
\mathcal{D}_{e_{\mu}}%
%TCIMACRO{\TeXButton{g}{\slg}}%
%BeginExpansion
\slg
%EndExpansion
\mathbf{\neq}0, \label{l17}%
\end{equation}
i.e., the gauge covariant derivative $\mathcal{D}_{e_{\mu}}$ has a non null
nonmetricity tensor relative to $%
%TCIMACRO{\TeXButton{g}{\slg}}%
%BeginExpansion
\slg
%EndExpansion
$. Indeed, we know from the results of Section 4.7.1 that if $\mathcal{D}%
_{e_{\mu}}%
%TCIMACRO{\TeXButton{eta}{\mbox{\boldmath{$\eta$}}}}%
%BeginExpansion
\mbox{\boldmath{$\eta$}}%
%EndExpansion
=0$ the gauge covariant derivative compatible with $%
%TCIMACRO{\TeXButton{g}{\slg}}%
%BeginExpansion
\slg
%EndExpansion
$ is $\mathcal{D}_{e_{\mu}}^{\prime}$, the $%
%TCIMACRO{\TeXButton{h}{\slh}}%
%BeginExpansion
\slh
%EndExpansion
$-deformation of $\mathcal{D}_{e_{\mu}}$ such that if $a,b\in\sec%
%TCIMACRO{\dbigwedge ^{1}}%
%BeginExpansion
{\displaystyle\bigwedge^{1}}
%EndExpansion
TM$,%
\begin{equation}
\mathcal{D}_{a}^{\prime}b=%
%TCIMACRO{\TeXButton{h}{\slh}}%
%BeginExpansion
\slh
%EndExpansion
^{-1}(\mathcal{D}_{a}%
%TCIMACRO{\TeXButton{h}{\slh}}%
%BeginExpansion
\slh
%EndExpansion
(b)). \label{lnew}%
\end{equation}

\textbf{Remark F3 \ }Another\ consequence of a non clear specification of the
relation between $\mathcal{D}_{e_{\mu}}$ and $D_{e_{\mu}}$ in
\cite{dola,hestenes2005} is the following. Those authors introduce a Dirac
like operator, here written as $\mathcal{D}=g^{\alpha}\mathcal{D}_{e_{\alpha}%
}$. Then, it is stated (see, e.g., Eq.(131) in \cite{hestenes2005}) that
\begin{equation}
\mathcal{D\wedge}g^{\alpha}=\frac{1}{2}(L_{\mu\nu}^{\alpha}-L_{\nu\mu}%
^{\alpha})g^{\nu}\wedge g^{\mu} \label{l18}%
\end{equation}
\textit{is} the torsion tensor (more precisely the torsion $2$-forms
$\Theta^{\alpha}$). However, since the structure coefficients of the basis
$\{g_{\mu}\}$ are non null, the correct expression for the torsion $2$-form
fields $\Theta^{\alpha}$ are, as well known (see. e.g., \cite{rodcap2007})%
\begin{equation}
\Theta^{\alpha}=\frac{1}{2}(L_{\mu\nu}^{\alpha}-L_{\nu\mu}^{\alpha}-c_{\mu\nu
}^{\alpha})g^{\nu}\wedge g^{\mu}. \label{l19}%
\end{equation}
It follows that even assuming $L_{\mu\nu}^{\alpha}=L_{\nu\mu}^{\alpha}$ the
covariant derivative $\mathcal{D}_{e_{\mu}}$ has torsion, contrary to what is
claimed in \cite{dola,hestenes2005}. Of course, a simple calculation shows
that the Riemann curvature tensor of $\mathcal{D}_{e_{\mu}}$ is also non
null$.\medskip$

\textbf{Remark F4 \ }The above comments written long ago clearly show that the
gravitational theory presented in \cite{dola,hestenes2005} without making
clear the relation of $\mathcal{D}_{e_{\mu}}$ with \ $D_{e_{\mu}}$ implies
that it is not possible to conclude that $\mathcal{D}_{e_{\mu}}$ is not
compatible with $%
%TCIMACRO{\TeXButton{g}{\slg}}%
%BeginExpansion
\slg
%EndExpansion
$ and without such a conclusion it cannot be claimed that the gauge theory of
\cite{dola,hestenes2005} reproduces the results of \textit{GRT}\footnote{At
http://www.mrao.cam.ac.uk/\symbol{126}clifford/publications/index.html there
is a list of articles (published in very good journals) written by the authors
of [4] and collaborators and based on their gravitational theory may be
found.}.\medskip\ 

\textbf{Remark F5 }On August 2014 W.A.R. meets A. Lasenby at ICCA 10 in
Tartu\footnote{See http://icca10.ut.ee/.} where they discussed the above
comments. As a result Lasenby agreed that the symbol $\mathcal{D}_{\mu}$ used
for the gauge covariant derivative must be interpreted as$\ \mathcal{D}%
_{e_{\mu}}$ and more important, besides the properties of $\mathcal{D}%
_{e_{\mu}}$ listen in [4]\ it becomes clear to W.A.R. that it is necessary to
make clear the following statement: $\mathcal{D}_{e_{\mu}}$ is to be taken
\ by \emph{definition} as metric compatible with the metric $%
%TCIMACRO{\TeXButton{g}{\slg}}%
%BeginExpansion
\slg
%EndExpansion
$, i.e.,
\begin{equation}
\mathcal{D}_{e_{\mu}}%
%TCIMACRO{\TeXButton{g}{\slg}}%
%BeginExpansion
\slg
%EndExpansion
=0
\end{equation}
and moreover it is necessary to make the identification%
\begin{equation}
\mathcal{D}_{e_{\mu}}e_{\nu}=\mathbf{L}_{\mu\nu}^{\alpha}e_{\alpha}=D_{e_{\mu
}}e_{\nu}=\Gamma_{\mu\nu}^{\alpha}e_{\alpha},\label{120}%
\end{equation}
where $\mathbf{L}_{\mu\nu}^{\alpha}=\Gamma_{\mu\nu}^{\alpha}$ are the
Christoffel symbols of the metric compatible connection $D$ of standard
differential geometry in the basis $\{e_{\mu}\}$, i.e.,
\begin{equation}
D_{e_{\mu}}%
%TCIMACRO{\TeXButton{g}{\slg}}%
%BeginExpansion
\slg
%EndExpansion
=0.\label{12a}%
\end{equation}
Under these conditions which do \emph{not} appear in any clear
way\footnote{Which cannot be deduced from the formalism as presented in
[4,5].} in \cite{dola,hestenes2005} we have that
\begin{equation}
\mathcal{D}_{e_{\mu}}g_{\nu}=%
%TCIMACRO{\TeXButton{h}{\slh}}%
%BeginExpansion
\slh
%EndExpansion
^{\dagger}(\mathcal{D}_{e_{\mu}}e_{\nu})=\Gamma_{\mu\nu}^{\alpha}g_{\alpha
}\label{121}%
\end{equation}
and then it follows that $\mathcal{D}_{e_{\mu}}%
%TCIMACRO{\TeXButton{eta}{\mbox{\boldmath{$\eta$}}}}%
%BeginExpansion
\mbox{\boldmath{$\eta$}}%
%EndExpansion
=0$ may express the metric compatibility condition $D_{e_{\mu}}%
%TCIMACRO{\TeXButton{g}{\slg}}%
%BeginExpansion
\slg
%EndExpansion
=0$ and also, of course the gauge covariant derivative $\mathcal{D}_{e_{\mu}}$
has no torsion. This may looks strange at first sight but may be justified if
we impose some constraints, to be discussed next. \medskip

\textbf{Remark F6 \ }The identifications recalled in Remark F5 implies the
following constraints which must also be specified in order for the gauge
theory described in \cite{dola,hestenes2005} to be coherent. Let
$\mathcal{\mathring{D}}_{e_{\mu}}$ be a gauge covariant derivative compatible
with the Minkowski metric\ $%
%TCIMACRO{\TeXButton{eta}{\mbox{\boldmath{$\eta$}}}}%
%BeginExpansion
\mbox{\boldmath{$\eta$}}%
%EndExpansion
$, i.e.,%
\begin{equation}
\mathcal{\mathring{D}}_{e_{\mu}}%
%TCIMACRO{\TeXButton{eta}{\mbox{\boldmath{$\eta$}}}}%
%BeginExpansion
\mbox{\boldmath{$\eta$}}%
%EndExpansion
=0
\end{equation}
and such that its torsion tensor is null but its Riemann curvature tensor is
not null.

Then writing%
\begin{equation}
\mathcal{\mathring{D}}_{e_{\mu}}g_{\nu}:=\mathring{\Gamma}_{\mu\nu}^{\alpha
}g_{\alpha},\text{ \ \ }\mathcal{\mathring{D}}_{e_{\mu}}g^{\alpha}%
=-\mathring{\Gamma}_{\mu\nu}^{\alpha}g^{\nu},\text{\ \ \ }\mathcal{\mathring
{D}}_{e_{\mu}}\zeta^{\alpha}=-\mathring{\Gamma}_{\mu\nu}^{\alpha}\zeta^{\nu
}\text{\ }%
\end{equation}%

\begin{equation}
\mathcal{\mathring{D}}_{e_{\mu}}e_{\nu}:=\mathbf{\mathring{\Gamma}}_{\mu\nu
}^{\alpha}e_{\alpha},\text{ \ \ }\mathcal{\mathring{D}}_{e_{\mu}}e^{\alpha
}=-\mathbf{\mathring{\Gamma}}_{\mu\nu}^{\alpha}e^{\nu},\text{ \ \ }%
\mathcal{\mathring{D}}_{e_{\mu}}\theta^{\alpha}=-\mathbf{\mathring{\Gamma}%
}_{\mu\nu}^{\alpha}\theta^{\nu}.
\end{equation}
we have%

\begin{align}
\mathcal{\mathring{D}}_{e_{\mu}}%
%TCIMACRO{\TeXButton{eta}{\mbox{\boldmath{$\eta$}}}}%
%BeginExpansion
\mbox{\boldmath{$\eta$}}%
%EndExpansion
&  =\mathcal{\mathring{D}}_{e_{\mu}}(g_{\alpha\beta}\zeta^{\alpha}\otimes
\zeta^{\beta})\nonumber\\
&  =(\partial_{\mu}g_{\alpha\beta}-g_{\rho\beta}\mathring{\Gamma}_{\mu\alpha
}^{\rho}-g_{\alpha\rho}\mathring{\Gamma}_{\mu\beta}^{\rho})\zeta^{\alpha
}\otimes\zeta^{\beta},
\end{align}

Now, taking into account the relation between the gauge covariant derivatives
$\mathcal{D}_{e_{\mu}}$ and $\mathcal{\mathring{D}}_{e_{\mu}}$ we get
\begin{equation}
\mathcal{D}_{e_{\mu}}g_{\nu}:=\boldsymbol{h}^{\dagger}(\mathcal{\mathring{D}%
}_{e_{\mu}}\boldsymbol{h}^{\clubsuit}(g_{\nu}))=\boldsymbol{h}^{\dagger
}(\mathcal{\mathring{D}}_{e_{\mu}}e_{\nu})=\mathcal{\mathring{D}}_{e_{\mu}%
}g_{\nu}=\mathring{\Gamma}_{\mu\nu}^{\alpha}g_{\alpha}%
\end{equation}
and then since $\mathcal{D}_{e_{\mu}}g_{\nu}=\Gamma_{\mu\nu}^{\alpha}%
g_{\alpha}$ we arrive at the conclusion that necessarily
\begin{equation}
\Gamma_{\mu\nu}^{\alpha}=\mathring{\Gamma}_{\mu\nu}^{\alpha},
\end{equation}
i.e, the connection coefficients $\mathring{\Gamma}_{\mu\nu}^{\alpha
}=\mathcal{\mathring{D}}_{e_{\mu}}g_{\nu}\cdot$ $g^{\alpha}$ in the hybrid
basis $(\{e_{\mu}\}$ and $\{g_{\mu}\})$ must be equal to the connection
coefficients $\Gamma_{\mu\nu}^{\alpha}=\mathcal{D}_{e_{\mu}}e_{\nu}\cdot$
$e^{\alpha}$ in the coordinate basis $\{e_{\mu}\}$ or in the hybrid basis
since $\Gamma_{\mu\nu}^{\alpha}=\mathcal{D}_{e_{\mu}}g_{\nu}\cdot$ $g^{\alpha
}$

\end{document}